\documentclass[epjST]{svjour}
\usepackage{amsmath,amssymb}
\usepackage{float}
\usepackage{imakeidx}
\usepackage[colorlinks=true,linkcolor=blue,citecolor=blue,urlcolor=magenta]{hyperref}
\usepackage[framemethod=TikZ]{mdframed}
\usepackage{footnote}
\usepackage{comment}
\usepackage{doi}
\usepackage{enumitem}
\usepackage{slashed}
\usepackage{xcolor}
\usepackage{xstring}
\usepackage{xparse}
\usepackage{graphicx}
\usepackage{rotating}
\usepackage{aastex_hack}
\usepackage{appendix}
\makeindex

\makesavenoteenv{tabular}
\makesavenoteenv{table}
\makesavenoteenv{figure}

\newcommand{\orcidicon}{\includegraphics[width=0.32cm]{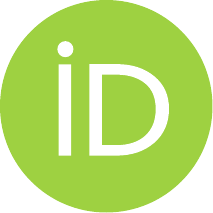}}
\newcommand{\orc}[1]{\href{https://orcid.org/#1}{\orcidicon}}

\newcommand{\orcA}{0000-0001-8217-1484}
\newcommand{\orcB}{0000-0001-5038-8427}
\newcommand{\orcC}{0000-0001-5474-2649}

\newcommand{\orcE}{0000-0002-2289-4856}
\newcommand{\orcF}{0000-0001-9985-1822}

\newcommand{\ts}{\textstyle}

\newcommand*{\GeV}{\text{\,GeV}}
\newcommand*{\MeV}{\text{\,MeV}}
\newcommand*{\keV}{\text{\,keV}}
\newcommand*{\eV}{\text{\,eV}}
\newcommand*{\meV}{\text{\,meV}}

\newcommand*{\bb}{\boldsymbol}
\newcommand*{\beqn}{\begin{equation}}
\newcommand*{\eeqn}{\end{equation}}
\newcommand{\beql}[1]{\begin{equation} \label{#1}}
\newcommand{\eeql}[1]{\label{#1} \end{equation} }  
\newcommand{\req}[1]{Eq.\,(\ref{#1})}
\newcommand{\rf}[1]{Fig.\,{\ref{#1}}}
\newcommand{\rt}[1]{Table~{\ref{#1}}}
\newcommand{\rsec}[1]{Sec.\,{\ref{#1}}}

\newcommand{\rapp}[1]{Appendix~{\ref{#1}}}

\newcommand{\ie}{{\em i.e.\/}}  
\newcommand{\eg}{{\em e.g.\/}}  

\newcommand{\grad}{\operatorname{grad}}
\newcommand{\diag}{\mathrm{diag}}

\newcommand{\cccite}[1]{Published in Ref.~\cite{#1} under the \href{https://creativecommons.org/licenses/by/4.0/}{CC BY 4.0} license}
\newcommand{\radapt}[1]{Adapted from Ref.~\cite{#1}}
\newcommand{\para}[1]{\paragraph{#1}\hfill\break\noindent}
\NewDocumentCommand{\allcite}{m o}{\nocite{#1}\hyperlink{cite.#1}{\StrBefore{#1}{:}\IfNoValueTF{#2}{ et. al. }{ and #2 }(\StrBehind{#1}{:}[\temp]\StrLeft{\temp}{4})}}
\NewDocumentCommand{\aucite}{m o}{\nocite{#1}\hyperlink{cite.#1}{\StrBefore{#1}{:}\IfNoValueTF{#2}{ }{ and #2 }(\StrBehind{#1}{:}[\temp]\StrLeft{\temp}{4})}}
\NewDocumentCommand{\allcitep}{m o}{\nocite{#1}\hyperlink{cite.#1}{\StrBefore{#1}{:}\IfNoValueTF{#2}{ et. al. }{ and #2 }(preprint \StrBehind{#1}{:}[\temp]\StrLeft{\temp}{4})}}

\newcommand{\eq}[1]{#1^{(\mathrm{eq})}}

\newcommand{\reqs}[2]{Eqs.\,({\ref{#1}}-{\ref{#2}})}

\newcommand{\ft}[1]{\widetilde{\boldsymbol{#1}}}
\newcommand{\hatv}[1]{\hat{\boldsymbol{#1}}}

\newcommand\Tstrut{\rule{0pt}{2.6ex}}
\newcommand\Bstrut{\rule[-0.9ex]{0pt}{0pt}}
\newcommand{\TBstrut}{\Tstrut\Bstrut}

\numberwithin{equation}{section}

\begin{document}
\title{Quarks to Cosmos: Particles and Plasma in\\ 
 Cosmological evolution}

\author{
Johann Rafelski${}^1$\orc{\orcA}\thanks{Corresponding author 
\email{johannr@arizona.edu}}, 
Jeremiah Birrell${}^{1,2}$\orc{\orcE}, 
Christopher Grayson${}^1$\orc{\orcF}, 
\newline Andrew Steinmetz${}^1$\orc{\orcC}, 
Cheng Tao Yang${}^1$\orc{\orcB}
}

\institute{${}^1$Department of Physics, The University of Arizona, Tucson, AZ, 85721, USA\\
${}^2$Department of Mathematics, Texas State University, San Marcos, TX, 78666, USA}

\abstract{We describe in the context of the particle physics (PP) standard model (SM) `PP-SM' the understanding of the primordial properties and composition of the Universe in the temperature range $130\,\mathrm{GeV}>T>20\,\mathrm{keV}$. The Universe evolution is described using FLRW cosmology. We present a global view on particle content across time and describe the different evolution eras using deceleration parameter $q$. In the considered temperature range the unknown cold dark matter and dark energy content of $\Lambda\mathrm{CDM}$ have a negligible influence allowing a reliable understanding of physical properties of the Universe based on PP-SM energy-momentum alone. We follow the arrow of time in the expanding and cooling Universe: After the PP-SM heavies $(t, h, W, Z)$ diminish in abundance below $T\simeq 50\,\mathrm{GeV}$, the PP-SM plasma in the Universe is governed by the strongly interacting Quark-Gluon content. Once the temperature drops below $T\simeq 150\,\mathrm{MeV}$, quarks and gluons hadronize into strongly interacting matter particles comprising a dense baryon-antibaryon content. Rapid disappearance of baryonic antimatter ensues, which adopting the present day photon-to-baryon ratio completes at $T_\mathrm{B}=38.2\,\mathrm{MeV}$. We study the ensuing disappearance of strangeness and mesons in general. We show that the different eras defined by particle populations are barely separated from each other with abundance of muons fading out just prior to $T=\mathcal{O}(2.5)\,\mathrm{MeV}$, the era of emergence of the free-streaming neutrinos. We develop methods allowing the study of the ensuing speed of the Universe expansion as a function of fundamental coupling parameters in the primordial epoch. We discuss the two relevant fundamental constants controlling the decoupling of neutrinos. We subsequently follow the primordial Universe as it passes through the hot dense electron-positron plasma epoch. The high density of positron antimatter disappears near $T=20.3\,\mathrm{keV}$, well after the Big-Bang Nucleosynthesis era: Nuclear reactions occur in the presence of a highly mobile and relatively strongly interacting electron-positron plasma phase. We apply plasma theory methods to describe the strong screening effects between heavy dust particle (nucleons). We analyze the paramagnetic characteristics of the electron-positron plasma when exposed to an external primordial magnetic field.}
\maketitle
\setcounter{tocdepth}{3}
\tableofcontents

\section{Introduction}
\subsection{Theoretical models of the primordial Universe}\label{ssec:UniLab}
In this report we explore the connection between particle, nuclear, and plasma physics in the evolution of the Universe. Our work concerns the domain described by the known laws of physics as determined by laboratory experiments.

Our journey in time through the expanding primordial Universe has the objective of understanding how different evolution eras impact each other. We are seeking to gain deeper insights into the fundamental processes that shaped our cosmos, by providing a clearer picture of the Universe's origin and its ongoing expansion. The question we address is how a very hot soup of elementary matter evolves and connects to the normal matter present today, indirectly observed by the elemental ashes of the Big-Bang nucleosynthesis (BBN)\index{Big-Bang!BBN}. 

{\color{black}We report on our  ongoing work on particle and plasma cosmology. Our aim is to create a continuous description of the primordial Universe spanning and connecting different particle and plasma epochs. This allows the reader to understand the physics spanning the entire temperature domain we explore, $130\GeV>T>20\,\mathrm{keV}$. We will define many topics deserving further study: There are many open questions to which this work provides an introduction.}

{\color{black}This manuscript is not a traditional review, as its focus is on our own ongoing work~\cite{Rafelski:2023emw,Birrell:2014ona,Grayson:2024okq,Steinmetz:2023ucp,Yang:2024ret}: We aim here to offer a readable report describing and explaining what can appear at times to be a fragmented body of work. A  guide- and time-line how our theoretical insights were gained over the past dozen years is presented in \rapp{list:Works}.}

{\color{black}As appropriate we highlight in \rapp{list:Works} the key results and provide fields of application not yet fully explored. This helps to create a backdrop of knowledge and introduces further areas of study potentially helping to understand the primordial Universe and sharpen prediction of observable quantities today.} 

\para{Dominance of visible radiation and matter in the primordial Universe}
We aim to connect various eras of cosmological evolution which can be addressed with some confidence in view of the already known particle and nuclear properties as measured experimentally. By analyzing the primordial Universe as a function of time in~\rf{fig:energy:frac} we are exploring the role of particle physics standard model (PP-SM) in the Universe evolution. We snapshot in this report specific epochs in the primordial Universe, and/or on specific physical conditions such as primordial magnetic fields.

\begin{figure}
\includegraphics[width=\linewidth]{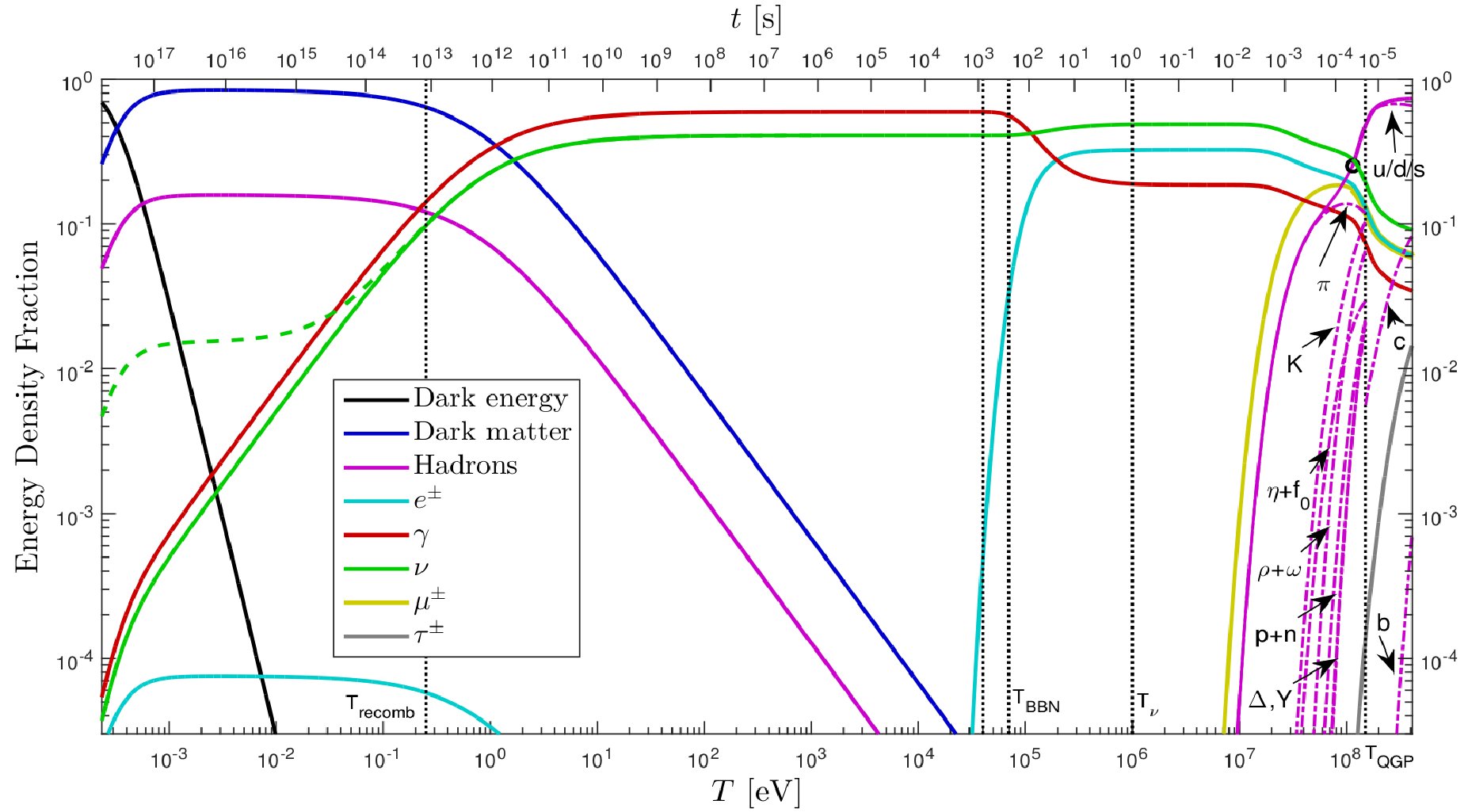}\label{fig:energy:frac}
\caption{Evolving in time fractional energy composition of the Universe. See text for discussion. \cccite{Rafelski:2023emw}. \radapt{Birrell:2014ona}.}
\end{figure}

In the cosmic epoch considered here with temperature above $kT=20\keV$, the present day dominant dark matter and dark energy played a negligible role in the cosmos. The changing energy component composition of the Universe\index{Universe!composition} is illustrated in~\rf{fig:energy:frac}. To create the figure we integrate the Universe backwards in time. The initial condition is the assumed composition of the Universe in the current era: $69\%$ dark energy, $26\%$ dark matter, $5\%$ baryons\index{baryon}, while photons and neutrinos comprise less than one percent. We further assumed one massless neutrino and two with $m_\nu=0.1$\,eV. Other neutrino mass values are possible; constraints remain weak\index{neutrino!mass}. How this solution is obtained will become evident at the end of~\rsec{sec:flrw} below. 

As described, there are two unknown dark components as one is able to disentangle these given two independent inputs in the cosmic energy-momentum tensor of homogeneous isotropic matter, pressure and energy density, which can be related by equations of state. The current epoch cosmic accelerated expansion (Nobel price 2011 to Saul Perlmutter, Adam Riess, and Brian P. Schmidt -- a graduate also in physics at The University of Arizona) creates the need for this two component ``darkness''.\index{Universe!darkness}

Dark energy in the conventional definition is akin to $\Lambda$=Einstein's cosmological term. $\Lambda$ is a fixed property of the Universe and does not scale with temperature. In comparison, radiation energy content scales with $T^4$ and is vastly dominant in the temperature range we explore; the dark energy (black line) emerges in a very recent past (on logarithmic time scale, see~\rf{fig:energy:frac}. Cold, {\it i.e.\/}, massive on temperature scale,  dark matter (CDM) content scales with $T^{3/2}$ for $m/T\gg 1$. In the temperature regime of interest to us CDM (blue line in~\rf{fig:energy:frac}) complements the invisible normal baryonic matter (purple line). Both are practically invisible in the Universe inventory in the epoch we explore, emerging just after as a $10^{-5}$ energy fraction shown~\rf{fig:energy:frac}. The further back we look at the hot Universe, the more irrelevant become all forms of matter, including the ``dark'' matter component.\index{Universe!radiation dominated} 

There is considerable tension\index{Hubble parameter!tension} between studies determining the present day speed of cosmic expansion (Hubble parameter)~\cite{DiValentino:2024spr,DiValentino:2021izs}: Extrapolation from more distant past, looking as far back as is possible, {\it i.e.\/}, the recombination epoch near redshift $z=1000$, are smaller than the Universe properties observed and studied in the current epoch. This result stated often asking the question ``$h_0=0.67$ or $0.73$?'' about contemporary Hubble parameter\index{Hubble parameter} $H_0(=100\,h_0$\,(km/Mpc)/s). This unresolved issue arises comparing diverse epochs when the Universe was in its atomic, molecular, stellar forms. One would think that therefore this discrepancy is in principle irrelevant to our particle and plasma study of the primordial Universe. 

However, as we will argue this separation of scales may not be complete. Depending on the details of PP-SM unobserved contents, {\it e.g.\/}, in the neutrino sector, free-streaming not quite massless quantum neutrinos contribute to darkness and may impact the result of extrapolation (``$h_0=0.67$ or $0.73$?'') of the Hubble expansion from recombination epoch to the current epoch. One could argue that the effort to study the ``Unknown'' darkness in cosmology suffers from the lack of the full understanding of the ``Known'' in the primordial cosmos which masquerades as darkness today. This is one of the many motivations for the research effort we pursue. 

\para{Cosmic plasma in the primordial Universe}
We use units in which the Boltzmann constant $k_\mathrm{B}=1$. In consequence, the temperature $T$ is discussed in this report in units of energy either MeV $\simeq 2 m_{\rm e} c^2$ ($m_{\rm e}$ is the electron mass) or GeV$= 1000$ MeV $\simeq m_{\mathrm N} c^2$ ($m_{\rm N}$ is the mass of a nucleon) or as the Universe cools in keV, one-thousandth of an MeV. The conversion of an MeV to temperature familiar units involves ten additional zeros. This means that when we explore hadronic matter at the `low' temperature: 
\begin{equation} \label{tempval}
100\MeV\equiv 116\times 10^{10}\,\mathrm{K}\, ,
\end{equation}
we exceed the conditions in the center of the Sun at $T=11\times 10^6$\,K by a factor of 100\,000.

The primordial hot Universe fireball underwent several nearly adiabatic phase changes that dramatically evolved its bulk plasma properties as it expanded and cooled. {\color{black} The periods near to a major structural change include the Electro-weak phase transition, QGP hadronization, neutrino decoupling, BBN nucleosynthesis, and development of cosmological scale magnetic fields. These transitions all require the development and/or application of nonequilibrium kinetic theory methods. These boundaries connect the different epochs in the history of the Universe. This report is dedicated to the study of these non-equilibrium transitions which require from the reader the full mastery of the thermal equilibrium conditions we also describe.}

We begin in the temperature range below temperature of electro-weak (EW) boundary at $T=130\GeV$, when massive elementary particles emerged in the electro-weak symmetry broken phase of matter. {\color{black}The cosmic plasma in the primordial Universe evolves within the first hour down to the temperature of about $T\simeq 10\keV$ where this report will end. We will address the four well separated domains of particle plasma all visible by inspection of~\rf{fig:energy:frac}, connect these and extend our study to the two topical cosmological challenges, the BBN and cosmological magnetic field development.}

Notable plasma epochs include:\index{Universe!plasma epochs} 
\begin{enumerate}
\item \textbf{Primordial quark-gluon plasma epoch:} 
At early times when the temperature was between $130\GeV>T>0.15\GeV$ we have in the primordial plasma in their thermal abundance all PP-SM building blocks of the Universe as we know them today, including the Higgs particle, the vector gauge electroweak and strong interaction bosons, all three families of leptons and free deconfined quarks: For most of the evolution of QGP all hadrons\index{QGP!hadronization} are dissolved into their constituents $u,d,s,t,b,c,g$. However, as temperature decreases below heavy particle mass, the thermal abundance is much reduced but is in general expected to remain in abundance (chemical) equilibrium due to the presence of strong interactions. \\[0.2cm]
However, we will show in~\rsec{Bottom} that near to the QGP phase transition $300\, \mathrm{MeV}>T>150\MeV$, the chemical equilibrium\index{chemical equilibrium} of the bottom quark\index{bottom quark} abundance is broken. The abundance described by the fugacity parameter relatively slowly diminishes, see~\rf{fugacity_bc}, with only a small deviations from stationary state detailed balance, see~\rf{NonFugacity}. The expansion of the Universe through the epoch of the bottom quark abundance disappearance from the particle inventory provides us the arrow of time often searched for, but never found in the current epoch\index{arrow of time}.\\[0.2cm]
For general reference we establish the energy density near to the end of the QGP epoch in the Universe by considering a benchmark value at $T\simeq 150\MeV$
\begin{equation} \label{endensval}
\epsilon=1\,\mathrm{GeV/fm}^3
= 1.8\times 10^{15}\,\mathrm{g\,cm^{-3}} 
=1.8\times 10^{18}\,\mathrm{kg m^{-3}}\,.
\end{equation}
The corresponding relativistic matter pressure converted into human environment unit is
\begin{equation} \label{presval}
P\simeq \ts\frac{1}{3} \epsilon=0.52\times 10^{30} \,\mathrm{bar}\,.
\end{equation}
\item \textbf{Hadronic epoch:} Near the Hagedorn temperature\index{Hagedorn!temperature} $T_H\approx 150\MeV$, a phase transformation occurred, forcing the free quarks and gluons to become confined within baryons\index{baryon} and mesons; experimental results confirming the universal nature of the hadronization process were described in Ref.\,\cite{Letessier:2005qe}. In the temperature range $ 150\MeV >T>20\MeV$, the Universe is rich in physical phenomena involving strange mesons and (anti)baryons including long lasting (anti)hyperon abundances~\cite{Fromerth:2012fe,Yang:2021bko}. The antibaryons\index{baryon!antibaryon} disappear from the Universe inventory at temperature $T=38.2\MeV$. However, strangeness remains in the inventory down to $T\approx13\MeV$. The detailed balance assures that the weak decay is compensated by inverse reactions, see~\rsec{Strangeness} for detailed discussion.
\item \textbf{Lepton-photon epoch:} For temperature $10\MeV >T>2\MeV$, massless leptons and photons controlled the fate of the Universe: The Universe contained relativistic electrons, positrons, photons, and three species of (anti)neutrinos. During this epoch Massive $\tau^\pm$ disappear from the plasma at high temperature via decay processes. However, $\mu^\pm$ leptons can persist in the primordial Universe until temperature $T=4.2\MeV$.\\[0.2cm]
In this temperature epoch neutrinos were still coupled to the charged leptons via the weak interaction~\cite{Birrell:2014ona,Birrell:2012gg}, they freeze-out in the temperature range $3\MeV >T>2\MeV$, exact value depends on the neutrino's flavors and the magnitude of the PP-SM parameters, see~\rsec{Neutrino} for detailed discussion. After neutrino freeze-out, they still play a important role in the Universe expansion via the effective number of neutrinos\index{neutrino!effective number} $N_{\nu}^{\mathrm{eff}}$, which relates to the Hubble parameter value in the current epoch.
\item \textbf{Electron-positron epoch:} After neutrinos freeze-out at $T=3\sim 2\MeV$ and become free-streaming in the primordial Universe, the cosmic plasma was dominated by electrons, positrons, and photons. In the $e^+e^-$ plasma, positrons $e^+$ persisted in similar to electron $e^-$ abundance until the temperature $T=20.3\keV$, see~\rsec{Electron} for detailed discussion. Properties of this plasma need to be studied in order to understand the behavior of the nucleon dust dynamics including:
\item \textbf{BBN in the midst of the ${\mathbf e^+e^-}$ plasma:}\index{Big-Bang!BBN} Contrary to what was the prevailing context only a few years ago, today it is understood that BBN occurred within a rich electron-positron $e^+e^-$ plasma environment. There are thousands if not millions of ${ e^+e^-}$-pairs for each nucleon undergoing nuclear fusion reactions during the BBN epoch. 
\item \textbf{Primordial magnetism:}\index{magnetic!primordial fields} The $e^{+}e^{-}$-pair plasma at temperatures reaching well below BBN epoch in the primordial Universe could be an origin of the present day intergalactic magnetic fields~\cite{Rafelski:2023emw,Steinmetz:2023nsc}. In~\rsec{Electron} we present a detailed discussion of our model: We explore Landau diamagnetic and magnetic dipole moment paramagnetic properties. A relatively small magnitude of the $e^{+}e^{-}$ magnetic moment polarization asymmetry suffices to produce a self-magnetization in the Universe consistent with present day observations. 
\end{enumerate}

{\color{black} However, the origin of primordial magnetic fields remains unknown. The further back in time that origin could be traced, the stronger the magnetic field will be due to cosmological scaling described in~\rsec{Electron} . It is well possible that the primordial origin could occur when the QGP filled the Universe. We prepare for this possibility that connects directly with the ongoing relativistic heavy ion collision studies by presenting in~\rsec{chap:QCD} the current understanding of the QGP phase in the presence of strong magnetic fields in a linear response model.}

After $e^+e^-$ annihilation finishes at a temperature near 20.3\,keV, the Universe was still opaque to photons due to large photon-electron scattering Thompson cross-section. Observational cosmology study of the Cosmic Microwave Background\index{CMB} (CMB)~\cite{Planck:2018vyg} addresses the visible epoch beginning after free electron binding into atoms -- a process referred to as recombination (clearly better called atom-formation). This is complete and the Universe becomes visible to optical experiments at $T_\mathrm{recomb}\approx 0.25\,\mathrm{eV}$. 

\para{Toward experimental study of primordial particle Universe} Just before quarks and gluons were adopted widely as elementary degrees of freedom in PP-SM, the so-called `Lee-Wick' model of dense primordial matter prompted a high level meeting: The Bear Mountain November 29-December 1, 1974 symposium had a decisive impact on the development of the research program leading to the understanding of primordial particles in the Universe. This meeting was not open to all interested researchers: Only a few dozen were invited to join the participant club, see last page of the meeting report:\index{Bear Mountain!1974 Symposium} \url{https://www.osti.gov/servlets/purl/4061527}. This is an unusual historical fact witnessed by one of us (JR), for further discussion see Ref.\,\cite{Rafelski:2019twp}.\index{Bear Mountain!symposium}

It is noteworthy that our report appears in essence on the 50th-year anniversary of this 1974 meeting and is accompanied by the passing of the arguably the most illustrious symposium participant, T.D.\,Lee (passed away August 4, 2024 at nearly 98\index{Lee-Wick!dense matter}). Within just half a century the newly developed PP-SM knowledge has rendered all but one insight of the 1974 meeting obsolete: The participating representatives of particle and nuclear physics elite of the epoch recognized the novel opportunity to experimentally explore hot and dense hadron (strongly interacting) matter by colliding high energy nuclei (heavy-ions); the initial objective was the discovery of the Lee-Wick super dense matter but the objectives evolved rapidly in following years. One of the symposium participants, Alfred Goldhaber, planted in the Nature magazine~\cite{Goldhaber:1978qp} the seed which grew into the RHIC collider at BNL-New York. 

\para{Phase transformation in the primordial Universe} Thanks to the tireless effort of Rolf Hagedorn~\cite{Rafelski:2016hnq}\index{Hagedorn}, the European laboratory CERN\index{CERN} was intellectually well positioned to embark on the rapid development of related physics ideas and the required experimental program. The preeminent physics motivation that soon emerged was the understanding of the primordial composition of the hot Universe. The pre-1970 idea advanced by Hagedorn, by Huang and Weinberg~\cite{Huang:1970iq} and in the following by many others, was that the Universe was bound to the maximum Hagedorn temperature of $kT\le kT_H=150-180\MeV$ at which the energy content diverged\index{Universe!maximum Hagedorn temperature}. In the following years and indeed by the time of the Bear Mountain meeting the idea that a symmetry restoring change in phase structure at finite temperature was already taking hold~\cite{Weinberg:1974hy,Harrington:1974fc}, unnoticed by the limited in scope Bear Mountain crowd.

Today we understand Hagedorn temperature $T_H$\index{Hagedorn!temperature} to be the phase transformation to the deconfined phase of matter where quarks and gluons can exist. The first clear statement about the existence of such a phase boundary connecting the Hagedorn hadron gas phase\index{hadrons!gas phase} with the constituent quarks and gluons, and invoking deconfinement at high temperature, was the 1975 work of Cabibbo and Parisi~\cite{Cabibbo:1975ig}. This was followed by a more quantitative characterization within the realm of the MIT bag model by~\cite{Chin:1978gj} and soon after by Rafelski and Hagedorn incorporating Hagedorn bootstrap model of hadronic matter with finite size hadrons melting into QGP, see Ref.\,\cite{Rafelski:2015cxa} and appendices A and B therein. This work implemented Cabibbo-Parisi proposal as well as it was at that time possible.

Could deconfined state of a hot phase of quarks and gluons we call QGP really exist beyond Hagedorn temperature? A broad acceptance of this new insight took decades to take hold. For some, this was natural. In 1992 Stefan Pokorski asked ``What else could be there?'' when one of us (JR) was struggling to convince the large and skeptical academic course crowd at the Heisenberg-MPI in Munich. Those who were like Pokorski convinced that QCD state of matter prevails in the 1970's and 1980's epoch missed the need to smoothly connect quarks to hadrons, or as we say in the title of this work, quarks to cosmos\index{Quarks!to cosmos}, and do this incorporating gluons. 

Neglecting or omitting the gluonic degrees of freedom pushed the transformation temperature in the Universe towards $T=400\MeV$, creating a glaring conflict with the well established Hagedorn hadronic phase temperature limit $T_H\simeq 160\pm 10\MeV$. Yet other large body of work in this epoch addressed the dissolution at ultra high density and {\bf zero temperature} of hadrons into quark constituents, a process of astrophysical interest, without relevance to the understanding of both the primordial Universe and of dynamic phenomena observed in relativistic heavy-ion collisions\index{heavy-ion!collisions}.

The present day understanding of the primordial QGP Universe was for some reason out of context for most nuclear scientists of the epoch, while to some of us the key issues became clear within less than a decade. Arguably the first Summer School connecting quarks to cosmos and relativistic heavy-ion laboratory experiments was held in the Summer 1992 under the leadership of Hans Gutbrod and one of us (JR) in the small Italian-Tuscan resort Il Ciocco. The following is the abstract of the forward article {\it Big-Bang in the Laboratory}\index{Big-Bang} of the proceedings volume presented more than 30 years ago~\cite{Gutbrod1993}: 

\begin{quote}
`Particle Production in Excited Matter' (the title of the proceeding volume, and of the meeting) happened at the beginning of our Universe. It is also happening in the laboratory when nuclei collide at highly relativistic energies. This topic is one of the fundamental research interests of nuclear physics of today and will continue to be the driving force behind the accelerators of tomorrow. In this work we are seeking to deepen the understanding of the history of time. Unlike other areas of Physics, Cosmology, the study of the birth and evolution of the Universe has only one event to study. But we hope to recreate in the laboratory a state of matter akin to what must have been a stage in the evolution when nucleons were formed. This occurred not too long after the Big-Bang birth of the Universe, when the disturbance of the vacuum made appear an extreme energy density leading to the creation of particles, nucleons, atoms and ultimately nebulas and stars. Figure 1 depicts the evolution of the Universe as we understand it today. On the left-hand scale is shown the decrease of the temperature as a function of time shown on the right side. The cosmological eras associated with the different temperatures and sizes of the Universe are described in between.
\end{quote}

Indeed! Today the ongoing laboratory work at CERN-LHC\index{CERN!LHC} and BNL-RHIC\index{BNL!RHIC} exploring the physics of QGP in the high temperature and high particle density regime reached in relativistic heavy-ion collisions allows us to study elementary strongly interacting matter connecting quarks to the cosmos. These two fields, primordial Universe and ultra relativistic heavy-ion collisions relate to each other very closely. There is little if any relation to the other, dense neutron matter research program. Such matter is found in compact stars; super-novae explosions create at much different matter density temperatures reaching 50\,MeV. 

\para{Comparing Big-Bang with laboratory micro-bang}
The heavy-ion collision micro-bang involves time scales many orders of magnitude shorter compared to the characteristic scale governing the Universe Big-Bang: The expansion time scale of the Universe is determined by the interplay of the gravitational force and the energy content of the hot matter, whereas in the micro-bangs there is no gravitation to slow the explosive expansion.\index{heavy-ion!micro-bang} The initial energy density is reflecting on the nature of strong interactions, and the lifespan of the micro-bang is a fraction of $\tau_\mathrm{MB}\le 10^{-22}$\,s, the time for particles to cross at the speed of light the localized fireball of matter generated in relativistic heavy-ion collision. 

It is convenient to represent the Universe expansion time constant\index{Universe!expansion at hadronization} $\tau_{\mathrm U}$ as the inverse of the Hubble parameter\index{Hubble parameter} at a typical ambient energy density $\rho_0$
\beql{taudef}
\tau_{\mathrm U} \equiv \frac{1}{H[\rho_0=1\,\mathrm{GeV/fm^3}]}=14\,\mu\mathrm{s}
\,.
\eeqn
The Universe is indeed expected to be about 15 orders of magnitude slower in its expansion compared to the exploding micro-bang fireball formed in collisions of relativistic heavy-ion laboratory experiments. 

Above, the value of $\rho_0$ is chosen in the context of the `freezing' of deconfined QGP into hadrons\index{QGP!hadronization} in primordial Universe near to $T_0\simeq 150\MeV$: The strongly interacting degrees of freedom contribute as measured in laboratory relativistic heavy-ion collisions about half to this value, $\rho_h\simeq 0.5\,\mathrm{GeV/fm^3}$; the other half is the contribution of neutrinos, charged leptons, and photons. The fact that these two energy density components are nearly equal is implicit in many results shown in the following, see for example~\rf{EntropyDOF:Fig}: At hadronization we have twice as many (entropic) degrees of freedom than will remain in the radiation dominated Universe once hadrons disappear.

We obtain the relation between $H$ and $\rho$ by remembering one of the fundamental relations in the Friedmann-Lema{\^i}tre-Robertson-Walker (FLRW)\index{cosmology!FLRW} cosmology, the so-called Hubble equation\index{Hubble!equation}
\beql{Hubble:eq}
H^2=\frac{8\pi G_N}{c^2}\,\frac{\rho}{3}=
c^2 \frac{\hbar c}{M_p^2c^4} \,\frac{\rho}{3}
\,.
\eeqn
We introduced here and will use often the Planck mass\index{Planck!mass} $M_p$, defined in terms of $G_N$
\beql{eq:GN}
\frac{1}{c^4}8\pi G_N\equiv \frac{\hbar c}{M_p^2c^4}\,, \qquad 
M_p c^2=2.4353\, 10^{18}\GeV\,.
\eeqn
This definition of $M_p$, while more convenient in cosmology, differs by the factor $1/\sqrt{8\pi}$ from the particle physics convention introduced by particle data group (PDG)~\cite{ParticleDataGroup:2022pth}
\beql{eq:MplPDG}
 \sqrt{8\pi} M_p c^2 \equiv M_p^\mathrm{PDG} c^2 =1.2209\, 10^{19}\GeV\,.
\eeqn

The difference between the ``two bangs'' (heavy-ion and the Universe) due to the very different time scales involved is difficult to resolve. The evolution of the Universe is slow on the hadronic reaction time scale. Given the value of characteristic $\tau_{\mathrm U}$ we obtained, we expect that practically all unstable hadronic particles evolve to fully attain equilibrium, with ample time available to develop a `mixed phase' of QGP and hadrons, and for electromagnetic and even weak interactions to take hold generating complete particle equilibrium. All this can not occur during the life span of the dense matter created in relativistic nuclear-collisions. To understand the Universe based on laboratory experiments running at a vastly different time scale, we must therefore use theoretical models as developed in this report.\index{Big-Bang!micro-bang comparison}

There are other notable differences between the laboratory fireball and the cosmic primordial plasma. The primordial quark-hadron Universe was practically baryon\index{baryon} free comparing the net (less antibaryon) baryon number to cosmic backgrounds of remnant particles. The residual asymmetry level was and remains at $10^{-9}$. In the laboratory micro-bang at the highest CERN-LHC energy,\index{CERN!LHC} we create a fireball of dense matter with a net baryon number per total final particle multiplicity at a fraction of a percent. This matter-antimatter-abundance asymmetry between laboratory and primordial Universe is easily overcome theoretically, since it implies a relatively minor extrapolation. Any small abundance of baryons can be an experimental diagnostic signal for QGP but not a key feature of the matter produced.

\para{Can QGP be discovered experimentally?}
This takes us right to the question: Can we really tell apart in these explosive ultra relativistic heavy-ion experiments the two different phases of strongly interacting matter, the deconfined quark gluon plasma and `normal', confined strongly interacting matter? Existence of these two distinct phases is a new paradigm that superseded the Hagedorn singularity at the Hagedorn temperature. In the laboratory, the outcome of ultra-relativistic heavy-ion collisions\index{heavy-ion!collisions} seems to be very much the same irrespective of the applicable paradigm, we achieve the conversion of the kinetic energy of colliding nuclei into many material particles. So is a transient deconfined QGP phase really formed in relativistic heavy collisions? This question haunted this field of research for decades~\cite{Rafelski:2015cxa,Harris:2024aov}, a topic which is not addressed in this work beyond the following few words: 

When one of us (JR) first arrived at CERN in 1977, he found himself immersed into ardent discussions about both what the structure of the hot primordial Universe could be, and if indeed we could figure out how to find the answer in an experiment. Was the Universe perhaps a dense baryon-antibaryon singular Hagedorn Universe? Or was indeed the confinement condition not really retained at high temperature~\cite{Weinberg:1974hy,Harrington:1974fc,Cabibbo:1975ig}? And, above all, how can we tell these models apart doing laboratory experiments? By 1979 it became clear that new experimental ideas and a new observable was needed, sensitive to specific properties of the dense deconfined hot matter if formed in experiments. Strange antibaryon\index{baryon!antibaryon} enhancement was one of the proposed novel approaches and in the opinion of one of us (JR); this was to be later the decisive QGP discovery evidence~\cite{Rafelski:2019twp}.\index{quark-gluon plasma!signature}

\subsection{Concepts in equilibrium statistical physics} \label{sec:statphys}
We now recall the fundamental statistical physics concepts necessary to explore the properties of the Universe during its 'first hour'. {\color{black} We begin by considering the laws of equilibrium thermodynamics which provide a framework for understanding the behavior of particle's energy density, pressure, number density and entropy in the Universe. These expressions presume local thermal equilibrium which prevailed in the relatively slowly expanding Universe.}

We will address the general Fermi and Bose distributions\index{Bose!distribution}\index{Fermi!distribution} and its application in the primordial Universe, as well as the cases of special interest to thermodynamics in the primordial Universe. We describe partial freeze-out conditions, {\it i.e.\/}, rise of the chemical nonequilibrium abundance, while kinetic scattering equilibrium is maintained, and the case of free-streaming particles, allowing for switching from radiation like to massive nonrelativistic condition. In the following we use natural units\index{natural units} $c=\hbar=k_\mathrm{B}=1$. While we have shown  $c$ and $\hbar$ explicitly before, we have measured temperature in units of energy, thus implicitly taking $k_\mathrm{B}T\to T$, {\it i.e.\/}, $k_\mathrm{B}=1$.

\para{Quantum statistical distributions}
\index{statistical distribution}
In the primordial Universe, the reaction rates of particles in the cosmic plasma were much greater than the Universe expansion rate $H$. Therefore, the local thermal equilibrium was in general maintained. Assuming the particles are in thermal equilibrium, the dynamical information about local energy density can be estimating using he single-particle quantum statistical distribution function. The general relativistic covariant Fermi/Bose momentum distribution can be written as
\begin{align}
f_{F/B}(\Upsilon_i,p_i)=\frac{1}{\Upsilon^{-1}_i\exp{\left[(u\cdot p_i-\mu_i)/T\right]}\pm1}
\,,
\end{align}
where the plus sign applies for fermions, and the minus sign for bosons. The Lorentz scalar $(u_i\cdot p_i)$ is a scalar product of the particle four momentum $p^\mu_i$ with the local four vector of velocity $u^\mu$. In the absence of local matter flow, the local rest frame is the laboratory frame 
\begin{align}
u^\mu=\left(1,\vec{0}\right),\,\,\,\,\,\,\,\,\, p^\mu_i=\left(E_i,\vec{p}_i\right)\,.
\end{align} 
The parameter\index{fugacity} $\Upsilon_i$\index{Bose!distribution} is the fugacity of a given particle characterizing the pair density and it is the same for both particles and antiparticles. For $\Upsilon_i=1$, the distribution maximizes the entropy content at a fixed particle energy, this maximum is not very pronounced~\cite{Letessier:1993qa}. The parameter $\mu_i$ is the chemical potential\index{chemical potential} for a given particle which is associated to the density difference between particles and antiparticles.

\para{Chemical equilibrium}
In general there are two types of chemical equilibrium\index{chemical equilibrium} associated with the chemical parameters $\Upsilon$ and $\mu$ each. We have:
\begin{itemize}
\item \underline{\it Absolute chemical equilibrium:\/} 
The absolute chemical equilibrium is the level to which energy is shared into accessible degrees of freedom, \eg\ the particles can be made as energy is converted into matter.
The absolute equilibrium is reached when the phase space occupancy approaches unity $\Upsilon\to1$. 
 \item \underline{\it Relative chemical equilibrium:\/}
 The relative chemical equilibrium is associated with the chemical potential $\mu$ which involves reactions that distribute a certain already existent element/property among different accessible compounds. 
 \end{itemize}
The dynamics of absolute chemical equilibrium, in which energy can be converted to and from particles and antiparticles, is especially important. The consequences for the energy conversion to from particles/antiparticle can be seen in the first law of thermodynamics by introducing the chemical potential $\mu_N$ for particle and $\mu_{\bar{N}}$ for antiparticle as follows:
\begin{align}
\mu_N\equiv\mu+T\ln\Upsilon,\qquad{\mu_{\bar{N}}}\equiv{-\mu}+T\ln\Upsilon.
\end{align}
Then the first law of thermodynamics can be written as
\begin{align}
dE&=-PdV+TdS+{\mu_N}dN+{\mu_{\bar{N}}}d{\bar{N}}
\\&=-PdV+TdS+{\mu}(dN-d{\bar{N}})+T\ln{\Upsilon}(dN+d{\bar{N}}).
\end{align}
Here the chemical potential\index{chemical potential}\index{fugacity} $\mu$ is the energy required to change the difference between particles and antiparticles, and $T\ln\Upsilon$ is the energy required to change the total number of particles and antiparticles; the fugacity $\Upsilon$ is the parameter allowing to adjust this energy.

\para{Boltzmann equation and particle freeze-out}
The Boltzmann equation describes the evolution of {\color{black} a (nonequilibrium)} distribution function $f$ in phase space. General properties of the Boltzmann-Einstein equation\index{Boltzmann-Einstein equation} in an arbitrary spacetime are explored in~\rsec{sec:BoltzmannEinstein}. The Boltzmann equation in the FLRW\index{cosmology!FLRW} Universe takes the Einstein-Vlasov form\index{Einstein-Vlasov equation}
\begin{align}\label{Hubble:Boltzmann}
\frac{\partial f}{\partial t}-\frac{\left(E^2-m^2\right)}{E}H\frac{\partial f}{\partial E}=\frac{1}{E}\sum_{i}\mathcal{C}_i[f]\,,
\end{align}
where $H=\dot{a}/a$\index{Einstein-Vlasov equation!Hubble expansion} is the Hubble parameter\index{Hubble parameter},~\req{dynamic}, see~\rsec{sec:flrw} below for a more detailed cosmology primer. Due to the homogeneity and isotropy of the Universe, the distribution function depends on time $t$ and energy $E=\sqrt{p^2+m^2}$ only. The collision term $\sum_i \mathcal{C}_i$ represents all elastic and inelastic interactions and the index $i$ labels the corresponding physical process. In general, the collision term is proportional to the relaxation time for a given collision as follows~\cite{Anderson:1974nyl}
\begin{align}
\frac{1}{E}\mathcal{C}_i[f]\propto\frac{1}{\tau_\mathrm{rel}}\,,
\end{align}
where $\tau_\mathrm{rel}$ is the relaxation time for the reaction, which characterizes the magnitude of the reaction time to reach chemical equilibrium. 

As the Universe expands, the collision term in the Boltzmann equation competes with the Hubble term. In general, a given particle freezes-out from the cosmic plasma when its interaction rate $\tau_\mathrm{rel}^{-1}$ becomes smaller than the Hubble expansion rate\index{freeze-out}
\begin{align}
H\geqslant\tau_\mathrm{rel}^{-1}.
\end{align}
When this happens, the particle's interactions are not rapid enough to maintain thermal distribution, either because the density of particles becomes so low that the chance of any two particles meeting each other becomes negligible, or because the particle energy becomes too low to interact. The freeze-out process can be categorized into three distinct stages based on the type of freeze-out interactions, we have~\cite{Birrell:2012gg,Rafelski:2023emw}:
\begin{itemize}
\item \underline{\it Chemical freeze-out:\/}\index{freeze-out!chemical} 
As the Universe expands and the temperature drops, the rate of the inelastic scattering (\eg\ production and annihilation reaction) that maintains the equilibrium density becomes smaller than the expansion rate. At this point, the inelastic scattering ceases, and a relic population of particles remain. Prior to the chemical freeze-out temperature, number changing processes are significant and keep the particle in thermal equilibrium, implying that the distribution function has the usual Fermi-Dirac\index{Fermi!distribution} form 
\begin{equation}\label{equilibrium}
f_\mathrm{ch}(t,E)=\frac{1}{\exp[(E-\mu)/T]+1},\qquad \text{ for } T(t)> T_\mathrm{ch}
\,,
\end{equation}
where $T_\mathrm{ch}$ represents the chemical freeze-out temperature. \\[-0.2cm]
\item \underline{\it Kinetic freeze-out:\/}\index{freeze-out!kinetic}
After chemical freeze-out, at yet lower temperature in an expanding Universe particles still scatter elastically from other particles and keep thermal equilibrium in the primordial plasma. As the temperature drops, the rate of elastic scattering reaction that maintain the thermal equilibrium become smaller than the expansion rate. At that time, elastic scattering processes cease, and the relic particles do not interact with other particles in the primordial plasma anymore; they free-stream. 

Once chemical freeze-out takes hold, the distribution function has the kinetic equilibrium\index{kinetic equilibrium} form with pair abundance typically below maximum yield~$\Upsilon \le 1$
\begin{equation}\label{kinetic:equilibrium}
f_\mathrm{k}(t,E)=\frac{1}{\Upsilon^{-1}\exp[(E-\mu)/T]+1},\qquad \text{ for }T_k< T(t)< T_\mathrm{ch},
\end{equation}
where $T_k$ represents the kinetic freeze-out temperature. The generalized fugacity\index{fugacity} $\Upsilon(t)$ controls the occupancy of phase space and is necessary once $T(t)<T_\mathrm{ch}$ in order to conserve the particle number. {\color{black} In general we encounter $T_k<T_{\mathrm{ch}}$ allowing us to denote the complete thermal and kinetic freeze-out temperature as $T_\mathrm{F}\simeq T_k$: for $T< T_\mathrm{F}$ all scattering becomes negligible for the considered particle species.}\\[-0.2cm]
\item \underline{\it Free streaming:\/}
{\color{black}After freeze-out at $T<T_\mathrm{F}$, all particles have fully decoupled from the primordial plasma and become free-streaming\index{free-streaming}. They ceased influencing the dynamics of the Universe  other than by their gravity contributing  to the energy-momentum tensor. Therefore, the free-streaming momentum distribution follows from the non-interacting  Einstein-Vlasov momentum evolution equation, see \req{VEeqFLR}.  This equation} can be solved~\cite{Choquet-Bruhat:2009xil} as we describe in \rsec{sec:model:ind},  and the free-streaming momentum distribution can be exactly given\index{Einstein-Vlasov equation} as~\cite{Birrell:2012gg}\index{free-streaming!quantum distribution}
\begin{equation}\label{freeStreamDist}
f_\mathrm{fs}(t,E)=\frac{1}{\Upsilon^{-1}\exp{\left[\sqrt{\frac{E^2-m^2}{T_\mathrm{fs}^2}+\frac{m^2}{T^2_\mathrm{F}}}-\frac{\mu}{T_\mathrm{F}}\right]+1}}\,,
\quad T_\mathrm{fs}(t)=\frac{a(t_F)}{a(t)}\,T_\mathrm{F}\,,
\end{equation}
{\color{black}where $t_F$ is the time at which full freeze-out occurs. The free-streaming effective temperature $T_\mathrm{fs}$ is obtained by redshifting the temperature at kinetic freeze-out as shown.}\\[0.2cm] 
Considering a massive particle (\eg\ dark matter) freeze-out process from cosmic plasma in the nonrelativistic regime, $m\gg T_\mathrm{F}$. We can use the
Boltzmann\index{Boltzmann!approximation} approximation, and the free-streaming distribution for nonrelativistic particle becomes
\begin{align}\label{freeStreamDistNR}
&f^B_\mathrm{fs}(t,p)=\Upsilon\,e^{-(m+\mu)/T_\mathrm{F}}\exp\left[-\frac{1}{ T_\mathrm{eff}}\frac{p^2}{2m}\right],\quad T_\mathrm{eff}=\left(\frac{a(t_\mathrm{F})}{a(t)}\right)^2T_\mathrm{F}\,,
\end{align}
where we note the effective temperature $T_\mathrm{eff}$ for massive free-streaming particles: The effective temperature for massive particles decreases faster than the Universe temperature cools. The difference in temperatures between cold free-streaming particles and hot cosmic plasma is worth emphasizing. How this would affect the evolution of the primordial Universe requires more detailed study. 
\end{itemize}

The division of the freeze-out process into these three regimes is a simplification of much more complex overlapping dynamical processes. It is, however, a very useful approximation in the study of cosmology~\cite{Rafelski:2023emw,Birrell:2012gg,Mangano:2005cc,Birrell:2014gea}.

\para{Particle content of the Universe} 
Our detailed understanding of the primordial Universe arises from half a century of research in the fields of cosmology, ultra relativistic heavy-ion collisions, particle, nuclear and plasma physics. Today we believe that the primordial deconfined matter we call quark-gluon plasma (QGP) filled the entire Universe and lasted for about first $20\,\mathrm{\mu s}$ after the Big-Bang~\req{taudef}\index{Big-Bang}. The deconfined condition allows free motion of quarks and gluons along with all other elementary particles. 

This hot primordial particle soup filled the expanding Universe as long as it was well above hadronization\index{QGP!hadronization} Hagedorn temperature\index{Hagedorn!temperature} $T_H\simeq 150\MeV$. Well below $T\ll T_H$, the Universe contained all the building blocks of the usual matter that surrounds us today. Depending on the temperature, many other elementary matter particles could also be present as seen in \rf{fig:energy:frac}. The total particle inventory thus includes:
\begin{itemize}
\item The up $u$ and down $d$ quarks now hidden in protons and neutrons;
\item Electrons, and  three flavors  of (Majorana) neutrinos;\\[-0.2cm]
\item[] There were also unstable particles present which can decay but are reformed in hot Universe:\\[-0.2cm]
\item Heavy unstable leptons muon $\mu$ and tauon $\tau$;
\item Unstable when bound in present day matter strange $s$, and heavy charm $c$ and bottom $b$ quarks;\\[-0.2cm]
\item[] At yet higher temperatures unreachable in laboratory experiments today we encounter all the remaining much heavier standard model particles:\\[-0.2cm]
\item Electroweak theory gauge bosons W$^\pm$ and Z$^0$, the top $t$ quark, and the Higgs particle H.
\item The QGP phase of matter contains also the gluons, particles mediating the strong interaction of deconfined quarks.
\end{itemize}

Using the relativistic covariant Fermi/Bose momentum distribution, the corresponding energy density $\rho_i$\index{particle!energy density}, pressure $P_i$\index{particle!pressure}, and number densities $n_i$\index{particle!density} for particle species $i$ and mass $m_i$ are given by
\begin{align}
\rho_i&=g_i\int\!\!\frac{d^3p}{(2\pi)^3}Ef_{F/B}=\frac{g_i}{2\pi^2}\!\int_{m_i}^\infty\!\!\!dE\,\frac{E^2\left(E^2-m_i^2\right)^{1/2}}{\Upsilon_i^{-1}e^{(E-\mu_i)/T}\pm 1}\,,
\label{energy_density}\\[0.2cm]
P_i&=g_i\int\!\!\frac{d^3p}{(2\pi)^3}\frac{p^2}{3E}f_{F/B}=\frac{g_i}{6\pi^2}\!\int_{m_i}^\infty\!\!\!dE\,\frac{\left(E^2-m_i^2\right)^{3/2}}{\Upsilon_i^{-1} e^{(E-\mu_i)/T}\pm 1}\,,
\label{Pressure_density}\\[0.2cm]
n_i&=g_i\int\!\!\frac{d^3p}{(2\pi)^3}f_{F/B}=\frac{g_i}{2\pi^2}\!\int_{m_i}^\infty\!\!\!dE\,\frac{E(E^2-m_i^2)^{1/2} }{\Upsilon_i^{-1}e^{(E-\mu_i)/T}\pm 1}\,,
\label{number_density}\\[0.2cm]
\sigma_i=&-g_i\int\!\! \frac{d^3p }{(2\pi)^3}(f_{F/B}\ln(f_{F/B})\pm(1\mp f_{F/B})\ln(1\mp f_{F/B}))
\,,
\end{align}
where $g_i$ is the degeneracy of the particle species `$i$'. Inclusion of the fugacity parameter $\Upsilon_i$ allows us to characterize particle properties in chemical nonequilibrium situations. 

{\color{black}Given the above relations for the energy density $\rho_i$, pressure $P_i$, and number densities $n_i$, the entropy density $\sigma_i$\label{entropy_density} for particle species $i$ of mass $m_i$ can be obtained by integration by parts leading to the usual Gibbs–Duhem form}\index{entropy!density}
\begin{align}\label{entropy}
\sigma_i=\frac{S_i}{V}=\frac{\partial P_i}{\partial T}=\left(\frac{\rho_i+P_i}{T}-\frac{\tilde \mu_i}{T}\,n_i\right)\,,\qquad
n_i=\frac{\partial P_i}{\partial \tilde{\mu}_i}\,,\qquad
\tilde \mu_i=T\ln \Upsilon_i +\mu_i\,,
\end{align}
where the chemical potential associated with charge, baryon number, etc., changes sign between particles and antiparticles, while $\Upsilon_i$ does not.

Once full decoupling is achieved, the corresponding free-streaming energy density, pressure, number density and entropy arising from the solution of the Boltzmann-Einstein equation\index{Boltzmann-Einstein equation} differ from the thermal equilibrium \req{energy_density}, \req{Pressure_density}, \req{number_density}, and~\req{entropy} by replacing the mass by a time dependent effective mass $m\,T_\mathrm{fs}(t)/T_\mathrm{F}$ in the exponential, and other related changes which will be derived in~\rsec{sec:model:ind}, see \req{eq:NeutrinoRho}, \req{eq:NeutrinoP}, \req{eq:NumDensity}, and \req{eq:EntropyIntegrand}. Once decoupled, the free-streaming particles maintain their comoving number and entropy density, see~\req{eq:ConstEntropy}.

In general the chemical potential is associated with the baryon number. The net baryon number density relative to the photon number density\index{baryon!per photon ratio} is near to $10^{-9}$. In many situations we can neglect the small chemical potential when calculating the total entropy density\index{entropy!density} in the Universe. The total entropy density in the primordial Universe can be written as
\begin{align}\label{eq:entg}
&\sigma=\sum_i\,\sigma_i=\frac{2\pi^2}{45}g^s_\ast\,T^3,\\
&g^s_\ast=\sum_{i=\mathrm{bosons}}g_i\left({\frac{T_i}{T_\gamma}}\right)^3B\left(\frac{m_i}{T_i}\right)+\frac{7}{8}\sum_{i=\mathrm{fermions}}g_i\left({\frac{T_i}{T_\gamma}}\right)^3F\left(\frac{m_i}{T_i}\right),\label{eq:entg002}
\end{align}
where $g^s_\ast$ counts the effective number of `entropy' degrees of freedom. The functions $B(m_i/T)$ and $F(m_i/T)$ are defined as \index{entropy!degrees of freedom}
\begin{align}
&B\left(\frac{m_i}{T}\right)=\frac{45}{12\pi^4}\int^\infty_{m_i/T}\,dx\sqrt{x^2-\left(\frac{m_i}{T}\right)^2}\left[4x^2-\left(\frac{m_i}{T}\right)^2\right]\frac{1}{\Upsilon^{-1}_ie^x-1},\\
&F\left(\frac{m_i}{T}\right)=\frac{45}{12\pi^4}\frac{8}{7}\int^\infty_{m_i/T}\,dx\sqrt{x^2-\left(\frac{m_i}{T}\right)^2}\left[4x^2-\left(\frac{m_i}{T}\right)^2\right]\frac{1}{\Upsilon^{-1}_ie^x+1}.
\end{align}
In~\rf{EntropyDOF:Fig} we show $g^s_\ast$ as a function of temperature, the effect of particle mass threshold~\cite{Coc:2006rt} is considered in the calculation for all considered particles. When $T$ decreases below the mass of particle $T\ll m_i$, this particle species becomes nonrelativistic and the contribution to $g^s_\ast$ becomes negligible, creating the smooth dependence on $T$ across mass threshold seen in~\rf{EntropyDOF:Fig}: The vertical lines identify particle mass thresholds on temperature axis, $m_e=0.511\MeV$, $m_\mu=105.6\MeV$, and pion average mass $m_\pi\approx138\MeV$.

\begin{figure} 
\centerline{\includegraphics[width=0.75\linewidth]
{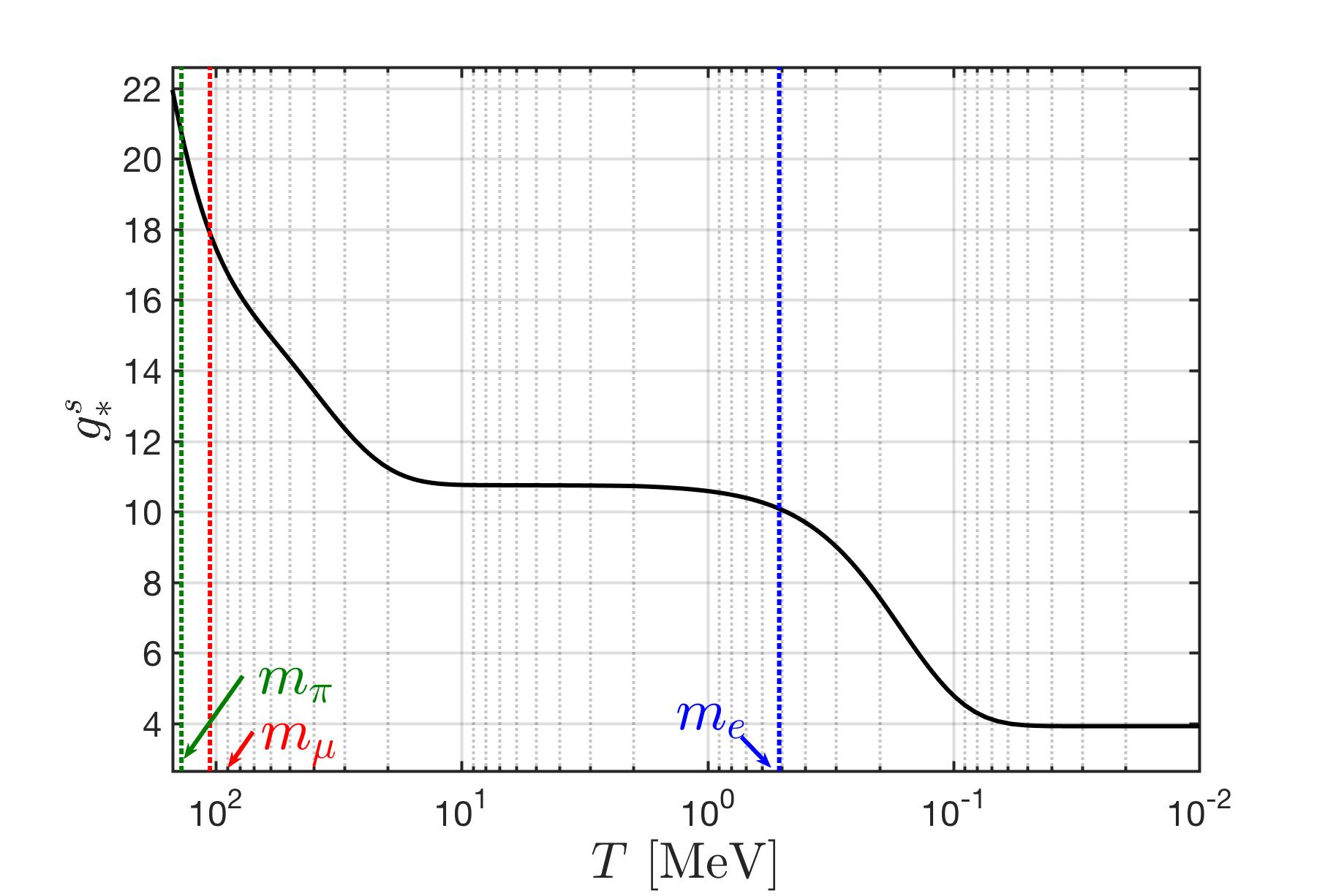}}
\caption{The entropy degrees of freedom as a function of $T$ in the primordial Universe epoch after hadronization $10^{-2}\MeV \leqslant T \leqslant 150 \MeV$. \radapt{Yang:2024ret}.}
\label{EntropyDOF:Fig} 
\end{figure}

\para{Departure from detailed balance}
A well-known textbook result for the case of two particle scattering is that the Boltzmann scattering term, the right-hand side in~\req{Hubble:Boltzmann}, vanishes when particles reach thermal equilibrium. The rates of the forward and reverse reactions are equal, resulting in a balance between production and annihilation of particles. Such a balance is called detailed balance\index{detailed balance}. Thermal equilibrium implies both chemical equilibrium\index{chemical equilibrium} (particle abundances are balanced) and kinetic equilibrium (equipartition of energy according to the equilibrium distributions).

Kinetic equilibrium\index{kinetic equilibrium} is usually established much quicker by means of scattering processes not capable of generating particles, thus the approach to kinetic equilibrium often has little impact on the actual particle abundances, that is, on chemical equilibrium. Chemical nonequilibrium is often driven by time dependence of the environment in which particles evolve, for example in~\req{Hubble:Boltzmann} by the Hubble parameter\index{Hubble parameter} $H(t)$ term. The well studied example is the emergence in the BBN era of light isotope abundances with many yields being dependent on the speed of Universe expansion~\cite{Pitrou:2018cgg,Kolb:1990vq,Dodelson:2003ft,Mukhanov:2005sc}. 

In elementary particle context the competition is often between elementary processes and not so much with the Hubble expansion. This can lead to stationary population in detailed balance not in chemical equilibrium, with the actual value of particle fugacity determined by reaction dynamics for a fixed ambient temperature. In the primordial Universe a particle abundance can be in detailed balance and yet not in chemical equilibrium. We will investigate this type of nonequilibrium situation in the primordial Universe for bottom quarks in~\rsec{Bottom} and strange quarks in~\rsec{Strangeness}.

There are thus two environments in the primordial Universe in which we can expect chemical nonequilibrium to arise:
\begin{enumerate}
\item The particle production rate becomes slower than the rate of Universe expansion and the production reaction freeze-out. Once the production reactions freeze-out from the cosmic plasma, the corresponding detailed balance is broken. In the case of unstable particles their abundance decreases via the decay/annihilation reactions.
\item The nonequilibrium can also be achieved when the production reaction slows down and is not able to keep up with decay/annihilation reaction. In this case, the Hubble expansion rate can be much longer than the decay and production rate and is not relevant to the nonequilibrium process. The key factor is competition between production and decay/annihilation which can result in chemical nonequilibrium in the primordial Universe in which detailed balance is maintained.
\end{enumerate}
The chemical nonequilibrium conditions in the primordial Universe are of general interest: they are understood to be prerequisite for the arrow of time to take hold in the expanding Universe.

\subsection{Cosmology Primer}
\label{sec:flrw}
We present now a short review of the Universe dynamics within the FLRW cosmology which will be useful throughout this work. Our objective is to recognize and identify markers clarifying and quantifying the different eras. This section unlike the remainder of the work relies on $\Lambda\mathrm{CDM}$ model of cosmology, which leads to the results seen in~\rf{fig:energy:frac} obtained with a pie-chart energy content of the contemporary Universe comprising: 69\% dark energy, 26\% dark matter, 5\% baryons\index{baryon}, and $<1$\% photons and neutrinos in energy density~\cite{Davis:2003ad,Planck:2018vyg}. 

As noted earlier, for most part our results will remain valid if one day this model evolves to account for tensions in modeling the current Universe Hubble expansion. This is so since our work applies to the primordial Universe period where neither dark energy nor dark matter is relevant; expansion of the Universe is driven nearly solely by radiation and matter-antimatter content and unknown properties of neutrinos do not contribute.\index{Universe!radiation dominated}

\para{About cosmological sign conventions}
There are several sign conventions\index{cosmology!sign conventions} in use in general relativity. As discussed by Hobson, Efstathiou and Lasenby~\cite{Hobson:2006se}, these conventions differ by three sign factors $S1$, $S2$, $S3$, which appear in the following objects:

\begin{subequations}
\vspace*{3mm}
\indent Metric Signature: 
\beql{conv:metric}\eta^{\mu\nu}=(S1)\text{Diag}(1,-1,-1,-1)
\vspace*{3mm}
\eeqn
\indent Riemann Tensor: 
\beql{conv:Riemann}
R^\mu_{\alpha\beta\gamma}=(S2)(\partial_{\beta}\Gamma^\mu_{\alpha\gamma}-\partial_{\gamma}\Gamma^\mu_{\alpha\beta}+\Gamma^\mu_{\sigma\beta}\Gamma^\sigma_{\gamma\alpha}-\Gamma^\mu_{\sigma\gamma}\Gamma^\sigma_{\beta\alpha})
\vspace*{3mm}
\eeqn
\indent Einstein Equation: 
\beql{conv:EinstEq}
G_{\mu\nu}=(S3)8\pi G_NT_{\mu\nu}
\vspace*{3mm}
\eeqn
\indent Ricci Tensor:
\beql{conv:RicciT}
R_{\mu\nu}=(S2)(S3)R^\alpha_{\mu\alpha\nu}
\vspace*{3mm}
\eeqn
\end{subequations}
\noindent The sign $S3$ comes from the choice of what index is contracted in forming the Ricci tensor. Since that sign factor appears in both $R_{\mu\nu}$ and $R$ it affects the overall sign of $G_{\mu\nu}$ and therefore Einstein's equation as shown above (here the cosmological constant is considered part of $T_{\mu\nu}$). In this work we will use the 
\beql{eq:3S}
\{(S_1), (S_2),(S_3)\}=(+,+,+)
\eeqn
convention.

\para{FLRW Cosmology} The Friedmann-Lema{\^i}tre-Robertson-Walker (FLRW)\index{cosmology!FLRW} line element and metric~\cite{Hobson:2006se,Hartle:2003yu,Misner:1973prb,Weinberg:1972kfs} in spherical coordinates is
\begin{gather}
 \label{FLRW} ds^2=dt^2-a^2(t)\left[\frac{dr^2}{1-kr^{2}}+r^{2}d\theta^2+r^{2}\sin\theta^{2}d\phi^2\right]\,,\\[0.3cm]
 g_{\alpha\beta}=
 \begin{pmatrix}
 1&0&0&0\\
 0&-\displaystyle\frac{a^{2}(t)}{1-kr^{2}}&0&0\\
 0&0&-a^{2}(t)r^{2}&0\\
 0&0&0&-a^{2}(t)r^{2}\sin\theta^{2}
 \end{pmatrix}\,.
\end{gather}
The Gaussian curvature $k$ informs the spatial hyper-surfaces defined by comoving observers. The spatial shape of the Universe has the following\index{Universe!geometry} possibilities~\cite{Planck:2018vyg}: infinite flat Euclidean $(k=0)$, finite spherical but unbounded $(k=+1)$, or infinite hyperbolic saddle-shaped $(k=-1)$. Observation indicates our Universe is flat or nearly so. Current observation of cosmic microwave background (CMB)\index{CMB} anisotropy imply the preferred value $k=0$~\cite{Planck:2018vyg,Planck:2015fie,Planck:2013pxb}.

In an expanding (or contracting) Universe which is both homogeneous and isotropic, the scale factor $a(t)$ denotes the change of proper distances $L(t)$ over time as
\begin{gather}
 L(t)=L_{0}\frac{a_{0}}{a(t)}\rightarrow L(z)=L_{0}(1+z)\,,
\end{gather}
where $z$ is the redshift and $L_{0}$ the comoving length. This implies volumes change with $V(t)=V_{0}/a^{3}(t)$ where $V_{0}=L_{0}^{3}$ is the comoving Cartesian volume. In terms of temperature, we can consider the expansion to be an adiabatic process~\cite{Abdalla:2022yfr} which results in a smooth shifting of the relevant dynamical quantities. As the Universe undergoes isotropic expansion, the temperature decreases as 
\begin{gather}
 \label{tscale}
 T(t)=T_{0}\frac{a_{0}}{a(t)}\rightarrow T(z)=T_{0}(1+z)\,,
\end{gather}
where $z$ is the redshift\index{redshift}. The entropy within a comoving volume is kept constant until gravitational collapse effects become relevant. The comoving temperature $T_{0}$ is given by the present CMB temperature $T_{0}=2.726{\rm\ K}\simeq 2.349\times10^{-4}\eV$~\cite{Planck:2018vyg}, with contemporary scale factor $a_{0}=1$.

The cosmological dynamical equations describing the evolution of the Universe follow from the Einstein equations. In general, the Einstein equation with cosmological constant $\Lambda$ can be written as:
\beqn\label{Einstine}
G^{\mu\nu} -\Lambda g^{\mu\nu}=\frac{\hbar c}{c^4M_p^2} T^{\mu\nu}\,, \quad G^{\mu\nu}=R^{\mu\nu}-\frac{R}{2} g^{\mu\nu}\,,
\quad R= g_{\mu\nu}R^{\mu\nu}\,.
\eeqn
{\color{black}The space curvature $R$ has dimension 1/Length$^2$; the energy momentum tensor $T^{\mu\nu}$ has dimension energy/Length$^3$. These units are maintained by factors $\hbar$ and $c$ shown above. In the following we will in general omit to state explicitly factors $\hbar$ or $c$. We note that given the definition of Planck mass in \req{eq:GN} we see that the appearance of $\hbar$ is purely decorative, allowing to represent Newton's constant  using Planck's mass, there is no quantum physics inherent to above equations.}

Recall that the Einstein tensor\index{Einstein tensor} $G^{\mu\nu}$ is divergence-free and so is the stress energy tensor, $T^{\mu\nu}$\index{stress-energy tensor}. In a homogeneous isotropic spacetime, the matter content is necessarily characterized by two quantities, the energy density $\rho$ and isotropic pressure~$P$
\begin{equation}
 T^\mu_\nu =\mathrm{diag}(\rho, -P, -P, -P)\,.
\end{equation}
 It is common to absorb the Einstein cosmological constant $\Lambda$ into $\rho$ and $P$ by defining dark energy components
\beqn\label{EpsLam}
\rho_\Lambda=M_p^2\Lambda\,, \qquad P_\Lambda=-M_p^2 \Lambda\,.
\eeqn
We implicitly consider this done from now on. 

As the Universe expands, redshift (referring verbally to the increase in de Broglie wavelength $\lambda_\mathrm{dB}=\hbar /p$) reduces the momenta $p$ of particles, thus lowering their contribution to the energy content of the Universe. This cosmic momentum redshift is written as
\begin{alignat}{1}
 \label{Redshift} p_{i}(t) = p_{i,0}\frac{a_{0}}{a(t)}\,.
\end{alignat}
Momentum (and the energy of massless particles $E=pc$) scales with the same factor as temperature. Since mass does not evolve in time, the energy of massive free particles in the Universe scales differently based on their momentum (and thus temperature). Only the energy of hot and relativistic particles decreases inversely with scale factor, like radiation. As the particles transition to nonrelativistic (NR) energies, they decrease with the inverse square of the scale factor, compare \req{freeStreamDist} and \req{freeStreamDistNR}
\begin{alignat}{1}
 \label{EScale} E(t) = E_{0}\frac{a_{0}}{a(t)}\xrightarrow{\mathrm{NR}}\ E_{0}\frac{a_{0}^{2}}{a(t)^{2}}\,.
\end{alignat}
This occurs because of the functional dependence of energy on momentum in the relativistic $E\sim p$ versus nonrelativistic $E\sim p^{2}$ cases.

\para{Hubble parameter and deceleration parameter}
The global Universe dynamics can be characterized by two quantities, the Hubble parameter $H(t)$, a strongly time dependent quantity on cosmological time scales, and the deceleration parameter\index{cosmology!deceleration parameter} $q$,\index{Hubble parameter}
\beqn\label{dynamic}
H(t)\equiv\frac{\dot a }{a} \,, 
\eeqn
\beqn\label{dynamic1}
q\equiv -\frac{a\ddot a}{\dot a^2}\,.
\eeqn
We note the resulting relations
\beqn\label{eq:Hdot1}
 \frac{\ddot a}{a}=-qH^2,
 \eeqn
\beqn\label{eq:Hdot}
 \dot H=-H^2(1+q)\,. 
\eeqn
\index{cosmology!deceleration parameter}

Two dynamically independent equations arise using the metric~\req{FLRW} in the Einstein equation~\req{Einstine}
\beqn\label{hubble}
\frac{8\pi G_N}{3} \rho = \frac{\dot a^2+k}{a^2}
=H^2\left( 1+\frac { k }{\dot a^2}\right),
\qquad
\frac{4\pi G_N}{3} (\rho+3P) =-\frac{\ddot a}{a}=qH^2.
\eeqn
These are also known as the Friedmann equations\index{cosmology!Friedmann equations}. 

There is a simple way to determine dependence of $q$ on Universe structure and dynamics: We can eliminate the strength of the interaction, $G_N$, by solving the equations~\req{hubble} for ${8\pi G_N}/{3}$ and equating the two results to find a relatively simple constraint for the deceleration parameter
\beqn\label{qparam0}
q=\frac 1 2 \left(1+3\frac{P}{\rho}\right)\left(1+\frac{k}{\dot a^2}\right).
\eeqn
From this point on, we work within the flat cosmological model with $k=0$. It is good to recall that one must always satisfy the constraint on $H$ introduced by the first of the Friedmann equations~\req{hubble}, which for $k$=0, flat Universe is the Hubble equation,~\req{Hubble:eq}.

The parameter $q$ and thus time evolution of $H$ according to~\req{eq:Hdot} is determined entirely within the FLRW cosmological model by the matter content of the Universe
\begin{equation}\label{qparam}
q=\frac 1 2 \left(1+3\frac{P}{\rho}\right)\,.
\end{equation}
We note that in FLRW Universe according to~\req{eq:Hdot1} the second derivative of scale parameter $a$ changes sign when the sign of $q$ changes: the Universe decelerates (hence name of $q>0$) initially slowing down due to gravity action. The Universe will reverse this and accelerate under influence of dark energy as $q$ changes sign. Even so, the Hubble parameter according to~\req{qparam} keeps its sign since even when dark energy dominates we asymptotically approach $q=-1$, that is according to~\req{EpsLam} $P=-\rho$. In the dark energy dominated Universe pressure approaches this condition without ever reaching it as normal matter remains within the Universe inventory: In the FLRW Universe $\dot H=0$ is impossible; $H(t)$ is continuously decreasing in its value, we cannot have a minimum in the value of $H$ as some have proposed to explain the Hubble $H_0$ tension\index{Hubble parameter!minimum}. 

\para{Conservation laws}
In a flat FLRW Universe, the spatial components of the divergence of the stress energy tensor automatically vanish, leaving the single condition
\begin{equation}\label{stress_energy_eq}
\nabla_\mu \mathcal{T}^{\mu 0}=\dot{\rho}+3\left(\rho+P\right)\frac{\dot{a}}{a}=0\,.
\end{equation}
If the set of particles can be portioned into subsets such that there is no interaction between the different subsets, then this condition applies independently to each and leads to an independent temperature for each such subset. We will focus on a single such group and use~\req{stress_energy_eq} to derive an equivalent condition involving entropy and particle number, which illustrates how the entropy of the Universe evolves in time. 

Consider a collection of particles with distributions $f_i$ that are in kinetic equilibrium\index{kinetic equilibrium} \req{kinetic:equilibrium} at a common temperature $T$, with zero chemical potentials and distinct fugacities\index{fugacity} $\Upsilon_i$, and which satisfy~\req{stress_energy_eq}. For the purpose of the following derivation, it is useful to rewrite the fugacities as if they came from chemical potentials, {\it i.e.\/}, define $\hat{\mu}_i$ by $\Upsilon_i=e^{\hat{\mu}_i/T}$. This gives the expressions a familiar thermodynamic form with $\hat{\mu}_i$ playing the role of chemical potential\index{chemical potential} and helps with the calculations, but should not be confused with a chemical potential as discussed above. We already know the particle energy density \req{energy_density}, particle pressure \req{Pressure_density}, particle number density \req{number_density}, and entropy density \req{entropy_density} of a particle of mass $m$ with momentum distribution $f_{F/B}$.

Combining~\req{stress_energy_eq} with the identities in~\req{entropy} we can obtain the rate of change of the total comoving entropy as follows. Letting $\sigma=\sum_i \sigma_i$ be the total entropy density, first compute
\begin{align}\frac{1}{a^3}\frac{d}{dt}(a^3\sigma T)&=\frac{1}{a^{3}}\frac{d}{dt}\left(a^3\left(P+\rho-\sum_i \hat{\mu}_i n_i\right)\right)\\
&=\dot{P}+\dot{\rho}-\sum_i \left(\dot{\hat{\mu}}_in_i+\hat{\mu}_i\dot{n}_i\right)+3\left(P+\rho-\sum_i \hat{\mu}_i n_i\right)\dot{a}/a\notag\\
&=\frac{\partial P}{\partial T} \dot{T}+\sum_i\frac{\partial P_i}{\partial \hat{\mu}_i} \dot{\hat{\mu}}_i-\sum_i \left(\dot{\hat{\mu}}_in_i+\hat{\mu}_i\dot{n}_i+3\hat{\mu}_i n_i \dot{a}/a\right)+\nabla_\mu \mathcal{T}^{\mu 0}\notag\\
&=\sigma\dot{T}-\sum_i \left(\hat{\mu}_i\dot{n}_i+3\hat{\mu}_i n_i \dot{a}/a\right)
=\sigma\dot{T}- a^{-3}\sum_i\hat{\mu}_i\frac{d}{dt}(a^3n_i)\,.\notag
\end{align}
Therefore we find
\begin{align}\label{S:n:eq}
\frac{d}{dt}(a^3\sigma )=&\frac{1}{T}\frac{d}{dt}(a^3\sigma T)-a^3\sigma \frac{\dot T}{T}=-\sum_i\ln(\Upsilon_i)\frac{d}{dt}(a^3n_i)\,.
\end{align}
From this we can conclude that comoving entropy is conserved as long as each particle satisfies one of the following conditions:
\begin{enumerate}
\item
The particle is in chemical equilibrium\index{chemical equilibrium}, {\it i.e.\/}, $\Upsilon_i= 1$;
\item
The particle has frozen out chemically and thus has conserved comoving particle number, {\it i.e.\/}, $\frac{d}{dt}(a^3n_i)$. 
\end{enumerate}
Therefore, under the instantaneous freeze-out assumptions, we can conclude conservation of comoving entropy during all three eras, \req{equilibrium}~-~\eqref{freeStreamDist}, of the evolution of a distribution. 

These observations provide an alternative characterization of the dynamics of a FLRW Universe that is composed of entirely of particles in chemical or kinetic equilibrium. The dynamical quantities are the scale factor $a(t)$; the common temperature $T(t)$, and the fugacity of each particle species $\Upsilon_i(t)$ that is not in chemical equilibrium. 

The dynamics are given by the Hubble equation \req{Hubble:eq}, conservation of the total comoving entropy of all particle species, and conservation of comoving particle number for each species not in chemical equilibrium (otherwise $\Upsilon_i=1$ is constant),
\begin{equation}\label{eq:dynamics}
H^2=\frac{\rho_{tot}}{3M_p^2}\,, \qquad \frac{d}{dt}(a^3\sigma )=0\,,\qquad \frac{d}{dt}(a^3n_i)=0 \,\text{ when } \Upsilon_i\neq 1\,.
\end{equation}
We emphasize here that $\rho_{tot}$ is the total energy density of the Universe, which may be composed of contributions from multiple particle groupings with cross group interactions being absent. In such case, each grouping has its own temperature and independently conserves its comoving entropy. 
%

\subsection{Dynamic Universe}\label{sec:dynamic}
\para{Eras of the Universe}
The dynamic Universe\index{Universe!eras} is governed by the total pressure and energy content: For the energy content $\rho=\rho_\mathrm{total}$ we have the sum of all contributions from any form of matter, radiation, particle or field. This includes but is not limited to: dark energy $(\Lambda)$, dark matter (DM), baryons\index{baryon} (B), leptons $(\ell,\nu)$ and photons $(\gamma)$. The same remark applies to pressure $P$. Depending on the age of the Universe, the relative importance of each particle group changes as each dilutes differently under expansion, with dark energy remaining constant, thus emerging in relative importance and accelerating the expansion of the aging Universe today. 

It turns out that $q$, the acceleration-deceleration\index{cosmology!deceleration parameter} parameter~\req{qparam} is a very convenient tool to characterize the different epochs of the Universe~\cite{Rafelski:2013yka}. $q$ is for historical reasons positive under deceleration $q>0$. Conversely, the accelerating Universe has $q<0$. This convention was chosen under the tacit assumption that the Universe should be decelerating, before the discovery of dark energy. The value of $q$ for different eras is found to be:
\begin{itemize}
\item Radiation-dominated\index{Universe!composition} Universe: \beql{Eq:radU}
P=\rho/3 \implies q=1\,.
\eeqn
\item (Nonrelativistic) Matter-dominated Universe: 
\beql{Eq:nonrmU}
P\ll\rho \implies q=1/2\,.
\eeqn
\item Dark energy $\Lambda$-dominated Universe:\index{Universe!dark energy dominated} 
\beql{Eq:darkU} 
P=-\rho \implies q=-1\,.
\eeqn
\end{itemize}
The value of the deceleration parameter is thus, according to~\req{qparam}, an indicator of the transition between different eras of the Universe's history: radiation dominated, matter dominated and dark energy dominated with the Universe switching to accelerating expansion when $q$ changes sign.

To illustrate the power of the era characterization in terms of the acceleration parameter\index{acceleration parameter} we survey its value considering the range of Universe evolution shown in~\rf{fig:today}. The time span covered is in essence the entire lifespan of the Universe, but on a logarithmic time scale there is a lot of room for interesting physics in the tiny blip that happened before neutrino decoupling where on the left the time axis begins. 

\begin{figure}
\centerline{\includegraphics[width=0.75\linewidth]{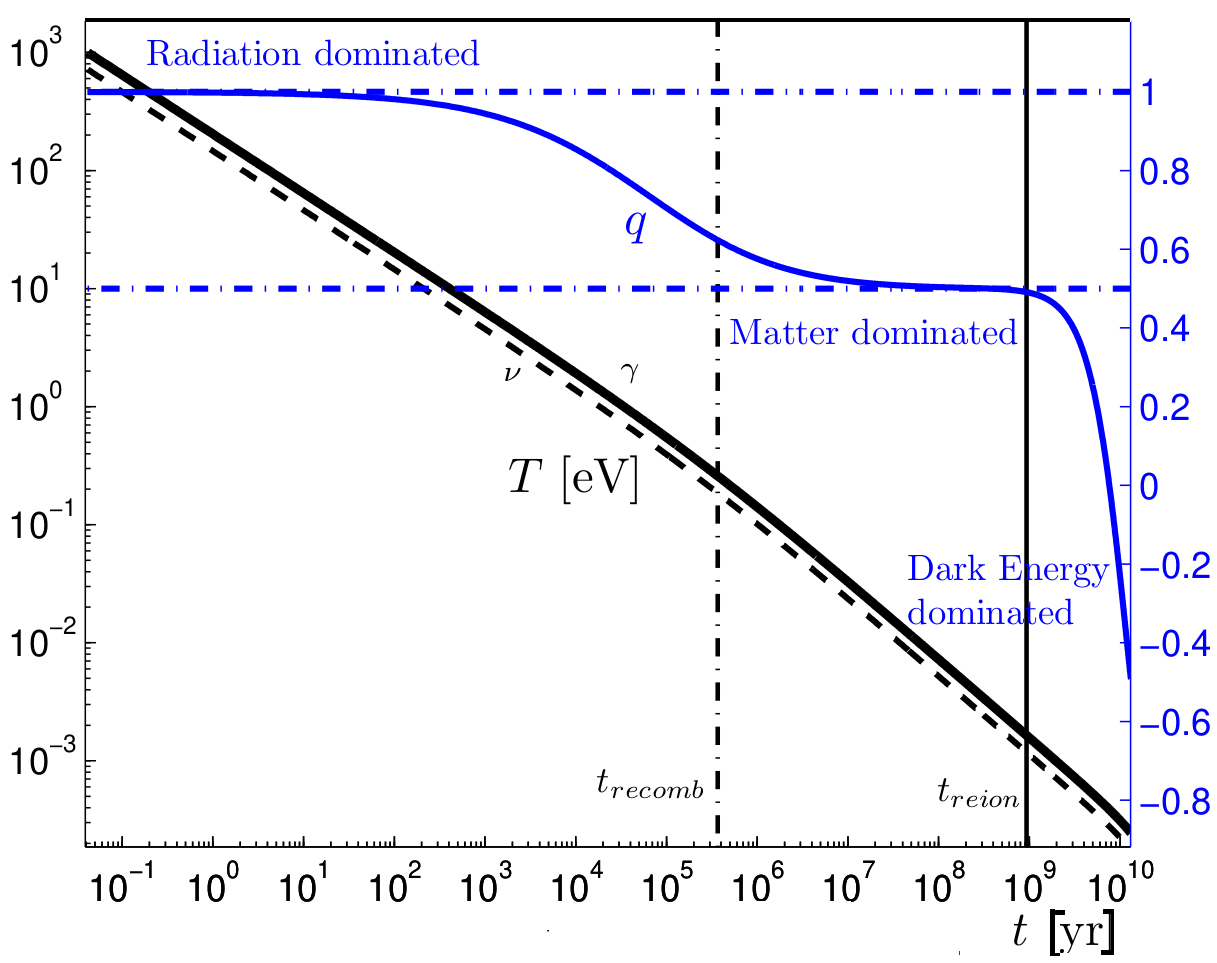}}
\caption{Deceleration parameter (blue lines, right-hand scale) shows transitions in the composition of the Universe as a function of time. The left-hand scale indicates the corresponding temperature $T$, dashed is the lower $T$-value for neutrinos. Vertical lines indicate recombination and reionization conditions. \radapt{Rafelski:2013yka}.
\label{fig:today} }
\end{figure}

On the left-axis in~\rf{fig:today} we see temperature $T$\,[eV] while on right-axis (blue) we see the deceleration parameter $q$. The horizontal dot-dashed lines show the pure radiation-dominated value of $q=1$ and the matter-dominated value of $q=1/2$. The expansion in this era starts off as radiation-dominated. We see relatively long transitions to matter-dominated domain starting around $T=\mathcal{O}(300\eV)$ and ending at $T=\mathcal{O}(10\eV)$. The matter-dominated Universe\index{Universe!matter dominated} begins near recombination and ends right at the edge of reionization. Thereafter begins the transition to a dark energy $\Lambda$-dominated era which is in full swing already at $T=\mathcal{O}(1\eV)$. $q$ changes sign near to $T=\mathcal{O}(200\meV)$. Today $q=-0.5$ indicates we are in the midst of a rapid transition to dark energy $\Lambda$-dominated regime. 

The vertical dot-dashed lines in~\rf{fig:today} show the time of recombination at $T\simeq0.25\eV$, when the Universe became transparent to photons, and reionization at $T\simeq {\cal O}(1\meV)$, when hydrogen in the Universe was again ionized due to light from the first galaxies~\cite{Zaroubi:2012in} is also shown. The usefulness of $q$ to predict the present day value of the Hubble parameter\index{Hubble parameter} is even better appreciated noting that we can easily integrate~\req{eq:Hdot} 
\beqn\label{eq:HdotInt}
H(t)=\frac{H_i}{1+H_i\int_{t_i}^{t}(1+q)dt}=
\frac{H_i}{1+1.5\,H_i\int_{t_i}^{t}(1+P/\rho)dt}\,.
\eeqn
Given an initial (measured) value $H_i$ in an epoch after free electrons disappeared (recombination epoch) the time dependence of $q$ or equivalently, $P/\rho$, see~\rf{fig:today} impacts the current epoch $H(t_0)=H_0$. The Hubble parameter 
$H$\,[s$^{-1}$] (left ordinate, black) and the redshift $z$ (right ordinate, blue) 
\beqn\label{eq:zdef}
z+1\equiv \frac{a_0}{a(t)}\,,
\eeqn
are shown in~\rf{fig:today1} spanning a wide ranging domain following on the domain of interest in this work. 

\begin{figure}
\centerline{\includegraphics[width=0.65\linewidth]{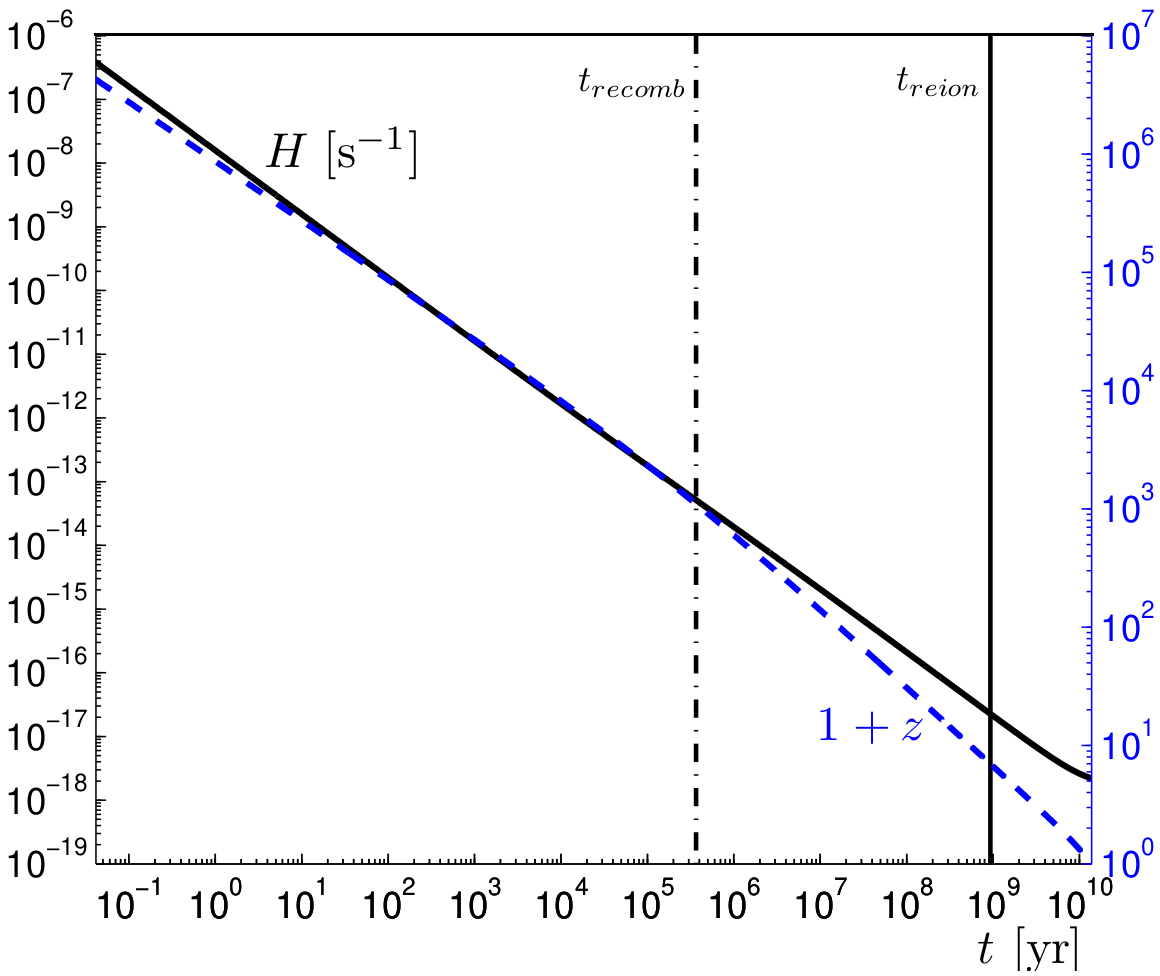}}
\caption{Temporal evolution of the Hubble parameter $H$ (in units 1/s] (left-hand scale) and of redshift $1+z$ (right-hand scale, blue). \radapt{Rafelski:2013yka}. 
\label{fig:today1} }
\end{figure}

There is a visible deviation from a power law behavior in~\rf{fig:today1} due to the transitions from radiation to matter dominated and from matter to dark energy dominated expansion we saw in~\rf{fig:today}. To achieve an increase $H$ in the current epoch beyond what is expected all it takes is to have the value of $q$ a bit more negative, said differently closer to being dark energy dominated altering the balance between matter, radiation (neutrinos, photons) and dark energy.\index{Universe!particle content} We conclude that it is important to understand the particle content of the Universe which we used to construct these results in order to understand the riddle of the Hubble value tension.\index{Hubble parameter!tension}

\para{Relation between time and temperature}
{\color{black} To determine the relation between time and temperature as the Universe evolves we first  consider} comoving entropy conservation\index{entropy!conservation}, 
\begin{align}
S=\sigma V\propto g^s_\ast T^3a^3=\mathrm{constant},
\end{align}
where $g^s_\ast$ is the entropy degree of freedom and $a$ is the scale factor. Differentiating the entropy with respect to time $t$ we obtain
\begin{align}
\left[\frac{\dot{T}}{g^s_\ast}\frac{dg^s_\ast}{dT}+3\frac{\dot{T}}{T}+3\frac{\dot{a}}{a}\right]g^s_\ast T^3a^3=0,\qquad \dot{T}=\frac{dT}{dt}.
\end{align}
The square bracket has to vanish. Solving for $\dot T $ we obtain
\begin{align}
\frac{dT}{dt}=-\frac{HT}{1+\frac{T}{3g^s_\ast}\frac{d\,g^s_\ast}{dT}}\,.
\end{align}
Taking the integral the relation between time and temperature in the primordial Universe is obtained\index{Universe!time-temperature relation}
\begin{align}\label{time}
t(T)=t_0-\int^T_{T_0} \frac{d\,T }{T\,H}\left[1+\frac{T}{3g^s_\ast}\frac{dg^s_\ast}{dT}\right],\qquad H=\sqrt{\frac{8\pi G_N}{3}\rho_{tot}(T)}
\,,
\end{align}
where $T_0$ and $t_0$ represent the initial temperature and time respectively. $H=\dot a/a$ is the Hubble parameter~\req{dynamic} related to the total energy density $\rho_{tot}$ in the Universe by the Hubble equation~\req{Hubble:eq} restated for convenience. The temperature derivative of the entropy degrees of freedom\index{entropy!degrees of freedom}, $g^\ast_s$ seen in~\rf{EntropyDOF:Fig} allows us to obtain a smooth time-temperature relation shown in~\rf{Fig:Overview}. We are using here the particle inventory in the Universe discussed earlier.

\begin{figure}
\centerline{\includegraphics[width=0.85\linewidth]{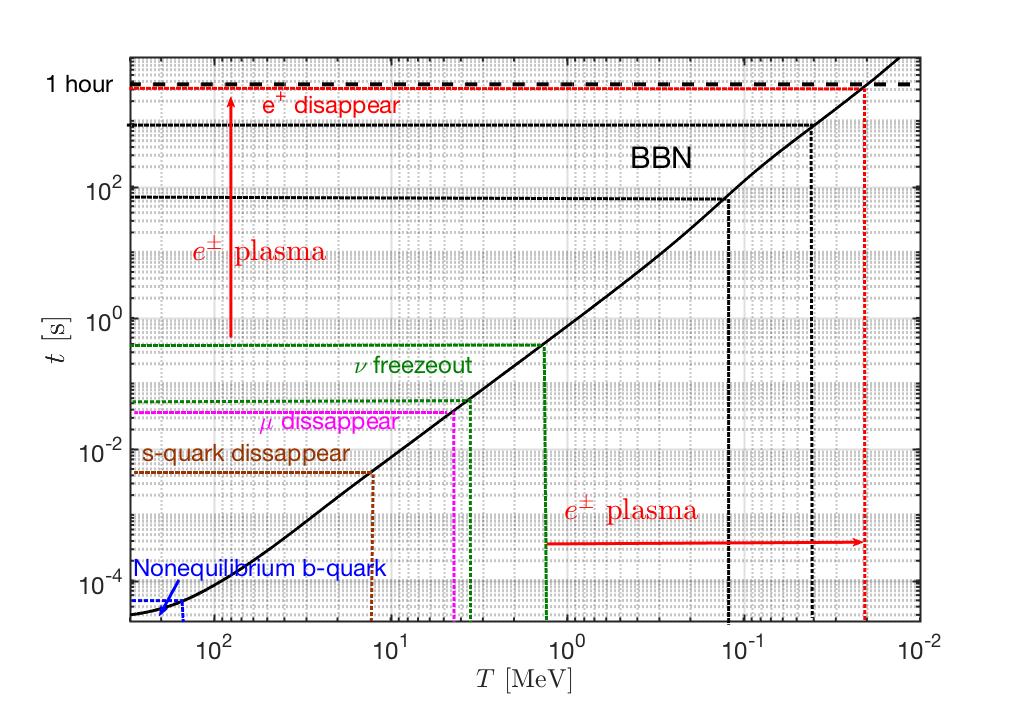}}
 \caption{The relation between time and temperature in the first hour of the Universe beginning shortly before QGP hadronization\index{QGP!hadronization} $300\MeV >T>0.02\MeV $ and ending with antimatter disappearance. Temperature/time range for several epochs is indicated. \radapt{Yang:2024ret}.}
 \label{Fig:Overview}
\end{figure}

In~\rf{Fig:Overview} the black line presents the computed relation between time $t$\,[s] (ordinate, increasing scale) and temperature $T$\,[MeV](abscissa, decreasing scale) during the first hour of the evolution of the Universe, reaching down to the temperature $T=10\keV$. Vertical and horizontal lines indicate some characteristic epochal events related to the Universe particle inventory, as marked. 

In the temperature range we consider in this work, $T>0.02\MeV$, particle-matter-radiation content of the Universe is relevant. There is vanishing dependence on $\Lambda$CDM model. However, in the contemporary Universe, the $\Lambda$CDM model uncertainties related to the lack of understanding of `darkness' and the need to know the pie-chart composition of the Universe at least at one `initial' time compound making in our view the direct measurements of $H_0$ a value that the extrapolations from recombination epoch should aim to resolve, eliminating the Hubble tension\index{Hubble parameter!tension}. Such a current epoch biased fit of data would provide as example the so called effective number of neutrino degrees of freedom that we address further below, see~\rsec{sec:model:ind}.

\para{Neutrinos in the cosmos}
In the primordial Universe the neutrinos are kept in equilibrium with cosmic plasma via the weak interaction processes, which at temperatures below $\mathcal{O}(20)\MeV$ involve predominantly the $e^+e^-$-pair plasma. However, as the Universe expands, these weak interactions gradually became too slow to maintain equilibrium, neutrinos ceased interacting and decouple from the cosmic background as we describe in this report in detail in the temperature range $T=2.5\pm1.5\MeV$.

According to theoretical models we and others have developed, at around $1\MeV$ all cosmic neutrinos have stopped interacting. Neutrinos evolve as free-streaming\index{free-streaming} particles in the Universe responding only to gravitational background they co-create, as individual particles they are unlikely to interact again in the rapidly expanding and diluting Universe. Today they are the relic neutrino background.\index{neutrino!relic background} We recall that photons become free-streaming much later, near to $0.25$\,eV and today they make up the Cosmic Microwave Background (CMB), currently at a temperature\index{CMB} $T_{\gamma,0} =2.726\,\mathrm{K}=0.2349\MeV$.

The relic neutrino background carries important information about our primordial Universe: If we ever achieve relic neutrino experimental observation we will be observing our Universe when it was about $1$ sec old. Since photons were reheated by ensuing electron-positron annihilation, the neutrino relic background should have a lower temperature and we show below $T_\nu^0\simeq 1.95\,\mathrm{K}\simeq 0.168\MeV$ in the present epoch.
The relic neutrinos have not been directly measured, but their impact on the speed of expansion of the Universe is imprinted on the CMB. Indirect measurements of the relic neutrino background, such as by the Planck satellite~\cite{Planck:2018vyg,Planck:2015fie,Planck:2013pxb}, constrain to some degree in model dependent analysis the neutrino properties such as number of massless degrees of freedom and a bound on mass.

We know that the neutrinos are not massless particles and we return to discuss how this insight was gained. Their square mass difference $\Delta m^2_{ij}$ has been\index{neutrino!mass} determined~\cite{ParticleDataGroup:2022pth}:
\begin{align}
&\Delta{m}_{21}^2=73.9\pm 2\MeV ^2,\\
&\Delta{m}_{32}^2=2450\pm 30\MeV ^2\,.
\end{align}
Thus neutrino mass\index{neutrino!mass} values can be ordered in the normal mass hierarchy ($m_1\ll m_2<m_3$) or inverted mass hierarchy ($m_3\ll m_1<m_2$). 

All three mass states remained relativistic until the temperature dropped below their rest mass. Today one of the neutrinos could be still relativistic. We will return in~\rsec{ch:nu:today} to discuss the relic massive neutrino flux in the Universe.
 
We will study the neutrino freeze-out\index{neutrino!freeze-out} temperature in the context of the kinetic Boltzmann-Einstein equation\index{Boltzmann-Einstein equation} for the three flavors, and refine the results by noting that there are three different freeze-out processes for neutrinos:
\begin{enumerate}
\item Neutrino chemical freeze-out: the temperature at which neutrino number changing processes such as $e^-e^+\to\nu\overline\nu$ effectively cease. After chemical freeze-out, there are no reactions that, in a noteworthy fashion, can change the neutrino abundance and so particle number is conserved.
\item Neutrino kinetic freeze-out: the temperature at which the neutrino momentum exchanging interactions such as $e^\pm\nu\to e^\pm\nu$ are no longer occurring rapidly enough to maintain an equilibrium momentum distribution. 
\item Collisions between neutrinos $\nu\nu\to\nu\nu$ are capable of re-equilibrating energy within and between neutrino flavor families. These processes end at a yet lower temperature and the neutrinos will be free-streaming from that point on.
\end{enumerate}

To obtain the freeze-out temperature $T=\mathcal{O}(2.5\pm1.5\mathrm{MeV})$, we solve the Boltzmann-Einstein equation including all required collision terms. To be able to do this
we developed a new method for analytically simplifying the collision integrals and showing that the neutrino freeze-out temperature is controlled by one fundamental coupling constants and particle masses. We give further discussion of these methods in~\rsec{ch:param:studies}. The required mathematical theory and numerical method is developed in Appendices~\ref{ch:vol:forms}, \ref{ch:boltz:orthopoly}, and \ref{ch:coll:simp}. Our report follows the comprehensive investigation of neutrino freeze-out found also in Jeremiah Birrell PhD thesis~\cite{Birrell:2014ona}.\index{neutrino!freeze-out}

The freeze-out temperature we obtain depends only on the magnitude of the symmetry breaking Weinberg angle $\sin^2(\theta_W)$, and a dimensionless relative interaction strength parameter $\eta$,
\begin{align}\label{etaCTY}
\eta\equiv M_p m_e^3 G_F^2, \qquad M_p\equiv\sqrt{\frac{1}{8\pi G_N}}\,, 
\end{align}
a combination of the electron mass $m_e$, Newton constant $G_N$ (expressed above in terms of Planck mass $M_p$,~\req{eq:GN}), and the Fermi constant $G_F$. These dimensionless strength parameters in the present-day vacuum state have the following values
\begin{align}\label{eta0CTY}
\eta_0\equiv \left.M_p m_e^3 G_F^2\right|_0 = 0.04421\,, \qquad \sin^2(\theta_W)=0.2312\,.
\end{align}

The magnitude of neither $\eta$ nor of the Weinberg angle is fixed by known phenomena. Therefore both the interaction strength $\eta$ and $\sin^2(\theta_W)$ could be subject to variation as a function of time or temperature. Therefore it is of interest to study the neutrino freeze-out as function of these parameters. The dependence of neutrino freeze-out temperatures on $\eta$ is shown in~\rf{fig:freeze-outT_eta} and the dependence on the Weinberg angle is shown in~\rf{fig:freeze-outT_B}. The present day vacuum value of Weinberg angle puts the $\nu_\mu,\nu_\tau$ freeze-out temperature, seen in the right-hand frame of \rf{fig:freeze-outT_B}, near its maximum value.
 
\begin{figure}
\centerline{\includegraphics[width=0.49\linewidth]{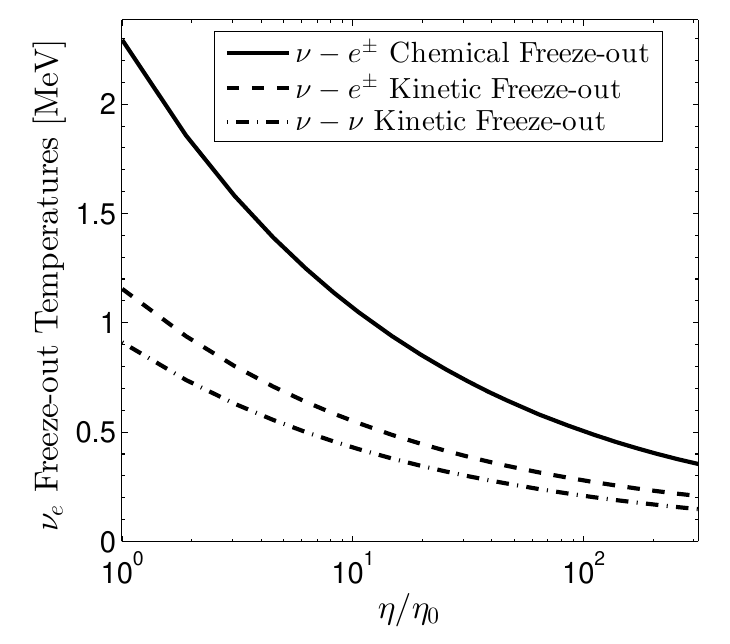} \includegraphics[width=0.49\linewidth]{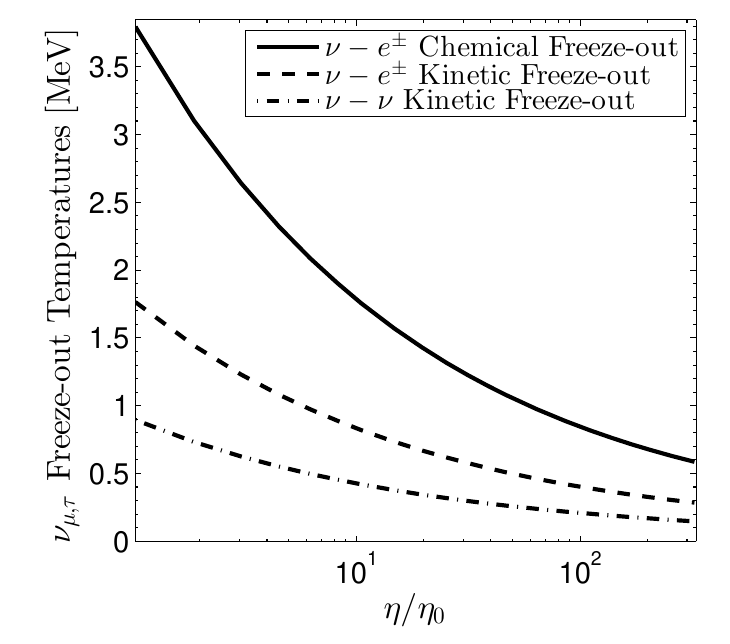}}
\caption{Freeze-out temperatures for electron neutrinos (left-hand frame) and $\mu$, $\tau$ neutrinos (right-hand frame) for the three types of processes, see insert, as functions of interaction strength $\eta>\eta_0$. \cccite{Birrell:2014uka}}
\label{fig:freeze-outT_eta}
 \end{figure}

\begin{figure}
\centerline{\includegraphics[width=0.49\linewidth]{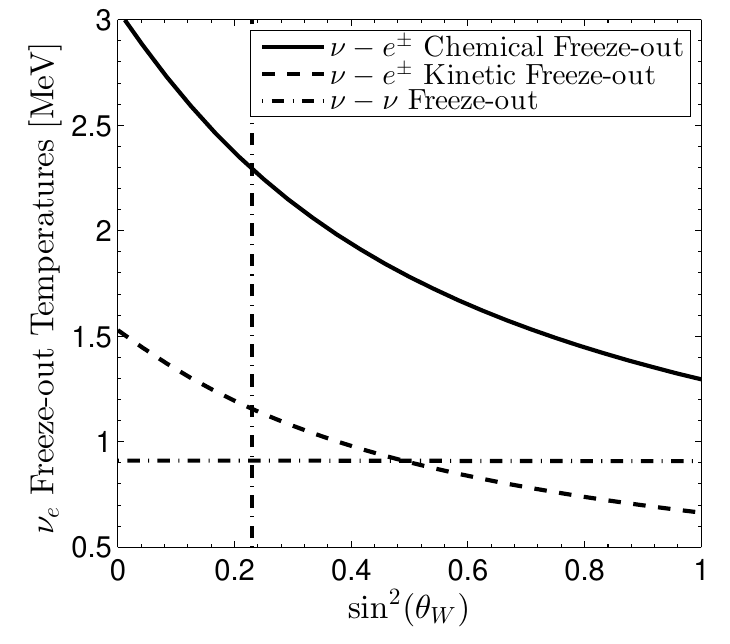}
\includegraphics[width=0.49\linewidth]{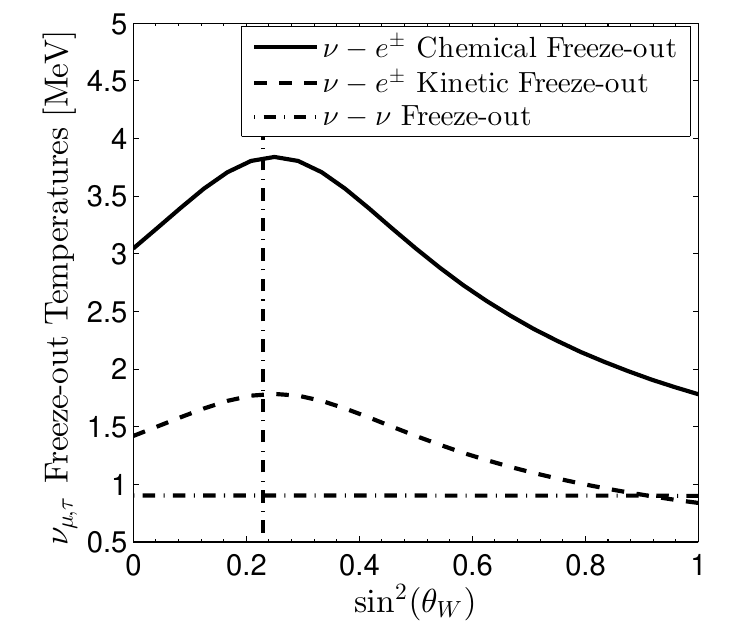}}
\caption{Freeze-out temperatures for electron neutrinos (left-hand frame) and $\mu$, $\tau$ neutrinos (right-hand frame) for three types of processes, see insert, as functions of the value of the Weinberg angle $\sin^2(\theta_W)$. Vertical line is at present epoch $\sin^2(\theta_W)=0.23$. \cccite{Birrell:2014uka}}
\label{fig:freeze-outT_B}
 \end{figure}

We do not explore here the pivotal insight that neutrinos in elementary processes are not produced in mass eigenstates but in flavor eigenstates. Due to the difference in the three neutrino masses the propagating flavor eigenstates contain three coherent amplitudes moving at different velocity. This leads to the experimentally observed oscillation of neutrino flavor as function of travel distance.\index{neutrino!flavor oscillation} This is also how the constraints on neutrino masses shown above were obtained.

How does this neutrino mixing impact neutrino freeze-out? We inspect our results to understand the hierarchy of freeze-out: Near to freeze-out temperature the electron-neutrino can still `annihilate' on electrons while the absence of muons and taus in the cosmic plasma at a temperature of a few MeV makes these two neutrino flavors less interactive and their freeze-out temperature is higher. Oscillation thus provides a mechanism in which the heavier flavors remain reactive in matter as they share in the more interactive electron-neutrino component. Conversely, electron-neutrino interaction is weakened since only a part of this flavor wave remains available to interact. The net effect was found negligible in the work of Mangano et. al.~\cite{Mangano:2005cc}. The impact of the cosmic magnetic field on neutrino oscillation is an avenue for further research~\cite{Rafelski:2023zgp}.

In regard to our results one can say that the differences in freeze-out between the three different flavors diminishes allowing for oscillations. We chose not to quantify this effect as the mixing of neutrino mass eigenstates into flavor eigenstates and neutrino masses remain a vibrant research field. Without knowing all the required input parameters the outcome is uncertain. Given the results we obtained and methods we developed we will be able, once the neutrino mixing and masses are well understood, to update our results. 

A discussion of the implications and connections of the results on neutrino freeze-out to other areas of physics, including BBN\index{Big-Bang!BBN} and dark radiation is described in more detail in~\cite{Birrell:2014uka,Dreiner:2011fp,Boehm:2012gr,Blennow:2012de}. 

We now characterize the era $30\MeV>T>0.01\MeV$. At the high end muons\index{muon} and pions are nonrelativistic and are disappearing from the Universe, we then pass through neutrino decoupling and the era where $e^+e^-$-pairs become nonrelativistic. In~\rf{fig:BBN} the black line refers to the left ordinate and shows the temperature as a function of time, dashed the lower value of $T$ for free-streaming neutrinos. We further indicate in~\rf{fig:BBN} the domain of Big-Bang Nucleosynthesis (BBN)~\cite{Iocco:2008va}\index{Big-Bang!BBN}, the period when the lighter elements were synthesized amidst of a $e^+e^-$-pair plasma, which is already reduced in abundance but not entirely eliminated. This insight will keep us very busy in this report.

\begin{figure}
\centerline{\includegraphics[width=0.60\linewidth]{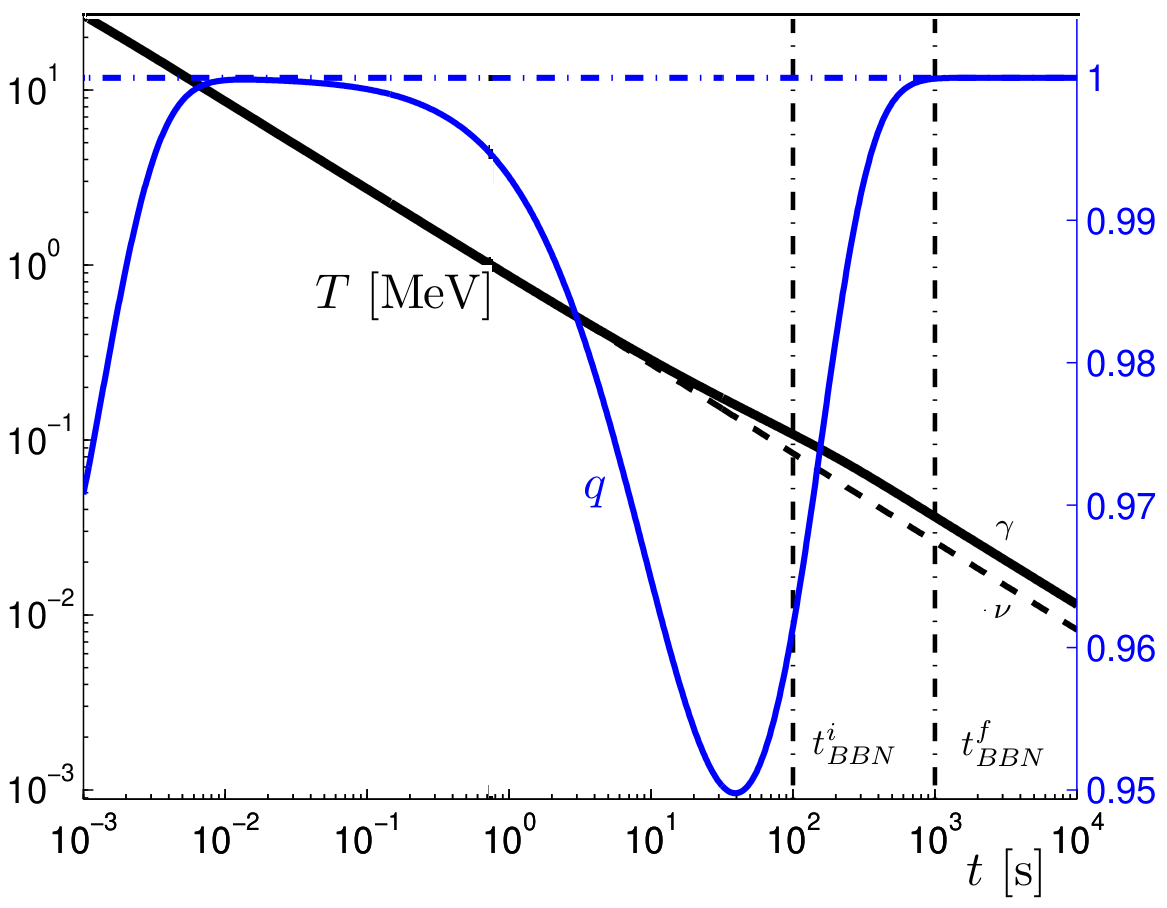}}
\caption{The first hours in the lifespan of the Universe from the end of baryon antimatter annihilation through BBN: Deceleration parameter\index{cosmology!deceleration parameter} $q$ (blue line, right-hand scale) shows impact of emerging antimatter components; at millisecond scale anti-baryonic matter and at 35\,sec. scale positronic nonrelativistic matter appears. The left-hand scale shows photon $\gamma$ temperature $T$ in\,eV, dashed is the emerging lower value for neutrino $\nu$ which are not reheated by $e^+e^-$ annihilation. Vertical lines bracket the BBN domain. \cccite{Birrell:2014uka}. \radapt{Rafelski:2013yka}} 
\label{fig:BBN}
\end{figure}

The blue lines in~\rf{fig:BBN} refer to the right ordinate: The horizontal dot-dashed line for $q=1$ shows the pure radiation dominated value with two exceptions. In~\rf{fig:BBN} the unit of time is seconds and the range spans the domain from fractions of a millisecond to a few hours.

The just noted presence of massive pions and muons reduces the value of $q$ towards matter dominated near to the maximal temperature shown. Second, when the temperature is near the value of the electron mass, the $e^+e^-$-pairs are not yet fully depleted but already sufficiently nonrelativistic to cause another dip in $q$ towards matter-dominated value. These dips in $q$ are not large; the Universe is still predominantly radiation-dominated.\index{acceleration parameter} But $q$ provides a sensitive measure of when various mass scales become relevant and is therefore a good indicator for the presence of a reheating period, where some particle population disappears and passes its entropy to the thermal background.

\para{Reheating history of the Universe}
At times where dimensional scales are irrelevant, entropy conservation\index{entropy!conservation} means that temperature scales inversely with the scale factor $a(t)$. This follows from the only contributing scale being $T$ and therefore by dimensional counting $ \rho\simeq 3P \propto T^4$. However, as the temperature drops and at their respective $m\simeq T$ scales, successively less massive particles annihilate and disappear from the thermal Universe. Their entropy reheats the other degrees of freedom and thus in the process, the entropy originating in a massive degree of freedom is shifted into the effectively massless degrees of freedom that still remain. 

This causes the $T\propto 1/a(t)$ scaling to break down; during each of these `reorganization' periods the drop in temperature is slowed by the concentration of entropy in fewer degrees of freedom, leading to a change in the reheating ratio, $R$, defined as\index{reheating}
\begin{equation}\label{redshiftratio}
R\equiv \frac{1+z}{ T_\gamma/T_{\gamma,0}}, \qquad 1+z\equiv \frac{a_{0}}{a(t)}.
\end{equation}
The reheating ratio connects the photon temperature redshift to the geometric redshift, where $a_0$ is the scale factor today (often normalized to $1$) and quantifies the deviation from the scaling relation between $a(t)$ and $T$. There is additional Universe expansion due to reheating of remaining degrees of freedom so that the total entropy is conserved as entropy in particles decreases. This is Universe reheating inflation.\index{Universe!reheating inflation}

The change in $R$ can be computed by the drop in the number of degrees of freedom and we learn from this actual redshift $1+z$. For the just discussed era $30\MeV>T>0.01\MeV$, we show in~\rf{fig:BBN1} in blue the value of $1+z$ as a function of time and in black (left ordinate) the value of $H$[s$^{-1}$]. It is interesting to observe that study of BBN\index{BBN!redshift} extends the range of redshift explored to $10^8<1+z_\mathrm{BBN}<10^9$.

\begin{figure}
\centerline{\includegraphics[width=0.60\linewidth]{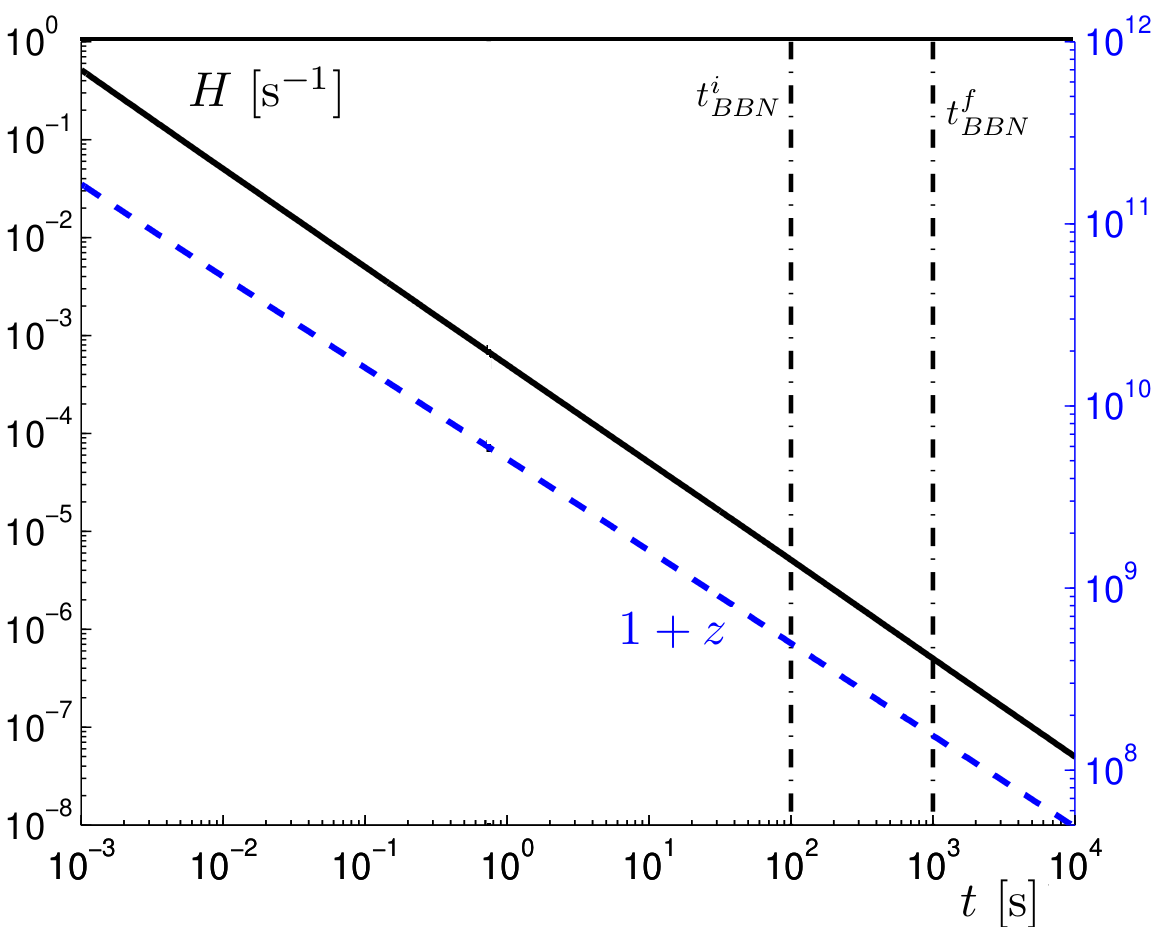}} 
\caption{First hours in the evolution of the Universe: Hubble parameter $H$ in units [1/s] (left-hand scale) and the redshift $1+z$ (right-hand scale, blue) spanning the epoch from well below the end of baryon antimatter annihilation through BBN, compare~\rf{fig:BBN}.\index{reheating!inflation} \radapt{Rafelski:2013yka}. \cccite{Birrell:2014uka}} \label{fig:BBN1}
\end{figure}

We are interested in determining by how much the Universe inflated in addition to its expected expansion in follow-up on particle disappearance from inventory. We begin at the highest temperature to count the particle degrees of freedom: At a temperature on the order of the top quark mass, when all standard model particles were in thermal equilibrium, the Universe was pushed apart by 28 bosonic and 90 fermionic degrees of freedom. The total number of degrees of freedom can be computed as discussed below. 

For bosons we have the following: the doublet of charged Higgs\index{Higgs} particles has $4=2\times2=1+3$ degrees of freedom -- three will migrate to the longitudinal components of $W^\pm, Z$ when the electro-weak vacuum freezes and the EW symmetry breaking arises\index{symmetry!breaking}, while one is retained in the one single dynamical charge-neutral Higgs particle component. In the massless stage, the SU(2)$\times$U(1) theory has 4$\times$2=8 gauge degrees of freedom where the first coefficient is the number of particles $(\gamma, Z, W^\pm)$ and each massless gauge boson has two transverse conditions of polarization. Adding in $8_c\times2_s=16$ gluonic degrees of freedom we obtain 4+8+16=28 bosonic degrees of freedom. 

The count of fermionic degrees of freedom includes three $f$ families, two spins $s$, another factor two for particle-antiparticle duality. We have in each family of flavors a doublet of $2\times 3_c$ quarks, 1-lepton and 1/2 neutrinos (due left-handedness which was not implemented counting spin). Thus we find that a total $3_f\times 2_p\times 2_s\times(2\times 3_c+1_l+1/2_\nu)=90$ fermionic degrees of freedom. We further recall that massless fermions contribute 7/8 of that of bosons in both pressure and energy density. Thus the total number of massless Standard Model particles at a temperature above the top quark mass scale, referring by convention to bosonic degrees of freedom, is $g_{\rm SM}=28+90\times 7/8=106.75$\,.

In~\rf{fig:dof} we show the reheating ratio $R$ \index{Universe!reheating inflation}~\req{redshiftratio} as a function of time beginning in the primordial elementary particle Universe epoch on the left, connecting to the present epoch on the right. The periods of change seen in~\rf{fig:dof} come when the evolution temperature crosses the mass of a particle species that is in equilibrium. One can see drops corresponding to the disappearance of thermal particle yields as indicated. After $e^+e^-$ annihilation on the right, there are no significant degrees of freedom remaining to annihilate and feed entropy into photons, and so $R$ remains constant until today. We do not model in detail the QGP phase transition and hadronization\index{QGP!hadronization} period near $T\simeq O(150\,\MeV), t\simeq 20\,\mu\mathrm{s}$ covering-up the resultant kinky connection. A more precise model using lattice QCD, see \eg~\cite{Borsanyi:2013bia}, together with a high temperature perturbative QCD expansion, see \eg~\cite{Letessier:2002ony}, can be considered. These complex details do not impact this study and so we do not consider these issues further here.

\begin{figure}
{\includegraphics[width=\linewidth]{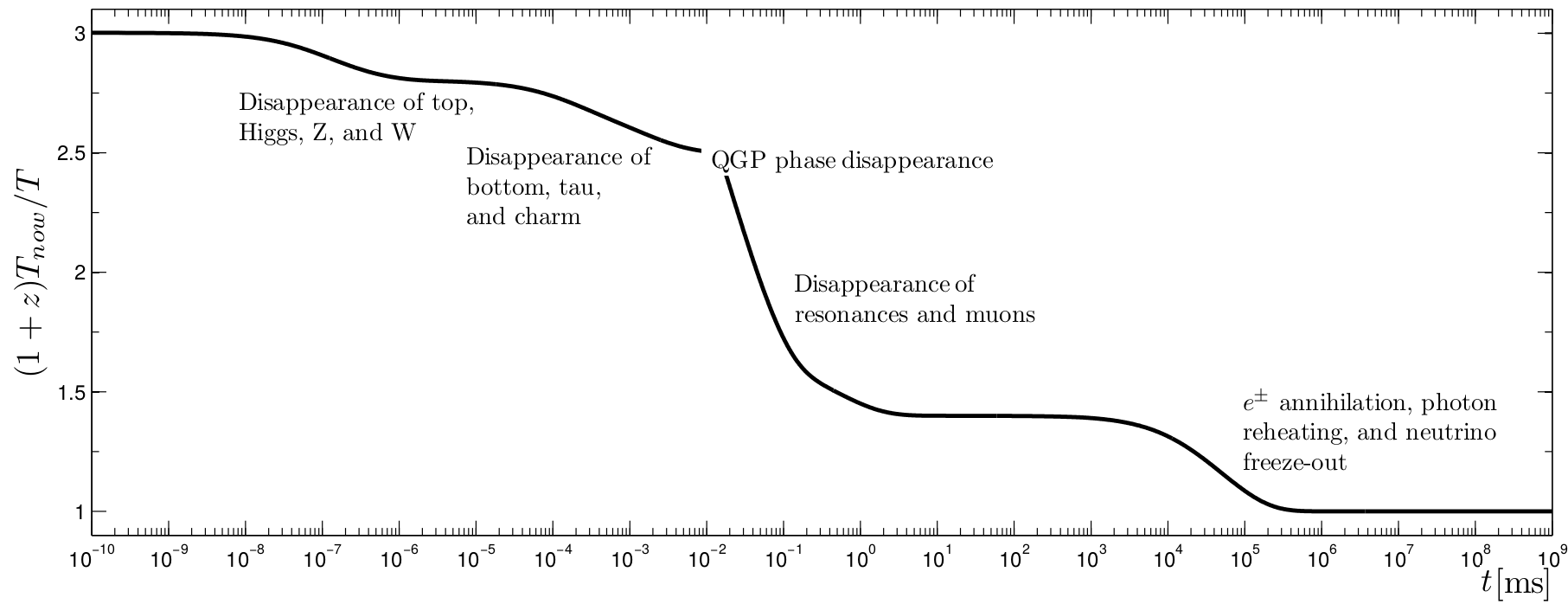}}
\caption{Universe inflation due to the disappearance of degrees of freedom as a function of time $t$\,[ms] (milliseconds). The Universe volume inflated by approximately a factor of 27 above the naive thermal redshift scale as massive particles disappeared successively from the inventory while entropy remained conserved. \radapt{Birrell:2014ona}}\label{fig:dof}
\end{figure}

As long as the microscopic local dynamics are at least approximately entropy conserving, the total drop in $R$ is entirely determined by the global entropy conservation governing expansion of the Universe based on FLRW\index{cosmology!FLRW} cosmology. Namely, the magnitude of the drop in $R$ seen in~\rf{fig:dof} is a measure of the number of degrees of freedom that have disappeared from the Universe. Consider two times $t_1$ and $t_2$ at which all particle species that have not yet annihilated are effectively massless. By conservation of comoving entropy and scaling $T\propto 1/a$ we have
\begin{equation}\label{r_ratio}
1=\frac{a_1^3S_{1}}{a_2^3 S_2}=\frac{a_1^3\sum_ig_i T_{i,1}^3}{a_2^3\sum_j g_j T_{j,2}^3},\qquad \left(\frac{R_1}{R_2}\right)^3=\frac{\sum_ig_i (T_{i,1}/T_{\gamma,1})^3}{\sum_j g_j (T_{j,2}/T_{\gamma,2})^3}
\,,
\end{equation}
where the sums are over the total number of degrees of freedom present at the indicated time and the degeneracy factors $g_i$ contain the $7/8$ factor for fermions. In the second form we divided the numerator and denominator by $a_{0}T_{\gamma,0}$. 

We distinguish between the temperature of each particle species and our reference temperature, the photon temperature. This is important since today neutrinos are colder than photons, due to photon reheating from $e^+e^-$ annihilation occurring after neutrinos decoupled (this is only an approximation, a point we will study in detail in subsequent chapters). By conservation of entropy one obtains the neutrino to photon temperature ratio of
\begin{equation}\label{T_nu_T_gamma}
T_\nu/T_\gamma=({4}/{11})^{1/3}.
\end{equation}
We will call this the reheating ratio in the decoupled limit. 

We now compute the total drop in $R$ shown in~\rf{fig:dof}. At $T=T_\gamma=\mathcal{O}(130\GeV)$ the number of active degrees of freedom is slightly below $g_{\rm SM}=106.75$ due to the partial disappearance of top quarks $t$ which have mass $174\,\GeV$, but this approximation will be good enough for our purposes. At this primordial time, all the species are in thermal equilibrium with photons.

Today we have $2$ photon and $7/8\times 6$ neutrino degrees of freedom and a neutrino-to-photon temperature ratio~\req{T_nu_T_gamma}. Therefore, for the overall reheating ratio since the primordial elementary particle Universe epoch we have\index{Universe!reheating}
\begin{equation}
\left(\frac{R_{100GeV}}{R_{now}}\right)^3= \frac{g_{SM}}{g_{\rm now}}=\frac{106.75}{2+\frac{7}{8}\times 6\times \frac{4}{11}}\approx 27.3
\end{equation}
which is the fractional change we see in~\rf{fig:dof}. The meaning of this factor is that the Universe approximately inflated by a factor 27 above the thermal redshift scale as massive particles disappeared successively from the inventory. 

Another view of the reheating is implicit in our presentation of particle energy inventory in~\rf{fig:energy:frac}. There the initial highest temperature is on the right at the end of the hadron era marked by the disappearance of muons and pions and other heavier particles as marked. This constitutes a major reheating period, with energy and entropy from these particles being transferred to the remaining $e^+e^-$, photon, neutrino plasma. Continuing to $T=\mathcal{O}(1)\MeV$, we come to the annihilation of $e^+e^-$ and the photon reheating period. Notice that only the photon energy density fraction increases here. As discussed above, a common simplifying assumption is that neutrinos are already decoupled at this time and hence do not share in the reheating process, leading to a difference in photon and neutrino temperatures~\req{T_nu_T_gamma}.

After passing through a long period, from $T=\mathcal{O}(1)\MeV$ until $T=\mathcal{O}(1)$\,eV, where the energy density is dominated by photons and free-streaming neutrinos, we then come to the beginning of the matter dominated regime, where the energy density is dominated by dark matter and baryonic\index{baryon} matter. This transition is the result of the redshifting of the photon and neutrino de Broglie wavelength and hence particle energy, for relativistic particles $\rho\propto T^4$, whereas for nonrelativistic matter $\rho\propto a^{-3}\propto T^3$. Note that our inclusion of neutrino mass\index{neutrino!mass} causes the leveling out of the neutrino energy density fraction during this period, as compared to the continued redshifting of the photon energy.

Finally, as we move towards the present day CMB\index{CMB} temperature of $T_{\gamma,0}=0.235$ meV on the left-hand side, we have entered the dark energy dominated regime. For the present day values, we have used the fits from the Planck data~\cite{Planck:2018vyg,Planck:2015fie,Planck:2013pxb} of $69\%$ dark energy, $26\%$ dark matter and $5\%$ baryons (and zero spatial curvature). The photon energy density is fixed by the CMB temperature $T_{\gamma,0}$ and the neutrino energy density is fixed by $T_{\gamma,0}$ along with the photon to neutrino temperature ratio. Both constitute $<1\%$ of the current energy budget in the pie chart of the Universe.

\para{The baryon-per-entropy density ratio}
An important\index{baryon!entropy ratio}
result of the FLRW cosmology is that following on the era of matter genesis both baryon and entropy content is conserved in the comoving volume, that is the volume where length scales account for the Universe $a(t)$ expansion scale parameter. Therefore the ratio of baryon number density to visible matter entropy density remains constant throughout the evolution of the thermally equilibrated Universe. 

Baryonic dust floating in the Universe dilutes due to volume growth with the $a(t)^3$ factor. The entropy described using the entropic degrees of freedom $g^\ast_s$ seen in~\rf{EntropyDOF:Fig} scales overall with the third power of Temperature and thus with the third power of the same expansion parameter, $a(t)^3$. During the short epochs when mass matters, scattering allows the disappearing massive particles to transfer their entropy to the remaining thermal background such that the scale parameter $a(t)$ inflates in each period of reheating, see prior discussion. 

{\color{black} By these considerations,} we have\index{baryon!entropy ratio}
\begin{align}
\frac{n_B-n_{\overline{B}}}{\sigma}= \left.\frac{n_B-n_{\overline{B}}}{ \sigma}\right|_{t_0}=\mathrm{Const.}\;
\end{align}
The subscript $t_0$ denotes the present day condition, and $\sigma$ is the total entropy density.
The observation gives the present baryon-to-photon\index{baryon!per photon ratio} ratio~\cite{ParticleDataGroup:2022pth} $5.8 \times 10^{-10} \leqslant(n_B-n_{\overline{B}})/n_\gamma\leqslant6.5\times10^{-10}$. This small value quantifies the matter-antimatter asymmetry in the present day Universe, and allows the determination of the present value of baryon per entropy ratio~\cite{Rafelski:2019twp,Fromerth:2002wb,Fromerth:2012fe}:
\begin{align}\label{BaryonEntropyRatio}
\left.\frac{n_B-n_{\overline{B}}}{ \sigma}\right|_{t_0}=\eta_\gamma\left(\frac{n_\gamma}{\sigma_\gamma+\sigma_\nu}\right)_{\!t_0}\!\!\!\!=(8.69\pm0.05)\!\!\times\!\!10^{-11},\qquad \eta_\gamma=\frac{(n_B-n_{\overline{B}})}{n_\gamma},
\end{align}
where the value $\eta_\gamma=(6.14\pm0.02)\times10^{-10}$~\cite{ParticleDataGroup:2022pth} is used in calculation. 

To obtain the above ratio, we have considered the Universe today to be containing photons and free-streaming massless neutrinos~\cite{Birrell:2012gg}, and $\sigma_\gamma$ and $\sigma_\nu$ are the entropy densities for photon and neutrino respectively. We have
\begin{align}
 \frac{\sigma_\nu}{\sigma_\gamma}=\frac{7}{8}\,\frac{g_\nu}{g_\gamma}\left(\frac{T_\nu}{T_\gamma}\right)^3\,,\qquad\frac{T_\nu}{T_\gamma}=\left(\frac{4}{11}\right)^{1/3}
 \,,
\end{align}
and the entropy-per-particle\index{entropy!per particle} for massless bosons and fermions are given by~\cite{Fromerth:2012fe}
\begin{align}
\sigma/n|_\mathrm{boson}\approx 3.60\,,\qquad
\sigma/n|_\mathrm{fermion}\approx 4.20\,.
\end{align}
The evaluation of entropy of free-streaming fluid in terms of effectively massless $m\,a_f/a(t)$ free-streaming particles (neutrinos) needs further consideration, as does the free-streaming particles entropy definition. We will return to consider these very important questions in the near future.
 
\section{Quark and Hadron Universe}\label{part2}
\subsection{Heavy particles in the QGP epoch}
\label{HiggsQGP}
\para{Matter phases in extreme conditions}
This section will be focused on a few examples of interest to cosmological context. In the temperature domain below electro-weak (EW) boundary near $T=130$\,GeV we explore in preliminary fashion novel and interesting physical processes. We will consider the Higgs\index{Higgs}, meson, and the heavy quarks $t,b,c$ with emphasis on bottom quarks\index{bottom quark}. We will show that the bottom quarks can deviate from chemical equilibrium\index{chemical equilibrium} $\Upsilon\neq 1$ by breaking the detailed balance between production and decay reactions. It is easy to see considering temperature scaling and additional degrees of freedom that the energy density of matter near to EW phase transition is a stunning 12 orders of magnitude greater compared to the benchmark we discussed for QGP-hadronization\index{hadrons!hadronization}, see~\req{endensval}.
 
The dynamical bottom $ b,\bar b$-quark pair abundance depends on the competition between the strong interaction two gluon fusion process into $b\bar b$-pair\index{quark!bottom} and weak interaction decay rate of these heavy quarks. This leads to the off-equilibrium phenomenon of the bottom quark freeze-out in abundance near the hadronization temperature as discussed in Ref.\,\cite{Yang:2020nne} and below. Here we further argue that the same unusual situation could exist for any other heavy particle in QGP at a temperature well below their mass scale $m_h\simeq 125\GeV$. {\color{black}Note that the `official' abbreviation for the Higgs-particle is $H^0$ or subscript `H'. This conflicts with: a) The Hubble parameter in current epoch $H_0$ and with: b) The hadron gas subscript which is either `HG' or `H' in literature. Therefore in this report the Higgs-particle is abbreviated using the (subscript) symbol $h$ which has a less conflicted meaning.} We study as an example the abundance of the Higgs-particle\index{Higgs-particle} at condition $m_h\gg T$. Higgs is a particularly interesting case due to its special position in the particle ZOO and a narrow width.

We also explore the properties of hadronic phase after hadronization with special emphasis on gaining an understanding about the strangeness $s,\bar s$ content of the Universe which persists to unexpectedly low temperature. Many of the methods we use in this context were developed in order to understand the properties of strongly interacting QGP formed in relativistic \ie\ high-energy heavy-ion\index{heavy-ion!collisions} \ie\ nuclear collision experiments. Such experimental program is in progress at the Relativistic heavy-ion Collider (RHIC) at BNL-New York\index{BNL!RHIC} and the Large Hadron Collider (LHC) at CERN\index{CERN!LHC}. 

Let us remind the reasons why the dynamics of particles and plasma in the primordial Universe differs greatly from the laboratory environment. We focus here on the case of QGP-hadron phase boundary but a similar list applies to other era boundaries:\index{Big-Bang!micro-bang comparison}
\begin{enumerate} 
\item The primordial Big-Bang QGP epoch lasts for about $20\,\mu$s. On the other hand, the QGP formed in collision micro-bangs has a lifespan of around $10^{-23}$\,s. 
\item In the primordial Universe the microscopic transformation of quarks into hadrons proceeded through creation of the so called mixed phase allowing for local equilibration and a full relaxation of strongly interacting degrees of freedom during about $10\,\mu$s~\cite{Fromerth:2002wb}. Contemporary lattice QCD model simulations predict a smooth transformation as well. The transformation in the laboratory is much closer to what can be called explosive and sudden conversion of quarks into hadronic (confined) degrees of freedom~\cite{Rafelski:2000by}. Such a situation can mimic phenomena usually observed in a true phase transition of first order.
\item Half of the degrees of freedom present in the Universe (charged leptons, photons, neutrinos) are not part of the thermal laboratory micro-bang.
\item Experimental reach today is at and below $T\simeq 0.5$\,GeV allowing us to explore the hadronization process of the QGP but not the heavy particle (h,W,Z,$t$) content, $b$ and $c$ quarks are difficult to study. 
\item Though the baryon content of the laboratory QGP is very low it is probably also much higher compared to the observed baryon asymmetry\index{baryon!asymmetry} in the Universe. 
\end{enumerate}
 
\para{Higgs equilibrium abundance in QGP} 
We would like to show that it is of interest to study the Higgs\index{Higgs} particle dynamics at a relatively late stage of Universe evolution. This is an ongoing project which is described here for the first time. We are now considering in the primordial Universe the temperature range $10\,\mathrm{GeV}>T>1\,\mathrm{GeV}$, and recall the mass of the Higgs-particle $m_h\simeq 125\GeV$. Therefore the number density of the Higgs can be written using the relativistic Boltzmann\index{Boltzmann!approximation} approximation
\begin{align}\label{DensityH}
n_{h}=\frac{\Upsilon_h}{2\pi^2}T^3W(m_h/T)\,\qquad W(m_h/T)=\left(\frac{m_h}{T}\right)^2 K_2(m_h/T)
\end{align}\index{Higgs!particle abundance}
where $K_2$ is the modified Bessel function of integer index '$2$'.

We are interested in comparing the abundance of the Higgs-particle to the net abundance of baryon excess over antibaryons\index{baryon!antibaryon} to determine at which temperature the Higgs-particle yield drops below this tiny Universe asymmetry. Our interest derives from the question how far down in temperature a baryon number breaking Higgs-particle decay could be of relevance. Clearly, once the Higgs-particle yield falls far below baryon asymmetry yield it would be difficult to argue that Higgs-particle sourced mechanism can contribute to a significant growth of the baryon asymmetry in the Universe. Moreover, comparing to baryon asymmetry provides in our opinion a good measure of general physical relevance. After all, our present Universe structure derives from this small asymmetry, probably developed in the primordial epoch we explore here. 
 
The density between Higgs and baryon asymmetry (quark-antiquark asymmetry) can be written normalizing to ambient entropy density
\begin{align}
\frac{n_h}{(n_B-n_{\bar{B}})}=\frac{n_{h}}{\sigma_{tot}}\,\left(\frac{\sigma_{tot}}{n_B-n_{\bar{B}}}\right)=
\frac{n_{h}}{\sigma_{tot}}\left[\frac{\sigma_{\gamma,\nu}}{n_B-n_{\bar{B}}}\right]_{t_0}\,.
\end{align}
Assuming there is no `late' baryon genesis and entropy conserving Universe expansion, we introduce in \req{BaryonEntropyRatio} in the last equality the present day value of the baryon per entropy ratio\index{baryon!entropy ratio}. The entropy density\index{entropy!density} $\sigma_{tot}$ in QGP can be obtained employing the entropic degrees of freedom $g^s_\ast$, \req{eq:entg} and~\rf{EntropyDOF:Fig}
\begin{align}
 &\sigma_{tot}=\frac{2\pi^2}{45}g^s_\ast T_\gamma^3,\qquad g^s_\ast=\sum_{i=\mathrm{g},\gamma}g_i\left({\frac{T_i}{T_\gamma}}\right)^3+\frac{7}{8}\sum_{i=l^\pm,\nu,u,d}g_i\left({\frac{T_i}{T_\gamma}}\right)^3\,.
\end{align}
The entropy content at a given temperature $T$, to a very good approximation, is dominated by all effectively massless $m_i<T$ particles. 

{\color{black}The baryon-to-photon density ratio $\eta_\gamma$ allows to quantify the matter-antimatter asymmetry in the Universe. $\eta_\gamma$ was bracketed early on  by the limits $5.8\times10^{-10} \leqslant\eta\leqslant6.5\times10^{-10}$,\index{baryon!per photon ratio} a more precise central 2022 PDG value $\eta_\gamma=(6.14\pm0.02)\times10^{-10}$~\cite{ParticleDataGroup:2022pth} is used in our study.} 

{\color{black}The density ratio between Higgs-particle\index{Higgs-particle} $h$ and baryon asymmetry for the case of chemical equilibrium $\Upsilon_h=1$ is seen in~\rf{HiggsDensity:fig}, solid (red) line. At temperature $T=5.7\GeV$ this ratio is equal to unity. This implies that Higgs-particle baryon number violating decay processes could populate and influence the baryon asymmetry down to this relatively low temperature scale, and even at lower values, due to recurrent production processes.} 

\begin{figure}
\centerline{\includegraphics[width=0.85\linewidth]{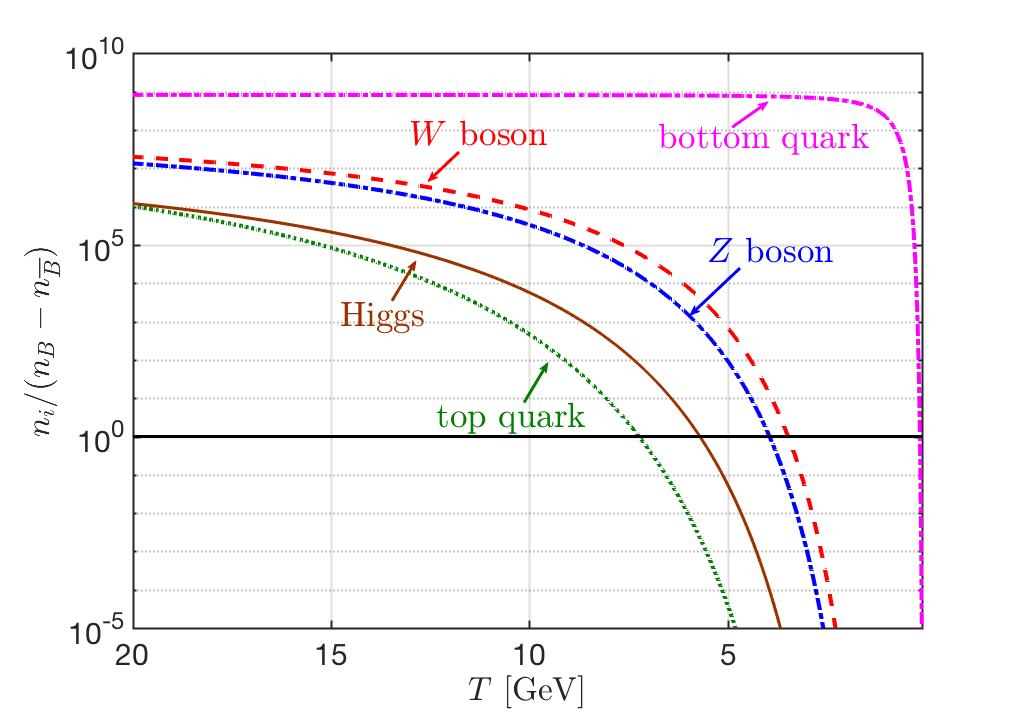}}
\caption{{\color{black} The ratio between thermal equilibrium density of heavy particles $t,h, \mathrm{Z,W}, b$ with the baryon excess density $n_B-n_{\bar B}$ as a function of temperature $20\GeV>T>0.1\GeV$. \radapt{Yang:2024ret}
}}
\label{HiggsDensity:fig} 
\end{figure}

{\color{black}Aside of the Higgs-particle we also see in \rf{HiggsDensity:fig} all other heavy particle thermal abundances normalized to baryon excess: the top quark with $m_t=172.7\GeV$ (dotted green), gauge bosons $m_\mathrm{W}=80.4\GeV$ (dashed red), and $m_\mathrm{Z}=91.19\GeV$ (dot-dashed blue). The relatively light $m_b=4.18\GeV$ bottom quark (dashed-dotted violet) reaches relatively quickly the asymptotic value of the ratio, $10^{9}$. We see that the these respective relative ratios reach unity $n_i/(n_B-n_{\bar{B}})\to 1$ at temperatures: $T_t=7.0$\,GeV , $T_h=5.7$ GeV, $T_\mathrm{Z}=3.9$\,GeV, $T_\mathrm{W}=3.3$\,GeV. The bottom quark remains of interest down to hadronization temperature $T_b\simeq 0.15$\,GeV. We will look at the bottom quark freeze-out in more detail below.}

\para{Baryon asymmetry and Sakhraov conditions}
The small value of the baryon asymmetry\index{Sakharov conditions!baryogenesis} in the Universe could be interpreted as being due to the initial conditions in the Universe. However, in the current standard cosmological model, it is believed that the inflation event can erase any pre-existing asymmetry between baryons and antibaryons. In this case, we need a dynamic baryogenesis process to generate excess of baryon number compared to antibaryon number in order to create the observed baryon number today.

The precise epoch responsible for the observed matter genesis $\eta_\gamma$ in the primordial Universe has not been established yet. Several mechanisms have been proposed to explain baryogenesis with investigations typically focusing on the temperature range between GUT (grand unified theory) phase transition $T_\mathrm{G}\simeq10^{16}\,\mathrm{GeV}$ and the EW phase transition near $T_\mathrm{W}\simeq130\,\mathrm{GeV}$ \cite{Kuzmin:1985mm,Kuzmin:1987wn,Arnold:1987mh,Kolb:1996jt,Riotto:1999yt,Nielsen:2001fy,Giudice:2003jh,Davidson:2008bu,Morrissey:2012db,Canetti:2012zc}. {\color{black}The EW phase transition temperature regime was favored before mass of the Higgs-particle became known to be too heavy to allow a strong EW phase transition.}

In 1967, Andrei Sakharov formulated the three conditions necessary to permit baryogenesis in the primordial Universe~\cite{Sakharov:1967dj} and in 1988 he refined the three conditions as follows~\cite{Sakharov:1988vdp}:
\begin{subequations}\label{Sakharov}
\begin{align}
 &\text{Absence of baryonic charge conservation}\\
 &\text{Violation of CP-invariance}\index{CP
 violation}\\
&\text{Non-stationary conditions in absence of local thermodynamic equilibrium}
\end{align}
\end{subequations}
{\color{black} In the following we discuss these three Sakharov conditions for matter asymmetry to form, and argue that these conditions  could be satisfied during the entire cosmic QGP era. We will study below in more detail the bottom quark case and argue for the Higgs-particle case. Other cases are possible. We begin our considerations with the second and third condition in \req{Sakharov} above as these can easily be recognized in laboratory and/or theoretical studies of the cosmic plasma dynamics. We discuss the first condition last.}

The second Sakharov \req{Sakharov} condition requiring $CP$ (charge conjugation $\times$ parity)  violation assures us that we can recognize, in universal manner, the difference between matter and antimatter. Clearly, we could not enhance one form with reference to the other without being able to tell matter from antimatter. $CP$ violation allows us to share with another distant civilization that we are made of matter. A nice textbook discussion showing how to do this using the Kaon system CP violation is offered by Perkins~\cite{Perkins:1982xb}.

The third Sakharov condition \req{Sakharov} is a requirement for departure from thermal equilibrium which allows the breaking of detailed balance\index{detailed balance} condition: It is evident that in thermal equilibrium, the net effect of baryogenesis processes is cancelled out by equality between the  forward and back-reactions. {\color{black}Breaking of detailed balance must be non-stationary in the sense that it does not disappear entirely slowing down the Universe expansion $H\to 0$. In other words for baryogenesis to work the breaking of detailed balance must rely on the competition between microscopic processes, and not on a competition of microscopic processes with the Hubble parameter appearing in the Einstein-Vlasov equation \req{Hubble:Boltzmann}. We fully agree here with Sakharov's more precise thinking presented in 1988 \cite{Sakharov:1988vdp}.} 

{\color{black}Space-time domains near to phase transitions harbor non-stationary nonequilibrium in the expanding Universe. However, proximity of phase transformation invoked usually is not an exclusive physical environment. We will demonstrate within PP-SM how another opportunity in the primordial plasma arises: A competition between strong and EW interaction. Strong interactions can be rendered relatively weak due to mass thresholds.} 

We distinguish kinetic (momentum distribution) and chemical (particle abundance) equilibrium. This is so since kinetic equilibrium\index{kinetic equilibrium} is usually established much more quickly, while abundance yields are more difficult to establish, especially so for particles with masses in excess, or at least similar to ambient temperatures~\cite{Koch:1986ud,Birrell:2014gea}. {\color{black}This distinction allows to recognize that detailed balance can arise without a strict chemical equilibrium condition. This is seen in other cosmic and astrophysical environments, including the nucleosynthesis processes in the Universe (BBN) and stars.}  

Specifically for all heavy primordial particles including the top $t$ and bottom $b$ quarks, W and Z gauge bosons, and the Higgs-particle $h$, we observe that when the Universe expands and temperature cools down well below the particle mass, the production process and decay processes create a stationary equilibrium with detailed balance outside of equilibrium. Higgs-particle is an excellent candidate for non-stationary effects due to its small and even vanishing coupling to the massless and low mass particle plasma. 

We believe that the presence of chemical (abundance) nonequilibrium
{\color{black}is a required but not sufficient condition for a baryogenesis environment. We need further either time dependence of chemical non-equilibrium processes or a sufficiently weak coupling of one of the involved particles to the thermal background in order to extend the potential for baryogenesis to a much wider temperature domain, beyond the proximity of the EW phase transition condition. Study of the Higgs-particle in this context is one of our ongoing research challenges. However, in this work
}
we use the case of bottom quarks to demonstrate the mechanism we are exploring. We interpret the third condition of Sakharov in our specific context as follows:
\begin{itemize}
\item Non-stationary conditions in absence of local thermodynamic equilibrium $\Longrightarrow$ Absence of detailed balance associated with nonequilibrium yields and non-stationary particle abundance evolution.
\end{itemize} 

{\color{black} We now return to consider the first Sakharov condition \req{Sakharov}.  So far all efforts to create consistent description of baryogenesis based on well studied EW phase transition near $T=130$\,GeV have not been able to generate the observed baryon asymmetry \cite{Kuzmin:1985mm,Kuzmin:1987wn,Arnold:1987mh,Kolb:1996jt,Riotto:1999yt,Nielsen:2001fy,Giudice:2003jh,Davidson:2008bu,Morrissey:2012db,Canetti:2012zc}. An ad-hoc `far'-primordial to the era of EW  phase transition baryon-antibaryon asymmetry creation seems less `attractive', in the sense that we push back baryogenesis to an unknown cosmic era with mechanisms well beyond the scope of known PP-SM. Even so this is the conclusion reached after several decades of study of the EW-baryogenesis. We refer to excellent review of these efforts as presented by Morrissey and Ramsey-Musolf~\cite{Morrissey:2012db}, which work invites new baryogenesis physics before EW phase transition, and to Canetti, Drewes,  and Shaposhnikov \cite{Canetti:2012zc}. We defer to these reviews in regard to intricate details of EW-baryogenesis  addressed in depth there: Within the PP-SM the sphaleron mechanism plays the pivotal role (see also \url{https://en.wikipedia.org/wiki/Sphaleron} sourced September 2024) both in erasing any baryon asymmetry in presence of 2nd order EW phase transition and in creating asymemtry in presence of 1st order transition. The EW phase transition is, considering PP-SM parameters, widely accepted to be 2nd order. This has stimulated a not yet conclusive search for BSM (beyond standard model) extension allowing a 1st order EW transition while not contradicting the established PP-SM properties.}

{\color{black}However, this  widely adopted conclusion to look `beyond', either in temperature or PP-SM is not fully compelling. We believe that the observed baryon-antibaryon  asymmetry can originate in the evolution of the cosmic QGP after EW transition down to hadronization condition provided that there is a  not yet discovered direct baryon conservation violating elementary process, which, for example, protects the $B-L$ (baryon minus lepton) quantum number. This is so since PP-SM does not contain baryon or lepton conservation and several minimal extensions that have been considered conserve $B-L$.  We refer to Nath and Perez \cite{Nath:2006ut} for a wider scope review of baryon number non-conservation. }

{\color{black}We note that the experimentally known limits on baryon number violating decay of protons were obtained at zero temperature, a very different environment compared to primordial QGP. This search could be extended to study of heavy particle decays: Of interest here would be baryon conservation violating decay of heavy elementary particles ($t$, $h$, W, Z, $b$). The current experimental data lack required precision to exclude  baryon number violation at the sub-nano-level required to assure that these processes do not participate in cosmic baryogenesis.} 
 
{\color{black}Another path to recognize a mechanism of baryon excess generation is the discovery of an (effective) force in primordial QGP that could create large domain separation of baryons and antibaryons. In general the elementary quark or lepton antimatter is electrically attracted to matter. However, in the strong interaction sector there could be hidden baryon number differentiating processes:  Interlocking flavor and color organization of quark matter in the so-called CFL-phase (color-flavor locked phase) \cite{Rajagopal:2000ff,Alford:2001zr,Kaplan:2001qk} is here of interest. Another context worth further exploration  are neutral composite anti-baryonic particles (\eg\ anti-neutrons, $\overline\Lambda(uds)$ and more exotic variants) present abundantly during the mixed phase QGP hadronization process. We will not further explore this alternative which has not attracted much attention so far. }

\para{Production and decay of Higgs-particle in QGP}
The  uniqueness of the Higgs-particle\index{Higgs-particle} among heavy PP-SM particles is also due to its stability: The total width is $\Gamma_h\simeq 2.5\times\,10^{-5}m_h$. This combines with the unexpected low value of $T=5.7\GeV$ of interest where the Higgs-particle yield is equal to the baryon asymmetry in the Universe. This motivates us to examine here, in a qualitative manner, the dynamical abundance of the Higgs-particle in the QGP epoch, seeking the eventual non-stationary condition needed for baryogenesis 

The Higgs-particle predominantly decays via the $W,Z$ decay channels as follows:
\begin{align}
h\longrightarrow WW^\ast\,. ZZ^\ast\longrightarrow\mathrm{anything}\,.
\end{align}
Here $W^\ast,Z^\ast$ represent the production of virtual off-mass-shell gauge bosons decaying rapidly into relevant particle pairs. Therefore, once Higgs-particle decays via this channel, at least four particles are ultimately formed and there is no path back for $T\ll m_h$. This is so since the spectral energy of the four produced particles, on average $31\GeV$, is epithermal compared to the ambient plasma at the low temperature of interest near to $T\simeq 6\GeV$. Therefore, a back-reaction production of Higgs-particle in this channel would be weaker destroying the detailed balance required for the chemical equilibrium yield. 

In the QGP epoch, the dominant production of the Higgs-particle is the bottom quark pair fusion reaction: 
\begin{align}
b+\overline{b}\longrightarrow h\,,
\end{align}
which is the inverse to the important but by far not dominant decay process of $h\to b+\overline{b}$. This means that in first approximation the detailed balance Higgs-particle yield is reached well below the chemical equilibrium.

However, there could be considerable deviation from kinetic momentum equilibrium as well. This is so since bottom fusion will in general produce a Higgs-particle out of kinetic momentum equilibrium. A heavy particle immersed into a plasma of lighter particles requires many, many collisions to equilibrate the momentum distribution. This is a well-known kinetic theory result. Moreover, the Higgs-particle interacts weakly with all lower mass particles in QGP present at $T<10\GeV$. 

The Higgs-particle is by far the best candidate to fulfill the Sakharov non-stationary condition in the primordial Universe at a temperature range of interest to baryogenesis. A full dynamic study leading to proper understanding of the off-chemical and off-kinetic equilibrium non-stationary abundance of Higgs-particle is one of the near future projects we consider, and is therefore today beyond the scope of this report. 

\subsection{Heavy quark production and decay}
\label{sec:heavyQ}
\index{quark!abundance}\index{QGP!quark abundance}
\para{Heavy quarks in the primordial QGP}
The primordial QGP refers to the state of matter that existed in the primordial Universe, specifically for time $t\approx 20\, \mathrm{\mu s}$ after the Big-Bang\index{Big-Bang}. At that time the Universe was controlled by the strongly interacting particles: quarks and gluons. In this chapter, we study the heavy bottom and charm flavor quarks near to the QGP hadronization\index{hadrons!hadronization} temperature, $0.3\,\mathrm{GeV}>T>0.15\,\mathrm{GeV}$, and examine the relaxation time for the production and decay of bottom/charm quarks. Then we show that the bottom quark nonequilibrium occur near to QGP–hadronization and creates the arrow in time in the primordial Universe.
 
In the QGP epoch, up and down $(u,d)$ (anti)quarks are effectively massless and provide, along with gluons, some leptons, and photons, the thermal bath defining the thermal temperature. Strange $(s)$ (anti)quarks are also found to be in equilibrium considering their weak, electromagnetic, and strong interactions; indeed this equilibrium continues in hadronic epoch until $T\approx13\MeV$~\cite{Yang:2021bko}. 

The massive top $(t)$ (anti)quarks couple to the plasma via the channel~\cite{ParticleDataGroup:2022pth} 
\begin{equation}
t\leftrightarrow W+b\,,\qquad \Gamma_t=1.4\pm0.2\,\mathrm{GeV}\,.
\end{equation}
As is well-known, the width prevents formation of bound toponium states. Given the large value of $\Gamma_t$ there is no freeze-out of top quarks until $W$ itself freezes out. To address the top quarks in QGP, a dynamic theory for $W$ abundance is needed, a topic we will embark on in the future. 
 
The semi-heavy bottom $(b)$ and charm $(c)$ quarks can be produced by strong interactions via quark-gluon pair fusion processes. These quarks decay via weak interaction decays and their abundance depends on the competition between the strong interaction fusion processes at low temperature inhibited by the mass threshold, and weak decay reaction rates.

In the following we consider the temperature near QGP hadronization $0.3\,\mathrm{GeV}>T>0.15\,\mathrm{GeV}$, and study the bottom and charm abundance by examining the relevant reaction rates of their production and decay.
In thermal equilibrium the number density of light quarks can be evaluated in the massless limit, and we have\index{number density of quark}
\begin{align}\label{FermiN}
n_q=\frac{g_{q}}{2\pi^2}\,T^3 F(\Upsilon_q)\;, \quad F=\int_0^\infty \frac{x^2dx}{1+\Upsilon_q^{-1}e^x}\;,
\end{align}
where $\Upsilon_q$ is the quark fugacity\index{fugacity!quark}. We have $ F(\Upsilon_q=1)=3\,\zeta(3)/2$ with the Riemann zeta function $\zeta(3)\approx1.202$.
The thermal equilibrium number density of heavy quarks with mass $m\gg T$ can be well described by the Boltzmann expansion of the Fermi distribution function, giving
\begin{align}\label{BoltzN}
n_{q}\!=\!\frac{g_{q}T^3}{2\pi^2}\sum_{n=1}^{\infty}\frac{(-1)^{n+1}\Upsilon_q^n}{n^4}\left(\frac{n\,m_{q}}{T}\right)^{\!2}\!K_2\left(\frac{n\,m_{q}}{T}\right),
\end{align} 
where $K_2$ is the modified Bessel\index{Bessel function} functions of integer order `$2$'. In the case of interest, when $m\gg T$, it suffices to consider the Boltzmann\index{Boltzmann!approximation} approximation and to keep the first term $n=1$ in the expansion. The first term $n=1$ also suffices for both charmed $c$-quarks and bottom $b$-quarks, giving
\begin{align}
&n_{b,c}={\Upsilon_{b,c}\,}n^{th}_{b,c},\qquad n^{th}_{b,c}=\frac{g_{b,c}}{2\pi^2}\,T^3\left(\frac{m_{b,c}}{T}\right)^2\,K_2(m_{b,c}/T).
\end{align}
However, for strange $s$ quarks, several terms are needed. 

In~\rf{number_entropy_b002} we show the equilibrium ($\Upsilon=1$) bottom and charm number density per entropy density ratio as a function of temperature $T$. The $b$-quark mass parameters shown are $m_b=4.2\,\mathrm{GeV}$ (blue) dotted line, $m_b=4.7\,\mathrm{GeV}$ (black) solid line, and $m_b=5.2\,\mathrm{GeV}$ (red) dashed line. For $c$-quark, $m_c=0.93\,\mathrm{GeV}$ (blue) dotted line, $m_c=1.04\,\mathrm{GeV}$ (black) solid line, and $m_c=1.15\,\mathrm{GeV}$ (red) dashed line. The entropy density\index{entropy!density} is given by~\req{entropy} and only light particles contribute significantly. Thus the result we consider is independent of the actual abundance of $c$, $b$ and other heavy particles. 

\begin{figure}
\centerline{\includegraphics[width=0.8\linewidth]{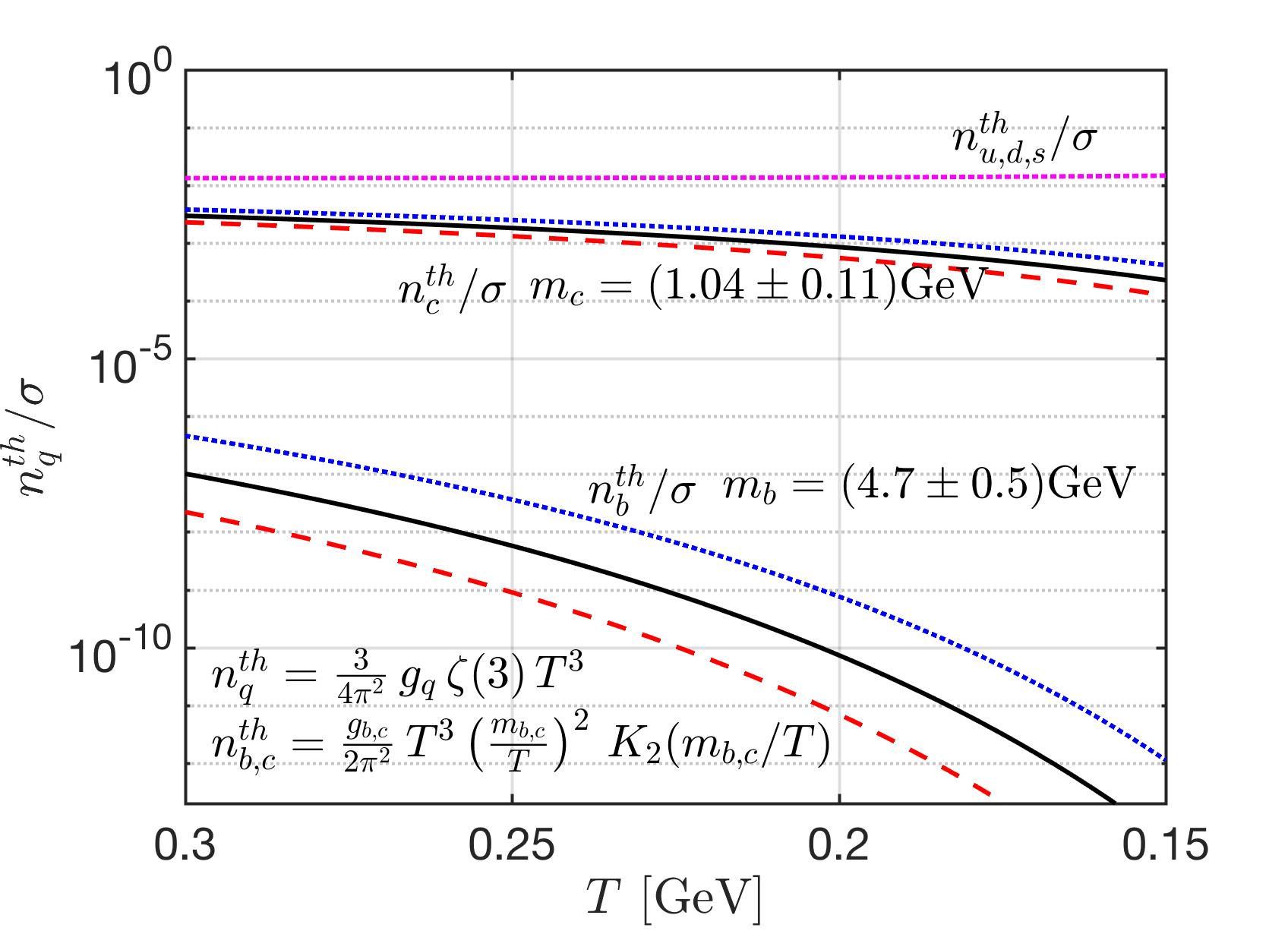}}
\caption{
The equilibrium charm and bottom quark number density normalized by entropy density, as a function of temperature in the primordial Universe, see text for discussion of different mass values. \radapt{Yang:2024ret}}
\label{number_entropy_b002} 
\end{figure}

The $m_b\simeq 5.2\,\mathrm{GeV}$ is a typical potential model mass used in modeling bound states of bottom, and $m_b=4.2,\,4.7\,\mathrm{GeV}$ is the current quark mass at low and high energy scales. In~\rf{number_entropy_b002} we see that the charm abundance in the domain of interest $0.3\,\mathrm{GeV}>T>0.15\,\mathrm{GeV}$ is about $10^4\sim 10^{9}$ times greater than the abundance of bottom quarks. This implies that the small $b$,$\bar b$ quark abundance is embedded in a large background comprising all lighter $u,d,s,c$ quarks and anti-quarks, as well as gluons $g$.

In the following we will calculate the production and decay rate for bottom and charm quarks and compare to the Universe expansion rate. We will show that in the epoch of interest to us the characteristic Universe expansion time $1/H$ is much longer than the lifespan and production time of the bottom/charm quark. In this case, the dilution of bottom/charm quark due to the Universe expansion is slow compare to the strong interaction production, and the weak interaction decay of the bottom/charm. Any abundance nonequilibrium will therefore be nearly stationary. 

It is important for following analysis to know that the expansion of the Universe is the slowest process, allowing many microscopic reactions at a `fixed' temperature range $T$ to proceed. To show this we evaluate the Hubble\index{Hubble!parameter} relation to obtain $1/H$\,[s]
\begin{align}
H^2=\frac{8\pi G_N}{3}\left(\rho_\gamma+\rho_{\mathrm{lepton}}+\rho_{\mathrm{quark}}+\rho_{g,{W^\pm},{Z^0}}\right)
\,.
\end{align}
The effectively massless particles and radiation dominate particle energy density $\rho_i$, defining the speed of expansion of the Universe within temperature range $130\, \mathrm{GeV}>T>0.15\,\mathrm{GeV}$. We have the following particles: photons, $8$ color charge gluons, $W^\pm$, $Z^0$, three generations of $3$ color charge quarks and leptons in the primordial QGP. The characteristic Universe expansion time constant $1/H$ is seen in~\rf{BCreaction:fig} below. In the epoch of interest to us $0.3\,\mathrm{GeV}>T>0.15\,\mathrm{GeV}$, the Hubble time $1/H\approx10^{-5}\,\mathrm{s}$, which is much longer than the microscopic lifespan and production time of the bottom and charm quarks we study 

\para{Quark production rate via strong interaction}
In primordial QGP\index{quark!production rate}, the bottom and charm quarks can be produced from strong interactions via quark-gluon pair fusion processes. For production, we have the following processes
\begin{align}
 q+\bar{q}&\longrightarrow b+\bar b,\qquad q+\bar{q}\longrightarrow c+\bar c,\\
 g+g&\longrightarrow b+\bar b,\qquad g+g\longrightarrow c+\bar c\,.
\end{align}

For the quark-gluon pair fusion processes\index{bottom quark!production rate}
the evaluation of the lowest-order Feynman diagrams yields the cross-sections~\cite{Letessier:2002ony}:
\begin{align}
&\sigma_{q\bar{q}\rightarrow b\bar{b},c\bar{c}}=\frac{8\pi\alpha_s^2}{27s}\left(1+\frac{2m_{b,c}^2}{s}\right)w(s),\,\qquad w(s)=\sqrt{1-{4m^2_{b,c}}/{s}},\\
&\sigma_{gg\rightarrow b\bar{b},c\bar{c}}=\!\frac{\pi\alpha_s^2}{3s}\bigg[\left(1\!+\!\frac{4m^2_{b,c}}{s}\!+\!\frac{m^4_{b,c}}{s^2}\right)\ln{\left(\frac{1+w(s)}{1-w(s)}\right)}\!-\!\left(\frac{7}{4}\!+\!\frac{31m^2_{b,c}}{4s}\right)w(s)\bigg],
\end{align} 
where $m_{b,c}$ represents the mass of bottom or charm quark, $s$ is the Mandelstam variable, and $\alpha_s$ is the QCD coupling constant. Considering the perturbation expansion of the coupling constant $\alpha_s$ for the two-loop approximation~\cite{Letessier:2002ony}, we have:
\begin{align}
\alpha_s(\mu^2)=\frac{4\pi}{\beta_0\ln({\mu^2/\Lambda^2})}\bigg[1-\frac{\beta_1}{\beta_0}\frac{\ln(\ln{(\mu^2/\Lambda^2)})}{\ln(\mu^2/\Lambda^2)}\bigg],
\end{align}
where $\mu$ is the renormalization energy scale and $\Lambda^2$ is a parameter that determines the strength of the interaction at a given energy scale in QCD. The energy scale we consider is based on required gluon/quark collisions above $b\bar b$ energy threshold, so we have $\mu=2m_b+T$. For the energy scale $\mu>2m_b$ we have $\Lambda=180\sim230\MeV$ ($\Lambda\approx205\MeV$ in our calculation), and the parameters $\beta_0=11-2n_f/3$, $\beta_1=102-38n_f/3$ with the number of active fermions $n_f=4$. 

\begin{figure} 
\centerline{\includegraphics[width=0.49\linewidth]{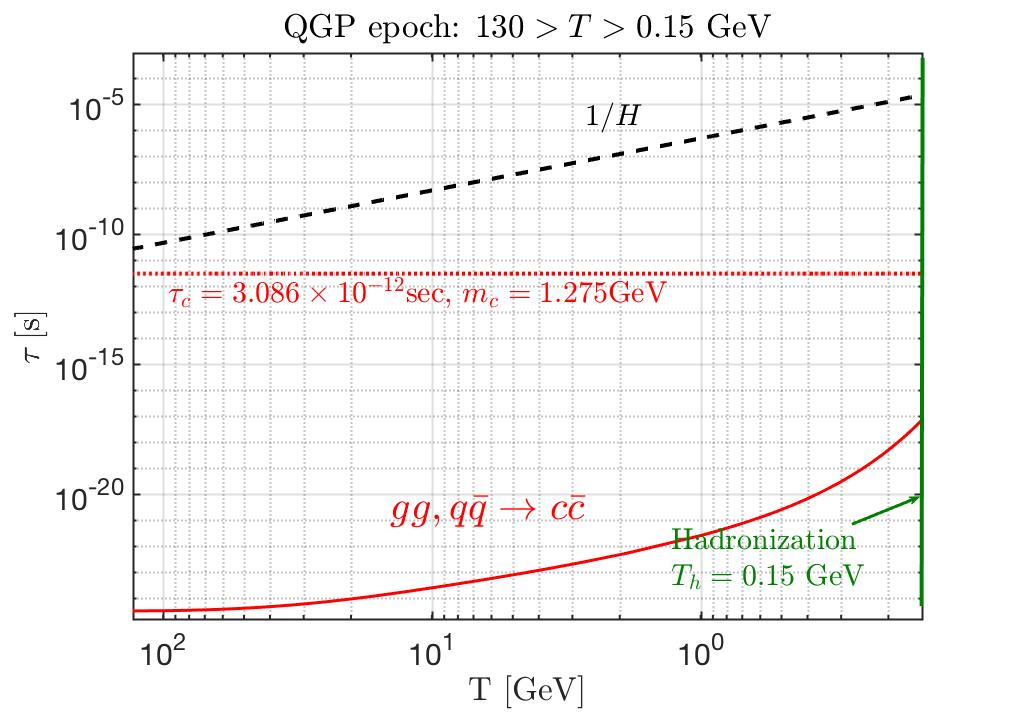}
\includegraphics[width=0.49\linewidth]{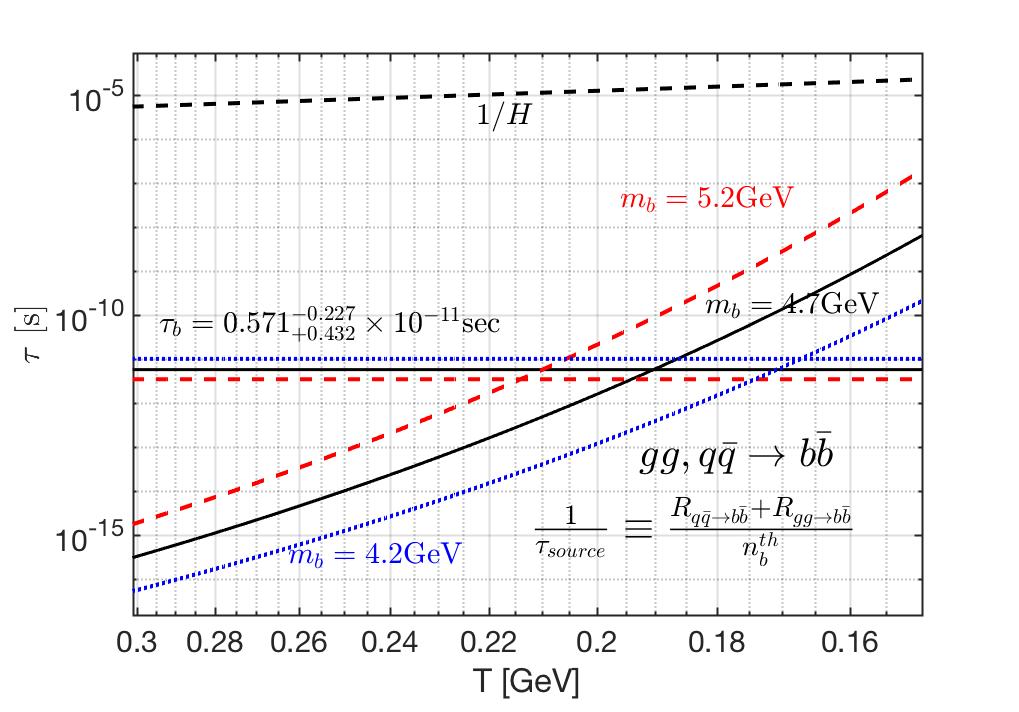}}
\caption{Comparison of Hubble\index{Hubble!time} time $1/H$, quark lifespan $\tau_{q}$, and characteristic time for production via quark, gluon pair fusion. The upper frame for charm $c$-quark in the entire QGP epoch $T$ rang; the lower frame for bottom $b$-quark amplifying the dynamic detail balance $T\simeq 200\MeV$. Both figures end at the hadronization temperature of $T_{H}\approx150\MeV$. See text for additional information. \cccite{Rafelski:2023emw}. \radapt{Yang:2024ret}}
\label{BCreaction:fig}
\end{figure}

In general the thermal reaction rate per unit time and volume $R$ can be written in terms of the scattering cross-section as follows~\cite{Letessier:2002ony}:
\begin{align}
R\equiv\sum_i\int_{s_{th}}^\infty\!ds\,\frac{dR_i}{ds}=\sum_i\int_{s_{th}}^\infty\!ds\,\sigma_i(s)\,P_i(s),
\end{align}
where $\sigma_i(s)$ is the cross-section of the reaction channel $i$, and $P_i(s)$ is the number of collisions per unit time and volume. Considering the quantum nature of the colliding particles (i.e., Fermi and Bose distribution)\index{Bose!distribution}\index{Fermi!distribution} with the massless limit and chemical equilibrium\index{chemical equilibrium} condition ($\Upsilon=1$), we obtain~\cite{Letessier:2002ony}
\begin{align}
&P_i(s)=\frac{g_1g_2}{32\pi^4}\,\frac{T}{1+I_{12}}\frac{\lambda_2}{\sqrt{s}}\!\sum_{l,n=1}^{\infty}\!(\pm)^{l+n}\frac{K_1(\sqrt{lns}/T)}{\sqrt{ln}},\\
&\lambda_2\equiv\left[s-\left(m_1+m_2\right)^2\right]\,\left[s-\left(m_1-m_2\right)^2\right],
\end{align}
where $+$ is for boson and $-$ is for fermions, and the factor $1/(1+I_{12})$ is introduced to avoid double counting of indistinguishable pairs of particles. $I_{12}=1$ for identical pair of particles, otherwise $I_{12}=0$. Hence the total thermal reaction rate per volume for bottom quark production can be written as
\begin{align}
\label{Bquark_Source}
R^{\mathrm{Source}}_{b,c}=\int^\infty_{s_{th}}ds\,\bigg[\sigma_{q\bar{q}\rightarrow b\bar{b},c\bar{c}}\,P_q+\sigma_{gg\rightarrow b\bar{b},c\bar{c}}\,P_g\bigg]
\,.
\end{align}
We introduce the bottom/charm quark relaxation time for the quark-gluon pair fusion as follows:
\begin{align}
\label{relaxation_time}
&{\tau_{b,c}^{\mathrm{Source}}}\equiv\frac{dn_{b,c}/d\Upsilon_{b,c}}{R^{\mathrm{Source}}_{b,c}}\;,
\end{align}
where $dn_{b,c}/d\Upsilon_{b,c}=n^{th}_{b,c}$ in the Boltzmann\index{Boltzmann!approximation} approximation. The relaxation time is on the order of magnitude of time needed to reach chemical equilibrium. 

In~\rf{BCreaction:fig} we show the characteristic time for $b$ and $c$ quark strong interaction production. The $c$ quark (upper frame) is shown in the entire QGP temperature range. We note the vast 15 orders of magnitude difference between the Hubble time and the rate of production. This means there will be very many microscopic cycles of charm production decay, erasing any non-stationary effect. For $b$ (lower frame) we restrict the view to temperature range in the domain of interest, $ 0.3\,\mathrm{GeV}>T> 0.15\,\mathrm{GeV}$. Three different masses $m_{b}=4.2\GeV$ (blue short dashes), $4.7\GeV$, (solid black), $5.2\GeV$ (red long dashes) for bottom quarks are shown. 
 
\para{Quark decay rate via weak interaction}
The bottom/charm quark decay\index{quark!weak decay rate} via the weak interaction \index{bottom quark!decay rate} 
\begin{align}
 &b\longrightarrow c+l+\overline{\nu_l}, \qquad b\longrightarrow c+q+\bar{q},\\
&c\longrightarrow s+l+\overline{\nu_l},\qquad c\longrightarrow s+q+\bar{q}\,.
\end{align}
The vacuum decay rate for $1\to2+3+4$ in vacuum can be evaluated via the weak interaction:
\begin{align}
\frac{1}{\tau_1}=&\frac{64G^2_F\,V^2_{12}\,V^2_{34}}{(4\pi)^3g_1}\,m^5_1\times\left[\frac{1}{2}{\left(1-\frac{m^2_2}{m^2_1}-\frac{m^2_3}{m^2_1}+\frac{m^2_4}{m^2_1}\right)}\mathcal{J}_1-\frac{2}{3}\mathcal{J}_2\right],
\end{align}
where the Fermi constant is $G_F=1.166\times10^{-5}\,\mathrm{GeV}^{-2}$, $V_{ij}$ is the element of the Cabibbo-Kobayashi-Maskawa (CKM) matrix~\cite{Czarnecki:2004cw}\index{CKM matrix} for quark channel and $V_{l\nu_l}=1$ for lepton channel. The functions $\mathcal{J}_1$, and $\mathcal{J}_2$ are given by
\begin{align}
&\mathcal{J}_1\!=\!\!\!\int_0^{(1-m^2_2/m^2_1)/2}\!\!\!\!\!\!\!\!dx\left(1\!-\!2x\!-\!\frac{m^2_2}{m_1^2}\right)^{\!\!2}\left[\frac{1}{(1-2x)^2}-1\right]\,,\\
&\mathcal{J}_2\!=\!\!\!\int_0^{(1-m^2_2/m^2_1)/2}\!\!\!\!\!\!\!\!dx\left(1\!-\!2x\!-\!\frac{m^2_2}{m_1^2}\right)^{\!\!3}\left[\frac{1}{(1-2x)^3}-1\right]
\,.
\end{align}
The modification due to the heat bath (plasma) is small because the bottom and charm mass $m_{b,c}\gg T$~\cite{Kuznetsova:2008jt}. In the temperature range we are interested in, the decay rate in the vacuum is a good approximation for our calculation. 

We show the lifespan for bottom and charm quarks in~\rf{BCreaction:fig}. For charm (upper frame) the decay is always much slower compared to production. This assures that the strong interaction processes can easily maintain equilibrium. Thus, during the entire era of QGP, charm quarks can be assumed to be in an equilibrium condition. 

After hadronization\index{hadrons!hadronization}, charm quarks form heavy mesons that decay into several hadronic particles. The daughter particles from charm meson decay can interact and re-equilibrate within the hadron plasma. There are very many branching reactions and some involve production of only light particles. In this case the energy required to drive the inverse reaction to produce heavy charm mesons is difficult to overcome. We believe this is causing the charm quark to vanish from the inventory shortly after hadronization, but a detailed study has not been carried out due to complexity of the situation. 

Looking at the lower frame in~\rf{BCreaction:fig}, we see that in the case of bottom quarks the decay crosses the production rate, and this happens within QGP near to $T=200\MeV$. The intersection implies that the bottom quark\index{bottom quark} freeze-out from the primordial plasma before hadronization as the production process slows down at low temperatures and the subsequent weak interaction decay leads to a dilution of the bottom quark content within the QGP plasma. All of this occurs with rates significantly faster than Hubble expansion and thus, as the Universe expands, the system departs from chemical equilibrium in near stationary manner, because of the competition between decay and production reactions in QGP. We will show how the dynamic equation cause the distribution to deviate from equilibrium with $\Upsilon\neq1$ in the temperature range below the crossing point but before the hadronization. 

\subsection{Is baryogenesis possible in QGP phase?}\label{Bottom}
\index{quark!bottom nonequilibrium}
\para{Bottom quark abundance nonequilibrium}
The competition between weak interaction decay and strong interaction production rates can lead to a nonequilibrium dynamic heavy quark abundance. We explore as example the case of bottom quarks in QGP. Similar considerations apply to all heavier PP-SM particles including in particular Higgs, W,Z gauge bosons, top $t$ quark. However, the case of $b$-quarks attracted our attention early on in context of baryogenesis since there is strong known CP violation\index{CP violation} also present.
 
The dynamic equation for bottom quark abundance in QGP can be written as \index{bottom quark!population equation}
\begin{align}
\label{Bquark_eq}
\frac{1}{V}\frac{dN_b}{dt}=\big(\,1-\Upsilon^2_{b}\,\big)\,R^{\mathrm{Source}}_{b}-\Upsilon_b\,R^{\mathrm{Decay}}_{b}\;,
\end{align}
where $R^{\mathrm{Source}}_{b}$ and $R^{\mathrm{Decay}}_{b}$ are the thermal reaction rates per volume of production and decay of bottom quark, respectively. The bottom source rates are the gluon and quark fusion rates \req{Bquark_Source}. The decay rate depends on whether the bottom quarks are freely present in the plasma or are bounded within mesons. We consider two extreme scenarios for the bottom quark population: 1.) all bottom flavor is free, and 2.) all bottom flavor is bounded into mesons in QGP. In~\rf{ReactionTime} we show the characteristic interaction times relevant to the abundance of bottom quarks, as well as the Hubble time $1/H$ for the temperature range of interest, $0.3\,\mathrm{GeV}> T> 0.15\,\mathrm{GeV}$.

\begin{figure} 
\centerline{\includegraphics[width=0.85\linewidth]{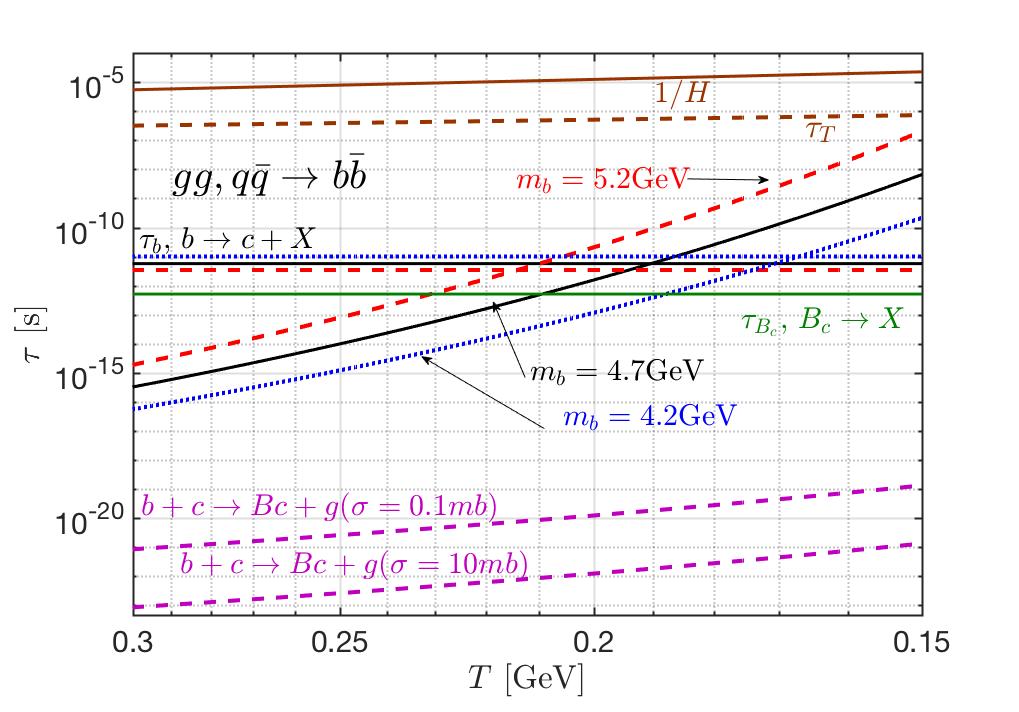}}
\caption{{\color{black}Top of figure: solid (brown) line is the characteristic $1/H$[s] cosmological time as a function of temperature $0.3\,\mathrm{GeV}>T> 0.15\MeV$. Just below it dashed (brown) dilution time $\tau_T$, \req{eq:tauT}. The total strong interaction bottom production $g+g\to b+\bar{b}$, $q+\bar q\to b+\bar{b}$ relaxation time is shown for three different bottom masses: $m_b=4.2\,\mathrm{GeV}$ (blue dotted line), $m_b=4.7\,\mathrm{GeV}$ (black solid line), $m_b=5.2\,\mathrm{GeV}$ (red dashed line). The nearly horizontal lines are bottom-quark (in QGP) EW decay lifetimes $\tau_b$ with the same color coding as the production processes. Horizontal solid (green) line is the vacuum lifespan $\tau_{B_c}$. At the bottom of the figure dashed lines (purple) are in QGP plasma pre-formation process $b+c\rightarrow \mathrm{B}_c+g$ with cross-section bracketing range $\sigma=\{0.1,10\} \,\mathrm{mb}$. \radapt{Yang:2024ret}.
}}
\label{ReactionTime}
\end{figure}

Considering all bottom flavor is free in QGP, the bottom decay rate per volume is the bottom lifespan weighted with density of particles \req{BoltzN}, see Ref.\,\cite{Kuznetsova:2008jt}. We have
\begin{align}\hspace{0.5cm}
R^{\mathrm{Decay}}_b=\frac{dn_b/d\Upsilon_b}{\tau_b},\,\,\,\,\, \tau_b\approx0.57\times10^{-11} \mathrm{s}.
\end{align}
On the other hand, $b$,\,$\bar b$ quark abundance is embedded in a large background comprising all lighter quarks and anti-quarks (see~\rf{number_entropy_b002}). After formation the heavy $b,\,\bar b$ quark can bind with any of the available lighter quarks, with the most likely outcome being a chain of reactions 
\begin{align}
&b+q\longrightarrow\mathrm{B}+g\;,\\
&\mathrm{B}+s\longrightarrow\mathrm{B}_s+q\;,\\
&\mathrm{B}_s+c\longrightarrow\mathrm{B}_c+s\;,
\end{align}
with each step providing a gain in binding energy and reduced speed due to the diminishing abundance of heavier quarks $s, c$. To capture the lower limit of the rate of $\mathrm{B}_c$ production we show in~\rf{ReactionTime} the expected formation rate by considering the direct process $b+\overline c\rightarrow \mathrm{B}_c+g$, considering the range of cross-section $\sigma=0.1\sim10\,\mathrm{mb}$ ~\cite{Schroedter:2000ek}. The rapid formation rate of B$_c(b\bar c)$ states in primordial plasma is shown by purple dashed lines at bottom in~\rf{ReactionTime}, we have
\begin{align}
\tau (b+\overline c\rightarrow \mathrm{B}_c+g)\approx(10^{-16}\sim10^{-14})\times\frac{1}{H} \;.
\end{align}

Despite the low abundance of charm, the rate of $\mathrm{B}_c$ formation is relatively fast, and that of lighter flavored B-mesons is substantially higher. Note that as long as we have bottom quarks made in gluon/quark fusion bound practically immediately with any quarks $u, d, s$ into B-mesons, we can use the production rate of $b, \bar b$ pairs as the rate of B-meson formation in the primordial-QGP, which all decay with lifespan of pico-seconds. We believe that this process is fast enough to allow consideration of bottom decay from the B$_c(b\bar c)$, $\overline{\mathrm{B}}_c(\bar b c)$ states~\cite{Yang:2020nne}. 
 
Based on the hypothesis that all bottom flavor is bound rapidly into $\mathrm{B}_c^\pm$ mesons, we have 
\begin{align}\label{Bc_source}
g+g, q+q \longleftrightarrow &b+\bar b\;[b(\bar{b})+\bar{c}(c)]\longrightarrow \mathrm{B}_c^\pm\longrightarrow\mathrm{anything}.
\end{align}
In this case, the decay rate per volume can be written as
\begin{align}\hspace{0.5cm}
 R^{\mathrm{Decay}}_b=\frac{dn_b/d\Upsilon_b}{\tau_{\mathrm{B}_c}},\,\,\,\,\, \tau_{\mathrm{B}_c}\approx0.51\times10^{-12} \mathrm{s}.
 \end{align}

\para{Stationary and non-stationary deviation from equilibrium}
To investigate the nonequilibrium phenomena of bottom quarks, we aim to replace the variation of particle abundance seen on LHS in \req{Bquark_eq} by the time variation of the abundance fugacity\index{fugacity} $\Upsilon$. This substitution allows us to derive the dynamic equation for the fugacity parameter and enables us to study the fugacity as a function of time. Considering the expansion of the Universe we have
\begin{align}\label{number_dilution}
\frac{1}{V}\frac{dN_b}{dt}=\frac{dn_b}{d\Upsilon_b}\frac{d\Upsilon_b}{dt}+\frac{dn_b}{dT}\frac{dT}{dt}+3Hn_b,\;
\end{align}
where we use $d\ln(V)/dt=3H$ for the Universe expansion. Substituting \req{number_dilution} into \req{Bquark_eq} and dividing both sides of equation by $dn_b/{d\Upsilon_b}=n^{th}_b$, the fugacity equation becomes
\begin{align}
\frac{d\Upsilon_b}{dt}+&3H\Upsilon_b+\Upsilon_b\frac{dn^{th}_b/dT}{n^{th}_b}\frac{dT}{dt}=\left(1-\Upsilon_b^2\right)\frac{1}{\tau_{b}^{\mathrm{Source}}}-\Upsilon_b\frac{1}{\tau^{\mathrm{Decay}}_b}\;,
\end{align}
where relaxation time for bottom production is obtained using \req{relaxation_time}. 

It is convenient to introduce the additional expansion dilution time $1/\tau_T$ as follows,
\begin{align}\label{eq:tauT}
\frac{1}{\tau_T}\equiv-\frac{dn^{th}_b/dT}{n^{th}_b}\frac{dT}{dt},
\end{align}
where we introduce '$-$' sign in the definition to have $\tau_T>0$.  Dilution time $\tau_T$ characterizes the bottom density dilution due to the Universe cooling. The fugacity equation can now be written as
\begin{align}\label{Fugacity_Eq0}
\frac{d\Upsilon_b}{dt}\!\!=&(1-\Upsilon_{b}^2)\frac{1}{\tau_{b}^{\mathrm{Source}}}
\!-\!\Upsilon_{b}\left(\frac{1}{\tau^{\mathrm{Decay}}_b}+3H\!-\!\frac{1}{\tau_T}\right).
\end{align}
In following sections we will solve the fugacity differential equation in two different scenarios: stationary and non-stationary Universe.

In~\rf{BCreaction:fig} (bottom) we show that the relaxation time for both production and decay are faster than the Hubble\index{Hubble!time} time $1/H$ for the duration of QGP, which implies that $H,1/\tau_T\ll1/\tau_{b}^{\mathrm{Source}},1/\tau^{\mathrm{Decay}}_b$. In this scenario, we can solve the fugacity equation by considering the stationary Universe first, i.e., the Universe is not expanding and we have
\begin{align}\label{stationary}
H=0,\qquad 1/\tau_T=0.
\end{align} 
In the stationary Universe at each given temperature we consider the dynamic equilibrium condition (detailed balance)\index{detailed balance} between production and decay reactions that keep
\begin{align}
\frac{d\Upsilon_b}{dt}=0.
\end{align}
Neglecting the time dependence of the fugacity $d\Upsilon_b/dt$ and substituting the condition \req{stationary} into the fugacity equation \req{Fugacity_Eq0}, then we can solve the quadratic equation to obtain the stationary fugacity as follows: 
\begin{align}
\label{Fugacity_Sol}
\Upsilon_{\mathrm{st}}&=\sqrt{1+\left(\frac{\tau_{source}}{2\tau_{decay}}\right)^2}-\left(\frac{\tau_{source}}{2\tau_{decay}}\right).
\end{align} 

\begin{figure} 
\centerline{\includegraphics[width=0.8\linewidth]{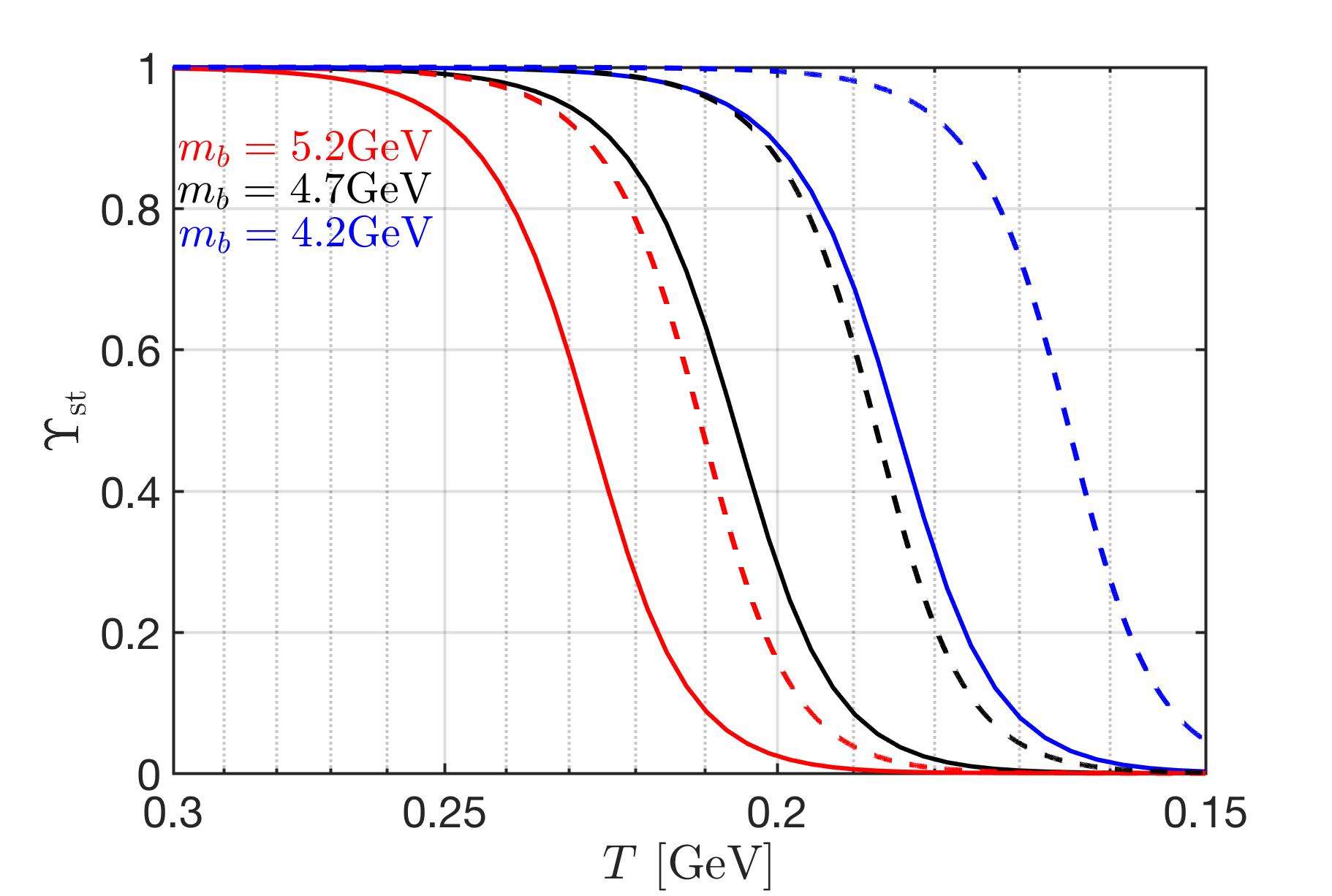}}
\caption{Dynamical fugacity of bottom quark as a function of temperature in primordial Universe. Solid line shows bottom quark bound into $B_c$, dashed lines the case of free bottom quark: $m_b=4.2\,\mathrm{GeV}$ (blue), $m_b=4.7\,\mathrm{GeV}$ (black), and $m_b=5.2\,\mathrm{GeV}$ (red). \cccite{Rafelski:2023emw}. \radapt{Yang:2024ret}}
\label{fugacity_bc}
\end{figure}

In~\rf{fugacity_bc} the fugacity of bottom quark\index{bottom quark} $\Upsilon_{\mathrm{st}}$ as a function of temperature, \req{Fugacity_Sol} is shown around the temperature $T=0.3\,\mathrm{GeV}>T>0.15\,\mathrm{GeV}$ for different masses of bottom quarks. In all cases we see prolonged nonequilibrium: This happens since the decay and reformation rates of bottom quarks are comparable to each other as we have noted in~\rf{ReactionTime} where both lines cross. One of the key results shown in~\rf{fugacity_bc} is that the smaller mass of bottom quark slows the strong interaction formation rate to the value of weak interaction decays just near the phase transformation of QGP to HG phase. Finally, the stationary fugacity corresponds to the reversible reactions in the stationary Universe. In this case, there is no arrow in time for bottom quark because of the detailed balance.

We now consider non-stationary correction in expanding Universe allowing for the Universe expanding and thus temperature being a function of time. This leads to a non-stationary correction related to time dependent fugacity in the expanding Universe. 

In general, the fugacity of bottom quark can be written as 
\begin{align}\label{Nonstationary_sol}
&\Upsilon_b=\Upsilon_{\mathrm{st}}+\Upsilon^{\mathrm{non}}_{\mathrm{st}}=\Upsilon_\mathrm{st}\left(1+x\right),\quad x\equiv{\Upsilon_\mathrm{st}^{\mathrm{non}}}/{\Upsilon_\mathrm{st}},
\end{align}
where the variable $x$ corresponds to the correction due to non-stationary Universe. Substituting the general solution \req{Nonstationary_sol} into differential equation \req{Fugacity_Eq0}, we obtain
\begin{align}\label{Nonstationary_eq}
\frac{dx}{dt}=-x^2\frac{\Upsilon_\mathrm{st}}{\tau_{source}}&-x\left[\frac{1}{\tau_{eff}}+3H-\frac{1}{\tau_T}\right]-\left[\frac{d\ln\Upsilon_\mathrm{st}}{dt}+3H-\frac{1}{\tau_T}\right],
\end{align}
where the effective relaxation time $1/\tau_{eff}$ is defined as
\begin{align}
\frac{1}{\tau_{eff}}\equiv\left[\frac{2\Upsilon_\mathrm{st}}{\tau_{source}}+\frac{1}{\tau_{decay}}+\frac{d\ln\Upsilon_\mathrm{st}}{dt}\right].
\end{align}

\begin{figure} 
\centerline{\includegraphics[width=0.8\linewidth]{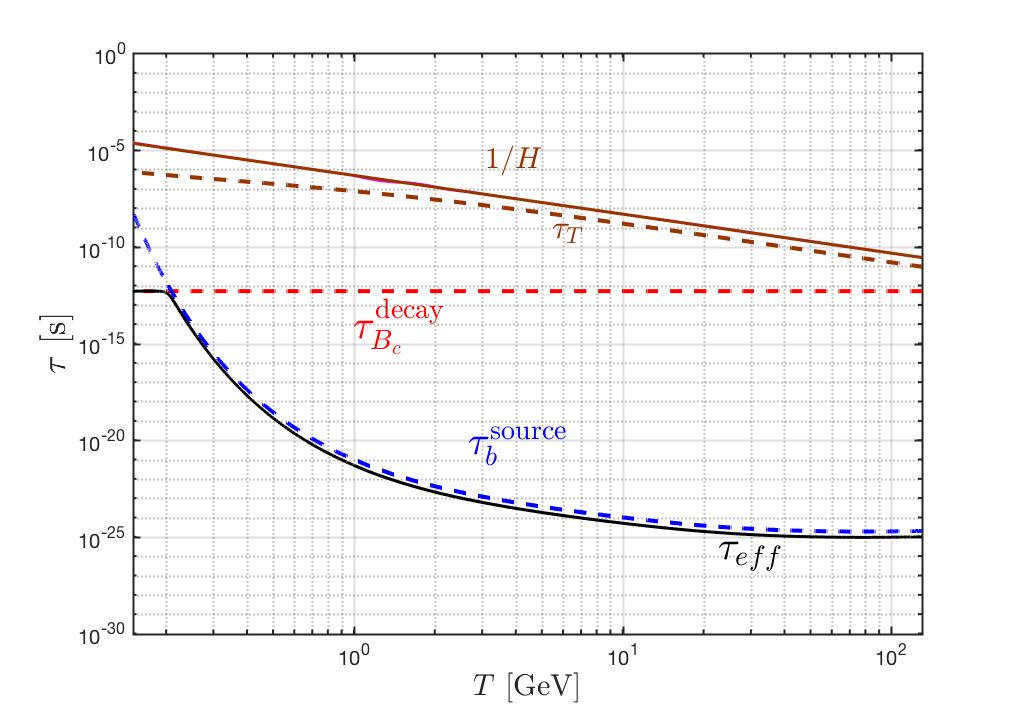}}
\caption{The effective relaxation time $\tau_{eff}$ as a function of temperature in the primordial Universe for bottom mass $m_b=4.7$\,GeV. For comparison, we also plot the vacuum lifespan of $B_c$ meson $\tau_{B_c}^{decay}$ (red dashed-line), the relaxation time for bottom production $\tau^b_{source}$ (blue dashed-line), Hubble expansion time $1/H$(brown solid line) and relaxation time for temperature cooling $\tau_T$ (brown dashed-line). \radapt{Yang:2024ret}}
\label{RelaxationTime_eff}
\end{figure}

In~\rf{RelaxationTime_eff} we see that when temperature is near to $T=0.2\GeV$, we have $1/\tau_{eff}\approx10^{7}H$, and $1/\tau_{eff}\approx10^5/\tau_T$. In this case, the last two dilution terms in \req{Nonstationary_eq}  are small compared to $1/\tau_{eff}$ and can be neglected. The differential equation becomes
\begin{align}\label{nonstationary_eq}
\frac{dx}{dt}=-\frac{x^2\,\Upsilon_\mathrm{st}}{\tau_{source}}&-\frac{x}{\tau_{eff}}-\left[\frac{d\ln\Upsilon_\mathrm{st}}{dt}+3H-\frac{1}{\tau_T}\right]
\,.
\end{align}

To solve the variable $x$ we consider the case $dx/dt,x^2\ll1$ first; we neglect the terms $dx/dt$ and $x^2$ in \req{nonstationary_eq}, then solve the linear fugacity equation. We will establish that these approximations are justified by checking the magnitude of the solution. Neglecting terms $dx/dt$ and $x^2$ in \req{nonstationary_eq}, we obtain
\begin{align}
x\approx\tau_{eff}\left[\frac{d\ln\Upsilon_\mathrm{st}}{dt}+3H-\frac{1}{\tau_T}\right].
\end{align}
It is convenient to change the variable from time to temperature. For an isentropically-expanding Universe, we have
\begin{align}\label{tau_H}
\frac{dt}{dT}=-\frac{\tau^\ast_H}{T},\qquad \tau^\ast_H=\frac{1}{H}\left(1+\frac{T}{3g^s_\ast}\frac{dg^s_\ast}{dT}\right).
\end{align}
In this case, we have
\begin{align}
x=\tau_{eff}\left[\frac{1}{\Upsilon_\mathrm{st}}\frac{d\Upsilon_\mathrm{st}}{dT}\frac{T}{\tau^\ast_H}+3H-\frac{1}{\tau_T}\right].
\end{align}
Finally, we can obtain the non-stationary fugacity\index{fugacity} by multiplying the fugacity ratio $x$ with $\Upsilon_\mathrm{st}$, giving \index{bottom quark! non-stationary fugacity}
\begin{align}
\Upsilon_{\mathrm{st}}^{\mathrm{non}}
&\approx\left(\frac{\tau_{eff}}{\tau^\ast_H}\right)\left[\frac{d\Upsilon_\mathrm{st}}{dT}T-\Upsilon_{\mathrm{st}}\left(3H\tau^\ast_H-\frac{\tau^\ast_H}{\tau_T}\right)\right].
\end{align}

In~\rf{NonFugacity} we plot the non-stationary $\Upsilon^{\mathrm{non}}_\mathrm{st}$ as a function of temperature. The non-stationary fugacity $\Upsilon^{\mathrm{non}}_\mathrm{st}$ follows the behavior of $d\Upsilon_{\mathrm{st}}/dT$, which corresponds to the irreversible process in the expanding Universe. 

The irreversible nonequilibrium process creates the arrow in time for bottom quarks evolution in the Universe. The relatively large value of Hubble time (relatively slow expansion) is suppressing the value of the non-stationary fugacity resulting in the small $\mathcal{O}\sim10^{-7}$ effect. On one hand this shows that  neglecting $dx/dt,x^2\ll1$ is a good approximation for obtaining the non-stationary fugacity in the primordial Universe. On the other hand this also means that chemical non-stationary effects near to hadronization of bottom quarks could be too weak to allow sufficient baryogenesis.

\begin{figure}
\centerline{\includegraphics[width=0.8\linewidth]{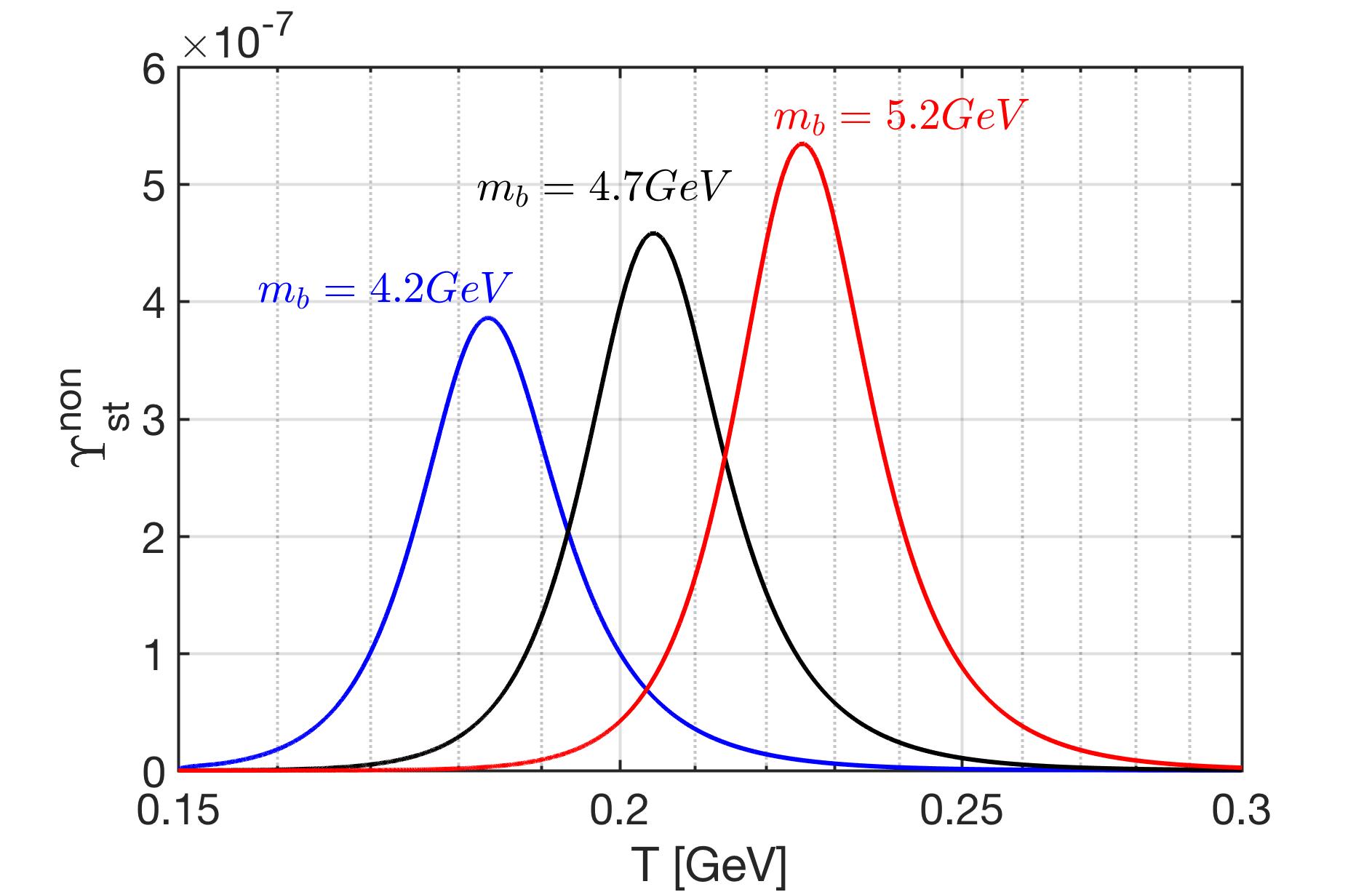}}
\caption{The non-stationary fugacity $\Upsilon_\mathrm{st}^{\mathrm{non}}$ as a function of the Universe $T$ for three different bottom masses $m_b=4.2\,\mathrm{GeV}$ (blue), $m_b=4.7\,\mathrm{GeV}$ (black), and $m_b=5.2\,\mathrm{GeV}$ (red). \radapt{Yang:2024ret}}
\label{NonFugacity}
\end{figure}

\para{Is there enough bottom flavor to matter?} Considering that FLRW-Universe evolves\index{cosmology!FLRW} conserving entropy, and that baryon and lepton numbers following on the era of matter genesis is conserved, the current day baryon $B$ to entropy $S$, $B/S$-ratio must be achieved during matter genesis. The estimates of present day baryon-to-photon density\index{baryon!per photon ratio} ratio $\eta_\gamma$ allows the determination of the present value of baryon per entropy\index{baryon!entropy ratio} ratio \cite{Fromerth:2012fe,Rafelski:2019twp,Letessier:2002ony,Fromerth:2002wb}:
\begin{align}
\left(\frac{B}{S}\right)_{t_0}\!\!\!\!=\eta_\gamma\left(\frac{n_\gamma}{\sigma_\gamma+\sigma_\nu}\right)_{\!t_0}\!\!\!\!=(8.69\pm0.05)\!\!\times\!\!10^{-11},
\end{align}
where the subscript $t_0$ denotes the present day value, where $\eta_\gamma=(6.14\pm0.02)\times10^{-10}$~\cite{ParticleDataGroup:2022pth}  is used here. Here we consider that the Universe today is dominated by photons and free-streaming  low mass neutrinos~\cite{Birrell:2012gg}, and $\sigma_\gamma$ and $\sigma_\nu$ are the entropy density for photons and neutrinos, respectively. 
 
In chemical equilibrium\index{chemical equilibrium} the ratio of bottom quark\index{bottom quark} (pair) density $n_b^{th}$ to entropy density\index{entropy!density} $\sigma=S/V$ just above the quark-gluon hadronization\index{hadrons!hadronization} temperature $T_\mathrm{H}=150\sim160\MeV$ is $n_b^{th}/\sigma=10^{-10}\sim 10^{-13}$, see~\rf{number_entropy_b002}. By studying the bottom density per entropy near to the hadronization temperature and comparing it to the baryon-per-entropy ratio $B/S$, we obtain the abundance of bottom quarks to consider in the proposed matter genesis mechanism.

\para{Example of bottom-catalyzed matter genesis}
Given that the nonequilibrium non-stationary component of bottom flavor arises at a relatively low QGP temperature, {\color{black}this Sakharov condition Eq.~(\ref{Sakharov}) is available around QGP hadronization.\index{Sakharov conditions!baryogenesis}} Let us now look back and see how different requirements are fulfilled
\begin{itemize}
\item
 We have demonstrated non-stationary conditions with absence of detailed balance: The competition between weak interaction decay and the strong interaction gluon fusion process is responsible for driving the bottom quark departure from the equilibrium in the primordial Universe near to QGP hadronization condition around the temperature $T=0.3\sim0.15\GeV$ as shown in~\rf{fugacity_bc}. Albeit small there is clear non-stationary component required for baryogenesis, see~\rf{NonFugacity}.
\item Violation of CP\index{CP violation} asymmetry was observed in the amplitudes of hadron decay including neutral B-mesons, see for example~\cite{LHCb:2019jta,LHCb:2020vut}. The weak interaction $CP$ violation\index{CP violation}\index{CKM matrix} arises from the components of Cabibbo-Kobayashi-Maskawa (CKM) matrix associated with quark-level transition amplitude and $CP$-violating phase. There is clear coincidence of non-stationary component of bottom yield with the bottom quark $CP$ violating decays of preformed $\mathrm{B}_x$ meson states, $x=u,d,s,c$~\cite{Karsch:1987pv,Brambilla:2010vq,Aarts:2011sm,Brambilla:2017zei,Bazavov:2018wmo,Offler:2019eij}. The exploration of the here interesting $CP$ symmetry breaking in B$_c(b\bar c)$ decay is in progress~\cite{ParticleDataGroup:2022pth,Tully:2019ltb,HFLAV:2019otj}.
\item
We do not know if there is baryon conservation\index{baryon!conservation} violating process in which one of the heavy QGP particles is participating. However, if such a process were to exist it is likely, considering mass thresholds, that it would be most active in the decays of heaviest standard model particles. It is thus of considerable interest to study in lepton colliders baryon number non conserving processes at resonance condition. Such a research program will additionally be motivated by our demonstration of an extended period of baryogenesis in the primordial Universe. 
\end{itemize}

\para{Circular Urca amplification}
The off equilibrium non-stationary behavior of bottom quarks near the temperature range $T=0.3\sim0.15\GeV$ can provide the environment for baryogenesis to occur in the primordial-QGP hadronization era. The non-stationary effects of interest, as we saw, are small. However there is an additional amplifying factor. 
 
 Let us consider the scenario where all bottom quarks are confined within $B_c^\pm$ meson. In this case, we know that the $B_c^\pm$ decay in the primordial-QGP has $CP$ violating component. We also know as shown above that there is a small  irreversible fraction of the evolving fugacity. Let us assume in addition that  there is a tiny baryon number breaking decay of the $B^\pm_c$ meson. A baryon symmetry will be accompanied by the asymmetry between leptons\index{lepton} and anti-leptons assuming $B-L=0$ baryon number violating processes.

The heavy $B_c^\pm$ meson decay into multi-particles in `cold' $T=0.3\sim0.15\GeV$ plasma is  naturally associated with the irreversible process. This is because after decay the daughter particles can interact with plasma and distribute their energy to other particles and reach equilibrium with the plasma quickly. In this case the energy required for the inverse reaction to produce $B_c^\pm$ meson is difficult to overcome and therefore we have an irreversible process for multi-particle decay in plasma.

This said we note that the rapid $B_c^\pm$ decay and bottom reformation speed at picosecond scale assures that there are millions of individual microscopic processes involving bottom quark production and decay before and during the hadronization epoch of QGP. In this case, we have a so-called Urca (after Urca Casino in Rio de Janeiro, Brazil) process for the bottom quark, i.e. a cycling reaction in which we have a multitude of circular production and decay processes involving $B_c^\pm$ meson. 

The Urca process is a fundamental physical process and has been studying the realms of in astrophysics and nuclear physics. In our case, for bottom quark as an example: at low temperature, the number of bottom quark cycling can be estimated as
\begin{align}
\left.\mathrm{C_{cycle}}\right|_{T=0.2\mathrm{GeV}}=\frac{\tau_H}{\tau_{B_c}}\approx2\times10^7,
\end{align}
where the lifespan of $B_c^\pm$ is $\tau_{\mathrm{B}_c}\approx0.51\times10^{-12}\,\mathrm{s}$ and at temperature $T=0.2\GeV$ the Hubble\index{Hubble!time} time is $\tau_H=1/H=1.272\times10^{-5}\,\mathrm{s}$. 

The Urca process\index{Urca process!baryogenesis} plays a significant role by potentially amplifying any small and currently unobserved violation of baryon number associated with the bottom quark. The small baryon asymmetry is enhanced by the Urca-like process with cycling ${\tau^\ast_H}/{\tau_\ast}$ in the primordial Universe. This amplification helps achieving the required baryogenesis strength.
  
\subsection{Electromagnetic plasma properties of QGP}\label{chap:QCD}
\para{The electromagnetic QGP medium}
{\color{black} The experimental study of QGP utilizing relativistic heavy-ion collisions provides the only known method to probe some key cosmic QGP properties. In these collisions, the strongly interacting quarks and gluons reach conditions that have governed the early Universe during its primordial quark-gluon plasma epoch, where the temperature exceeded the hadronization condition $T>150\MeV$.} 

{\color{black} Aside from strongly screened color-strong interactions, the QGP phase is governed by the electromagnetic(EM) properties of quarks. EM response of QGP is of particular interest considering possible cosmic magnetic fields that may have been generated and/or are present during the QGP epoch; we return to this topic in quantitative detail in \rsec{sec:mag:universe}. Novel EM mechanisms in QGP, such as the chiral magnetic effect \cite{Kharzeev:2007jp}, could also have an interesting impact. Long range cosmic magnetic fields could influence particle interactions and plasma dynamics, similar to how strong magnetic fields play a role in relativistic heavy-ion collisions. The transport properties, such as the electromagnetic conductivity of QGP, are essential for understanding how electromagnetic fields evolve and how particles interact within the plasma.}

{\color{black} The heavy-ion collision experiments allow us to explore how cosmic-scale magnetic fields might have affected the QGP's behavior. This insight directly connects laboratory studies to cosmological phenomena, offering a window into the Universe’s primordial state. In this section, we will review a semi-analytic study of the electromagnetic field of heavy ions in QGP following Ref.\,\cite{Grayson:2022asf}. This method may lead to the experimental determination of the electromagnetic conductivity of QGP which influences decisively the freeze-out magnetic field in heavy ion collisions~\cite{STAR:2023jdd}. This would provide the required insight into the EM properties of strongly interacting primordial QGP, such as the electric conductivity, present during the early Universe.}

{\color{black} The semi-analytic framework discussed below builds the groundwork for understanding the electromagnetic properties of QGP in heavy ion collisions allowing us to describe the strongly interacting particle plasma in the early Universe. By studying QGP experimentally in the collisions of heavy ions, we aim to determine its electromagnetic transport properties, such as its static conductivity, which can then be used to infer the properties of QGP in the early Universe.}

\para{Transport theory methods}
We consider the ultra-relativistic limit of the polarization tensor seen in Chapter \ref{chap:PlasmaSF} to study the electromagnetic properties of quark-gluon plasma (QGP) \cite{Grayson:2022asf}. We are interested in understanding  electromagnetic fields generated by colliding relativistic heavy-ion collisions which arguably  the largest in the known Universe, on the order of 
\begin{equation}\label{eq:Bcol}
ec|B| \approx m_\pi^2,
\end{equation}
but exist for very short times 
\begin{equation}\label{eq:tcol}
t_{\text{coll}}= 2 R/\gamma \sim 10^{-25}\,\textrm{s}\,,
\end{equation}
due to the Lorentz contraction by the Lorentz-factor $\gamma$ of the colliding nuclei. The magnetic field\index{magnetic!field} generated in these collisions is interesting due to its role in separating electric charge in the QGP through the chiral magnetic effect (CME) \cite{Kharzeev:2007jp}.

The electric current generated by the CME could lead to a charge separation along magnetic field lines. If a magnetic field survives in QGP until the time of hadronization\index{QGP!hadronization} of the QGP, which we will refer to as the freeze-out time $t_f$, it could also lead to a difference in the observed global polarization of $\Lambda$ hyperons and anti-hyperons\index{hyperon} \cite{Muller:2018ibh}. Charge separation in the hadron was recently studied experimentally \cite{STAR:2023jdd}. 

The distribution of the vacuum magnetic field\index{magnetic!fields} of two colliding nuclei in the center of momentum frame (CM-frame) is given by the Li\'enard-Wiechert fields shown in \rf{fig:vacmag}. This is the same magnetic field found by Lorentz boosting the Coulomb field of a nucleus at rest. We neglect the portion of the field that depends on acceleration arising in the actual collision since it is small for in vacuum scattering of heavy nuclei, as compared to the field that depends only on velocity.

\begin{figure}
\centering \includegraphics[width=0.85\linewidth]{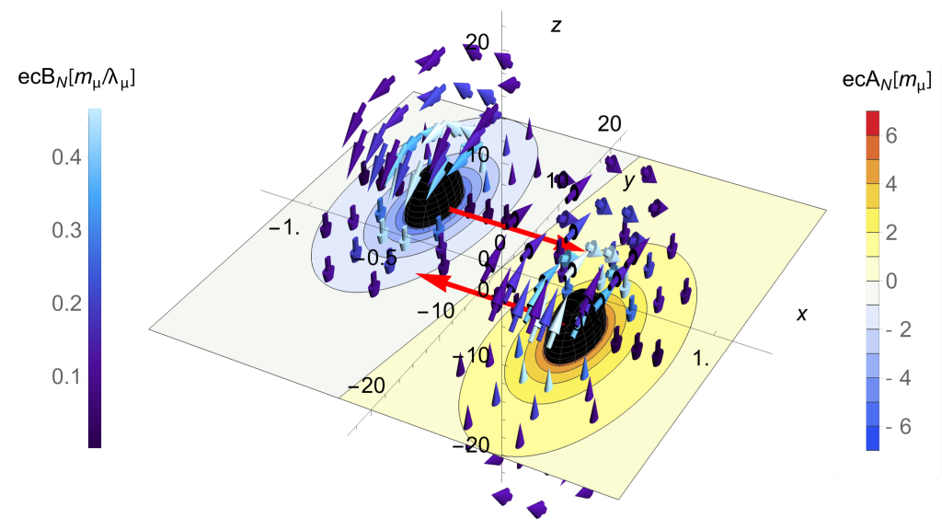}
\caption{The vacuum magnetic field and vector potential for two colliding lead Pb nuclei. See text for further discussion. \radapt{Grayson:2024okq}.}
    \label{fig:vacmag}
\end{figure}

{\color{black}The vacuum magnetic field  and the vacuum vector potential is depicted in \rf{fig:vacmag}  for the case of two colliding lead Pb nuclei in CM-frame. The vector potential is plotted in the collision plane in the right-half of figure, the magnetic field on left-half.  Red arrows indicate the direction of the moving nuclei with an impact parameter $b=3R$, $R$ is nuclear radius, and the Lorentz-factor $\gamma =37$.  At a larger Lorentz-factor, a graphical representation is difficult to visualize without scaling the fields with $\gamma$.  This plot shows how the magnetic field distribution is Lorentz contracted along the direction of motion. The magnetic field lines circulate out of the collision plane perpendicular to the velocity, adding together at the collision center.} 

This magnetic field is treated as an external perturbation on the quark-gluon plasma, filling the overlap region between the two nuclei after they collide. For simplicity, the QGP is modeled as an infinite medium to avoid edge complications at the plasma boundary. The temperature of QGP depends strongly on the collision energy of the nuclei. In \cite{Grayson:2022asf} we study Au+Au collisions at $\sqrt{s_{\text{NN}}}=200\,$GeV for QGP temperature $T=300$\,MeV.  After Heavy Ions collide, the conducting QGP medium generates long-range decaying tails or wake-fields in the magnetic field that extend far beyond the collision time \cite{Tuchin:2010vs}. The conductivity of QGP determines the strength of these wake-fields. We aim to model these fields in QGP using the formulation discussed in Chapter \ref{chap:PlasmaSF}.

\para{EM conductivity of quark-gluon plasma}
Past analytic calculations \cite{Tuchin:2010vs,Deng:2012pc,McLerran:2013hla,Tuchin:2013apa,Gursoy:2014aka,Li:2016tel,Roy:2015kma} solve Maxwell's equations in the presence of static electric conductivity 
\begin{equation}
   \sigma_0 = \frac{m_D^2}{3\kappa}\,,
\end{equation} 
in a  hydrodynamically evolving QGP. For a collisionless plasma $\kappa\rightarrow0$, the conductivity is infinite, and the medium behaves as a perfect conductor. Our work introduces the frequency and wave-vector dependence of the QGP analytically using the polarization tensor previously obtained in Ref.\,\cite{Formanek:2021blc}.

Prior work \cite{Inghirami:2016iru,Inghirami:2019mkc} incorporated the dynamical response of QGP by numerically solving the coupled magneto-hydrodynamic equations for a conducting quark-gluon plasma in the presence of the colliding nuclear charges. More recent calculations \cite{Yan:2021zjc,Wang:2021oqq} also incorporate the frequency and wave-vector dependence of the QGP response to electromagnetic fields by solving the coupled Vlasov-Boltzmann--Maxwell equations\index{Vlasov-Boltzmann equation}  numerically.

\para{The Ultrarelativistic EM polarization tensor in QGP}
Here we review the ultra-relativistic polarization tensor, including damping, for the idealized case where the QGP is infinite, homogeneous, and stationary. This calculation differs from our earlier work \cite{Formanek:2021blc} only in that we consider three quark species: up, down, and strange in thermal equilibrium abundance. We start with the Vlasov-Boltzmann equation for each quark flavor \req{eq:boltzmanncov}, where we assume all quarks collide on a momentum-averaged time scale $\tau_{\text{rel}} = \kappa^{-1}$. The induced current $ j_{\mathrm{ind}}^\mu$ can be written in terms of the phase-space distribution of quarks and anti-quarks as
\begin{equation}\label{eq:current}
   j_{\mathrm{ind}}^\mu(x) = 2 N_c \int (dp)p^\mu \\ \times \sum_{u,d,s} q_f (f_{f}(x,p) - f_{\bar{f}}(x,p))=  4 N_Q e^2 \int (dp)p^\mu \delta f(x,p)\,,
\end{equation}
where  $N_c$ is the number of colors. We sum over the quark flavors with charges $q_f$, and in the final result, we replace $q_f \delta f = \delta f_f$. The result \req{eq:current} differs from that found in the case of an electron-positron plasma by the factor
\begin{equation}
N_Q \equiv N_c\sum_f (q_f/e)^2 = 2\,,
\end{equation}
for three light quark flavors ($u,d,s$).

In the ultrarelativistic limit, neglecting quark masses, one finds the polarization functions \cite{Formanek:2021blc}:
\begin{align}\label{eq:polfuncsUltra}
&\Pi_{\parallel}(\omega,|\boldsymbol{k}|) = m_D^2\frac{\omega^2}{\boldsymbol{k}^2}\left(1 - \frac{\omega \Lambda}{2|\boldsymbol{k}|-i\kappa \Lambda}\right)\,,\\
&\Pi_{\perp}(\omega,|\boldsymbol{k}|) = \frac{m_D^2\,\omega}{4 |\boldsymbol{k}|}\left( \Lambda \left(\frac{\omega'^2}{\boldsymbol{k}^2} - 1\right) - \frac{2\omega'}{ |\boldsymbol{k}|}\right)\,,
\end{align}
where $\Lambda(\omega,\boldsymbol{k})$ is defined as
\begin{align}\label{eq:definitions}
 \Lambda \equiv \ln \frac{\omega'+  |\boldsymbol{k}|}{\omega'- |\boldsymbol{k}|}\,, \quad \text{with} \quad \omega' = \omega+i\kappa\,.
\end{align}
The parallel and transverse polarization functions have the same form as in \cite{Formanek:2021blc} except for an overall factor $N_Q$  as found in \cite{Kapusta:1992fm,Grayson:2022asf}:
\begin{equation}\label{eq:DebyemQCD}
    {m_D}^2_{(\text{EM})} = \sum_{u,d,s} q^2_f T^2 \frac{N_c}{3} = N_Q\frac{e^2T^2}{3} \equiv C_{\text{em}}T^2\,,
\end{equation}
where $C_{\text{em}} =  2e^2/3$. In the following, we will use $m_D$ as short-hand notation for the electromagnetic screening mass since we do not discuss color screening here.
The transverse conductivity $\sigma_{\perp}$, which controls the response of the plasma to magnetic fields\index{magnetic!fields}, is related to the imaginary part of the transverse polarization function as in \req{eq:sigmaperp}.

\para{QCD Damping rate and simple models of conductivity in QGP}
The strength of the plasma response to an external magnetic field depends on the quark damping rate $\kappa$ and the electromagnetic screening mass $m_D$. The scale of the collisional quark damping $\kappa$ is much larger than the electromagnetic Debye mass $m_D$ and electromagnetic damping $\kappa_{\text{EM}}$ because it depends on the strong coupling constant $\alpha_s$, not the electromagnetic coupling $\alpha$.

In \cite{Grayson:2022asf}, we use the first-order electromagnetic Debye mass \req{eq:DebyemQCD} to estimate the electromagnetic screening mass $m_D$. The collision rate $\kappa$ is related to the inverse of the mean-free time of quarks in QGP. We adopt a value for $\kappa$ from \cite{Mrowczynski:1988xu} where the mean-free time is given by the product of the parton density in the QGP and the quark-parton transport cross-section, leading to the expression 
\begin{equation}\label{eq:kappadef}
    \kappa(T) = \frac{10}{17\pi} (9 N_f +16) \zeta(3) \alpha_s^2 \ln\left(\frac{1}{\alpha_s}\right) T\,,
\end{equation}
where $N_f$ is the number of flavors, $\zeta(x)$ denotes the Riemann zeta function, and $\alpha_s(T)$ is the running QCD coupling.  We model the running of the QCD coupling constant as a function of temperature in the range $T<5T_c$ using a fit provided in \cite{Letessier:2002ony}:
\begin{equation}\label{eq:alphas}
    \alpha_s(T) \approx \frac{\alpha_s(T_c)}{1+C \ln(T/T_c)}\,,
\end{equation}
where $C=0.760 \pm 0.002$. For the QCD (pseudo-)critical temperature we use $T_c = 160\,$MeV. The  electromagnetic QED Debye mass is compared to $\kappa(T)$ in \rf{fig:kappaDebye}. We  expect the electromagnetic response of QGP to be over-damped since $\kappa> \frac{2}{\sqrt{3} m_D}$ giving a plasma frequency \req{eq:plasmafreq} which is imaginary over the range of temperatures relevant for QGP. We note that at the QGP temperature $T=300\,$MeV we use as example, $\kappa = 4.86\, m_D$.    

\begin{figure}
    \centering
    \includegraphics[width=0.7\linewidth]{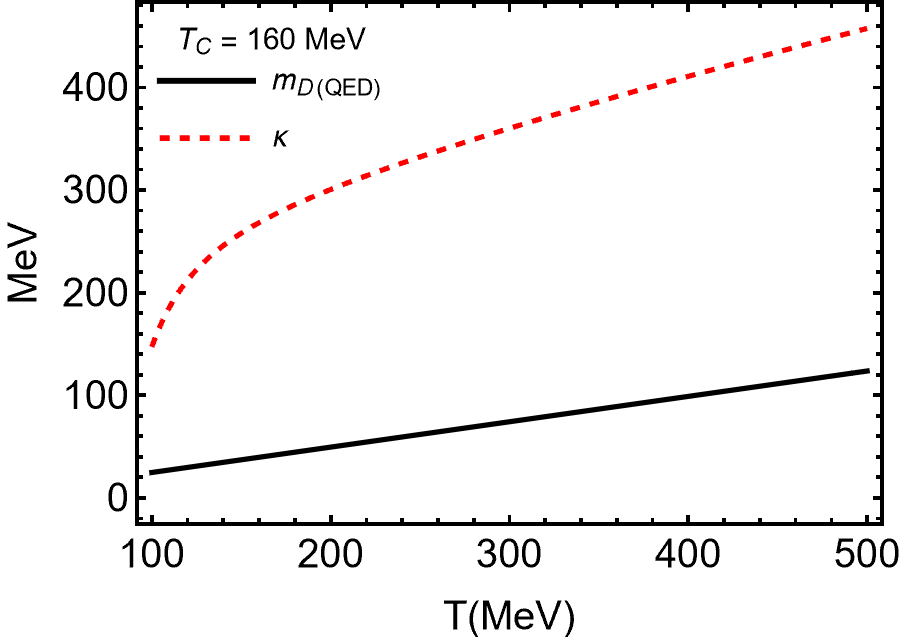}
    \caption{The electromagnetic Debye mass $m_D$, solid (black) line, and the QCD dampening rate $\kappa$, dashed (red) line, as a function of temperature.  \cccite{Grayson:2022asf}}
    \label{fig:kappaDebye}
\end{figure}

A  well studied model of conductivity is the Drude model~\cite{Drude:1900}, the long wavelength limit $k \to 0$ of the conductivity, which of course is isotropic as it depends only of $\omega$
\begin{equation}\label{eq:drude}
    \sigma_\parallel(\omega,0) = \sigma_\perp(\omega,0) = \frac{\sigma_0}{1-i \omega/\kappa} \,,
\end{equation}
with the static conductivity given by
\begin{equation}\label{eq:condstat}
   \sigma_0 = \frac{m_D^2}{3\kappa}\,.
\end{equation} 
The Drude conductivity can be obtained from the polarization tensor \req{eq:sigmaperp} determined within linear response method, see \rsec{chap:PlasmaSF}.

We can then use the Debye mass \req{eq:DebyemQCD} and the damping rate \req{eq:kappadef} to calculate the static conductivity \req{eq:condstat}, shown as a black line in \rf{fig:lattice comp}, which we then compare to Lattice calculations of the conductivity in QGP.  The factor of $C_{\text{em}}$, defined in \req{eq:DebyemQCD}, normalizes the conductivity by the charge of the plasma constituents, such that results using different numbers of dynamical quark flavors can be compared.

\begin{figure}
    \centering
    \includegraphics[width=0.75\linewidth]{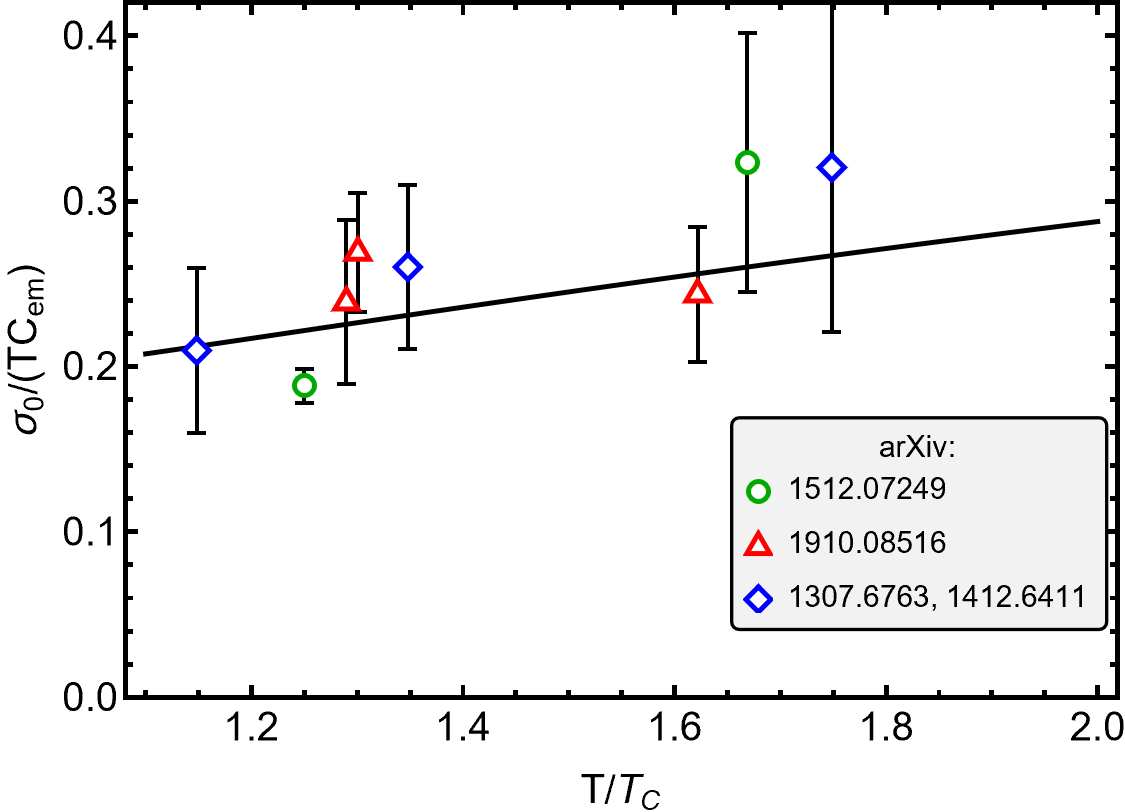}
    \caption{The static conductivity $\sigma_0$ scaled with $T$ as a function of temperature $T$ scaled with the `critical' value $T_c\simeq 150\MeV$ as predicted by \req{eq:condstat}. `Experimental' data are lattice results adapted from \cite{Aarts:2020dda} for $T>T_c$.  We indicate each set of points by its arXiv reference: blue diamonds \cite{Amato:2013naa,Aarts:2014nba}, green circles \cite{Brandt:2015aqk}, and red triangles \cite{Astrakhantsev:2019zkr}. \radapt{Grayson:2024okq}.}
    \label{fig:lattice comp}
\end{figure}

The  lattice-QCD results seen in  \rf{fig:lattice comp} \cite{Amato:2013naa,Aarts:2014nba,Brandt:2015aqk,Astrakhantsev:2019zkr} are scaled with temperature $T$ to remove the linear temperature dependence. One can see that the conductivity value predicted by \req{eq:kappadef}, plotted in \rf{fig:lattice comp} as a black line, lies well within the lattice-QCD results. We will use the value predicted by \rf{fig:lattice comp}, $\sigma = 5.01\,$MeV at $T=300\,$GeV, in the next section to compute the screened heavy-ion fields\index{heavy-ion!fields} in QGP.

One of the important situations to note in context of highly relativistic heavy-ion collisions  is that the fields of the ions, traveling near the speed of light, probe the polarization tensor, see \rsec{chap:PlasmaSF}, \req{eq:sigmaperp}, near the light cone. The transverse light-cone conductivity is given by
\begin{equation}\label{eq:lightcone}
    \sigma_\perp (\omega = |\boldsymbol{k}|)  =  i \frac{m_D^2}{4 \omega}\left( \frac{\kappa^2}{\omega^2} \xi \ln\xi +\frac{i\kappa}{\omega}\left(\xi+1\right)\right)\,,
\end{equation}
where $\xi$ is defined as
\begin{equation}\label{eq:xidef}
    \xi \equiv 1- 2i \frac{\omega}{\kappa}\,.
\end{equation}

\para{Magnetic field in QGP during a nuclear collision}
Assuming that the QGP is an infinite homogeneous and stationary medium near equilibrium, we can solve Maxwell's equations for the self-consistent fields, see \rsec{sec:Maxwell}. Then the magnetic field\index{QGP!magnetic fields} is obtained by  Fourier transforming the momentum space expressions given in \reqs{eq:aperp}{eq:ftfields} to the position space
\begin{equation}\label{eq:magorgin}
   \boldsymbol{B}(t, z) = \int \frac{d^4k}{(2\pi)^4}  e^{-i\omega t+ik_z z}
 \frac{\mu_0 i \boldsymbol{k} \times\ft{j}_{\perp \text{ext}}(\omega, \boldsymbol{k})}{\boldsymbol{k}^2 - \omega^2 - \mu_0 \Pi_{\perp}(\omega, \boldsymbol{k})}\,.
\end{equation}
We choose the collision center as the origin of our spatial coordinate system and align the spatial $z$-axis with the beam direction. Due to the symmetry of the colliding ions, the only nonzero component of the magnetic field along the $z$-axis points out of the collision plane ($x-y$ plane). In our coordinate system used in \cite{Grayson:2022asf}, this corresponds to the $y$-component of the magnetic field. 

For ease of calculation, we specify the external 4-current using two colliding Gaussian charge distributions normalized to the nuclear root mean square radius $R$ and charge $Z$:
\begin{equation}\label{eq:rhoext}
\rho_{\text{ext}\pm }(t,\boldsymbol{x}) = \frac{Zq\gamma}{\pi^{3/2}R^3}e^{-\frac{1}{R^2}(x\mp b/2)^2}e^{-\frac{1}{R^2}y^2}
\times e^{-\frac{\gamma^2}{R^2}(z\mp \beta t)^2}\,,
\end{equation}
where $\gamma$ is  as before the Lorentz-factor, $\beta\to 1$ is the ratio of the ion speed to the speed of light, respectively, and $b$ is the impact parameter of the collision. The plus and minus signs indicate motion in the $\pm \hat{z}$-direction (beam-axis). This charge distribution corresponds to the vector current
\begin{equation}\label{eq:jext}
\boldsymbol{j}_{\text{ext}\pm}(t, \boldsymbol{x}) = \pm \beta \hatv{z} \rho_{\text{ext}\pm}(t, \boldsymbol{x})\,.
\end{equation}
Further details of the external charge distribution for two colliding nuclei are presented in Appendix B  of Ref.\,\cite{Grayson:2022asf}.

The numerical result for the position-space magnetic field found by Fourier transforming \req{eq:magorgin} using the full transverse polarization function \req{eq:polfuncsUltra} is shown as a red dashed line in \rf{fig:bfcomp} and compared with various models of conductivity. These other models and their connections to published works are discussed in detail in  Ref.\,\cite{Grayson:2022asf}.

\begin{figure}
\includegraphics[width=0.50\linewidth]{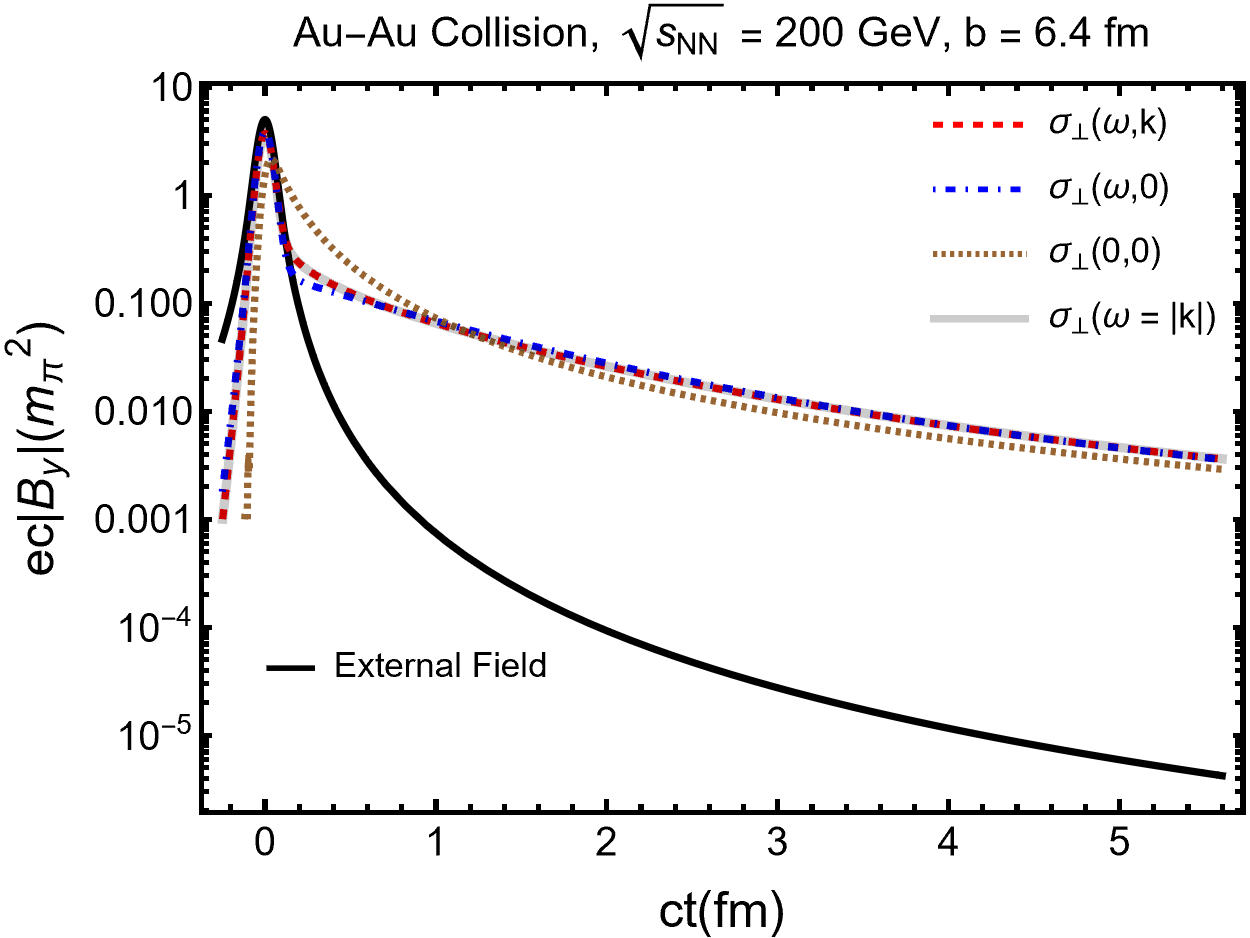}
\includegraphics[width=0.47\linewidth]{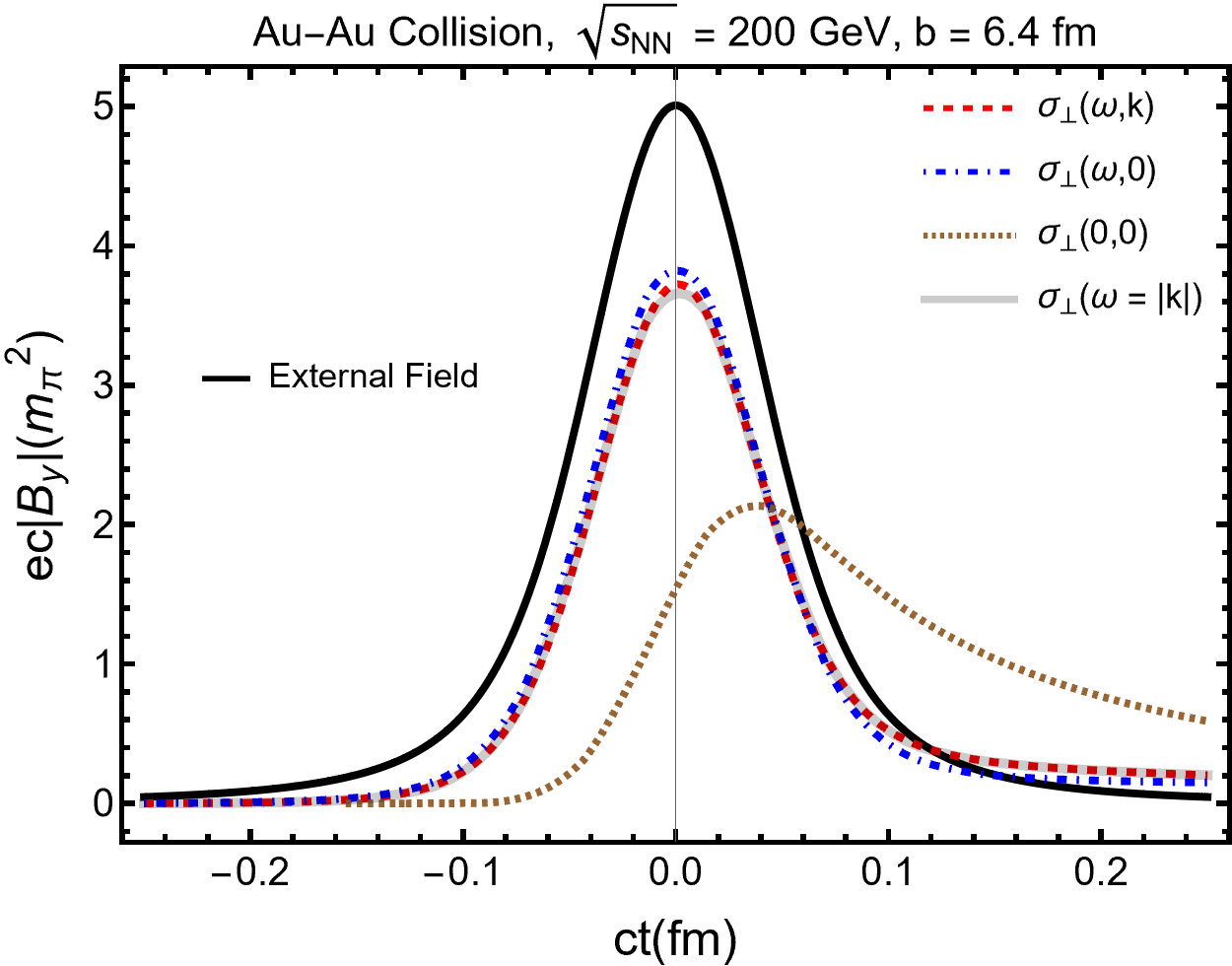}
\caption{The magnetic field at the collision center as a function of time, for QGP at $T = 300$\,MeV formed in Au-Au collisions ($Z=79$) at $\sqrt{s_\text{NN}} = 200$\,GeV and impact parameter $b = 6.4\,$fm. The left-hand frame shows the magnetic field\index{QGP!magnetic field} on a semi-logarithmic scale up to $ct = 5$\,fm. The right-hand frame shows the early-time magnetic field on a linear scale. \cccite{Grayson:2022asf}\label{fig:bfcomp}}
\end{figure}

Note that the QGP plasma considered in \rf{fig:bfcomp} is homogeneous and stationary. In a more realistic situation, the field would become screened only after the QGP is formed in the collision. At the considered temperature $T = 300$\,MeV, the electromagnetic Debye mass is $m_D = 74\,$MeV, and the quark damping rate is $\kappa = 4.86\,m_D$. This gives a static conductivity of $\sigma_0 = 5.01\,$MeV. 

Comparing the different approximations   \rf{fig:bfcomp}, we see  all three approximations using static conductivity $\sigma_\bot(0,0)$, \req{eq:condstat}, the Drude conductivity $\sigma_\bot(\omega,0)$, \req{eq:drude},  and the light-cone conductivity $\sigma_\bot(\omega=\|\boldsymbol{k}|)$, \req{eq:lightcone},  have similar asymptotic behavior for large times and agree in outcome with the full conductivity $\sigma_\perp(\omega,\boldsymbol{k})$. However, the static conductivity fails to describe the field at the early times as seen very clearly in the right-hand frame of \rf{fig:bfcomp}. One subject of future study of the heavy-ion experimental environment is to use the light-cone conductivity to attain analytical formulas for electromagnetic fields in position space in light-cone coordinates.

The light-cone conductivity \req{eq:lightcone} simplifies the calculation of plasma response since it only depends on a single variable ($\omega = |\boldsymbol{k}|$). One can see in the right-hand frame of \rf{fig:bfcomp} that results obtained using the light-cone conductivity,\req{eq:lightcone}, shown as an opaque (grey) line, is hardly visible under the  dashed (red) line showing the full numerical solution \req{eq:magorgin}. The light-cone conductivity accurately models the early magnetic field in QGP formed in ion collisions since the ions traveling near the light's speed only sample the polarization tensor near the light-cone. 

The simplest method to calculate the late-time magnetic field of colliding nuclei is to assume a static conductivity \cite{Tuchin:2013apa}. In this case, the magnetic field\index{heavy-ion!magnetic fields} in Fourier space has the form
\begin{equation}\label{eq:bstat}
    \ft{B}(\omega,\boldsymbol{k}) = \frac{ \mu_0 i\boldsymbol{k} \times \ft{j}_{\perp \text{ext}}}{\boldsymbol{k}^2 - \omega^2 - i\omega\sigma_0}\,,
\end{equation}
which is Fourier transformed using contour integration in the appendix of \cite{Grayson:2022asf} to
\begin{equation}\label{eq:banalyticapp}
   B_y(t) = -\mu_0 \frac{ Zq \beta }{(2\pi)} \frac{ b\sigma_0}{4t^2} e^{\frac{-b^2 \sigma_0}{16 t}}\,.
\end{equation}

Looking at the left-hand frame of \rf{fig:bfcomp}, the static conductivity, dotted line, initially overestimates the magnetic field after the external field begins to disappear since the effect of dynamic screening is not captured. Approaching the freeze-out time $t_f \approx 5\,$fm/c every model of the response function predicts similar values for the magnetic field \cite{Song:2007ux}. This is because the static conductivity determines the dependence of the magnetic field at times later than $t>1/\sigma \approx 59$\,fm/c after which damping of the initial magnetic field pulse is irrelevant. 

Alternatively, by assuming a point-like charge distribution $R\rightarrow 0$ and approximating the magnetic field for $ 1/\sigma_0 > t\gg 1/\kappa$, one can derive the late-time magnetic field using the Drude conductivity \req{eq:drude}
\begin{equation}\label{eq:latetimeB}
   B_y(t) \approx  \mu_0 \frac{ Ze \beta b \kappa \omega_p }{8\pi}\bigg[ \frac{1- e^{-\kappa t}}{\kappa t} - e^{-\kappa t} \text{Ei}\left(t\kappa\right)\bigg]\,.
\end{equation}
This result has the advantage of accurately describing the late-time magnetic field $t>t_f$  at large $\gamma$ as shown in \rf{fig:bcolcomp}. Both these results illustrate that the late-time magnetic field has a finite limit when $\gamma\rightarrow\infty$ as it depends only on $\beta\to 1$, but not on $\gamma$. The approximation used to derive this solution holds for $\gamma\beta \gg \sqrt{ \kappa/\sigma_0} \approx 12$. 

\begin{figure}
\centering
\includegraphics[width=0.8\linewidth]{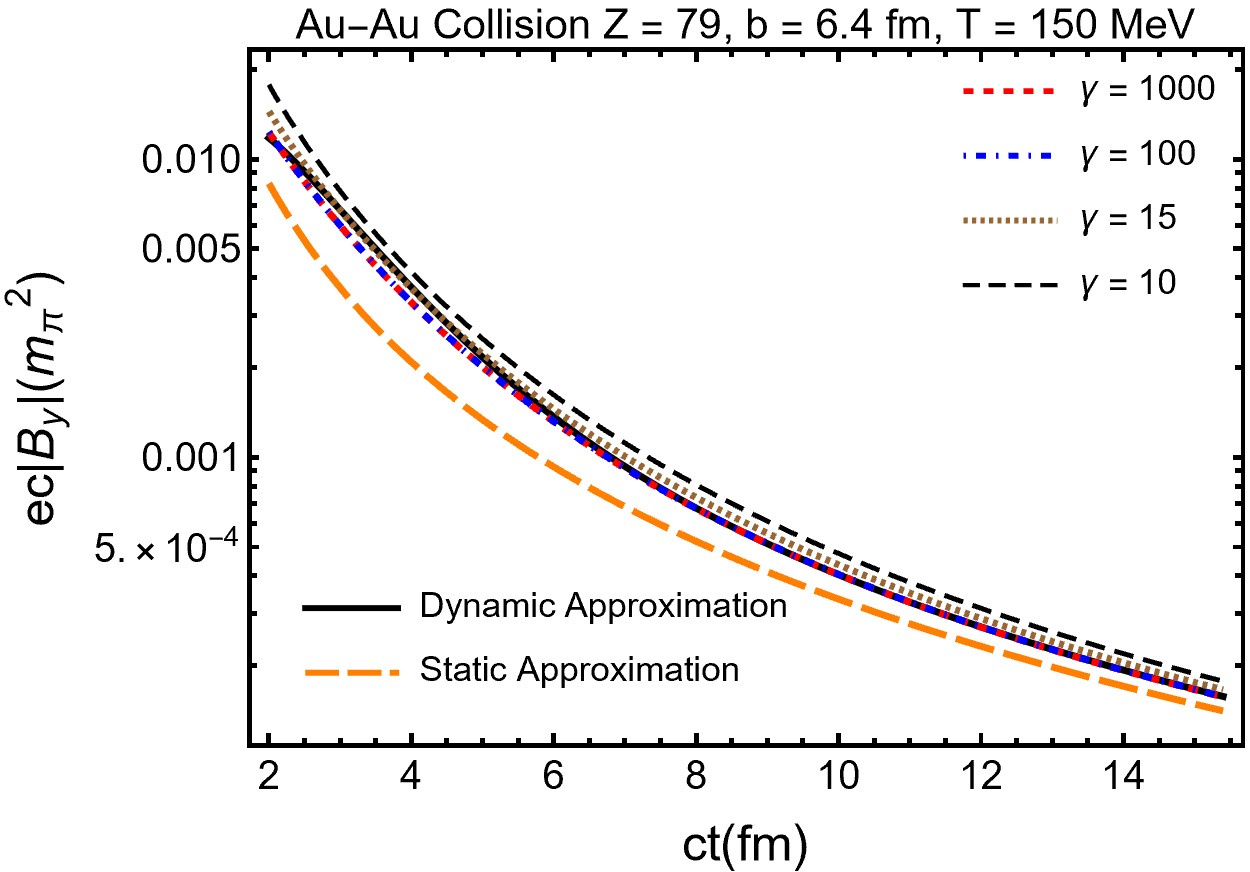}
    \caption{The freeze-out magnetic field for $T= 150$\,MeV for collisions with colliding heavy-ions at different value of Lorentz-factor, $\gamma=10,15,100,1000$ as function of freeze-out time.   \radapt{Grayson:2022asf}\label{fig:bcolcomp}}
\end{figure}

In \rf{fig:bcolcomp} we study the freeze-out magnetic field for $T= 150$\,MeV for collisions with colliding heavy-ions. We expect that around this temperature, QGP will hadronize \cite{Letessier:1992xd}.  We compare \req{eq:banalyticapp} to the full numerical result using the full polarization tensor \req{eq:magorgin} and explore dependence on the Lorentz-factor $\gamma\in\{10,1000\}$. The approximate static late time solution \req{eq:banalyticapp} shown as an orange dashed line is compared to numerical calculations  and to the late time analytic expansion \req{eq:latetimeB}. The approximate solution does not fully match the ultrarelativistic limit until times $t > t_{\sigma} \approx 59$\,fm/c. The freeze-out magnetic field\index{freeze-out!magnetic field} at fixed freeze-out time and temperature $T= 150$\,MeV \req{eq:banalyticapp} (black solid line) is independent of the beam energy over a wide range of $\gamma$. It begins to diverge slowly from the ultra-relativistic case at around $\gamma \leq 15$ for the time window shown in the figure. Lower beam energies result in a somewhat larger field at late times.  

As seen in \rf{fig:bcolcomp} the late-time magnetic field has a very soft dependence on the heavy-ion collision energy. However, the time $t_f$ at which QGP hadronization\index{QGP!hadronization} occurs, varies with collision energy; this has a much stronger effect on the magnitude of the freeze-out field.  As the QGP begins to hadronize at time $t_f$, one may expect hadrons to be statistically polarized with respect to the magnetic field. In \cite{Muller:2018ibh} the measured difference in global polarization of hyperons\index{hyperon!polarization} and anti-hyperons is used to give an upper bound on the magnetic field at QGP freeze-out, $B \sim 2.7\times 10^{-3}\,m_{\pi}^2$ for Au+Au collisions at $\sqrt{s_\text{NN}} = 200$\,GeV. Looking at \rf{fig:bcolcomp} the magnetic field for $\gamma = 100$ at QGP freeze-out $t_f \approx 5 $\,fm/c is predicted to be $B \approx 1.2\times 10^{-3}\,m_{\pi}^2$, somewhat below this upper bound. Note that this assumes the polarization rapidly equilibrates in the plasma. It also neglects any interactions during the hadron gas stage\index{hadrons gas}  of the collision. 

Since the remnant magnetic field at hadronization does not depend directly on the collision energy, an experimental measurement of the magnetic field at different collision energies could permit the determination of the electrical conductivity of the QGP or a determination of the freeze-out time of QGP if the conductivity is assumed to be known. As discussed at the beginning we are particularly interested in the primordial cosmic QGP conductivity which allows us to model the response of primordial plasma to strong magnetic fields. 
 
\subsection{Strange hadron abundance in cosmic plasma}
\label{Strangeness}\index{strangeness}
\para{Hadron populations in equilibrium}
As the Universe expanded and cooled down to the QGP Hagedorn temperature\index{Hagedorn!temperature} $T_H\approx150\MeV$, the primordial QGP underwent a phase transformation called hadronization. Quarks and gluons fragmented, combined and formed matter and antimatter we are familiar with. After hadronization\index{hadrons!hadronization}, one may think that all relatively short lived massive hadrons decay rapidly and disappear from the Universe. However, the most abundant hadrons\index{hadrons}, pions $\pi(q\bar q)$, can be produced via their inverse decay process $\gamma\gamma\rightarrow\pi^0$. Therefore they retain their chemical equilibrium down to $T=3\sim5\MeV$~\cite{Kuznetsova:2008jt}. 

{\color{black} This result demonstrates a key difference between laboratory heavy-ion experiments and the hot hadronic Universe: In the evolving Universe we allow for a full adjustment in the particle inventory to the ambient temperature, and we follow this inventory as a function of temperature with chemical potential(s) constrained by the baryon asymmetry, this is how the overview \rf{fig:energy:frac} was obtained. The key dynamic difference between Universe and laboratory is that in the former each particle has individual decoupling conditions while in the latter we study only strongly interacting particles, which all freeze-out near to phase cross-over from QGP to the hadron phase: Laboratory experimental data provide snap-shot image of QGP explosion into hadrons. For our study of laboratory environment see Ref.\,\cite{Letessier:2005qe,Petran:2013qla,Rafelski:2014cqa}. It is the mastery of the methods developed in the study of relativistic heavy-ion collisions that makes precise modeling of the primordial Universe possible.}

We begin by determining the Universe particle population composition assuming both kinetic and particle abundance equilibrium (chemical equilibrium) of non-interacting bosons and fermions. By considering the charge neutrality\index{charge neutrality} and a prescribed conserved baryon-per-entropy-ratio\index{baryon!entropy ratio} ${(n_B-n_{\overline{B}})}/{\sigma}$ we can determine the baryon chemical potential\index{chemical potential!baryon} $\mu_B$~\cite{Fromerth:2002wb,Fromerth:2012fe,Rafelski:2013yka}. We extend this approach allowing for the presence of strange hadrons, and imposing conservation of strangeness in the primordial Universe -- the strange quark content in hadrons must equal the anti-strange quark content in statistical average $\langle s-\bar s \rangle=0$. 

Given $\mu_B(T)$, $\mu_s(T)$ the baryon and strangeness chemical potentials as a function of temperature, we can obtain the particle number densities for different strange and non-strange species and study their population in the primordial Universe. Our approach prioritizes strangeness pair production into bound hadron states by either strong or electromagnetic interactions, or both, over the also possible weak interaction strangeness changing processes, which are capable to amplify the effect of baryon asymmetry. This is another topic beyond the scope of this work deserving further attention.

To characterize the baryon and strangeness content of a hadron we employ the chemical fugacity\index{fugacity!strangeness} for strangeness $\lambda_s$ and for light quarks $\lambda_q$ \index{quark!fugacity}
\begin{align}
\lambda_s=\exp(\mu_s/T)\,\quad \lambda_q=\exp(\mu_B/3T)\,.
\end{align}
Here $\mu_s$ and $\mu_B$ are the chemical potential of strangeness and baryon, respectively. To obtain quark fugacity\index{fugacity!quark} $\lambda_q$, we divide the baryo-chemical potential of baryons by quark content in the baryon, \ie\ three.

When the baryon chemical potential does not vanish, the chemical potential of strangeness in the primordial Universe is obtained by imposing the conservation of strangeness constraint $\langle s-\bar s \rangle=0$, see Section 11.5 in Ref.\,\cite{Letessier:2002ony}\index{strangeness! chemical potential} which leads considering only the dominant single strange hadrons to the condition
\begin{align}\label{museq}
\lambda_s=\lambda_q\sqrt{\frac{F_K+\lambda^{-3}_q\,F_Y}{F_K+\lambda^3_q\,F_Y}}\,.
\end{align}
Here we employ the phase-space function $F_i$ for sets of nucleon $N$, kaons $K$, and hyperon\index{hyperon} $Y$ particles 
\begin{align}
&F_N=\sum_{N_i}\,g_{N_i}W(m_{N_i}/T)\;, \quad N_i=n, p, \Delta(1232),\\
&F_K=\sum_{K_i}\,g_{K_i}W(m_{K_i}/T)\;, \quad K_i=K^0, \overline{K^0}, K^\pm, K^\ast(892),\\
&F_Y=\sum_{Y_i}\,g_{Y_i}W(m_{Y_i}/T)\;, \quad Y_i=\Lambda, \Sigma^0,\Sigma^\pm, \Sigma(1385),
\end{align}
$g_{N_i,K_i,Y_i}$ are the degeneracy factors, and the function $W(x)=x^2K_2(x)$ was introduced in \req{DensityH}.

Considering the massive particle number density in the Boltzmann\index{Boltzmann!approximation} approximation, we obtain
\begin{align}
\label{Density_N}
&n_N=\frac{T^3}{2\pi^2}\lambda_q^3F_N,\quad\qquad\qquad n_{\overline N}=\frac{T^3}{2\pi^2}\lambda^{-3}_qF_N,\\
\label{Density_K}
&n_K=\frac{T^3}{2\pi^2}\left(\lambda_s\lambda_q^{-1}\right)F_K,\,\qquad n_{\overline{K}}=\frac{T^3}{2\pi^2}\left(\lambda_s^{-1}\lambda_q\right)F_K,\\
\label{Density_Y}
&n_Y=\frac{T^3}{2\pi^2}\left(\lambda_q^2\lambda_s\right)F_Y,\quad\qquad n_{\overline Y}=\frac{T^3}{2\pi^2}\left(\lambda^{-2}_q\lambda_s^{-1}\right)F_Y.
\end{align}
In this case, the net baryon density in the primordial Universe with temperature range $150\MeV> T>10\MeV$ can be written as 
\begin{align}
\frac{\left(n_B-n_{\overline{B}}\right)}{\sigma}&=\frac{1}{\sigma}\left[\left(n_p-n_{\overline{p}}\right)+\left(n_n-n_{\overline{n}}\right)+\left(n_Y-n_{\overline{Y}}\right)\right]\notag\\
&=\frac{T^3}{2\pi^2\,\sigma}\left[\left(\lambda_q^3-\lambda^{-3}_q\right)F_N+\left(\lambda_q^2\lambda_s-\lambda^{-2}_q\lambda_s^{-1}\right)F_Y\right]\notag\\
&=\frac{T^3}{2\pi^2\sigma}\left(\lambda_q^3-\lambda_q^{-3}\right)F_N\left[1+\frac{\lambda_s}{\lambda_q}\left(\frac{\lambda_q^3-\lambda^{-1}_q\lambda_s^{-2}}{\lambda^3_q-\lambda^{-3}_q}\right)\,\frac{F_Y}{F_N}\right]\notag\\
&\approx\frac{T^3}{2\pi^2\sigma}\left(\lambda_q^3-\lambda_q^{-3}\right)F_N\left[1+\frac{\lambda_s}{\lambda_q}\,\frac{F_Y}{F_N}\right],
\end{align}
where we can neglect the term $F_Y/F_K$ in the expansion of \req{museq} in our temperature range. 

Introducing the strangeness conservation $\langle s-\bar s\rangle=0$ constraint and using the entropy density\index{entropy!density} in the primordial Universe, the explicit relation for baryon to entropy ratio becomes
\begin{align}\label{muBeq}
\frac{n_B-n_{\overline{B}}}{\sigma}&=\frac{45}{2\pi^4g^s_\ast}\sinh\left[\frac{\mu_B}{T}\right]F_N\times\left[1+\frac{F_Y}{F_N}\sqrt{\frac{1+e^{-\mu_B/T}\,F_Y/F_K}{1+e^{\mu_B/T}\,F_Y/F_K}}\right].
\end{align}
The present-day baryon-per-entropy-ratio is needed in \req{muBeq}, see \req{BaryonEntropyRatio}. It is convenient to also remember the quantum value of entropy per particle, for a massless bosons $\sigma/n|_\mathrm{boson}\approx 3.60$, and for a massless fermions $\sigma/n|_\mathrm{fermion}\approx 4.20$. We solve \req{museq}) and \req{muBeq} numerically to obtain baryon and strangeness chemical potentials as functions of temperature shown in the left-hand side frame of~\rf{Baryon:fig}.

\begin{figure}
\centerline{\includegraphics[width=0.50\linewidth]{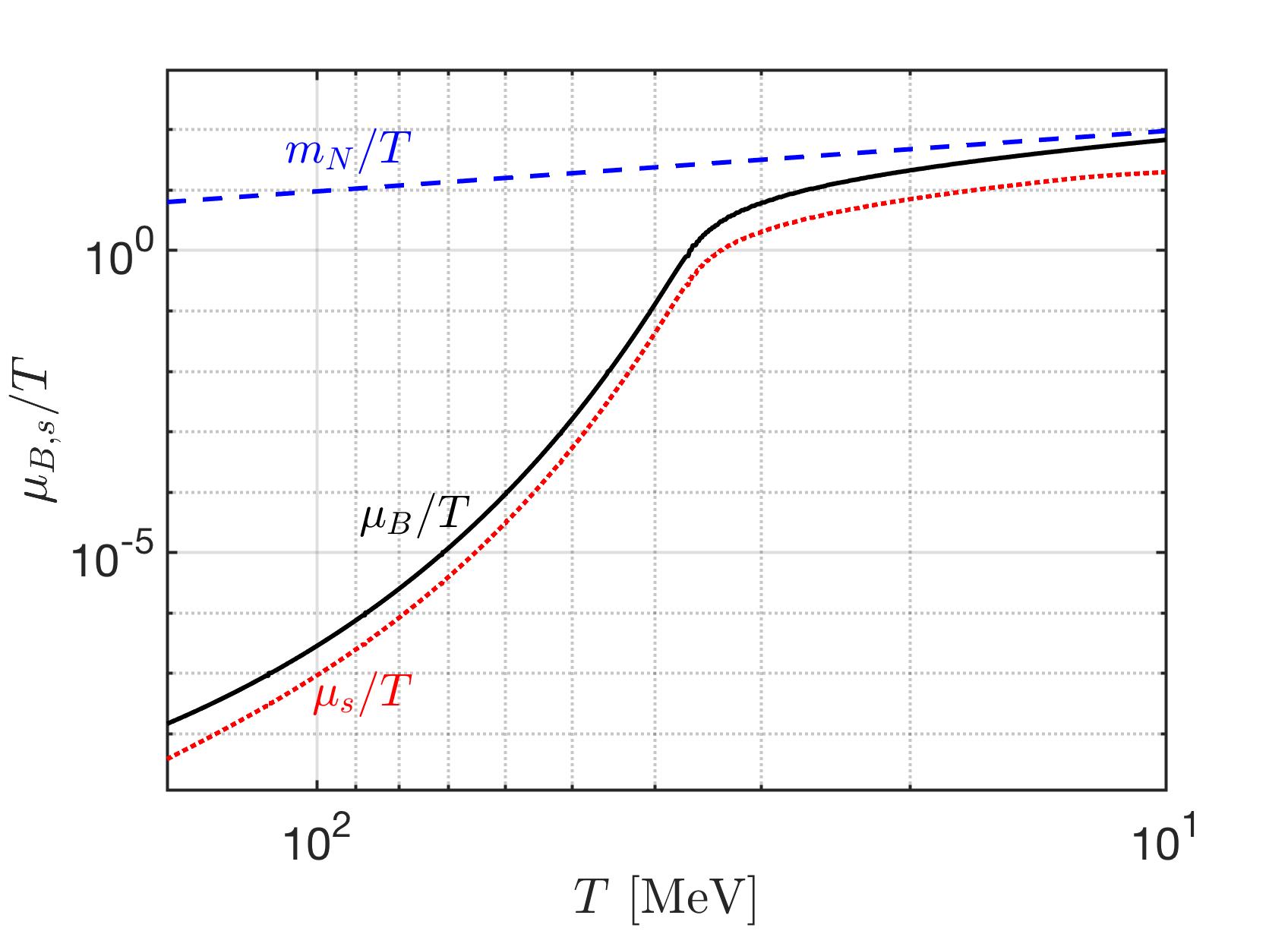}
\includegraphics[width=0.505\linewidth]{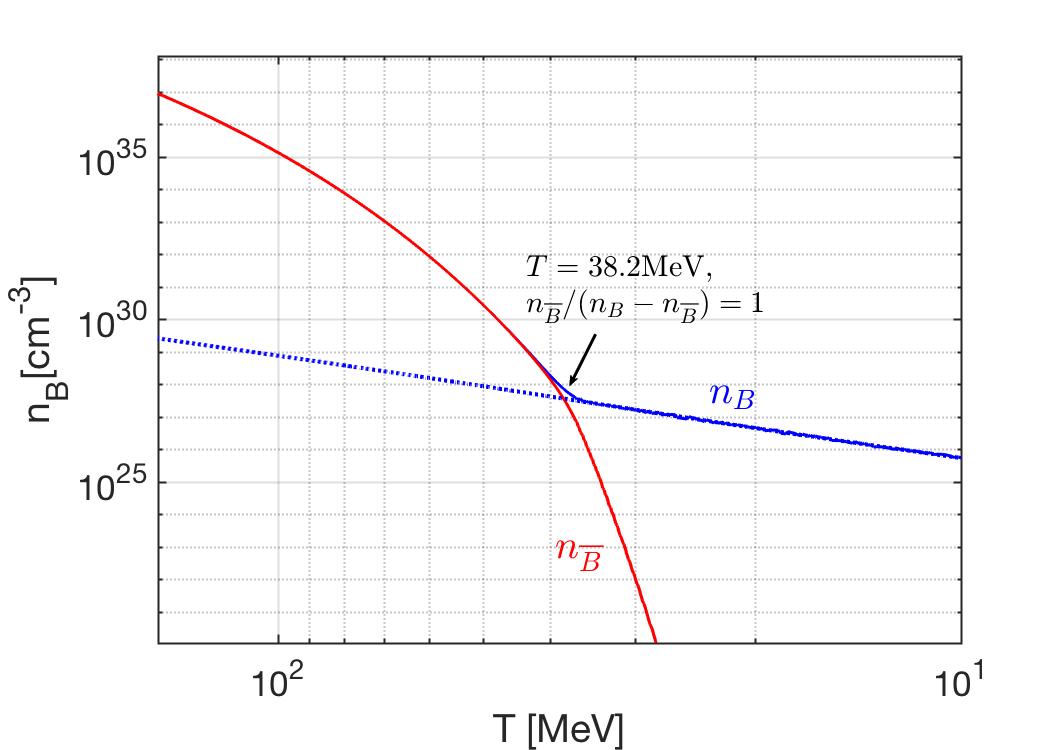}}
\caption{Left frame: Solid (black) line baryon number chemical potential $\mu_B/T$; dotted (red) line strangeness chemical potential $\mu_s/T$, as a function of temperature $150\MeV> T>10\MeV$ in the primordial Universe; for comparison we show dashed (blue) $m_N/T $ with $m_N=938.92\MeV$, the average nucleon mass. \cccite{Yang:2021bko}. \radapt{Yang:2024ret}
\index{strangeness!chemical potential} \index{chemical potential!baryon}
{\color{black} Right frame: Net baryon number $n_{B}-n_{\overline{B}}$ in the temperature range $150\MeV>T>5\MeV$, the dashed (blue) line turns into solid (blue) line depicting baryon density $n_{B}$ once the antibaryon $n_{\overline B}$ number density, solid line (red), diminishes; the transition point $n_{\overline B}/(n_B-n_{\overline B})=1$ is at $T=38.2\MeV$.\index{Antibaryons!in Universe} \cccite{Rafelski:2023emw}. \radapt{Yang:2024ret}}
}
\label{Baryon:fig}
\end{figure}

The shape of chemical potentials in the left-hand frame of ~\rf{Baryon:fig} changes dramatically, showing a knee in the temperature window $50\MeV\le T\le 30\MeV$. This behavior is describing the antibaryon\index{baryon!antibaryon} disappearance from the Universe inventory. Substituting the chemical potential $\lambda_q$ and $\lambda_s$ into particle density \req{Density_N}, \req{Density_K}, and \req{Density_Y}, we can obtain the particle number densities for different species as functions of temperature. 

The net baryon $n_B-n_{\overline B}$ number density is compared to the antibaryon number density in the right-hand frame of~\rf{Baryon:fig}. We note the transition point $n_{\overline B}\ll(n_B-n_{\overline B})=1$ at temperature $T=38.2\MeV$ marking the effective antibaryon disappearance temperature from the Universe inventory. This precisely computed result is in good agreement with the qualitative result presented in 1990 by Kolb and Turner~\cite{Kolb:1990vq}. Below this temperature, antibaryons rapidly disappear, while the baryon density dilutes due to Universe expansion. The baryon to entropy ratio remains constant.

In~\rf{EquilibPartRatiosFig} we present examples of hadronic particle abundance ratios of interest\index{hadrons!abundance ratios}. Pions $\pi(q\bar q)$ are the most abundant hadrons $n_\pi/n_B\gg1$, because of their low mass and the reaction $\gamma\gamma\rightarrow\pi^0$, which assures chemical yield equilibrium~\cite{Kuznetsova:2008jt} in the era of interest here. For $150\MeV>T>20.8\MeV$, we see the ratio $n_{{\overline K}(\bar q s)}/n_B\gg1$, which implies pair abundance of strangeness is more abundant than baryons, and is dominantly present in mesons, since $n_{\overline K}/n_Y\gg1$. This also implies that the exchange reaction $\overline{K}+N\rightarrow \Lambda+\pi$ can re-equilibrate kaons and hyperons; therefore strangeness symmetry $s=\bar s$ can be maintained. Considering $n_Y/n_B$ we see that hyperons\index{hyperon} $Y(sqq)$ remain a noticeable 1\% component in the baryon yield through the domain of antibaryon decoupling. Below $T=12.9\MeV$ we have $n_Y/n_B>n_{\overline K}/n_B$, now the still existent tiny abundance of strangeness is found predominantly in hyperons.

\begin{figure} 
\centerline{
\includegraphics[width=0.8\linewidth]{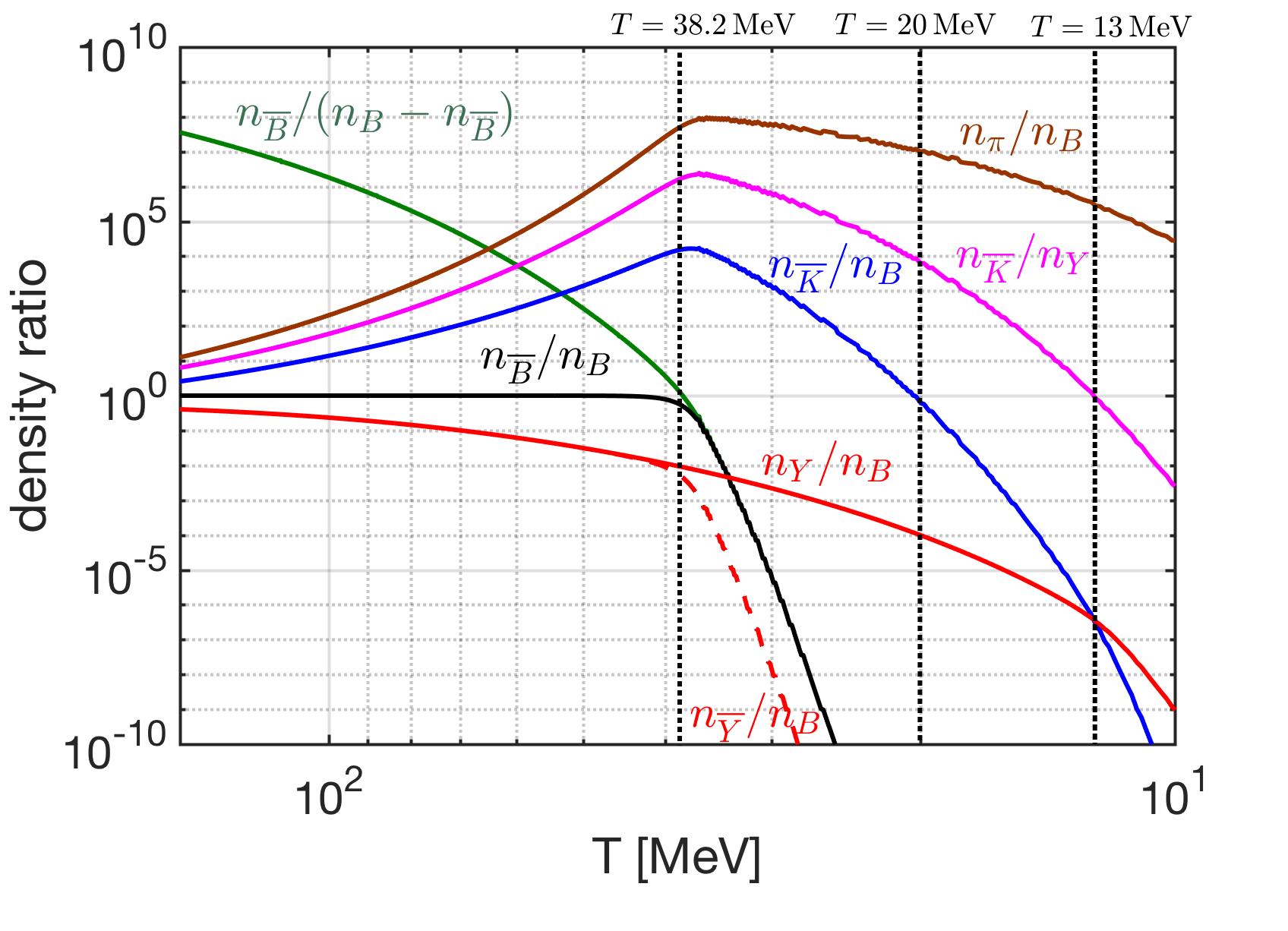}}
\caption{Ratios of hadronic particle number densities with baryon $B$ yields as a function of temperature $150\MeV> T>10\MeV$: Pions $\pi$ (brown line), kaons $K( q\bar s)$ (blue), antibaryon $\overline B$ (black), hyperon $Y$ (red) and anti-hyperons $\overline Y$ (dashed red). Also shown $\overline K/Y$(purple). Vertical lines highlight events described in text. \index{strange!hadrons} 
\cccite{Rafelski:2023emw}. \radapt{Yang:2021bko}}
\label{EquilibPartRatiosFig} 
\end{figure}

\para{Strange hadron dynamic population}
Given the equilibrium abundances\index{strangeness!dynamic population} of hadrons in the epoch of interest $150\MeV\ge T\ge 10\MeV$, we turn now to study the freeze-out temperature for different particles and strangeness by comparing the relevant reaction rates with each other and with the Hubble expansion rate. We will need to explore a large number of reactions, going well beyond the relative simplicity of the case of QGP phase of matter studied in relativistic heavy ion collisions. We find that strangeness in hadronic particles is kept in equilibrium in the primordial Universe down until $T\approx 13\MeV$. This study addresses non-interacting particles; nuclear interactions can be many times greater compared to this temperature. Thus further exploration of this result seems necessary in the future.

Let us first consider an unstable strange particle $S$ decaying into two particles $1$ and $2$, which themselves have no strangeness content. In a dense and high-temperature plasma with particles $1$ and $2$ in thermal equilibrium, the inverse reaction populates the system with particle $S$. This is written schematically as
\begin{align}
 S\Longleftrightarrow1+2,\qquad \mathrm{Example}: K^0\Longleftrightarrow\pi+\pi\,.
\end{align}
As long as both decay and production reactions are possible, particle $S$ abundance remains in thermal equilibrium; as already discussed this balance between production and decay rates is the `detailed balance'.

Once the primordial Universe expansion rate $1/H$ overwhelms the strongly temperature dependent back-reaction and the back-reaction freeze-out, then the decay $S\rightarrow 1+2$ occurs out of balance and particle $S$ disappears rather rapidly from the inventory. 

Next on our list are the two-to-two strangeness producing and burn-up reactions. These have significantly higher strangeness production reaction thresholds, thus especially near to strangeness decoupling, their influence is negligible. Such reactions are more important near the QGP hadronization temperature $T_H\simeq 150\MeV$. A typical strangeness exchange reaction is $\mathrm{K}+N\leftrightarrow \Lambda+\pi$, (see Chapter 18 in~Ref.\,\cite{Letessier:2002ony}).

In~\rf{Strangeness_map2} we show some reactions relevant to strangeness evolution in the considered Universe evolution epoch $150\MeV\ge T\ge 10\MeV$ and their pertinent reaction strength. Specifically:
\begin{itemize}
\item
We study strange quark abundance in baryons and mesons, considering both open and hidden strangeness (hidden: $s\bar s$-content). Important source reactions are $l^-+l^+\rightarrow\phi$, $\rho+\pi\rightarrow\phi$, $\pi+\pi\rightarrow K_\mathrm{S}$, $\Lambda \leftrightarrow \pi+ N$, and $\mu^\pm+\nu\rightarrow K^\pm$. 
\item
Muons\index{muon} and pions are coupled through electromagnetic reactions $\mu^++\mu^-\leftrightarrow\gamma+\gamma$ and $\pi\leftrightarrow\gamma+\gamma$ to the photon background and retain their chemical equilibrium\index{chemical equilibrium} until the temperature $T =4$\, MeV and $T=5\MeV$, respectively~\cite{Rafelski:2021aey,Kuznetsova:2008jt}. The large $\phi\leftrightarrow K+K$ rate assures $\phi$ and $K$ are in relative chemical equilibrium.
\end{itemize}

\begin{figure} 
\centerline{\includegraphics[width=0.85\linewidth]{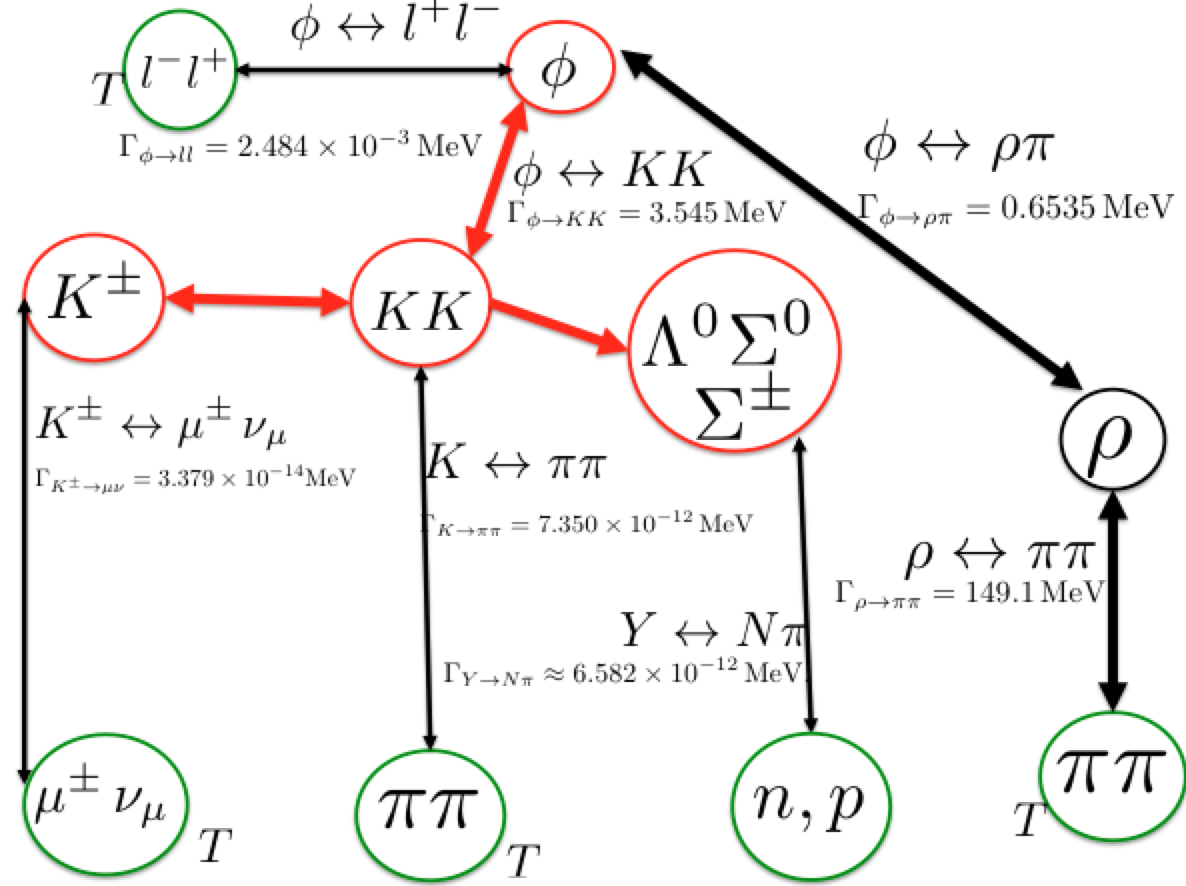}}
\caption{The strangeness abundance changing reactions in the primordial Universe. The central circles (red) show strangeness carrying hadronic particles; red thick lines denote effectively instantaneous reactions. Black thick lines show relatively strong hadronic reactions. \cccite{Rafelski:2023emw}. \radapt{Yang:2024ret,Yang:2021bko}}
\label{Strangeness_map2} 
\end{figure}

In order to determine where exactly strangeness disappears from the Universe inventory, we explore the magnitudes of different rates of production and decay processes in mesons and hyperons and present results in figures. The reaction rates required to describe strangeness time evolution are presented in Ref.~\cite{Rafelski:2020ajx}. 
 
\para{Strangeness creation and annihilation rates in mesons}
From~\rf{Strangeness_map2} in the meson domain, the relevant interaction rates competing with Hubble time\index{Hubble!time} are the reactions\index{strangeness!mesons production rate}
\begin{align}
 &\pi+\pi\leftrightarrow K\,,\quad\mu^\pm+\nu\leftrightarrow K^\pm\,,\quad l^++l^-\leftrightarrow\phi\,,\\
 &\rho+\pi\leftrightarrow\phi\,,\quad \pi+\pi\leftrightarrow\rho\,.
\end{align}
The thermal reaction rate per time and volume for two body-to-one particle reactions $1+2\rightarrow 3$ has been presented before~\cite{Koch:1986ud,Kuznetsova:2008jt,Kuznetsova:2010pi}. 

In full kinetic and chemical equilibrium, the reaction rate per time per volume can be written as~\cite{Kuznetsova:2010pi} :\index{inverse decay rate}
\begin{align}
&R_{12\to 3}=\frac{g_3}{(2\pi)^2}\,\frac{m_3}{\tau^0_3}\,\int^\infty_0\frac{p^2_3dp_3}{E_3}\frac{e^{E_3/T}}{e^{E_3/T}\pm1}\Phi(p_3)\;,
\end{align}
where $\tau^0_3$ is the vacuum lifetime of particle $3$. The positive sign `$+$' is for the case when particle $3$ is a boson, and negative sign `$-$' for a fermion. The function $\Phi(p_3)$ for the nonrelativistic limit $m_3\gg p_3,T$ can be written as 
\begin{align}
\Phi(p_3\to0)=2\frac{1}{(e^{E_1/T}\pm1)(e^{E_2/T}\pm1)}.
\end{align}

Considering the Boltzmann\index{Boltzmann!approximation} limit, the thermal reaction rate per unit time and volume becomes
\begin{align}
\label{Thermal_Rate}
R_{12\rightarrow3}=\frac{g_3}{2\pi^2}\left(\frac{T^3}{\tau^0_3}\right)\left(\frac{m_3}{T}\right)^2\,K_1(m_3/T),
\end{align}
where $K_1$ is the modified Bessel\index{Bessel function} functions of integer order `$1$'. 

In order to compare the reaction time with Hubble time $1/H$, it is convenient to define the relaxation time for the process $1+2\rightarrow 3$ as follows:
\begin{align}
\label{Reaction_Time}
\tau_{12\rightarrow 3}\equiv\frac{n^{eq}_{1}}{R_{12\rightarrow n}}\,,\quad
n^{eq}_1=\frac{g_1}{2\pi^2}\int_{m_1}^\infty\!\!\!\!dE\,\frac{E\,\sqrt{E^2-m_1^2}}{\exp{\left(E/T\right)}\pm1}\;, 
\end{align}
where $n^{eq}_1$\,is the thermal equilibrium number density of particle\,$1$ with the `heavy' mass $m_1>T$. Combining \req{Thermal_Rate} with \req{Reaction_Time} we obtain
\begin{align}\label{RelaxationTime}
&\frac{\tau_{12\rightarrow3}}{ \tau^0_3}= 
\frac{2\pi^2 n^{eq}_1/T^3}{g_3(m_3/T)^2\,K_1(m_3/T)}\,, \quad 
n^{eq}_1\simeq g_1\left(\frac{m_1 T}{2\pi}\right)^{3/2}e^{-m_1/T}\,,
\end{align}
where, conveniently, the relaxation time does not depend on the abundant and often relativistic heat bath component $2$, \eg\ $l^\pm,\pi,\nu,\gamma$. The density of heavy particles\,$1$\,and\,$3$ can in general be well approximated using the leading and usually nonrelativistic Boltzmann term as shown above.

In general, the reaction rates for an inelastic collision process capable of changing particle number, for example $\pi\pi\to K^0$, are suppressed by the factor $\exp{(-m_{K^0}/T)}$. On the other hand, there is no suppression for the elastic momentum and energy exchanging particle collisions in plasma. In general for the case $m\gg T$, the dominant collision term in the relativistic Boltzmann equation is the elastic collision term, keeping all heavy particles in kinetic energy equilibrium with the plasma. This allows us to study the particle abundance in plasma presuming the energy-momentum statistical distribution equilibrium shape exists. This insight was discussed in detail in the preparatory phase of laboratory exploration of hot hadron and quark matter, see~\cite{Koch:1986ud}. 

In order to study the particle abundance in the Universe when $m\gg T$, instead of solving the exact Boltzmann equation, we can separate the fast energy-momentum equilibrating collisions from the slow particle number changing inelastic collisions. This approach makes it possible to explore the rates of inelastic collision and compare the relaxation times of particle production in all relevant reactions with the Universe expansion rate at a fixed temperature which governs the shape of particle distributions.

It is common to refer to particle freeze-out as the epoch where a given type of particle ceases to interact with other particles. In this situation the particle abundance decouples from the cosmic plasma, a chemical nonequilibrium and even complete abundance disappearance of this particle can accompany this; the condition for the given reaction $1+2\rightarrow 3$ to decouple is
\begin{align}
\tau_{12\rightarrow 3}(T_f)=1/H(T_f),
\end{align}
where $T_f$ is the freeze-out temperature.

In the epoch of interest, $150\MeV>T>10\MeV$, the Universe is dominated by radiation and effectively massless matter behaving like radiation. The Hubble parameter can be obtained from the Hubble equation\index{Hubble!equation}
\req{Hubble:eq} and written as~\cite{Kolb:1990vq}
\begin{align}\label{H2g}
H^2=H^2_{rad}\left(1+\frac{\rho_{\pi,\,\mu,\,\rho}}{\rho_\mathrm{rad}}+\frac{\rho_\mathrm{strange}}{\rho_\mathrm{rad}}\right)=\frac{8\pi^3G_\mathrm{N}}{90}g^e_\ast T^4,\qquad H^2_\mathrm{rad}=\frac{8\pi G_\mathrm{N}\,\rho_\mathrm{rad}}{3},
\end{align}
where: $g^e_\ast$ is the total number of effective relativistic `energy' degrees of freedom; $G_\mathrm{N}$ is the Newtonian constant of gravitation; the `radiation' energy density includes $\rho_\mathrm{rad}=\rho_\gamma+\rho_\nu+\rho_{e^\pm}$ for photons, neutrinos, and massless electrons(positrons). The massive-particle correction is $\rho_{\pi,\,\mu,\,\rho}=\rho_\pi+\rho_\mu+\rho_\rho$; and at highest $T$ of interest, also of (minor) relevance, $\rho_\mathrm{strange}=\rho_{K^0}+\rho_{K^\pm}+\rho_{K^\ast}+\rho_{\eta}+\rho_{\eta^\prime}$.
Equating $1/H$ to the computed reaction rate we obtain the freeze-out temperature $T_f$. 

When considering the reaction rates and quoting $T_f$, we must check allowing for a finite reaction time how sudden the freeze-out happens. We refer to this temperature uncertainty as $\Delta T_f$, which by a simple scale consideration can be defined by
\begin{equation}\label{eq:DeltaT}
\Delta T_f\simeq \frac{1}{R(T_f)}\times \frac{dT}{dt}\,. 
\end{equation}
$R$\,[MeV] is the value of reaction rate at freeze-out. The greater is the rate $R_f$ the sharper is the freeze-out, thus smaller $\Delta T_f$.\index{freeze-out!uncertainty}

For the temperature range $50\MeV>T>5\MeV$, we have $10^{-1}<dT/dt<10^{-4}\MeV$/$\mu$s. We estimate the width of the freeze-out temperature interval $\Delta T_f$ using reaction rates for $dt$ as follows
\begin{align}
\frac{1}{\Delta T_f}\equiv \left[\frac{1}{(\Gamma_{12\to3}/H)}\frac{d(\Gamma_{12\to3}/H)}{dT}\right]_{T_f},\quad \Gamma_{12\to3}\equiv\frac{1}{\tau_{12\to3}}.
\end{align}
Using \req{H2g} and \req{RelaxationTime} and considering the temperature range $50\MeV>T>5\MeV$ with $g^e_\ast\approx\mathrm{constant}$ we obtain, using the Boltzmann approximation to describe the massive particles\,$1$\,and\,$3$,
\begin{align}\label{DeltaFreezeout}
 \frac{\Delta T_f}{ T_f} \approx\frac{T_f }{ m_3 - m_1 -2T_f}\,,\quad m_3 - m_1>> T_f\,.
\end{align}
The width of the freeze-out domain is shown in the right column in Table~\ref{FreezeoutTemperature_table}. We see a range of $2$-$10\%$. Therefore it is nearly justified to consider as a decoupling condition in time the value of temperature at which the pertinent rate crosses the Hubble expansion rate, see~\rf{reaction_time_tot}.\index{freeze-out!duration}
 
\begin{table} 
\centering
\begin{tabular}{c| c| c}
\hline\hline
Reactions &Freeze-out $T_f$\,[MeV] & {Uncertainty $\Delta T_f$\,[MeV]} \\
\hline
$\mu^\pm\nu\rightarrow K^\pm$ & $T_f=33.8\MeV$ & {$3.5$ \,MeV}\\ 
\hline
$e^+e^-\rightarrow \phi$ & $T_f=24.9\MeV$ &{$0.6\MeV$}\\
$\mu^+\mu^-\rightarrow\phi$ & $T_f=23.5\MeV$ &{$0.6\MeV$}\\
\hline
 $\pi\pi\rightarrow K$ & $T_f=19.8\MeV$&{$1.2\MeV$}\\
\hline
$\pi\pi\rightarrow\rho$ & $T_f=12.3\MeV$&{$0.2\MeV$}\\
\hline\hline
\end{tabular}
\caption{Strangeness producing reactions in the primordial Universe, their freeze-out temperature $T_f$; and temperature uncertainty $\Delta T_f$}
\label{FreezeoutTemperature_table} 
\end{table}

\begin{figure}
\centerline{\includegraphics[width=0.5\linewidth]{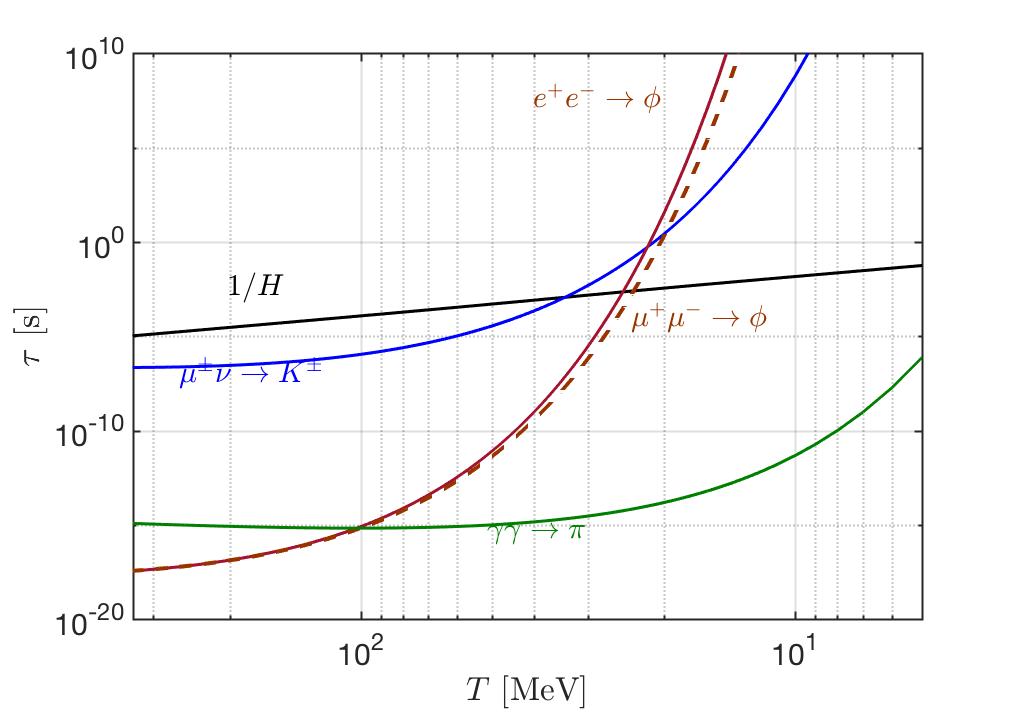}
\includegraphics[width=0.5\linewidth]{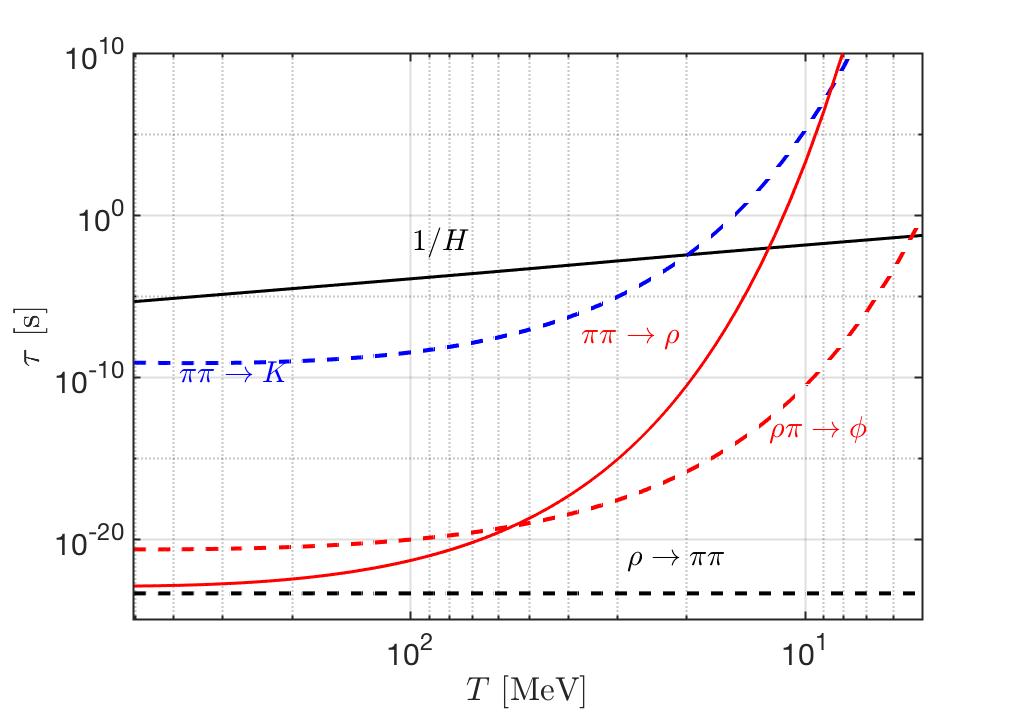}}
\caption{Hadronic relaxation reaction times, see \req{Reaction_Time}, as a function of temperature $T$, are compared to Hubble time $1/H$ (black solid line). On right the bottom horizontal black-dashed line is the natural (vacuum) lifespan of $\rho$. \cccite{Rafelski:2023emw}. \radapt{Yang:2024ret,Yang:2021bko}}
\label{reaction_time_tot} 
\end{figure}

In~\rf{reaction_time_tot} we plot the hadronic reaction relaxation times $\tau_{i}$ in the meson sector as a function of temperature compared to Hubble time $1/H$. We note that the weak interaction reaction $\mu^\pm+\nu_{\mu}\rightarrow K^\pm$ becomes slower compared to the Universe expansion near temperature $T_f^{K^\pm}=33.8\MeV$, signaling the onset of abundance nonequilibrium for $K^\pm$. For $T<T_f^{K^\pm}$, the reactions $\mu^\pm+\nu_{\mu}\rightarrow K^\pm$ decouples from the cosmic plasma; the corresponding detailed balance\index{detailed balance} can be broken and the decay reactions $K^\pm\rightarrow\mu^\pm+\nu_{\mu}$ are acting like a (small) ``hole'' in the strangeness abundance ``pot''. If other strangeness production reactions did not exist, strangeness would disappear as the Universe cools below $T_f^{K^\pm}$. However, there are other reactions: $l^++l^-\leftrightarrow\phi$, $\pi+\pi\leftrightarrow K$, and $\rho+\pi\leftrightarrow\phi$ can still produce the strangeness in cosmic plasma and the rate is very large compared to the weak interaction decay.

In Table~\ref{FreezeoutTemperature_table} we also show the characteristic strangeness reactions and their freeze-out temperatures in the primordial Universe. The intersection of strangeness reaction times with $1/H$ occurs for $l^-+l^+\rightarrow\phi$ at $T_f^\phi=25\sim23\MeV$, and for $\pi+\pi\rightarrow K$ at $T_f^K=19.8\MeV$, for $\pi+\pi\rightarrow\rho$ at $T_f^\rho=12.3\MeV$. The reactions $\gamma+\gamma\rightarrow\pi$ and $\rho+\pi\leftrightarrow\phi$ are faster compared to $1/H$. However, the $\rho\to\pi+\pi$ lifetime (black dashed line in~\rf{reaction_time_tot}) is smaller than the reaction $\rho+\pi\leftrightarrow\phi$; in this case, most the of $\rho$-mesons decays faster, thus are absent and cannot contribute to the strangeness creation in the meson sector. Below the temperature $T<20\MeV$, all the detail balances in the strange meson reactions are broken and the strangeness in the meson sector should disappear rapidly, were it not for the small number of baryons present in the Universe.

\para{Strangeness production and exchange rates involving hyperons}
In order to understand strangeness\index{strangeness!hyperons} in hyperons in the baryonic domain, we now consider the strangeness production reaction $\pi +N\rightarrow K+\Lambda$, the strangeness exchange reaction $\overline{K}+N\rightarrow \Lambda+\pi$; and the strangeness decay $\Lambda\rightarrow N+\pi$. The competition between different strangeness reactions allows strange hyperons and anti-hyperons to influence the dynamic nonequilibrium condition, including development of $\langle s-\bar s\rangle \ne 0$.

To evaluate the reaction rate in two-body reaction $1+2\rightarrow3+4$ in the Boltzmann approximation\index{Boltzmann!approximation} we can use the reaction cross-section $\sigma(s)$ and the relation~\cite{Letessier:2002ony}:\index{hyperon!production rate}
\begin{align}
R_{12\rightarrow34}=\frac{g_1g_2}{32\pi^4}\frac{T}{1+I_{12}}\!\!\int^\infty_{s_{th}}\!\!\!\!ds\,\sigma(s)\frac{\lambda_2(s)}{\sqrt{s}}\!K_1\!\!\left({\sqrt{s}}/{T}\right),
\end{align}
where $K_1$ is the Bessel\index{Bessel function} function of order $1$ and the function $\lambda_2(s)$ is defined as
\begin{align}
\lambda_2(s)=\left[s-(m_1+m_2)^2\right]\left[s-(m_1-m_2)^2\right],
\end{align}
with $m_1$ and $m_2$, $g_1$ and $g_2$ as the masses and degeneracy of the initial interacting particle. The factor $1/(1+I_{12})$ is introduced to avoid double counting of indistinguishable pairs of particles; we have $I_{12}=1$ for identical particles and $I_{12}=0$ for others. 

The thermal averaged cross-sections for the strangeness production and exchange processes are about $\sigma_{\pi N\rightarrow K\Lambda}\sim0.1\,\mathrm{mb}$ and $\sigma_{\overline{K}N\rightarrow \Lambda\pi}=1\sim3\,\mathrm{mb}$ in the energy range in which we are interested~\cite{Koch:1986ud}. The cross-section can be parameterized as follows:\\
1) For the cross-section $\sigma_{\overline{K}N\rightarrow \Lambda\pi}$ we use~\cite{Koch:1986ud}
 \begin{align}
 \sigma_{\overline{K}N\rightarrow \Lambda\pi}=\frac{1}{2}\left(\sigma_{K^-p\rightarrow \Lambda\pi^0}+\sigma_{K^-n\rightarrow \Lambda\pi^-}\right)\,.
\end{align}
Here the experimental cross-sections can be parameterized as 
\begin{align}
&\sigma_{K^-p\rightarrow \Lambda\pi^0}\!\!=\!\!\left\{\begin{array}{l}\!\!1479.53\mathrm{mb}\!\cdot\!\displaystyle e^{\left(\frac{-3.377\sqrt{s}}{\mathrm{GeV}}\right)}\qquad   \mathrm{for\ } \sqrt{s_m}\!\!<\!\!\sqrt{s}\!<\!3.2\mathrm{GeV}\,, \\[0.5cm]
0.3\mathrm{mb}\!\cdot\!\displaystyle e^{\left(\frac{-0.72\sqrt{s}}{\mathrm{GeV}}\right)}\qquad\qquad   \mathrm{for\ }\sqrt{s}>3.2\mathrm{GeV}\,,\end{array}\right.
\\[0.4cm]
&\sigma_{K^-n\rightarrow \Lambda\pi^-}\!\!=\!\!1132.27\mathrm{mb}\!\cdot\!e^{\left(\frac{-3.063\sqrt{s}}{\mathrm{GeV}}\right)} \qquad\   \mathrm{for\ }\sqrt{s}>1.699\mathrm{GeV}\,,
\end{align}
where $\sqrt{s_m}=1.473\GeV$.\\
2) For the cross-section $\sigma_{\pi N\rightarrow K\Lambda}$ we use~\cite{Cugnon:1984pm}
\begin{align}
&\sigma_{\pi N\rightarrow K\Lambda}=\frac{1}{4}\times\sigma_{\pi p\rightarrow K^0\Lambda}\,.
\end{align}
The experimental $\sigma_{\pi p\rightarrow K^0\Lambda}$ can be approximated as follows
\begin{align}
\sigma_{\pi p\rightarrow K^0\Lambda}=\left\{
\begin{array}{l}
\displaystyle\frac{0.9\mathrm{mb}\cdot\left(\sqrt{s}-\sqrt{s_0}\right)}{0.091\mathrm{GeV}}\qquad  \mathrm{for\ } \sqrt{s_0}<\sqrt{s}<1.7\mathrm{GeV}\,, \\[0.5cm]
\displaystyle\frac{90\mathrm{MeV\cdot mb}}{\sqrt{s}-1.6\mathrm{GeV}}\qquad\qquad \quad  \mathrm{for\
 }\sqrt{s}>1.7\mathrm{GeV}\,,\end{array}\right.
 \,,
 \end{align}
 with $ \sqrt{s_0}=m_\Lambda+m_K$. 

Given the cross-sections, we obtain the thermal reaction rate per volume for strangeness exchange reaction seen in~\rf{Lambda_Rate_volume.fig}. We see that near to $T=20\MeV$, the dominant reactions for the hyperon\index{hyperon} $\Lambda$ production is $\overline{K}+N\leftrightarrow\Lambda+\pi$. At the same time, the $\pi+\pi\to K$ reaction becomes slower than Hubble time and kaon $K$ decay rapidly in the primordial Universe. However, the anti-kaons $\overline K$ produce the hyperon $\Lambda$ because of the strangeness exchange reaction $\overline{K}+N\rightarrow\Lambda+\pi$ in the baryon-dominated Universe. We have strangeness in $\Lambda$ and it disappears from the Universe via the decay $\Lambda\rightarrow N+\pi$. Both strangeness and anti-strangeness disappear because of the $K\rightarrow\pi+\pi$ and $\Lambda\rightarrow N+\pi$, while the strangeness abundance $s = \bar{s}$ in the primordial Universe remains.

\begin{figure} 
\centerline{\includegraphics[width=0.7\linewidth]{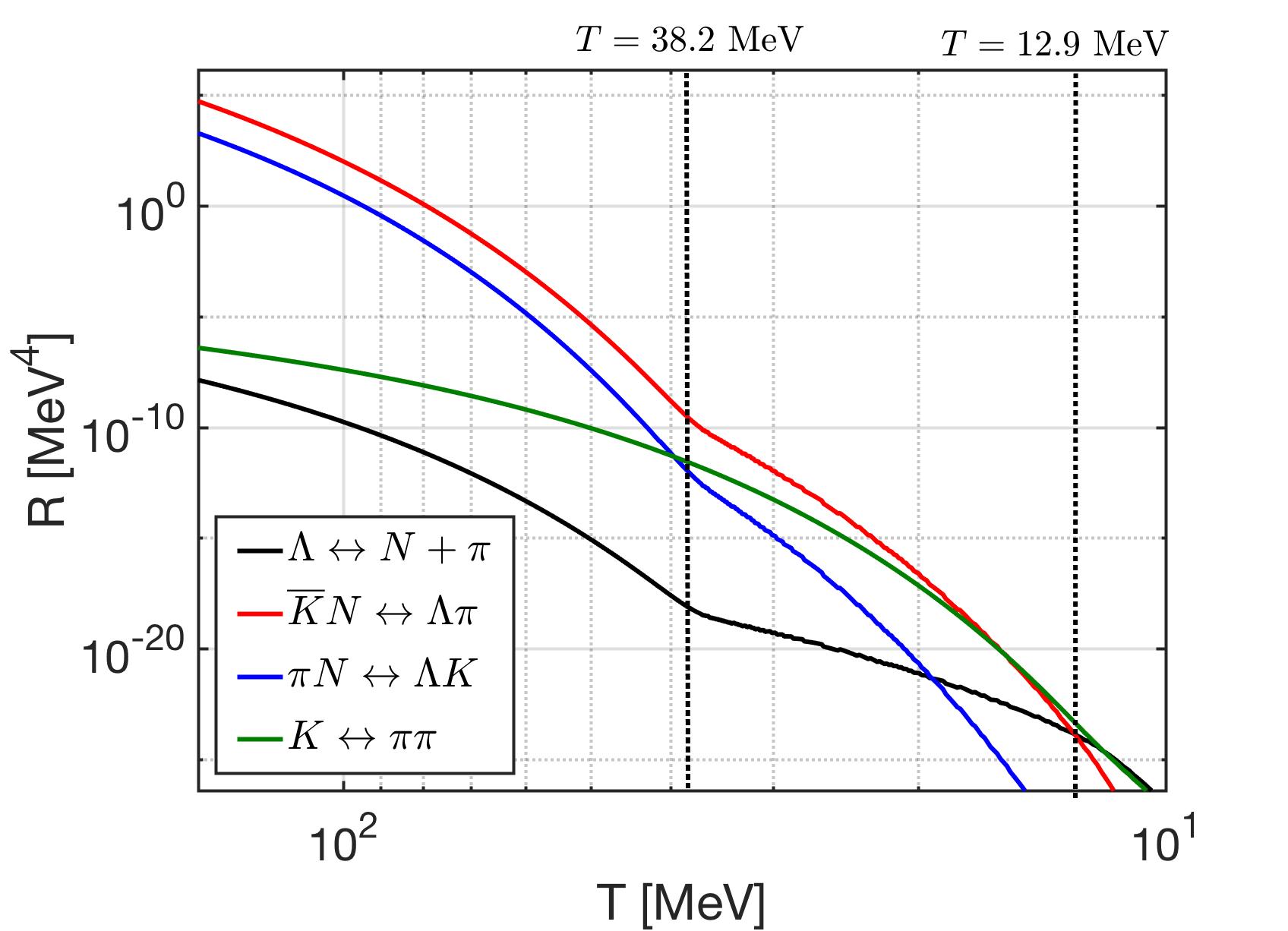}}
\caption{Thermal reaction rate $R$ per volume and time for important hadronic strangeness production and exchange processes as a function of temperature $150\MeV> T>10\MeV$ in the primordial Universe. \cccite{Rafelski:2023emw}. \radapt{Yang:2024ret,Yang:2021bko}}
\label{Lambda_Rate_volume.fig} 
\end{figure}

Near to $T=12.9\MeV$ the reaction $\Lambda+\pi\rightarrow\overline{K}+N$ becomes slower than the strangeness decay $\Lambda\leftrightarrow N+\pi$ and shows that at the low temperature the $\Lambda$ particles are still in equilibrium via the reaction $\Lambda\leftrightarrow N+\pi$ and little strangeness remains in the $\Lambda$. Then strangeness abundance becomes asymmetric $s\gg \bar{s}$, which implies that the assumption for strangeness conservation can only be valid until the temperature $T\sim13\MeV$. Below this temperature a new regime opens up in which the tiny residual strangeness abundance is governed by weak decays with no re-equilibration with mesons. Also, in view of baron asymmetry, $\langle s-\bar s\rangle \ne 0$.

{\color{black}
\para{Strangeness comparison: laboratory with the Universe}
In laboratory experiments we study abundance of hadrons present at the instant of chemical freeze-out measuring the value of $T_\mathrm{ch}$, \req{equilibrium}. The dynamic explosive QGP fireball disintegration, often referred to as `sudden hadronization' spans a temperature range which is smaller than the range $155\MeV >T_\mathrm{ch}>140\MeV$, the observed production domain which populates particle abundances near to the QGP breakup condition. There is no background of photons, leptons, or neutrinos. 

Said differently, in the laboratory heavy-ion experiments: a) Only strongly interacting degrees of freedom are normally present; b) Particle yields follow from a single chemical freeze-out near to QGP-hadron gas phase cross-over; c) Detailed particle yields allow us to measure the dynamically created chemical potentials; d) An analysis of laboratory experimental data creates a snap-shot image of the abundance freeze-out instant.

In the evolving Universe  we allow for a full adjustment of particle yields to the ambient temperature of the dynamically expanding Universe, and we follow these yields as a function of temperature with chemical potential(s) constrained by the Universe baryon asymmetry. Each particle
has an individual decoupling temperature, with unstable particles usually rapidly disappearing after decoupling, while stable decoupled particles are free-streaming. At low temperature we must consider reactions involving thermal photon, lepton, neutrino background.

A comparison between the early Universe particle inventory and laboratory experiments is implicit in yields seen on the left margin of \rf{EquilibPartRatiosFig}, except for $n_{\overline{B}}/(n_B-n_{\overline{B}})$, which depends on baryon content of the Universe. In this figure the ordinate scale changes by many orders of magnitude since we track the yields over a wide range of temperatures. This prevents us from reading off relevant to heavy-ion collisions results, published in many  other works in past decades.

What this means is that our study of evolving abundances in the Universe is entirely different from laboratory experiments  using same theoretical methods. A detailed comparison between the early Universe particle inventory and laboratory experiments is not insightful.
} \section{Neutrino Plasma}\label{Neutrino}
\subsection{Neutrino properties and reactions}\label{ssec:nuproperties}
Neutrinos are fundamental particles which play an important role in the evolution of the Universe. In the early Universe the neutrinos are kept in equilibrium with cosmic plasma via the weak interaction. The neutrino-matter interactions play a crucial role in our understanding of neutrinos evolution in the early Universe (such as neutrino freeze-out) and the later Universe (the property of today's neutrino background). In this chapter, we will examine the neutrino coherent and incoherent scattering with matter and their application in cosmology. The investigation of the relation between the effective number of neutrinos\index{neutrino!effective number} $N^{\mathrm{eff}}_\nu$ and lepton asymmetry\index{lepton!asymmetry} $L$ after neutrino freeze-out\index{neutrino!freeze-out} and its impact on Universe expansion is also discussed in this chapter. 

\para{Matrix elements for neutrino coherent \& incoherent scattering}
 According to the standard model, neutrinos interact with other particles via the Charged-Current(CC) and Neutral-Current(NC) interactions. Their Lagrangian can be written as~\cite{Giunti:2007ry}
\begin{align}
&\mathcal{L}^{CC}=\frac{g}{2\sqrt{2}}\left(j^\mu_W\,W_\mu+{j^\mu_W}^\dagger\,W^\dagger_\mu\right),\qquad\mathcal{L}^{NC}=-\frac{g}{2\cos{\theta_W}}\,j^\mu_Z\,Z_\mu,
\end{align}
where $g=e\sin\theta_W$, $W^\mu$ and $Z^\mu$ are W and Z boson gauge fields, and $j^\mu_W$ and $j^\mu_Z$ are the charged-current and neutral-current, respectively. In the limit of energies lower than the $W(m_W=80\,\mathrm{GeV})$ and $Z(m_z=91\,\mathrm{GeV})$ gauge bosons, the effective Lagrangians are given by
\begin{align}\label{L_low}
\mathcal{L}^{CC}_{eff}=-\frac{G_F}{\sqrt{2}}\,j^\dagger_{W\,\mu}\,j^\mu_W,\qquad
\mathcal{L}^{NC}_{eff}=-\frac{G_F}{\sqrt{2}}\,j^\dagger_{Z\,\mu}\,j^\mu_Z,\qquad \frac{G_F}{\sqrt{2}}=\frac{g^2}{8m^2_W},
\end{align}
where $G_F=1.1664\times10^{-5}\,\mathrm{GeV}^{-2}$ is the Fermi constant, which is one of the important parameters that determine the strength of the weak interaction rate. When neutrinos interact with matter, based on the neutrino's wavelength, they can undergo two types of scattering processes: coherent scattering and incoherent scattering with the particles in the medium. 

With coherent scattering, neutrinos interact with the entire composite system rather than individual particles within the system. The coherent scattering is particularly relevant for low-energy neutrinos when the wavelength of neutrino is much larger than the size of system. In $1978$, Lincoln Wolfenstein pointed out that the coherent forward scattering of neutrinos off matter could be very important in studying the behavior of neutrino flavor oscillation\index{neutrino!flavor oscillation} in a dense medium~\cite{Wolfenstein:1977ue}. The fact that neutrinos propagating in matter may interact with the background particles can be described by the picture of free neutrinos traveling in an effective potential.

For incoherent scattering, neutrinos interact with particles in the medium individually. Incoherent scattering is typically more prominent for high-energy neutrinos, where the wavelength of neutrino is smaller compared to the spacing between particles. Study of incoherent scattering of high-energy neutrinos is important for understanding the physics in various astrophysical systems (e.g. supernova, stellar formation) and the evolution of the early Universe.

In this section, we discuss the coherent scattering between long wavelength neutrinos and atoms, and study the effective potential for neutrino coherent interaction. Then we present the matrix elements that describe the incoherent interaction between high energy neutrinos and other fundamental particles in the early Universe. Understanding these matrix elements is crucial for comprehending the process of neutrino freeze-out in the early Universe.

\para{Long wavelength limit of neutrino-atom coherent scattering}
According to the standard cosmological model, the Universe today is filled with the cosmic neutrinos with temperature $T_{\nu}^0=1.9 \,\mathrm{K}=1.7\times10^{-4}\,\mathrm{eV}$. The average momentum of present-day relic neutrinos is given by $\langle p_\nu^0\rangle\approx3.15\,T_\nu^0$ and the typical wavelength $\lambda_{\nu}^{0}={2\pi}/{\langle p_{\nu}^{0}\rangle}\approx2.3\times10^5\,$\AA, which is much larger than the radius at the atomic scale, such as the Bohr radius $R_{\mathrm{atom}}=0.529\,$\AA. In this case we have the long wavelength condition $\lambda_\nu\gg\,R_{\mathrm{atom}}$ for cosmic neutrino background today. 

Under the condition $\lambda_\nu\gg\,R_{\mathrm{atom}}$, when the neutrino is scattering off an atom, the interaction can be coherent scattering~\cite{PhysRevD.38.32,Lewis:1979mu,Papavassiliou:2005cs}. According to the principles of quantum mechanics, with neutrino scattering it is impossible to identify which scatters the neutrino interacts with and thus it is necessary to sum over all possible contributions. In such circumstances, it is appropriate to view the scattering reaction as taking place on the atom as a whole, i.e.,\index{neutrino!coherent scattering}
\begin{align}
\nu+\mathrm{Atom}\longrightarrow\nu+\mathrm{Atom}.
\end{align}

Considering a neutrino elastic scattering off an atom which is composed of $Z$ protons, $N$ neutrons and $Z$ electrons. For the elastic neutrino atom scattering, the low-energy neutrinos scatter off both atomic electrons and nucleus. For nucleus parts, we consider that the neutrinos interact via the $Z^0$ boson with a nucleus as
\begin{align}
\nu+A^{Z}_N\longrightarrow\nu+A^{Z}_N.
\end{align}
In this process a neutrino of any flavor scatters off a nucleus with the same strength. Therefore, the scattering will be insensitive to neutrino flavor. On the other hand, the neutrons can also interact via the $W^\pm$ with nucleus as 
\begin{align}
\nu_l+A^{Z}_N\longrightarrow\,l^-+A^{{Z}+1}_N,
\end{align}
which is a quasi-elastic process for neutrino scattering with the nucleus; we have $A^{Z_e}_N\rightarrow\,A^{{Z_e}+1}_N$. Since this process will change the nucleus state into an excited one, we will not consider its effect here. For detail discussion pf quasi-elastic scattering see ~\cite{SajjadAthar:2022pjt}.

For atomic electrons, the neutrinos can interact via the $Z^0$ and $W^\pm$ bosons with electrons for different flavors, we have
\begin{align}
&\nu_e+e^-\longrightarrow\nu_e+e^-\,\,\,(\mathrm{Z^0,\,W^\pm\,exchange}),\\
&\nu_{\mu,\tau}+e^-\longrightarrow\nu_{\mu,\tau}+e^-\,\,\,(\mathrm{Z^0\,exchange}).
\end{align}
Because of the fact that the coupling of $\nu_e$ to electrons is quite different from that of $\nu_{\mu,\tau}$, one may expect large differences in the behavior of $\nu_e$ scattering compared to the other neutrino types.

\para{Neutrino-atom coherent scattering amplitude \& matrix element} 
This section considers how a neutrino scatters from a composite system, assumed to consist of $N$ individual constituents at positions $x_i,\,i=1,2,....N$. Due to the superposition principle, the scattering amplitude $\mathcal{M}_\mathrm{sys}(\mathbf{p}^\prime,\mathbf{p})$ for scattering from an incoming momentum $\mathbf{p}$ to an outgoing momentum $\mathbf{p}^\prime$ is given as the sum of the contributions from each constituent~\cite{Freedman:1977xn,Papavassiliou:2005cs}:
\begin{align}
\mathcal{M}_\mathrm{sys}(\mathbf{p}^\prime,\mathbf{p})=\sum_i^N\,\mathcal{M}_i(\mathbf{p}^\prime,\mathbf{p})\,e^{i\mathbf{q}\cdot\mathbf{x}_i},
\end{align}
where $\mathbf{q}=\mathbf{p}^\prime-\mathbf{p}$ is the momentum transfer and the individual amplitudes $\mathcal{M}_i(\mathbf{p}^\prime,\mathbf{p})$ are added with a relative phase factor determined by the corresponding wave function. 
In principle, due to the presence of the phase factors, major cancellation may take place among the terms for the condition $|\mathbf{q}|R\gg1$, where $R$ is the size of the composite system, and the scattering would be incoherent. However, for the momentum small compared to the inverse target size, i.e., $|\mathbf{q}|R\ll1$, then all phase factors may be approximated by unity and contributions from individual scatters add coherently.

In the case of neutrino coherent scattering with an atom: If we consider sufficiently small momentum transfer to an atom from a neutrino which satisfies the coherence condition, i.e., $|\mathbf{q}|R_{\mathrm{atom}}\ll1$, then the relevant phase factors have little effect, allowing us to write the transition amplitude as \cite{Nicolescu:2013rxa}
\begin{align}
\label{M_atom}
\mathcal{M}_\mathrm{atom}=\sum_t\frac{G_F}{\sqrt{2}}\left[\overline{u}(p^\prime_\nu)\gamma_\mu\left(1-\gamma_5\right)u(p_\nu)\right]\left[\overline{u}(p^\prime_t)\gamma^\mu\left(c^t_V-c^t_A\gamma^5\right)u(p_t)\right],
\end{align}
where $t$ is all the target constituents (Z protons, N neutrons and Z electrons). The transition amplitude includes contributions from both charged and neutral currents, with
\begin{align}\label{CC_int}
&\mathrm{Charged\,\,Current}: c^t_V=c^t_A=1\,,\\
\label{NC_int}
&\mathrm{Neutral\,\, Current}: c^t_V=I_3-2\mathcal{Q}\sin^2\theta_W,\qquad c^t_A=I_3\,,
\end{align}
where $I_3$ is the weak isospin, $\theta_W$ is the Weinberg angle, and $\mathcal{Q}$ is the particle electric charge. 

Considering the target can be regarded as an equal mixture of spin states $s_z=\pm1/2$, and we can simplify the transition amplitude by summing the coupling constants of the constituents \cite{Lewis:1979mu,Sehgal:1986gn}. We have
\begin{align}
\label{Transition}
\mathcal{M}_\mathrm{atom}=&\frac{G_F}{2\sqrt{2}}\left[\overline{u}(p^\prime_\nu)\gamma_\mu\left(1-\gamma_5\right)u(p_\nu)\right]\notag\\&\bigg[\overline{u}(p^\prime_{a})\sum_t\left(C_L+C_R\right)_t\gamma^\mu\,u(p_{a})-\overline{u}(p^\prime_{a})\sum_t\left(C_L-C_R\right)_t\gamma^\mu\gamma^5u(p_{a})\bigg],
\end{align}
where the $u(p_\nu)$, $u(p^\prime_\nu)$ are the initial and final neutrino states and $u(p_a)$, $u_(p^\prime_a)$ are the initial and final states of the target atom. 
The coupling coefficients $C_L$ and $C_V$ are defined as
\begin{align}
&C_L=c_V+c_A,\,\,\,\,\,C_R=c_V-c_A,
\end{align}
where the coupling constants for neutrino scattering with proton, neutron, and electron are given by \rt{Table_coupling}. The coupling constants for $\nu_{\mu,\tau}$ are the same as for the $\nu_e$, excepting the absence of a charged current in neutrino-electron scattering.

\begin{table}
\centering
\begin{tabular}[c]{c|c|c|c|c}
\hline\hline
& Electron ($Z^0$ boson) & Electron ($W^\pm$ boson) & Proton (uud) & Neutron (udd)\\
\hline
$C_L$ & $-1+2\sin^2\theta_W$ & $2$ & $1-2\sin^2\theta_W$ & $-1$ \\
\hline
$C_R$ & $2\sin^2\theta_W$ & $0$ &$-2\sin^2\theta_W$ & $0$ \\
\hline\hline
\end{tabular}
\caption{The coupling constants for neutrino scattering with proton, neutron, and electron.}
\label{Table_coupling} 
\end{table}

Given the neutrino-atom coherent scattering amplitude \req{Transition}, the transition matrix element can be written as
\begin{align}
\label{scattering_matrix}
|\mathcal{M}_{\mathrm{atom}}|^2=\frac{G^2_F}{8}L_{\alpha\beta}^{\mathrm{neutrino}}\,\Gamma^{\alpha\beta}_{\mathrm{atom}},
\end{align}
where the neutrino tensor $L_{\alpha\beta}^{\mathrm{neutrino}}$ is given by
\begin{align}
\label{neutrino_tensor}
L_{\alpha\beta}^{\mathrm{neutrino}}
&=\mathrm{Tr}\left[\gamma_\alpha\left(1-\gamma_5\right)(\slashed{p}_\nu+m_\nu)\gamma_\beta\left(1-\gamma_5\right)(\slashed{p}^\prime_\nu+m_\nu)\right]\notag\\
&=8\left[(p_\nu)_\alpha\,(p^\prime_{\nu})_\beta+(p_\nu)^\prime_\alpha\,(p_\nu)_\beta-g_{\alpha\beta}(p_\nu\cdot\,p_\nu^\prime)+i\epsilon_{\alpha\sigma\beta\lambda}(p_\nu)^\sigma(p^\prime_\nu)^\lambda\right],
\end{align}
and the atomic tensor $\Gamma^{\alpha\beta}_\mathrm{atom}$ can be written as
\begin{align}
\label{atomic_tensor}
\Gamma^{\alpha\beta}_\mathrm{atom}
&=\mathrm{Tr}\bigg[(C_{LR}\gamma^\alpha-C^\prime_{LR}\gamma^\alpha\gamma^5)(\slashed{p}_a+M_a)(C_{LR}\gamma^\beta-C^\prime_{LR}\gamma^\beta\gamma^5)(\slashed{p}^\prime_a+M_a)\bigg]\notag\\
&=4\bigg\{(C^2_{LR}+C^{\prime2}_{LR})\left[(p_a)^\alpha\,(p^\prime_a)^\beta+(p_a)^{\prime\alpha}\,(p_a)^\beta\right]
+2iC_{LR}C^\prime_{LR}\epsilon^{\alpha\sigma^\prime\beta\lambda^\prime}(p_a)_{\sigma^\prime}(p^\prime_a)^{\lambda^\prime}\notag\\
&\qquad-g^{\alpha\beta}\bigg[(C^2_{LR}-C^{\prime2}_{LR})(p_a\cdot\,p_a^\prime)-(C^2_{LR}-C^{\prime2}_{LR})M^2_a\bigg]
\bigg\},
\end{align}
where $M_a$ is the target atom's mass $(M_a = AM_\mathrm{nucleon}, A=Z+N)$, and the coupling constants $C_{LR}$ and $C^\prime_{LR}$ are defined by
\begin{align}
C_{LR}=\sum_t(C_L+C_R)_t,\,\,\,\,\,\,C^\prime_{LR}=\sum_t(C_L-C_R)_t.
\end{align}
Substituting \req{neutrino_tensor} and \req{atomic_tensor} into \req{scattering_matrix}, the transition matrix element for coherent elastic neutrino atom scattering is given by:\index{neutrino!scattering matrix element}
\begin{align}
|\mathcal{M}_{\mathrm{atom}}|^2&=\frac{G^2_F}{8}L_{\alpha\beta}^{\mathrm{neutrino}}\,\Gamma^{\alpha\beta}_{\mathrm{atom}}
=8G^2_F\bigg[(C_{LR}+C^\prime_{LR})^2\,(p_\nu\cdot\,p_a)(p^\prime_\nu\cdot\,p^\prime_a)
\notag\\&\,\,\,\,\,\,
+(C_{LR}-C^\prime_{LR})^2\,(p_\nu\cdot\,p^\prime_a)(p^\prime_\nu\cdot\,p_a)
 -(C^2_{LR}-C^{\prime2}_{LR})M^2_a(p_\nu\cdot\,p_\nu^\prime)\bigg].
\end{align}
Taking the atom at rest in the laboratory frame, and considering small momentum transfer to an atom from a neutrino, i.e., $q^2=(p_\nu-p^\prime_\nu)^2=(p_a^\prime-p_a)^2\ll\,M^2_a$, we have
\begin{align}
&p_\nu\cdot\,p_a=E_\nu\,M_a,\\
&p_\nu^\prime\cdot\,p_a=E_\nu^\prime\,M_a\approx\,E_\nu\,M_a,\\
&p^\prime_\nu\cdot\,p^\prime_a=p^\prime_\nu\cdot(p_a+q)=E^\prime_\nu\,M_a\left[\left(1+\frac{q_0}{M_a}\right)-\frac{|p^\prime_\nu||q|}{M_a}\cos\theta\right]\approx\,E_\nu\,M_a,\\
&p_\nu\cdot\,p^\prime_a=p_\nu\cdot(p_a+q)=E_\nu\,M_a\left[\left(1+\frac{q_0}{M_a}\right)-\frac{|p^\prime_\nu||q|}{M_a}\cos\theta\right]\approx\,E_\nu\,M_a.
\end{align}
Then the transition matrix element for neutrino coherent elastic scattering off a rest atom can be written as
\begin{align}\label{M_general}
|\mathcal{M}_{\mathrm{atom}}|^2&=8\,G^2_F\,M_a\,E_\nu^2\left[C^2_{LR}\left(1+\frac{|p_\nu|^2}{E^2_\nu}\cos\theta\right)+3C^{\prime2}_{LR}\left(1-\frac{|p_\nu|^2}{3E_\nu^2}\cos\theta\right)\right],
\end{align}
which is consistent with the results in papers \cite{PhysRevD.38.32,Lewis:1979mu,Papavassiliou:2005cs,Smith:1984gym}.
From the above formula we found that the scattering matrix neatly divides into two distinct components: a vector-like component (first term) and an axial-vector like component (second term). They have different angular dependencies: the vector part has a $\left({|p_\nu|^2}/{E^2_\nu}\cos\theta\right)$ dependence, while the axial part has a $\left(-{|p_\nu|^2}/{3E_\nu^2}\cos\theta\right)$ behavior. However, in the case of the nonrelativistic neutrino, both angular dependencies can be neglected because of the limit $p_\nu\ll\,m_\nu$. 

Next, we consider the nonrelativistic electron neutrino $\nu_e$ scattering off an general atom with $Z$ protons, $N$ neutrons and $Z$ electrons. Then from Eq.~(\ref{M_general}), the matrix element can be written as
\begin{align}
\label{Probability_e}
|\mathcal{M}_{\mathrm{atom}}|^2&=8\,G^2_F\,M_a\,E_\nu^2\left[\left(3Z-A\right)^2\left(1+\frac{|p_\nu|^2}{E^2_\nu}\cos\theta\right)+3\left(3Z-A\right)^2\left(1-\frac{|p_\nu|^2}{3E_\nu^2}\cos\theta\right)\right]\notag\\
&\approx32\,G^2_F\,M_a\,E_\nu^2\left(3Z-A\right)^2,
\end{align}
where we neglect the angular dependence because of the nonrelativistic limit, and the coefficient $\left(3Z-A\right)^2$ for different target atoms are given in \rt{Table001}. 

\begin{table}[ht]
\centering
\begin{tabular}{c|c|c}
\hline\hline
 Neutrino Flavor:&$\nu_e$ &$\nu_{\mu,\tau}$\\
\hline\hline
Target Atom & $(3Z-A)^2$ & $(Z-A)^2$\\
\hline
$H_2(A=2, Z=2)$ & $16$ & $0$\\
\hline
${}^{3}H_e(A=3, Z=2)$ & $9$ & $1$\\
\hline
$HD(A=3, Z=2)$ & $9$ & $1$\\
\hline
${}^{4}_2H_e(A=4, Z=2)$ &$4$ & $4$\\
\hline
$DD(A=4, Z=2)$ & $4$ & $4$\\
\hline
${}^{12}_{{}6}C(A=12, Z=6)$ & $36$& $36$\\
\hline\hline
\end{tabular}
\caption{The coefficients for transition amplitude and scattering probability of $\nu_e$ and $\nu_{\mu,\tau}$ coherent elastic scattering off different target atoms. The definition of atomic mass is $A=Z+N$, where $Z$ and $N$ are the number of protons and neutrons, respectively.}
\label{Table001} 
\end{table}%

For nonrelativistic $\nu_{\mu,\tau}$, the scattering matrix is given by
\begin{align}
\label{Probability_mt}
|\mathcal{M}_{\mathrm{atom}}|^2&=8\,G^2_F\,M_a\,E_\nu^2\left[\left(A-Z\right)^2\left(1+\frac{|p_\nu|^2}{E^2_\nu}\cos\theta\right)+3\left(A-Z\right)^2\left(1-\frac{|p_\nu|^2}{3E_\nu^2}\cos\theta\right)\right]\notag\\
&\approx32\,G^2_F\,M_a\,E_\nu^2\left(Z-A\right)^2,
\end{align}
where the coefficient $\left(Z-A\right)^2$ for different target atoms are given in \rt{Table001}. The transition matrix for $\nu_e$ differs from that of $\nu_{\mu,\tau}$; this is due to the charged current reaction with the atomic electrons. Furthermore, the neutral current interaction for the electron and proton will cancel each other because of the opposite weak isospin $I_3$ and charge $\mathcal{Q}$. As a result, the coherent neutrino scattering from an atom is sensitive to the method of the neutrino-electron coupling.

\para{Mean field potential for neutrino coherent scattering}
When neutrinos are propagating in matter and interacting with the background particles, they can be described by the picture of free neutrinos traveling in an effective potential~\cite{Wolfenstein:1977ue}. In the following we describe the effective potential between neutrinos and the target atom, and generalize the potential to the case of neutrino coherent scattering with a multi-atom system.

Let us consider a neutrino elastic scattering off an atom which is composed of Z protons, N neutrons and Z electrons. For the elastic neutrino atom scattering, the low-energy neutrinos are scattering off both atomic electrons and the nucleus. Considering the effective low-energy CC and NC interactions, the effective Hamiltonian in current-current interaction form can be written as 
\begin{align}
\label{H_atom}
\mathcal{H}_I^{\mathrm{atom}}&=\mathcal{H}^\mathrm{electron}_I+\mathcal{H}^\mathrm{nucleon}_I=\frac{G_F}{\sqrt{2}}\,\left(j_\mu\,\mathcal{J}^\mu_{\mathrm{electron}}+j_\mu\,\mathcal{J}^\mu_\mathrm{nucleon}\right),
\end{align}
where $\mathcal{J}^\mu_{\mathrm{nucleon}}$ denotes the hadronic current for the nucleus, and $j^\mu$ and $\mathcal{J}^\mu_{\mathrm{electron}}$ are the lepton\index{lepton} currents for neutrino and electron, respectively. According to the weak interaction theory, the lepton current for the neutrino and electron can be written as
\begin{align}
&j_\mu=\overline{\psi_{\nu}}\,\gamma_\mu\,\left(1-\gamma_5\right)\,\psi_\nu,\\
\label{Current_e}
&\mathcal{J}^\mu_{\mathrm{electron}}=\overline{\psi_{e}}\,\gamma_\mu\,\left(1-\gamma_5\right)\,\psi_e\,\,\,\,\,(\mathrm{W^\pm\,exchange}),\\
&\mathcal{J}^\mu_{\mathrm{electron}}=\overline{\psi_{e}}\,\gamma_\mu\,\left(c_V^e-c_A^e\gamma_5\right)\,\psi_e\,\,\,\,\,(\mathrm{Z^0\,exchange}),
\end{align}
where $\psi_\nu$ and $\psi_e$ represent the spinor for the neutrino and electron, respectively. From Eq.~(\ref{NC_int}) the coupling coefficient for electrons are $c^e_V=-1/2+2\sin^2\theta_W$ and $c^e_A=-1/2$. The hadronic current is given by the expression~\cite{Giunti:2007ry}
\begin{align}
\label{Current_h}
\mathcal{J}^\mu_\mathrm{nucleon}\equiv\overline{\psi_t}\,\gamma^\mu\left(c^t_V-c^t_A\gamma^5\right)\psi_t,
\end{align}
where subscript $t$ means the target constituents (protons and neutrons). From Eq.~(\ref{NC_int}) the coupling constants for proton(uud) and neutron(udd) are given by
\begin{align}
&c^p_V=\frac{1}{2}-2\sin^2\theta_W,\,\,\,\,c^p_A=\frac{1}{2},\,\,\,\,\,\mathrm{proton}\,,\\
&c^n_V=-\frac{1}{2}\,\,\,\,c^n_A=-\frac{1}{2},\,\,\,\,\,\mathrm{neutron}\,.
\end{align}

To obtain the effective potential for an atom, we need to average the effective Hamiltonian over the electron and nucleon background. For the neutrino-nucleon (proton, neutron) interaction, we only have the neutral current interaction via $Z^0$ boson. However, for the neutrino-electron interaction, we can have charged-current or neutral current interaction depending on the flavor or neutrino. In following, we consider interaction between $\nu_e$ and electrons first which includes both charged and neutral-currents interaction for general discussion.

Considering atomic electrons as a gas of unpolarized electrons with a statistical distribution function $f(E_e)$, the effective potential for neutrino-electron interaction can be obtained by averaging the effective Hamiltonian over the electron background~\cite{Giunti:2007ry}
\begin{align}
\langle{\mathcal{H}^\mathrm{electron}_{I}}\rangle&=\frac{G_F}{\sqrt{2}}\int\,\frac{d^3p_e}{(2\pi)^32E_e}\,f(E_e,T)\left[\overline{\psi_\nu}(x)\,\gamma_\mu\left(1-\gamma_5\right)\,\psi_\nu(x)\right]\notag\\&\times\frac{1}{2}\!\sum_{h_e=\pm1}\!\!\langle\,e^-(p_e,h_e)|\overline{\psi_e}\,\gamma^\mu\big((1+c^e_V)\!-\!(1+c^e_A)\gamma_5\big)\,\psi_e|e^-(p_e,h_e)\rangle,
\end{align}
where $h_e$ denotes the helicity of the electron. The average over helicity of the electron matrix element can be calculated with Dirac spinor and gamma matrix traces ~\cite{Giunti:2007ry}. Then the average effective Lagrangian can be written as
\begin{align}
\langle{\mathcal{H}^\mathrm{electron}_{I}}\rangle&=\frac{G_F}{\sqrt{2}}(1+c^e_V)\int\,\frac{d^3p_e}{(2\pi)^3}f(E_e)\left[\overline{\psi_\nu}(x)\,\frac{\gamma^\mu{p_e}_\mu}{E_e}\left(1-\gamma_5\right)\,\psi_\nu(x)\right]\notag\\
&=\frac{G_F}{\sqrt{2}}\,(1+c^e_V)\,\left[\int\,\frac{d^3p_e}{(2\pi)^3}f(E_e)\left(\gamma^0-\frac{\vec{\gamma}\cdot\vec{{p}}_e}{E_e}\right)\right]\overline{\psi_\nu}(x)\left(1-\gamma_5\right)\psi_\nu(x)\notag\\
&=\left[\frac{G_F}{\sqrt{2}}\left(1+c^e_V\right)n_{e}\right]\,\overline{\psi_\nu}(x)\gamma^0\left(1-\gamma_5\right)\psi_\nu(x),
\end{align}
where $n_e$ is the number density of the electron. In this case, the effective potential for the case of neutrino-atomic electron interaction can be written as\index{neutrino!effective potential}
\begin{align}
V^{\mathrm{electron}}_{I}=\frac{G_F}{\sqrt{2}}\left(1+c^e_V\right)n_{e}=\frac{G_F}{\sqrt{2}}\left(4\sin^2\theta_W+1\right)n_{e}.
\end{align}
The same method can be applied to the neutrino-nuclear interactions. Following the same approach and averaging the effective neutrino-nuclear Hamiltonian over the nuclear background, the effective potential experienced by a neutrino in a background of neutron/proton is given by~\cite{Giunti:2007ry} 
\begin{align}
&V_{I}^{\mathrm{proton}}=\frac{G_F}{\sqrt{2}}\left(1-4\sin^2\theta_W\right)n_{p},\qquad V_{I}^{\mathrm{neutron}}=-\frac{G_F}{\sqrt{2}}\,n_{n},
\end{align}
where $n_p$ and $n_n$ represent the number density of proton and neutron.
Combining the neutron and proton potential together, we define the effective nucleon potential experienced by a neutrino as 
\begin{align}
V_I^{\mathrm{nucleon}}\equiv-\frac{G_F}{\sqrt{2}}\bigg[1-\left(1-4\sin^2\theta_W\right)\xi\bigg]n_{n},\qquad\xi=n_{p}/n_{n},
\end{align}
where $\xi$ is the ratio between proton and neutron number density.

In our study, we generalize the effective potential to the case of neutrino coherent scattering with multi-atom system, we consider a neutrino coherent forward scatters from a spherical symmetric system which is composed by atoms. In this case, the neutrino scatters off every atom, and it is impossible to identify which scatterer the neutrino interacts with and thus it is necessary to sum over all possible contributions from each atom. In such circumstances, it is appropriate to assume that the number density of electrons and neutrons can be written as
\begin{align}
&n_e=Z_e\,\left(\frac{N_\mathrm{atom}}{V}\right),\,\,\,\,\mathrm{and}\,\,\,\,n_n=N\,\,\left(\frac{N_\mathrm{atom}}{V}\right),
\end{align}
where $N_\mathrm{atom}$ is the number of atoms inside the system, $V$ is the volume of system, $Z$ is the number of electrons, and $N$ is the number of neutrons.
Then the effective potential is given by
\begin{align}
\label{Potential}
V_{I}&=V_I^{\mathrm{electron}}+V_I^{\mathrm{nucleon}}\notag\\&=\frac{G_F}{\sqrt{2}}\left(\frac{N_\mathrm{atom}}{V}\right)\bigg\{\left(4\sin^2\theta_W\pm1\right)\,Z_e-\bigg[1-\left(1-4\sin^2\theta_W\right)\xi\bigg]\,N\bigg\},
\end{align}
where the $+$ sign is for electron neutrinos $\nu_e$ and the $-$ sign is for muon(tau)\index{muon} neutrinos $\nu_{\mu,\tau}$, separately. 
Equation~(\ref{Potential})  shows that the effective potential depends on the number density of electrons and nucleons contained within the wavelength. 
Thus by increasing the atoms contained in the wavelength or selecting different atoms as targets, we can enhance the effective potential and may be able to provide a sensitive way to detect the cosmic neutrino background. Beside the detection of the cosmic neutrino background, the effective potential for multi-atom can also provide new approaches for studying other aspects of neutrino physics in the future.

\para{Matrix elements of incoherent neutrino scattering}
To determine the freeze-out temperature (chemical/kinetic freeze-out) for a given flavor of neutrinos, we need to know all the elastic and inelastic interaction processes in the early Universe and compare their interaction rates with the Hubble expansion rate. In this section we summarize the matrix elements for the neutrino annihilation/production processes and elastic scattering processes that are relevant for investigating neutrino freeze-out. These matrix elements serve as one of the fundamental ingredients for solving the Boltzmann equation ~\cite{Birrell:2014uka}.

Considering the Universe with temperature $T\approx\mathcal{O}$(MeV), the particle species in cosmic plasma are given by:
\begin{align}
\mathrm{Particle\,\,species\,\, in \,\,plasma:}
\left\{\gamma,\, l^-,\,l^+,\, \nu_e,\, \nu_\mu,\, \nu_\tau,\, \bar{\nu}_e,\, \bar{\nu}_\mu,\, \bar{\nu}_\tau\right\},
\end{align}
where $l^\pm$ represents the charged leptons. In this case, neutrinos can interact with all these particles via weak interactions and remain in equilibrium. In~\rt{T005} and \rt{T006} we present the matrix elements $|M|^2$ for different weak interaction processes in the early Universe.

\begin{table} 
\centering
\begin{tabular}{p{0.91\textwidth}}
\hline\hline
\vskip  -0.1cm 
Annihilation \& Production and Transition Amplitude $|M|^2$ \\[0.1cm]
\hline
\vskip -0.1cm 
$l^-+l^+\longrightarrow\nu_l+\bar{\nu}_l$\\[0.1cm]
$ 32G^2_F\left[\left(1+2\sin^2\theta_W\right)^2\left(p_1\cdot p_4\right)\left(p_2\cdot p_3\right)+\,\left(2\sin^2\theta_W\right)^2\left(p_1\cdot p_3\right)\left(p_2\cdot p_4\right)+2\sin^2\theta_W\left(1+2\sin^2\theta_W\right)m^2_l\left(p_3\cdot p_4\right)\right]$ \\[0.1cm]
\hline
\vskip -0.1cm 
$l^{\prime-}+l^{\prime+}\longrightarrow\nu_l+\bar{\nu}_l$\\
$32G^2_F\left[\left(1-2\sin^2\theta_W\right)^2\left(p_1\cdot p_4\right)\left(p_2\cdot p_3\right)+\,\left(2\sin^2\theta_W\right)^2\left(p_1\cdot p_3\right)\left(p_2\cdot p_4\right)-2\sin^2\theta_W\left(1-2\sin^2\theta_W\right)m^2_{l^\prime}\left(p_3\cdot p_4\right)\right]$ \\[0.1cm]
\hline
\vskip -0.1cm 
$\nu_l+\bar{\nu}_l\longrightarrow\nu_l+\bar{\nu}_l$ 
\hspace{1cm}
$32G^2_F\left[\left(p_1\cdot p_4\right)\left(p_2\cdot p_3\right)\right]$ \\[0.1cm]
\hline
\vskip -0.1cm 
$\nu_{l^\prime}+\bar{\nu}_{l^\prime}\longrightarrow\nu_l+\bar{\nu}_l$ 
\hspace{1cm}
$32G^2_F\left[\left(p_1\cdot p_4\right)\left(p_2\cdot p_3\right)\right]$ \\[0.1cm]
\hline\hline
\end{tabular}
\caption{The transition amplitude for different annihilation and production processes. The definition of particle number is given by $1+2\leftrightarrow3+4$, where $l,\,l^\prime=e,\,\mu,\,\tau\,(l\neq\,l^\prime)$.}
\label{T005}
\end{table}

In the calculation of transition amplitude, we use the low energy approximation for $W^\pm$ and $Z^0$ massive propagators, i.e.
\begin{align}
&\mbox{$Z^0$ boson}:\frac{-i\left[g_{\mu\nu}-\frac{q_\mu q_\nu}{M^2_z}\right]}{q^2-M^2_z}\approx\frac{ig_{\mu\nu}}{M^2_z},\quad
&\mbox{$W^\pm$ boson}:\frac{-i\left[g_{\mu\nu}-\frac{q_\mu q_\nu}{M^2_W}\right]}{q^2-M^2_W}\approx\frac{ig_{\mu\nu}}{M^2_W},
\end{align}
and consider the tree-level Feynman diagram contributions only. Then, following the Feynman rules of weak interaction~\cite{Griffiths:2008zz}, we obtain the matrix elements $|M|^2$ for different interaction processes.\index{neutrino!incoherent scattering} For a detail discussion please \rsec{ch:coll:simp} in Appendix.

\begin{table} 
\centering
\begin{tabular}{p{0.91\textwidth}}
\hline\hline
Elastic ($\nu_e$) 
Scattering Process and  Transition Amplitude $|M|^2$\\
\hline \vskip -0.1cm
$\nu_l+l^-\longrightarrow\nu_l+l^-$\\  
$ 32G^2_F\left[
\left(1+2\sin^2\theta_W\right)^2\left(p_1\cdot p_2\right)\left(p_3\cdot p_4\right)+\,\left(2\sin^2\theta_W\right)^2\left(p_1\cdot p_4\right)\left(p_2\cdot p_3\right)-2\sin^2\theta_W\left(1+2\sin^2\theta_W\right)m^2_l\left(p_1\cdot p_3\right)\right]$ \\[0.1cm]
\hline\vskip -0.1cm
$\nu_l+l^+\longrightarrow\nu_l+l^+$\\  
$ 32G^2_F\left[
 \left(1+2\sin^2\theta_W\right)^2\left(p_1\cdot p_4\right)\left(p_2\cdot p_3\right) +\,\left(2\sin^2\theta_W\right)^2\left(p_1\cdot p_2\right)\left(p_3\cdot p_4\right) -2\sin^2\theta_W\left(1+2\sin^2\theta_W\right)m^2_l\left(p_1\cdot p_3\right)\right]$ \\[0.1cm]
\hline\vskip -0.1cm
$\nu_l+l^{\prime-}\longrightarrow\nu_l+l^{\prime-}$\\ 
$ 32G^2_F\left[
 \left(1-2\sin^2\theta_W\right)^2\left(p_1\cdot p_2\right)\left(p_3\cdot p_4\right)+\,\left(2\sin^2\theta_W\right)^2\left(p_1\cdot p_4\right)\left(p_2\cdot p_3\right) +2\sin^2\theta_W\left(1-2\sin^2\theta_W\right)m^2_{l^\prime}\left(p_1\cdot p_3\right)\right]$ \\[0.1cm]
\hline\vskip -0.1cm
$\nu_l+l^{\prime+}\longrightarrow\nu_l+l^{\prime+}$\\ 
$ 32G^2_F\left[
 \left(1-2\sin^2\theta_W\right)^2\left(p_1\cdot p_4\right)\left(p_2\cdot p_3\right)+\,\left(2\sin^2\theta_W\right)^2\left(p_1\cdot p_2\right)\left(p_3\cdot p_4\right) +2\sin^2\theta_W\left(1-2\sin^2\theta_W\right)m^2_{l^\prime}\left(p_1\cdot p_3\right)\right]$ \\[0.1cm]
\hline\vskip -0.1cm
$\nu_l+\nu_l\longrightarrow\nu_l+\nu_l$ 
\hspace*{1cm}
$\frac{1}{2!}\frac{1}{2!}\times32G^2_F\left[4\left(p_1\cdot p_2\right)\left(p_3\cdot p_4\right)\right]$ \\[0.1cm]
\hline\vskip -0.1cm
$\nu_l+\bar{\nu}_l\longrightarrow\nu_l+\bar{\nu}_l$
\hspace*{1cm}
$32G^2_F\left[4\left(p_1\cdot p_4\right)\left(p_2\cdot p_3\right)\right]$ \\[0.1cm]
\hline\vskip -0.1cm
$\nu_l+\nu_{l^\prime}\longrightarrow\nu_l+\nu_{l^\prime}$ 
\hspace*{1cm}
$32G^2_F\left[\left(p_1\cdot p_2\right)\left(p_3\cdot p_4\right)\right]$ \\[0.1cm]
\hline\vskip -0.1cm
$\nu_l+\bar{\nu}_{l^\prime}\longrightarrow\nu_l+\bar{\nu}_{l^\prime}$ 
\hspace*{1cm}
$32G^2_F\left[\left(p_1\cdot p_4\right)\left(p_2\cdot p_3\right)\right]$ \\[0.1cm]
\hline\hline
\end{tabular}
\caption{The transition amplitude for different elastic scattering processes. The definition of particle number is given by $1+2\leftrightarrow3+4$, where $l,\,l^\prime=e,\,\mu,\,\tau\,(l\neq\,l^\prime)$.}
\label{T006}
\end{table}
\subsection{Boltzmann-Einstein equation}\label{sec:BoltzmannEinstein}
We now begin a detailed study of the nonequilibrium properties of the neutrino freeze-out\index{neutrino!freeze-out} and its impact on the effective number of neutrinos\index{neutrino!effective number}, an important cosmological observable. We model the dynamics of the neutrino freeze-out using the Boltzmann-Einstein equation\index{Boltzmann-Einstein equation}, also called the general relativistic Boltzmann equation, which describes the dynamics of a gas of particles that travel on geodesics in an general spacetime, with the only interactions being point collisions~\cite{Andreasson:2011ng,cercignani,Choquet-Bruhat:2009xil,ehlers},
\begin{equation}\label{eq:BoltzmannEinstein}
p^\alpha\partial_{x^\alpha}f-\sum_{j=1}^3\Gamma^j_{\mu\nu}p^\mu p^\nu\partial_{p^j}f=C[f]\,.
\end{equation}
Here $ \Gamma^\alpha_{\mu\nu}$ is the affine connection (Christoffel symbols) corresponding to a metric $g_{\alpha\beta}$, and the distribution function $f$ is a function of four-momentum on the mass shell, \ie, that satisfies
 \begin{equation}
g_{\alpha\beta}p^\alpha p^\beta=m^2\,.
\end{equation}
Here and in the following, repeated Greek indices are summed from $0$ to $3$. $C[f]$ is the collision operator\index{collision operator} and encodes all information about point interactions between particles. 

If $C[f]$ vanishes, then the equation is called the Vlasov equation and describes particles that move on geodesics (or free stream). At this point, we are not invoking the assumption that the distribution function has a kinetic equilibrium form\index{kinetic equilibrium}, nor are we assuming a FLRW\index{cosmology!FLRW} Universe; in this section we will discuss general properties of \req{eq:BoltzmannEinstein} before turning to the study of neutrino freeze-out in subsequent sections. We will need the following definitions of entropy current $S^\mu$, stress-energy tensor ${T}^{\mu\nu}$, and number current $N^\mu$,\index{entropy!current}\index{stress-energy tensor}\index{number current}
\begin{align}
\label{smdef} S^\mu&=-\int \left(f\ln(f)\pm(1\mp f)\ln(1\mp f)\right)p^\mu d\pi\,,\\
\label{Tmndef}{T}^{\mu\nu}&=\int p^\mu p^\nu f d\pi\,,\\
\label{nmdef} N^\nu&=\int f p^\nu d\pi\,,\\
d\pi&=\frac{\sqrt{-g}}{p_0}\frac{g_pd^3{\bf p}}{8\pi^3}\,,
\end{align}
where $d\pi$ is the volume element on the future mass shell; here $g$ denotes the determinant of the metric tensor, $p_0=g_{0\alpha} p^\alpha$; non-bold $p$ are four-momenta while bold ${\bf p}$ denotes the spacial components; the upper signs are for fermions and the lower signs for bosons. See \rapp{ch:vol:forms} for the derivation of the form of the volume element.

\para{Collision operator}
We now elaborate on the form of the collision operator. Our presentation is an expanded version of the survey in \cite{ehlers}. Suppose we have a collection of distinct particle and antiparticle types $\mathcal{C}$ with distribution functions $f_{C}$, $C\in\mathcal{C}$, and they partake in some number of reactions or interactions $I=n_{B_1} B_1, n_{B_2}B_2 \ldots \longrightarrow n_{A_1} A_1,n_{A_2}A_2 \ldots $, $A_i\in\mathcal{C}$ distinct and $B_j\in\mathcal{C}$ distinct, where $n_{A_i}$ is the number of particles of type $A_i$ occurring in the interaction (all nonzero) and similarly for $n_{B_i}$. Given an interaction, $I$, we let $r(I)$ be the collection of particle types that are reactants in the interaction, $p(I)$ be the collection of particle types that are products, and we let $\overleftarrow{I}$ denote the reverse reaction, \ie, with reactants and products reversed. We let $int$ denote the set of all interactions and, for any given species $A$, $int(A)$ be the set of all interactions involving $A$ as a product. We will assume that $\overleftarrow{I}\in int$ whenever $I\in int$. With these conventions, the collision operator for particle type $A$ takes the form\index{collision operator}
\begin{align}\label{collision:operator}
C[f_A]=&\sum_{I\in int(A)} \frac{n_A}{\prod_i n_{A_i}!\prod _j n_{B_j}!}\int\left[\left(\prod_j \prod_{l=1}^{n_{B_j}}f_{B_j}(p_{B_j}^l)\right)\left(\prod_i \prod_{k=1}^{n_{A_i}}f^{A_i}(p_{A_i}^k)\right)W^I(p_{B_j}^l,p_{A_i}^k) \right.\notag\\[0.3cm]
 -& \left.\left(\prod_i \prod_{k=1}^{n_{A_i}}f_{A_i}(p_{A_i}^k)\right)\left(\prod_j \prod_{l=1}^{n_{B_j}}f^{B_j}(p_{B_j}^l)\right)W^{\overleftarrow{I}}(p_{A_i}^k,p_{B_j}^l) \right] \delta(\Delta p)\prod_i \widehat{dV}_{A_i}\prod_j dV_{B_j}\,
\end{align}
where
\begin{align}
f^C=&1\mp f_C, \hspace{2mm} 
\Delta p=\sum_i \sum_{k=1}^{n_{A_i}}p^k_{A_i}-\sum_j \sum_{l=1}^{n_{B_j}}p^l_{B_j}\,,  \hspace{2mm}
\widehat{dV}_{A_i}=\tilde{\pi}_{A_i}\prod_{k=2}^{n_{A_i}}\frac{1}{2}d\pi^k_{A_i}\,, \hspace{2mm} 
dV_{B_j}=(2\pi)^4\prod_{l=1}^{n_{B_j}}\frac{1}{2}d\pi^l_{B_j}\,,\notag\\
 \tilde{\pi}_{A_i}=&\frac{1}{2} \text{ if } A_i=A \text{ and } \tilde{\pi}_{A_i}=\frac{1}{2}d\pi^1_{A_i} \text{ otherwise\,,}\hspace{2mm} 
d\pi_{C}^r=\frac{\sqrt{-g}}{(p_{C}^r)_0}\frac{g_{C}d^3{\bf p}_{C}^r}{8\pi^3}, p_0=g_{0\alpha}p^\alpha.\notag
\end{align}
 The integrations are over the future mass shells of all the particles, so the $p$ are related by $g_{\alpha \beta}p^\alpha p^\beta=m^2$. 
 
 The factorials in the collision term take into account the indistinguishability of the particles and prevent one from overcounting the independent ways a reaction can happen when integrating over momentum. The terms $f^A$ are due to quantum statistics and account for Fermi repulsion or Bose attraction (again, upper signs are for fermions and lower signs for bosons). $W^I(p_{B_j}^l,p_{A_i}^k)$, an abbreviation for $W^I(p_{B_1}^1,p_{B_1}^2,\ldots ,p_{B_1}^{n_{B_1}},p_{B_2}^1,\ldots ,p_{A_1}^1,\ldots )$, is the scattering kernel that encodes the probability of $n_{B_j}$ particles of types $B_j$ with momenta $p_{B_j}^l$ interacting to form $n_{A_i}$ particles of types $A_i$ with momenta $p_{A_i}^k$ in the process $I=n_{B_1}B_1,n_{B_2}B_2,\ldots\longrightarrow n_{A_1}A_1,n_{A_1}A_1,\ldots $, and so it is non-negative. 
 
 The delta function in the collision term enforces conservation of four-momentum. The factors of $(2\pi^4)$ and $\frac{1}{2}$ in the definitions of the volume elements come from normalization of the transition functions from quantum scattering calculations. For computational purposes, the expression \eqref{collision:operator} must be further simplified, taking into account the structure of each interaction. For example, see \rapp{ch:coll:simp} for a detailed study of the collision operator in the case of neutrino freeze-out.

 As defined, $C[f_A]$ is a function of $p_{A_i}^1$ where $A=A_i$. The choice not to integrate over $p_{A_i}^1$ rather than any of the other $p_{A_i}^k$ is completely arbitrary, but makes no difference in the result since the interaction does not depend on how we number the participating particles. In terms of the scattering kernels, this means we assume $W^I$ has the property
\begin{equation}\label{reorderProperty}
W^I(p_{A_1}^{\sigma_1},p^{\sigma 2}_{A_1},\ldots )=W^I(p_{A_1}^1,p_{A_1}^2,\ldots )\,,
\end{equation}
for any permutation $\sigma$, and similarly for any other permutation with one of the collections $p_{A_i}^k$ or $p_{B_j}^l$ for any choice of $i$ or $j$. For economy of notation in these derivations, we will employ the additional abbreviations for a given interaction $I=n_{B_i}B_i\longrightarrow n_{A_i}A_i$:
\begin{align}
f_{p,I}(p^k_{A_i})&\equiv f_{p,I}(p^1_{A_i},p^2_{A_i},\ldots ,p^{n_{A_i}}_{A_i})\equiv \prod_i \prod_{k=1}^{n_{A_i}}f_{A_i}(p_{A_i}^k)\,,\\
f^{p,I}(p^k_{A_i})&=f^{p,I}(p^1_{A_i},p^2_{A_i},\ldots ,p^{n_{A_i}}_{A_i})=\prod_i \prod_{k=1}^{n_{A_i}}f^{A_i}(p_{A_i}^k)\,,\notag\\
f_{r,I}(p^l_{B_j})&\equiv f_{r,I}(p^1_{B_j},p^2_{B_j},\ldots ,p^{n_{B_j}}_{B_j})\equiv \prod_j \prod_{l=1}^{n_{B_j}}f_{B_j}(p_{B_j}^l)\,,\notag\\
f^{r,I}(p^l_{B_j})&=f^{r,I}(p^1_{B_j},p^2_{B_j},\ldots ,p^{n_{B_j}}_{B_j})=\prod_j \prod_{l=1}^{n_{B_j}}f^{B_j}(p_{B_j}^l)\,,\notag\\
n_I&=\prod_i n_{A_i}!\prod _j n_{B_j}!\,,\notag\\
\widehat{dV}_I&=\delta(\Delta p)\prod_i\widehat{dV}_{A_i}\prod_jdV_{B_j}\,,\notag\\
dV_I&=\delta(\Delta p)\prod_idV_{A_i}\prod_jdV_{B_j}\,,\notag
\end{align}
where the $r$ and $p$ sub and superscripts stand for reactants and products respectively. See \rapp{ch:vol:forms} for more information on the precise meaning and properties of the delta function factors.

In the following subsections we derive several important properties of the equation \eqref{eq:BoltzmannEinstein}. While in principle these properties are well known~\cite{Andreasson:2011ng,cercignani,Choquet-Bruhat:2009xil,ehlers}, here we prove them at a level of generality that, to the authors knowledge, is not available in other references, \ie, for a general collection of interactions as encapsulated in \req{collision:operator}. We note that Riemannian normal coordinates will a key tool in these derivations. These are coordinates centered at a chosen point, $x$, in spacetime wherein the geodesics through $x$ are straight lines in the coordinate system and the derivatives of the metric in the coordinate system vanish at $x$. In particular, the Christoffel symbols vanish at $x$; see, e.g., page 42 in \cite{Wald:1984rg} or pages 72-73 of \cite{o1983semi}. 

\para{Conserved currents}
Suppose all the interactions of interest conserve some charge $b_A$, \ie,
 \begin{align}\label{eq:conservedCharge}
\sum_{A\in p(I)} n_Ab_A=\sum_{A\in r(I)} n_Ab_A
\end{align}
for all $I\in int$. We can construct a $4$-vector current corresponding to this charge as follows:
\begin{equation}
B^\mu=\sum_A b_A N_A^\mu\,,
\end{equation}
where $N^\mu_A$ are the number currents of the particle species \req{nmdef}. In this section we show that $B^\mu$ has vanishing divergence, \ie, a $B^\mu$ satisfies a conservation law.

For any point $x$ in spacetime, by transforming to Riemannian normal coordinates at $x$ and using \eqref{eq:BoltzmannEinstein} along with the fact that the first derivatives of the metric vanish at $x$, one can compute
\begin{equation}\label{useNormalCoords}
\nabla_\mu N_A^\mu=\int p^\mu \partial_{x^\mu} f d\pi_A=\int C[f_A] d\pi_A
\end{equation}
at $x$. The left and right-hand sides are scalars and therefore they are equal in any coordinate system. Noting this, we can then calculate
\begin{align}\label{eq:charge_current_div}
\nabla_\mu B^\mu=&\sum_A b_A\int C[f_A]d\pi_A\notag\\
=&\sum_A\sum_{I\in int(A)} \frac{n_Ab_A}{n_I}\int\int\left(f_{r,I}(p_{B_j}^l)f^{p,I}(p_{A_i}^k)W^I(p_{B_j}^l,p_{A_i}^k) 
-f_{p,I}(p_{A_i}^k)f^{r,I}(p_{B_j}^l)W^{\overleftarrow{I}}(p_{A_i}^k,p_{B_j}^l)\right)\widehat{dV}_I d\pi_A\notag\\
=&\sum_A\sum_{I\in int(A)} \frac{n_Ab_A}{n_I}\int\left(f_{r,I}(p_{B_j}^l)f^{p,I}(p_{A_i}^k)W^I(p_{B_j}^l,p_{A_i}^k) 
-f_{p,I}(p_{A_i}^k)f^{r,I}(p_{B_j}^l)W^{\overleftarrow{I}}(p_{A_i}^k,p_{B_j}^l)\right) dV_I\;. 
\end{align}
Now observe that, for any collection of finite sets $D_j$ indexed by a finite set $J$ with $\bigcup_{j\in J}D_j=D$ and any function $h:J\times D\rightarrow \mathbb{R}^m$, we have
\begin{equation}\label{sumLemma}
\sum_{j\in J}\sum_{x\in D_j} h(j,x)=\sum_{x\in D}\sum_{\{j:x\in D_j\}}h(j,x)\,.
\end{equation}
Using this fact, we can switch the order of the sums to obtain
\begin{align}\label{delB}
&\nabla_\mu B^\mu=\sum_{I\in int}\sum_{A\in p(I)} n_Ab_A R_I \,,\\
&R_I\equiv \frac{1}{n_I}\int\left(f_{r,I}(p_{B_j}^l)f^{p,I}(p_{A_i}^k)W^I(p_{B_j}^l,p_{A_i}^k) -f_{p,I}(p_{A_i}^k)f^{r,I}(p_{B_j}^l)W^{\overleftarrow{I}}(p_{A_i}^k,p_{B_j}^l)\right) dV_I\,.\notag
\end{align}
The sum over all interactions splits over a sum over symmetric interactions, $int_{s}$, and a sum over asymmetric interactions. For each asymmetric interaction, pair it up with its reverse and arbitrarily choose one of them to call the forward direction. Let the set of these forward interactions be denoted $\overrightarrow{int}$. Then the sum in \req{delB} splits as follows
\begin{equation}
\nabla_\mu B^\mu=\sum_{I\in int_s}R_I\sum_{A\in p(I)} n_A b_A+\sum_{I\in\overrightarrow{int}}R_I\sum_{A\in p(I)} n_Ab_A+\sum_{I\in\overrightarrow{int}}R_{\overleftarrow{I}}\sum_{A\in p(\overleftarrow{I})} n_Ab_A\,.
\end{equation}
For every $I\in int_s$ we have $W^I=W^{\overleftarrow{I}}$, $f_{A_i}=f_{B_i}$, and $f^{A_i}=f^{B_i}$, and therefore
\begin{align}
R_I=&\frac{1}{n_I}\left(\int f_{r,I}(p_{B_j}^l)f^{p,I}(p_{A_i}^k)W^I(p_{B_j}^l,p_{A_i}^k) dV_I 
-\int f_{p,I}(p_{A_i}^k)f^{r,I}(p_{B_j}^l)W^{\overleftarrow{I}}(p_{A_i}^k,p_{B_j}^l) dV_I\right)\notag\\
=&\frac{1}{n_I}\left(\int f_{r,I}(p_{B_j}^l)f^{p,I}(p_{A_i}^k)W^I(p_{B_j}^l,p_{A_i}^k) dV_I 
-\int f_{r,I}(p_{A_i}^k)f^{p,I}(p_{B_j}^l)W^{I}(p_{A_i}^k,p_{B_j}^l) dV_I\right)
=0\,,
\end{align}
as the two integrals differ only by a relabeling of integration variables. Asymmetric interactions satisfy
\begin{align}
R_{\overleftarrow{I}}=&\frac{1}{n_I}\int\left(f_{p,I}(p_{A_i}^k)f^{r,I}(p_{B_j}^l)W^{\overleftarrow{I}}(p_{A_i}^k,p_{B_j}^l) 
-f_{r,I}(p_{B_j}^l)f^{p,I}(p_{A_i}^k)W^I(p_{B_j}^l,p_{A_i}^k)\right) dV_I\notag\\
=&-R_I.
\end{align}
Combining this with \req{eq:conservedCharge}, we find
\begin{align}
\nabla_\mu B^\mu=\sum_{I\in\overrightarrow{int}}R_I\left(\sum_{A\in p(I)} n_Ab_A-\sum_{A\in p(\overleftarrow{I})} n_Ab_A\right)
=\sum_{I\in\overrightarrow{int}}R_I\left(\sum_{A\in p(I)} n_Ab_A-\sum_{A\in r(I)} n_Ab_A\right)=0\,.
\end{align}
Therefore $B^\mu$ is a conserved current, as claimed.

\para{Divergence freedom of stress energy tensor}
The Einstein equation implies that the total stress energy tensor\index{stress-energy tensor} of all matter coupled to gravity is divergence free. Here we show that the relativistic Boltzmann stress energy tensor \req{Tmndef} has this property, and is therefore a natural candidate matter model for coupling to gravity. 

First use Riemannian normal coordinates to compute
\begin{align}
\nabla_\mu T^{\mu\nu}=&\sum_A \int p_A^\nu C[f_A]d\pi_A \\
=&\sum_A\sum_{I\in int(A)}\frac{n_A}{n_I}\int (p^1_{A_\ell})^\nu\left(f_{r,I}(p_{B_j}^l)f^{p,I}(p_{A_i}^k)W^I(p_{B_j}^l,p_{A_i}^k) 
-f_{p,I}(p_{A_i}^k)f^{r,I}(p_{B_j}^l)W^{\overleftarrow{I}}(p_{A_i}^k,p_{B_j}^l)\right) dV_I\,,\notag
\end{align}
where $\ell$ is the unique index such that $A_\ell=A$ ($\ell$ depends on $A$ and $I$, but we suppress this dependence for simplicity of notation). Using \req{sumLemma} we can switch the summation order to get
\begin{align}
\nabla_\mu T^{\mu\nu}=&\sum_{I\in int}\sum_{A\in p(I)} \frac{n_A}{n_I}\int (p^1_{A_\ell})^{\nu}\left(f_{r,I}(p_{B_j}^l)f^{p,I}(p_{A_i}^k)W^I(p_{B_j}^l,p_{A_i}^k)
-f_{p,I}(p_{A_i}^k)f^{r,I}(p_{B_j}^l)W^{\overleftarrow{I}}(p_{A_i}^k,p_{B_j}^l)\right) dV_I\,.
\end{align}
By \req{reorderProperty} and the surrounding remarks, we can rewrite this as
\begin{align}\label{delTSum}
\nabla_\mu T^{\mu\nu}=&\sum_{I\in int}\sum_{A\in p(I)} \frac{1}{n_I}\sum_{a=1}^{n_A}\int (p^a_{A_\ell})^{\nu}\left(f_{r,I}(p_{B_j}^l)f^{p,I}(p_{A_i}^k)W^I(p_{B_j}^l,p_{A_i}^k) 
-f_{p,I}(p_{A_i}^k)f^{r,I}(p_{B_j}^l)W^{\overleftarrow{I}}(p_{A_i}^k,p_{B_j}^l)\right) dV_I\notag\\
=&\sum_{I\in int} \frac{1}{n_I}\sum_{\ell}\sum_{a=1}^{n_{A_\ell}}\int (p_{A_\ell}^a)^{\nu}\left(f_{r,I}(p_{B_j}^l)f^{p,I}(p_{A_i}^k)W^I(p_{B_j}^l,p_{A_i}^k) 
-f_{p,I}(p_{A_i}^k)f^{r,I}(p_{B_j}^l)W^{\overleftarrow{I}}(p_{A_i}^k,p_{B_j}^l)\right) dV_I\,.
\end{align}
As before, we can break the sum over $I$ into a sum over symmetric processes and two other sums over forward and backward asymmetric processes, respectively. For a symmetric interaction $I=\overleftarrow{I}$ and $f_{A_i}=f_{B_i}$ for all $i$, hence
\begin{align}
&\sum_\ell\sum_{a=1}^{n_{A_\ell}}\int (p_{A_\ell}^a)^{\nu}\left(f_{r,I}(p_{B_j}^l)f^{p,I}(p_{A_i}^k)W^I(p_{B_j}^l,p_{A_i}^k) 
-f_{p,I}(p_{A_i}^k)f^{r,I}(p_{B_j}^l)W^{\overleftarrow{I}}(p_{A_i}^k,p_{B_j}^l)\right) dV_I\notag\\
&\hspace{2cm}=\int\sum_\ell\sum_{a=1}^{n_{A_\ell}}\left((p_{A_\ell}^a)^\nu- (p_{B_\ell}^a)^{\nu}\right) f_{r,I}(p_{B_j}^l)f^{p,I}(p_{A_i}^k)W^I(p_{B_j}^l,p_{A_i}^k) dV_I
=0\,,
\end{align}
due to the delta function $\delta(\Delta p)$ in the volume form $dV_I$. Therefore the terms in the sum \req{delTSum} corresponding to symmetric interactions vanish. For every pair of forward and backward asymmetric interactions we obtain
\begin{align}
&\sum_\ell\sum_{a=1}^{n_{A_\ell}}\int (p_{A_\ell}^a)^{\nu}\left(f_{r,I}(p_{B_j}^l)f^{p,I}(p_{A_i}^k)W^I(p_{B_j}^l,p_{A_i}^k) 
-f_{p,I}(p_{A_i}^k)f^{r,I}(p_{B_j}^l)W^{\overleftarrow{I}}(p_{A_i}^k,p_{B_j}^l)\right) dV_I\notag\\
&+\sum_{\tilde\ell}\sum_{c=1}^{n_{B_{\tilde\ell}}} \int (p_{B_{\tilde\ell}}^c)^{\nu}\left(f_{p,I}(p_{A_i}^k)f^{r,I}(p_{B_j}^l)W^{\overleftarrow{I}}(p_{A_i}^k,p_{B_j}^l) 
-f_{r,I}(p_{B_j}^l)f^{p,I}(p_{A_i}^k)W^I(p_{B_j}^l,p_{A_i}^k)\right) dV_I\notag\\
=&\int\left( \sum_\ell\sum_{a=1}^{n_{A_\ell}}(p_{A_\ell}^a)^{\nu} -\sum_{\tilde \ell}\sum_{c=1}^{n_{B_{\tilde\ell}}} (p_{B_{\tilde\ell}}^c)^{\nu}\right)f_{r,I}(p_{B_j}^l)f^{p,I}(p_{A_i}^k)W^I(p_{B_j}^l,p_{A_i}^k) dV_I\notag\\
&+\int \left(\sum_{\tilde\ell}\sum_{c=1}^{n_{B_{\tilde\ell}}}(p_{B_{\tilde\ell}}^c)^{\nu}-\sum_\ell\sum_{a=1}^{n_{A_\ell}}(p_{A_\ell}^a)^{\nu}\right)f_{p,I}(p_{A_i}^k)f^{r,I}(p_{B_j}^l)W^{\overleftarrow{I}}(p_{A_i}^k,p_{B_j}^l) dV_I
=0\,,
\end{align}
again because of $\delta(\Delta p)$ in the volume forms. This shows $\nabla_\mu T^{\mu\nu}=0$, as claimed.

\index{Boltzmann!H-theorem}
\para{Entropy and Boltzmann's H-Theorem}
Finally, we prove that the entropy four-current satisfies $\nabla_\mu S^\mu\geq 0$, known as Boltzmann's H-theorem. This result requires the additional assumption that the interactions are time-reversal symmetric, \ie,
\begin{equation}\label{timeSymmetry}
W^I(p_{B_j}^l,p_{A_i}^k)=W^{\overleftarrow{I}}(p_{A_i}^k,p_{B_j}^l)
\end{equation}
for all $I$. Working in Riemannian normal coordinates once again, we can compute
\begin{align}
\nabla_\mu S^\mu=&-\sum_A\int p^\mu \partial_{x^\mu}\left(f_A\ln\left(f_A\right)\pm\left(1\mp f_A\right)\ln\left(1\mp f_A\right)\right)d\pi_A
=\sum_A\int\ln\left(1/f_A\mp 1\right)C[f_A]d\pi_A\,. 
\end{align}
Similar reasoning to the above two subsections then gives
\begin{align}
\nabla_\mu S^\mu=&\sum_{I\in int}\frac{1}{n_I}\sum_\ell\sum_{a=1}^{n_{A_\ell}}\int\ln\left(1/f_{A_\ell}(p_{A_\ell}^a)\mp1\right)\left(f_{r,I}(p_{B_j}^l)f^{p,I}(p_{A_i}^k)W^I(p_{B_j}^l,p_{A_i}^k) \right.\\
&\left. -f_{p,I}(p_{A_i}^k)f^{r,I}(p_{B_j}^l)W^{\overleftarrow{I}}(p_{A_i}^k,p_{B_j}^l)\right) dV_I\,.\notag
\end{align}
Once again, we break the summation into a sum over symmetric processes and two other sums over forward and backward asymmetric processes respectively. Each symmetric process contributes a term of the form
\begin{align}
&\int\sum_\ell\sum_{a=1}^{n_{A_\ell}}f_{r,I}(p_{B_j}^l)f^{p,I}(p_{A_i}^k)\left(\ln\left(1/f_{A_\ell}(p_{A_\ell}^a)\mp 1\right)
-\ln\left(1/f_{B_\ell}(p_{B_\ell}^a)\mp 1\right)\right) W^I(p_{B_j}^l,p_{A_i}^k) dV_I\notag\\
=& \int\ln\left(\frac{f^{p,I}(p_{A_i}^k)f_{r,I}(p_{B_j}^l)}{f_{p,I}(p_{A_i}^k)f^{r,I}(p_{B_j}^l)}\right) f_{r,I}(p_{B_j}^l)f^{p,I}(p_{A_i}^k)W^I(p_{B_j}^l,p_{A_i}^k) dV_I\notag\\
=& \frac{1}{2}\int\ln\left(\frac{f^{p,I}(p_{A_i}^k)f_{r,I}(p_{B_j}^l)}{f_{p,I}(p_{A_i}^k)f^{r,I}(p_{B_j}^l)}\right)\left( f_{r,I}(p_{B_j}^l)f^{p,I}(p_{A_i}^k)
- f_{p,I}(p_{A_j}^l)f^{r,I}(p_{B_i}^k)\right)W^I(p_{B_j}^l,p_{A_i}^k) dV_I\,,
\end{align}
where to obtain the last line we used the time-reversal property \eqref{timeSymmetry}.

A pair of forward and backward asymmetric interactions combine to give a term of the form
\begin{align}
&\sum_\ell\sum_{a=1}^{n_{A_b}}\int\ln\left(1/f_{A_\ell}(p_{A_\ell}^a)\mp 1\right)\left(f_{r,I}(p_{B_j}^l)f^{p,I}(p_{A_i}^k)W^I(p_{B_j}^l,p_{A_i}^k)
-f_{p,I}(p_{A_i}^k)f^{r,I}(p_{B_j}^l)W^{\overleftarrow{I}}(p_{A_i}^k,p_{B_j}^l)\right) dV_I\notag\\
&+\sum_{\tilde\ell}\sum_{c=1}^{n_{B_{\tilde\ell}}} \int\ln\left(1/f_{B_{\tilde\ell}}(p_{B_{\tilde\ell}}^c)\mp 1\right)\left(f_{p,I}(p_{A_i}^k)f^{r,I}(p_{B_j}^l)W^{\overleftarrow{I}}(p_{A_i}^k,p_{B_j}^l) 
-f_{r,I}(p_{B_j}^l)f^{p,I}(p_{A_i}^k)W^I(p_{B_j}^l,p_{A_i}^k)\right) dV_I\notag\\
=&\int\left( \sum_\ell\sum_{a=1}^{n_{A_\ell}}\ln\left(1/f_{A_\ell}(p_{A_\ell}^a)\mp 1\right) 
-\sum_{\tilde\ell}\sum_{c=1}^{n_{B_{\tilde\ell}}} \ln\left(1/f_{B_{\tilde\ell}}(p_{B_{\tilde\ell}}^c)\mp 1\right)\right)f_{r,I}(p_{B_j}^l)f^{p,I}(p_{A_i}^k)W^I(p_{B_j}^l,p_{A_i}^k) dV_I\notag\\
&-\int \left(\sum_\ell\sum_{a=1}^{n_{A_\ell}}\ln\left(1/f_{A_\ell}(p_{A_\ell}^a)\mp 1\right) 
-\sum_{\tilde\ell}\sum_{c=1}^{n_{B_{\tilde\ell}}} \ln\left(1/f_{B_{\tilde\ell}}(p_{B_{\tilde\ell}}^c)\mp 1\right)\right)f_{p,I}(p_{A_i}^k)f^{r,I}(p_{B_j}^l)W^{I}(p_{B_j}^l,p_{A_i}^k) dV_I\,,
\end{align}
where to obtain the first equality we used the time-reversal property \eqref{timeSymmetry}. Combining the symmetric and asymmetric cases we find
\begin{align}\label{eq:entropy_current}
\nabla_\mu S^\mu=&\sum_{I\in int_s} \frac{1}{2n_I}\int\ln\left(\frac{f^{p,I}(p_{A_i}^k)f_{r,I}(p_{B_j}^l)}{f_{p,I}(p_{A_i}^k)f^{r,I}(p_{B_j}^l)}\right)\left( f_{r,I}(p_{B_j}^l)f^{p,I}(p_{A_i}^k)
- f_{p,I}(p_{A_j}^l)f^{r,I}(p_{B_i}^k)\right)W^I(p_{B_j}^l,p_{A_i}^k) dV_I\notag\\
&+\sum_{I\in \overrightarrow{int}}\frac{1}{n_I}\int\ln\left(\frac{f_{r,I}(p_{B_j}^l)f^{p,I}(p_{A_i}^k)}{f_{p,I}(p_{A_i}^k)f^{r,I}(p_{B_j}^l)}\right)\left(f_{r,I}(p_{B_j}^l)f^{p,I}(p_{A_i}^k)
-f_{p,I}(p_{A_i}^k)f^{r,I}(p_{B_j}^l)\right)W^I(p_{B_j}^l,p_{A_i}^k) dV_I\,.
\end{align}
Each term in either sum is the integral of a non-negative quantity $W^{I}$ times a quantity of the form $(a-b)\ln(a/b)$, $a,b>0$, which is easily seen to be non-negative. Therefore we obtain the claimed result $\nabla_\mu S^\mu\geq 0$. The entropy four-current is future directed, due to the volume element being supported on the future mass shell. Therefore, given any splitting of spacetime into space and time $M=S\times T$, Boltzmann's H-theorem implies that the total entropy is non-decreasing on $T$.

\subsection{Neutrinos in the early Universe}
\label{sec:model:ind}

\para{Instantaneous freeze-out model}
Neutrino freeze-out\index{neutrino!freeze-out} is, as far as we know, the unique era in the history of the Universe when a significant matter fraction froze out at the same time that a reheating period was beginning, due to the onset of the $e^+e^-$ annihilation process. It is this coincidence involving the last reheating period that makes neutrino freeze-out a rich and complicated period to study as compared to the many other reheating periods in the history of the Universe.

A key cosmological observable that is impacted by the dynamics of neutrino freeze-out is the effective number of neutrinos\index{neutrino!effective number}, $N^{\text{eff}}_\nu$, which quantifies the amount of radiation energy density, $\rho_r$, in the Universe prior to photon freeze-out and after $e^\pm$ annihilation. $N^{\text{eff}}_\nu$ can be measured by fitting to the distribution of CMB\index{CMB} temperature fluctuations. The early Planck~\cite{Planck:2013pxb} analysis found $N^{\text{eff}}_{\nu}=3.36\pm 0.34$ (CMB only) and $N^{\text{eff}}_{\nu}=3.62\pm 0.25$ (CMB+$H_0$) ($68\%$ confidence levels), indicating a possible tension in the current understanding of $N^{\text{eff}}_\nu$ though this tension has lessened with further analysis from Planck~\cite{Planck:2015fie,Planck:2018vyg}. This section, as well as \rsec{ch:param:studies}, works towards a detailed understanding of $N^{\text{eff}}_{\nu}$ with an eye towards this old tension.

Mathematically, $N^{\text{eff}}_\nu$ is defined by the relation\index{neutrino!effective number}
\begin{align}\label{eq:NeffDef}
\rho_r=\left(1+(7/8)R_\nu^{4}N^{\text{eff}}_\nu\right)\rho_\gamma\,,
\end{align}
where $\rho_r$ is the radiation component of the Universe energy density, $\rho_\gamma$ is the photon energy density and $R_\nu\equiv T_\nu/T_\gamma=({4}/{11})^{1/3}$ is the photon to neutrino temperature ratio in the limit where the annihilating $e^\pm$ pairs do not transfer any entropy to Standard Model (SM) left-handed neutrinos, i.e., under the assumption that neutrinos have completely frozen out at the time of $e^\pm$ annihilation. The factor 7/8 is the ratio of Fermi to Bose reference normalization in $\rho$ and the neutrino to photon temperature ratio $R_\nu$ is the result of the transfer of $e^\pm$ entropy into photons after neutrino freeze-out\index{neutrino!freeze-out}. 

The definition \req{eq:NeffDef} is constructed such that if photons and SM left-handed neutrinos are the only significant massless particle species in the Universe between the freeze-out of left-handed neutrinos at $T_\gamma=\mathcal{O}(1)$ MeV and photon freeze-out at $T_\gamma=0.25$ eV, and assuming zero reheating of neutrinos, then $N^{\text{eff}}_{\nu}=3$, corresponding to the number of SM neutrino flavors. Detailed numerical study of the neutrino freeze-out process within the SM gives $N^{\text{eff}}_{\nu}=3.046$~\cite{Mangano:2005cc}, a value close to the number of flavors, indicating only a small amount of neutrino reheating. 
 We emphasize that $N^{\text{eff}}_\nu$ is named after neutrinos as they are the only significant contributor in SM cosmology. However, $N^{\text{eff}}_\nu$ could be impacted by non SM particles.

First we study how $N^{\text{eff}}_{\nu}$ is impacted by non-SM neutrino dynamics by characterizing its dependence on the neutrino freeze-out temperature within an instantaneous freeze-out model. This model, based on the work in \cite{Birrell:2013gpa,Birrell:2012gg}, allows us to study $N^{\text{eff}}_{\nu}$ without requiring a detailed description of the underlying non-SM interactions; the latter will be considered later in \rsec{ch:param:studies}. In addition, we explore the possibility of non-SM neutrino contributions to $N^{\text{eff}}_\nu$; the latter is based on \cite{Birrell:2014cja}.

\para{Chemical and kinetic equilibrium}
As the Universe expands and cools, the various components of the Universe transition from equilibrium to non-interacting free-streaming. This process is governed by two key temperatures: 1) The chemical freeze-out temperature, $T_{ch}$, above which the particles are kept in chemical equilibrium\index{chemical equilibrium} by number changing interactions. 2) The kinetic freeze-out temperature $T_k$, above which the particles are kept in thermal equilibrium, i.e., equilibrium momentum distribution. In reality, these are not sharp transitions, but we approximate them as such in this section. Our arguments follow the discussion presented earlier, see \req{equilibrium} and \req{kinetic:equilibrium}. The insights gained here will be important when studying the more detailed model of neutrino freeze-out in later sections.

At sufficiently high temperatures, such as existed in the early Universe, both particle creation and annihilation (i.e., chemical) processes and momentum exchanging (i.e., kinetic) scattering processes can occur sufficiently rapidly to establish complete thermal equilibrium of a given particle species. The distribution function $f_{ch}^\pm$ of fermions (+) and bosons (-) in both chemical and kinetic equilibrium\index{kinetic equilibrium} is found by maximizing entropy subject to energy being conserved and is given by \req{equilibrium}.
 
As temperature decreases, there will be a period where the temperature is greater than the kinetic freeze-out temperature, $T_k$, but below chemical freeze-out. During this period, momentum exchanging processes continue to maintain an equilibrium distribution of energy among the available particles, which we call kinetic equilibrium, but particle number changing processes no longer occur rapidly enough to keep the equilibrium particle number yield, i.e., for $T<T_{ch}$ the particle number changing processes have `frozen-out'. In this condition the momentum distribution, which is in kinetic equilibrium but chemical nonequilibrium, is obtained by maximizing entropy subject to particle number and energy constraints and thus two parameters appear\index{kinetic equilibrium}\index{fugacity}, see \req{kinetic:equilibrium}. The need to preserve the total particle number within the distribution introduces an additional parameter $\Upsilon$ called fugacity. 

The time dependent fugacity, $\Upsilon(t)$, controls the occupancy of phase space and is necessary once $T(t)<T_{ch}$ in order to conserve particle number. A fugacity different from $1$ implies an over-abundance ($\Upsilon>1$) or under-abundance ($\Upsilon<1$) of particles compared to chemical equilibrium and in either of these situations one speaks of chemical nonequilibrium. 

The effect of $\ln(\Upsilon)$ is similar after that of chemical potential\index{chemical potential} $\mu$, except that $\ln(\Upsilon)$ is equal for particles and antiparticles, and not opposite. This means $\Upsilon>1$ increases the density of both particles and antiparticles, rather than increasing one and decreasing the other as is common when the chemical potential is associated with conserved quantum numbers. Similarly, $\Upsilon<1$ decreases both. The fact that $\ln(\Upsilon)$ is not opposite for particles and antiparticles reflects the fact that both the number of particles and the number of antiparticles are conserved after chemical freeze-out, and not just their difference. 

Ignoring the small particle antiparticle asymmetry, their equality reflects the fact that any process that modifies the distribution would affect both particle and antiparticle distributions in the same fashion. Such an asymmetry would be incorporated by replacing $\Upsilon\rightarrow \Upsilon e^{\pm\mu/T}$ where $\mu$ is the chemical potential, but we ignore it in this work as the matter antimatter asymmetry is on the order of $1$ part in $10^9$.

We also emphasize that the fugacity is time dependent and not just an initial condition. At high temperatures $\Upsilon=1$ and we will find that $\Upsilon<1$ emerges dynamically as a result of the freeze-out process. The importance of fugacity was first introduced in \cite{Rafelski:1982pu} in the context of quark-gluon plasma. Its presence in cosmology was noted in \cite{Bernstein:1985th,Dolgov:1992wf} but its importance has been largely forgotten and the consequences unexplored in the literature. 

\para{Comoving entropy and particle number in an FLRW Spacetime}
In an FLRW spacetime, and assuming the particle distributions have the kinetic equilibrium form \req{kinetic:equilibrium} with fugacities $\Upsilon_{A}$, we can use the general formalism from \rsec{sec:BoltzmannEinstein} to compute the rate of change of comoving particle number and entropy; later we will apply these results to study neutrino freeze-out.

The number density of particle species $A$ in an FRW Universe is given by $N_A^0$ and so the comoving particle number can be written in terms of the divergence of the number current, \req{nmdef}, as follows
\begin{align}
\frac{d}{dt}(a^3N^0_A)=a^{3} \nabla_\mu N_A^\mu\,. 
\end{align}
By a similar computation to \req{eq:charge_current_div} we obtain
\begin{align}\label{eq:FRWcomovingNumGen}
a^{-3}\frac{d}{dt}(a^3N^0_A)=&\sum_{I\in int(A)} \frac{n_A}{n_I}\int\left(f_{r,I}(p_{B_j}^l)f^{p,I}(p_{A_i}^k)W^I(p_{B_j}^l,p_{A_i}^k) 
-f_{p,I}(p_{A_i}^k)f^{r,I}(p_{B_j}^l)W^{\overleftarrow{I}}(p_{A_i}^k,p_{B_j}^l)\right) dV.
\end{align}
The kinetic equilibrium distribution satisfies
\begin{equation}
f_A=\Upsilon_A e^{-E_A/T}f^A\,,
\end{equation}
where $E_A=p_A^0$. 

{\color{black}
Substituting this into \req{eq:FRWcomovingNumGen} and simplifying we find
\begin{align}\label{eq:comovingPartNumFRWfinal}
a^{-3}\frac{d}{dt}(a^3N_A^0)
=&\sum_{I\in int(A)}\frac{n_A}{n_I}\left(\int \prod_j\left(\Upsilon_{B_j}^{n_{B_j}}e^{-\sum_{l=1}^{n_{B_j}}E_{B_j}^l/T}\right)f^{p,I}(p_{A_i}^k)f^{r,I}(p_{B_j}^l)W^I(p_{B_j}^l,p_{A_i}^k)dV\right.\notag\\
&\left. -\int \prod_i\left(\Upsilon_{A_i}^{n_{A_i}}e^{-\sum_{k=1}^{n_{A_i}}E_{A_i}^k/T}\right)f^{p,I}(p_{A_i}^k)f^{r,I}(p_{B_j}^l)W^{\overleftarrow{I}}(p_{A_i}^k,p_{B_j}^l)dV\right)\notag\\
=&\sum_{I\in int(A)}\left(\prod_j\Upsilon_{B_j}^{n_{B_j}}- \prod_i\Upsilon_{A_i}^{n_{A_i}}\right){n_A}R_I\,,
\\
&R_I\equiv \frac{1}{n_I}\int\left(e^{-\sum_j\sum_{l=1}^{n_{B_j}}E_{B_j}^l/T}\right) f^{p,I}(p_{A_i}^k)f^{r,I}(p_{B_j}^l)W^I(p_{B_j}^l,p_{A_i}^k)dV\notag\,,
\end{align}
where $j$ indexes the reactants of $I$ and $i$ the products. To obtain this result we have assumed time-reversal symmetry of the scatting kernel \req{timeSymmetry} and used conservation of 4-momentum, $\Delta p=0$. In particular, if all particle species are in chemical equilibrium then the rate of change of comoving particle number is zero. The non-negative quantity $R_I$ can be viewed as quantifying the strength of the interaction $I$. Finally, note the terms corresponding to symmetric interactions vanish. 

We can also use \req{eq:comovingPartNumFRWfinal} to obtain a formula for the comoving total entropy of a collection of particles in kinetic equilibrium in an FRW Universe. Denoting the total entropy current $S^\mu\equiv\sum_A S_A^\mu$, with $S_A^\mu$ defined as in \req{smdef}, and noting that the entropy density in an FRW Universe is given by $\sigma=S^0$, we can combine \req{S:n:eq} with \req{eq:comovingPartNumFRWfinal} to compute
\begin{align}
&a^{-3}\frac{d}{dt}(a^3\sigma)=-\sum_A \ln(\Upsilon_A) \frac{d}{dt}(a^3N_A^0)
=-\sum_A \ln(\Upsilon_A)\sum_{I\in int(A)} \left(\prod_j\Upsilon_{B_j}^{n_{B_j}}- \prod_i\Upsilon_{A_i}^{n_{A_i}}\right)n_A R_I\notag\\
&=\sum_{I\in int} \left( \prod_i\Upsilon_{A_i}^{n_{A_i}}-\prod_j\Upsilon_{B_j}^{n_{B_j}}\right)R_I\sum_{A\in p(I)}n_A \ln(\Upsilon_A)\notag\\
&=\sum_{I\in \overrightarrow{int}} \left[\left( \prod_i\Upsilon_{A_i}^{n_{A_i}}-\prod_j\Upsilon_{B_j}^{n_{B_j}}\right)R_I\sum_{A\in p(I)}n_A \ln(\Upsilon_A)
+\left( \prod_j\Upsilon_{B_j}^{n_{B_j}}-\prod_j\Upsilon_{A_i}^{n_{A_i}}\right)R_I\sum_{B\in r(I)}n_B \ln(\Upsilon_B)\right]\notag\\
&=\sum_{I\in \overrightarrow{int}} \left( \prod_i\Upsilon_{A_i}^{n_{A_i}}-\prod_j\Upsilon_{B_j}^{n_{B_j}}\right)\left(\sum_i n_{A_i}\ln(\Upsilon_{A_i})-\sum_jn_{B_j}\ln(\Upsilon_{B_j})\right)R_I\,, 
\end{align}
where $j$ indexes the reactants of $I$ and $i$ the products. Again, there is no contribution from symmetric interactions. Observing that 
 \begin{align}
 \sum_in_{A_i}\ln(\Upsilon_{A_i})\geq \sum_jn_{B_j} \ln(\Upsilon_{B_j}) \text{ if and only if }\prod_i \Upsilon^{n_{A_i}}_{A_i}\geq\prod_j\Upsilon^{n_{B_j}}_{B_j}
 \,,
 \end{align} 
 we can conclude 
\begin{equation}
\frac{d}{dt}(a^3\sigma)\geq 0
\end{equation}
with strict inequality if the particle species are not all in chemical equilibrium. In particular, we have re-derived the H-Theorem in the case where all the distributions have the kinetic equilibrium form \req{kinetic:equilibrium}.
}

\para{Free-streaming in FLRW spacetime}
Once\index{cosmology!free-streaming}\index{Einstein-Vlasov equation} the temperature drops below the kinetic freeze-out temperature $T_k$ of a particular component of the Universe, we reach the free-streaming period where all particle scattering processes have completely frozen out. The dynamics are therefore determined by the free-streaming\index{free-streaming} Boltzmann-Einstein\index{Boltzmann-Einstein equation} equation, \req{eq:BoltzmannEinstein} with $C[f]=0$, known as the Einstein-Vlasov equation, in a spatially flat FLRW Universe.\index{Einstein-Vlasov equation}

Due to the assumed homogeneity and isotropy, the particle distribution function depends on $t$ and $p^0=E$ only and so the Einstein-Vlasov equation becomes
\begin{equation}\label{VEeqFLR}
E\partial_tf_\nu+(m^2-E^2)\frac{\partial_ta}{a}\partial_{E}f_\nu=0\,.
\end{equation}
Henceforth when discussing free-streaming distributions we replace subscript `$\nu$' by `fs' for free-streaming, as the results we obtain are of wider relevance. 

The general solution to \req{VEeqFLR} can be found in, \eg,~\cite{Choquet-Bruhat:2009xil,Wong:2011ip}:
\begin{equation}
f_\mathrm{fs}(t,E)=K(x)\,,\qquad
x\equiv\frac{a(t)^2}{D^2}(E^2-m^2)\,,
\end{equation}
where $K$ is an arbitrary smooth function and $D$ is an arbitrary constant with units of mass. To continue the evolution beyond thermal freeze-out, we choose $K$ to match the kinetic equilibrium\index{kinetic equilibrium} distribution \req{kinetic:equilibrium} at the freeze-out time $t_k$. This is accomplished by setting
\begin{equation}
K(x)=\frac{1}{\Upsilon_\mathrm{fs}^{-1}e^{\sqrt{x+m^2/T_k^2}}+ 1}
\end{equation}
and $D=T_k a(t_k)$. 

The free-streaming Fermi-Dirac-Einstein-Vlasov (FDEV) distribution function after freeze-out is then\index{Fermi!Einstein-Vlasov distribution}\index{neutrino!free-streaming}
\begin{equation}\label{eq:NeutrinoDist}
f_\mathrm{fs}(t,E)=\frac{1}{\Upsilon_\mathrm{fs}^{-1}e^{\sqrt{(E^2-m^2)/T_\mathrm{fs}^2+m_\mathrm{fs}^2 /T_k^2}}+ 1}\,.
\end{equation}
where
\begin{equation}\label{eq:TneutrinoDist}
T_\mathrm{fs}(t)=\frac{T_ka(t_k)}{a(t)}\,. 
\end{equation}
We will call $T_\mathrm{fs}$ in \req{eq:TneutrinoDist} the cosmic free-streaming (neutrino) background temperature, even though $T_\mathrm{fs}$ will in general differ from the temperature of the photon background and the distribution of free-streaming particles has a thermal shape only for $m=0$. The shape of momentum distribution for massive particles seen in \req{eq:NeutrinoDist} describes a Fermion quantum gas free-streaming in an expanding Universe below the freeze-out temperature where $T_\mathrm{fs}(t_k)=T_k$. 

The energy, pressure, number, and entropy densities\index{entropy!density}, denoted by $\rho$, $P$, $n$, and $\sigma$ respectively for the free-streaming distribution \req{eq:NeutrinoDist}, can be computed using \req{Tmndef}, \req{nmdef}, and \req{smdef}\index{free-streaming!energy density}\index{free-streaming!pressure}\index{free-streaming!number density}
\begin{align}
\rho_\mathrm{fs}&=\frac{g_\mathrm{fs}}{2\pi^2}\!\int_0^\infty\!\!\!\frac{\left(m_\mathrm{fs}^2+p^2\right)^{1/2}p^2dp }{\Upsilon_\mathrm{fs}^{-1}e^{\sqrt{p^2/T_\mathrm{fs}^2+m_\mathrm{fs}^2/T_k^2}}+ 1}\,,\label{eq:NeutrinoRho}\\[0.2cm]
P_\mathrm{fs}&=\frac{g_\mathrm{fs}}{6\pi^2}\!\int_0^\infty\!\!\!\frac{\left(m_\mathrm{fs}^2+p^2\right)^{-1/2}p^4dp }{\Upsilon_\mathrm{fs}^{-1} e^{\sqrt{p^2/T_\mathrm{fs}^2+m_\mathrm{fs}^2/T_k^2}}+ 1}\,,\label{eq:NeutrinoP}\\[0.2cm]
n_\mathrm{fs}&=\frac{g_\mathrm{fs}}{2\pi^2}\!\int_0^\infty\!\!\!\frac{p^2dp }{\Upsilon_\mathrm{fs}^{-1}e^{\sqrt{p^2/T_\mathrm{fs}^2+m_\mathrm{fs}^2/T_k^2}}+ 1}\,,\label{eq:NumDensity}\\[0.3cm]
\sigma_\mathrm{fs}&=-\frac{g_\mathrm{fs}}{2\pi^2}\!\int_0^\infty\!\!\!H(p^2/T_\mathrm{fs}^2)p^2dp\,,\,\,\,H\equiv K\ln K +(1-K)\ln(1-K)\,,\label{eq:EntropyIntegrand}
\end{align}
where $g_\mathrm{fs}$ is the degeneracy (not to be confused with the metric factor $\sqrt{-g}=a^3)$.

Comparing these results \req{eq:NeutrinoRho}, \req{eq:NeutrinoP}, \req{eq:NumDensity}, \req{eq:EntropyIntegrand}, to the formulas for the corresponding quantities in kinetic equilibrium, we see that they differ by the replacement $m\rightarrow m T_\mathrm{fs}(t)/T_k$ in the exponential factor {\em only}. Changing variables to $u=p/T_\mathrm{fs}$, one sees that both $n$ and $s$ are proportional to $T_\mathrm{fs}^3$. The  free-streaming temperature, $T_\mathrm{fs}$, is inversely proportional to $a$, hence we see that
\begin{equation}\label{eq:ConstEntropy}
a^3n=\text{constant}\text{ and } a^3\sigma=\text{constant}\,.
\end{equation}
This proves that the particle number and entropy in a comoving volume are conserved, irrespective of the form of $K$ that defines the shape of the momentum distribution at freeze-out. It should be noted that this conservation of entropy in free-streaming neutrinos relies on the Boltzmann equation model, and its corresponding entropy current\index{entropy!current} \req{smdef}, an approximation which may break down in later epochs of the evolution of the Universe when the quantum fluid nature of the neutrino emerges. 

\para{Neutrino fugacity and photon to neutrino temperature ratio}
We assume instantaneous freeze-out allowing the use of conservation laws in \req{eq:dynamics} to characterize the neutrino fugacity\index{fugacity!neutrino} and temperature in terms of the (kinetic) freeze-out temperature $T_k$. We first outline the physics of the situation qualitatively. For $T_k<T<T_{ch}$, the evolution of the temperature of the common $e^\pm,\gamma,\nu$ plasma and the neutrino fugacity are determined by conservation of comoving neutrino number (since $T<T_{ch}$) and conservation of entropy. As shown above, after thermal freeze-out the neutrinos begin to free-stream and therefore $\Upsilon_\mathrm{fs}$ is constant, the neutrino temperature evolves as $1/a$, and the comoving neutrino entropy and neutrino number are exactly conserved. 

The photon temperature then evolves to conserve the comoving entropy within the coupled system of photons, electrons, and positrons. As annihilation occurs, entropy from $e^+e^-$ is fed into photons, leading to reheating\index{reheating}. We now make this analysis quantitative in order to derive a relation between the reheating temperature ratio and neutrino fugacity\index{fugacity}.

Assuming $T_{ch}\gg m_e$, the entropy, $S$, in a given comoving volume, $V$, at chemical freeze-out is the sum of relativistic (turning to free-streaming) neutrinos (with $\Upsilon_\nu=1$), electrons, positrons, and photons
\begin{equation}
S(T_{ch})=\left(\frac{7}{8}g_\nu+\frac{7}{8}g_{e^\pm} +g_\gamma \right)\frac{2\pi^2}{45} T_{ch}^3V(t_{ch})\,,
\end{equation}
where $T_{ch}$ is the common neutrino, $e^+e^-$, and $\gamma$ temperature at chemical freeze-out and $t_{ch}$ is the time at which chemical freeze-out occurs. The number of neutrinos and anti-neutrinos at chemical freeze-out in this same volume is
\begin{equation}
\mathcal{N}_\nu(T_{ch})=\frac{3g_\nu}{4\pi^2}\zeta(3)T_{ch}^3V(t_{ch})\,.
\end{equation}
The particle-antiparticle, flavor, and spin-helicity statistical factors are $g_\nu=6$, $g_{e^\pm}=4$, $g_\gamma=2$.

Distinct chemical and thermal freeze-out temperatures lead to a nonequilibrium modification of the neutrino distribution in the form of a fugacity factor $\Upsilon_\nu$ when $T_k<T<T_{ch}$. This leads to the following expressions for, respectively, neutrino entropy and number at $T=T_k$ in the comoving volume
\begin{align}
S(T_k)=&\left(\frac{2\pi^2}{45}g_\gamma T_k^3+S_{e^\pm}(T_k)+S_{\nu}(T_k)\right)V(t_k)\,,\\
\mathcal{N}_{\nu}(T_k)=&\frac{g_\nu}{2\pi^2}\int_0^\infty \frac{u^2 du}{\Upsilon_\nu^{-1}(T_k)e^u+1}T_k^3V(t_k)\,,\notag
\end{align}
where $t_k$ is the time at which kinetic freeze-out occurs.

After neutrino freeze-out\index{neutrino!freeze-out} and when $T_{\gamma}\ll m_e$, the entropy in neutrinos is conserved independently of the other particle species and the $e^+e^-$ entropy is nearly all transferred to photons:
\begin{equation}
S_{\gamma}(T_\gamma)=\frac{2 \pi^2}{45}g_\gamma T_{\gamma}^3 V\,.
\end{equation}
 Note that we must now distinguish between the neutrino and photon temperatures.

The conservation laws \req{eq:dynamics} and \req{eq:ConstEntropy} then imply the following relations.
\begin{enumerate}
\item Conservation of comoving neutrino number between chemical and kinetic freeze-out:
\begin{equation}\label{modindeq1}
\frac{T_{ch}^3V(t_{ch})}{T_k^3V(t_k)}=\frac{2}{3\zeta(3)}\int_0^\infty \frac{u^2 du}{\Upsilon_\nu^{-1}(T_k)e^u+1}.
\end{equation}
\item Conservation of entropy in $e^\pm$, $\gamma$, and neutrinos prior to neutrino freeze-out:
\begin{align}\label{modindeq2}
&\left(\frac{7}{8}g_\nu+\frac{7}{8}g_{e^\pm} +g_\gamma \right)\frac{2\pi^2}{45} T_{ch}^3V(t_{ch})=\left(S_{\nu}(T_k)+S_{e^\pm}(T_k)+\frac{2\pi^2}{45}g_\gamma T_k^3\right)V(t_k)\,.
\end{align}
\item Conservation of the entropy in $e^\pm$ and $\gamma$ between neutrino freeze-out and $e^\pm$ annihilation:
\begin{equation}\label{modindeq3}
\frac{2 \pi^2}{45}g_\gamma T_{\gamma}^3 V(t)=\left(\frac{2\pi^2}{45}g_\gamma T_k^3+S_{e^\pm}(T_k)\right)V(t_k)\,, \,\,\, T_\gamma\ll \min\{m_e, T_k\}\,.
\end{equation}
\end{enumerate}

These relations allow one to solve for the fugacity, reheating ratio, and effective number of neutrinos\index{neutrino!effective number} in terms of the kinetic freeze-out temperature, irrespective of the details of the dynamics that leads to a particular freeze-out temperature. Specifically, combining \req{modindeq1} and \req{modindeq2} one obtains
\begin{align}
 \frac{S_{\nu}(T_k)/T_k^3+S_{e^\pm}(T_k)/T_k^3+\frac{2\pi^2}{45}g_\gamma }{\left(\frac{7}{8}g_\nu+\frac{7}{8}g_{e^\pm} +g_\gamma \right)\frac{2\pi^2}{45} }=\frac{2}{3\zeta(3)}\int_0^\infty \frac{u^2 du}{\Upsilon_\nu^{-1}(T_k)e^u+1}\,.
\end{align}

This can be solved numerically to compute $\Upsilon_\nu(T_k)$. One can also use these relations to analytically derive the following expansion for the photon to neutrino temperature ratio after $e^\pm$ annihilation (see \cite{Birrell:2012gg}), where here we define $\delta=\ln(\Upsilon)$:
\begin{align}\label{eq:UpsilonRatio}
&\frac{T_\gamma}{T_\nu}=a\Upsilon^{b}\left(1+c\delta^2+O(\delta^3)\right),\\
\label{value_a}
&a=\left(1+\frac{7}{8}\frac{g_{e^\pm}}{g_\gamma}\right)^{1/3}=\left(\frac{11}{4}\right)^{1/3}\,,\,\,\,b\approx 0.367\,, \,\,\,c\approx -0.0209\,.\notag
\end{align}

An approximate power law fit was first obtained numerically in \cite{Birrell:2013gpa}. A relation between the fugacity\index{fugacity} $\Upsilon=e^\delta$ and the effective number of neutrinos \req{eq:NeffDef} was also derived in \cite{Birrell:2012gg} using these methods:
\begin{equation}\label{eq:NnuApprox}
N^{\mathrm{eff}}_\nu=\frac{360}{7\pi^4}\frac{e^{-4b\delta}}{(1+c\delta^2)^4}\int_0^\infty \frac{u^3}{e^{u-\delta}+1}du\left(1+O(\delta^3)\right)\,.
\end{equation}

\begin{figure}
\centerline{\includegraphics[width=0.51\linewidth]{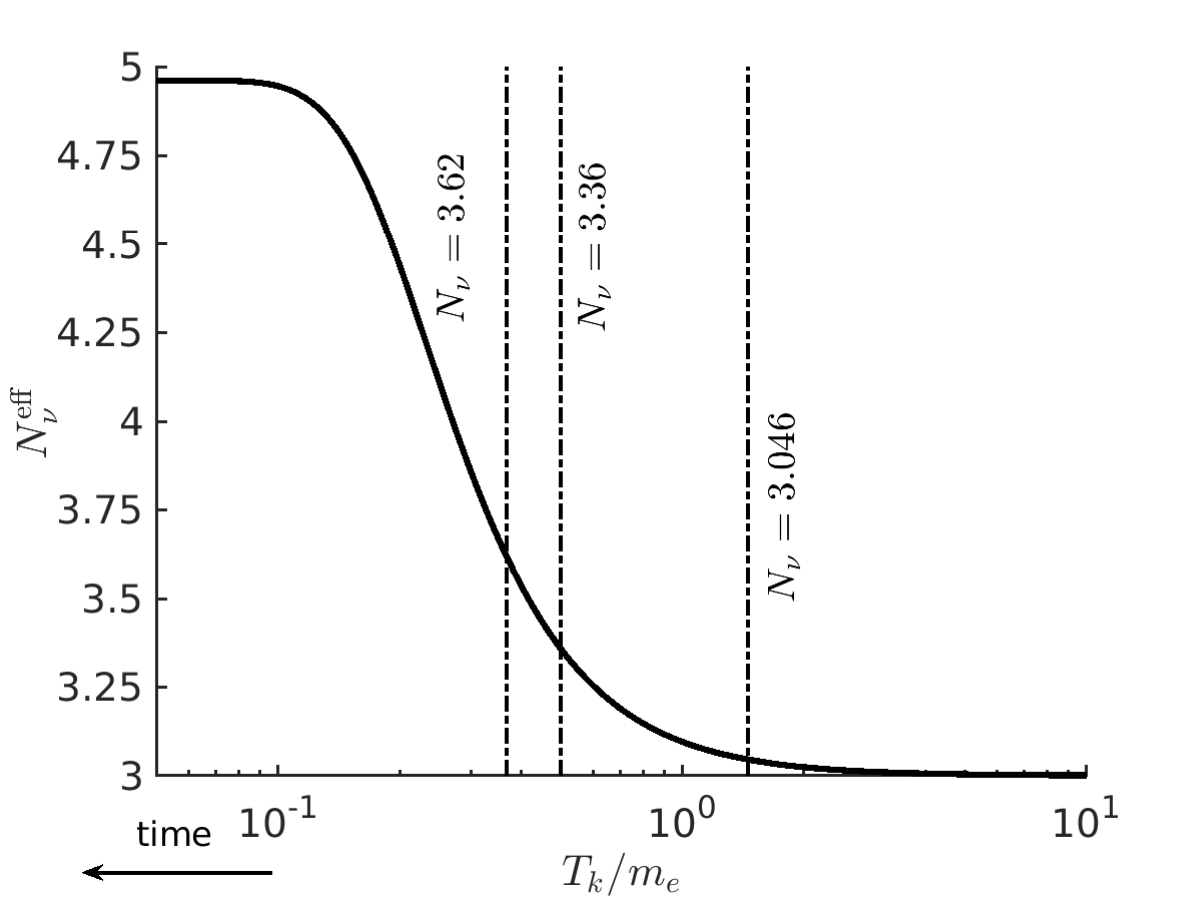}\hspace{-0.95cm}
\includegraphics[width=0.50\linewidth]{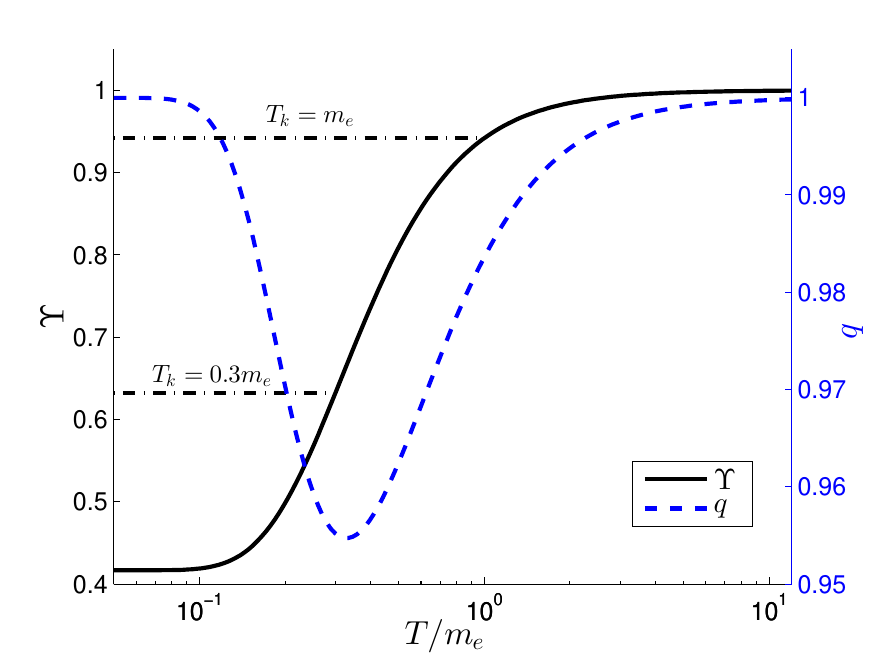}}
\caption{Dependence of the effective number of neutrinos (left-hand frame) and neutrino fugacity\index{fugacity!neutrino} (right-hand frame) on the neutrino kinetic freeze-out temperature. We also show the evolution of the deceleration parameter through the freeze-out period (right-hand frame)}
\label{fig:Tk_dependence}
\end{figure}

In \rf{fig:Tk_dependence} we plot that dependence of $N^{\mathrm{eff}}_\nu$ and $\Upsilon$ on $T_k$ that is implied by these calculations. In particular, the fugacity evolves following the solid black curve in the bottom plot until it reaches the kinetic freeze-out temperature, at which point the neutrinos decouple and $\Upsilon$ remains constant thereafter, as shown in the dashed black curves for two sample values of $T_k$. 
 
The early Planck CMB\index{CMB} results~\cite{Planck:2013pxb} contain several fits based on different data sets which suggest that $N^{\mathrm{eff}}_\nu$ is in the range $3.30\pm 0.27$ to $3.62\pm0.25$ ($68\%$ confidence level). We note the more recent Planck CMB analysis~\cite{Planck:2018vyg} which chooses to sidestep these considerations, which in part gives rise to the Hubble-tension issues.

A numerical computation based on the Boltzmann equation with two body scattering~\cite{Mangano:2005cc} gives to $N_{\nu}^{\rm eff}=3.046$. These values are shown in the vertical lines in the left figure. The tension between the Planck results and theoretical reheating studies motivated some of our work.

\para{Contribution to effective neutrino number from sub-eV mass sterile particles}
We  briefly explore  the effect on $N_\nu^{\text{eff}}$ by non-SM light weakly coupled particle species \cite{Birrell:2014cja}, referred to here as a sterile particles (SP)\index{sterile particles}. Independent of their source, once the SPs decouple from the particle inventory at a photon temperature of $T_{d,s}$, a difference in their temperature from that of photons will build up during subsequent photon reheating periods, similarly to earlier computations we presented.

Such hypothetical SPs would behave as `dark radiation'~\cite{Steigman:2013yua} rather than cold dark matter and would therefore impact $N_\nu^{\text{eff}}$ in a similar manner to neutrinos, though with a different freeze-out temperature not described by PP-SM.  The possibility that Goldstone bosons, one candidate for SPs, could be such dark radiation was identified in \cite{Weinberg:2013kea}. Another viable candidate for SPs are sterile neutrinos. In fact three sterile right-handed neutrinos could contribute with needed strength to the effective number of neutrinos\index{neutrino!effective number}, $N^{\text{eff}}_{\nu}$, if their freeze-out temperature is in the vicinity of the quark gluon plasma (QGP) phase transition~\cite{Anchordoqui:2011nh,Anchordoqui:2012qu}. 

If SPs originating in the QGP phase transition are interpreted as Goldstone bosons, this would imply that in the deconfined phase there is an additional hidden symmetry, weakly broken at hadronization. For example, if this symmetry were to be part of the baryon conservation riddle, then we can expect that these Goldstone bosons will couple to particles with baryon number, and possibly only in the domain where the vacuum is modified from its present day condition. These consideration motivate our study~\cite{Birrell:2014cja} of the contribution to 
$N^{\text{eff}}_{\nu}$ of boson or fermion degrees of freedom (DoF) that froze out near to the QGP phase transformation. 

We use the lattice-QCD derived QGP EoS from~\cite{Borsanyi:2013bia} to characterize the relation between $N^{\text{eff}}_{\nu}$ and the number of DoF that froze out at the time that the quark-gluon deconfined phase froze into hadrons near $T=150\MeV$. We work within the instantaneous freeze-out approximation, using the same reasoning that was applied to neutrinos, {\it i.e.\/}, we employ comoving entropy conservation\index{entropy!conservation} along with the facts that frozen-out particle species undergo temperature scaling with $1/a(t)$ and the remaining coupled particles undergo reheating at each $T\simeq m$ threshold, caused by a disappearing particle species transfer entropy into the remaining particles.

We denote by $S$ the conserved `comoving' entropy in a volume element $dV$, which scales with the factor $a(t)^3$. As we are no longer only considering just the neutrino freeze-out\index{neutrino!freeze-out}, here we employ the definition of the effective number of entropy DoF, $g_*^S$, given by
\begin{equation}
S=\frac{2\pi^2}{45}g^S_*T_\gamma^3 a^3\,.
\end{equation} 
For ideal fermion and boson gases, 
\begin{equation}
g_*^S=\!\!\!\!\sum_{i=\text{bosons}}\!\!\!\!g_i \left(\frac{T_i}{T_\gamma}\right)^3\!\!\!f_i^-+\frac{7}{8}\!\!\!\sum_{i=\text{fermions}}\!\!\!\! g_i \left(\frac{T_i}{T_\gamma}\right)^3\!\!\!f_i^+\,.
\end{equation}
The $g_i$ are degeneracies, $f_i^\pm$ are known functions, valued in $(0,1)$ (see \req{eq:entg002} for details), that turn off the various species as the temperature drops below their mass; compare to the analogous Eqs.\,(2.3) and (2.4) in~\cite{Blennow:2012de}. 

\begin{figure}
\centerline{\includegraphics[width=0.8\linewidth]{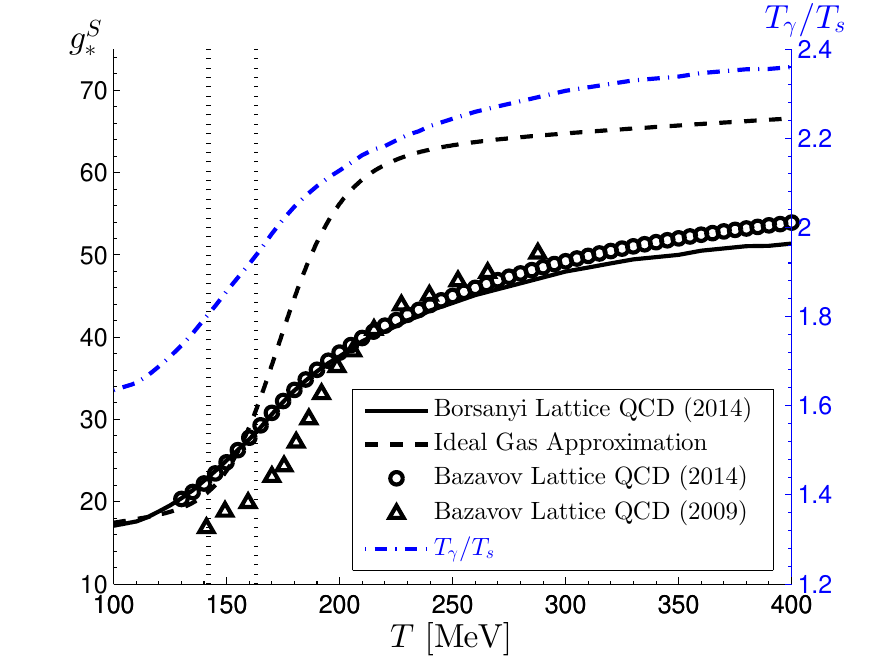}}
\caption{Left axis: Effective number of entropy-DoF, including lattice QCD effects applying~\cite{Borsanyi:2013bia} (solid line) and~\cite{HotQCD:2014kol} (circles), compared to the earlier results~\cite{Bazavov:2009zn} (triangles) used by~\cite{Anchordoqui:2011nh}, and the ideal gas model of~\cite{Coleman:2003hs} (dashed line) as functions of temperature $T$. Right axis: Photon to SP temperature ratio, $T_\gamma/T_s$, as a function of SP decoupling temperature (dash-dotted (blue) line). The vertical dotted lines at $T=142\MeV$ and $T=163\MeV$ delimit the QGP transformation region. \cccite{Birrell:2014cja}\label{fig:gS}}
 \end{figure}

Such a simple characterization does not hold in the vicinity of the QGP phase transformation where quark-hadron degrees of freedom are strongly coupled and the system must be studied using lattice QCD. A computation of $g_*^S$ that incorporates the lattice QCD results is shown in the solid line in \rf{fig:gS} (left axis). Specifically, we used the table of entropy density\index{entropy!density} values through the QGP phase transition presented by Borsanyi et al.~\cite{Borsanyi:2013bia}, while circles show recent results from Bazavov et al.~\cite{HotQCD:2014kol} correcting their earlier 2009 data\cite{Bazavov:2009zn} (triangles), also shown. This should be compared to the use of the ideal gas approximation (dashed) from~\cite{Coleman:2003hs} together with the fit in~\cite{Wantz:2009it} to interpolate though the QGP phase transition. The free gas approximation has a maximum error of $10\%$ in the QGP phase transition temperature range $T\simeq 150$\,MeV. The early 2009 lattice data, used in~\cite{Anchordoqui:2011nh} has a maximum error on the order of $25\%$, which leads to a non-negligible difference in the relation between freeze-out temperature and $N^{\text{eff}}_{\nu}$. 

Conservation of entropy leads to a temperature ratio at $T_\gamma<T_{d,s}$, shown in the dot-dashed line in Figure \ref{fig:gS} (right axis), given by
\begin{equation}\label{eq:TRatio}
R_s\equiv T_{s}/T_{\gamma}=\left(\frac{g_*^S(T_\gamma)}{g_*^S(T_{d,s})}\right)^{1/3}\,.
\end{equation}
Evolving the Universe through neutrino freeze-out\index{neutrino!freeze-out}, if $T_s$ and $T_\gamma$ are the light SP and photon temperatures, both after $e^\pm$ annihilation, and $g_s$ is the number of DoF of the SPs normalized to bosons (i.e., for fermions it includes an additional factor of $7/8$) then this leads to the following change in the effective number of neutrinos\index{neutrino!effective number} in excess of the SM value:
\begin{equation}\label{Neff1}
\delta N_{\text{eff}}\equiv N^{\text{eff}}_{\nu}-3.046=\frac{4g_s}{7}\left(\frac{T_s}{R_s T_{\gamma}}\right)^4\,,
\end{equation}
where $3.046$ is the SM neutrino contribution. Using \req{eq:TRatio} we can rewrite $\delta N_{\text{eff}}$ as
\begin{equation}\label{eq:deltaN}
\delta N_{\text{eff}}=\frac{4g_s}{7R_\nu^4}\left(\frac{g_*^S(T_{\gamma})}{g_*^S(T_{d,s})}\right)^{4/3}\,,
\end{equation}
where $T_{d,s}$ is the decoupling temperature of the SP and $T_{\gamma}$ is any photon temperature in the regime $T_{\gamma}\ll m_e$. The SM particles remaining (in relevant amounts) at such $T_{\gamma}$ are photons and SM neutrinos, the latter with temperature $R_\nu T_{\gamma}$, and so $g_*^S(T_{\gamma})=2+7/8\times 6\times 4/11$ and (see also Eq.(2.7) in~\cite{Blennow:2012de})
\begin{align}\label{eq:deltaN2}
\delta N_{\text{eff}}\approx&g_s\left(\frac{7.06}{g_*^S(T_{d,s})}\right)^{4/3}\,.
\end{align}

\begin{figure}
\centerline{\includegraphics[width=0.75\linewidth]{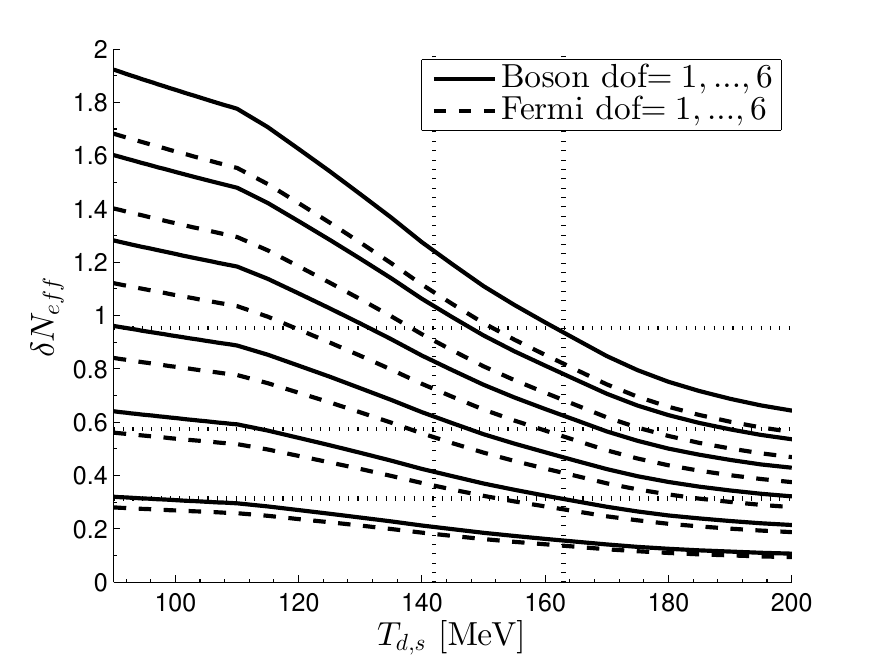}}
\caption{Solid lines: Increase in $\delta N_{\text{eff}}$ due to the effect of $1,\dots,6$ light sterile boson DoF ($g_s=1,\dots,6$, bottom to top curves) as a function of freeze-out temperature $T_{d,s}$. Dashed lines: Increase in $\delta N_{\text{eff}}$ due to the effect of $1,\dots,6$ light sterile fermion DoF ($g_s=7/8\times 1,\dots,7/8\times 6$, bottom to top curves) as a function of freeze-out temperature $T_{d,s}$. The horizontal dotted lines correspond to $\delta N_{\text{eff}}+0.046=0.36,0.62,1$. The vertical dotted lines show the reported range of QGP transformation temperatures $T_c=142-163\MeV$. \cccite{Birrell:2014cja}\label{fig:NeffTdZoom}}
\end{figure}

In Figure \ref{fig:NeffTdZoom} we plot $\delta N_{\text{eff}}$ as a function of $T_{d,s}$ for $1,\dots,6$ boson (solid lines) and fermion (dashed lines) DoF. For a low decoupling temperature $T_{d,s}<100$\,MeV; a single bose or fermi SP can help alleviate the observed tension in $N^{\text{eff}}_{\nu}$. Within the QGP hadronization\index{hadrons!hadronization} temperature range $T_c=142-163\MeV$ (marked by vertical dotted lines) we see that three boson degrees of freedom or four fermion degrees of freedom are the most likely cases to resolve the tension. If the SPs froze out in the QGP phase at $T_{d,s}\gg 163\MeV$ then a significantly larger number of SPs would be required. While such a scenario cannot be excluded, such a large number undiscovered weakly broken symmetries, or/and sterile neutrino-like particles, seems unlikely. Therefore we suggest that Figure \ref{fig:NeffTdZoom} pinpoints the QGP temperature range and below as the primary domain of interest for the freeze-out of a small to moderate number of hypothetical degrees of freedom, should an excess in $N_\nu^{\text{eff}}$ above the SM value be needed~\cite{Birrell:2014cja}.

\subsection{Neutrino freeze-out solving the Boltzmann-Einstein equation}\label{ch:param:studies}
In this section we remove the instantaneous freeze-out assumption and present results of a more precise study of neutrino freeze-out\index{neutrino!freeze-out}: We do not assume that the distribution is either in chemical or kinetic equilibrium\index{kinetic equilibrium} or is free-streaming. The required mathematical theory and numerical method is developed in Appendices \ref{ch:vol:forms}, \ref{ch:boltz:orthopoly}, and \ref{ch:coll:simp}. Here we focus our attention on the physical implications, in particular the dependence of the freeze-out process on natural constants\index{natural constants!variation}. This allows us to identify potential avenues by which the tension between observed in terms of the present day value of Hubble parameter\index{Hubble!parameter} $H_0$ and the related theoretical value of $N^{\mathrm{eff}}_\nu$, the key feature of the invisible today neutrino background, may be alleviated. 

Our study also constrains the time and/or temperature variation of certain natural constants by comparing the results with measurements of $N_\nu^{\mathrm{eff}}$. Further details on this work were presented in~\rsec{sec:model:ind}, more details can be found in Ref.\,\cite{Birrell:2014uka}. The topic of the time variation of natural constants is a very active field with a long history; for a comprehensive review of this area, with which we make only slight contact, see {\it e.g.\/} Ref.\,\cite{Uzan:2010pm}. 

\para{Neutrino freeze-out temperature and relaxation time} 
To connect with the instantaneous freeze-out model from \rf{sec:model:ind}, we now give a definition of the kinetic freeze-out temperature that is applicable to the Boltzmann-Einstein equation model and use this to calculate the neutrino freeze-out temperature. Any such definition will be only approximate, as the freeze-out process is not a sharp transition. Our definition is motivated in part the treatment in \cite{Kolb:1990vq}. 

We first define a characteristic length between scatterings. Using the formula \req{n:div}, we obtain the fractional rate of change of comoving particle number
\begin{align}
\frac{\frac{d}{dt}(a^3 n)}{a^3n}=\frac{g_\nu}{2\pi^2n}\int C[f]p^2/Edp\,.
\end{align}
Here we don't want the net change, but rather to count the number of interactions. For that reason, we imagine that only one direction of the process is operational and define the relaxation rate\index{relaxation rate}
\begin{align}
\Gamma\equiv\frac{g_\nu}{2\pi^2n}T^2\int \tilde C[f]zdz\,,
\end{align}
where the one-way directional collision term $\tilde C[f]$ is computed as in \req{coll} except with $F$ replaced by 
\begin{equation}
\tilde F=f_1(p_1)f_2(p_2)f^3(p_3)f^4(p_4)\,.
\end{equation}
When the particle type being considered appears in both directions of a reaction, the corresponding term for the other reaction direction must be added. We emphasize that the key difference here, there is no minus sign; here we are counting reactions, not net particle number change.

\begin{figure} 
\centerline{\includegraphics[width=0.5\linewidth]{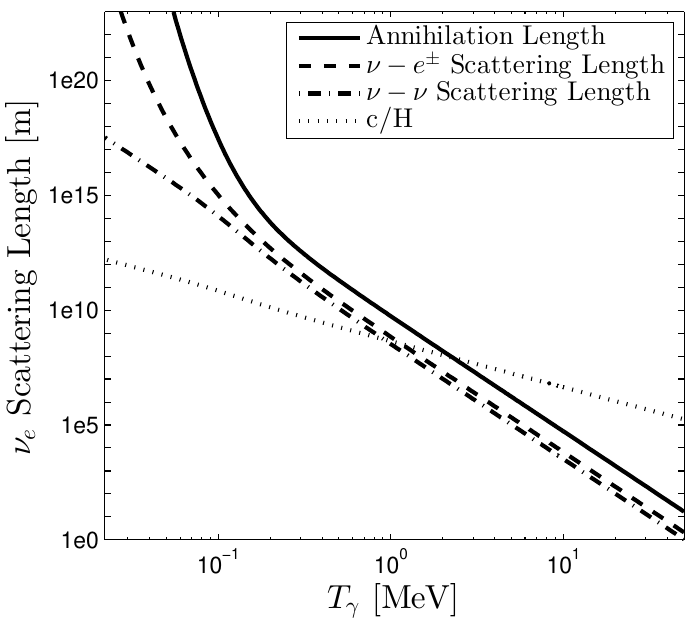}
\includegraphics[width=0.5\linewidth]{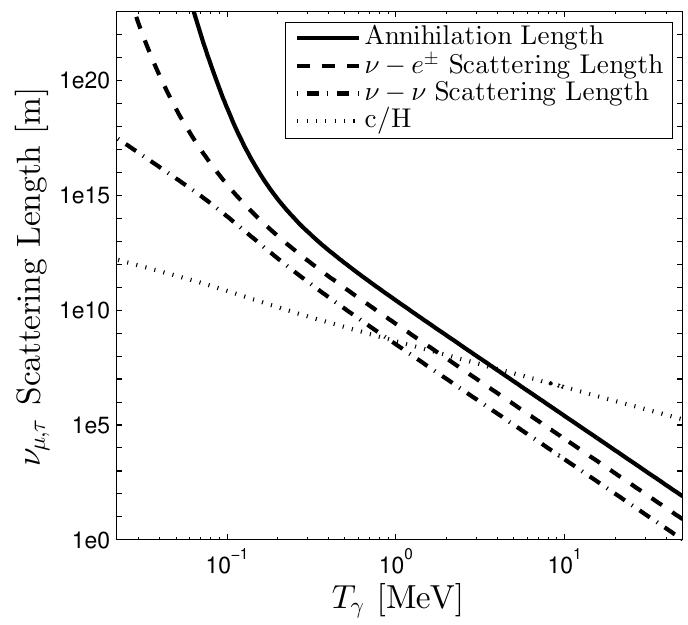}}
\caption{Comparison of Hubble parameter to neutrino scattering length for various types of PP-SM processes, left-hand frame for electron neutrino $\nu_e$ and right-hand frame for the other two flavors $\nu_\mu$, $\nu_\tau$. \cccite{Birrell:2014uka} }\label{fig:scattLength}
\end{figure}

Using the average velocity, which for (effectively massless) neutrinos is $\bar v=c=1$, we obtain what we call the scattering length\index{scattering length}
\begin{align}
L_\Gamma&\equiv\displaystyle\frac{\bar v}{\Gamma}=\displaystyle\frac{\int_0^\infty\displaystyle\frac{1}{\Upsilon^{-1}e^z+1}z^2dz}{\int_0^\infty \tilde C[f] z^2/E dz}\,.
\end{align}
This can be compared to the Hubble length $L_H=c/H$\index{Hubble!length} and the temperature at which $L_\Gamma=L_H$ we call the freeze-out temperature\index{freeze-out!temperature} for that reaction. Figure \ref{fig:scattLength} shows the scattering length and $L_H$ for various types of neutrino reactions. The solid line corresponds to the annihilation process $e^+e^-\rightarrow \nu\bar\nu$, the dashed line corresponds to the scattering $\nu e^\pm\rightarrow \nu e^\pm$, and the dot-dashed line corresponds to the combination of all processes involving only neutrinos. The freeze-out temperatures in MeV are given in Table \ref{table:freezeoutTemp}.

\begin{table}[ht]
\centering 
\begin{tabular}{|c|c|c|c|}
\hline
 & $e^+e^-\rightarrow \nu\bar\nu$ & $\nu e^\pm\rightarrow \nu e^\pm$ & $\nu$-only processes\\
\hline
$\nu_e$ &2.29 & 1.15&0.910\\
\hline
$\nu_{\mu,\tau}$ &3.83 & 1.78& 0.903\\
\hline
\end{tabular}
\caption{Freeze-out temperatures in MeV for electron neutrinos and for $\mu$,$\tau$ neutrinos.}
\label{table:freezeoutTemp}
\end{table}

We now consider the relaxation time\index{relaxation time} for a given reaction, defined by $\tau=1/\Gamma$. Suppose we have a time interval $t_f>t_i$ and corresponding temperature interval $T_f<T_i$ during which there is no reheating and the Universe is radiation dominated. Normalizing time so $t=0$ corresponds to the temperature $T_i$ we have
\begin{equation}\label{ch6:Heq}
\dot a/a=-\dot T/T\,,\hspace{2mm} H=\frac{C}{2Ct+T_i^2}\propto T^2
\,,
\end{equation}
where $C$ is a constant that depends on the energy density and the Planck mass. Its precise form will not be significant for us. Note that \req{ch6:Heq} implies
\begin{equation}
1/H(t)-1/H(0)=2t\,.
\end{equation}

At $T\gg m_e$, the rates for reactions under consideration from Tables \ref{table:nu:e:reac} and \ref{table:nu:mu:reac} scale as $\Gamma\propto T^5$. Therefore, supposing $H(T_f)/\Gamma(T_f)=1$ (which occurs at $T_f=O(1\MeV)$ as seen in the above figures), at any time $t_f>t>t_i$ we find 
\begin{align}\label{relaxTime}
\tau(t)/t=&\frac{2}{\Gamma(t)}\left(\frac{1}{H(t)}-\frac{1}{H(0)}\right)^{-1}=\frac{2T_f^5}{\Gamma(T_f)T^5}\left(\frac{T_f^2}{H(T_f)T^2}-\frac{T_f^2}{H(T_f)T_i^2}\right)^{-1}\\
=&\frac{2T_f^3}{T^3}\left(1-\frac{T^2}{T_i^2}\right)^{-1}\,.
\end{align}
Therefore, given any time $t_i<t_0<t_f$ we have
\begin{equation}\label{ch6:tauEq}
\tau(t)<\tau(t_0)=\frac{2T_f^3}{T_0^3}\left(1-\frac{T_0^2}{T_i^2}\right)^{-1}\Delta t \text{ \,for all } t<t_0\,,
\end{equation}
where $\Delta t=t_0-t_i=t_0$.

\begin{figure} 
\centerline{\includegraphics[width=0.52\linewidth]{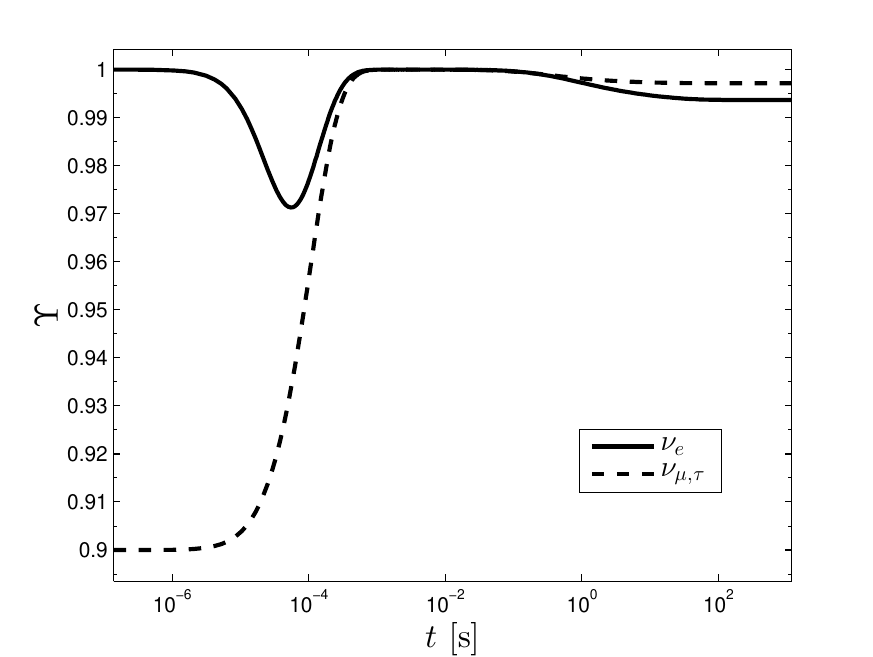}\hspace*{-0.7cm}
\includegraphics[width=0.52\linewidth]{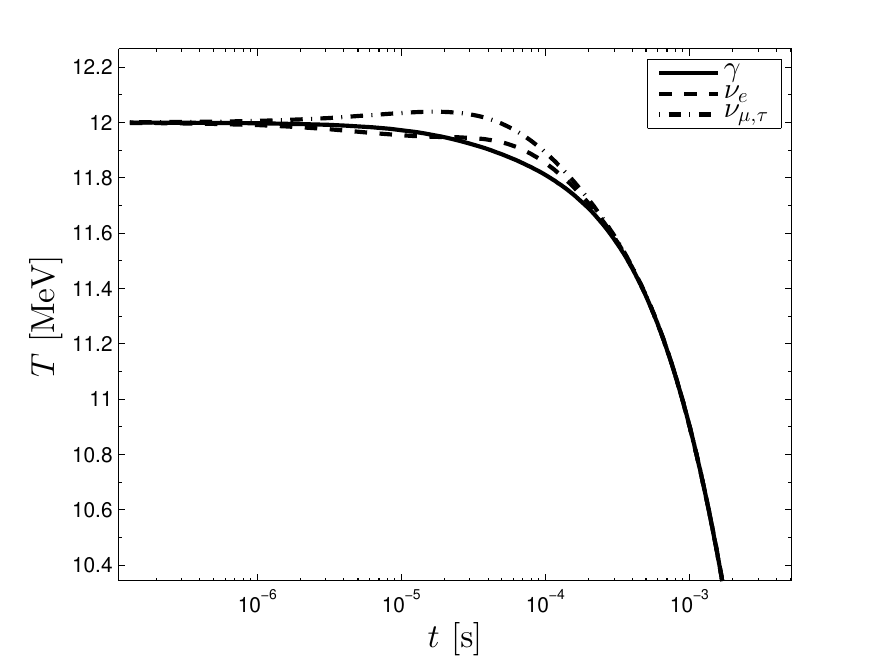}}
\caption{Starting at $12\MeV$, this figure shows the relaxation of a nonequilibrium $\mu,\tau$-neutrino distribution towards equilibrium. The fugacities are shown in the left-hand frame while the temperatures are shown in the right-hand frame}
\label{fig:relax}
 \end{figure}

 The first reheating period that precedes neutrino freeze-out\index{neutrino!freeze-out} is the disappearance of muons\index{muon} and pions around $O(100\MeV)$, as seen in Figure \ref{fig:energy:frac}, and so we let $T_i=100\MeV$. \req{ch6:tauEq} is minimized at $T_0\approx 77.5\MeV$ at which point we have 
\begin{equation}
\tau(t)<10^{-5} \Delta t_0 \text{ for } t<t_0.
\end{equation}
This shows that the relaxation time during the period between $100\MeV$ and $77.5\MeV$ is at least five orders of magnitude smaller than the corresponding time interval. Therefore the system has sufficient time to relax back to equilibrium after any potential nonequilibrium aspects developed during the reheating period. This justifies our assumption that the neutrino distribution has the equilibrium Fermi Dirac form at $T=O(10 \MeV)$ when we begin our numerical simulation. This can also be demonstrated numerically in Figure \ref{fig:relax}, where we have initialized the system at $T_\gamma=12\MeV$ with a nonequilibrium distribution of $\mu$ and $\tau$ neutrinos, giving them $\Upsilon=0.9$, and let them evolve under the Boltzmann-Einstein equation. We see that after approximately $10^{-3}$ seconds the system relaxes back to equilibrium, well before neutrino freeze-out near $t=1$s.

\para{Dependence of effective neutrino number on PP-SM parameters}
Only two key PP-SM parameters influence the effective number of neutrinos;\index{neutrino!effective number} this is the Weinberg angle and the generalized interaction strength $\eta$. We explore in the following how $N_\nu^{\mathrm{eff}}$ depends on these parameters.

The Weinberg angle\index{Weinberg angle} is one of the key standard model parameters that impacts the neutrino freeze-out process. More specifically, it is found in the matrix elements of weak force processes, including the reactions $e^+e^-\rightarrow \nu\bar\nu$ and $\nu e^\pm\rightarrow \nu e^\pm$ as found in Tables \ref{table:nu:e:reac} and \ref{table:nu:mu:reac}. It is determined by the $SU(2)\times U(1)$ coupling constants $g$, $g^{'}$ by
\begin{equation}
\sin(\theta_W)=\frac{g^{'}}{\sqrt{g^2+(g^{'})^2}}\,.
\end{equation}
It is also related to the mass of the $W$ and $Z$ bosons and the Higgs vacuum expectation value\index{Higgs!vacuum expectation value} $v$ by
\begin{equation}\label{eq:MWMZ}
M_Z=\frac{1}{2}\sqrt{g^2+(g^{'})^2}v\,,\hspace{2mm} M_W=\frac{1}{2}gv\,,\hspace{2mm} \cos(\theta_W)=\frac{M_W}{M_Z}\,,
\end{equation}
as well as the electromagnetic coupling strength
\begin{equation}\label{eq:egg}
e=\frac{2M_W}{v}\sin(\theta_W) =\frac{gg^{'}}{\sqrt{g^2+(g^{'})^2}}\,.
\end{equation}
It has a measured value in vacuum $\theta_W\approx 30^\circ$, giving $\sin(\theta_W)\approx 1/2$, but its value is not fixed within the Standard Model. For this reason, a time or temperature variation can be envisioned and this would have an observable impact on the neutrino freeze-out process, as measured by $N_\nu^{\mathrm{eff}}$.

In letting $\sin(\theta_W)$, and hence $g$ and $g^{'}$, vary, we must fix the electromagnetic coupling $e$ so as not to impact sensitive cosmological observables such as Big-Bang Nucleosynthesis. Another argument to keep the value of electric charge fixed is the gauge invariance symmetry which assures charge conservation. 

 According to \req{eq:MWMZ}, $M_Z>M_W\gg |p|$ for neutrino momentum $p$ in the energy range of neutrino freeze-out, around $1\MeV$, even as we vary $\sin(\theta_W)$. This approximation is inherent in the formulas for the matrix elements in Tables~\ref{table:nu:e:reac} and \ref{table:nu:mu:reac} and continues to be valid here. We will characterize the dependence of $N_\nu^{\mathrm{eff}}$ on $\sin(\theta_W)$ in following, but first we identify the remaining parameter dependence in the Boltzmann-Einstein system

Beyond the Weinberg angle, the remaining dependence of the Boltzmann-Einstein system on dimensioned quantities during neutrino freeze-out can be combined into one overall interaction strength factor. To show this, we now convert the system to dimensionless form. Letting $m_e$ be the mass scale and $M_p/m_e^2$ be the time scale the Einstein equations take the form
\begin{equation}
H^2=\frac{\rho}{3}\,,\hspace{2mm}\dot\rho=-3H(\rho+P)\,.
\end{equation}
Since $e^\pm$ are the only (effectively) massive particles in the system, by scaling all energies, momenta, energy densities, pressures, and temperatures by $m_e$ we have removed all scale dependent parameters from the Einstein equations. The Boltzmann-Einstein equation becomes
\begin{equation}\label{etaDef}
\partial_tf-pH\partial_pf=\eta\frac{C[f]}{E}\,,\hspace{2mm}\eta\equiv M_p m_e^3G_F^2\,,
\end{equation}
where we have also factored out of $C[f]$ the $G_F^2$ term that is common to all of the neutrino interaction matrix elements. 

Aside from the $\theta_W$ dependence of the matrix elements seen in Tables \ref{table:nu:e:reac} and \ref{table:nu:mu:reac}, the complete dependence on natural constants is now contained in a single dimensionless neutrino interaction strength parameter\index{neutrino!interaction strength} $\eta$ with the vacuum present day value\index{neutrino!decoupling strength $\eta$} presented in \req{eta0CTY}
\begin{equation}\label{eta0Def}
\eta\to \eta_0 = 0.04421\, .
\end{equation}

\para{Impact of QED corrections to equation of state}
At the time of neutrino freeze-out, the Universe is at a sufficiently high temperature for photons and $e^\pm$ to be in chemical and kinetic equilibrium.\index{QED!Corrections EOS} The temperature is also sufficiently high for QED corrections to the photon and $e^\pm$ equation of state to be non-negligible. Therefore, in our study here we use the results given in \cite{Heckler:1994tv,Mangano:2001iu} to include these in our computation by modifying the combined photon, $e^\pm$ equation of state
\begin{align}
P=P^0+P^{int},\hspace{2mm} \rho=-P+T\frac{dP}{dT}\,,
\end{align}
where
\begin{align}
P^{int}=&-\frac{1}{2\pi^2}\int_0^\infty\left[\frac{k^2}{E_k}\frac{\delta m_e^2}{e^{E_k/T}+1}+\frac{k}{2}\frac{\delta m_\gamma^2}{e^{k/T}-1}\right]dk\,,\hspace{2mm} E_k=\sqrt{k^2+m_e^2}\,,\\
\delta m_e^2=&\frac{2\pi\alpha^2}{3}+\frac{4\alpha}{\pi}\int_0^\infty \frac{k^2}{E_k}\frac{1}{e^{E_k/T}+1}dk\,,\hspace{2mm} \delta m_\gamma^2=\frac{8\alpha}{\pi}\int_0^\infty \frac{k^2}{E_k}\frac{1}{e^{E_k/T}+1}dk\,,
\end{align}
and $P^0$ is the pressure of a non-interacting gas of photons and $e^\pm$ in chemical equilibrium\index{chemical equilibrium}.

\para{Freeze-out T and effective neutrino number dependence on PP-SM}
We now present the dependence of the effective number of neutrinos\index{neutrino!effective number}, $N_\nu^{\mathrm{eff}}$, on the SM parameters $\sin^2(\theta_W)$ and $\eta$, as computed using the Boltzmann-Einstein equation method developed in Appendices \ref{ch:vol:forms}, \ref{ch:boltz:orthopoly}, and \ref{ch:coll:simp}. These results are shown in \rf{NnuParams}, presented as a function of Weinberg angle $\sin^2(\theta_W) $ for $\eta/\eta_0=1,2,5,10$. The effects of an increase in both parameters above the vacuum values can generate a significant increase in $N_\nu^{\mathrm{eff}}\to 3.5$.
\begin{figure}
\centerline{\includegraphics[width=0.80\linewidth]{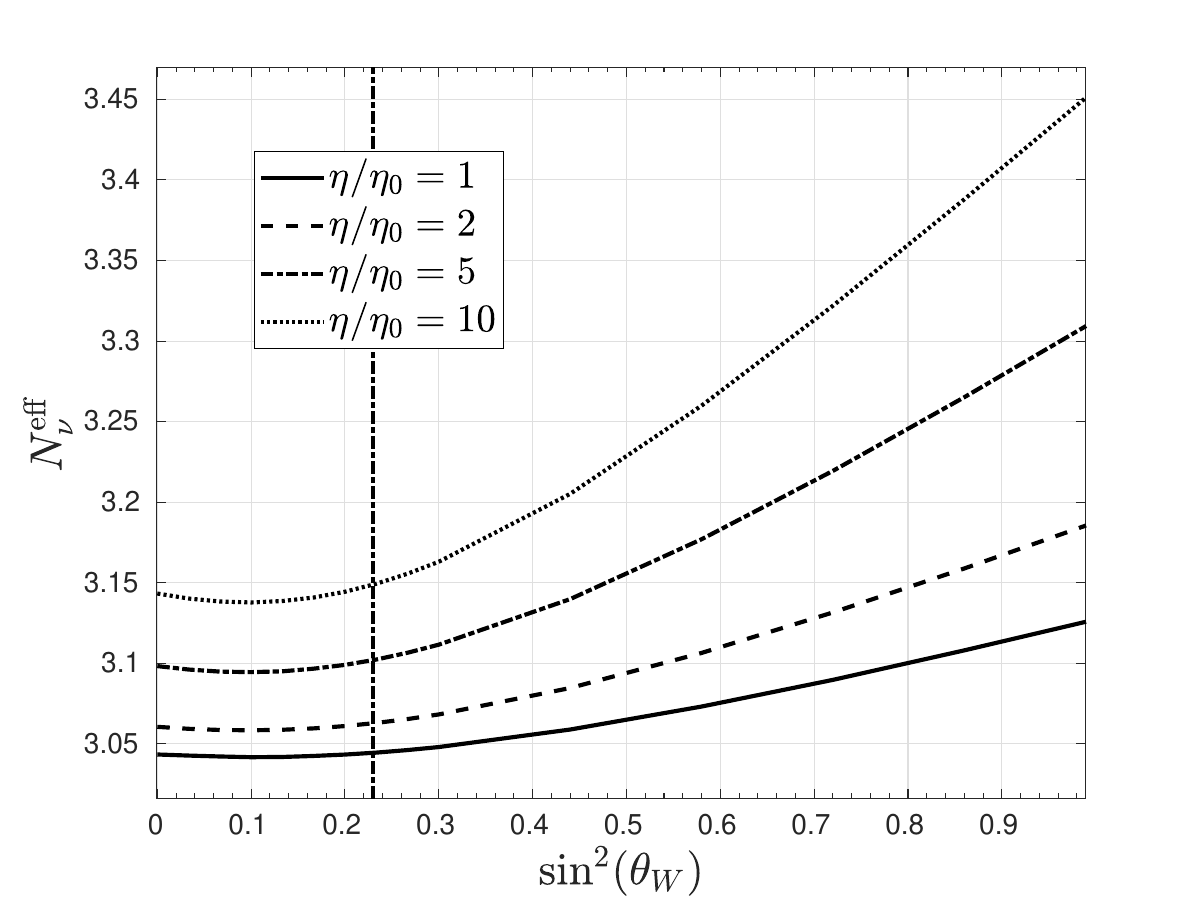}}
\caption{Change in the effective number of neutrinos $N_\nu^{\mathrm{eff}}$ as a function of the Weinberg angle for several values of $\eta/\eta_0=1,2,5,10$. Vertical line is $\sin^2(\theta_W)=0.23$. \radapt{Birrell:2014uka}}
\label{NnuParams} 
 \end{figure}

We performed a least-squares fit of $N_\nu^{\mathrm{eff}}$ over the range $0\leq \sin^2(\theta_W)\leq 1$, $1\leq \eta/\eta_0\leq 10$ shown in figure \ref{NnuParams}, obtaining a result with relative error less than $0.2\%$,
\begin{align}
N_\nu^{\mathrm{eff}}=&3.003-0.095\sin^2(\theta_W) +0.222\sin^4(\theta_W ) -0.164\sin^6(\theta_W )\notag\\
+&\sqrt{\frac{\eta}{\eta_0}}\left(0.043+0.011\sin^2(\theta_W) +0.103\sin^4(\theta_W)\right)\,.
\end{align}
$N_\nu^{\mathrm{eff}}$ is monotonically increasing in $\eta/\eta_0$ with dominant behavior scaling as $\sqrt{ \eta/\eta_0}$. Monotonicity is to be expected, as increasing $\eta$ decreases the freeze-out temperature and the longer neutrinos are able to remain coupled to $e^\pm$, the more energy and entropy from annihilation is transferred to neutrinos.

We complement this with fits to the photon to neutrino temperature ratios $ T_\gamma / T_{\nu_e}$, $T_\gamma / T_{\nu_\mu}= T_\gamma / T_{\nu_\tau} $, and the neutrino fugacities, $\Upsilon_{\nu_e}, \Upsilon_{\nu_\mu}=\Upsilon_{\nu_\tau}$, again with relative error less than $0.2\%$, 
\begin{align}
\frac{T_\gamma}{T_{\nu_\mu}}=&1.401+0.015x-0.040x^2+0.029x^3-0.0065y+0.0040xy-0.017x^2y\,, \notag\\
\Upsilon_{\nu_e}=&1.001+0.011x-0.024x^2+0.013x^3-0.005y-0.016xy+0.0006x^2y\,,\notag\\ 
\frac{T_\gamma}{T_{\nu_e}}=&1.401+0.015x-0.034x^2+0.021x^3-0.0066y-0.015xy-0.0045x^2y\,,\notag\\
\Upsilon_{\nu_\mu}=&1.001+0.011x-0.032x^2+0.023x^3-0.0052y+0.0057xy-0.014x^2y\,,
\end{align}
where
\begin{equation}
x\equiv \sin^2(\theta_W)\, ,\qquad
y\equiv \sqrt{\frac{\eta}{\eta_0}}\,.
\end{equation}

\begin{figure}
\centerline{\includegraphics[width=0.95\linewidth]{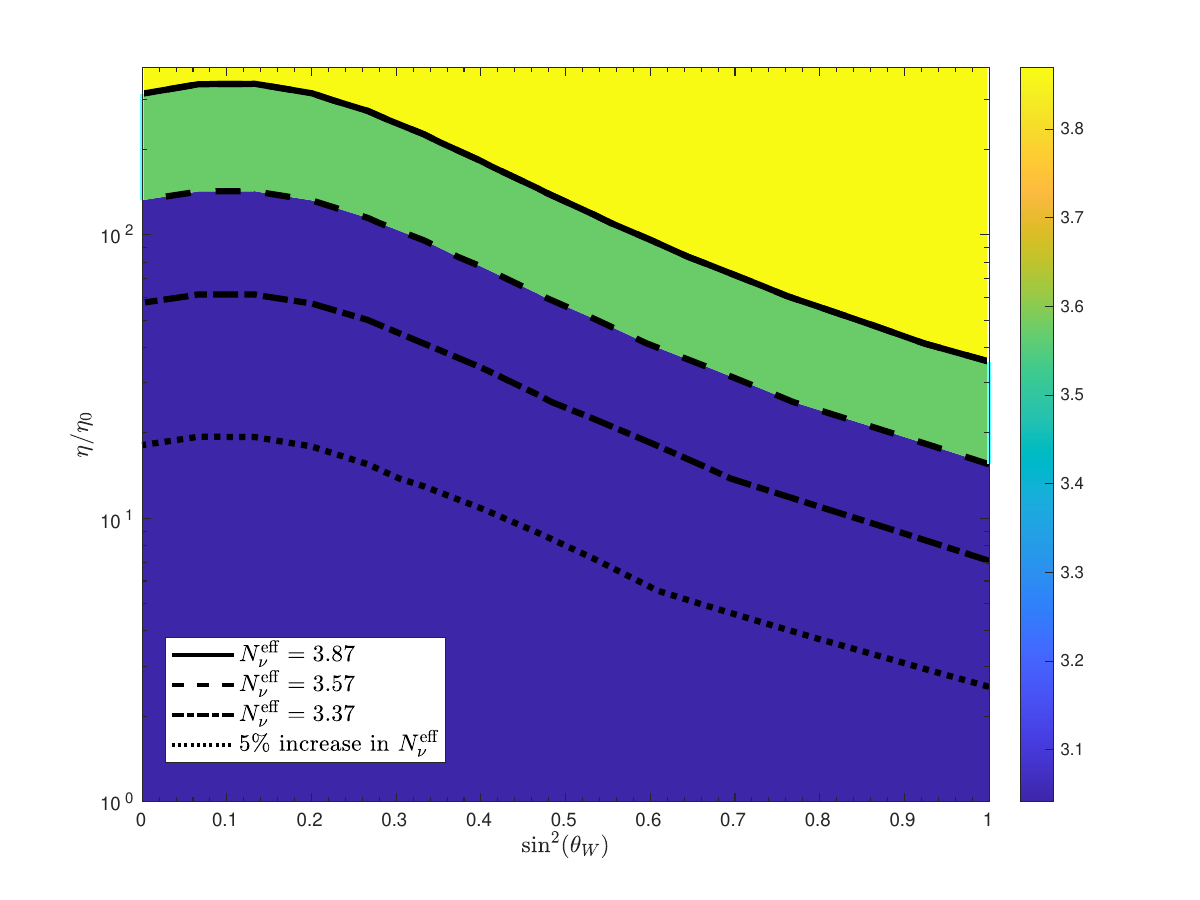}
}
\caption{$N_\nu^{\mathrm{eff}}$ bounds in the $\eta/\eta_0, \sin^2(\theta_W)$ plane. Blue for $N_\nu^{\mathrm{eff}}\in (3.03,3.57)$ corresponding to Ref.~\cite{Planck:2013pxb} CMB+BAO analysis\index{CMB} and green extends the region to $N_\nu^{\mathrm{eff}}<3.87$ i.e. to CMB+$H_0$. Dot-dashed line delimits the 1 standard-deviation lower boundary of the second analysis. \radapt{Birrell:2014uka}}
\label{NnuDomain}
 \end{figure}

The bounds on $N_\nu^{\mathrm{eff}}$ from the Planck analysis \cite{Planck:2013pxb} can be used to constrain time or temperature variation of $\sin^2(\theta_W)$ and $\eta$. 
In Figure \ref{NnuDomain} the blue region shows the combined range of variation of natural constants\index{natural constants!variation} compatible with CMB+BAO and the green region shows the extension in the range of variation of natural constants for CMB+$H_0$, both at a $68\%$ confidence level. The dot-dashed line within the blue region delimits this latter domain. The dotted line shows the limit of a 5\% change in $N_\nu^{\mathrm{eff}}$. Any increase in $\eta/\eta_0$ and/or $\sin^2(\theta_W)$ moves the value of $N_\nu^{\mathrm{eff}}$ into the domain favored by current experimental results.

We have omitted here a discussion of flavor neutrino oscillations. If it weren't for the differences between the matrix elements for the interactions between $e^\pm$ and $\nu_e$ on one hand and $e^\pm$ and $\nu_\mu,\nu_\tau$ on the other, oscillations would have no effect on the flow of entropy into neutrinos and hence no effect on $N_\nu^{\mathrm{eff}}$, but these differences do lead to a modification of $N_\nu^{\mathrm{eff}}$. In \cite{Mangano:2005cc} the impact of oscillations on neutrino freeze-out\index{neutrino!freeze-out} for the present day measured values of $\theta_W$ and $\eta$ was investigated. It was found that while oscillations redistributed energy amongst the neutrino flavors, the impact on $N_\nu^{\mathrm{eff}}$ was negligible. We have neglected oscillations in our study.

\para{Primordial variation of natural constants}
We end our study of neutrino freeze-out by exploring what neutrino decoupling in the early Universe can tell us about the values of natural constants when the Universe was about one second old\index{natural constants!variation} and at an ambient temperature near to 1 MeV (11.6 billion degrees K). Our results were presented assuming that the Universe contains no other effectively massless particles but the three left-handed neutrinos and three corresponding right-handed anti-neutrinos. 

In \rf{NnuParams} we see that, near the present day value of the Weinberg angle $\sin^2(\theta_W)\simeq 0.23$, the effect of changing $\sin^2(\theta_W)$ on the decoupling of neutrinos is relatively small. The dominant variance is due to the change in the coupling strength $\eta/\eta_0$, \req{etaDef} and \req{eta0Def}. The dotted line in Figure \ref{NnuDomain} shows that in order to achieve a change in $N_\nu^{\mathrm{eff}}$ at the level of up to 5\%, i.e., $N_\nu^{\mathrm{eff}}\lesssim 3.2 $, $\eta/\eta_0$ must change significantly, e.g., increasing by an order of magnitude.

It is not possible to exclude with certainty such a large scale in the primordial Universe as we will now argue considering the natural constants contributing to $\eta$ and their required
modification:
\begin{itemize}
\item
In models of emergent gravity we can imagine a `melting' of gravity in the hot primordial Universe, just like we see the vacuum structure and quark confinement melt. Conversely, and perhaps more attractive in light of the present day interest in the so-called Hubble tension\index{Hubble!tension}, there could be present era weakening of gravity which would allow the Universe expansion to accelerate and more generally could also modify the dark energy input into Universe dynamics. Whether such a variable gravity model can be realized will be a topic for future consideration. Considering that $\eta\propto M_p\propto G_N^{-1/2}$ the value of $\eta$ will change in the opposite to the strength of gravity: An order magnitude change in $\eta$ at the time of neutrino decoupling translates into two orders of magnitude (inverse) change in the strength of gravity. One would not think this is a possible scenario mainly because neutrino decoupling occurs at a scale so much different from gravity. The question about temporal variation of gravity strength, along with temperature dependence cannot be as yet addressed in the absence of fundamental gravity theory. 
\item
Compared to all other elementary particles the electron mass has an unusually low value. This could imply a more complicated mass origin of the electron when compared to other elementary particles which are drawing their mass by the minimal coupling from the Higgs field\index{Higgs!field}. We studied a strong field mechanism for electron mass melting recently~\cite{Evans:2019zyk}. Since $\eta\propto m_e^3$, electron mass would need to change at the time of decoupling of neutrinos by `only' a factor 2.15 to create an order of magnitude impact on $\eta$. This seems not entirely impossible.
\item
A modification by `only' a factor of 1.8 in the vacuum expectation value (VEV) of the Higgs field $v_0\simeq 246$ GeV controlling the weak interaction coupling $G_\mathrm{F}\propto 1/v^4$ would suffice to alter $\eta$ by an order of magnitude. However, if we allow electron mass to be also Higgs controlled, three powers of $v$ would cancel and a change in $v$ by an order of magnitude near to $T\simeq m_e$ would be required. In either case, given our good understanding of the standard model of particle physics we do not believe that the VEV of the Higgs field could be impacted by the MeV-scale temperature prevailing at the time of neutrino decoupling.
\end{itemize}
To summarize: Gravity, even though it is an effective theory poorly understood at a fundamental level, is governed by the Planck mass scale which is many, many orders of magnitude above the scales we are exploring in the epoch of neutrino decoupling. Similarly, the Higgs VEV which controls $G_F$ seems also immutable at the neutrino decoupling temperature, considering the relevant scale being different by a factor of about 500,000. On the other hand, electron mass $m_e$ is `anomalously' small, it is the only elementary scale below the temperature scale of neutrino decoupling, hence it is prone to be modifiable in primordial hot Universe. One can wonder if its small mass is due to an interplay between quantum effects, Higgs coupling and QED interaction. If so the mass would be modifiable at a temperature that is larger than the mass value which is the condition for neutrino decoupling. This therefore could be the cause of a substantial primordial increase in $\eta$, impacting the present day Universe expansion speed through the value of $N_\nu^\mathrm{eff}$.
 
One could further argue that any value of $\sin^2(\theta_W)$ is possible at time of neutrino decoupling, as there is no rational for the vacuum observed symmetry breaking mixing value of $\sin^2(\theta_W)$. However, in the SU(5) model unifying quarks and leptons\index{lepton} a natural value $\sin^2(\theta_W)=1/4$ appears. Since this model has been discredited by baryon stability, we could still admit any temperature and/or time dependence of $\sin^2(\theta_W)$. Even so the appearance of a natural $\sin^2(\theta_W)=1/4$ value in the framework of one model could imply that a more realistic model will lead to a similar value.

\subsection{Lepton number and effective number of neutrinos}\label{sec:NeffIntro}
\para{Invisible lepton number: relic neutrinos}
As discussed in the preceding sections, neutrinos\index{neutrino!effective number} decoupled from the cosmic plasma in the early Universe at a temperature of $T=\mathcal{O}(2\mathrm{MeV})$ and became free-streaming\index{free-streaming}. However, after freeze-out neutrinos still continue to play a significant role in the evolution of the Universe and have an impact on cosmological observations involving BBN, CMB, and the matter spectrum for large scale structure. This is due to the sensitivity of the Hubble parameter\index{Hubble!parameter} to the total energy density in the Universe. Besides photons, neutrinos are the most abundant species and contribute significantly to the relativistic energy density throughout the early Universe, affecting the Hubble expansion rate significantly. 

The contribution of energy density from the neutrino sector can be described by the effective number of neutrinos $N_{\nu}^{\mathrm{eff}}$, which captures the number of relativistic degrees of freedom for neutrinos as well as any reheating that occurred in the sector after freeze-out. The effective number of neutrino is defined as 
\begin{align}\label{Neff}
N_\nu^{\mathrm{eff}}\equiv\frac{\rho^{\mathrm{tot}}_\nu}{\frac{7\pi^2}{120}\left(\frac{4}{11}\right)^{4/3}T_\gamma^4}\;,
\end{align}
where $\rho_\nu^{\mathrm{tot}}$ is the total energy density in neutrinos and $T_\gamma$ is the photon temperature. $N_\nu^{\mathrm{eff}}$ is defined such that three neutrino flavors with zero participation of neutrinos in reheating during $e^+e^-$ annihilation results in $N_\nu^{\mathrm{eff}}=3$. The factor of $\left(4/11\right)^{1/3}$ relates the photon temperature to the free-streaming neutrinos temperature, under the assumption of zero neutrino reheating after $e^+e^-$ annihilation. The currently accepted theoretical value is $N_\nu^{\mathrm{eff}}=3.046$, after including the slight effect of neutrino reheating \cite{Mangano:2005cc,Birrell:2014uka}. The favored value of $N_\nu^{\mathrm{eff}}$ can be found by fitting to CMB data. In 2013 the Planck collaboration found $N_\nu^{\mathrm{eff}}=3.36\pm0.34$ (CMB only) and $N_\nu^{\mathrm{eff}}= 3.62\pm0.25$ (CMB and $H_0$)~\cite{Planck:2013pxb}.

To explain the experimental value of $N_\nu^{\mathrm{eff}}$, many studies aim to improve the calculation of neutrino decoupling in the early Universe, including exploring the dependence of freeze-out on natural constants~\cite{Birrell:2014uka}, the entropy transfer from $e^+e^-$ annihilation and finite temperature correction~\cite{Dicus:1982bz,Heckler:1994tv,Fornengo:1997wa}, neutrino decoupling with flavor oscillations~\cite{Mangano:2001iu,Mangano:2005cc}\index{neutrino!flavor oscillation}, and investigating nonstandard neutrino interactions \cite{Morgan:1981zy,Fukugita:1987uy,Elmfors:1997tt,Vogel:1989iv,Mangano:2006ar,Giunti:2008ve,Mangano:2006ar}.

The standard cosmological model assumes that the lepton asymmetry\index{lepton!asymmetry} $L\equiv [N_\mathrm{L}-N_{\overline{\mathrm{L}}}] /N_\gamma $ (normalized with the photon number) 
between leptons and anti-leptons is small, similar to the \index{baryon!asymmetry} $\eta_\gamma=[N_\mathrm{B}-N_{\overline{\mathrm{B}}}]/N_\gamma $; most often it is assumed $L=B$. Barenboim, Kinney, and Park~\cite{Barenboim:2016shh,Barenboim:2017dfq} noted that the lepton asymmetry of the Universe is one of the most weakly constrained parameters is cosmology and they propose that models with leptogenesis are able to accommodate a large lepton number asymmetry surviving up to today. Moreover, the discrepancy between $H_\mathrm{CMB}$ and $H_0$ has increased~\cite{riess2018new,Riess:2018byc,Planck:2018vyg}. The Hubble tension and the possibility that leptogenesis in the early Universe resulted in neutrino asymmetry motivate our study of the dependence of $N_\nu^{\mathrm{eff}}$ on lepton asymmetry, $L$. In our work~\cite{Yang:2018oqg} we consider $L\simeq 1$ and explore how this large cosmological lepton yield relates to the effective number of (Dirac) neutrinos $N^{\mathrm{eff}}_\nu$. 

\para{Relation between the effective number of neutrinos and chemical potential} We consider how neutrinos decouple~\cite{Birrell:2014gea} at a temperature of $T_f\simeq 2\,\mathrm{MeV}$ and are subsequently free-streaming. Assuming exact thermal equilibrium at the time of decoupling, the resulting free-streaming FDEV distribution was obtained in \rsec{sec:model:ind}, \req{eq:NeutrinoDist} 
\begin{align}
\label{fnudef}
&f_\nu=\frac{1}{\exp{\left(\sqrt{\frac{E^2-m_\nu^2}{T_\nu^2}+\frac{m^2_\nu}{T^2_f}}-\sigma\frac{\mu_\nu}{T_f}\right)+1}}\;,\qquad T_\nu\equiv\frac{a(t_f)}{a(t)}T_f\,,
\end{align}
where $\sigma=+1(-1)$ denotes particles (antiparticles) and the effective neutrino temperature $T_\nu$ is understood in terms of red-shifting the momentum in the comoving volume element of the Universe.

Since the freeze-out temperature $T_f\gg m_\nu$ and also neutrino temperature $T_\nu\gg m_\nu$ in the domain of our analysis, we can consider the massless limit in Eq.\;(\ref{fnudef}). Under this approximation, the total neutrino energy density can be written as
\begin{align}
\label{EnergyDensity}
\rho_\nu^{\mathrm{tot}}
&=\frac{g_\nu\,T_\nu^4}{2\pi^2}\left[\frac{7\pi^4}{60}+\frac{\pi^2}{2}\left(\frac{\mu_\nu}{T_f}\right)^{\!\!2}+\frac{1}{4}\left(\frac{\mu_\nu}{T_f}\right)^{\!\!4}\right].
\end{align}
Substituting Eq.\;(\ref{EnergyDensity}) into the definition of the effective number of neutrinos\index{neutrino!effective number} Eq.~(\ref{Neff}), we obtain 
\begin{align}
\label{Neff002}
N_\nu^{\mathrm{eff}}\!\!
=\!3\!\left(\frac{11}{4}\right)^{\!\!\frac{4}{3}}\!\!\left(\frac{T_\nu}{T_\gamma}\right)^{\!\!4}\!
\left[1\!+\!\frac{30}{7\pi^2}\!\!\left(\frac{\mu_\nu}{T_f}\right)^{\!\!2} 
\!\!+\frac{15}{7\pi^4}\!\!\left(\frac{\mu_\nu}{T_f}\right)^{\!\!4}\right].
\end{align}
From Eq.\;(\ref{Neff002}) we have for the standard photon reheating ratio $T_\nu/T_\gamma=(4/11)^{1/3}$ \cite{Kolb:1990vq} and degeneracy $g_\nu=3$ (flavor), the relation between the effective number of neutrinos and the chemical potential\index{chemical potential} at freeze-out
\begin{align}
\label{NeffPotential}
N_\nu^{\mathrm{eff}}=3\left[1+\frac{30}{7\pi^2}\left(\frac{\mu_\nu}{T_f}\right)^{\!\!2}+ \frac{15}{7\pi^4} \left(\frac{\mu_\nu}{T_f}\right)^{\!\!4}\right].
\end{align}
To solve the neutrino chemical potential $\mu_\nu/T_f$ as a function of the effective number of neutrinos, we can neglect the $(\mu_\nu/T_f)^4$ term in Eq.\;(\ref{NeffPotential}) because $m_\nu\ll T_f$ and obtain
\begin{align}\label{Solution}
\frac{\mu_\nu}{T_f}=\pm\sqrt{\frac{7\pi^2}{30}\left(\frac{N_\nu^{\mathrm{eff}}}{3}-1\right)}\,.
\end{align}

\begin{figure}[t]
\begin{center}
\includegraphics[width=0.8\linewidth]{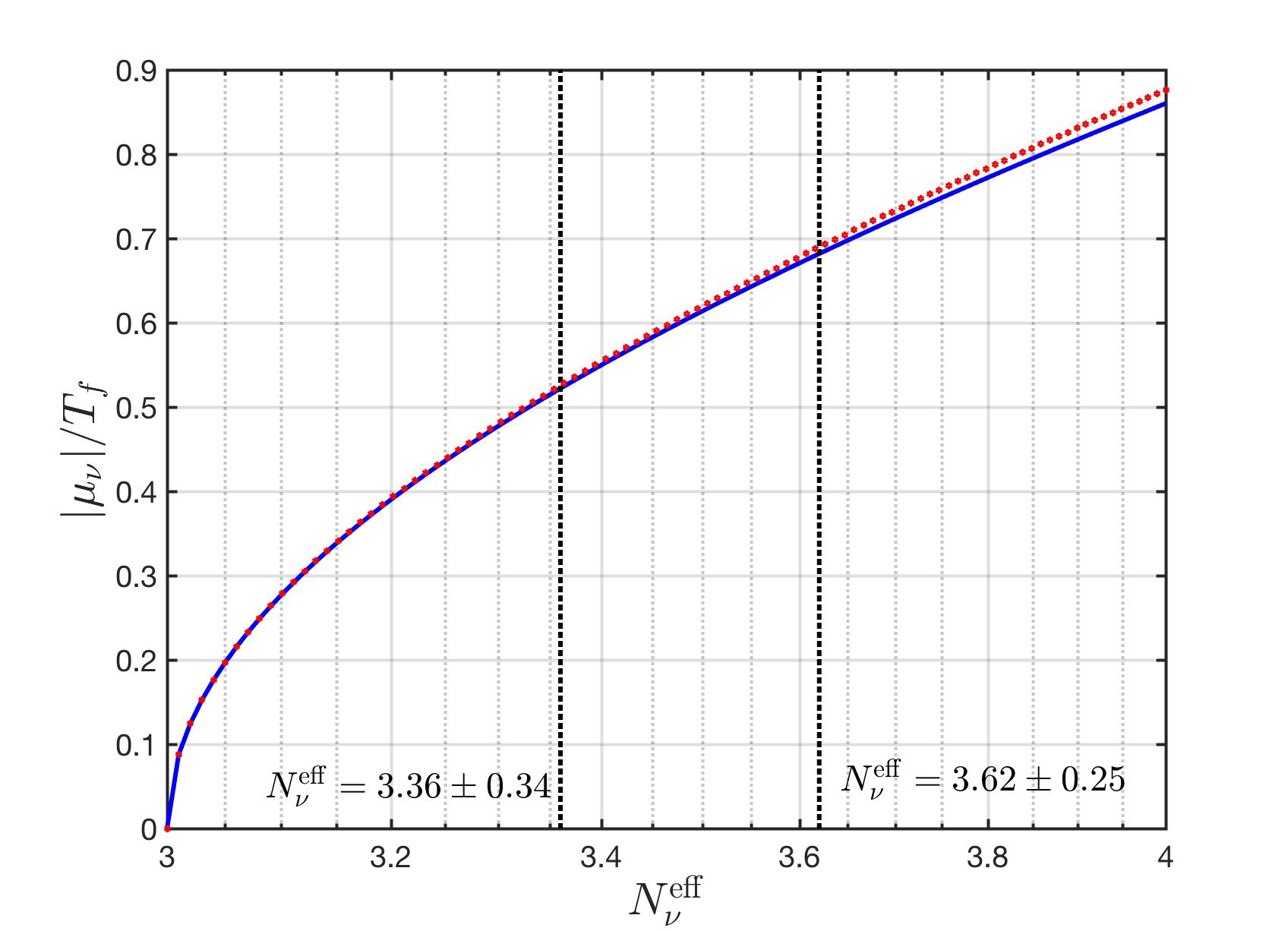}
\caption{The free-streaming neutrino chemical potential $|\mu_\nu|/T_f$ as a function of the effective number of neutrinos $N_\nu^{\mathrm{eff}}$. The solid (blue) line is the exact solution and the (red) dashed line is the approximate solution neglecting the $(\mu_\nu/T_f)^4$ term; the maximum difference in the domain shown is about $2\%$. \radapt{Yang:2024ret}}
\label{ChemicalPotentialNeff}
\end{center}
\end{figure}
In Fig.\;\ref{ChemicalPotentialNeff} we plot the free-streaming\index{free-streaming} neutrino chemical potential $|\mu_\nu|/T_f$ as a function of the effective number of neutrinos $N_\nu^{\mathrm{eff}}$. For comparison, the solid (blue) line is the exact solution of $|\mu_\nu|/T_f$ by solving Eq.~(\ref{NeffPotential}) numerically, and the (red) dashed line is the approximate solution Eq.~(\ref{Solution}) by neglecting the $(\mu_\nu/T_f)^4$ in calculation. In the parameter range of interest, we show that the term $(\mu_\nu/T_f)^4$ only contributes $\approx 2\%$ to the calculation and henceforth we neglect it, and use the approximation Eq.\;(\ref{Solution}). 

The SM value of the effective number of neutrinos, $N_\nu^{\mathrm{eff}}=3$, is obtained under the assumption that the neutrino chemical potentials are not essential, {\it i.e.\/}, $\mu_\nu\ll T_f$. From Fig.\;\ref{ChemicalPotentialNeff}, to interpret the literature values $N_\nu^{\mathrm{eff}}=3.36\pm0.34$ (CMB only) and $N_\nu^{\mathrm{eff}}= 3.62\pm0.25$ (CMB and $H_0$), we require $0.52\leqslant\mu_\nu/T_f\leqslant0.69$. These values suggest a possible neutrino-antineutrino asymmetry at freeze-out, {\it i.e.\/} a difference between the number densities of neutrinos and antineutrinos.

\index{lepton!asymmetry}
\para{Dependence of effective number of neutrinos on lepton asymmetry}
We now obtain the relation between neutrino chemical potential and the lepton-to-baryon\index{lepton!per baryon ratio} ratio. Let us consider the neutrino freeze-out\index{neutrino!freeze-out} temperature $T_f\simeq 2.0$\,MeV; here we treat neutrino freeze-out as occurring instantaneously and prior to $e^+e^-$ annihilation (implying zero neutrino reheating). Comoving lepton (and baryon) number is conserved after the epoch of leptogenesis (baryogenesis, respectively) which precedes the epoch under consideration in this work ($T\lesssim 2$\;MeV). 

The lepton-density asymmetry $\ell $ at neutrino freeze-out can be written as\index{lepton!asymmetry}
\begin{align}
\ell_f \equiv\big(n_e-n_{\overline{e}}\big)_f+\sum_{i=e,\mu, \tau}\big(n_{\nu_i}-n_{\overline{\nu}_i}\big)_f,
\end{align}
where we use the subscript $f$ to indicate that the quantities should be evaluated at the neutrino freeze-out temperature. As a first approximation, here we assume that all neutrinos freeze-out at the same temperature and their chemical potentials are the same; {\it i.e.\/},
\begin{align}
\mu_\nu=\mu_{\nu_e}=\mu_{\nu_\mu}=\mu_{\nu_\tau}.
\end{align}
Furthermore, neutrino oscillation implies that neutrino number is freely exchanged between flavors; {\it i.e.\/}, $\nu_e\rightleftharpoons\nu_\mu\rightleftharpoons\nu_\tau$, and we can assume that all neutrino flavors share the same population. Under these assumptions, the lepton-density asymmetry can be written as
\begin{align}
\label{Lasymmetry} 
\ell_f=\big(n_e-n_{\overline{e}}\big)_f+\big(n_{\nu}-n_{\overline{\nu}}\big)_f,
\end{align}
where the three flavors are accounted for by taking the degeneracy $g_\nu=3$ in the last term. The difference in yield of neutrinos and antineutrinos can be written as
\begin{align}
\label{ExcessNeutrino}
\left(n_\nu-n_{\overline{\nu}}\right)_f=\frac{g_\nu}{6\pi^2}T^3_f\bigg[\pi^2\left(\frac{\mu_\nu}{T_f}\right)+\left(\frac{\mu_\nu}{T_f}\right)^{\!\!3}\bigg].
\end{align}

On the other hand, the baryon-density asymmetry $b$ at neutrino freeze-out is given by
\begin{align}
\label{Basymmetry}
b_f \equiv\big(n_p-n_{\overline{p}}\big)_f+\big(n_n-n_{\overline{n}}\big)_f \approx \big(n_p+n_n\big)_f,
\end{align}
where $n_{\overline{n}}$ and $n_{\overline{p}}$ are negligible in the temperature range we consider here. Taking the ratio $\ell_f/b_f$, using charge neutrality\index{charge neutrality}, and introducing the entropy density\index{entropy!density} at freeze-out we obtain
\begin{align}\label{LfBf}
\left(\frac{\ell_f}{b_f}\right) 
\approx\left(\frac{n_p}{n_B} \right)_f+\left(n_{\nu}-n_{\overline{\nu}}\right)_f \left(\frac{\sigma}{n_B}\right)_f \frac{1}{\sigma_f},\qquad n_B=(n_p+n_n),
\end{align}
where we introduce the notation $n_B$ for the baryon number density. The proton concentration at neutrino freeze-out is given by
\begin{align}
\label{X:proton}
\left(\frac{n_p}{n_B}\right)_f&=\frac{1}{1+(n_n/n_p)_f}=\frac{1}{1+\exp{\big[-\left(Q+\mu_\nu\right)/T_f\big]}},
\end{align}
with $Q=m_n-m_p=1.293\,\mathrm{MeV}$. We neglect the electron chemical potential\index{chemical potential!electron} in the last step because the $e^+e^-$ asymmetry is determined by the proton density, and at energies of order a few MeV, the proton density is small, {\it i.e.\/}, $\mu_e\ll T_f$. 

However, as we will see, for our study of $N_\nu^{\mathrm{eff}}$ we will be interested in the case of a large lepton-to-baryon ratio\index{lepton!per baryon ratio}. From Eq.\;(\ref{X:proton}) it is apparent that this can only be achieved through the second term in Eq.\;(\ref{LfBf}), with the first term then being negligible, as it is smaller than $1$. So we further approximate
\begin{align}\label{LB:ratio}
\left(\frac{\ell_f}{b_f}\right) 
\approx\left(n_{\nu}-n_{\overline{\nu}}\right)_f \left(\frac{\sigma}{n_B}\right)_f \frac{1}{\sigma_f}.
\end{align}
We retained the full expression Eq.\;(\ref{X:proton}) in our above discussion to show that the presence of a chemical potential $\mu_\nu\simeq 0.2\,Q$ could lead to small, perhaps noticeable, effects on pre-BBN proton and neutron abundance. We defer this unrelated discussion to a separate future work. Note that for large $|\mu_\nu|$, Eq.\;(\ref{LB:ratio}) implies that the signs of $\mu_\nu$ and $\ell_f$ are the same. However, for very small $\mu_\nu$ the sign of $\ell_f$ is determined by the interplay between (anti)electrons and (anti)neutrinos; {\it i.e.\/}, there is competition between the two terms in Eq.\;(\ref{Lasymmetry} ).

In general, the total entropy density at freeze-out can be written
\begin{align}
\label{eq:EntropyDensity}
\sigma_f=\frac{2\pi^2}{45}g^s_\ast(T_f)\,T_f^3,
\end{align}
where the $g^s_\ast$ counts the degree of freedom for relativistic particles~\cite{Kolb:1990vq}. At $T_f\simeq 2\mathrm{MeV}$, the relativistic species in the early Universe are photons, electron/positrons, and $3$ neutrino species. We have
\begin{align}
g^s_{\ast}&= g_\gamma+\frac{7}{8}\,g_{e^\pm}+\frac{7}{8}\,g_{\nu\bar{\nu}}\left(\frac{T_\nu}{T_\gamma}\right)^{\!\!3}\bigg[1+\frac{15}{7\pi^2}\left(\frac{\mu_\nu}{T_f}\right)^{\!\!2}\bigg]=10.75+\frac{45}{4\pi^2}\left(\frac{\mu_\nu}{T_f}\right)^{\!\!2}\;,
\end{align}
where the degrees of freedom are given by $g_\gamma=2$, $g_{e^\pm}=4$, and $g_{\nu\bar{\nu}}=6$, and we have $T_\nu=T_\gamma=T_f$ at neutrino freeze-out.

\begin{figure}
\begin{center}
\includegraphics[width=0.8\linewidth]{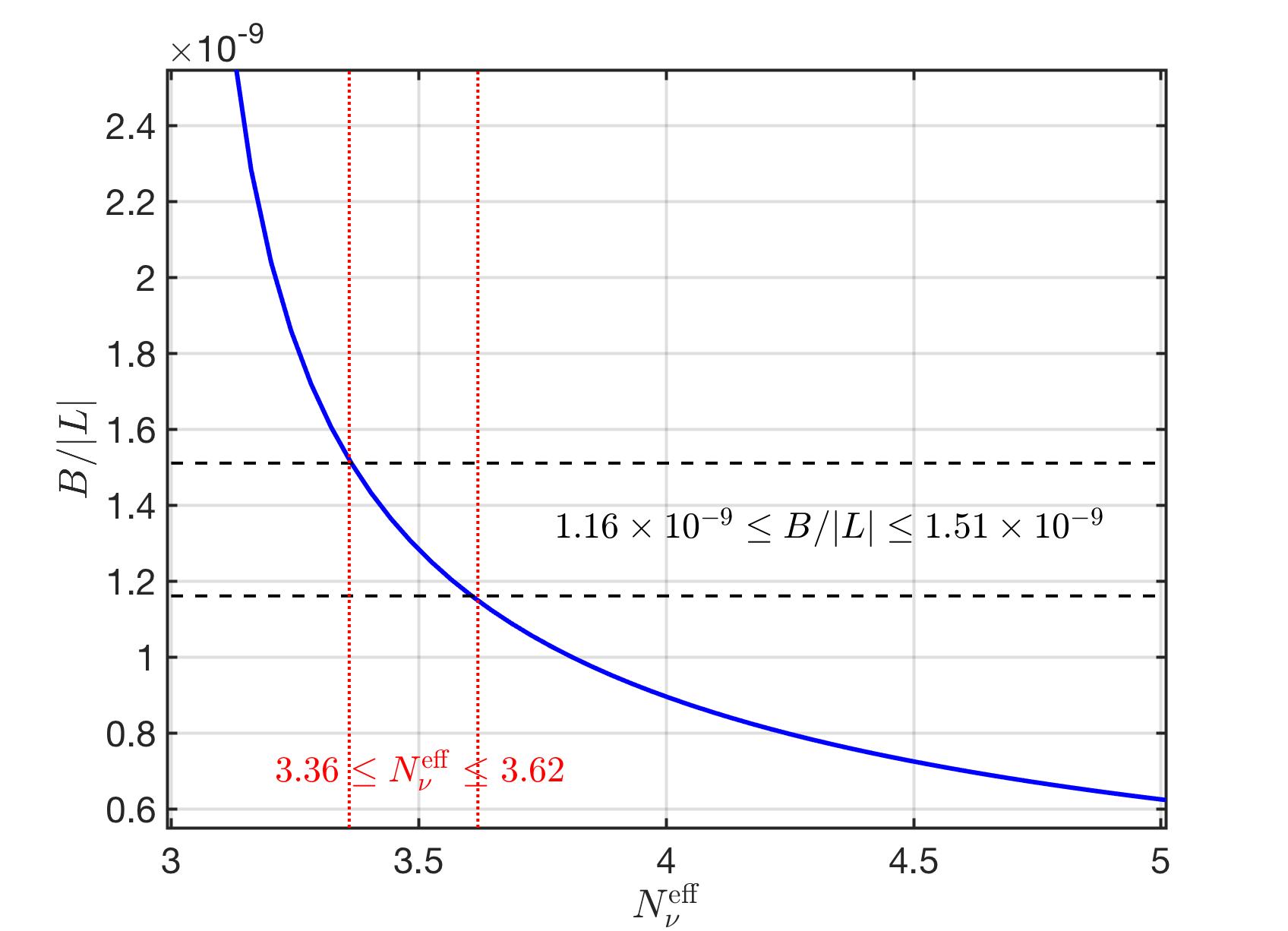}
\caption{The ratio $B/|L|$ between the net baryon number and the net lepton number as a function of $N^{\mathrm{eff}}_\nu$: The solid blue line shows $B/|L|$. The vertical (red) dotted lines represent the values $3.36\leqslant N_\nu^{\mathrm{eff}}\leqslant3.62$, which correspond to $1.16 \times 10^{-9}\leqslant B/|L|\leqslant 1.51 \times 10^{-9}$ (horizontal dashed lines). \radapt{Yang:2024ret}}
\label{fig:BLRatio}
\end{center}
\end{figure}

Finally, since the entropy-per-baryon from neutrino freeze-out up to the present epoch is constant, we can obtain this value by considering the Universe's entropy content today~\cite{Fromerth:2012fe}. For $T\ll1\,\mathrm{MeV}$, the entropy content today is carried by photons and neutrinos, yielding
\begin{align}
\label{NbS}
\left(\frac{\sigma}{n_B}\right)_{t_0}&=\frac{\sum_i\,\sigma_i}{n_B}=\frac{n_\gamma}{n_B}\,\bigg(\frac{\sigma_\gamma}{n_\gamma}+\frac{\sigma_\nu}{n_\gamma}+\frac{\sigma_{\bar{\nu}}}{n_\gamma}\bigg)\;\\
&=\left(\frac{1}{\eta_\gamma}\right)_{\!\!t_0}\!\!\left[\frac{\sigma_\gamma}{n_\gamma}+\frac{4}{3T_\nu}\frac{\rho_\nu^{\mathrm{tot}}}{n_\gamma}-\frac{\mu_\nu}{T_f}\left(\frac{n_\nu-n_{\bar{\nu}}}{n_\gamma}\right)\right]_{t_0}\;,
\end{align}
where $t_0$ denotes the present day values, and we have $\eta_\gamma\equiv n_B/n_\gamma= 0.614\pm0.002\times10^{-9}$ 
(CMB)~\cite{ParticleDataGroup:2022pth} \index{CMB!baryon to photon ratio}. The entropy per particle for a massless boson at zero chemical potential\index{chemical potential} is $(s/n)_{\mathrm{boson}}\approx 3.602$.

Substituting Eq.\;(\ref{ExcessNeutrino}) and Eq.\;(\ref{eq:EntropyDensity}) into Eq.\;(\ref{LB:ratio}) yields the lepton-to-baryon ratio\index{lepton!per baryon ratio}
\begin{align}\label{LBratioFinal}
&\frac{L}{B}=\frac{45}{4\pi^4}\frac{\pi^2(\mu_\nu/T_f)+(\mu_\nu/T_f)^3}{10.75+{45}(\mu_\nu/T_f)^2/{4\pi^2}}\left(\frac{\sigma}{n_B}\right)_{\!\!t_0}\;,
\end{align}
in terms of $\mu_\nu/T_f$ which is given by Eq.(\ref{Solution}) and the present day entropy-per-baryon ratio. In Fig.\;\ref{fig:BLRatio} we show the ratio between the net baryon number and the net lepton number as a function of the effective number of neutrino species $N^{\mathrm{eff}}_\nu$ with the parameter $ \eta_\gamma|_{t_0} =0.614\times 10^{-9}$(CMB). We find that the values $N_\nu^{\mathrm{eff}}=3.36\pm0.34$ and $N_\nu^{\mathrm{eff}}= 3.62\pm0.25$ require the ratio between baryon number and lepton number to be $1.16 \times 10^{-9} \leqslant\, B/|L| \leqslant 1.51\times 10^{-9}$. These values are close to the baryon-to-photon ratio $0.57 \times 10^{-9} \leqslant \eta_\gamma \leqslant 0.67\times 10^{-9}$.

A large cosmic neutrino lepton asymmetry\index{lepton!asymmetry} can affect the neutron lifespan in cosmic plasma which is one of the important parameter controlling the BBN sourced light element abundances\index{BBN}. 
In general the neutron lifespan depends on the temperature of the cosmic medium as we will discuss in \rsec{sec:neutron}. This is so since electrons and neutrinos in the background plasma can reduce the neutron decay rate by Fermi blocking. A large neutrino asymmetry chemical potential thus could have an observable impact on light element formation in BBN.

\para{Extra background neutrinos from microscopic primordial processes}
We are interested in improving the understanding of the abundance of neutrinos produced by secondary processes after neutrino  freeze-out. We recall that neutrinos decouple at $T_f\simeq 2$\,MeV and become free streaming after freeze-out. The presence of electron-positron plasma until $T=20$\,keV, see \rsec{section:electron} allows the following weak reaction to occur 
\begin{align}
\gamma+\gamma\longrightarrow e^-+e^+\longrightarrow \nu+\bar{\nu}\,,
\end{align}
adding neutrinos to the neutrino background inventory. This process continues until last positrons disappear from Universe inventory at the temperature $T\simeq 20$\,keV. It could modify the free streaming distribution and the effective number of neutrinos\index{neutrino!effective number}. We examine this electron-positron plasma source of extra neutrinos and develop methods for a
future more detailed study.

 Given the reaction rates per volume $R_{\gamma\gamma\to e\overline{e}}$ for the reaction $\gamma\gamma\to e\overline{e}$, and $R_{e\overline{e}\to\nu\overline{\nu}}$ for the reaction $e\overline{e}\to\nu\overline{\nu}$, the combined thermal reaction rate per volume for $\gamma\gamma\to e^-e^+\to\nu\bar{\nu}$ is
\begin{align}
R_{\gamma\to e\to\nu}=R_{\gamma\gamma\to e\overline{e}}\left(\frac{R_{e\overline{e}\to\nu\overline{\nu}}}{R_{\gamma\gamma\to e\overline{e}}+R_{e\overline{e}\to\nu\overline{\nu}}}\right)\approx R_{e\overline{e}\to\nu\overline{\nu}}\,.
\end{align}
In \rf{ExtraNeutrinoRate} we show the thermal reaction rate per volume for relevant reactions as a function of temperature $2\MeV>T>0.05\MeV$. We see that the bottleneck for the process $\gamma\gamma\to e^-e^+\to\nu\bar{\nu}$ is the $e\overline{e}\to\nu\overline{\nu}$, hence $R_{\gamma\to e\to\nu}=R_{e\overline{e}\to\nu\overline{\nu}}$ in the temperature regime in which we are interested.
\begin{figure}
\begin{center}
\includegraphics[width=0.8\linewidth]{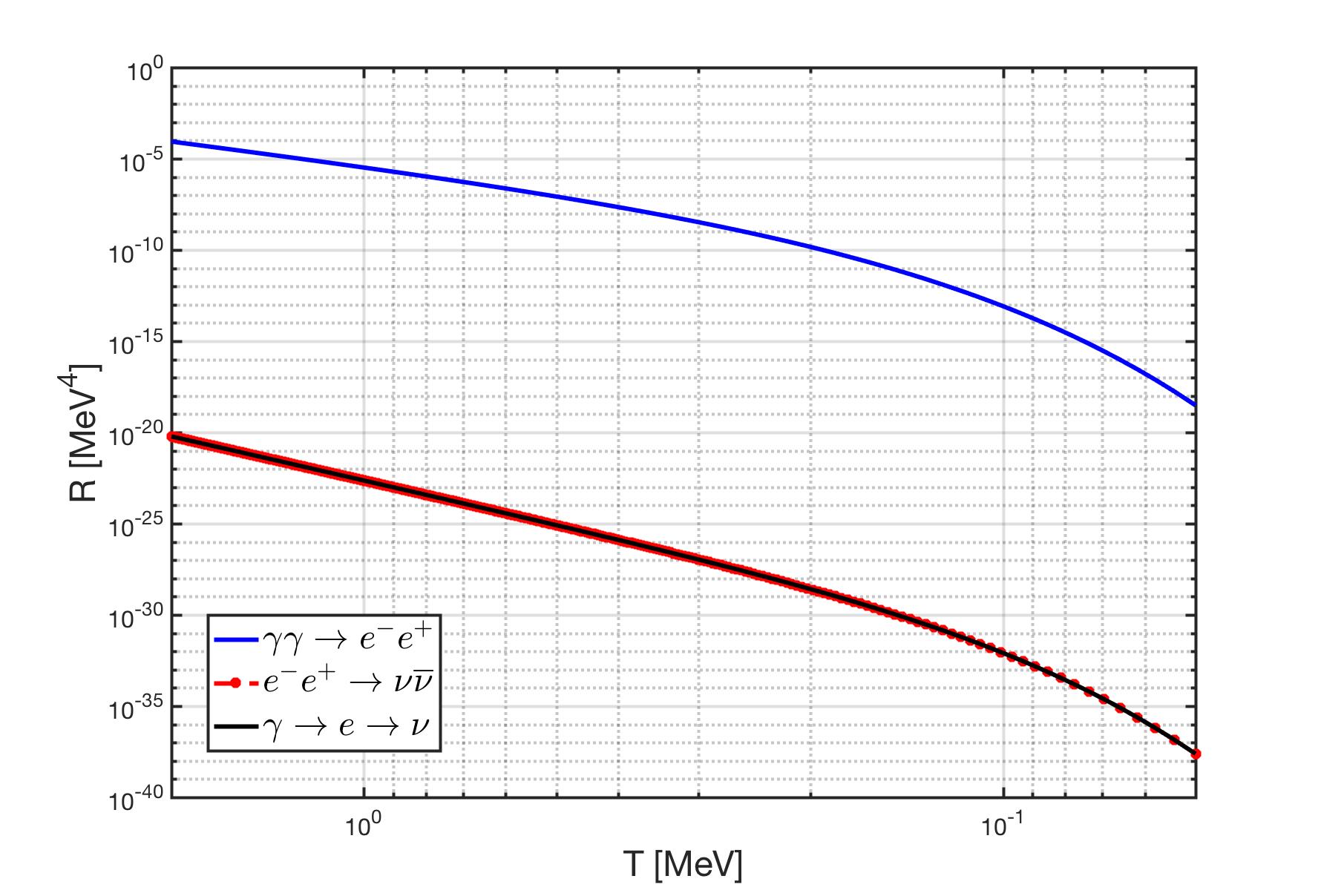}
\caption{The thermal reaction rates per volume for $\gamma\gamma\to e^-e^+$ top line (blue) and $e^-e^+\to\nu\bar{\nu}$ bottom line (red) as a function of temperature $2\MeV>T>0.05\MeV$. . \radapt{Yang:2024ret}.}
\label{ExtraNeutrinoRate}
\end{center}
\end{figure}

Given the thermal reaction rate, the dynamic equation describing the relic neutrino abundance after freeze-out is
\begin{align}\label{ExtraNeutrioEq}
\frac{dn_\nu}{dt}+3Hn_\nu=R_{e\overline{e}\to\nu\overline{\nu}}(T_{\gamma,e^\pm})-R_{\nu\overline{\nu}\to e\overline{e}}(T_\nu),
\end{align}
where $n_\nu$ is the number density of neutrinos and $H$ is the Hubble parameter. 

The parameter $T_{\gamma,e^\pm}$ is the equilibrium temperature between photons and $e^\pm$ which undergo annihilation, while   $T_\nu$ describes the back-reaction which has as temperature the free-streaming\index{free-streaming} neutrinos: 
\begin{align}
T_\nu=\frac{a(t_f)}{a(t)}T_f,
\end{align}
where $T_f$ is the neutrino freeze-out\index{neutrino!freeze-out} temperature. After neutrinos have decoupled from the cosmic plasma, we have $T_\nu < T_{\gamma,e^\pm}$. This is because after freeze-out, the relic neutrino entropy is conserved independently and the entropy from $e^+e^-$ annihilation flows solely into photons and reheats  photon's temperature. If  the considered process is significant than after neutrino freeze-out, extra entropy from electron-positron plasma can still flow into the free-streaming neutrino sector via the reaction $\gamma\gamma\to e^-e^+\to\nu\bar{\nu}$. To describe this novel situation, kinetic theory for entropy production needs to be adapted, a topic we will address in the future. Here we neglect this extra entropy and explore the magnitude of the expected effect.  

In  \rf{DimensionlessRatio} we show the temperature ratio $T_\nu/T_{\gamma,e^\pm}$, the rate ratio $R_{\nu\overline{\nu}\rightarrow e\overline{e}}/ R_{e\overline{e}\rightarrow\nu\overline{\nu}}$ and $(R_{e\overline{e}\rightarrow\nu\overline{\nu}}-R_{\nu\overline{\nu}\rightarrow e\overline{e}})/ R_{e\overline{e}\rightarrow\nu\overline{\nu}}$ as a function of temperature. We see that after neutrino freeze-out, the back reaction $\nu\overline{\nu}\rightarrow e\overline{e}$ becomes smaller compared to the reaction $e\overline{e}\rightarrow\nu\overline{\nu}$ as the neutrino temperature cools down. This is because as $T_\nu$ cools down, the density of relic neutrinos becomes so low and their energy becomes too small to interact. However, the relatively hot and massive electron-positron plasma can still annihilate into neutrino pairs.

\begin{figure} 
\begin{center}
\includegraphics[width=0.8\linewidth]{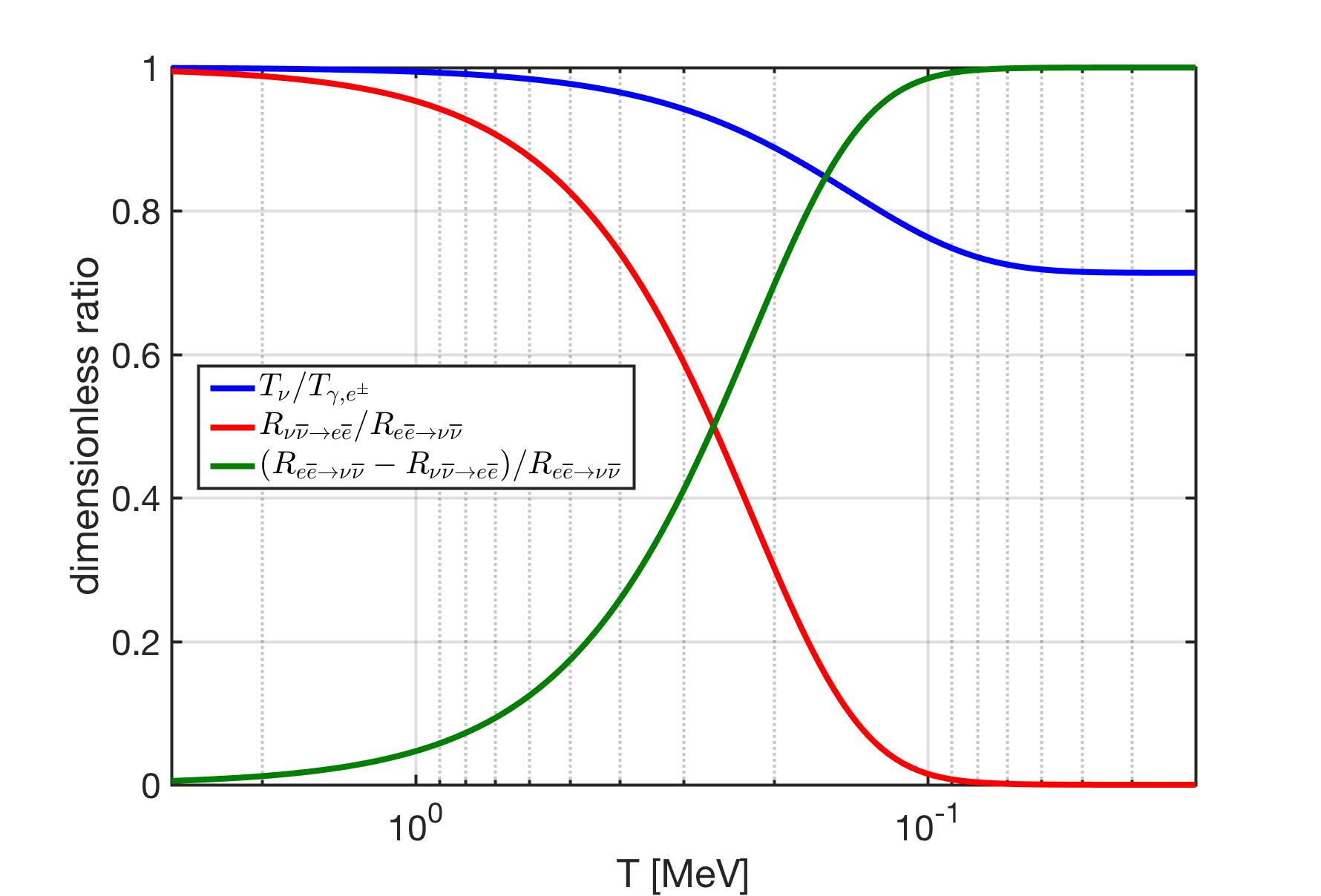}
\caption{The temperature ratio $T_\nu/T_{\gamma,e^\pm}$ (blue line), the rate ratio $R_{\nu\overline{\nu}\rightarrow e\overline{e}}/ R_{e\overline{e}\rightarrow\nu\overline{\nu}}$ (red line) and $(R_{e\overline{e}\rightarrow\nu\overline{\nu}}-R_{\nu\overline{\nu}\rightarrow e\overline{e}})/ R_{e\overline{e}\rightarrow\nu\overline{\nu}}$ (green line) as a function of temperature. It shows that the reaction $\nu\overline{\nu}\rightarrow e\overline{e}$ is small compare to the reaction $e\overline{e}\rightarrow\nu\overline{\nu}$ as temperature cooling down. \radapt{Yang:2024ret}.}
\label{DimensionlessRatio}
\end{center}
\end{figure}

The general solution of the dynamic \req{ExtraNeutrioEq} can be written as
\begin{align}
n_\nu(T)=n_\mathrm{relic}(T)+n_\mathrm{extra}(T),\qquad T=T_{\gamma,e^\pm},
\end{align}
where $n_\mathrm{relic}$ represents the relic neutrino number density and $n_\mathrm{extra}$ is the extra number density from the $e^\pm$ annihilation. The relic neutrino density is given by
\begin{align} &n_\mathrm{relic}=n_\nu^0\exp\left(-3\int_{t_i}^t{dt^\prime}H(t^\prime)\right)=n_\nu^0\exp\left(3\int_{T_i}^T\frac{dT^\prime}{T^\prime}(1+\mathcal{F})\right),\\
&n^0_\nu=g_\nu\frac{3\zeta(3)}{4\pi^2}T^3_i,\qquad \mathcal{F}=\frac{T}{3g^\ast_s}\frac{dg^\ast_s}{dT},
\end{align}
where $T_i$ is the initial temperature and $g^\ast_s$ is the entropy degrees of freedom\index{entropy!degrees of freedom}. The extra neutrino density can be written as
\begin{align}
n_\mathrm{extra}=-&\exp\left(3\int_{T_i}^T\frac{dT^\prime}{T^\prime}(1+\mathcal{F})\right)\notag\\
\times&\int_{T_i}^T\frac{dT^\prime}{T^\prime}\frac{R_{e\overline{e}}(T^\prime)-R_{\nu\overline{\nu}}(T^\prime_\nu)}{H(T^\prime)}\left(1+\mathcal{F}\right)\exp\left(-3\int_{T_i}^{T^{\prime}}\frac{dT^{\prime\prime}}{T^{\prime\prime}}(1+\mathcal{F})\right).
\end{align}

\begin{figure}
\begin{center}
\includegraphics[width=0.8\linewidth]{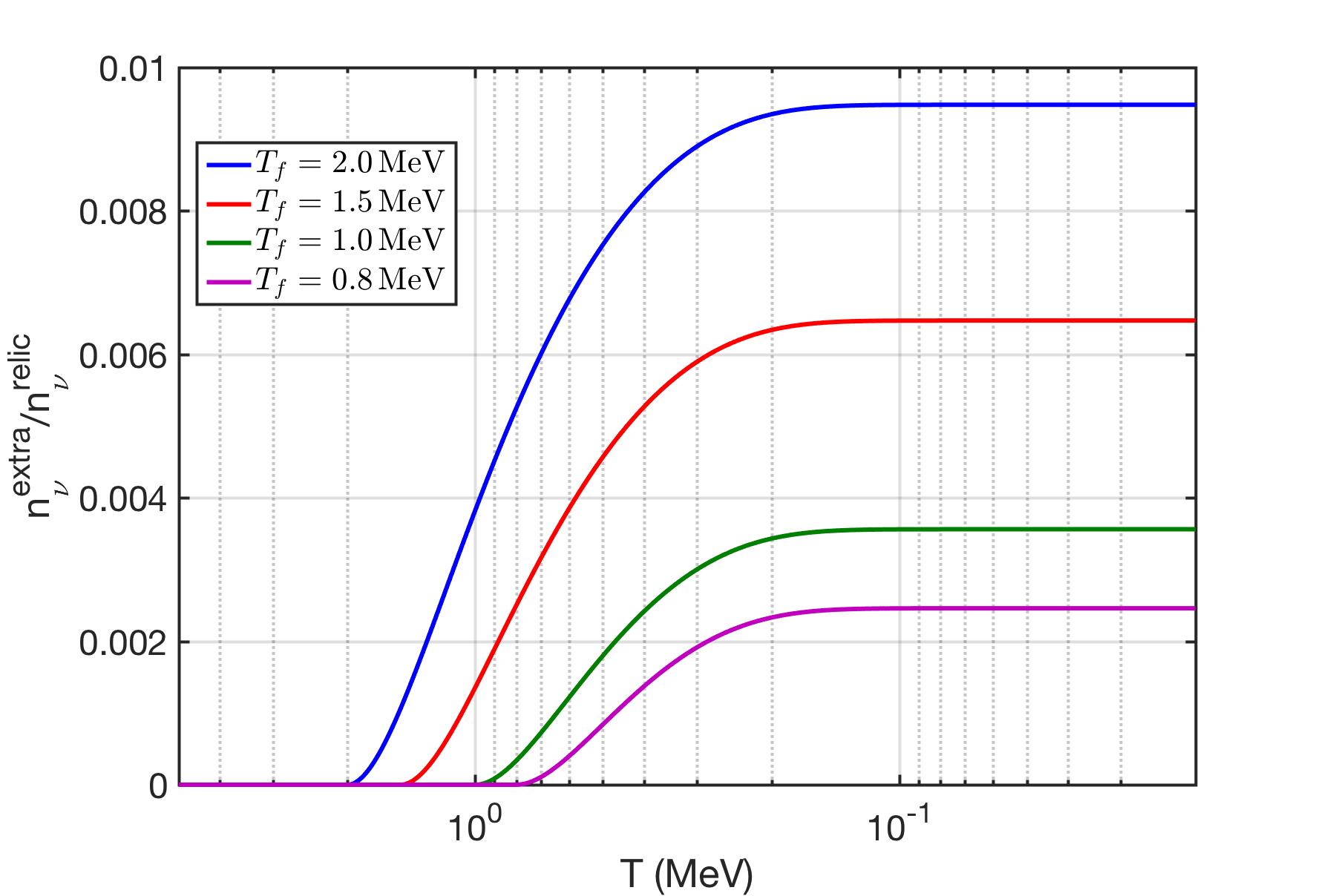}
\caption{the ratio between $n_{extra}/n_{relic}$ as a function of temperature with different neutrino freeze-out temperature $T_f$. \radapt{Yang:2024ret}.}
\label{ExtraNeutrinoRatio}
\end{center}
\end{figure}

In \rf{ExtraNeutrinoRatio} we show the ratio between $n_\mathrm{extra}/n_\mathrm{relic}$ as a function of temperature varying the neutrino freeze-out temperature $T_f$, with top line (blue) for $T_f=2\MeV$ and other lines for $T_f=1.5\MeV$ (red), $T_f=1 \MeV$ (green), and  $T_f=0.8 \MeV$ (violet) . The number of extra neutrinos depends strongly on the parameter $T_f$ because the freeze-out temperature determines the timing of the entropy transfer between $e^\pm$ and photon, which subsequently affects the evolution of temperature ratio between neutrinos and photons in the early Universe. The temperature ratio affects the rate ratio between $\nu\overline{\nu}\to e\overline{e}$ and $ e\overline{e}\to\nu\overline{\nu}$, because once the neutrino is too cold and the back reaction $\nu\overline{\nu}\to e\overline{e}$ can not maintain the balance, the $e^\pm$ annihilation starts to feed the extra neutrinos to the relic neutrino background. These results indicate  that the higher realistic freeze-out temperature $T_f$ leads to an added  number of extra neutrinos, in our example up to 1\%. 

In addition to the annihilation of electron-positron pairs process, there are other sources that can contribute to the presence of extra neutrinos in the early Universe. These additional sources include particle physics phenomena and plasma effects: neutrinos from charged leptons $\mu^\pm,\tau^\pm$ decay, neutrinos from the $\pi^\pm$ decay, and neutrino radiation from massive in plasma photon decay via virtual $e^+e^-$ pair, $\gamma_M\to e^+e^-\to nu\bar\nu$. 

All of these potential sources of extra neutrinos can impact the distribution of freely streaming neutrinos and the effective number of neutrinos. Understanding these effects is crucial to comprehending how the neutrino component influences the expansion of the Universe, as well as the potential implications for large-scale structure formation and the spectrum of relic neutrinos.

\subsection{Neutrinos today}\label{ch:nu:today}
We conclude our comprehensive exploration of neutrino freeze-out considering the distribution of free-streaming\index{relic neutrino!background} relic neutrinos in the present day, as seen from the reference frame of the moving Earth~\cite{Birrell:2014qna}. Such a study is prequel to the future experimental detection of the cosmic background neutrinos,  a challenge of great interest \cite{Stodolsky:1974aq,Cabibbo:1982bb,Shvartsman,Langacker:1982ih,Smith:1983jj,Tupper:1987sf,Ferreras:1995wf,Hagmann:1999kf,Duda:2001hd,Gelmini:2004hg,Ringwald:2009bg,Liao:2012wb,Hedman:2013hha}. With the  recently proposed PTOLEMY experiment, which aims to detect relic  electron-neutrino capture by tritium \cite{Betts:2013uya,PTOLEMY:2019hkd}, the characterization of the relic neutrino background is increasingly relevant.  Using our  characterization of the neutrino distribution after freeze-out and the subsequent free-streaming dynamics from  \rsec{sec:model:ind} and \cite{Birrell:2012gg}, we lay groundwork for a characterization of the present day relic neutrino spectrum, which we explore  from the  perspective of an observer moving relative to the neutrino background, including the dependence on neutrino mass\index{neutrino!mass} and effective number of neutrinos, $N_\nu^{\mathrm{eff}}$. Beyond consideration of the observable neutrino distributions, we evaluate the ${\cal O}(G_F^2)$ mechanical drag force acting on the moving observer. This section is adapted from the work in \cite{Birrell:2014qna}.

\para{Neutrino distribution in a moving frame}
The cosmic neutrino background (C$\nu$B)\index{C$\nu$B} and the cosmic microwave background (CMB)\index{CMB} were in equilibrium until decoupling (freeze-out) at $T_k\simeq {\cal O}{\rm (MeV)}$, hence one surmises that an observer would have the same relative velocity relative to the relic neutrino background  as with CMB. As a particular example in considering the spectrum, we present in more detail the case of an observer comoving with  Earth velocity $v_\oplus=300$\,km/s relative to the CMB,  modulated by orbital velocity ($\pm29.8$\,km/s).  We will write velocities in units of $c$, though our specific results will be presented in km/s.

In the cosmological setting, for $T<T_k$ the neutrino spectrum evolves according to the well known Fermi-Dirac-Einstein-Vlasov\index{Fermi!Einstein-Vlasov distribution} (FDEV) free-streaming\index{free-streaming} distribution~\cite{Langacker:1982ih,Choquet-Bruhat:2009xil,Wong:2011ip,Birrell:2012gg}.  By casting it in a relativistically invariant form we can then make a transformation to the rest frame of an observer moving with relative velocity $v_{\text{rel}}$ and obtain
\begin{align}\label{eq:neutrinoDistB}
f(p^\mu)=&\frac{1}{\Upsilon^{-1} e^{\sqrt{(p^\mu U_\mu)^2-m_\nu^2}/T_\nu}+1}\,.
\end{align}
The 4-vector characterizing the rest frame of the neutrino FDEV distribution is
\begin{equation}
U^\mu=(\gamma,0,0,v_{\text{rel}}\gamma)\,,\hspace{2mm} \gamma={1}/{\sqrt{1-v_{\text{rel}}^2}}\,,
\end{equation} 
where we have chosen coordinates so that the relative motion is in the $z$-direction. 

The neutrino effective temperature $T_\nu(t)= T_k\,(a(t_k)/a(t))$ is the scale-shifted freeze-out temperature $T_k$. Here $a(t)$ is the cosmological scale factor where $\dot a(t)/a(t)\equiv H$ is the observable Hubble parameter\index{Hubble!parameter}. $\Upsilon$ is the fugacity\index{fugacity} factor, here describing the underpopulation of neutrino phase space that was frozen into the neutrino FDEV distribution in the process of decoupling from the $e^\pm,\gamma$-QED background  plasma.

There are several available bounds on neutrino masses\index{neutrino!mass}. Neutrino energy and pressure components are important before photon freeze-out and thus $m_\nu$ impacts Universe dynamics. The analysis of CMB data alone leads to $\sum_i m_\nu^i<0.66$eV ($i=e,\mu,\tau$) and including Baryon Acoustic Oscillation (BAO) gives $\sum m_\nu<0.23$eV~\cite{Planck:2013pxb}.  {\small PLANCK CMB} with lensing observations~\cite{Battye:2013xqa} lead to  $\sum m_{\nu}=0.32\pm0.081$ eV. Upper bounds have been placed on the electron neutrino mass in direct laboratory measurements  $m_{\bar\nu_e}<2.05$eV~\cite{Troitsk:2011cvm}.   In the subsequent analysis we will focus on the neutrino mass range $0.05$eV to $2$eV in order to show that direct measurement sensitivity allows the exploration of a wide mass range. 

 The relations in \req{modindeq1}\,-\,\req{modindeq3}, see also~\cite{Birrell:2012gg}, determine $T_\nu/T_\gamma$ and  $\Upsilon$ in terms of the measured  value of  $N_\nu^{\mathrm{eff}}$ under the assumption of a strictly SM-particle inventory.  In the following we treat $N_\nu^{\mathrm{eff}}$  as a variable model parameter and use the above mentioned relations to characterize our results in terms of $N_\nu^{\mathrm{eff}}$.

\para{Velocity, energy, and wavelength distributions}
Using \req{eq:neutrinoDistB}, the normalized FDEV velocity distribution for an observer in relative motion has the form
\begin{align} \label{fvdistrib}
&f_v=\frac{g_\nu}{n_\nu 4\pi^2}\!\!\!\int_0^\pi \!\!\!\!\frac{ p^2dp/dv\sin(\phi) d\phi}{\Upsilon^{-1}e^{\sqrt{( E-v_{\text{rel}} p \cos(\phi))^2\gamma^2-m_\nu^2}/T_\nu}+1}\,,\notag\\
&p(v)=\frac{m_\nu v}{\sqrt{1-v^2}}\,,\qquad \frac{dp}{dv}=\frac{m_\nu}{(1-v^2)^{3/2}}\,.
\end{align}
The normalization $n_\nu$ depends on $N_\nu^{\mathrm{eff}}$ but not on $m_\nu$ since decoupling occurred at $T_k\gg m_\nu$. For each neutrino flavor (all flavors are equilibrated by oscillations) we have, per neutrino or antineutrino and at nonrelativistic relative velocity,
\begin{equation}\label{nnu}
n_\nu=[-0.3517(\delta N_\nu^{\mathrm{eff}})^2+6.717\delta N_\nu^{\mathrm{eff}}+56.06]\,{\rm cm}^{-3}
\end{equation}
($\delta N_\nu^{\mathrm{eff}}\equiv N_\nu^{\mathrm{eff}}-3$), compare to Eq.(55) in Ref.~\cite{Birrell:2012gg}.

We show $f_v$ in \rf{fig:RelvDist300} left-hand panel for several values of the neutrino mass\index{neutrino!mass}, $v_{\text{rel}}=300$ km/s, and $N_\nu^{\mathrm{eff}}=3.046$ (solid lines) and $N_\nu^{\mathrm{eff}}=3.62$ (dashed lines). As expected, the lighter the neutrino, the more $f_v$  is weighted towards higher velocities with the velocity becoming visibly peaked about $v_{\text{rel}}$ for $m_\nu=2$ eV. A similar procedure produces the normalized FDEV energy distribution $f_E$.  In \req{fvdistrib} we replace $dp/dv\to dp/dE$ where it is understood that 
\begin{equation}
p(E)=\sqrt{E^2-m_\nu^2},\qquad \frac{dp}{dE}=\frac{E}{p}.
\end{equation}
We show $f_E$ in \rf{fig:RelvDist300} right-hand frame for several values of the neutrino mass, $v_{\text{rel}}=300$ km/s, and $N_\nu^{\mathrm{eff}}=3.046$ (solid lines) and $N_\nu^{\mathrm{eff}}=3.62$ (dashed lines). The width of the FDEV energy distribution is on the micro-eV scale and the kinetic energy $T=E-m_\nu$ is peaked about $T=\frac{1}{2}m_\nu v_{\text{rel}}^2$, implying that the relative velocity between the Earth and the CMB\index{CMB} is the dominant factor for $m_\nu>0.1$ eV.

\begin{figure}
\centerline{\includegraphics[width=0.5\linewidth]{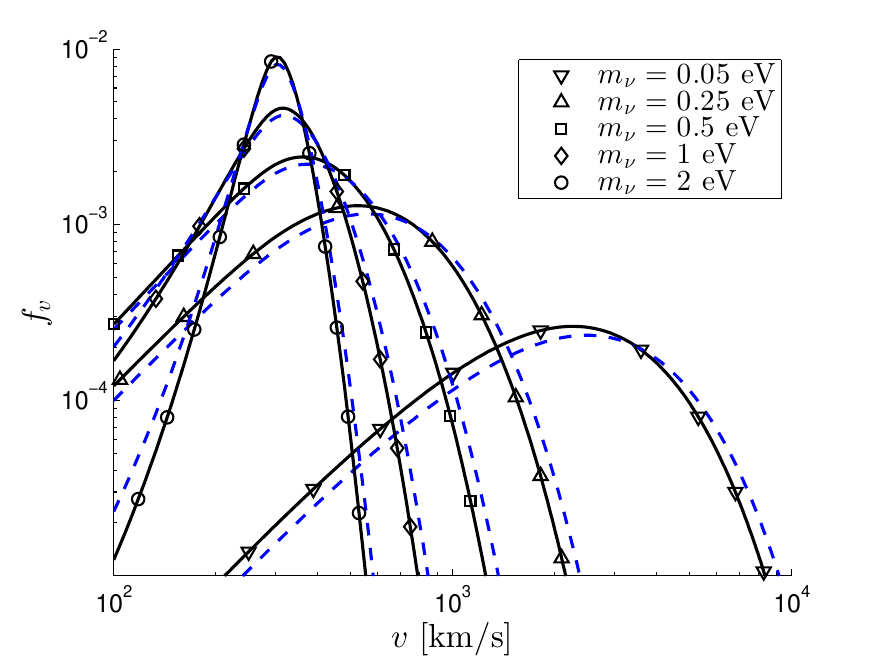}\hspace*{-0.5cm}
\includegraphics[width=0.5\linewidth]{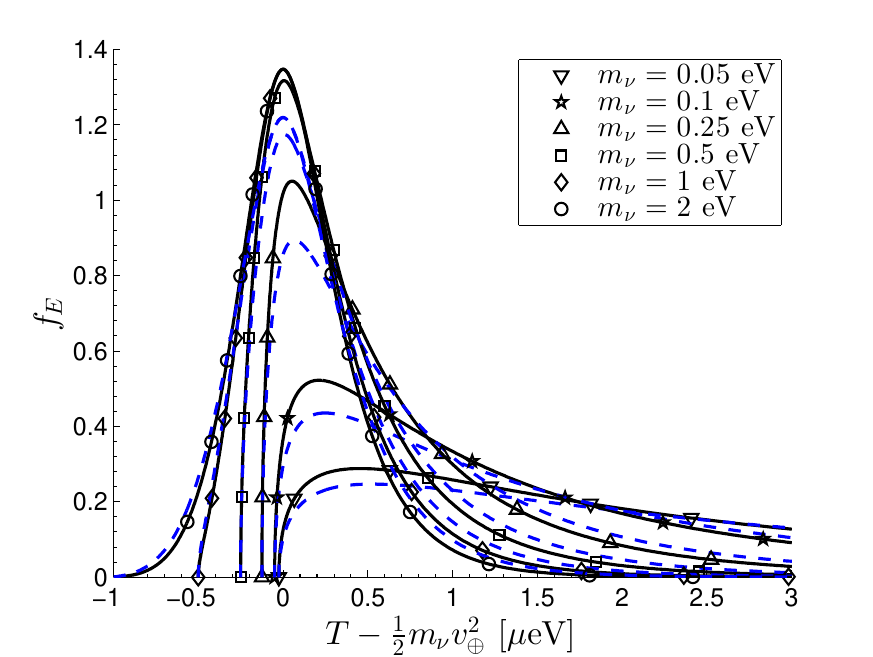}}
\caption{Normalized neutrino FDEV  for velocity on left and energy on right. We show the distribution for $N_\nu^{\mathrm{eff}}=3.046$ (solid lines) and $N_\nu^{\mathrm{eff}}=3.62$ (dashed lines). \cccite{Birrell:2014qna}}
\label{fig:RelvDist300}
 \end{figure}

\begin{figure}
\centerline{\includegraphics[width=0.7\linewidth]{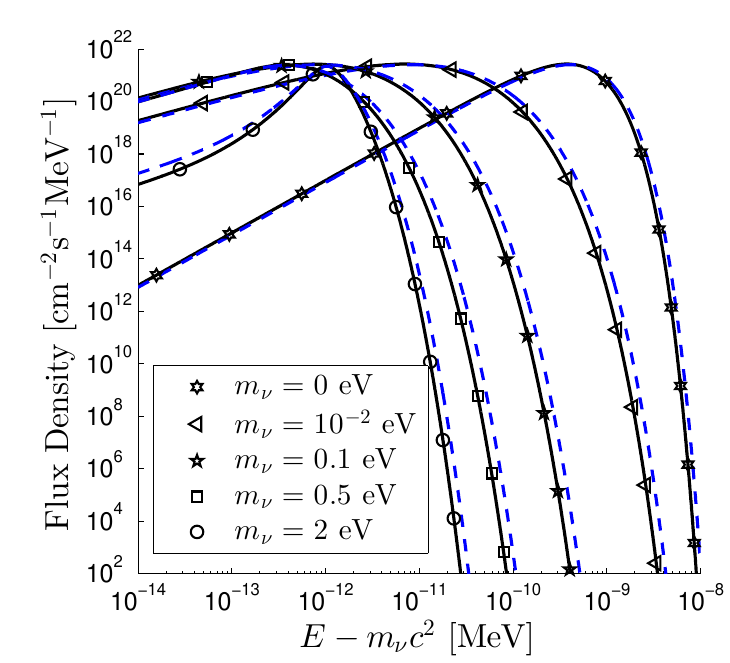}}
\caption{Neutrino flux density in the Earth frame. We show the result for $N_\nu^{\mathrm{eff}}=3.046$ (solid lines) and $N_\nu^{\mathrm{eff}}=3.62$ (dashed lines) for an observer moving with $v_\oplus=300$\,km/s. \cccite{Birrell:2014qna}}
\label{fig:fluxDist}
 \end{figure}

By multiplying $f_E$ by the neutrino velocity and number density for a single neutrino flavor (without anti-neutrinos) we obtain the particle flux density,
 \begin{equation}
 \frac{dJ}{dE}\equiv\frac{dn}{dAdtdE}\,,
\end{equation} 
shown in Figure \ref{fig:fluxDist}. We show the result for $N_\nu^{\mathrm{eff}}=3.046$ (solid lines) and $N_\nu^{\mathrm{eff}}=3.62$ (dashed lines). The flux is normalized in these cases to a local density $56.36$~cm${}^{-3}$ and $60.10$~cm${}^{-3}$, respectively. 

The precise neutrino flux in the Earth frame is significant for efforts to detect relic neutrinos, such as the PTOLEMY experiment~\cite{Betts:2013uya}. The energy dependence of the flux shows a large sensitivity to the mass. However, the maximal fluxes do not vary significantly with $m$. In fact the maximum values are independent of $m$ when $v_{\text{rel}}=0$, as follows from the fact that $v=p/E=dE/dp$.  In the Earth frame, where $0<v_\oplus\ll c$, this translates into only a small variation in the maximal flux.

\begin{figure} 
\centerline{\includegraphics[width=0.52\linewidth]{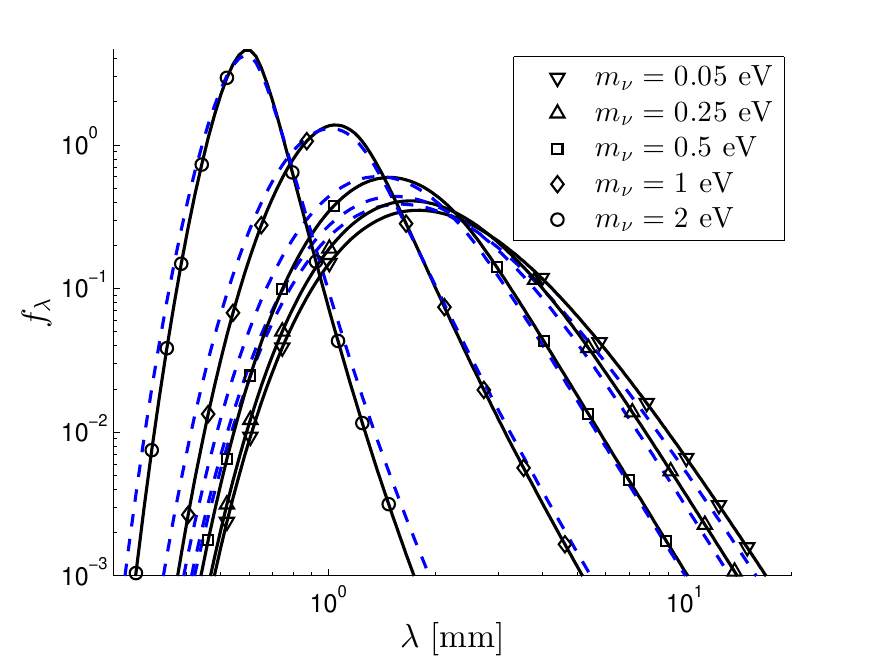} \hspace*{-0.8cm}
\includegraphics[width=0.52\linewidth]{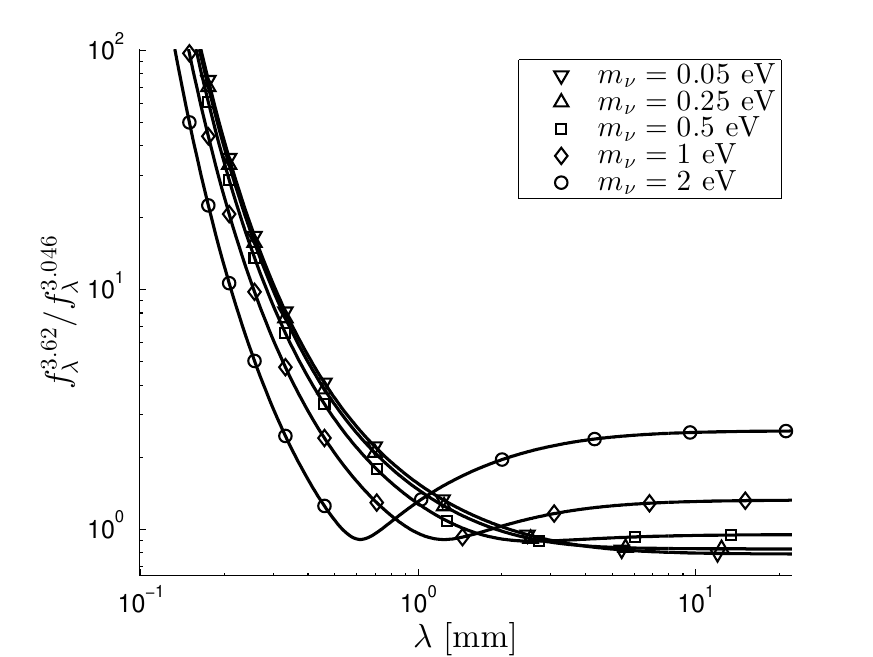}}
\caption{Neutrino FDEV de Broglie wavelength  distribution in the Earth frame. We show in left panel the distribution for $N_\nu^{\mathrm{eff}}=3.046$ (solid lines) and $N_\nu^{\mathrm{eff}}=3.62$ (dashed lines) and in right panel their ratio. \cccite{Birrell:2014qna}.  }\label{fig:deBrogle300}
 \end{figure}

Using $\lambda=2\pi/p$, we find  the normalized FDEV de Broglie wavelength distribution
\begin{equation}
f_\lambda=\frac{ 2\pi g_\nu}{n_\nu\lambda^4}\!\!\int_0^\pi\!\!\! \!\frac{\sin(\phi) d\phi}{\Upsilon^{-1}e^{\sqrt{( E-v_{\text{rel}} p \cos(\phi))^2\gamma^2-m_\nu^2}/T_\nu}\!\!+\!1}\,,
\end{equation}
shown in Figure \ref{fig:deBrogle300} for $v_{\text{rel}}=300$ km/s and for several values $m_\nu$ comparing  $N_\nu^{\mathrm{eff}}=3.046$ with $N_\nu^{\mathrm{eff}}=3.62$. 

\para{Drag force}
Given the neutrino distribution, we evaluate the drag force due to the anisotropy of the neutrino distribution in the rest frame of the moving object for $N_\nu^{\mathrm{eff}}=3.046$. The relic neutrinos will undergo potential scattering with the scale of the potential strength being
\begin{equation}\label{V0}
V_0=CG_F\rho_{N_c},\hspace{2mm} \rho_N\equiv N_c/R^3
\end{equation}
where $R$ is the linear size of the detector.  

When the detector size is smaller than the quantum de Broglie wavelength of the neutrino, all scattering centers are added coherently to for the target effective `charge' $N_c$.  $\rho_{N_c}$ is the charge density, and C=O(1) and is depending on material composition of the object. Such considerations are of interest both for scattering from terrestrial detectors, as well as for ultra-dense objects of neutron star matter density, e.g.   strangelet  CUDOS~\cite{Rafelski:2011bby} - recall that such nuclear matter fragments with $R<\lambda$  despite their small size would have a mass rivaling that of large meteors. We find $V_0\simeq 10^{-13}$ eV for normal matter densities, but for nuclear target density a potential well with $V_0\simeq {\cal O}(10 {\rm eV})$.  

We consider relic neutrino potential scattering to obtain the average momentum transfer to the target and hence the drag force. The particle flux per unit volume in momentum space is
\begin{equation}\label{dnQuantum}
\frac{dn}{dtd Ad^3{\bf p}}({\bf p})=\frac{2}{(2\pi)^3}f({\bf p})p/m_\nu\,,\hspace{2mm} p\equiv |{\bf p}|\,,
\end{equation}
where the factor of two comes from combining neutrinos and anti-neutrinos of a given flavor. 

Our use of nonrelativistic velocity is justified by  \rf{fig:RelvDist300}.  The recoil change in detector momentum per unit time is  
\begin{align}
\frac{d{\bf p}}{dt}=& \int  {\bf q}A \frac{dn}{dtdAd^3p}({\bf p})d^3p\,,\\
{\bf q}A\equiv &\int ({\bf p}-{\bf p_f})\frac{d\sigma}{ d\Omega}({\bf p_f},{\bf p})d\Omega\,.
\end{align}
Here ${\bf p}$ and ${\bf p}_f$, the incoming and outgoing momenta respectively, have the same magnitude. $qA$ is the momentum transfer times area, averaged over outgoing momenta, and $d\Omega$ is the solid angle for to ${\bf p}_f$.  

For a spherically symmetric potential, the differential cross-section depends only on the incoming energy and the angle $\phi$ between ${\bf p}$ and ${\bf p_f}$.  Therefore, for each ${\bf p}$ the integral over $d\Omega$ of the components orthogonal to ${\bf p}$ is zero by symmetry.  This implies
\begin{align}
{\bf q}A\equiv &2\pi{\bf p}\int(1-\cos(\phi))\frac{d\sigma}{ d\Omega}(p,\phi)\sin(\phi)d\phi\,.
\end{align}
The only angular dependence in the neutrino distribution is in ${\bf p}\cdot {\bf\hat z}$ and therefore the components of the force orthogonal to ${\bf \hat z}$ integrate to zero, giving
\begin{align}\label{drag}
\frac{d{\bf p}}{dt}=&\frac{{\bf\hat z}}{\pi m_\nu } \int p^4g(p) f(p,\tilde\phi) \cos(\tilde\phi) \sin(\tilde\phi) dpd\tilde\phi\,,\\
g(p)\equiv& \int_0^\pi(1-\cos(\phi))\frac{d\sigma}{ d\Omega}(p,\phi)\sin(\phi)d\phi\,.\label{geq}
\end{align}

For the case of normal density matter, the Born approximation is valid due to the weakness of the potential compared to the neutrino energy seen in \rf{fig:RelvDist300} right-hand panel. To obtain an order of magnitude estimate, we take a Gaussian potential \begin{equation}
V(r)=V_0e^{-r^2/R^2}\,,
\end{equation}
for which the differential cross-section in the Born approximation can be analytically evaluated
\begin{align}
&\frac{d\sigma}{ d\Omega}(p,\phi)=\frac{\pi m_\nu^2V_0^2R^6}{4}e^{-q^2R^2/2}\,,\notag\\
&q=|{\bf p}-{\bf p}_f|=2p\sin(\phi/2)\,.
\end{align}

The integral over $\phi$ in \req{geq} can also be done analytically, giving
\begin{align}
g(p)=&\pi m_\nu^2V_0^2R^6\frac{1-(2R^2p^2+1)e^{-2R^2p^2}}{4R^4p^4}\,.
\end{align}
 In the long and short wavelength limit we have 
\begin{align}
\label{longWaveDrag}
&g(p)\simeq\frac{\pi}{2} m_\nu^2V_0^2R^6\,,\quad pR\ll 1\,,\\
&F_L\simeq \frac{m_\nu V_0^2R^6}{2} \int p^4 f(p,\tilde\phi) \cos(\tilde\phi) \sin(\tilde\phi) dpd\tilde\phi\,,\notag\\
\label{shortWaveDrag}
&g(p)\simeq\frac{\pi m_\nu^2V_0^2R^2}{4p^4}\,, 
\quad pR\gg 1\,,\\
&F_S\simeq \frac{ m_\nu V_0^2R^2}{4} \int f(p,\tilde\phi) \cos(\tilde\phi) \sin(\tilde\phi) dpd\tilde\phi\notag\,.
\end{align}
 We also note that in the short wavelength limit, our coherent scattering treatment is only applicable to properly prepared structured targets \cite{Liao:2012wb}.

Inserting \req{V0} we see that this force is $O(G_F^2)$, see also \cite{Shvartsman,Smith:1983jj,Gelmini:2004hg}, as compared to the $O(G_F)$ effects debated in  \cite{Opher:1974drq,Lewis:1979mu,Opher2,Cabibbo:1982bb,Langacker:1982ih,Smith:1983jj,Ferreras:1995wf}. In long wavelength limit the size $R$ cancels, in favor of $N_c^2$ which explicitly shows that scattering is on the square of the charges of the target. 

This results in an enhancement of the force by a factor of $N_c$ over the incoherent scattering case, due to $V_0^2$ scaling with $N_c^2$. This effect exactly parallels the proposed detection of supernovae MeV energy scale neutrinos by means of collisions with the entire atomic nucleus~\cite{Divari:2012zz}.  

Fits to the integrals in the above force formulas \req{longWaveDrag} and \req{shortWaveDrag} can be obtained in the region $0.005 \text{eV}\leq m_\nu\leq 0.25\text{eV}$, $v_\text{rel}\leq 300$km/s, yielding
\begin{align}\label{FL}
F_L\!=&8\,10^{-34}{\rm N}\!\left(\!\frac{m_\nu}{0.1 \text{eV}}\!\right)^{\!\!2}\!\! \left(\!\frac{V_0}{1\text{peV}}\!\right)^{\!\!2}\!\!\left(\!\frac{R}{1\text{mm}}\!\right)^{\!\! 6} \!\!\frac{v_{\text{rel}}}{v_\oplus}\,,\\[0.2cm]
F_S=&2\, 10^{-35}{\rm N}\!\left(\!\frac{m_\nu}{0.1\text{eV}}\!\right)^{\!\!2}\! \left(\!\frac{V_0}{1\text{peV}}\!\right)^{\!\!2}\!\!\left(\!\frac{R}{1\text{mm}}\!\right)^{\!\!2}
\frac{v_{\text{rel}}}{v_\oplus}\!\left(\!1\!-\!0.2\frac{m_\nu}{0.1\text{eV}}\frac{v_{\text{rel}}}{v_\oplus}\right)\,.
\end{align}
We emphasize that they are not valid in the limit as $m_\nu\rightarrow 0$. Considering that the current frontier of precision force measurements at the level of individual ions is on the order of $10^{-24}$N \cite{Biercuk}, the ${\cal O}(G_F^2)$ force on a coherent mm-sized terrestrial detector is negligible, despite the factor of $N_c$ enhancement. 

We now consider scattering from nuclear matter density $\rho_N\simeq 3 \, 10^{8}{\rm kg/mm}^3$ objects where $V_0={\cal O}(10\text{eV})$ is effectively infinite compared to the neutrino energy unless the object velocity relative to the neutrino background is ultra-relativistic.  Therefore we are in the hard `ball' scattering limit. As with the analysis for normal matter density, we will investigate both the long and short wavelength limits. 

In the long wavelength limit, only the S-wave contributes to hard sphere scattering and $d\sigma/d\Omega=R^2$, independent of angle. Using \req{drag} and a similar fit to \req{FL} gives
\begin{align}\label{FLHard}
F_L=&\frac{2\pi^2R^2}{\pi m_\nu} \int p^4 f(p,\tilde\phi) \cos(\tilde\phi) \sin(\tilde\phi) dpd\tilde\phi\notag\\
\simeq &\, 2\, 10^{-22}{\rm N}\left(\frac{R}{1\text{mm}}\right)^2\frac{v_{\text{rel}}}{v_\oplus}\,.
\end{align}
In particular the force is independent of $m_\nu$.  We also note that at high velocity, \req{FLHard}  underestimates the drag force. The resulting acceleration is
\begin{equation}
a=4\, 10^{-31}\frac{m}{s^2}\frac{v_{\text{rel}} }{v_\oplus}\! \left(\frac{R}{1\text{mm}}\right)^{-1}\!\!\left(\frac{\rho}{\rho_N}\right)^{-1}\,.\!\!
\end{equation}

The Newtonian drag time constant, $v_{\rm rel}/a$, is
\begin{equation}
\tau= 2\,10^{28}\text{yr}\frac{R}{1\text{mm}}\,\frac{\rho}{\rho_N}\,,
\end{equation}
which suggests that the compact object produced early on in stellar evolution remain largely unaltered.

The last case to consider is the short wavelength hard sphere scattering limit.  This limit is classical and so we no longer treat it as quantum mechanical potential scattering, but rather as elastic scattering of point particle neutrinos from a hard sphere of radius $R$.  

For a single scattering event where the component of the momentum normal to the sphere is ${\bf p}^\perp=({\bf p}\cdot \hat {\bf r}) \hat {\bf r}$, the change in particle momentum is  $\Delta {\bf p}=-2{\bf p}^\perp$. The particle flux per unit volume in momentum space at a point ${\bf r}$ on a radius $R$ sphere $S_R^2$ and inward pointing momentum ${\bf p}$ (i.e. ${\bf p}\cdot \hat{\bf  r}<0$) is
\begin{equation}\label{dnClassical}
\frac{dn}{dtd Ad^3{\bf p}}({\bf x},{\bf p})=\frac{2}{(2\pi)^3}f({\bf p})|{\bf v}\cdot \hat{\bf r}|\,,
\end{equation}
where the factor of two comes from combining neutrinos and anti-neutrinos of a given flavor.  

Note that for point particles the flux is proportional to the normal component of the velocity, as opposed to wave scattering where it is proportional to the magnitude of the velocity, seen in \req{dnQuantum}.

Using \req{dnClassical}, the recoil change in momentum per unit time is  
\begin{equation}
\frac{d{\bf p}}{dt}= -\frac{2}{(2\pi)^3}\int_{{\bf p}\cdot \hat{\bf r}<0} \!\!\!\!\!\!\Delta {\bf p}  f({\bf p}) \frac{1}{m_\nu}|{\bf p}\cdot \hat{\bf r}| d^3{\bf p}R^2d\Omega\,.
\end{equation}
The only angular dependence in $f$ is through ${\bf p}\cdot \hat {\bf z}$ so by symmetry, the ${\bf \hat x}$ and ${\bf \hat y}$ components integrate to $0$.  Therefore we have
\begin{equation}
\frac{d{\bf p}}{dt}=-\frac{R^2\hat{\bf z}}{2\pi^3m_\nu}\int_{{\bf p}\cdot \hat{\bf r}<0}\!\!\!   f({\bf p}) ({\bf p}\cdot \hat{\bf r})^2 \hat {\bf r}\cdot\hat{\bf z}\, d^3{\bf p}d\Omega\,.  
\end{equation}

We perform this integration in spherical coordinates for ${\bf r}$ and in the spherical coordinate vector field basis for ${\bf p}=p_r\hat{\bf r} +p_\theta\hat{\bf r}_\theta+p_\phi\hat{\bf r}_\phi,\hspace{2mm} p_r<0$, where we recall
\begin{align}
&\hat{\bf r}=\cos  \theta \sin \phi \, \hat{\bf x}+\sin \theta \sin \phi \hat{\bf y}+\cos  \phi\,\hat{\bf z}\notag\,,\\
&\hat{\bf r}_\theta=-\sin \theta \hat{\bf x}+\cos \theta \hat{\bf y}\,,\\
&\hat{\bf r}_\phi=\cos \theta \cos \phi\,\hat{\bf x}+\sin \theta \cos \phi \,\hat{\bf y}-\sin \phi \, \hat{\bf z}\notag\,.
\end{align}
Therefore the force per unit surface area is
\begin{align}
\frac{1}{A}\frac{d{\bf p}}{dt}=&-\frac{1}{4\pi^3 m_\nu}\int_0^\pi\!\!\int_{p_r<0}\!\!\!\!\!\!\!\! f({\bf p}) p_r^2  d^3{\bf p}\cos \phi \sin \phi  d\phi\hat{\bf z}\,,\notag\\
f({\bf}p)=&\frac{1}{  \Upsilon^{-1}e^{\sqrt{( E-V_\oplus {\bf p}\cdot \hat {\bf z})^2\gamma^2-m_\nu^2}/T_\nu}+1 }\,,\\
          &{\bf p}\cdot\hat{\bf z}=p_r\cos \phi-p_\phi\sin\phi \,.\notag
\end{align}

We obtain an approximation over the range $v_{\text{rel}}\leq v_\oplus ;\ 0.05\text{eV}\leq m_\nu\leq 0.25\text{eV}$ given by
\begin{align}
&F_S =  4\,10^{-23} \text{N}\left(\frac{R}{1{\rm mm}}\right)^2\frac{v_{\text{rel}}}{ v_\oplus}\,.
\end{align}
This is a result similar to the long wavelength hard sphere limit \req{FLHard}, but the fact that it is only applicable to objects larger than the neutrino wavelength means that the acceleration it generates is negligible on the timescale of the Universe.

\para{Prospects for detecting relic neutrinos}
In this section we characterized the relic cosmic neutrinos and their velocity, energy, and de Broglie wavelength distributions in a frame of reference moving relative to the neutrino background. We have shown explicitly the mass $m_\nu$ dependence and the dependence on neutrino reheating expressed by $N_\nu^{\mathrm{eff}}$, choosing a range within the experimental constraints. This is a necessary input for the measurement of $N_\nu^{\mathrm{eff}}$ and neutrino mass by future detection efforts.  

Finally, we have discussed in detail the $O(G_F^2)$ mechanical drag force  originating in the dipole anisotropy induced by motion relative to the neutrino background.  Despite enhancement with the total target charge found within the massive neutrino wavelength, the magnitude of the force is found to be well below the reach of current  precision force measurements.

Our results are derived under the assumption that $N_\nu^{\mathrm{eff}}$ is due entirely to SM neutrinos, with no contribution from new particle species. In principle future relic neutrino detectors, such as PTOLEMY~\cite{Betts:2013uya,PTOLEMY:2019hkd}, will be able to distinguish between these alternatives since the effect of $N_\nu^{\mathrm{eff}}$ as presented here is to increase neutrino flux~\cite{Birrell:2012gg}, see \req{nnu}. However, to this end one must gain precise control over the enhancement of neutrino galactic relic density due to  gravitational effects~\cite{Ringwald:2004np} as well as the annual modulation~\cite{Safdi:2014rza}. 

\section{Plasma physics methods: Electromagnetic fields,  BBN}\label{part4}
\subsection{Plasma response to electromagnetic fields}
\label{chap:PlasmaSF}
The interaction of electromagnetic fields within relativistic plasma is of interest in wide area of cosmology, astrophysics, and laboratory environments involving  intense laser interactions with matter, and  relativistic heavy-ion collisions forming the new phase of matter, quark-gluon plasma, in relativistic heavy-ion collisions, which discovery allows one to probe in the laboratory some aspects of the primordial Universe as we have discussed before, see \rsec{chap:QCD}. Several methods have been introduced to study the linear response of a collisionless ultrarelativistic QGP following the seminal work by~\cite{Weldon:1982aq} by using semiclassical transport theory based on the Boltzmann equation~\cite{Mrowczynski:1987jr,Mrowczynski:1989np,Blaizot:1993zk,Kelly:1994ig,Kelly:1994dh}. However, applications of this formalism are restricted to dilute plasmas where collisions can be neglected~\cite{Blaizot:2001nr}. 

Here, we will study semi-classical transport using the Vlasov-Boltzmann equation\index{Vlasov-Boltzmann equation} with momentum-averaged quantum collisions between particles, a topic discussed in numerous other works, such as \cite{DeGroot:1980dk,Cercignani:2002bk,Hakim:2011bk,Carrington:2003je,Schenke:2006xu}.
Previously, the effects of collisions within the plasma were mainly studied to derive transport coefficients, such as the electrical conductivity, of interest to the study of plasma response to long-wavelength perturbations~\cite{Mrowczynski:1988xu,Heiselberg:1993cr,Ahonen:1996nq,Baym:1997gq,Ahonen:1998iz}. In quantum field theory, transport coefficients have also been calculated using effective propagators that re-sum thermal modifications to avoid infrared divergences~\cite{Heiselberg:1994ms,Arnold:2002zm,Arnold:2003zc}.

The theoretical description of relativistic interacting  plasma is based on transport theory, i.e., the relativistic form of Liouville's equation. The one-particle phases space distribution function $f(x,p)$ undergoes Liouville flow,
\begin{align}
    \frac{d f(x,p)}{d\tau} = \{H(x,p), f(x,p)\} = 0\,,
\end{align}
where $p$ is the canonical four-momentum, and $x$ is the canonical position. The collision term $C[f]$ represents elastic/inelastic interactions and gives deviations away from Liouville's theorem
\begin{align}\label{eq:LpC}
    \frac{d f(x,p)}{d\tau} = C[f]\,,
\end{align}
or equivalently, entropy generation. The collision term is necessary to describe systems where the mean free path of plasma constituents is less than or equal to the characteristic length scale of the plasma or when the mean free time $\tau$ is smaller than the characteristic oscillation time of the plasma. This pertains to systems with high density, low temperature, or strongly coupled systems.

The Boltzmann-Einstein equation, see  \rsec{sec:BoltzmannEinstein}, with a realistic collision operator, i.e., modeling scattering among neutrinos and $e^\pm$, was used in \rsec{ch:param:studies} to study the cosmological neutrino freeze-out. However, in many cases a detailed treatment of the microscopic collision term \req{eq:collisionMicro} is computationally prohibitive.  In this section our focus is on the interaction of electromagnetic fields within relativistic plasmas and so in place of the microscopic collision term we employ the relaxation-time approximation (RTA) technique, as proposed by~\cite{Anderson:1974nyl}. 

RTA is a commonly made simplification to the Boltzmann equation, reducing it from an integrodifferential equation to a differential equation. The relativistic form of this collision term takes the form
\begin{equation}\label{eq:lincoll}
C[f] = (p^\mu u_\mu) \kappa [ f_\mathrm{eq}(p) - f(x,p) ] \,,
\end{equation}
where $\kappa=1/\tau$ is the relaxation rate, $f(x,p)$ is the phase space distribution of charged particles in the plasma, $f_\mathrm{eq}(p)$ is their equilibrium distribution, and $u_\mu$ is the four-velocity of the plasma rest frame.

The RTA collision term assumes the nonequilibrium distribution $f$ returns to the equilibrium distribution in some characteristic time $\tau$, which is evident when writing \req{eq:LpC} in the form
\begin{equation}
    \frac{d f(x,p)}{dt} = \frac{f_\mathrm{eq}(p) - f(x,p) }{\tau}\,.
\end{equation}
The relaxation time $\tau$ can be computed using the schematic relaxation time approximation where an average relaxation time is introduced~\cite{Mrowczynski:1988xu,Satow:2014lia} or by calculating the momentum-dependent relaxation rate $\kappa(p)$ with the input of perturbative matrix elements~\cite{Ahonen:1996nq}. We use the average relaxation time approximation with momentum averaged $\kappa$ to make all calculations analytically tractable.

The well-known disadvantage of the RTA is that it forces all quantities, even conserved ones, to return to their equilibrium value at a rate $\tau$. This can cause the dynamics derived from this collision term to violate current and energy-momentum conservation. The violation of energy conservation is similar to introducing frictional damping into one particle Newtonian dynamics where energy is lost to the environment.

Correcting for current and energy-momentum conservation is possible by adding terms that ensure that conserved quantities are unaffected~\cite{Bhatnagar:1954zz,Greene1973,Rocha:2021zcw,Singha:2023eia}. It is worth noting that this breaking of conservation law does not always affect the physical behavior of the plasma. For instance, the behavior of transverse waves in an infinite homogeneous plasma is unaffected by the addition of current conservation~\cite{Formanek:2021blc}.

We generalize the BGK modification of the linearized collision term to relativistic plasmas using the Anderson-Witting form Eq.\,(\ref{eq:lincoll}), ensuring current conservation \req{eq:collision} but not energy-momentum conservation. In \cite{Formanek:2021blc} we have shown that the resulting linear response functions satisfy current conservation and gauge invariance constraints. 

The preceding sections will discuss obtaining exact solutions for the covariant polarization tensor in linear response limit via Fourier transform with the BGK collision term \req{eq:collision}. We will present the plasma's electromagnetic properties by using the polarization tensor to derive the electromagnetic fields.

\para{Covariant kinetic theory}
A full microscopic picture of plasma kinematics, useful in numerical simulations, is often more involved than what is required to understand changes in the macroscopic quantities of plasmas. A conventional simplification to the microscopic picture is to average over the discrete states to yield a distribution function $f(x,\boldsymbol{p})$, which describes the probability of finding some number of particles $dN$ in a small range of position $d\mathbf{r}^3$ and momentum $d\boldsymbol{p}^3$ or relativistically~\cite{Hakim:2011bk}
\begin{equation}
    \int_{\Sigma}d\Sigma_{\mu}\int  d^4p\frac{p^\mu}{m}f(x,p) = N,
\end{equation}
where $d\Sigma_\mu$ is the surface element on $\Sigma$
\begin{equation}
    d\Sigma_\mu = \frac{1}{3!}\epsilon_{\mu \nu \alpha\beta} dx^\nu \times dx^\alpha\times dx^\beta\,
\end{equation}
and where the covariant integration measure can be written as
\begin{equation}\label{eq:measure} 
 \frac{d^4p}{(2\pi)^4}4\pi \delta_+(p^2-m^2) = \left.\frac{d^3p}{(2\pi)^3p^0}\right|_{p^0 = \sqrt{|\boldsymbol{p}|^2 + m^2}} \,,
\end{equation}
where $p^0 = p \cdot u$ in the rest frame of the plasma; see \rapp{ch:vol:forms} for a detailed discussion of the relativistic volume element. The one particle distribution function is effectively the phase space density of the system. We will always refer to the four-momentum as $p = (p_0, \, \boldsymbol{p})$ and the 3-momentum as $\boldsymbol{p}$.

The kinetic equation describing the evolution of this distribution is the Vlasov-Boltzmann equation\index{Vlasov-Boltzmann equation} (VBE). The VBE is often derived in detail from heuristic arguments see \cite{DeGroot:1980dk,Cercignani:2002bk}. Here, we will outline how it relates to Liouville's theorem. A similar derivation of the equilibrium distribution in the presence of electromagnetic fields is found in \cite{Hakim:2011bk}.
We derive the classical one-species Vlasov-Boltzmann equation from the Liouville theorem
\begin{equation}
    \frac{d f(Q,P)}{d\tau} = \{H(Q,P), f(Q,P)\} = 0\,,
\end{equation}
where $P^{\mu}$ and $Q^{\mu}$ are the canonical coordinates. 
This theorem states that the canonical phase space density is conserved or the one particle phase space density $f(Q,P)$ satisfies the above continuity equation.  The Poisson bracket is explicitly written as 
\begin{equation}
    \frac{d f(Q,P)}{d\tau} = \frac{\partial Q^{\mu}}{\partial \tau}\partial_\mu f(Q,P) + \frac{\partial P^{\mu}}{\partial \tau}\frac{\partial f(Q,P)}{\partial P^{\mu}}\,.
\end{equation}
Since we consider these particles in the presence of electromagnetic fields, we use the relativistic EM Hamiltonian in the Bergmann form
\begin{equation}
    H(Q,P) = \sqrt{(P-q A(Q))_\mu(P-q A(Q))^\mu}\,,
\end{equation}
which contracts the kinetic momentum to give the relativistic energy of a particle in an electromagnetic field. The equations of motion are
\begin{align}
    \frac{\partial Q^{\mu}}{\partial \tau} &= \frac{\partial H(Q,P)}{\partial P_{\mu}}= \frac{(P-q A(Q))^{\mu}}{H(Q,P)}\,,\\
   -\frac{\partial P^{\mu}}{\partial \tau} &= \frac{\partial H(Q,P)}{\partial Q^{\mu}}= - \frac{(P-q A(Q))^{\nu}q \partial_\mu A_\nu(Q)}{H(Q,P)}\,.
\end{align}
If a canonical transformation is applied to our coordinates, the Liouville theorem states that the phase space density remains unchanged. 
The transformation we would like to consider is the transition from kinetic to canonical coordinates where $Q^{\mu}\rightarrow x^{\mu}$ and  $P^{\nu} \rightarrow P^{\nu} - q A^{\nu}(x)$. This new momentum is related to the actual velocity of the particle $P^{\nu} - q A^{\nu}(x) = p^{\mu} = m\frac{d x^{\mu}}{d \tau}$.  We then consider the Liouville theorem for the shifted function,  
\begin{equation}
 \frac{d x^{\mu}}{d \tau}\partial_\mu f(x,P-q A(x)) + \frac{d (P-q A(x))^{\mu}}{d \tau}\frac{\partial f(x,P-q A(x))}{\partial (P-q A(x))^{\mu}}\,.
\end{equation}
Then, we use the equations of motion to write
\begin{equation}
    \frac{(P-q A(x))^{\mu}}{H(x,P)} \partial_\mu f(x,P-q A(x)) + q\frac{(P-q A(x))_{\nu}}{H(x,P)} F^{\mu \nu}(x) \frac{\partial f(x,P-q A(x))}{\partial (P-q A(x))^{\mu}}\,,
\end{equation}
where the electromagnetic tensor is $F^{\mu \nu} = \partial^{\mu} A^{\nu}  - \partial^{\nu}A^{\mu}$. 
Since the canonical momentum is related to the kinetic momentum by $ P^{\mu}  = m\frac{d x^{\mu}}{d \tau} + q A^{\mu}(x)$, we rewrite the Liouville flow in terms of kinetic momentum $p^\mu = m \frac{dx^\mu}{d\tau}$. Applying Liouville's theorem allows us to set the whole expression to zero to recover the collisionless Vlasov-Boltzmann equation
\begin{equation}
    p^{\mu} \partial_\mu f(x, p) + q  p_{\nu} F^{\mu \nu}(x) \frac{\partial f(x, p )}{\partial  p^{\mu} }=0\,,
\end{equation}
where $ p^{\mu} = m\frac{d x^{\mu}}{d \tau}$. The collision term is then added to allow for deviations from constant phase space density flow
\begin{equation}\label{eq:VBE}
(p_k \cdot \partial) f_k(x,p_k) + q_k F^{\mu\nu} p^k_\nu \frac{\partial f_k(x,p_k)}{\partial p_k^\mu} =\sum_l \, (p_k\cdot u)C_{kl}(x,p_k)\,,
\end{equation}
where there are $k$ equations for each particle species and index $l$ allows to sum over all possible collisions with particle $k$. Usually, we drop the subscript $k$ on momentum if there is no ambiguity. The first term describes the flow or diffusion of particles in the medium, the second term generates an electromagnetic force on particles, and the collision term is on the right-hand side. Generally, each plasma constituent will have a Boltzmann equation and collisions between each species. The collision term represents the detailed microscopic scattering between the plasma constituents. The collision term for the reaction $k+l\rightarrow i+j$ is defined as
\begin{equation}\label{eq:collisionMicro}
    C_{kl}(x,p_k) = \frac{1}{2}\sum^N_{i=1}\sum^N_{j=1}\int \frac{d^3p_l}{(2 \pi)^3p_l^0}\frac{d^3p_i}{(2 \pi)^3p_i^0}\frac{d^3p_j}{(2 \pi)^3p_j^0}\left[f_if_j -f_k f_l
    \right]W_{kl|ij}\,,
\end{equation}
where 
$k,l = 1,2,...,N$ and $W_{ij|kl}$ is the transition rate for the respective collision.
It is important to note that in this framework for a plasma forced by external fields, the collision term is the only way a particle species can impact the dynamics of the phase space distribution of another species.

\para{The BGK collision term}
As discussed previously, the integral in \req{eq:collisionMicro} vastly complicates solving the Vlasov-Boltzmann equation\index{Vlasov-Boltzmann equation}. Instead, we will use a simplified collision term that returns the distribution $f(x,p)$ to equilibrium at some characteristic rate $\kappa = 1/\tau$, reducing \req{eq:VBE} from an integro-differential equation to a differential equation. The relaxation rate or damping rate $\kappa$ is the sum of all possible collisions~\cite{Das:2021bkz}
\begin{equation}
    \kappa_k(p) = \sum^N_{i=1}\sum^N_{j=1}\sum^N_{l=1} \frac{1}{2}\int\frac{d^3p_l}{(2 \pi)^3p_l^0}\frac{d^3p_i}{(2 \pi)^3p_i^0}\frac{d^3p_j}{(2 \pi)^3p_j^0}f_l^{\text{eq}}W_{kl|ij}\,.
\end{equation}
In \cite{Formanek:2021blc} we utilize the simplified collision term proposed by Ref.~\cite{Bhatnagar:1954zz} (BGK), which is amended to conserve the current 
\begin{equation}\label{eq:collision}
    C(x,p) =\kappa\left(f_{\text{eq}}(p)\frac{n(x)}{{n_{\text{eq}}}} - f(x,p)\right)\,.
\end{equation}
The nonequilibrium and equilibrium densities are defined covariantly as
\begin{align}
\label{eq:ndef1}n(x) &\equiv 2 \int \frac{d^3p}{(2\pi)^3p^0}(p \cdot u)f(x,p)\,,\\
\label{eq:ndef2}n_\mathrm{eq} &\equiv 2\int \frac{d^3p}{(2\pi)^3p^0}(p \cdot u) f_\mathrm{eq}(p)\,.
\end{align}
The factor of two accounts for the spin degrees of freedom. This correction is also proposed in \cite{Rocha:2021zcw} where they treat the collision term as an operator adding counterterms to ensure that when acting on conserved quantities like energy, momentum, and particle number, the modified collision operator yields zero, thereby respecting the fundamental conservation laws. We can see that \req{eq:collision} explicitly conserves the four-current~\cite{Formanek:2021blc}
\begin{equation}\label{eq:jmudef}
j_{\mathrm{ind}}^\mu (x)= 2q \int \frac{d^3p}{(2\pi)^3p^0}p^\mu f(x,p)\,,
\end{equation}
by applying $\partial_\mu$ on this expression and substituting back from the Boltzmann equation \req{eq:boltzmanncov}
\begin{equation}
\partial_\mu j^\mu = 2q \int \frac{d^3p}{(2\pi)^3p^0} \left\{-q F^{\mu\nu}p_\nu \frac{\partial f(x,p)}{\partial p^\mu}\right. 
\left. + (p \cdot u)\kappa \left[f_\mathrm{eq}(p) \frac{n(x)}{n_\mathrm{eq}}-f(x,p) \right] \right\}\,.
\end{equation} 
The first term should naturally vanish because the collisionless Vlasov equation preserves the four-current. This can be seen upon integration by parts and use of the antisymmetry of $F^{\mu\nu}$. On the other hand, the collision term vanishes by design - see definitions (\ref{eq:ndef1},\ref{eq:ndef2}). This is in contrast to the Anderson-witting collision term, which does not conserve current \req{eq:lincoll}.

\subsection{Linear response: electron-positron plasma}
The transport properties of electron-positron plasma are governed by the Vlasov-Boltzmann equations \cite{Grayson:2023flr}
\begin{align}\label{eq:VBEf}
(p \cdot \partial) f_\pm(x,p) + &q F^{\mu\nu} p_\nu \frac{\partial f_\pm(x,p)}{\partial p^\mu} = C_\pm(x,p)\,,\\
\label{eq:VBEg}(p \cdot \partial) f_\gamma(x,p) &= C_\gamma(x,p)\,.
\end{align}
The subscripts $-$, $+$, and $\gamma$ indicate the transport equation for electrons, positrons, and photons. These form a system of differential equations for each distribution function $f_i(x,p)$. We suppress the four-momentum subscript for each species $f_i(x,p) = f_i(x,p_i)$ to simplify notation. 

Since photons cannot couple directly to the electromagnetic field, they do not contribute to the dynamics of the electromagnetic field at first-order polarization response as indicated in Eq.\,(\ref{eq:VBEg}). This is not true for a QCD plasma where gluons could couple directly to an external gluon field.

To find the effect of electrons and positrons on the electromagnetic fields, we use the transport equations \req{eq:VBEf} to find the induced current in the plasma
\begin{equation}
j_{\mathrm{ind}}^\mu(x) = 2\int \frac{d^3 p}{(2 \pi)^3 p^0}p^\mu \left[f_+(x,p)-f_-(x,p)\right]\,,
\end{equation}
found via Fourier transformation and related to the induced current in the linear response equation
\begin{equation}
    \widetilde{j}_{\mathrm{ind}}^{\mu}(k) = {\Pi^{\mu}}_{\nu}(k) \widetilde{A}^{\nu}(k)\,,
\end{equation}
to identify the polarization tensor $\Pi^{\mu}_{\nu}$. To begin, we solve the Vlasov-Boltzmann equation with the BGK collision term
\begin{equation}\label{eq:boltzmanncov}
(p \cdot \partial) f_\pm(x,p) + q F^{\mu\nu} p_\nu \frac{\partial f_\pm(x,p)}{\partial p^\mu} = (p \cdot u)\kappa_\pm\left[f^\mathrm{eq}_\pm(p)\frac{n_\pm(x)}{n^\mathrm{eq}_\pm} - f_\pm(x,p)\right]\,.
\end{equation}
Since the solutions for these equations will differ only by the sign of charge, we need only solve one to understand dynamics. The $\pm$, which indicates electrons or positrons, may be dropped when unnecessary in the equations below.

We assume for the equilibrium distribution the covariant Fermi-Dirac distribution function~\cite{DeGroot:1980dk,Hakim:1967prd}:
\begin{equation}\label{eq:fb}
f^\mathrm{eq}_\pm(x,p) \equiv \frac{1}{e^{([p^{\mu} +q A^\mu (x) ] u_\mu\pm \mu_q)/T} + 1}\,,
\end{equation}
where $p^\mu+q A^\mu (x)$ is the canonical momentum in the presence of an electromagnetic four-potential, $u^\mu$ is the global four-velocity of the medium, $T$ denotes the temperature in the medium rest frame, and $\mu_q$ is the chemical potential related to charge. 

The linear response approximation assumes the distribution function can be written as a sum of the equilibrium distribution $f_\mathrm{eq}(x,p)$ plus a small perturbation away from the equilibrium $\delta f(x,p)$
\begin{equation}\label{eq:perturbation}
f(x,p) = f_\mathrm{eq}(x,p) + \delta f(x,p)\,.
\end{equation}
Here the small perturbation $\delta f(x,p)$ is induced by an external electromagnetic field. We expand \req{eq:boltzmanncov} in equilibrium and perturbation terms \cite{melrose2008quantum}
\begin{align}
    (p \cdot \partial)\left(f_\mathrm{eq}+ \delta f(x,p)\right) +  q \left(F_{\mathrm{eq}}^{\mu\nu} +\delta F^{\mu\nu}\right)p_\nu &\frac{\partial (f_\mathrm{eq}+\delta f(x,p))}{\partial p^\mu} 
    = \kappa (p\cdot u)\left(f_\mathrm{eq} \frac{\delta n(x)}{{n_\mathrm{eq}(x)}} - \delta f(x,p)\right)\,.
\end{align}
Since the equilibrium expressions are a solution to the collisionless Boltzmann equation, all the equilibrium terms combined are zero. The collision term is constructed to be zero at equilibrium. We will neglect the Lorentz force due to the induced field on the perturbation since it is second order in the perturbation
\begin{equation}
    (p \cdot \partial) \delta f(x,p)+ q \delta F^{\mu\nu}p_\nu \frac{\partial f(x,p)}{\partial p^\mu} = \kappa (p\cdot u)\left(f_{\text{eq}} (x,p)\frac{\delta n(x)}{{n_{\text{eq}}(x)}} - \delta f(x,p)\right)\,,
\end{equation}
where the quantity $\delta n(x)$ is defined following the definitions(\ref{eq:ndef1},\ref{eq:ndef2}) as
\begin{equation}
\delta n (x) \equiv 2 \int \frac{d^3p}{(2\pi)^3p^0} (p \cdot u)\delta f(x,p)\,.
\end{equation}
At this point, we will take the weak field limit of the equilibrium distribution, which assumes the change in energy of a particle due to the electromagnetic field is small in comparison to the thermal energy
\begin{equation}
    \frac{ qA(x)\cdot u}{T}\ll 1\,.
\end{equation}
In this case, the equilibrium distribution becomes the usual
\begin{equation}\label{eq:equilibriumFD}
f^\mathrm{eq}_\pm(x,p) \equiv \frac{1}{e^{(p^{\mu}  u_\mu\pm \mu_q)/T} + 1}\,.
\end{equation}
An explicit solution of the Vlasov-Boltzmann\index{Vlasov-Boltzmann equation}  equation can be obtained more easily in momentum space after a Fourier transformation.  We define the Fourier transform $\widetilde{g}(k^\mu)$ of a general function $g(x^\mu)$ of space-time coordinates as 

\begin{equation}\label{eq:ftdef}
g(x) = \int \frac{d^4k}{(2\pi)^4} \, e^{-i k \cdot x} \, \widetilde{g}(k)\,.
\end{equation} 
The Fourier transformation replaces partial derivatives $\partial_\mu$ with the four-momentum $k_\mu$:
\begin{equation}
\partial_\mu \rightarrow - i k_\mu \,.
\end{equation}
The four-vector $k^\mu = (\omega,\mathbf{k})$ represents the momentum and energy in the electromagnetic field. In contrast, $p^{\mu}  = (E,\boldsymbol{p})$ represents the momentum and energy of plasma constituents.

Using these definitions, the Fourier-transformed Boltzmann equation reads \cite{Formanek:2021blc}
\begin{equation}\label{eq:boltzfourier}
-i (p \cdot k) \widetilde{\delta f}(k,p) + q\widetilde{F}^{\mu\nu}p_\nu \frac{\partial f_\mathrm{eq}(p)}{\partial p^\mu} 
= (p \cdot u)\kappa \left[\frac{f_\mathrm{eq}(p)}{n_\mathrm{eq}}\widetilde{\delta n}(k) - \widetilde{\delta f}(k,p) \right]\,.
\end{equation}
In the following, we simplify the notation of derivatives of the equilibrium function with respect to momentum as
\begin{equation}
\frac{\partial f_\mathrm{eq}(p)}{\partial p^\mu} = \frac{d f_\mathrm{eq}(p)}{d (p \cdot u)} u_\mu \equiv f'_\mathrm{eq}(p) u_\mu \,.
\end{equation}
We solve \req{eq:boltzfourier} for the perturbation $\widetilde{\delta f}(k,p)$, which describes fluctuations away from equilibrium due to the electromagnetic field
\begin{equation}\label{eq:deltaftilde}
\widetilde{\delta f}(k,p) = \frac{i}{p \cdot k + i (p \cdot u) \kappa}\bigg[-q (u \cdot \widetilde{F} \cdot p)f'_\mathrm{eq}(p) 
\left.+ (p \cdot u) \kappa \frac{f_\mathrm{eq}(p)}{n_\mathrm{eq}}\widetilde{\delta n}(k)\right]\,.
\end{equation}
This can be readily integrated to obtain an equation for $\widetilde{\delta n}(k)$
\begin{equation}
\widetilde{\delta n}(k) = R(k) - Q(k)\widetilde{\delta n}(k)\,,
\end{equation}
where the integrals are defined as
\begin{align}\label{eq:R}
R(k)  \equiv -2i \int \frac{d^3p}{(2\pi)^3p^0}(p \cdot u) \frac{q(u \cdot \widetilde{F} \cdot p)f'_\mathrm{eq}}{p \cdot k + i (p \cdot u)\kappa}\,,\\
\label{eq:Q}Q(k) \equiv -2i \frac{\kappa}{n_\mathrm{eq}}\int \frac{d^3p}{(2\pi)^3p^0}(p \cdot u)^2 \frac{f_\mathrm{eq}(p)}{p\cdot k + i(p \cdot u)\kappa}\,.
\end{align}
The solution for $\widetilde{\delta n}(k)$ in terms of the external fields is simply 
\begin{equation}
\widetilde{\delta n}(k) = \frac{R(k)}{1+Q(k)}\,.
\end{equation}
We can substitute this result back into (\ref{eq:deltaftilde}) to obtain an explicit expression for $\widetilde{\delta f}(k,p)$ found in \cite{Formanek:2021blc}
\begin{equation}\label{eq:deltafsolution}
	\widetilde{\delta f}(k,p) = \frac{i}{p \cdot k + i (p \cdot u) \kappa}\bigg[-q (u \cdot \widetilde{F} \cdot p)f'_\mathrm{eq}(p) 
	\left.+ (p \cdot u) \kappa \frac{f_\mathrm{eq}(p)}{n_\mathrm{eq}} \frac{R(k)}{1+Q(k)}\right]\,.
\end{equation}
The right-hand side contains only known quantities. In the next section, we will use \req{eq:deltafsolution} to calculate the induced current in the plasma. Adding additional conservation laws requires further integrals to solve the Vlasov-Boltzmann equation involving higher moments of the fluctuation $\delta f$ as discussed in \cite{Rocha:2021zcw,Singha:2023eia}.

\para{Induced current}
The induced charge current is the sum of the antiparticle distribution $\widetilde{f}_-$ and the particle distribution $\widetilde{f}_+$
\begin{equation}\label{eq:perturbation1}
\tilde{j}_{\mathrm{ind}}^\mu(k) = 2\int \frac{d^3 p}{(2 \pi)^3 p^0}p^\mu 
\sum_{i = \, +, \, -} q_i \tilde{f}_{i}(k,p)\,,
\end{equation}
with the factor of two accounting for spin. Sometimes, this is referred to as the first moment of $\delta f$.
After expanding in linear response \req{eq:perturbation}, and specifying $q_\pm = \pm e$ the induced current is a function of the perturbation
\begin{align}\label{eq:perturbation2}
\tilde{j}_{\mathrm{ind}}^\mu(k) &= 2\int \frac{d^3 p}{(2 \pi)^3 p^0}p^\mu \Big( e \left[\tilde{f}^{\mathrm{eq}}_+(k,p)-\tilde{f}^{\mathrm{eq}}_-(k,p)\right]
+ e\left[\delta\tilde{f}_+(k,p)-\delta\tilde{f}_-(k,p)\right]
\Big) \notag \\
&=4 e\int \frac{d^3 p}{(2 \pi)^3 p^0}p^\mu \delta\tilde{f}(k,p)
\,.
\end{align}
The equilibrium currents cancel in the weak field limit for zero chemical potential\index{chemical potential}, and the perturbations add since they differ by the charge $\delta f_\pm=\pm e \delta f' $. For finite chemical potential $\mu_q$, the equilibrium terms can be combined with hyperbolic trig-identities
\begin{equation}
 \tilde{j}_{\mathrm{ind}}^\mu(k) 
=2 e\int \frac{d^3 p}{(2 \pi)^3 p^0}p^\mu \Big(-\frac{\sinh{(\mu_q)}}{\cosh{(p \cdot u)}+\cosh{(\mu_q)}} +  \left[\delta\tilde{f}_+(k,p)-\delta\tilde{f}_-(k,p)\right]
 \Big)
\,.
\end{equation}
For now, we will focus on the case of zero chemical potential, $\mu_q=0$, where the first term vanishes.
We can express the induced current in terms of defined integrals \cite{Formanek:2021blc} resulting from inserting \req{eq:deltafsolution} into the induced current
\begin{equation}\label{eq:jmu}
\widetilde{j}_{\mathrm{ind}}^\mu(k) = R^\mu(k) - \frac{R(k)}{1+Q(k)} Q^\mu(k)\,,
\end{equation}
where the integrals $R^\mu(k)$ and $Q^\mu(k)$ are defined analogously to (\ref{eq:R},\ref{eq:Q}) as
\begin{align}
\label{eq:Rmu}R^\mu(k)  \equiv -4q^2i \int \frac{d^3p}{(2\pi)^3p^0} p^\mu \frac{(u \cdot \widetilde{F} \cdot p)f'_\mathrm{eq}}{p \cdot k + i (p \cdot u)\kappa}\,,\\
\label{eq:Qmu}Q^\mu(k) \equiv -4qi \frac{\kappa}{n_\mathrm{eq}}\int \frac{d^3p}{(2\pi)^3p^0}(p\cdot u) p^\mu \frac{f_\mathrm{eq}(p)}{p\cdot k + i(p \cdot u)\kappa}\,.
\end{align} 
Note that we absorbed the factor $4e$ from the current (\ref{eq:perturbation2}) into the definition of these integrals. The $R^{\mu}$ term is what one would find from the collisionless case $\kappa \rightarrow 0^+$. The induced current for the normal RTA collision term, which does not conserve current, is obtained by setting $\delta n \rightarrow n_{eq}$, or equivalently,
\begin{equation}\label{eq:jRTA}
\widetilde{j}_{\mathrm{AW}}^\mu(k) = R^\mu(k) - Q^\mu(k)\,.
\end{equation}

\para{Covariant polarization tensor}
To find the polarization tensor, we compare our result (\ref{eq:jmu}) to the covariant formulation of Ohm's law~\cite{Starke:2014tfa} both of which describe the induced current in the momentum space
\begin{equation}\label{eq:ohm}
\widetilde{j}^\mu(k) = \Pi^\mu_\nu(k) \widetilde{A}^\nu(k)\,.
\end{equation}
To perform this comparison and extract the polarization tensor we must rewrite the Fourier transform of the electromagnetic tensor in terms of the four-vector potential in momentum space $\widetilde{A}^\mu(k)$
\begin{equation}\label{eq:ftfmunu}
\widetilde{F}^{\mu\nu}(k) = -i k^\mu \widetilde{A}^\nu(k) + i k^\nu \widetilde{A}^\mu(k)\,.
\end{equation}
We then substitute this into the definition of $R^\mu(k)$ (\ref{eq:Rmu}) and isolate $\widetilde{A}^\mu$ as so it is in the form of \req{eq:ohm} to obtain \cite{Formanek:2021blc}
\begin{equation}
R^\mu(k) = - 4q^2 \int \frac{d^3p}{(2\pi)^3p^0} f'_\mathrm{eq}(p)
\;
\frac{(u\cdot k)p^\mu p_\nu - (k \cdot p)p^\mu u_\nu}{p\cdot k + i (p \cdot u) \kappa} \widetilde{A}^\nu(k)\,,
\end{equation}
from which we see that the contribution of $R^\mu$ to the polarization tensor is
\begin{equation}\label{eq:Rmunu}
R^\mu_\nu(k) \equiv - 4q^2 \int \frac{d^3p}{(2\pi)^3p^0} f'_\mathrm{eq}(p)
\;
\frac{(u\cdot k)p^\mu p_\nu - (k \cdot p)p^\mu u_\nu}{p\cdot k + i (p \cdot u) \kappa}.
\end{equation}
The contribution of the second term is hidden in the $R(k)$ scalar. In terms of the four-vector potential in the momentum space $\widetilde{A}^\nu$, we have
\begin{equation}
R(k) = - 2q \int \frac{d^3p}{(2\pi)^3p^0}(p \cdot u)f'_\mathrm{eq}(p)
\;
\frac{(u\cdot k)p_\nu - (k \cdot p)u_\nu}{p\cdot k + i (p \cdot u) \kappa}\widetilde{A}^\nu(k)\,.
\end{equation}
We can identify in this expression a four-vector $H_\nu(k)$ defined as
\begin{equation}\label{eq:Hnu}
H_\nu(k) \equiv - 2q \int \frac{d^3p}{(2\pi)^3p^0}(p \cdot u)f'_\mathrm{eq}(p)
\;
\frac{(u\cdot k)p_\nu - (k \cdot p)u_\nu}{p\cdot k + i (p \cdot u) \kappa}
\end{equation}
so that the polarization tensor is given by
\begin{equation}\label{eq:pimunu}
\Pi^\mu_\nu(k) = R^\mu_\nu(k) - \frac{Q^\mu(k) H_\nu(k)}{1+Q(k)},
\end{equation}
where the covariant quantities $R^\mu_\nu$, $Q^\mu$, $H_\nu$, and $Q$ are given by the integrals (\ref{eq:Rmunu}, \ref{eq:Qmu}, \ref{eq:Hnu}, \ref{eq:Q}) respectively. 
This is the final covariant form of the current conserving covariant polarization tensor for an infinite homogeneous plasma. The bulk of the work in applying \req{eq:pimunu} to a specific scenario is choosing an equilibrium distribution and evaluating the integrals. Explicit expressions for the components of this tensor in the rest frame of the plasma are found in the ultrarelativistic limit \req{eq:polfuncsUltra} and in the nonrelativistic limit \req{eq:polfuncs} in \cite{Formanek:2021blc}.
This polarization tensor is also derived in \cite{Carrington:2003je} and \cite{Schenke:2006xu}. The correction to the polarization tensor found by using the collision term with current conservation \req{eq:collision} is given by the second term in \req{eq:pimunu}. The current conserving correction modifies the longitudinal polarization properties of the tensor related to charge fluctuations but not the transverse properties related to electromagnetic waves. 

The Anderson-Witting form of the polarization tensor found using the collision term \req{eq:lincoll} is equivalent to $R^\mu_\nu$ and the polarization tensor for a collisionless plasma is $R^\mu_\nu$ with $\kappa \rightarrow 0^+$. {\color{black}We note that a gauge transformation of the linear response relation \req{eq:ohm} results in a shift of the Fourier transformed four-potential
\begin{equation}
    \widetilde{j}'^\mu(k) = \Pi^\mu_\nu(k) \widetilde{A}'^\nu(k) = \Pi^\mu_\nu(k)\left(\widetilde{A}^\nu(k) -i k^\nu \widetilde{\chi}(k)\right)
\end{equation}
where $-i k^\nu \widetilde{\chi}(k)$ is the gradient of an arbitrary gauge potential. 

Gauge invariance requires then that
\begin{equation}
\Pi^\mu_\nu(k) k^\nu = 0
\end{equation}
this requirement is satisfied for the general polarization tensor of an isotropic plasma by the longitudinal \req{eq:Long} and transverse \req{eq:transverse} projection tensors and is satisfied for our specific polarization tensor due to the form of $H_\nu(k)$ in \req{eq:Hnu} and $R^\mu_\nu(k)$ \req{eq:Rmunu}.
}

\subsection{Self-consistent electromagnetic fields in a medium}\label{sec:Maxwell}
To find the electromagnetic field in a plasma, we solve Maxwell's equations self-consistently in an infinite homogeneous and stationary polarizable medium. In this medium, Maxwell's equations take on the usual form \cite{melrose2008quantum}
\begin{equation}
\partial^{[\mu}F^{\nu \rho]}(x) =0, \quad \partial_{\mu}F^{\mu \nu}(x) = \mu_0 J^{\nu}(x)\,.
\end{equation}
Using the Fourier transform defined as in equation \req{eq:ftdef} we replace partial derivatives $\partial_\mu$ with the four-momentum $-i k_\mu$. Then Maxwell's equations in Fourier space are
\begin{equation}
-i k^{[\mu}\widetilde{F}^{\nu \rho]}(k) =0, \quad -i k_{\mu}\widetilde{F}^{\mu \nu}(k) = \mu_0 \widetilde{J}^{\nu}(k)\,,
\end{equation}
where $k=(\omega, \boldsymbol{k})$ is the four-wavevector of the electromagnetic field. The properties of the medium are introduced by writing the four-current $\widetilde{J}^{\mu}$ in terms of its induced and external parts
\begin{equation}
 \widetilde{J}^{\mu}(k) = \widetilde{j}_{\mathrm{ext}}^{\mu}(k)+ \widetilde{j}_{\mathrm{ind}}^{\mu}(k)\,.
\end{equation}
The induced current $\widetilde{j}_{\mathrm{ind}}^{\mu}$, to leading order, is given by the polarization tensor through \req{eq:ohm}. Though the induced current is linear with respect to the self-consistent field $\widetilde{A}^{\nu}$, the field itself is intrinsically nonlinear regarding plasma response as we shall see when solving for the self-consistent fields \reqs{eq:phi}{eq:aperp}. Nonlinear response comes from higher-order terms involving nested convolution integrals of the polarization tensor and the self-consistent potential and is required when the polarization current is on the order of the external current.

Solving Maxwell's equations in the Lorentz gauge $k \cdot \widetilde{A}=0$, one finds the usual expression
\begin{equation}\label{eq:Amu}
\begin{split}
\widetilde{A}^{\mu}(k)&= -\frac{\mu_0}{k^2}\left(\widetilde{j}_{\mathrm{ext}}^{\mu}(k)+ \widetilde{j}_{\mathrm{ind}}^{\mu}(k)\right)\\
&= -\frac{\mu_0}{k^2}\left(\widetilde{j}_{\mathrm{ext}}^{\mu}(k)+  \Pi^\mu_\nu(k) \widetilde{A}^\nu(k)\right)\,,
\end{split}
\end{equation}
where $\mu_0$ denotes the magnetic permittivity of the vacuum, and we have used \req{eq:ohm} to express the induced current. 

\para{Projection of plasma polarization tensor} We proceed by algebraically solving for the self-consistent potential. To do this, we first note that in a homogeneous medium, the response depends only on two independent scalar polarization functions $\Pi_\parallel$ and $\Pi_\perp$ describing polarization in the parallel and transverse directions relative to the wave-vector $\boldsymbol{k}$ \cite{Weldon:1982aq}. The polarization tensor may be written in terms of these polarization functions as
\begin{equation}\label{eq:poltensgen}
 \Pi^{\mu \nu}(k,u) = \Pi_\parallel(k) L^{\mu \nu}(k,u) + \Pi_\perp(k) S^{\mu \nu}(k,u)\,,
\end{equation}
where $k^\mu$ is the four-momentum of the field and $u^\mu$ is the four-velocity of the medium. The polarization tensor represents the electromagnetic response of the medium to the electromagnetic field. $\Pi_\parallel$ usually describes charge fluctuations and $\Pi_\perp$ describes the properties of electromagnetic waves. For optically active or chiral mediums there is also a rotational portion of the polarization tensor $\Pi_R$. Since we neglect spin, our derivation of the polarization tensor is not sensitive to $\Pi_R$. Conventions for the longitudinal and transverse projection tensors, $L^{\mu \nu}$ and  $S^{\mu \nu}$, may be found in \cite{melrose2008quantum}. These tensors are reproduced here for convenience
\begin{equation}\label{eq:Long}
     L^{\mu \nu} \equiv \frac{k^2}{(k\cdot u)^2-k^2}\bigg[ \frac{ k^{\mu}u^{\nu}}{(k\cdot u)}+ \frac{ k^{\nu}u^{\mu}}{(k\cdot u)} -\frac{k^2u^{\mu}u^{\nu}}{(k\cdot u)^2}  -\frac{k^{\mu}k^{\nu}}{k^2} \bigg]\,, 
\end{equation}
\begin{equation}\label{eq:transverse}
     S^{\mu \nu} \equiv g^{\mu \nu} +\frac{1}{(k\cdot u)^2-k^2}\bigg[ k^{\mu}k^{\nu} 
     -(k\cdot u)( k^{\mu}u^{\nu}+k^{\nu}u^{\mu})+k^2u^{\mu}u^{\nu}\bigg]\,.
\end{equation}
These projections are equivalent to ones defined in \cite{Weldon:1982aq} up to an overall normalization. To simplify the calculation, the wave-vector $\boldsymbol{k}$ is chosen, without loss of generality, to point along the third spatial direction ($\mu=3$):
 \begin{equation}\label{eq:poltenmat}
    \Pi^{\mu}_{\nu}(\omega,\boldsymbol{k}) = \left[
    \begin{array}{cccc}
-\frac{|\mathbf{k}|^2}{\omega^2}\Pi_{\parallel}& 0 & 0 & \frac{|\mathbf{k}|}{\omega}\Pi_{\parallel} \\
 0 & \Pi_{\perp} & 0 & 0 \\
 0 & 0 & \Pi_{\perp} & 0 \\
 -\frac{|\mathbf{k}|}{\omega}\Pi_{\parallel} & 0 & 0 & \Pi_{\parallel} \\ 
\end{array}
\right]\,.
\end{equation}
Utilizing this decomposition, we can immediately see that the transverse polarization function will be related to the $\Pi^1_1 = \Pi^2_2 = \Pi_\perp$ component of the polarization tensor defined in \req{eq:pimunu}. Analogously, the longitudinal portion of the polarization tensor is given by calculating the $\Pi^3_3 = \Pi_\parallel$ component. The spatial component of the potential $\widetilde{\boldsymbol{A}}$  in these coordinates can be expressed as, compare \rf{fig:project}
\begin{equation}
\widetilde{\boldsymbol{A}} = \widetilde{A}_\parallel \hat{\boldsymbol{k}} + \widetilde{\boldsymbol{A}}_\perp\,,
\end{equation}
which implies
\begin{equation}
 \widetilde{A}_{\parallel} = \frac{\boldsymbol{k} \cdot  \widetilde{\boldsymbol{A}}}{|\boldsymbol{k}|}, \quad   \widetilde{\boldsymbol{A}}_{\perp} = \widetilde{\boldsymbol{A}} -  \widetilde{A}_{\parallel}\hat{\boldsymbol{k}}\,,
\end{equation}
with analogous definitions for the current, $\widetilde{j}_{\parallel}$ and $\widetilde{j}_{\perp}$. 

\begin{figure} 
    \centering
    \includegraphics[width=0.55\linewidth]{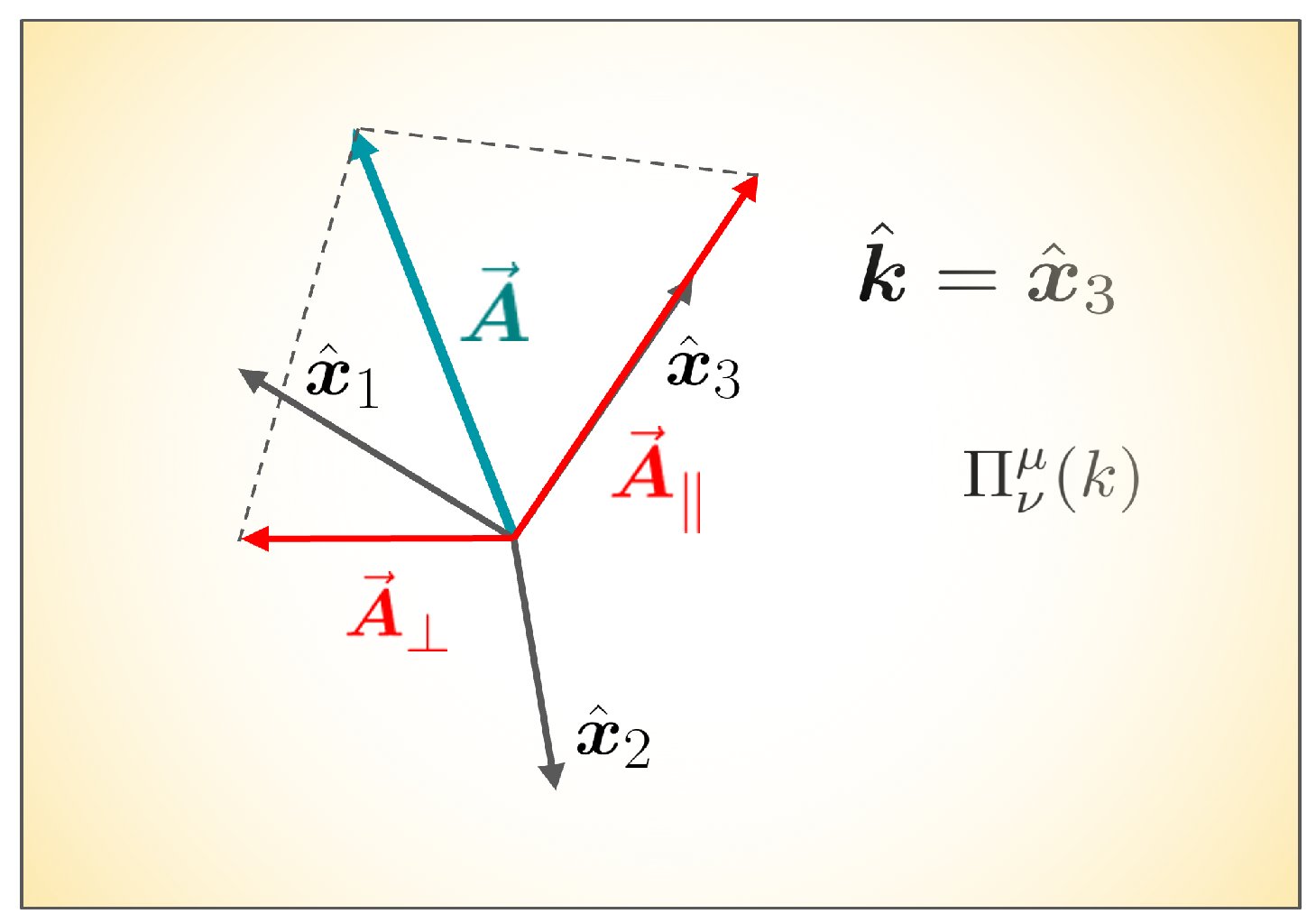}
    \caption{Vector potential is projected onto $\boldsymbol{\hat{k}} = \boldsymbol{\hat{x_3}} =\boldsymbol{\hat{z}}$. \radapt{Grayson:2024okq}.}
    \label{fig:project}
\end{figure}

Note that the Lorentz gauge condition $\partial_\mu A^\mu = 0$ implies
\begin{equation}\label{eq:apar}
\widetilde{A}_\parallel = \frac{\omega}{ |\boldsymbol{k}|}\widetilde{\phi}\, ,
\end{equation} 
with $\phi=A^0$. The induced charge is calculated using the projected polarization tensor \req{eq:poltenmat}:
\begin{equation}
    \widetilde{\rho}_\text{ind}(\omega,\boldsymbol{k})  = \Pi^0_\nu \widetilde{A}^\nu = -\frac{|\boldsymbol{k}|^2}{\omega^2} \Pi_{\parallel}\widetilde{\phi} +  \frac{|\boldsymbol{k}|}{\omega}\Pi_{\parallel} \widetilde{A}_{\parallel}\,.
\end{equation}
For the Lorentz gauge condition \req{eq:apar}, one finds
\begin{equation}\label{eq:indch}
    \widetilde{\rho}_\text{ind}(\omega,\boldsymbol{k})  = \Pi_{\parallel}\widetilde{\phi} \left(1 -\frac{|\boldsymbol{k}|^2}{\omega^2}\right)\,.
\end{equation}
The longitudinal current is,
\begin{equation}\label{eq:indjpar}
\widetilde{j}_{\parallel\text{ind}}(\omega,\boldsymbol{k})  =  \Pi^z_\nu \widetilde{A}^\nu  = \Pi_{\parallel}  \frac{\omega}{|\boldsymbol{k}|}\widetilde{\phi}\left(1-\frac{|\boldsymbol{k}|^2}{\omega^2} \right)\,,
\end{equation}
as expected from current conservation $\partial^\mu j_\mu(x) =0$.
The induced transverse current is
\begin{equation}\label{eq:indjperp}
   \boldsymbol{j}_{\perp\text{ind}}(\omega,\boldsymbol{k})  =  \Pi_{\perp} \widetilde{A}_\perp\,.
\end{equation}
Solving for the potential on both sides of \req{eq:Amu} with the help of \reqs{eq:indch}{eq:indjperp} gives the self-consistent solutions \cite{Grayson:2022asf}
\begin{align}\label{eq:phi}
&\widetilde{\phi}(\omega,\boldsymbol{k}) = \frac{\widetilde{\rho}_\text{ext}(\omega,\boldsymbol{k})}{\varepsilon_0(\boldsymbol{k}^2-\omega^2) \left(\Pi_{\parallel}/( \omega^2\varepsilon_0)+1\right) }\,, \\\label{eq:aperp}
&\widetilde{\boldsymbol{A}}_\perp(\omega,\boldsymbol{k}) = \frac{\mu_0 \widetilde{\boldsymbol{j}}_{\perp \text{ext}}(\omega,\boldsymbol{k})}{\boldsymbol{k}^2 - \omega^2 - \mu_0 \Pi_{\perp}}\,.
\end{align}
The gauge condition \req{eq:apar} gives the self-consistent potential $\widetilde{A}_\parallel$. These self-consistent potentials determine the electric and magnetic fields via the usual relations
\begin{equation}\label{eq:ftfields}
\widetilde{\boldsymbol{B}}(\omega,\boldsymbol{k}) = i\boldsymbol{k} \times \widetilde{\boldsymbol{A}}_\perp\,, \quad \widetilde{\boldsymbol{E}}(\omega,\boldsymbol{k}) = -i \boldsymbol{k} \widetilde{\phi} + i \omega \widetilde{\boldsymbol{A}}\,.
\end{equation}
To obtain the electromagnetic fields in position space, one must Fourier transform \reqs{eq:phi}{eq:aperp}. If done analytically, this usually requires finding the poles in the denominator of these expressions, which equates to finding the poles of the thermal photon propagator. These poles represent propagating modes in the plasma. Modes will often be located at complex values in the $\omega, \mathbf{k}$ plane, leading to finite lifetimes and spatial dispersion. 

\para{Small back-reaction limit}
Here, we briefly mention an alternative to the self-consistent fields, which comes from assuming that the back reaction of the plasma due to the external fields is small compared to the external field. In this case, one can use the external field in the linear response equation instead of the total field
\begin{equation}\label{eq:pert}
    \widetilde{j}_{\mathrm{ind}}^{\mu}(k) = {\Pi^{\mu}}_{\nu}(k) \widetilde{A}_\text{ext}^{\nu}(k)\,.
\end{equation}
Inserting this into \req{eq:Amu} successively to find a series expansion yields the same expression as expanding \reqs{eq:phi}{eq:aperp} in the polarization functions
\begin{align}\label{eq:phipert}
&\widetilde{\phi}(\omega,\boldsymbol{k}) = \sum_{n=0}^\infty\frac{\widetilde{\rho}_\text{ext}(\omega,\boldsymbol{k})}{\varepsilon_0(\boldsymbol{k}^2-\omega^2)}\left(-\frac{\Pi_{\parallel}}{ \omega^2\varepsilon_0}\right)^n\,, \\\label{eq:aperppert}
&\widetilde{\boldsymbol{A}}_\perp(\omega,\boldsymbol{k}) = \sum_{n=0}^\infty\frac{\mu_0 \widetilde{\boldsymbol{j}}_{\perp \text{ext}}(\omega,\boldsymbol{k})}{(\boldsymbol{k}^2 - \omega^2)^{n+1}}(\mu_0 \Pi_{\perp})^{n}\,.
\end{align}
The first term $n=0$ is the vacuum field, and higher-order terms describe the back reaction of the induced current on the external field. Notably, the series expansion of \req{eq:aperp} does not accurately represent the late-time magnetic field in QGP during heavy-ion collisions\index{heavy-ion!collisions}. This is because the infinite series of \reqs{eq:phi}{eq:aperp} must be performed to capture the pole structure of the field. 

Electromagnetic fields in a polarizable medium are often described using the electric displacement field $\mathbf{D}$, the magnetic fields $\mathbf{H}$, the polarization $\mathbf{P}$, and the magnetization $\mathbf{M}$. This formulation is only useful when the field or the medium's response is static or time-dependent. When introducing spatial and temporal dispersion, these definitions are no longer unique \cite{melrose2008quantum}. For instance, if the magnetization depends on space and time $\mathbf{M}(t,x)$ the time dependence of the magnetic field generated will lead to electric fields through Faraday's Law leading to ambiguity since the displacement field no longer depends on just polarization field $\mathbf{P}$.

\subsection{General properties of EM fields in a plasma}
In the case of an infinite homogeneous plasma, its properties are completely described by two independent polarization functions $\Pi_\parallel(k)$ and $\Pi_\perp(k)$. In the framework presented here, the properties of these scalar functions are imparted on the electromagnetic fields via the poles in the Fourier transform of the propagators in \reqs{eq:phi}{eq:aperp}. After contour integration, one effectively gets a sum of different electromagnetic fields at each pole, the amplitude of which depends on the residue of the pole and a spacetime dependence, leading to growth attenuation or propagation depending on the pole's location. An example of this process in done in \cite{Grayson:2022asf}, where we Fourier transform the magnetic field in the center of heavy-ion collisions. 

\para{Dispersion relation}
We can find the poles of the propagator or, equivalently, the zeros of the dispersion relation by inverting Maxwell's equations
\begin{equation}
    -ik_{\mu}\widetilde{F}^{\mu \nu} = \mu_0( \widetilde{j}_{\mathrm{ind}}^{\nu}+\widetilde{j}_{\mathrm{ext}}^{\nu})\,.
\end{equation}
Including the induced current on the left-hand side of the equation and writing the expression in terms of $A^{\mu}$ one finds
\begin{equation}
    (k^2g^{\mu \nu} - k^{\mu} k^{\nu} + \mu_0\Pi^{\mu \nu})\widetilde{A}_{\nu} = - \mu_0\widetilde{j}_{\mathrm{ext}}^{\nu} \,.
\end{equation}
The propagator $D^\mu_\nu(k)$ is obtained by inverting the previous equation
\begin{equation}
    \widetilde{A}_{\nu}(k) = -D^{\mu}_{\nu}(k) \,\widetilde{j}_{\mathrm{ext}}^{\nu}(k) \,.
\end{equation}
The poles of $D^\mu_\nu(k)$ are given by the dispersion equation~\cite{melrose2008quantum}:
\begin{equation}\label{eq:disp}
 \frac{1}{(k\cdot u)^2}\left[(k\cdot u)^2+ \mu_0\Pi_\parallel(k)\right]\left[k^2 + \mu_0 \Pi_\perp(k)\right]^2=0 \,.
\end{equation}
The transverse mode has duplicate solutions as it describes modes in a plane perpendicular to $\boldsymbol{k}$.

The dispersion \req{eq:disp} can be solved for numerous choices of variables describing the modes, such as frequency, phase velocity, or wavevector. We chose to solve for the modes of the plasma in terms of frequency $\omega_m (\mathbf{k})$, which can be thought of as a quasi-particle $m$ with energy $\omega$ and momentum $\mathbf{k}$ analogous to the usual momentum energy relation 
\begin{equation}
    E^2 = \boldsymbol{p}^2 +m^2\, ,
\end{equation}
with $c=1$. This is not always the best choice for simplifying the solutions of \req{eq:disp}, but these modes are often the easiest to interpret. A study of the modes for the general polarization tensor is not the most informative process unless one is looking for general behavior, which can be found in most plasma physics textbooks. Usually, in looking at these modes $\omega_m(\mathbf{k})$, one must first assume the external field's shape or some flow distribution in the plasma by specifying the equilibrium momentum distribution to yield interesting effects in the modes such as plasma instabilities.

When the plasma is perturbed in time in a way that doesn't depend on space, such as for a plane wave, one can take $k \to 0$ for both the transverse and longitudinal roots of the dispersion relation, which reduces the frequency of plasma oscillations \cite{Formanek:2021blc,Grayson:2022asf}
\begin{equation}\label{plasmafreq}
    \omega_{\pm} = -\frac{i\kappa}{2} \pm \sqrt{\omega_p^2 - \frac{\kappa}{4}^2}\,,
\end{equation}
the plasma frequency $\omega_p$ is explicitly given in the ultrarelativistic and nonrelativistic limits, respectively, by \cite{Formanek:2021blc}:
\begin{equation}
\omega_p^2 = \frac{1}{3} m_D^2 \quad (\mathrm{UR})\,, \qquad \omega_p^2 = m_L^2 \quad (\mathrm{NR}) \,,
\end{equation}
with
\begin{equation}
    m_D^2 = \frac{e^2 T}{3}\,.
\end{equation}
The Debye screening mass $m_D$ describes the strength of polarization in the plasma. The plasma frequency $\omega_p$ is the characteristic response frequency of the plasma. For an external field, which is an oscillatory wave of the form $E=E_0e^{-i\omega t}$, one would find that the response is weakly-damped or over-damped depending on the size of $\kappa$ according to \req{eq:plasmafreq}. Waves are weakly damped for $\kappa \ll \omega_p$, and since the square root is imaginary for $\kappa > 2\omega_p$, waves become over-damped. These general statements are subject to the spacetime dependence of the external perturbation. For instance, if a particle moves through the plasma at a constant velocity, the field will not experience much damping if the velocity is much less than the speed of sound in the plasma.

In the static limit $\omega \rightarrow 0$ the zeros in the longitudinal dispersion relation take on the form
\begin{equation}
    |\mathbf{k}| = \pm  i m_D \,.
\end{equation}
Fourier transforming using the positive root in \req{eq:phi} gives the Debye-H{\"u}ckel screening of a stationary charge within the plasma \cite{Debye_1923a,Debye_1923b}
\begin{equation}
    \phi(r) = \frac{Z \alpha \hbar c \, e^{-r/\lambda_D}}{r}\,, \quad \text{with} \quad  \lambda_D = \frac{m_D}{\hbar c}\,.
\end{equation}
The Debye length $\lambda_D$ describes the size of the polarization cloud around a charge generated by the plasma.

\para{Permittivity, susceptibility, and conductivity}
In most fields of applied physics, the effects of a polarizable medium on electromagnetic fields are not described by the polarization functions $\Pi_\parallel$ and $\Pi_\perp$. It is instructive to connect these quantities to more commonplace definitions such as relative permittivity $\epsilon$, susceptibility $\chi$, and conductivity $\sigma$.

The dielectric and susceptibility tensors are related to the spatial portion of the polarization tensor $\Pi^i_j$ ~\cite{Starke:2014tfa,melrose2008quantum},
\begin{equation}\label{dielten}
     \boldsymbol{K}^i_j(\omega,\boldsymbol{k}) = \boldsymbol{\varepsilon}^i_j/\varepsilon_0 = 1+\frac{\boldsymbol{\Pi}^i_j(\omega,\boldsymbol{k})}{\omega^2} = 1+\boldsymbol{\chi}^i_j(\omega,\boldsymbol{k})\,.
\end{equation}
When we project on the axis $\mu =3$, the spatial portion of the polarization tensor is
 \begin{equation}
    \boldsymbol{\Pi}^{i}_{j}(\omega,\boldsymbol{k}) = \left[
    \begin{array}{ccc}
  \Pi_{\perp} & 0 & 0 \\
  0 & \Pi_{\perp} & 0 \\
  0& 0 & \Pi_{\parallel} \\ 
\end{array}
\right]\,.
\end{equation}
It is then natural to discuss transverse and longitudinal susceptibilities,
\begin{equation}\label{eq:chi}
    \chi_\parallel(\omega,\boldsymbol{k}) =\frac{\Pi_\parallel(\omega,\boldsymbol{k})}{\omega^2}, \quad \text{and} \quad \chi_\perp(\omega,\boldsymbol{k}) = \frac{\Pi_\perp(\omega,\boldsymbol{k})}{\omega^2}\,,
\end{equation}
and their associated permeability $K_\parallel$ and $K_\perp$. These quantities are useful for studying the attenuation of electromagnetic fields by looking at light absorption.

The conductivity tensor is found by taking the spatial part of the linear response equation \req{eq:ohm} and expressing the vector potential in terms of the electric field $i \omega \widetilde{A^i} = \widetilde{E^i}$~\cite{Starke:2014tfa,melrose2008quantum}
\begin{align}\label{eq:sigmaperp}
    \sigma_\perp(\omega,\boldsymbol{k}) &\equiv - i \omega \chi_\perp(\omega,\boldsymbol{k}) =- i \frac{\Pi_\perp(\omega,\boldsymbol{k})}{\omega} \,,\\
    \sigma_\parallel(\omega,\boldsymbol{k}) &\equiv - i \omega \chi_\parallel(\omega,\boldsymbol{k}) =- i \frac{\Pi_\parallel(\omega,\boldsymbol{k})}{\omega} \,.
\end{align} 
In the long wavelength limit $k \to 0$ the conductivity \req{eq:sigmaperp} reduces to \req{eq:drude}, the Drude model of conductivity~\cite{Drude:1900}.
The Drude model is equivalent to solving the Vlasov-Boltzmann equation\index{Vlasov-Boltzmann equation}  using the Anderson-Witting collision term \req{eq:lincoll} and neglecting spatial dispersion.

These quantities and results are detailed and plotted in \cite{Formanek:2021blc}. While these quantities are useful for communicating the physics of plasma response, the limits of these quantities must be taken carefully to retain the causal properties of the field. Specifically, tacitly expanding these quantities in either $\omega$ and $\mathbf{k}$ and then inserting them into the self-consistent potentials \reqs{eq:phi}{eq:aperp} will not necessarily generate causal solutions. Instead of carefully expanding and taking limits of these quantities to ensure analytic behavior, it's often easier to expand the electromagnetic fields within their Fourier transforms as in Appendix B of \cite{Grayson:2022asf}.

\section{Charged Leptons and Neutrons before BBN} 
\subsection{Timeline for charged leptons in the primordial Universe}\label{Electron}
Charged leptons\index{lepton} $\tau^\pm,\mu^\pm,e^\pm$ played significant roles in the dynamics and evolution of the primordial Universe. They were kept in equilibrium via electromagnetic and weak interactions. In this chapter, we examine a dynamical model of the abundance of charged leptons $\mu^\pm$ and $e^\pm$ in the primordial Universe. Of particular interest in this work is the dense electron-positron plasma present during the primordial Universe evolution. We study the damping rate and the magnetization\index{magnetization} process in this dense $e^\pm$ plasma in the primordial Universe.

We comment briefly on the case of $\tau^\pm$-particle which is different as their mass $m_\tau=1776.86$\,MeV is above a threshold allowing the $\tau^\pm$-particle to decay into hadrons in about 2/3 of their decays mediated by the W-gauge boson; the vacuum lifespan for $\tau^\pm$-particle is~\cite{ParticleDataGroup:2022pth}
\begin{align}
&\tau_{\tau}=(290.3\pm0.5)\times10^{-15}\,\mathrm{sec}\,.
\end{align}
$\tau^\pm$-particle disappears at a temperature $T\simeq 75$-$50\MeV$ from the Universe inventory via multi-particle decay processes. This heavy lepton provides an interesting bridge between EM particles, and hadrons, near to QGP hadronization. While the full understanding of $\tau^\pm$-particle dynamics in the primordial Universe would be an interesting topic of future research it is not relevant to the regime we are considering here, with temperature $T<10\MeV$. 

On the other hand, the understanding the $\mu^\pm$-lepton abundance is required for the understanding of several fundamental questions concerning properties of the primordial Universe after emergence of residual baryon asymmetry\index{baryon!asymmetry} below $T=38$\,MeV. Muons play also an important role in the dynamics of the ensuing disappearance of strangeness flavor in the primordial Universe. We recall that the strangeness decay often proceeds into muons, energy thresholds permitting; the charged kaons K$^\pm$ have a 63\% branching into $\mu+\bar \nu_\mu$ decay channel. 

The disappearance of muons\index{muon} has therefore direct impact on strangeness flavor inventory in the primordial Universe. Muons are relatively strongly connected to charged pions through the decay and production reaction 
\begin{align}
&\pi^\pm\leftrightarrow \mu^\pm+\nu_\mu\,.
\end{align}
The decay process is nearly exclusive. The back reaction remains active down to relatively low temperature of a few MeV, as long as muons remain in the primordial Universe thermal population inventory. We conclude that if and when muons fall out of their thermal abundance equilibrium this would directly impact the detailed balance back-reaction processes involving strangeness. 

The lightest charged leptons $e^\pm$ can persist via the reaction $\gamma\gamma\to e^-e^+$ until below $T\simeq 20.3$\,keV any remaining positron rapidly disappears through annihilation, leaving only residual electrons required to maintain the primordial Universe's charge neutrality\index{charge neutrality} considering the baryon\index{baryon} (proton) abundance. The long lasting existence of an electron-positron plasma down to temperature range just above $T=20$\,keV plays a pivotal role in several aspects of the primordial Universe: 

1. The primordial electron-positron plasma\index{plasma!electron-positron} has not received the appropriate attention in the context of precision BBN studies\index{Big-Bang!BBN}. However, the presence of dense $e\bar e$-pair plasma before and during BBN has been recognized already a decade ago by Wang, Bertulani and Balantekin~\cite{Wang:2010px}. The primordial synthesis of light elements is found~\cite{Pitrou:2018cgg} to typically takes place in the temperature range $86\,\mathrm{keV}>T_{BBN}>50\,\mathrm{keV}$. Within this temperature range we show below presence of millions of electron-positron pairs per every charged nucleon and plasma densities which reach millions of times normal atomic particle density~\cite{Yang:2024ret,Grayson:2023flr}. Given that the BBN nucleosynthesis processes occur in an electron-positron-rich plasma environment we explore in this work the effect of modifications in the nuclear repulsive Coulomb potential due to the in plasma screening effects on BBN nuclear reactions~\cite{Grayson:2024okq,Grayson:2024uwg}. 

2. The Universe today is filled with magnetic fields at various scales and strengths, both within galaxies, and in deep extra-galactic space. The origin of these magnetic fields\index{magnetic!fields} is currently unknown. In the primordial Universe, above temperature $T>20$\,keV, we have a dense nonrelativistic $e^\pm$ plasma which could prove to be primordial origin of cosmic magnetism as we describe below~\cite{Steinmetz:2023ucp,Rafelski:2023emw,Steinmetz:2023nsc} and \rsec{sec:mag:universe}. We will show that beyond electric currents the magnetic moments of electrons can contribute to spin based magnetization process.

Understanding the abundances of $\mu^+\mu^-$ and $e^+e^-$-pair plasma provides essential insights into the evolution of the primordial Universe. In the following we discuss the muon density down to their persistence temperature and explore the electron/positron plasma properties, including the QED plasma damping rate and damped dynamic screening in \rsec{section:electron}.

\para{Muon pairs in the primordial Universe}
Our interest in strangeness flavor freeze-out in the primordial Universe requires the understanding of the abundance of muons in the primordial Universe. The specific question needing an answer is at which temperature muons remain in abundance (chemical) equilibrium established predominantly by electromagnetic and weak interaction processes, allowing diverse detailed-balance back-reactions to influence the primordial strangeness abundance.

In the primordial Universe in the cosmic plasma muons of mass $m_\mu=105.66$\,MeV can be produced by the following interaction processes~\cite{Yang:2024ret,Rafelski:2021aey}\index{muon!production}
\begin{align} 
\gamma+\gamma&\longrightarrow\mu^++\mu^-,\qquad &e^++e^-\longrightarrow \mu^++\mu^-\;,\\
\pi^-&\longrightarrow\mu^-+\bar{\nu}_\mu,\qquad &\pi^+ \longrightarrow\mu^++\nu_\mu\;.
\end{align}
The back reactions for all above processes are in detailed balance, provided all particles shown on the right-hand side (RHS) exist in chemical abundance equilibrium in the primordial Universe. We recall the empty space (no plasma) at rest lifetime of charged pions $\tau_\pi=2.6033\times 10^{-8}$\,s. We note that neutral pions decay much faster $\tau_{\pi^0}=8.43\times 10^{-17}$\,s.\index{pion!decay}

Any of the produced muons can decay via the well known reactions
\begin{equation}
\mu^-\rightarrow\nu_\mu+e^-+\bar{\nu}_e,\qquad \mu^+\rightarrow\bar{\nu}_\mu+e^++\nu_e\,,
\end{equation} 
with the empty space (no plasma) at rest lifetime $\tau_{\mu}=2.197 \times 10^{-6}\,\mathrm{s}$. 
 
The temperature range of our interests is the Universe when $m_\mu\gg T$. In this case the Boltzmann approximation\index{Boltzmann!approximation} is appropriate for studying massive particles such as muons and pions. The thermal decay rate per volume and time for muons $\mu^\pm$ (and pions $\pi^\pm$) in the Boltzmann limit are given by~\cite{Kuznetsova:2010pi}:\index{muon!decay rate}
\begin{align}
&R_\mu=\frac{g_\mu}{2\pi^2}\left(\frac{T^3}{\tau_\mu}\right)\left(\frac{m_\mu}{T}\right)^2K_1(m_\mu/T)\;,\\
&R_\pi=\frac{g_\pi}{2\pi^2}\left(\frac{T^3}{\tau_\pi}\right)\left(\frac{m_\pi}{T}\right)^2K_1(m_\pi/T)\;, 
\end{align}
where the lifespan of $\mu^\pm$ and $\pi^\pm$ in the vacuum were given above. This rate accounts for both the density of particles in chemical abundance equilibrium and the effect of time dilation present when particles are in thermal motion with respect to observer at rest in the local reference frame. The quantum effects of Fermi blocking or boson stimulated emission have been neglected using Boltzmann statistics.

\para{Muon production processes}
The thermal averaged reaction rate per volume for the reaction $a\overline{a}\rightarrow b\overline{b}$ in Boltzmann approximation is given by~\cite{Letessier:2002ony}\index{muon!production rate}
\begin{align}\label{pairR}
R_{a\overline{a}\rightarrow b\overline{b}}=\frac{g_ag_{\overline{a}}}{1+I}\frac{T}{32\pi^4}\int_{s_{th}}^\infty ds\frac{s(s-4m^2_a)}{\sqrt{s}}\sigma_{a\overline{a}\rightarrow b\overline{b}}~K_1(\sqrt{s}/T),
\end{align}
where $s_{th}$ is the threshold energy for the reaction, $\sigma_{a\overline{a}\rightarrow b\overline{b}}$ is the cross-section for the given reaction, and $K_1$ is the modified
Bessel\index{Bessel function} function of integer order ``$1$". We introduce the factor $1/1+I$ to avoid the double counting of indistinguishable pairs of particles; we have $I=1$ for an identical pair and $I=0$ for a distinguishable pair.

The leading order invariant matrix elements for the reactions $e^++e^-\to\mu^++\mu^-$ and $\gamma+\gamma\to\mu^++\mu^-$, are introduced in this work by~\cite{Kuznetsova:2008jt}
\begin{align}\label{Mee}
|M_{e\bar e\to\mu\bar\mu}|^2=&32\pi^2\alpha^2\frac{(m_\mu^2-t)^2+(m_\mu^2-u)^2+2m_\mu^2s}{s^2},\quad m_\mu\gg m_e\;,\\[0.2cm]
\label{Mgg}
|M_{\gamma\gamma\to\mu\bar\mu}|^2=&32\pi^2\alpha^2\left[\left(\frac{m_\mu^2-u}{m_\mu^2-t}+\frac{m_\mu^2-t}{m_\mu^2-u}\right)+4\left(\frac{m_\mu^2}{m_\mu^2-t}+\frac{m_\mu^2}{m^2_\mu-u}\right) 
-4\left(\frac{m_\mu^2}{m^2_\mu-t}+\frac{m^2_\mu}{m^2_\mu-u}\right)^{\!\!2}\;\right]\;,
\end{align}
 where $s, t, u$ are the Mandelstam variables\index{Mandelstam variables}. The cross-section required in \req{pairR} can be obtained by integrating the matrix elements \req{Mee} and \req{Mgg} over the Mandelstam variable $t$~\cite{Kuznetsova:2010pi}. We have
\begin{align}
\sigma_{e\bar e\to\mu\bar\mu} 
&=\frac{64\pi\alpha^2}{48\pi}\left(\frac{1+2m^2_\mu/s}{s-4m_e^2}\right)\beta\,,\qquad 
\beta =\sqrt{1-\frac{4m^2_\mu}{s}}\, ,\\
\sigma_{\gamma\gamma\to\mu\bar\mu}
 &=\frac{\pi}{2}\left(\frac{\alpha}{m_\mu}\right)^2(1-\beta^2)\left[(3-\beta^4)\ln\frac{1+\beta}{1-\beta}-2\beta(2-\beta^2)\right] \,.
\end{align}
Substituting the cross-sections into \req{pairR} we obtain the production rates for $e\bar e\to\mu\bar\mu$ and $\gamma\gamma\to\mu\bar\mu$, respectively.

As the temperature decreases in the expanding Universe, the initially dominant production rates ($e\bar e,\gamma\gamma\to\mu\bar\mu$) decrease with decreasing temperature, and eventually cross the $\mu^\pm$ decay rates. The muon abundance disappears as soon as any known decay rate is faster than the fastest production rate. In \rf{MuonRatenew:fig} we show the invariant thermal reaction rates per volume and time for rates of relevance, as a function of temperature $T$. It is important to first note that the pion decay rate (short-long dashed, black) is smaller compared to the other rates in the domain of temperatures we are interested. The weak decay rate of the muon is shown by dotted line (green)\index{muon!decay}. The dominant reactions for $\mu^\pm$ production are ${\gamma+\gamma\to\mu^++\mu^-}$ (dashed, blue) and $e^++e^-\to\mu^++\mu^-$ (dashed, red), the solid line (violet) shows the sum of both rates.

\begin{figure}
\centerline{\includegraphics[width=0.8\linewidth]{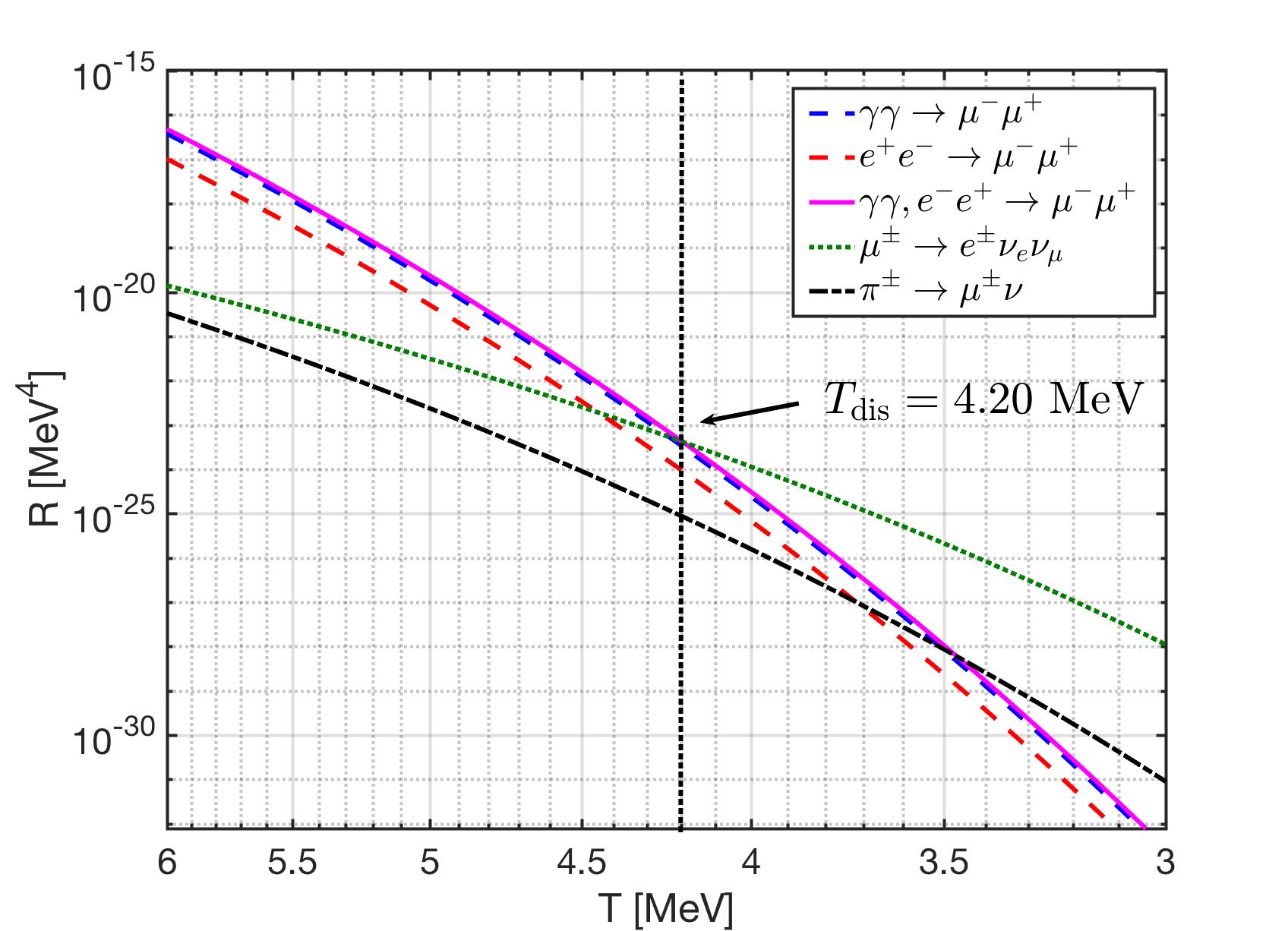}}
\caption{The thermal reaction rate per time and volume for different muon reactions as a function of temperature. See text.\cccite{Rafelski:2023emw}. \radapt{Yang:2024ret,Rafelski:2021aey}}
\label{MuonRatenew:fig} 
\end{figure}

The total $\mu^\pm$ production rate crosses the decay rate at temperature $T_{dissapear}\approx 4.195$\,MeV. Due to the relatively slow expansion of the Universe, the disappearance of muons\index{muon!disappearance} is sudden, and the abundance of muons vanishes as soon as a fast microscopic decay rate surpasses the total production rate. We see that irrespective of charged pion abundance, muons persist until the Universe cools below the temperature $T_\mathrm{disappear}=4.195$\,MeV, below that temperature the dominant reaction is the muon decay.

\para{Comparison of muon and baryon abundance}
It is of interest to explore and to compare the muon inventory in the Universe with the baryon number inventory. Considering the number density for nonrelativistic $\mu^\pm$ in the Boltzmann approximation, we obtain
\begin{align}\label{nmupm}
n_{\mu^\pm}=\frac{g_{\mu^\pm}}{2\pi^2}T^3\left(\frac{m_\mu}{T}\right)^2 K_2(m_\mu/T)=g_{\mu^\pm}\left(\frac{m_\mu T}{2\pi}\right)^{3/2}e^{-{m_\mu}/{T}}\;. 
\end{align}
The ratio of the number density between $n_{\mu^\pm}$ and baryons\index{baryon} $n_B$ can be written as follows
\begin{align}
\frac{n_{\mu^\pm}}{n_\mathrm{B}}=\frac{n_{\mu^\pm}}{\sigma}\frac{\sigma}{n_\mathrm{B}}=
\frac{n_{\mu^\pm}}{\sigma}\left[\frac{\sigma}{n_\mathrm{B}}\right]_{t_0},
\end{align}
where we assume that $\sigma/n_\mathrm{B}$ the ratio of entropy to baryon number remains constant and $t_0$ represent present day value. The present value is given by $(n_B/\sigma)_{t_0}\approx8.69\times10^{-11}$\index{baryon!entropy ratio}. We recall, see \rf{EntropyDOF:Fig}, that the entropy density\index{entropy!density} $\sigma$ can be characterized introducing $g^s_\ast$, the total number of \lq entropic\rq\ degrees of freedom, see \req{eq:entg}. For temperature $10\,\mathrm{MeV} >T>3 $\,MeV, the massless photons, nearly relativistic electrons and positrons, and practically massless neutrinos contribute to the count of degrees of freedom $g^s_\ast$. In this case, the number density between $n_{\mu^\pm}$ and baryon $n_B$ in the temperature interval we consider $10\,\mathrm{MeV} >T>3 $\,MeV is given by
\begin{align}\label{nmuperbF} 
\frac{n_{\mu^\pm}}{n_\mathrm{B}}=\frac{45}{2\pi^2}\frac{g_{\mu^\pm}}{g^s_\ast}\left(\frac{m_\mu}{2\pi T}\right)^{3/2}e^{-{m_\mu}/{T}}\;\left(\frac{\sigma}{n_\mathrm{B}}\right)_{\!t_0}.
\end{align}

In \rf{fig:DensityRatio} we show the muon to baryon density ratio \req{nmuperbF} as a function of $T$\index{muon!to baryon ratio}. We see that the very small muon pair abundance at $T=10$\,MeV exceeds that of residual baryons by a large factor 500,000, while at muon disappearance temperature $n_{\mu^\pm}/n_\mathrm{B}(T_\mathrm{disappear})\approx0.911$. The number density $n_{\mu^\pm}$ and $n_\mathrm{B}$ abundances are equal at the temperature $T_\mathrm{equal}\approx4.212\,\mathrm{MeV} > T_\mathrm{disappear}$. 

\begin{figure}
\centerline{\includegraphics[width=0.8\linewidth]{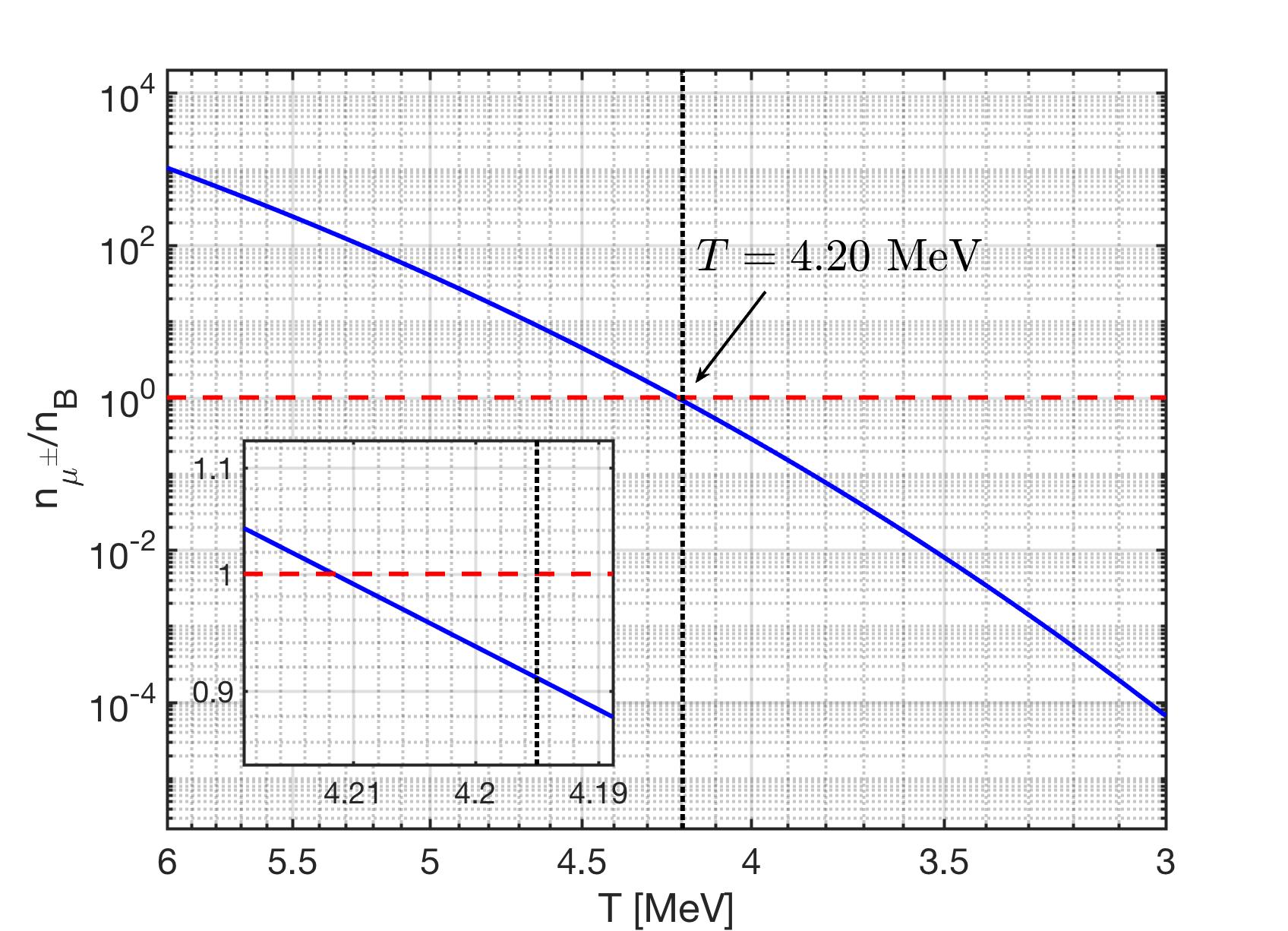}}
\caption{The density ratio between $\mu^\pm$ and baryons as a function of temperature. The density ratio at muon disappearance temperature is $n_{\mu^\pm}/n_\mathrm{B}(T_\mathrm{disappear})\approx0.911$, and around the temperature $T\approx4.212$\,MeV the computed density ratio $n_{\mu^\pm}/n_\mathrm{B}\approx1$. \cccite{Rafelski:2023emw}. \radapt{Yang:2024ret,Rafelski:2021aey}}
\label{fig:DensityRatio}
\end{figure}

\index{muon!baryon ratio}
The primary insight is that aside of protons, neutrons and other nonrelativistic particles, both positively and negatively charged muons $\mu^\pm$ are present in thermal equilibrium and in non-negligible abundance exceeding baryon abundance down to $T>T_\mathrm{dissapear}\approx 4.195$\,MeV. This coincidence, muon and baryon abundance being practically equal at temperature of muon disappearance, surprised us. We do not know if there is a physics content in this observation. 

\subsection{Cosmic electron-positron plasma and BBN}
\label{section:electron}
Following on the neutrino freeze-out at $T\approx 2$\,MeV, the Universe is dominated by the electron-positron-photon QED plasma. In this section, we derive the electron-positron density and chemical potential\index{chemical potential!charge neutrality} required for local charge neutrality\index{Universe!charge neutrality} of the Universe to show that during the normal BBN\index{Big-Bang!BBN} temperature range $86.7\,\mathrm{keV}>\mathrm{T_{BBN}}>50\,\mathrm{keV}$~\cite{Pitrou:2018cgg} the Universe was filled with a dense electron-positron pair-plasma dotted with a dispersed baryonic matter dust. 

We examine the microscopic collision properties of the electron-positron plasma in the primordial Universe allowing us to use appropriately generalized methods of plasma physics in a study of the role of the $e^+e^-$ plasma in the primordial Universe. The time scale of Universe expansion $H^{-1}$ is orders of magnitude larger than the microscopic reaction time scales of interest for all processes we consider, the dynamical processes we consider are thus occurring in expanding, but stationary Universe\index{Universe!stationray}.

\para{Electron chemical potential and number density}
We obtain the dependence of electron chemical potential\index{chemical potential!electron}, and hence $e^+e^-$ density, as a function of the photon background temperature $T$ by employing the following physical principles
\begin{enumerate}
\item Charge neutrality of the Universe:
\begin{align}\label{neutrality}
n_{e^-}-n_{{e^+}}=n_p-n_{\overline{p}}\approx\,n_p,
\end{align}
where $n_\ell$ denotes the number density of particle type $\ell$.
\item Neutrinos decouple (freeze-out) at a temperature $T_f\simeq 2$\,MeV, after which they free stream through the Universe with an effective temperature~\cite{Birrell:2012gg}
\begin{align}
 T_\nu(t)=T_f\,\frac{a(t_f)}{a(t)},
\end{align}
where $a(t)$ is the Friedmann-Lema\^{i}tre-Robertson-Walker (FLRW)\index{cosmology!FLRW} Universe scale factor (see cosmology primer \rsec{sec:flrw}) which is a function of cosmic time $t$, and $t_f$ represents the cosmic time when neutrino freezes out.
\item The total comoving entropy is conserved. At $T\leq T_f$, the dominant contributors to entropy are photons, $e^+e^-$, and neutrinos. In addition, after neutrino freeze out, neutrino comoving entropy is independently conserved~\cite{Birrell:2012gg}. This implies that the combined comoving entropy in $e^+e^-\gamma$ is also conserved for $T\leq T_f$.\index{entropy!conservation}
\end{enumerate} 
Motivated by the fact that comoving entropy in $\gamma$, $e^+e^-$ is conserved after neutrino freeze-out, we rewrite the charge neutrality condition, \req{neutrality}, in the form
\begin{align}\label{charge_neutral_cond2}
n_{e^-}-n_{{e^+}}=X_p\frac{n_B}{\sigma_{\gamma,e^\pm}} \sigma_{\gamma,e^\pm},\qquad X_p\equiv\frac{n_p}{n_B},
\end{align}
where $n_B$ is the number density of baryons, $s_{\gamma,e^\pm}$ is the combined entropy density\index{entropy!density} in photons, electrons, and positrons. During the Universe expansion, the comoving entropy and baryon number are conserved quantities; hence the ratio $n_B/\sigma_{\gamma,e^\pm}$ is conserved. We have
\begin{align}
\frac{n_B}{\sigma_{\gamma,e^\pm,}}=\left(\frac{n_B}{\sigma_{\gamma,e^\pm}}\right)_{t_0}\!\!\!\!=\left(\frac{n_B}{\sigma_{\gamma}}\right)_{t_0}\!\!\!\!=\left(\frac{n_B}{n_\gamma}\right)_{t_0}\left(\frac{n_\gamma}{s_{\gamma}}\right)_{t_0},
\end{align}
where the subscript $t_0$ denotes the present day value, and the second equality is obtained by observing that the present day $e^+e^-$-entropy density is negligible compared to the photon entropy density. We can evaluate the ratio introducing the present day baryon-to-photon ratio: $B/N_\gamma =n_B/n_\gamma= 0.605\times10^{-9}$ as obtained from the Cosmic Microwave Background (CMB)~\cite{ParticleDataGroup:2022pth}\index{CMB}, and the entropy per particle for a massless boson: $(s/n)_{\mathrm{boson}}\approx 3.602$.

The total entropy density of photons, electrons, and positrons can be written as
\begin{align}\label{entropy_per_baryon}
\sigma_{\gamma,e^\pm}=\frac{2\pi^2}{45}g_\gamma\,T^3+\frac{\rho_{e^\pm}+P_{e^\pm}}{T}-\frac{\mu_e}{T}(n_{e^-}-n_{{e^+}}),
 \end{align}
where $ \rho_{e^\pm}=\rho_{e^-}+\rho_{e^+}$ and $P_{e^\pm}=P_{e^-}+P_{{e^+}}$ are the total energy density and pressure of electrons and positron respectively.

By incorporating \req{charge_neutral_cond2} and \req{entropy_per_baryon}, the charge neutrality condition can be expressed as
\begin{align}\label{charge_neutral_cond3}
 &\left[1+X_p\left(\frac{n_B}{n_\gamma}\right)_{\!t_0}\!\!\left(\frac{n_\gamma}{\sigma_{\gamma}}\right)_{\!t_0}\!\!\frac{\mu_e}{T}\right]\frac{n_{e^-}-n_{{e^+}}}{T^3}=X_p\left(\frac{n_B}{n_\gamma}\right)_{\!t_0}\!\!\left(\frac{n_\gamma}{\sigma_{\gamma}}\right)_{\!t_0}\!\!\left(\frac{2\pi^2}{45}g_\gamma+\frac{\rho_{e^\pm}+P_{e^\pm}}{T^4}\right).
\end{align}

Using the Fermi distribution, the number density of electrons over positrons in the primordial Universe is given by
\begin{align}\label{ee_density}
n_{e^-}-n_{{e^+}}&=\frac{g_e}{2\pi^2}\left[\int_0^\infty\frac{p^2dp}{\exp{\left((E-\mu_e)\right)/T}+1}\right.\left.-\int_0^\infty\frac{p^2dp}{\exp{\left((E+\mu_e)/T\right)}+1}\right]\notag\\[0.3cm]
&=\frac{g_e}{2\pi^2}\,{T^3}\,\tanh(b_e)M_e^3\int_{1}^\infty \!\!\!\!\frac{ \zeta \sqrt{\zeta^2-1} d\zeta}{1+\cosh(M_e\zeta)/\cosh(b_e)},
\end{align}
where we have introduced the dimensionless variables as follows: 
\begin{align}\label{Variables}
\zeta=\frac{E}{m_e},\qquad M_e=\frac{m_e}{T},\qquad b_e=\frac{\mu_e}{T}.
\end{align}
Substituting \req{ee_density} into \req{charge_neutral_cond3} and giving the value of $X_p$, then the charge neutrality condition can be solved to determine $\mu_e/T$ as a function of $M_e$ and $T$.

In \rf{BBN:Electron} (left axis), we show (left axis, brown line) the electron chemical potential\index{chemical potential} as a function of temperature we obtain solving \req{charge_neutral_cond3} numerically employing the following parameters: proton concentration $X_p=0.878$ as derived from observation~\cite{ParticleDataGroup:2022pth} and $n_B/n_\gamma=6.05\times10^{-10}$ from CMB. We can see the value of chemical potential is comparatively small $\mu_e/T\approx10^{-6}\sim10^{-7}$ during the BBN epoch temperature range, implying a very small asymmetry in the number of electrons and positrons in plasma is needed to neutralize proton charge. 

\begin{figure}
\centerline{\includegraphics[width=0.8\linewidth]{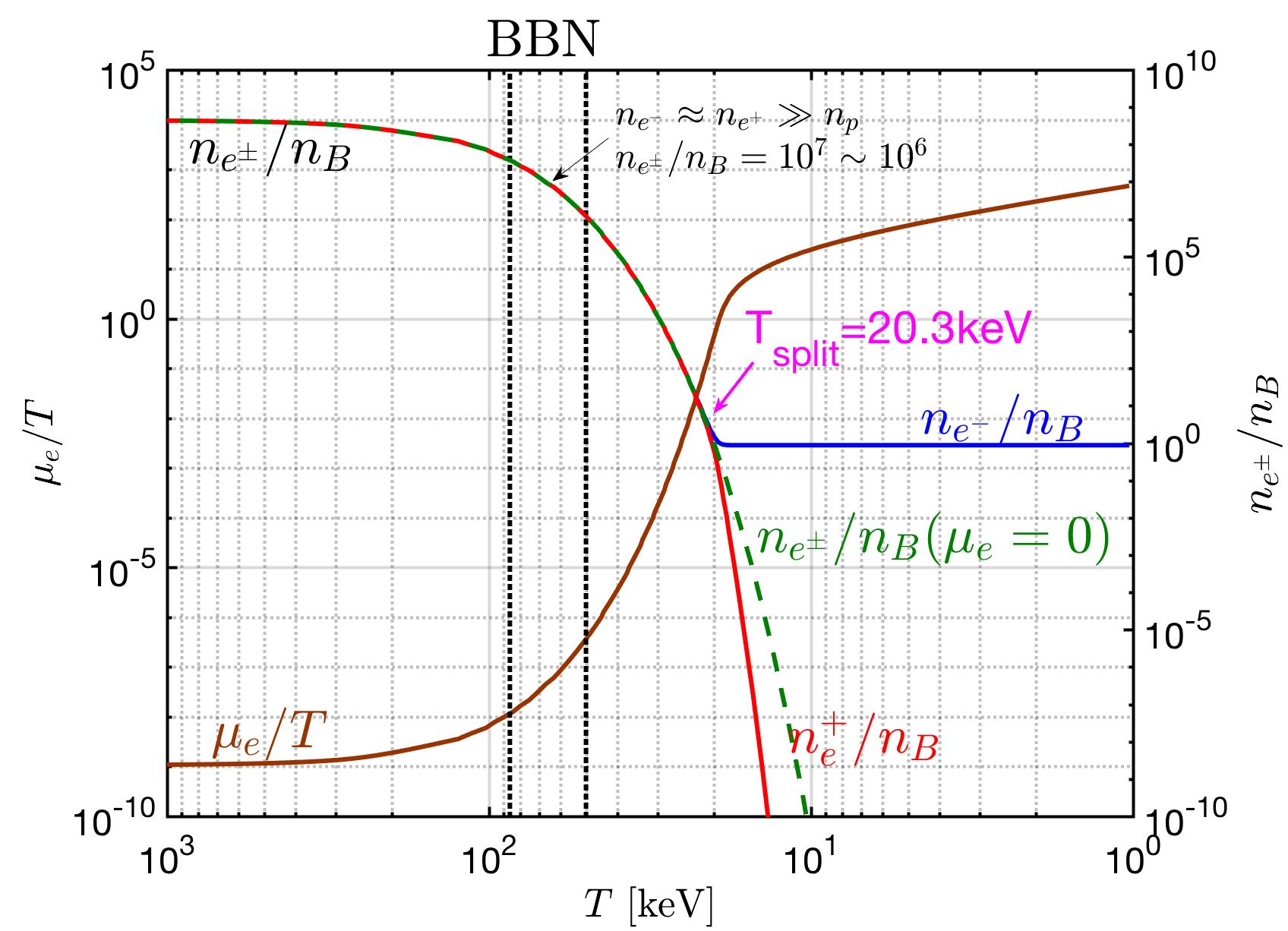}}
\caption{Left axis: The chemical potential of electrons $\mu_e/T$ as a function of temperature (solid (brown) line). Right axis: the ratio of electron (positron) number density to baryon density as a function of temperature. The solid blue line is the electron density, the red line is the positron density, and the green dashed line is obtained setting for comparison $\mu_e=0$. The vertical black dotted lines are bounds of BBN epoch. \cccite{Grayson:2023flr}. \radapt{Yang:2024ret}}
\label{BBN:Electron}
\end{figure}

The ratio of electron (positron) number density to baryon density,right axis in \rf{BBN:Electron}, shows that the Universe was filled with an electron-positron rich plasma during the BBN temperature range epoch in the (approximate) temperature range $86\,\mathrm{keV}>\mathrm{T_{BBN}}>50\,\mathrm{keV}$. When the temperature is \eg\ around $T=70\,\mathrm{keV}$, the density of electrons and positrons is comparatively large $n_{e^\pm}\approx10^7\,n_B$. At $90$\,keV, the electron and positron density is near the solar core density, compare Fig.~19 in Ref.~\cite{Rafelski:2023emw}. Near and below the temperature $T=20.3\,\mathrm{keV}$, the positron density decreases rapidly, transforming the pair-plasma into an electron-baryon plasma.

\para{QED plasma damping rate}
\index{plasma!QED damping}
The reactions of interest for the evaluation of the QED plasma damping are the (inverse) Compton scattering\index{Compton scattering}, the M{\o}ller\index{M{\o}ller scattering} scattering, and the Bhabha\index{Bhabha scattering} scattering, respectively
\begin{align}
e^\pm+\gamma\longrightarrow e^\pm+\gamma,\qquad e^\pm+e^\pm\longrightarrow e^\pm+e^\pm,\qquad e^\pm+e^\mp\longrightarrow e^\pm+e^\mp.
\end{align}
The general formula for thermal reaction rate per volume is discussed in~\cite{Letessier:2002ony} (Eq.(17.16), Chapter 17). For inverse Compton scattering we have
\begin{align}
R_{e^{\pm}\gamma}=\frac{g_eg_\gamma}{16\left(2\pi\right)^5}T\int_{m_e^2}^\infty\!\!\!\!ds\frac{K_1(\sqrt{s}/T)}{\sqrt{s}}\int^0_{-(s-m_e^2)^2/s}\!\!\!\!\!\!dt\, |M_{e^{\pm}\gamma}|^2,
\end{align} 
and for M{\o}ller and Bhabha reactions we have
\begin{align}
&R_{e^\pm e^\pm}=\frac{g_eg_e}{16\left(2\pi\right)^5}T\!\!\int_{4m_e^2}^\infty\!\!\!\!ds\frac{K_1(\sqrt{s}/T)}{\sqrt{s}}\int^0_{-(s-4m_e^2)}\!\!\!\!\!\!dt\,|M_{e^\pm e^\pm}|^2,\\[0.3cm]
&R_{e^\pm e^\mp}=\frac{g_eg_e}{16\left(2\pi\right)^5}T\!\!\int_{4m_e^2}^\infty\!\!\!\!ds\frac{K_1(\sqrt{s}/T)}{\sqrt{s}}\int^0_{-(s-4m_e^2)}\!\!\!\!\!\!dt\,|M_{e^\pm e^\mp}|^2,
\end{align}
where $g_i$ is the degeneracy of particle $i$, $|M|^2$ is the matrix element for a given reaction, $K_1$ is the Bessel\index{Bessel function} function of order $1$, and $s,t,u$ are Mandelstam variables\index{Mandelstam variables}. The leading order matrix element associated with inverse Compton scattering can be expressed in the Mandelstam variables~\cite{Kuznetsova:2011wt,Kuznetsova:2009bq} we have\index{Compton scattering}
\begin{align}
|M_{e^\pm\gamma}|^2\!=32 \pi^2\alpha^2\bigg[&4\left(\frac{m_e^2}{m_e^2-s}+\frac{m_e^2}{m_e^2-u}\right)^2
-\frac{4m_e^2}{m_e^2-s}-\frac{4m_e^2}{m_e^2-u} -
 \frac{m_e^2-u}{m_e^2-s} -\frac{m_e^2-s}{m_e^2-u}\bigg],
\end{align}
and for M{\o}ller\index{M{\o}ller scattering} and Bhabha scattering we have \index{Bhabha scattering}
\begin{align}
|M_{e^{\pm}e^{\pm}}|^{2}\!=64\pi^{2}\alpha^{2}\bigg[&
\frac{s^{2}+u^{2}+8m_e^{2}(t-m_e^{2})}{2(t-m^2_{\gamma})^{2}}
+\frac{{s^{2}+t^{2}}+8m_e^{2}
(u-m_e^{2})}{2(u-m_{\gamma}^2)^{2}} + \frac{\left( {s}-2m_e^{2}\right)\left({s}-6m_e^{2}\right)}
{(t-m_{\gamma}^2)(u-m_{\gamma}^2)} \bigg],
\end{align}
and
\begin{align}
|M_{e^\pm e^\mp}|^{2}=64\pi^{2}\alpha^{2}
\bigg[&\frac{s^{2}+u^{2}+8m_e^{2}(t-m_e^{2})}{2(t-m^2_{\gamma})^{2}}
+\frac{u^{2}+t^{2}+8m_e^{2}
(s-m_e^{2})}{2(s-m^2_{\gamma})^{2}} + \frac{\left({u}-2m_e^{2}\right)\left({u}-6m_e^{2}\right)}
 {(t-m^2_{\gamma})(s-m^2_{\gamma})} \bigg],
\label{M_fi_b}
\end{align}
where we introduce the photon mass $m_\gamma$ to account the plasma effect and avoid singularity in reaction matrix elements. 

The photon mass $m_\gamma$ in plasma is equal to the plasma frequency $\omega_p$, where we have~\cite{Kislinger:1975uy}\index{photon!plasma mass}
\begin{align}
m^2_\gamma=\omega^2_{p}=8\pi\alpha\int\frac{d^3p_e}{(2\pi)^3}\left(1-\frac{p_e^2}{3E_e^2}\right)\frac{f_e+f_{\bar e}}{E_e},
\end{align}
where $E_e=\sqrt{p_e^2+m^2_e}$. In the BBN temperature range $86\,\mathrm{keV}>T_{BBN}>50\,\mathrm{keV}$ we have $m_e\gg T$ and considering the nonrelativistic limit for electron-positron plasma, we obtain
\begin{align}
m^2_\gamma=\frac{4\pi\alpha}{2m_e}\left(\frac{2m_eT}{\pi}\right)^{3/2}e^{-m_e/T}\cosh\left(\frac{\mu_e}{T}\right).
\end{align}
In the BBN temperature range, we have $\mu_e/T\ll1$, which implies the equal number of electrons and positrons in plasma.

To discuss the collisions plasma by the linear response theory, it is convenient to define the average relaxation rate for the electron-positron plasma as follows:\index{plasma!electron-positron}
\begin{align}\label{Kappa}
\kappa=\frac{R_{e^\pm e^\pm}+R_{e^\pm e^\mp}+R_{e^\pm\gamma}}{\sqrt{n_{e^-}n_{e^+}}}\approx\frac{R_{e^\pm e^\pm}+R_{e^\pm e^\mp}}{\sqrt{n_{e^-}n_{e^+}}},
\end{align}
where the density function ${\sqrt{n_{e^-}n_{e^+}}}$ in the Boltzmann limit is given by
\begin{align}
{\sqrt{n_{e^-}n_{e^+}}}=\frac{g_e}{2\pi^3}T^3\left(\frac{m_e}{T}\right)^2K_2(m_e/T).
\end{align}

In \rf{RelaxationRate:fig}, we show the reaction rates for M{\o}ller reaction, Bhabha reaction, and inverse Compton scattering as a function of temperature. For temperatures $T>12.0$\,keV, the dominant reactions in plasma are M{\o}ller and Bhabha scatterings between electrons and positrons. Thus in the BBN temperature range, we can neglect the inverse Compton scattering. The total relaxation rate $\kappa$ (black line) is approximately constant, $\kappa=10\sim12$\,keV, during the BBN. However, at $T<20.3$\,keV the relaxation rate $\kappa$ decreases rapidly because the plasma changes its nature when positrons disappear.

\begin{figure}
\centering
\includegraphics[width=0.66\textwidth]{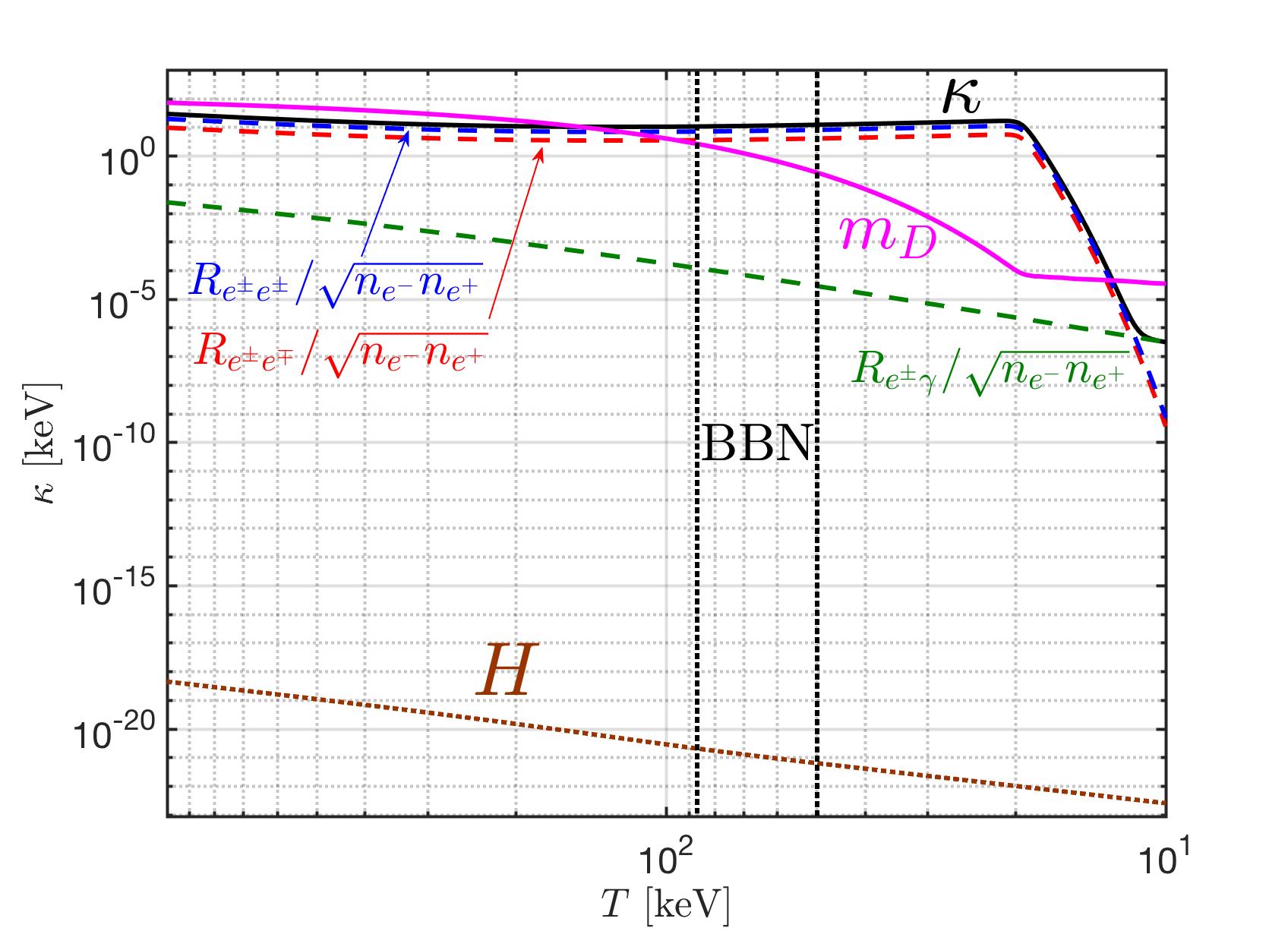}
\caption{The relaxation rate $\kappa$ (black line) as a function of temperature in the nonrelativistic electron-positron plasma, compared to reaction rates for M{\o}ller reaction $e^-+e^-\to e^-+e^-$ (blue dashed line), Bhabha reaction $e^-+e^+\to e^-+e^+$ (red dashed line), and inverse Compton scattering $e^-+\gamma\to e^-+\gamma$ (green dashed line) respectively. The Debye mass $m_D=\omega_{p}\sqrt{m_e/T}$ (purple line) is also shown. \cccite{Grayson:2023flr}. \radapt{Yang:2024ret}}
\label{RelaxationRate:fig}
\end{figure}

\para{Self-consistent damping rate}
In electron-positron plasma, the photon mass\index{photon!plasma mass} $m_\gamma^2$ appears in the transition matrices for M{\o}ller and Bhabha reactions, which is one of important parameters in the calculation of the relaxation rate in $e^\pm$ plasma. When evaluating M{\o}ller and Bhabha scattering, we included as is common practice the temperature-dependent mass of the photon obtained in plasma theory without damping. However, in general, the effective mass of the photon depends at a given temperature on all properties of the QED plasma. 

Considering the linear response theory, the dispersion relation for the photon in nonrelativistic $e^\pm$ plasma is given by~\cite{Formanek:2021blc}
\begin{align}\label{dispersion_damping}
w^2=|k|^2+\frac{w}{w+i\kappa}w_{pl}^2,
\end{align}
where $w_{pl}$ is the plasma frequency and $\kappa$ is the average collision rate of $e^\pm$ plasma. The effective plasma frequency in damped plasma can be solved by considering the case $|k|^2=0$~\cite{Formanek:2021blc}
\begin{align}\label{plasmafrequency_damped}
w_{\pm}=-i\frac{\kappa}{2}\pm\sqrt{w^2_{pl}-\frac{\kappa^2}{4}}.
\end{align}
The result shows that the plasma frequency in damped plasma $w_\pm$ is a function of $\kappa$ which we are computing using the effective plasma photon mass. 

The effective photon mass in damped plasma is also a function of the scattering rate. We have
\begin{align}\label{PhotonMass:self}
m_\gamma=w_\pm(w_{pl},\kappa)=m_\gamma(w_{pl},\kappa),
\end{align}
where the photon mass $m_\gamma=w_+$ for the under-damped plasma $w_{pl}>\kappa/2$, and $m_\gamma=w_-$ for over-damped plasma $w_{pl}<\kappa/2$. \req{PhotonMass:self} shows that computed damping strength $\kappa$ is the dominant scale for collisional plasma and it is also the main parameter determining the photon mass in plasma. 

Substituting the effective photon mass \req{PhotonMass:self} into the definition of the average relaxation rate \req{Kappa}, we obtain a self-consistent equation for damping rate $\kappa$ 
\begin{align}\label{RealaxtionSelf}
\kappa\,\left[\frac{g_e}{2\pi^3}T^3\left(\frac{m_e}{T}\right)^2K_2(m_e/T)\right]=\frac{g_eg_e}{32\pi^4}T\!\! \int_{4m_e^2}^\infty\!\!\!\!ds&
\frac{s(s-4m^2_e)}{\sqrt{s}}K_1(\sqrt{s}/T)
\bigg[\sigma_{e^\pm e^\pm}(s,w_{pl},\kappa)+\sigma_{e^\pm e^\mp}(s,w_{pl},\kappa)\bigg],
\end{align}
where the cross-sections depend on the parameter $w_{pl}$ and $\kappa$, and the variable $\kappa$ appears on both sides of the equation so we need to solve the equation numerically to determine the $\kappa$ value that satisfies this condition.

Depending on the nature of the plasma (overdamped or underdamped plasma), we can establish the photon mass in collision plasma based on two distinct conditions as follows:
\begin{itemize}
\item Case 1. The plasma frequency is larger than the collision rate $w_{pl}>\kappa/2$, we have
\begin{align}
m_\gamma=w_+=-i\frac{\kappa}{2}+\sqrt{w^2_{pl}-\frac{\kappa^2}{4}}.
\end{align}
\item Case 2. The plasma frequency is smaller than the collision rate $w_{pl}<\kappa/2$, we have
\begin{align}\label{PhotonMassPlasma}
m_\gamma=w_-=-i\left(\frac{\kappa}{2}+\sqrt{\frac{\kappa^2}{4}-w^2_{pl}}\right).
\end{align}
\end{itemize}
In \rf{RelaxationRate002:fig} we see that during the BBN epoch $50\leqslant T\leqslant 86$\,keV, the plasma frequency is smaller than the collision rate $w_{pl}<\kappa/2$. In this case, the effective photon mass in collision plasma is given by the overdamped relation \req{PhotonMassPlasma}. This result is valid for $T>20.3$\,keV. For temperature $T<20.3$\,keV, the composition turns into electron and proton plasma, which is beyond our current study because of assumed (for simplicity) equal numbers of electrons and positrons.

\begin{figure} 
\centerline{\includegraphics[width=0.8\linewidth]{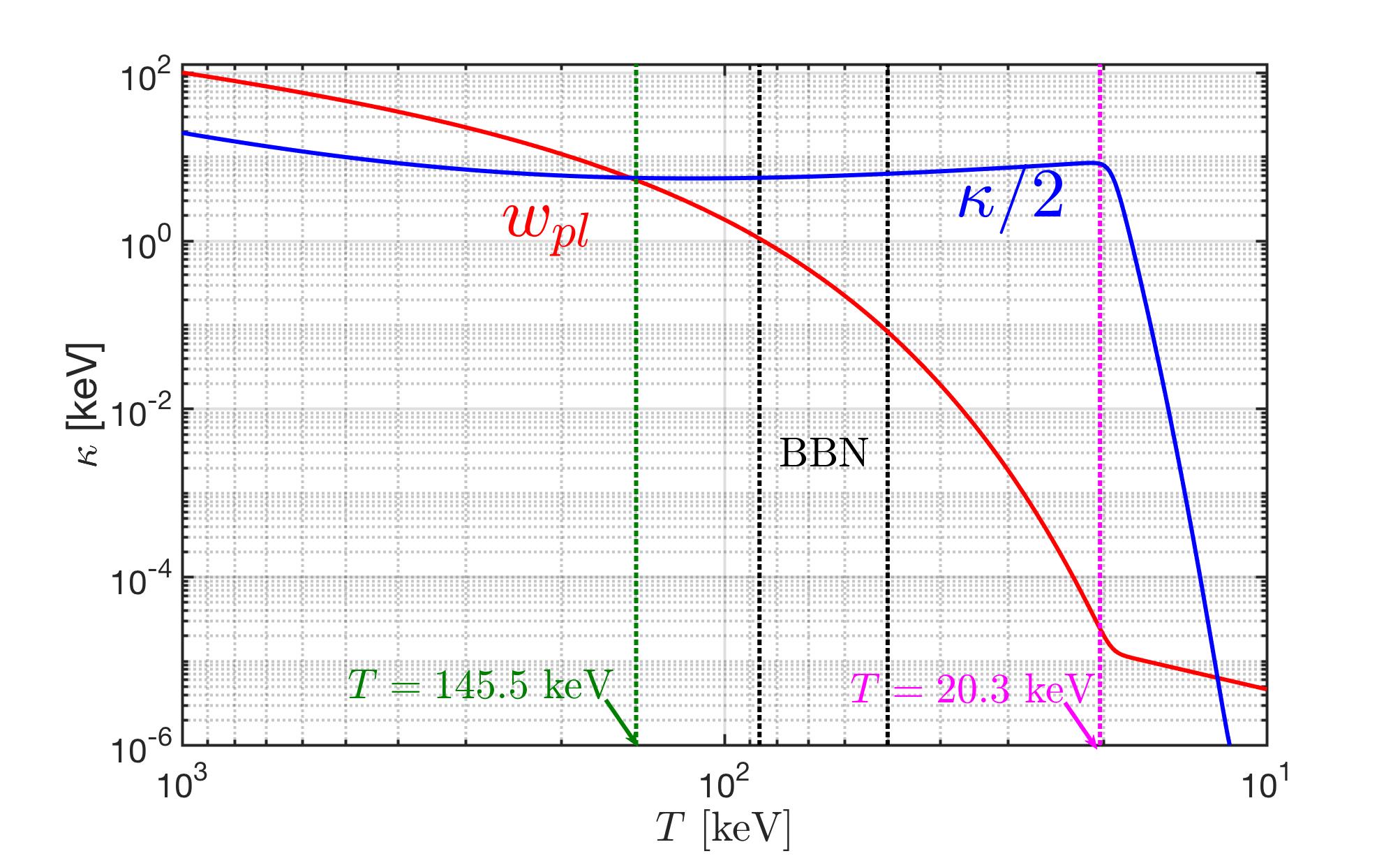}}
\caption{The relaxation rate $\kappa/2$ (blue line) and plasma frequency $\omega_{pl}$ (red line) as a function of temperature in nonrelativistic electron-positron plasma. Vertical green dashed line indicates the boundary between over- and under-damped plasma at $T<145.5$\,keV which is before the BBN epoch (vertical black lines). Temperature domain of validity is above disappearance of positrons (vertical line at 20.3\,keV). \radapt{Yang:2024ret}}
\label{RelaxationRate002:fig} 
\end{figure}

To calculate the effective cross-sections for M{\o}ller\index{M{\o}ller scattering} and Bhabha\index{Bhabha scattering} scattering we need in the overdamped regime to account for the imaginary photon mass in the calculation of reaction matrix elements. This imaginary part of the photon mass accounts for the decay in sense of propagation range of the massive photon in plasma. We now make a first estimate of the effect of self-consistent real part of the photon mass on the damping rate $\kappa$, we leave the photon decay to a future study.

For BBN\index{Big-Bang!BBN} temperature $50\leqslant T\leqslant 86$\,keV,
we have $w_{pl}<\kappa$ and the effective photon mass can be approximated as
\begin{align}
m^2_\gamma=w_-w_-^\ast&=\left(\frac{\kappa}{2}+\sqrt{\frac{\kappa^2}{4}-w^2_{pl}}\right)^2
=\frac{\kappa^2}{2}\left[\left(1-\frac{2w^2_{pl}}{\kappa^2}\right)+\sqrt{1-\frac{4w^2_{pl}}{\kappa^2}}\right]\notag\\
&=\frac{\kappa^2}{2}\left[\left(1-\frac{2w^2_{pl}}{\kappa^2}\right)+\left(1-\frac{2w^2_{pl}}{\kappa^2}+\cdots\right)\right]\approx\kappa^2\,,
\label{PhotonMassPlasma002}
\end{align}
where we consider the limit $w^2_{pl}/\kappa^2\ll 1$ and effective photon mass is equal to the average collision rate in plasma $m^2_\gamma\approx\kappa$.

Substituting the photon mass $m^2_\gamma=\kappa^2$ for overdamped plasma into the relaxation rate of electron-positron \req{RealaxtionSelf}, and introducing the following dimensionless variables
\begin{align}
x=\sqrt{s}/T,\qquad a=m_\gamma/T=\kappa/T,\qquad b=m_e/T,
\end{align}
the relaxation rate of electron-positron can be written as
\begin{align}\label{Numerical_eq}
&\left[\frac{g_e}{2\pi^2}T^4\left(\frac{m_e}{T}\right)^{\!2}\!K_2(m_e/T)\right]\,\left(\frac{\kappa}{T}\right)\notag\\
&\qquad\qquad\qquad=\frac{g^2_e\alpha^2}{8\pi^3}T^4\!\!\int_{2b}^\infty\!dxK_1(x)\left[\mathcal{F}_{e^\pm e^\pm}(x,\kappa/T)+\mathcal{F}_{e^\pm e^\mp}(x,\kappa/T)\right],
\end{align}
where the functions $\mathcal{F}_{e^\pm e^\pm}$ and $\mathcal{F}_{e^\pm e^\mp}$ are given by
\begin{align}
\mathcal{F}_{e^\pm e^\pm}(x,a=\kappa/T)&=\left\{2\left[3a^2+4b^2+\frac{4(b^4-a^4)}{x^2-4b^2+2a^2}\right]\ln\left(\frac{a^2}{x^2-4b^2+a^2}\right)\right.\notag\\
&\left.+\frac{(x^2-4b^2)(8b^4+2a^4+3a^2x^2+2x^4-4b^2(2x^2+a^2))}{a^2(x^2-4b^2+a^2)}\right\}
\end{align}
and 
\begin{align}
\mathcal{F}_{e^\pm e^\mp}(x,a=\kappa/T)&=\left\{\frac{2x^2(a^2+x^2)-4b^4}{x^2-a^2}\ln\left(\frac{a^2}{x^2-4b^2+a^2}\right)\right.\notag\\
&+\frac{(x^2-4b^2)(3x^2+4b^2+2a^2)}{(x^2-a^2)}+\frac{x^6-12b^4x^2-16b^6}{3(x^2-a^2)^2}\notag\\
&\left.+\frac{(x^2-4b^2)(8b^4+2a^4+3a^2x^2+2x^4-4b^2(2x^2+a^2))}{a^2(x^2-4b^2+a^2)}\!\right\}.
\end{align}

We solve \req{Numerical_eq} numerically. In \rf{KappaSol:fig}, we plot the resultant relaxation rate $\kappa$ that satisfies \req{Numerical_eq} as a function of temperature $50\,\mathrm{keV} \leqslant T\leqslant 86$\,keV. The result shows that in the BBN temperature range, the overdamping is considerably reduced: We remember that we started with $w_{pl}<\kappa$, and the effective photon mass $m^2_\gamma=\kappa^2$. Now we obtain a relaxation rate $\kappa=1.832\sim 0.350$\,keV during BBN epoch, which is smaller than the relaxation rate without damping effect on the photon mass, compare \rf{RelaxationRate:fig}, where the relaxation rate $\kappa=10\sim12$\,keV during the BBN epoch is shown.\index{plasma damping}

\begin{figure}
\centering
\includegraphics[width=0.66\linewidth]{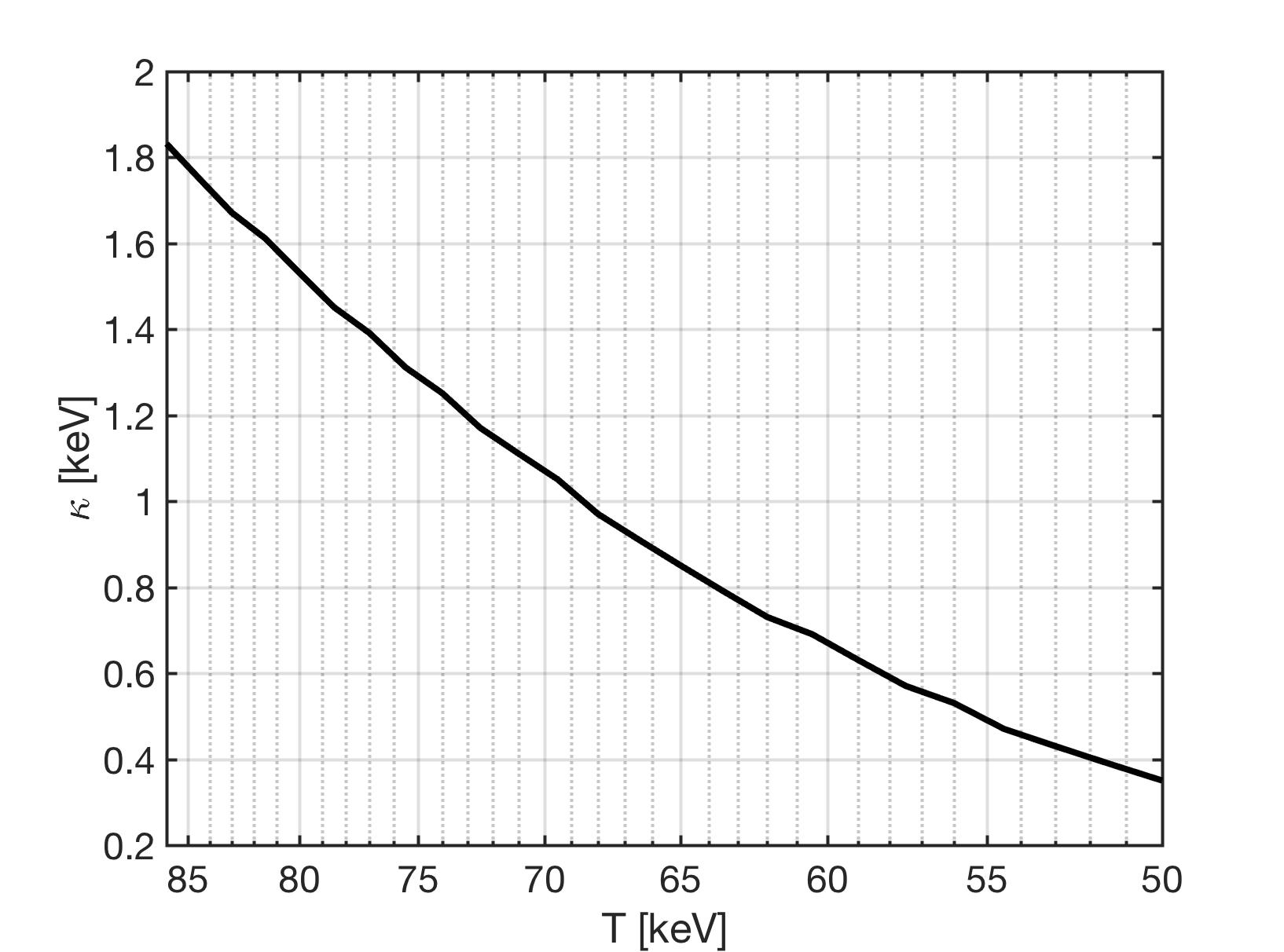}
\caption{The relaxation rate $\kappa$ that satisfies \req{Numerical_eq} self-consistently as a function of temperature $50\leqslant T\leqslant 86$\,keV. The minor fluctuations are due to limited numerical precision. \radapt{Yang:2024ret}}
\label{KappaSol:fig} 
\end{figure}

This first estimate of self-consistent plasma damping shows high sensitivity, demonstrating the need for a full self-consistent evaluation of damping rate in plasma within the context of a well-defined, self-consistent approach, where both damping and photon properties in plasma are determined in a mutually consistent manner, a project which is well ahead of the current state of the art and which is well beyond the scope of this report.

\subsection{Temperature dependence of the neutron lifespan}\label{sec:neutron}
\para{Understanding neutrons}
Element production during BBN is influenced by several parameters, e.g. baryon to photon ratio $\eta_\gamma$, number of neutrino species $N_\nu$, and neutron to proton ratio $X_n/X_p$, as controlled by both the dynamics of neutron\index{neutron} freeze-out at temperature $T_f\approx 0.8\,\mathrm{MeV}$ and neutron lifetime.

Since about 200 seconds, about 25\% of neutron lifespan, pass between neutron freeze-out, and BBN\index{Big-Bang!BBN} neutron burn at $T\approx0.07\,\mathrm{MeV}$, the in plasma neutron lifetime is one of the important parameters controlling BBN element yields~\cite{Pitrou:2018cgg}. However, the neutron population dynamics and decay within the cosmic plasma medium with large abundances of neutrinos and $e^+e^-$-pairs is not the same as in effective vacuum laboratory environment. The medium influence on particle decay was discussed for example by Kuznetsova et al~\cite{Kuznetsova:2010pi}, we will further develop and use this method in order to explore how cosmic primordial plasma influences neutron population dynamics.
 
After freeze-out when weak interaction scattering processes slow down to allow neutron abundance to free-stream, neutron abundance remains subject to natural decay
\begin{align}\label{Ndec}
n\longrightarrow p+e+\overline{\nu}_e\;.
\end{align}
The current experimental neutron lifetime remains method dependent, with a few seconds discrepancy, we adopt here the value $\tau_n^0=880.2\pm1.0\,\mathrm{sec}$. However measurements using magneto-gravitational traps unlike beam experiments find a bit shorter value, $877.7\pm0.7\,\mathrm{sec}$~\cite{Pattie:2017vsj}. In the standard BBN the neutron abundance when nucleosynthesis begins is assumed to be~\cite{Pitrou:2018cgg}
\begin{align}
\label{Xn_abundance}
X_n(T_{BBN})=X_n^f\exp\left(-\frac{t_{BBN}-t_f}{\tau_n^0}\right)\approx0.13\;.
\end{align}
The normalizing neutron freeze-out yield $X_n^f$ is in terms of abundances 
\begin{align}
\label{Xn_abundance2}
X_n^f \equiv \frac{n_n^f}{n_n^f+n_p^f}= \frac{n_n^f/n_p^f}{1+n_n^f/n_p^f}\;,
\end{align}
where $n_n^f$ and $n_p^f$ are the neutron and proton densities, respectively. The thermal equilibrium yield ratio is, assuming an instantaneous freeze-out
\begin{align}
\label{Xn_abundance3}
 \frac{n_n^f}{n_p^f}= \exp\left(-Q/T_f\right)\;,\qquad Q=m_n-m_p\;.
\end{align}
This value depends on the temperature $T_f$ at which neutrons decouple from the heat bath, and the neutron-proton mass difference (in medium). The values considered in literature are in the range $X_n^f=0.15\sim0.17$~\cite{Pitrou:2018cgg}. A dynamical approach to neutron freeze-out is necessary to fully understand $X_n^f$, we hope to return to this challenge in the near future.

Following freeze-out the neutron is subject to natural decay and normally the neutron lifetime in vacuum $\tau_n^0$ is used to calculate the neutron abundance resulting in the \lq desired\rq\ value $X_n(T_{BBN})\approx0.13$, \req{Xn_abundance} when BBN starts. To improve precision, a dynamically evolving neutron yield needs to be studied and for this purpose we explore here the neutron decay which occurs in medium, not vacuum. This leads to neutron lifespan dependence on temperature of the cosmic medium as the decay occurs for a particle emerged in electron-positron pair plasma and a background of free-streaming neutrinos.

Two physical effects of the medium influence the neutron lifetime in the primordial Universe noticeably:
\begin{itemize}
\item Fermi suppression factors from the medium: 
During the temperature range $T_f\geqslant T\geqslant T_{BBN}$, electrons and neutrinos in the background plasma can reduce the neutron decay rate by Fermi suppression to the neutron decay rate. Furthermore, the neutrino background can still provide the suppression after electron/positron pair annihilation becomes nearly complete.
\item Photon reheating:
When $T\ll m_e$ the electron/positron annihilation occurs, the entropy from $e^\pm$ is fed into photons, leading to photon reheating. The already decoupled (frozen-out) neutrinos remain undisturbed. Therefore, after annihilation we have two different temperatures in cosmic plasma: neutrino temperature $T_\nu$ and the photon and proton temperature $T$ respectively.
\end{itemize}
These two effects will be included in the following exploration of the neutron lifetime in the primordial Universe as a function of $T$. We show how these effects alter the neutron lifespan and obtain the modification of the neutron yield at the time of BBN. Yet another effect was considered by Kuznetsova et al~\cite{Kuznetsova:2010pi} which is due to time dilation originating in particle thermal motion. In our case for neutrons with $T/m<10^{-3}$ this effect is negligible. Below we will explicitly assume that the neutron decay is studied in the neutron rest frame.

\para{Decay rate in medium}
The invariant matrix element for the neutron decay \req{Ndec} for nonrelativistic neutron and proton is given by
\begin{align}
\langle|\mathcal{M}|^2\rangle\approx16\,G^2_FV^2_{ud}\,m_nm_p(1+3g^2_A)(1+RC)E_{\bar{\nu}}E_e,
\end{align}
where the Fermi constant is $G_F=1.1663787\times10^{-5}\,\mathrm{GeV}^{-2}$, $V_{ud}=0.97420$ is an element of the Cabibbo-Kobayashi-Maskawa (CKM)\index{CKM matrix} matrix~\cite{Czarnecki:2018okw,Marciano:2005ec,Czarnecki:2004cw}, and $g_A=1.2755$ is the axial current constant for the nucleons~\cite{Czarnecki:2018okw,Marciano:2014ria}. We also consider the total effect of all radiative corrections relative to muon decay that have not been absorbed into Fermi constant $G_F$. The most precise calculation of this correction~\cite{Marciano:2014ria,Marciano:2005ec} gives $(1+RC)=1.03886$. 

In the primordial Universe the neutron decay rate\index{neutron!lifespan} in medium, at finite temperature can be written as~\cite{Kuznetsova:2010pi}\index{neutron!decay rate in medium}
\begin{align}
\frac{1}{\tau^\prime_n}=\frac{1}{2m_n}\int&\frac{d^3p_{\bar{\nu}}}{(2\pi)^32E_{\bar{\nu}}}\frac{d^3p_p}{(2\pi)^32E_p}\frac{d^3p_e}{(2\pi)^32E_e}(2\pi)^4\delta^4\left(p_n-p_p-p_e-p_{\bar{\nu}}\right)\langle|\mathcal{M}|^2\rangle\notag\\
&\big[1-f_p(p_p)\big]\big[1-f_e(p_e)\big]\big[1-f_{\bar{\nu}}(p_{\bar{\nu}})\big]\;,
\end{align}
where we consider this expression in the rest frame of neutron, \ie\ $p_n=(m_n,0)$. The phase-space factors $(1-f_i)$ are Fermi suppression factors in the medium. The Fermi-Dirac\index{Fermi!distribution} distributions for electron and nonrelativistic proton are given by
\begin{align}
&f_e=\frac{1}{e^{E_e/T}+1},\\
&f_p=e^{-E_p/T}=e^{-m_p/T}\,e^{-p_p^2/2m_pT}.
\end{align}
For neutrinos, after neutrino/antineutrino kinetic freeze-out they become free streaming particles. If we assume that kinetic freeze out occurs at some time $t_k$ and temperature $T_k$, then for $t>t_k$ the free streaming distribution function \req{eq:NeutrinoDist} for antineutrinos is
\begin{align}
f_{\bar{\nu}}=\frac{1}{\exp{\left(\sqrt{\frac{E^2-m_\nu^2}{T_\nu^2}+\frac{m^2_\nu}{T^2_k}}+\frac{\mu_{\bar{\nu}}}{T_k}\right)+1}}\,,\qquad
T_\nu\equiv\frac{a(t_k)}{a(t)}T_k\,,
\end{align}
with the effective neutrino temperature $T_\nu$. In the following calculation, we assume the condition $T_k\gg\mu_{\bar{\nu}},\,m_\nu$, {\it i.e.\/} we consider the massless neutrino in plasma. Substituting the distributions into the decay rate formula and using the neutron rest frame, the decay rate can be written as 
\begin{align}
\label{Decay:Rate_01}
\frac{1}{\tau_n^\prime}&=\frac{G^2_FQ^5V^2_{ud}}{2\pi^3}\,(1+3g^2_A)\,(1+RC)
\int^1_{m_e/Q}d\xi\,\frac{\xi(1-\xi)^2}{\exp\left(-Q\xi/{T}\right)+1}\frac{\sqrt{\xi^2-(m_e/Q)^2}}{\exp\left(-Q(1-\xi)/T_\nu\right)+1}\,,
\end{align} 
where $Q$ was defined in Eq.\;(\ref{Xn_abundance3}) and we integrate using dimensionless variable $\xi=E_e/Q$. From \req{Decay:Rate_01}, the decay rate in vacuum can be written as
\begin{align}
&\frac{1}{\tau_n^0}=\frac{G^2_Fm_e^5V^2_{ud}}{2\pi^3}(1+3g^2_A)\,(1+RC)\,f^\prime,
\end{align}
where the phase space factor $f^\prime$ is given by
\begin{align}
f^\prime&\equiv\left(\frac{Q}{m_e}\right)^5\int^1_{m_e/Q}d\xi\,{\xi(1-\xi)^2}\sqrt{\xi^2-(m_e/Q)^2}=1.6360\,.
\end{align}

The phase space factor is also modified by the Coulomb correction between electron and proton, proton recoil, nucleon size correction etc. This has been studied by Wilkinson~\cite{Wilkinson:1982hu}, and the phase space factor is given by~\cite{Czarnecki:2018okw,Czarnecki:2004cw,Wilkinson:1982hu}
\begin{align}
f=1.6887.
\end{align}
These effects amount to adding the factor $\mathcal{F}$ to our calculation
\begin{align}
\mathcal{F}=\frac{f}{f^\prime}=1.0322,
\end{align}
then the neutron lifespan can be written as \index{neutron!lifespan in vacuum}
\begin{align}
\tau^{\mathrm{Vacuum}}_n=\frac{\tau^0_n}{\mathcal{F}}=879.481\,\mathrm{sec},
\end{align}
which compares well to the experimental result, $877.7\pm0.7\,\mathrm{sec}$~\cite{Pattie:2017vsj}. 

In the case of plasma medium, we do not expect that these effects (Coulomb correction between electron and proton, proton recoil, nucleon size correction etc) are modified in the cosmic plasma. Thus we adapt the factor into our calculation and the neutron decay rate in the cosmic plasma is given by
\begin{align}
\label{Decay:Rate_02}
\frac{1}{\tau_n^{\mathrm{Medium}}}=&\frac{G^2_FQ^5V^2_{ud}}{2\pi^3}\,(1+3g^2_A)\,(1+RC)\,\mathcal{F}
\int^1_{m_e/Q}d\xi\,\frac{\xi(1-\xi)^2}{\exp\left(-Q\xi/{T}\right)+1}\frac{\sqrt{\xi^2-(m_e/Q)^2}}{\exp\left(-Q(1-\xi)/T_\nu\right)+1}.
\end{align}
From \req{Decay:Rate_02} we see that the neutron decay rate in the primordial Universe depends on both the photon temperature $T$ and the neutrino effective temperature $T_\nu$.

\para{Photon reheating}
After neutrino freeze-out and when $m_e\gg T$, the $e^{\pm}$ becomes nonrelativistic and annihilate. In this case, their entropy is transferred to the other relativistic particles still present in the cosmic plasma, {\it i.e.\/} photons, resulting in an increase in photon temperature as compared to the free-streaming neutrinos. From entropy conservation\index{entropy!conservation} we have
\begin{align}
\label{Entropy}
\frac{2\pi}{45}g^s_\ast(T_k)T^3_kV_k+S_{\nu}(T_k)=\frac{2\pi}{45}g^s_\ast(T)T^3V+S_{\nu}(T),
\end{align}
where we use the subscripts $k$ to denote quantities for neutrino freeze-out\index{neutrino!freeze-out} and $g^s_\ast$ counts the degree of freedom for relativistic species in primordial Universe. After neutrino freeze-out, their entropy is conserved independently and the temperature can be written as
\begin{align}
T_\nu\equiv\frac{a(t_k)}{a(t)}T_k=\left(\frac{V_k}{V}\right)^{1/3}T_k.
\end{align}
In this case, from entropy conservation, \req{Entropy}, we obtain
\begin{align}
\label{Neutrino_Photon}
T_\nu=\frac{T}{\kappa},\,\,\,\,\,\,\kappa\equiv\left[\frac{g^s_\ast(T_k)}{g^s_\ast(T)}\right]^{1/3}.
\end{align}
From \req{Neutrino_Photon} the neutron decay rate in a heat bath can be written as
\begin{align}
\label{Decay:Rate_03}
\frac{1}{\tau_n^\mathrm{Medium}}=&\frac{G^2_FQ^5V^2_{ud}}{2\pi^3}(1+3g^2_A)\,(1+RC)\,\mathcal{F}
\int^1_{m_e/Q}d\xi\,\frac{\xi(1-\xi)^2}{\exp\left(-Q\xi/{T}\right)+1}\frac{\sqrt{\xi^2-(m_e/Q)^2}}{\exp\left(-Q(1-\xi)\kappa/T\right)+1}. 
\end{align}

In the high temperature regime, $T\gg Q$, the exponential terms in the Fermi distribution becomes $1$ and the decay rate is given by
\begin{align}
&\frac{1}{\tau_n^\mathrm{Medium}}=\frac{1}{4}\left(\frac{1}{\tau_n^\mathrm{Vacuum}}\right)\;,
\qquad
T\gg Q\;.
\end{align}
In \rf{Decay:Rate} on the left-hand side we show the  neutron lifetime $\tau^\mathrm{Medium}_n$ in plasma as a function of temperature. Fermi-suppression from electron and neutrino increases the neutron lifetime as compared to value in vacuum. At low temperature, $T<0.2\,m_e$, most of the electrons and positrons have annihilated and the main Fermi-blocking comes from the cosmic neutrino background. In this regime, the neutron lifetime depends also on the neutrino temperature, $T_\nu$. For cold neutrinos $T_\nu<T$, the Fermi suppression is smaller than the hot one $T_\nu=T$, the effect is made more visible in the insert.

\begin{figure} 
\centerline{\includegraphics[width=0.5\linewidth]{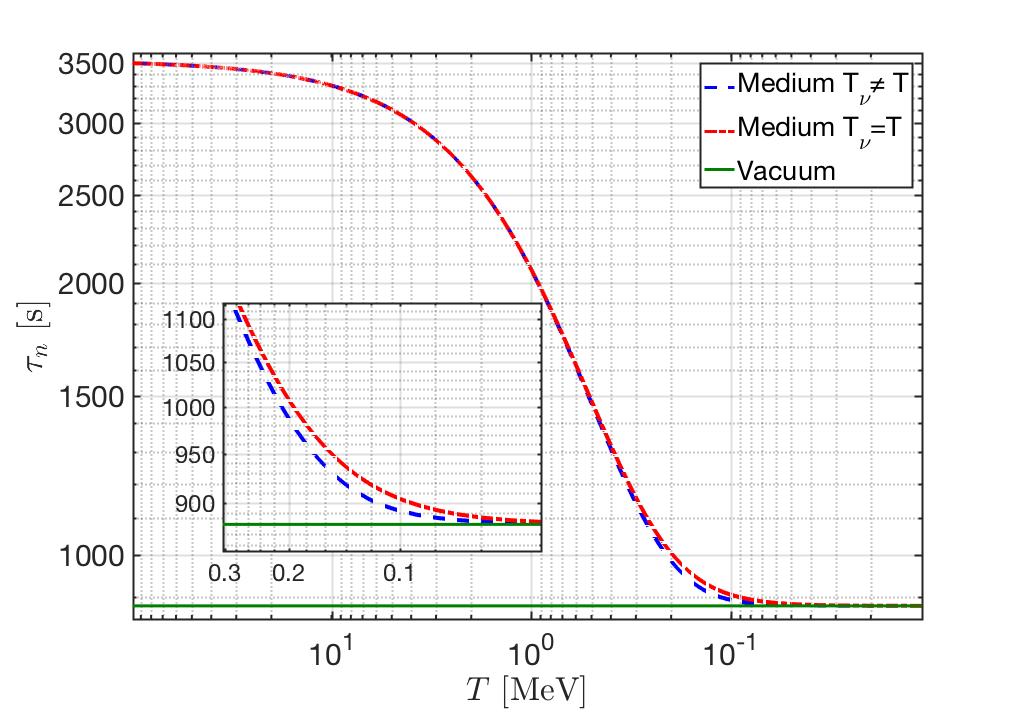}\hspace*{-0.5cm}
\includegraphics[width=0.5\linewidth]{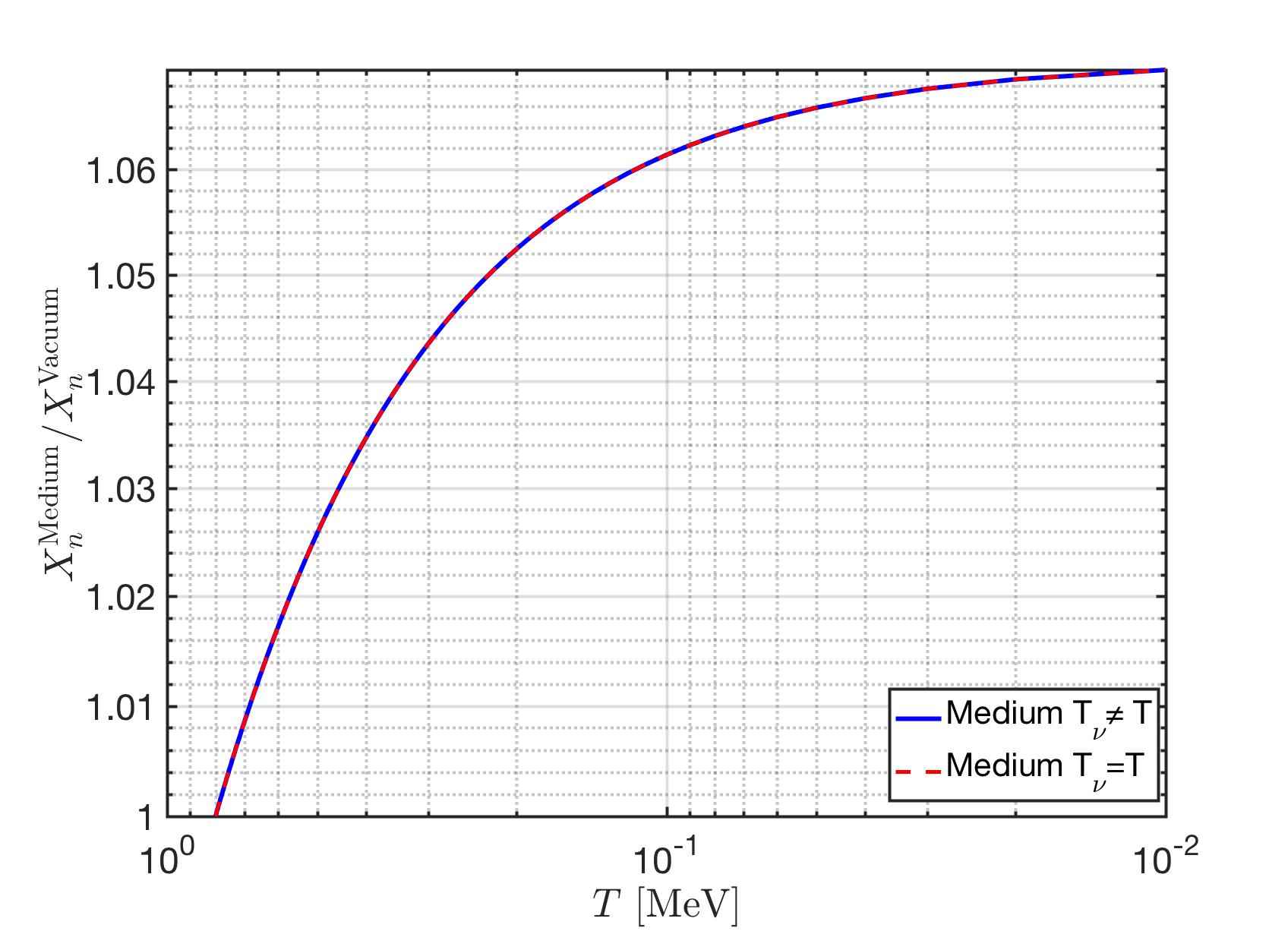}}
\caption{On left: The neutron lifetime $\tau_n^\mathrm{Medium}$ in the cosmic plasma; On right: The neutron abundance in the cosmic plasma, ratio to vacuum  abundance, as a function of temperature\cccite{Yang:2018qrr}}
\label{Decay:Rate}
\end{figure}

\para{Neutron abundance}
After the neutron freeze-out, the neutron abundance is only affected by the neutron decay. The neutron concentration can be written as \index{neutron!particle abundance}
\begin{align}
\label{Abundance}
X_n=X_n^f\,\exp\bigg[-\int^t_{t_f}\frac{dt^\prime}{\tau_n}\bigg],
\end{align}
where we use the subscripts $f$ to denote quantities at neutron freeze-out. Using \req{Decay:Rate_03} and \req{Abundance}, the neutron abundance ratio between plasma medium and vacuum is given by
\begin{align}
\label{Abundance_Ratio}
\frac{X_n^{\mathrm{Meduim}}}{X_n^{\mathrm{Vacuum}}}=\exp\bigg[-\int^t_{t_f}dt^\prime\left(\frac{1}{\tau^\prime_n}-\frac{1}{\tau^0_n}\right)\bigg].
\end{align}

On the right-hand side in \rf{Decay:Rate}  we show this neutron abundance ratio as a function of temperature. We consider as typical the neutron freeze-out temperature $T_f=0.08\mathrm{MeV}$ and the BBN  temperature $T_{BBN}\approx0.07\mathrm{MeV}$, which yields ${X_n^{\mathrm{Meduim}}}/{X_n^{\mathrm{Vacuum}}}=1.064$. Then from \req{Xn_abundance} the neutron abundance in plasma medium is given by
\begin{align}
X_n^{\mathrm{Meduim}}=1.064\,X_n^{\mathrm{Vacuum}}\approx0.138.
\end{align}
In this case, the neutron abundance will increase about $6.4\%$ in the cosmic plasma which should affect the final abundances of the Helium-4 and other light elements in BBN. A further study of neutron decoupling and decay in medium will be needed. 
 
\para{How is BBN impacted?}
One of the important parameters of standard BBN is the neutron lifetime, as it affects the neutron abundance after neutron freeze-out at temperature $T_f\approx 0.8 \mathrm{MeV}$ and before the BBN $T\approx0.07 \mathrm{MeV}$. 

In the standard BBN model, it is necessary to have a neutron-to-proton ratio $n/p\approx1/7$ when BBN begins in order to explain the observed values of hydrogen and helium abundance~\cite{Pitrou:2018cgg}. We have evaluated the effect of Fermi suppression on the neutron lifetime due to the background electron-positron  and free-streaming neutrino plasma. We found that in a very hot medium ($T>10$\,MeV) the neutron lifetime is lengthened by up to a factor 4. Our method should in principle also be considered in the study of medium modification of just about any of the relevant weak interaction rates which in this report we used as given in vacuum,  this remains a task for another day.

In the temperature range between neutron freeze-out just below $T=1$\;MeV and BBN conditions the effect of neutron lifespan is smaller but still quite noticeable. Near neutron freeze-out both decay electron and neutrino are blocked. However, after $e^\pm$ annihilation is nearly complete  Fermi-blocking comes predominantly from the cosmic neutrino background and the neutron lifetime depends on the temperature $T_\nu<T$.

We found that the increased neutron lifetime results in an increased neutron abundance ${X_n^{\mathrm{Meduim}}}/{X_n^{\mathrm{vacuum}}}=1.064$ at $T_{BBN}\approx0.07 \mathrm{MeV}$ \ie\ we find a $6.4\%$ \emph{increase} in neutron abundance due to the medium effect at the time of BBN. We believe that this effect needs to be accounted for in the precision BBN study of the final abundances of heavy hydrogen, helium and other light elements produced in BBN.
  
\subsection{Effective inter-nuclear potential in BBN}\label{sec:potential}

A few seconds after the Big-Bang\index{Big-Bang}, when the Universe was filled with electron-positron plasma\index{plasma!electron-positron} the assumption of homogeneity and stationary of the medium on the scale of the relevant plasma parameters, (Debye mass $m_D$, and damping $\kappa$, \rsec{section:electron}) is well justified. This allows us to use the methods of plasma theory obtained in \rsec{chap:PlasmaSF} to develop an in depth understanding of the BBN reactions in the presence of relevant and high density electron-positron plasma.

\para{Screening in BBN}
At present, the observation of light element 
(e.g. D, $^3$He, $^4$He, and $^7$Li) cosmic abundances predominantly produced in BBN allows us to probe the primordial Universe before the recombination era. Much effort in the BBN study is currently being made to reconcile the discrepancies and tensions between theoretical predictions and observations, \eg, $^7$Li problem~\cite{Pitrou:2018cgg}. Current BBN models assume that the Universe was essentially void of anything but reacting nucleons and light nuclei, and electrons needed to keep the local baryon density charge-neutral, a situation similar to the experimental environment where empirical nuclear reaction rates are obtained.

The electron-positron plasma we have shown to be present at BBN epoch can influence through electromagnetic screening of the nuclear potential the nuclear reaction rates. The electron cloud surrounding the charge of an ion screens other nuclear charges far from its radius and reduces the Coulomb barrier. In nuclear reactions, the reduction of the Coulomb barrier makes the penetration probability easier and enhances the thermonuclear reaction rates. In this case, the modification of the nuclei interaction due to the plasma screening effect may play a key role in the formation of light elements in the BBN. 

The enhancement factor of thermonuclear reaction rates by a static screening potential was calculated by Salpeter in 1954~\cite{Salpeter:1954nc} allowing one electron per proton in BBN epoch. In an isotropic and homogeneous plasma, the Coulomb potential of a point-like particle with charge $Ze$ at rest is modified~\cite{Salpeter:1954nc} according to 
\begin{align}
\phi_\text{stat}(r)=\frac{Ze}{4\pi\epsilon_0 r}e^{-m_Dr},
\end{align}
where $m_D$ is the Debye mass. After that, it has been exploited widely in BBN for static screening ~\cite{1969ApJ...155..183S,Famiano:2016hhs}. Subsequently, the study of dynamical screening for moving ions has been taken into account~\cite{Carraro:1988apj,Gruzinov:1997as,Hwang:2021kno}.  

In this section, we review the effect of  the nonrelativistic longitudinal polarization function to study the dynamics of the electron-positron plasma in the primordial Universe\cite{Grayson:2023flr}. In particular, we discussed the damping rate, the electron-positron to baryon density ratio, and their potential implications for Big-Bang Nucleosynthesis (BBN) through screening within linear response theory. We derived an approximate analytic formula for the potential of a moving heavy charge in a collisional plasma in \req{eq:pos_point_DDS} describing screening effects previously found only numerically \cite{Hwang:2021kno}. Our analytic formula can be readily used to estimate the effect of screening on thermonuclear reactions using \req{eq:DDSenhance}. 

The correction to thermonuclear reactions due to damped-dynamic screening is found to be small due to the low velocity of nuclei and a large amount of collisional scattering. This is in line with the findings of \cite{Hwang:2021kno}, who concludes that even though the densities are large, they are not enough to modify the potential at short distances related to screening. The analytic expression we find for the nuclear reaction rate enhancement \req{eq:DDSenhance} in a collisional plasma could be useful in other fusion environments such as stellar fusion and laboratory fusion experiments, such as those discussed in ~\cite{Labaune:2013dla,Margarone:2022mdpi}.

Overall, we were very surprised to find that the screening effects in BBN were relatively small even in the static case, considering that the number densities of free charges (electron-positron pairs) present during BBN epoch are $\sim 10^4$ times normal matter. If we compare this to screening effects on Earth, we can see that although plasma state occurs in much colder environments at lower densities. The strength of the screening effect is related to the Debye mass
\begin{equation}
m_D^2 \sim \frac{n_\text{eq} }{T}\,,
\end{equation}
which is on the order of a few keV during BBN. On Earth, $n_\text{eq}$ is decreased by $\sim 10^4$, but $T$ is decreased by $\sim 10^6$. Thus, we would expect to see similar, if not larger, screening effects on Earth. For instance, the Debye screening length in extracellular fluid in the body is 8 \AA ngstrom \cite{Wennerstrom:2020}, only a factor of $\sim 20$ times larger than the Debye length during BBN.

\para{The short-range screening potential}
In \cite{Grayson:2023flr}, a proposal is made to study the short-range potential relevant to quantum tunneling in thermonuclear reactions. Since the Gamow energy at which nuclei are most likely to tunnel is above the thermal energy, the portion of the screening potential relevant for tunneling does not satisfy the `weak-field' limit where the electromagnetic energy is small compared to the thermal energy
\begin{equation}
 \frac{q \phi(x)}{T} \ll 1\,,
\end{equation}
{\color{black}where we have chosen the Lorentz gauge}. When this condition is not satisfied, one must consider the full equilibrium distribution when calculating the short-range potential \cite{Hakim:1967prd,DeGroot:1980dk}
\begin{equation}\label{eq:Boltz}
 f_B^\pm(x,p) = e^{-(p_0\pm e\phi(x))/T}\,.
\end{equation}
The $e\phi$ term in the exponential accounts for the change in energy of a charge in the plasma due to its presence in an external field. For this equilibrium distribution, a linear response is no longer possible since the equilibrium distribution depends on the external electromagnetic field. 

In equilibrium, one can find the static screening potential for strong electromagnetic fields using the nonlinear Poisson-Boltzmann equation\index{Poisson-Boltzmann equation},
\begin{equation}\label{eq:Poisson-Boltz}
 -\nabla^2 e\phi_{(\text{eq})}(x)/T +m_D^2\sinh\left[e\phi_{(\text{eq})}(x)/T\right] =e\rho_\mathrm{ext}(x)/T\,.
 \end{equation}
This equation has a well-known solution for an infinite sheet, which we used to argue the importance of strong screening in BBN. We hope that, in a future publication, we will solve the Poisson-Boltzmann equation with strong screening to calculate the short-range screening potential in BBN. 

We note that the toy model in \cite{Grayson:2023flr} overestimates strong screening effects for two reasons: an infinite sheet has a constant electric field requiring more polarizing charge density to screen the field, and the Boltzmann distribution\index{Boltzmann!distribution} in \req{eq:Boltz} does not account for the stacking of electron-positron states when the density of electrons and positrons becomes very large near the nucleus. Both of these effects significantly reduce the effect of strong screening on reaction rates, but at the time of writing, it seems that strong screening will create a larger effect on nuclear reaction rates than damped-dynamic screening. Predicting enhanced screening may be relevant for the anomalous screening observed in the measurements of astrophysical S(E) factors \cite{Zhang:2020nuc}.
 
\para{Primordial cosmic plasma: nonrelativistic polarization tensor}
The properties of the BBN plasma are described by the relativistic Vlasov-Boltzmann transport equations\index{Vlasov-Boltzmann equation} \req{eq:VBEf}. Since photons do not couple directly to the electromagnetic field, they do not contribute to the polarization tensor at first order in $\delta f$ as indicated in Eq.\,(\ref{eq:VBEg}). We neglect photon influence on the electron-positron distribution through the scattering term since the rate of inverse Compton scattering\index{Compton scattering} $R_{e^{\pm}\gamma }$ shown in green in \rf{RelaxationRate:fig} is much smaller, in the BBN temperature range, than the total rate $\kappa$ shown as a black line. Each fermion Boltzmann equation \req{eq:VBEf} can be solved independently. Since the equations for electrons and positrons are equivalent, except for the charge sign, only one needs to be solved to understand the dynamics.

We take the equilibrium one particle distribution function $\eq{f_\pm}$ of electrons and positrons to be the relativistic Fermi-distribution
\begin{equation}\label{eq:equildist}
\eq{f}_\pm(p) = \frac{1}{\exp{\left(\frac{\sqrt{\boldsymbol{p}^2 + m^2}}{T}\right)}
+1}\,,
\end{equation}
with chemical potential\index{chemical potential} $\mu = 0 $. The electron and positron mass will be indicated by $m$ unless otherwise stated. At temperatures interesting for nucleosynthesis $T = 50-86$\,keV, we expect the plasma temperature to be much less than the mass of the plasma constituents. Only the nonrelativistic form of Eq.\,(\ref{eq:equildist}) will be relevant at these temperature scales
\begin{equation}
\eq{f}_\pm(p) \approx \exp\left(- \frac{m}{T}\left(1+\frac{|\pmb{p}|^2}{2m^2}\right)\right)\,.
\end{equation}
Keeping terms up to quadratic order in $|\boldsymbol{p}|/m$ we solve the Vlasov-Boltzmann equation Eq.\,(\ref{eq:VBEf}) for the induced current and identify the polarization tensor. This is done in detail in our previous work in~\cite{Formanek:2021blc}.

In the infinite homogeneous plasma filling the primordial Universe, the polarization tensor only has two independent components: the longitudinal polarization function $\Pi_{\parallel}$ parallel to field wave-vector $\boldsymbol{k}$ in the rest frame of the plasma and the transverse polarization function $\Pi_{\perp}$ perpendicular to $\boldsymbol{k}$~\cite{melrose2008quantum}. In the nonrelativistic limit, these functions are~\cite{Formanek:2021blc}
\begin{align}\label{eq:polfuncs}
	\Pi_\parallel(\omega,\boldsymbol{k}) &= -\omega_p^2\frac{\omega^2}{(\omega+ i \kappa)^2} \frac{1}{1-\frac{i\kappa}{\omega+ i \kappa}\left(1+\frac{ T |\boldsymbol{k}|^2}{m (\omega+ i \kappa)^2} \right)}\,,\\
	\Pi_{\perp}(\omega) &= -\omega_p^2 \frac{\omega}{\omega+ i \kappa}\,.
\end{align}
In these expressions, the plasma frequency $\omega_p$ (defined as $m_L$ in~\cite{Formanek:2021blc}) is related to the Debye screening mass in the nonrelativistic limit as
\begin{equation}\label{eq:plasmafreq}
 \omega_p^2 = m_D^2\frac{T}{m}\,.
\end{equation}

\begin{figure} 
\centerline{\includegraphics[width=0.80\linewidth]{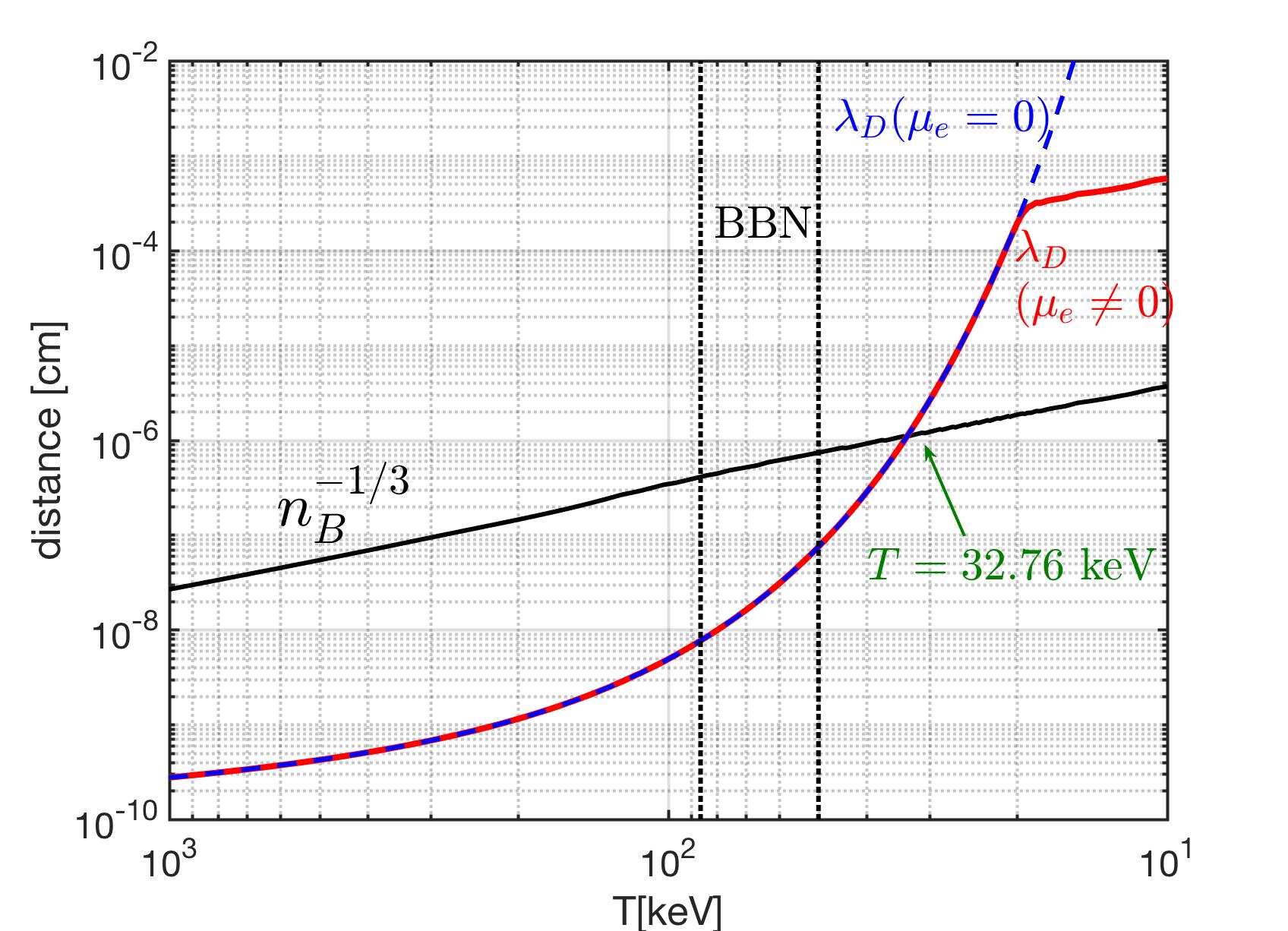}}
\caption{ The average distance between baryons $n_B^{-1/3}$ and the Debye length $\lambda_D$ ($\mu_e \neq 0$) as a function of temperature (red solid line). During the BBN epoch (vertical dotted lines) $n_B^{-1/3}>\lambda_D$. For temperature below $T<32.76$ keV we have $n_B^{-1/3}<\lambda_D$. For comparison, the Debye length for zero chemical potential $\mu_e=0$ is also plotted as a blue dashed line. \cccite{Grayson:2023flr}}
\label{MeanFreePath_fig} 
\end{figure}

The transverse response $\Pi_{\perp}$ relates to the dispersion of photons in the plasma. Here we need only consider $\Pi_\parallel$ since the vector potential $\boldsymbol{A}(t,\boldsymbol{x})$ of the traveling ion will be small in the nonrelativistic limit. This work does not consider the effect of a primordial magnetic field discussed in~\cite{Steinmetz:2023nsc} and \rsec{sec:mag:universe}. We note that Debye mass $m_D$ is related to the usual Debye screening length of the field in the plasma as
\begin{equation}\label{eq:mL}
	1/\lambda_D^{2} = m_D^2= 4 \pi \alpha \left(\frac{2mT}{\pi}\right)^{3/2}\frac{e^{-m/T}}{2T}\,.
\end{equation}
This formula describes the characteristic length scale of screening in the plasma.

\para{Longitudinal dispersion relation}
As discussed in Chapter \ref{chap:PlasmaSF} the poles in the propagator or roots of the dispersion equation represent the plasma's propagating modes, often called `quasi-particles' or `plasmons.' In the nonrelativistic limit, one can solve the longitudinal part of the dispersion equation \req{eq:disp}, which is relevant for finding charge oscillation modes in the plasma
\begin{equation}
    1+ \frac{\Pi_\parallel( k)}{(p\cdot u)^2}= 1+ \frac{\Pi_\parallel(\omega, \boldsymbol{k})}{\omega^2}=\varepsilon_\parallel(\omega,\boldsymbol{k}) =0 \,,
\end{equation}
evaluated in the rest frame. Then we insert \req{eq:polfuncs} to find
\begin{equation}
   1- \frac{\omega_p^2}{(\omega+ i \kappa)^2} \frac{1}{1-\frac{i\kappa}{\omega+ i \kappa}\left(1+\frac{T |\boldsymbol{k}|^2}{m(\omega+ i \kappa)^2} \right)}=0 \,.
\end{equation}
We can simplify the above expression since this is only a function of $\omega' =\omega+i\kappa$
\begin{equation}
   1- \frac{\omega_p^2}{\omega'^2-i\kappa\omega'+\frac{i\kappa T |\boldsymbol{k}|^2}{m \omega'} }=0 \,.
\end{equation}
Then we get a cubic equation for $\omega'(|\boldsymbol{k}|)$
\begin{equation}\label{eq:dispfact}
   \frac{1}{\omega'^3-i\kappa\omega'^2+\frac{i\kappa T |\boldsymbol{k}|^2}{m} }
    \left(\omega'^3-i\kappa\omega'^2 - \omega_p^2\omega'+\frac{i\kappa T |\boldsymbol{k}|^2}{m} \right)=0 \,.
\end{equation}
Cardano's formula gives the solutions to this cubic equation
\begin{equation}\label{eq:cardano}
\omega_n(\boldsymbol{k}) = \frac{1}{3}\left(i\kappa-\xi^n C-\frac{\Delta_0}{\xi^n C}\right), \qquad n \in \{0,1,2\} \,,
\end{equation}
with the quantities:
\begin{align}\label{eq:delta}
  \xi &=\frac{i\sqrt{3}-1}{2}\,,\\
    C &= \sqrt[3]{\frac{\Delta_1 \pm \sqrt{\Delta_1^2 - 4 \Delta_0^3}}2}\,,\\
    \Delta_0 &= -\kappa^2 + 3 \omega_p^2\,,\\
\Delta_1 &= 2i\kappa^3 - 9 i\kappa \omega_p^2 + 27\frac{i\kappa T |\boldsymbol{k}|^2}{m}.
\end{align}

Since the longitudinal dispersion relation is analytically solvable, the full nonrelativistic potential can be found in position space using contour integration. The residue of each pole will lead to the strength of that mode, and the location of the pole will lead to space and time dependence, which in simple cases is exponential. In practice, factoring out these roots in the Fourier transform of the potential leads to five poles, which do not seem to lead to simple expressions in position space. We found using the approximate expression derived in \req{sec:potential} was more practical. Deriving the full expression is the subject of future work.

\para{Damped-dynamic screening} 
We discuss the application of the nonrelativistic limit of the polarization tensor \rsec{chap:PlasmaSF} to the electron-positron plasma which existed during Big-Bang nucleosynthesis (BBN)~\cite{Grayson:2023flr}. The BBN Epoch occurred within the first 20 min after the Big-Bang when the Universe was hot and dense enough for nuclear reactions to produce light elements up to lithium \cite{Pitrou:2018cgg}. 

The BBN nuclear reactions typically take place within the temperature interval $86\, \mathrm{keV}>\mathrm{T_{BBN}}>50\, \mathrm{keV}$~\cite{Pitrou:2018cgg}. We refer to these elements produced in BBN\index{Big-Bang!BBN} as primordial light elements to distinguish them from those made later in the Universe's history. Primordial light element abundances are the most accessible probes of the primordial Universe before recombination. Though the current BBN model successfully predicts D, $^3$He, $^4$He abundances, well-documented discrepancies, such as $^7$Li, remain. Efforts to resolve the theoretical BBN model with present-day observations are discussed in detail in \cite{Pitrou:2021vqr,Bertulani:2022qly}.

A rather large electron-positron $e^-e^+$- number densities existed in the primordial Universe, \rsec{section:electron}, work is in progress to understand how this plasma impacts BBN~\cite{Wang:2010px,Hwang:2021kno,Rafelski:2023emw}. The charge particle densities in the BBN Universe are $10^2$ times larger than those present in the Sun \cite{Bahcall:2001smc} and $10^4$ times normal atomic densities \cite{Grayson:2023flr}. Charge screening is an essential collective plasma effect that modifies the inter-nuclear potential $\phi(r)$ changing thermonuclear reaction rates during BBN. An electron cloud around an ion's charge effectively diminishes the influence of nuclear charges beyond their immediate vicinity, lowering the Coulomb barrier. 

In the context of nuclear reactions, a reduced Coulomb barrier leads to a higher likelihood of penetration, boosting thermonuclear reaction rates. Consequently, this process influences the abundance of light elements in the primordial Universe by modifying their formation rates. Since the BBN temperature range is much less than the electron mass, we will use the nonrelativistic limit of the polarization tensor derived in \rsec{chap:PlasmaSF}. The screened potential relevant for thermonuclear reactions will be given by the longitudinal polarization function \req{eq:phi}.

The influence of screening on nuclear reactions is a well-established field of study. The concept of plasma screening effects on nuclear reactions was initially introduced in~\cite{Salpeter:1954nc}, who suggested determining the increase in nuclear reaction rates through the use of the static Debye-H{\"u}ckel potential~\cite{Debye_1923a,Debye_1923b,Salpeter:1969apj,Famiano:2016hhs}. Subsequent research expanded this framework to account for the thermal velocity of nuclei traversing the plasma~\cite{Hwang:2021kno,Carraro:1988apj,Gruzinov:1997as,Opher:1999jh,Yao:2016cjs}, introducing the concept of `dynamic' screening. 

In our current study, we address the high density of the $e^-e^+\gamma$ plasma by including collisional damping using the current conserving collision term developed in \cite{Formanek:2021blc} shown in \req{eq:collision}. The dense aspect of the BBN plasma has only recently been acknowledged by incorporating collision effects into numerical models \cite{Sasankan:2019oee,Kedia:2020xdc}. We will refer to this model of screening as 'damped-dynamic' screening. In \cite{Grayson:2023flr}, we find an analytic formula for the induced screening potential, which allows for estimating the enhancement of thermonuclear reaction rates.

\para{Screened nuclear potential}
We consider the effective nuclear potential for a light nucleus moving in the plasma at a constant velocity. This is done by Fourier transforming \req{eq:potentk}. The velocity of the nucleus is assumed to be the most probable velocity given by a Boltzmann distribution\index{Boltzmann!distribution}
\begin{equation}\label{eq:vel}
 \beta_{\text{N}} = \sqrt{\frac{2T}{m_N}}\,. 
\end{equation}
Since the poles of the \req{eq:potent} can be solved analytically, ideally, one would perform contour integration to get the position space field. Due to the intricacy of these poles $\omega_n(\boldsymbol{k})$, we find it insightful to look at the field in a series expansion around velocities of the light nuclei smaller than the thermal velocity of electrons and positrons and large damping.
\begin{equation}\label{eq:expansion}
\frac{(\boldsymbol{k}\cdot\boldsymbol{\beta}_{\text{N}})^2}{\omega_p^2} \ll \frac{\boldsymbol{k}^2}{m_D^2} \ll \frac{\kappa^2}{\omega_p^2}\, .
\end{equation}
This expansion is useful during BBN since the temperature is much lower than the mass of light nuclei and the damping rate $\kappa$ is approximately twice the Debye mass $m_D$, as seen in Fig.~\ref{RelaxationRate:fig}. Applying this expansion to \req{eq:potentk} and evaluating this expression for a point charge $r \rightarrow 0$ we find
\begin{equation}\label{eq:ddsint}
\phi(t,\boldsymbol{x}) =\phi_{\text{stat}}(t,\boldsymbol{x})-Ze\int \frac{d^3\boldsymbol{k}}{(2\pi)^3} e^{ i\boldsymbol{k}\cdot(\boldsymbol{x}-\boldsymbol{\beta_{\text{N}}} t)}\frac{i \boldsymbol{k}\cdot \boldsymbol{\beta_{\text{N}}} m_D^4 (\frac{\boldsymbol{k}^2}{m_D^2} - \frac{\kappa^2}{\omega_p^2})}{\boldsymbol{k}^2(\boldsymbol{k}^2+m_D^2)^2\kappa}\,.
\end{equation}
The second term is the damped-dynamic screening correction, which we refer to as $\Delta \phi$, where
\begin{equation}\label{eq:pos_point}
\phi(t,\boldsymbol{x}) = \phi_{\text{stat}}(t,\boldsymbol{x}) +\Delta \phi(t,\boldsymbol{x}) \,,
\end{equation}
and $\phi_{\text{stat}}$ is the standard static screening potential. The details of the integration of \req{eq:ddsint} can be found in \cite{Grayson:2023flr}, the result is
\begin{equation}\label{eq:pos_point_DDS}
\Delta \phi(t,\boldsymbol{x}) = \frac{Ze \beta_N \cos (\psi) m_D^2}{4 \pi \varepsilon_0 \kappa} \Bigg[\left(\frac{\nu_\tau^2}{m_D^2 r(t)^2} + \frac{\nu_\tau^2}{m_D r(t)}+\frac{1 + \nu_\tau^2}{2}\right)e^{-m_D r(t)}   -\frac{\nu_\tau^2}{m_D^2 r(t)^2}\Bigg]\,,
\end{equation}
where $\psi$ is the angle between $\boldsymbol{x}-\boldsymbol{\beta}_N t$ and $\boldsymbol{\beta}_N$ and $r(t) = |\boldsymbol{x}-\boldsymbol{\beta}_N t|$.
We introduce the ratio of the damping rate to the rate of oscillations in the plasma $\nu_\tau = \kappa/\omega_p$.  

\begin{figure} 
 \centerline{
\includegraphics[width=.80\linewidth]{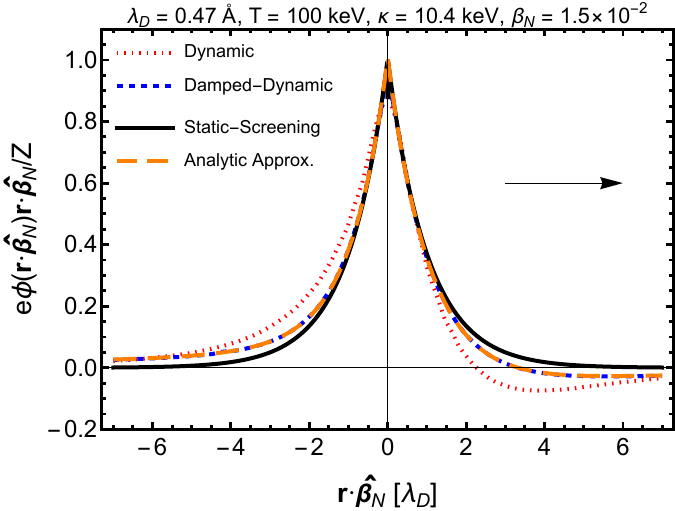}}
 \caption{Total screening potential scaled with charge Z and distance along the direction of motion. We show a comparison of the following screening models plotted along the direction of motion of a nucleus $\boldsymbol{r}\cdot\hat{\boldsymbol{\beta}_{\text{N}}}$: static screening (black), dynamic screening (red dotted) from \cite{Hwang:2021kno}, damped-dynamic screening (blue dashed), and the approximate analytic solution of \req{eq:pos_point} (orange dashed). A black arrow indicates the direction of motion of the nucleus $\hat{\boldsymbol{\beta}_{\text{N}}}$. \cccite{Grayson:2023flr}}
 \label{fig:dynamiclinear}
\end{figure} 

This expression \req{eq:pos_point_DDS} is valid for large damping and slow motion of the nucleus or if the velocity of the nuclei is small. A similar result valid at large distances, which only includes the last term, was previously derived in~\cite{Stenflo:1973} for dusty (complex) plasmas. For large distances and large $\nu_\tau$, the last term in the second line is dominant, indicating that the overall potential would be over-damped. In this regime, the potential is heavily screened in the forward direction and unscreened in the backward direction relative to the motion of the nucleus. As $\nu_\tau$ becomes small, the $1/2$ in the first portion of the third term, proportional to $m_D^2/\kappa$, dominates. This flips the sign of the damped-dynamic screening contribution, causing a wake potential to form behind the nuclei. This shift indicates the change from damped to undamped screening where \req{eq:pos_point_DDS} is no longer valid. 

\begin{figure} 
\centerline{\includegraphics[width=.80\linewidth]{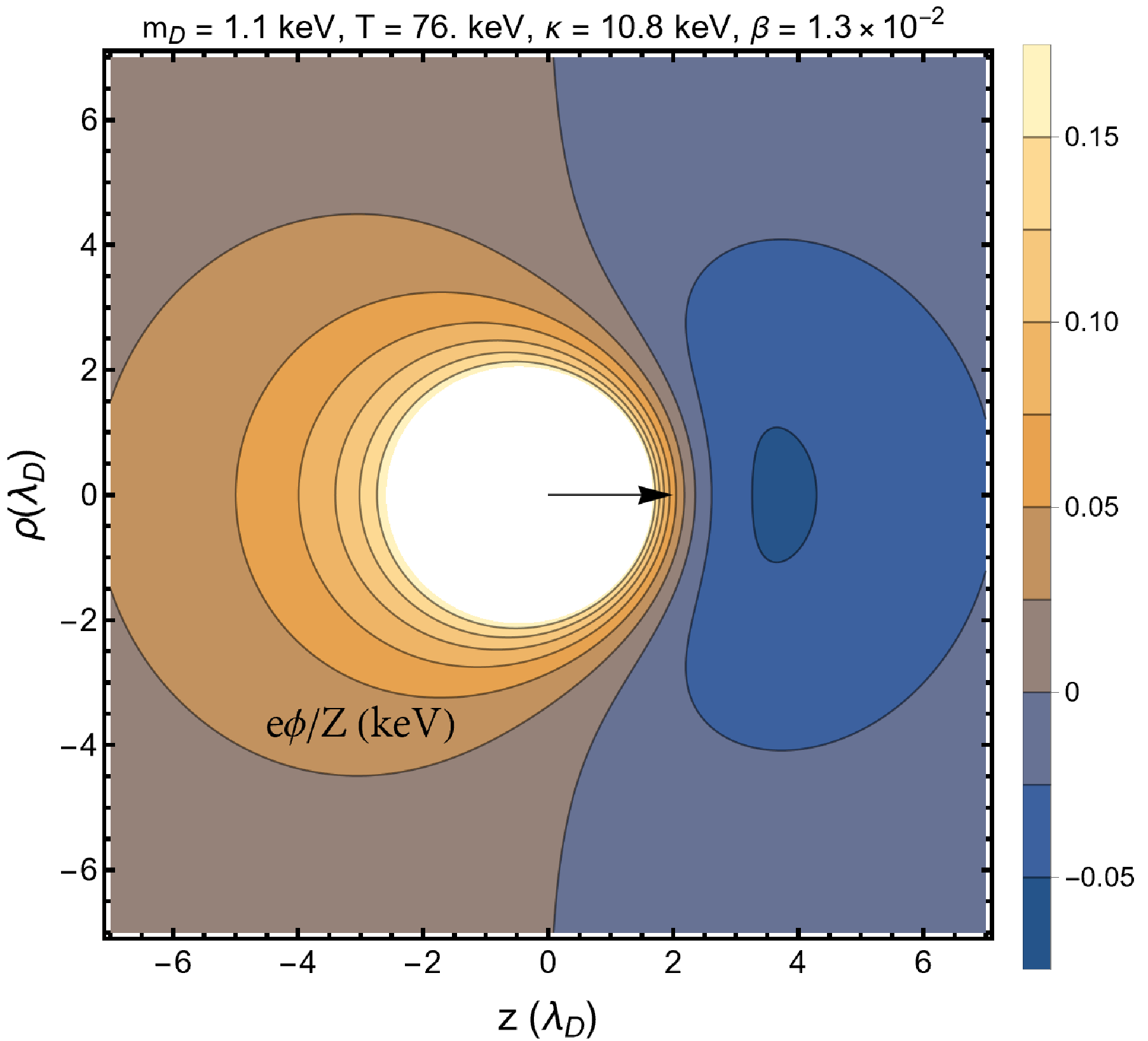}}
 \caption{Two dimensional plot of the total potential \req{eq:pos_point} scaled with Z, at $T=74\,$keV. The potential is cylindrically symmetric about the direction of motion $\boldsymbol{\hat{z}}$, which is indicated by a black arrow. The direction transverse to the motion is $\rho$. The sign of the damped-dynamic correction \req{eq:pos_point_DDS} changes sign due to the cosine term. \radapt{Grayson:2024okq}}
 \label{fig:numericalComp}
\end{figure} 

Figure~\ref{fig:dynamiclinear} demonstrates that the damped-dynamic response in the analytic approximation \req{eq:pos_point_DDS} (shown as orange dashed line) is sufficient to approximate the full numerical solution (blue dashed line) found by numerical integration of \req{eq:potent}. The temperature $T = 100$\, keV, above our upper limit of BBN temperatures, is chosen to relate to the dynamic screening result found in~\cite{Hwang:2021kno}. Our analytic solution differs from the numerical result in Fig.~4 of \cite{Hwang:2021kno} by a factor of $\sqrt{2}$ and is horizontally flipped. This reflection is due to a difference in convention in the permittivity, as seen in \req{eq:potentk}. We can see that dynamic screening is slightly stronger at large distances than damped screening, as expected. Damped and undamped screening are very similar at short distances, which is relevant to thermonuclear reaction rates. 

Dynamic screening in both the damped and undamped cases predicts less screening behind and more in front of the moving nucleus than static screening. This is shown in the two-dimensional plot \rf{fig:numericalComp}, of the total potential in plasma at $T=76\,$keV. This effect was previously observed for subsonic screening in electron-ion-dust plasmas ~\cite{Stenflo:1973,Shukla:2002ppcf,Lampe:2000pop}. As a result, a negative polarization charge builds up in front of the nucleus. The small negative potential in front alters the potential energy between light nuclei, possibly changing the equilibrium distribution of light elements in the primordial Universe plasma. This effect is much larger in the undamped case and is known in some cases to lead to the formation of dust crystals~\cite{Shukla:1996ccc}. 

We calculate the potential of light nuclei in the primordial Universe electron-positron plasma by Fourier transforming the screened scalar potential $\phi$ of a single traveling nuclei \req{eq:phi}
\begin{equation}\label{eq:potent}
 \phi(t,\boldsymbol{x}) = \int \frac{d^4k}{(2\pi)^4} e^{-i\omega t+ i\boldsymbol{k}\cdot\boldsymbol{x}} \frac{\widetilde{\rho}_\text{ext}(\omega,\boldsymbol{k})}{\varepsilon_\parallel(\omega,\boldsymbol{k})(\boldsymbol{k}^2-\omega^2) }\,,
\end{equation}
where $\widetilde{\rho}_{\text{ext}}(\omega,\boldsymbol{k})$ is the Fourier-transformed charge distribution of nuclei traveling at a constant velocity, and $\varepsilon_\parallel(\omega,\boldsymbol{k})$ is the longitudinal relative permittivity. The relative permittivity can be written in terms of the polarization tensor as
\begin{equation}\label{eq:epsilon}
 \varepsilon_\parallel(\omega,\boldsymbol{k})= \left(\frac{\Pi_{\parallel}(\omega,\boldsymbol{k})}{ \omega^2}+1\right)\,.
\end{equation}

In the linear response framework \req{eq:ohm}, the electromagnetic field still obeys the principle of superposition, so the potential between two nuclei can be inferred simply from the potential of a single nucleus. 

We can perform the $\omega$ integration in \req{eq:potent} using the delta function in the definition of the external charge distribution, whose effect is to set $\omega = \boldsymbol{\beta_{\text{N}}}\cdot \boldsymbol{k}$ where $\boldsymbol{\beta}_N = \boldsymbol{v}_N/c$ is the nuclei velocity. Then we have
\begin{equation}\label{eq:potentk}
 \phi(t,\boldsymbol{x}) = Ze\int \frac{d^3\boldsymbol{k}}{(2\pi)^3} e^{ i\boldsymbol{k}\cdot(\boldsymbol{x}-\boldsymbol{\beta_{\text{N}}} t)} \frac{ e^{-\boldsymbol{k}^2\frac{R^2}{4}}}{\boldsymbol{k}^2\varepsilon_\parallel(-\boldsymbol{\beta_{\text{N}}} \cdot \boldsymbol{k},\boldsymbol{k}) }\,,
\end{equation}
where $R$ is the Gaussian radius parameter.
In nonrelativistic approximation the Lorentz factor $\gamma \approx 1$ and we use the convention $\varepsilon_\parallel(-\boldsymbol{\beta_{\text{N}}} \cdot \boldsymbol{k},\boldsymbol{k})$ used in~\cite{Montgomery:1970jpp,Stenflo:1973,Shukla:2002ppcf,Shukla:1996ccc} which gives the correct causality for the potential. This ensures that, without damping, the wake field occurs behind the moving nucleus.

\para{Reaction rate enhancement}
We use the same argument as \cite{Salpeter:1954nc} to find the enhancement factor due to damped-dynamic screening. The enhancement of a nuclear reaction process by screening is related to the WKB probability of tunneling through the Coulomb barrier
\begin{equation} \label{eq:penprob}
    P(E) = \exp{\left( - \frac{2\sqrt{2 \mu_r}}{\hbar c}\int_{R}^{r_c}dr \sqrt{U(r)-E}\right)}\,,
\end{equation}
often referred to as the penetration factor. $U(r)$ is the potential energy of the two colliding nuclei, $\mu_r$ is their reduced mass, $E$ is the relative energy of the collision, $R$ is the radius of the nucleus, and $r_c$ is the classical turning point. 

In the weak screening limit, the screening charge density varies on the scale of $\lambda_D$, which is here on the order of \AA ngstrom. The distance scales relevant for tunneling are between $R$ and $r_c$, on the order of $10\,$fm. This allows us to approximate the contribution to the potential energy from screening, $H(r)$ defined as
\begin{equation}
    H(r) \equiv U(r) - U_\text{vac}(r)\,,
\end{equation}
as constant over the integral in \req{eq:penprob} taking the value of \req{eq:pos_point_DDS} at the origin,
\begin{equation}
     H(0) = Z_1\phi_2(0) = Z_1 Z_2 \alpha \left(m_D - \frac{\beta_N m_D^2}{2 \kappa}\right)\,.
\end{equation}
Then, the screening effect reduces to a constant shift in the relative energy $E \rightarrow E+H(0)$. In this approximation, the enhancement to reaction rates can be represented by a single factor \cite{Salpeter:1954nc,Kravchuk:2014sps}
\begin{equation}\label{eq:DDSenhance}
   \mathcal{F} = \exp\left[\frac{H(0)}{T} \right]=\exp\left[\frac{Z_1 Z_2 \alpha}{T} \left(m_D - \frac{\beta_N m_D^2}{2 \kappa}\right)\right]\,.
\end{equation}
This result is only valid in the weak damping limit $\omega_p<\kappa$. The first term is the weak field screening result, and the second is the contribution of damped-dynamic screening. Due to the large damping rate in comparison to the Debye mass and the small velocities of nuclei \req{eq:vel} during BBN\index{Big-Bang!BBN}, the correction due to damped dynamic screening is small, changing $H(0)$ by $10^{-5}$. 
  
\section{Magnetism in the Plasma Universe}
\label{sec:mag:universe}
\subsection{Towards a theory of primordial pair magnetism}
\label{sec:theory}
\para{Overview of primordial magnetism}
Macroscopic domains of magnetic fields have been found in all astrophysical environments from compact objects (stars, planets, etc.); interstellar and intergalactic space; and surprisingly in deep extra-galactic void spaces. Considering the ubiquity of magnetic fields in the\index{magnetic!intergalactic fields} Universe~\cite{Giovannini:2017rbc,Giovannini:2003yn,Kronberg:1993vk}, we search for a common primordial mechanism for the origin of the diversity of magnetism observed today. In this chapter, IGMF will refer to experimentally observed intergalactic fields of any origin while primordial magnetic fields\index{magnetic!primordial fields} (PMF) refers to fields generated via primordial Universe processes possibly as far back as inflation.

IGMF are notably difficult to measure and difficult to explain. The bounds for IGMF at a length scale of $1{\rm\ Mpc}$ are today~\cite{Neronov:2010gir,Taylor:2011bn,Pshirkov:2015tua,Jedamzik:2018itu,Vernstrom:2021hru}
\begin{gather}
 \label{igmf}
 10^{-8}{\rm\ G}>B_\mathrm{IGMF}>10^{-16}{\rm\ G}\,.
\end{gather}
We note that generating PMFs with such large coherent length scales is nontrivial~\cite{Giovannini:2022rrl} though currently the length scales for PMFs are not well constrained~\cite{AlvesBatista:2021sln}. Faraday rotation from distant radio active galaxy nuclei (AGN)~\cite{Pomakov:2022cem} suggest that neither dynamo nor astrophysical processes would sufficiently account for the presence of magnetic fields in the Universe today if the IGMF strength was around the upper bound of $B_\mathrm{IGMF}\simeq30-60{\rm\ nG}$ as found in Ref.~\cite{Vernstrom:2021hru}. Such strong magnetic fields would then require that at least some portion of the IGMF arise from primordial sources that predate the formation of stars. The conventional elaboration of the origins for cosmic PMFs are detailed\index{Universe!primordial magnetic field} in~\cite{Gaensler:2004gk,Durrer:2013pga,AlvesBatista:2021sln}.

Magnetized baryon inhomogeneities which in turn could produce anisotropies in the cosmic microwave background (CMB)~\cite{Jedamzik:2013gua,Abdalla:2022yfr}\index{CMB}. We note that according to Jedamzik~\cite{Jedamzik:2020krr} the presence of a intergalactic magnetic field of $B_\mathrm{PMF}\simeq0.1{\rm\ nG}$ could be sufficient to explain the Hubble tension\index{Hubble!tension}.

\begin{figure}
 \centering
 \includegraphics[width=0.99\linewidth]{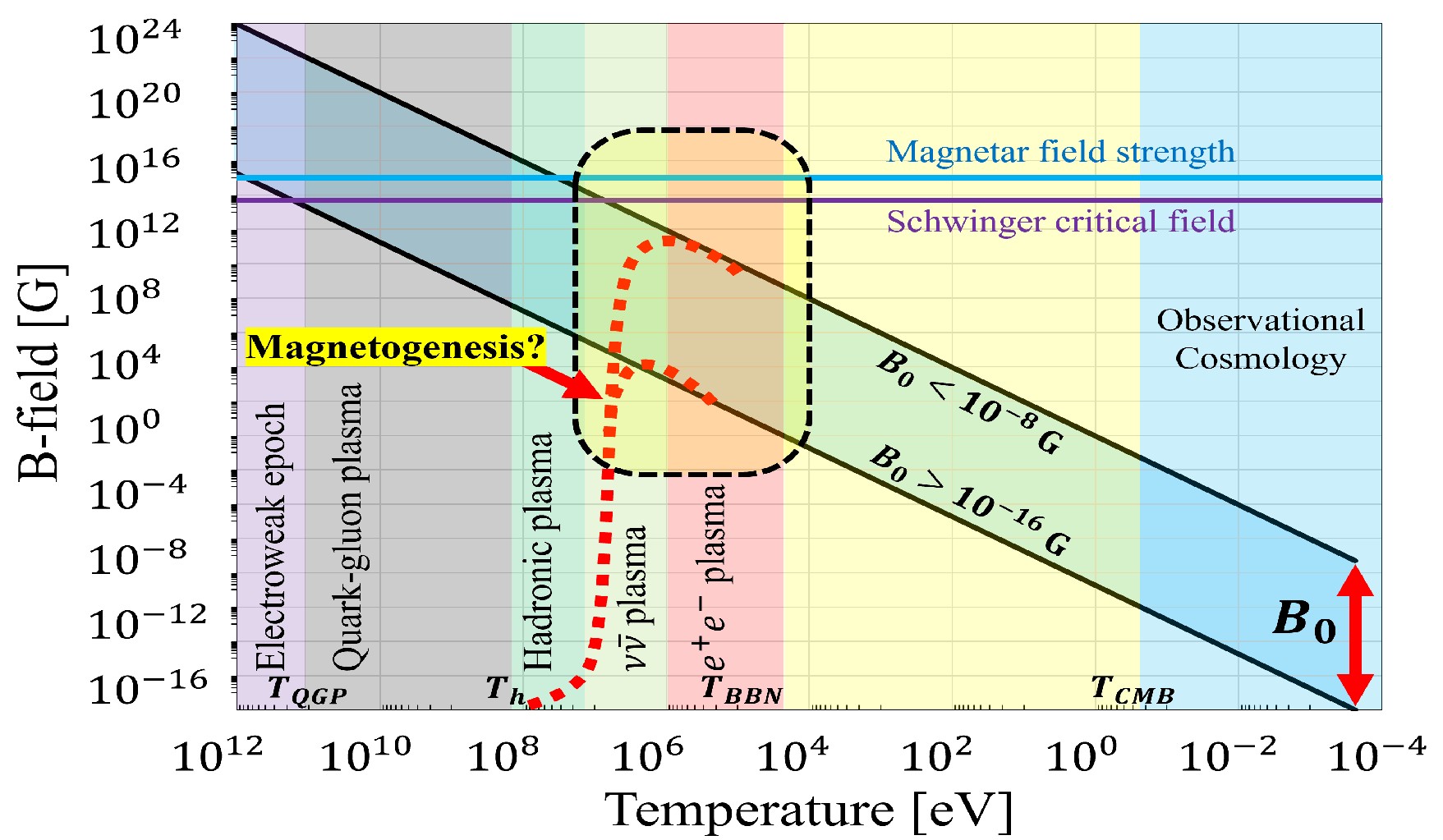}
 \caption{Qualitative plot of the primordial magnetic field strength over cosmic time. All figures are printed in temporal sequence in the expanding Universe beginning with high temperatures (and early times) on the left and lower temperatures (and later times) on the right. \cccite{Steinmetz:2023ucp}. \radapt{Rafelski:2023emw}}\index{primordial!magnetic field}
 \label{fig:pmf}
\end{figure}

Our motivating hypothesis is outlined qualitatively in \rf{fig:pmf} where PMF evolution is plotted over the temperature history of the Universe. The descending blue band indicates the range of possible PMF strengths. The different epochs of the primordial Universe according to $\Lambda\mathrm{CDM}$ are delineated by temperature. The horizontal lines mark two important scales: (a) the Schwinger critical field strength given by\index{Schwinger!critical magnetic field}
\begin{align}
 \label{crit:1}
 B_\mathrm{C} = \frac{m_{e}^{2}}{e}\simeq4.41\times10^{13}\,\mathrm{G}\,,
\end{align}
where electrodynamics is expected to display nonlinear characteristics and (b) the upper field strength seen in magnetars of $\sim10^{15}\,\mathrm{G}$. A schematic of magnetogenesis is drawn with the dashed red lines indicating spontaneous formation of the PMF within the primordial Universe plasma itself. The $e^{+}e^{-}$ era is notably the final epoch where antimatter exists in large quantities in the cosmos~\cite{Rafelski:2023emw}. We demonstrate that fundamental quantum statistical analysis can lead to further insights on the behavior of magnetized plasma, and show that the $e^\pm$ plasma is overall paramagnetic and yields a positive overall magnetization\index{magnetization}, which is contrary to the traditional assumption that matter-antimatter plasma lack significant magnetic responses.\index{plasma!magnetized}

As the primordial Universe cooled below temperature $T\!=\!m_{e}$ (the electron mass), the thermal electron and positron\index{plasma!electron-positron} comoving density depleted by over eight orders of magnitude. At $T_\mathrm{split}=20.3\keV$, the charged lepton asymmetry\index{lepton!asymmetry} (mirrored by baryon asymmetry\index{baryon!asymmetry} and enforced by charge neutrality)\index{charge neutrality} became evident as the surviving excess electrons persisted while positrons vanished entirely from the particle inventory of the primordial Universe due to annihilation.

\begin{figure}
 \centering
\includegraphics[width=0.80\linewidth]{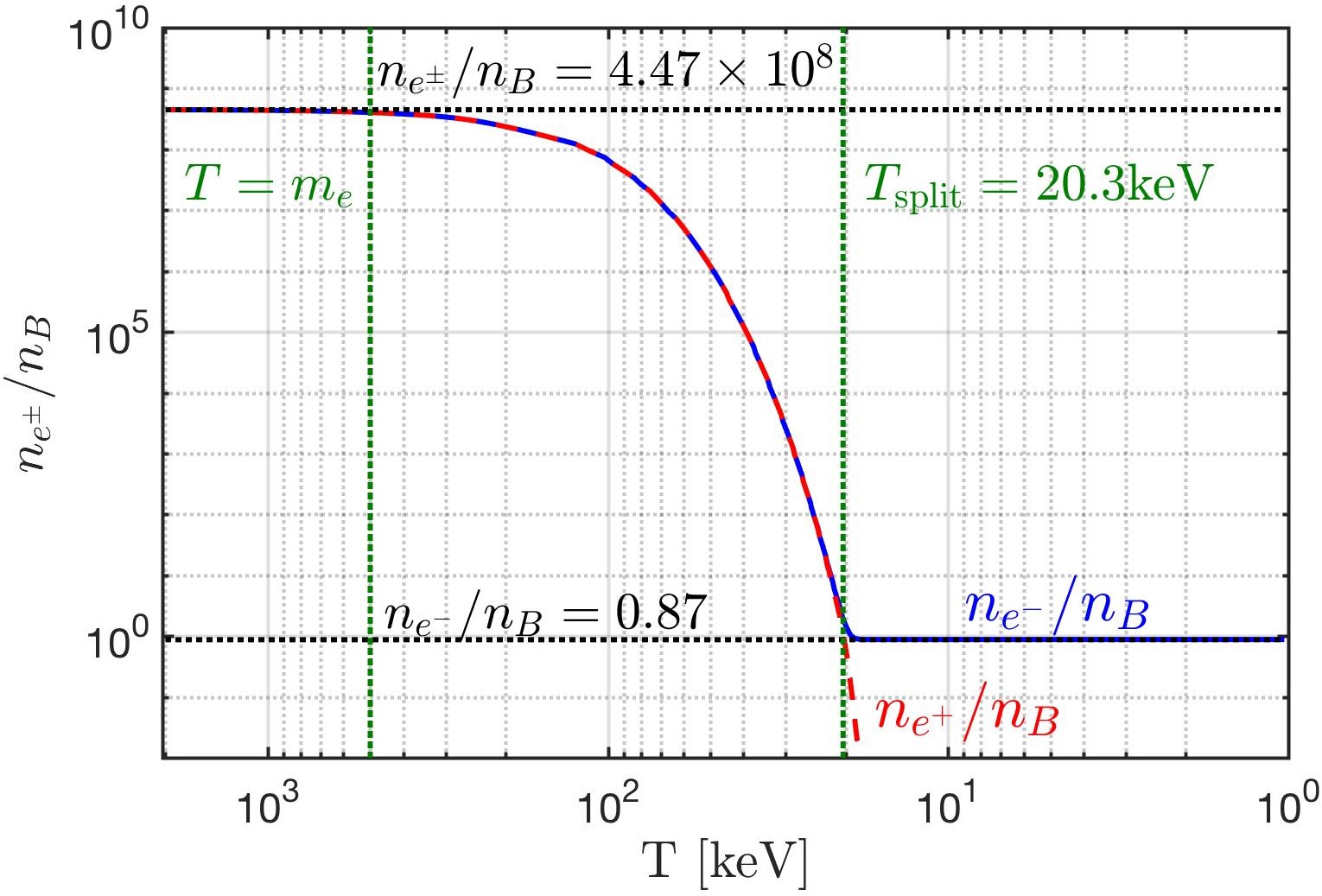}
 \caption{Number density of electron $e^{-}$ and positron $e^{+}$ to baryon ratio $n_{e^{\pm}}/n_{B}$ as a function of photon temperature in the primordial Universe. See \rsec{section:electron} for further details. In this work we measure temperature in units of energy (keV) thus we set the Boltzmann constant to $k_{B}=1$. \cccite{Steinmetz:2023nsc}}
 \label{fig:densityratio} \index{positron!abundance}
\end{figure}

The electron-to-baryon density\index{baryon!per electron ratio} ratio $n_{e^{-}}/n_{B}$ is shown in \rf{fig:densityratio} as the solid blue line while the positron-to-baryon ratio $n_{e^{+}}/n_{B}$ is represented by the dashed red line. These two lines overlap until the temperature drops below $T_\mathrm{split}=20.3\keV$ as positrons vanish from the primordial Universe, marking the end of the $e^{+}e^{-}$ plasma and the dominance of the electron-proton $(e^{-}p)$ plasma. 

The two vertical dashed green lines in \rf{fig:densityratio} denote temperatures $T\!=\!m_{e}\simeq511\keV$ and $T_\mathrm{split}=20.3\keV$. These results were obtained using charge neutrality and the baryon-to-photon content (entropy) of the primordial Universe; see details in~\cite{Rafelski:2023emw}, see also \rsec{section:electron}. The two horizontal black dashed lines denote the relativistic $T\gg m_e$ abundance of $n_{e^{\pm}}/n_{B}=4.47\times10^{8}$ and post-annihilation abundance of $n_{e^{-}}/n_{B}=0.87$. Above temperature $T\simeq85\keV$, the $e^{+}e^{-}$ primordial plasma density exceeded that of the Sun's core density $n_{e}\simeq6\times10^{26}{\rm\ cm}^{-3}$~\cite{Bahcall:2001smc}. 

Conversion of the dense $e^{+}e^{-}$ pair plasma into photons reheated the photon background~\cite{Birrell:2014uka} separating the photon and neutrino temperatures. The $e^{+}e^{-}$ annihilation and photon reheating period lasted no longer than an afternoon lunch break. Because of charge neutrality, the post-annihilation comoving ratio $n_{e^{-}}/n_{B}=0.87$~\cite{Rafelski:2023emw} is slightly offset from unity in~\rf{fig:densityratio} by the presence of bound neutrons in $\alpha$ particles and other neutron containing light elements produced during BBN epoch.

The abundance of baryons is itself fixed by the known abundance relative to photons~\cite{ParticleDataGroup:2022pth} and we employed the contemporary recommended value $n_B/n_\gamma=6.09\times 10^{-10}$. The resulting chemical potential\index{chemical potential!electron} needs to be evaluated carefully to obtain the behavior near to $T_\mathrm{split}=20.3\keV$ where the relatively small value of chemical potential $\mu$ rises rapidly so that positrons vanish from the particle inventory of the primordial Universe while nearly one electron per baryon remains. The detailed solution of this problem is found in \cite{Fromerth:2012fe,Rafelski:2023emw} leading to the results shown in \rf{fig:densityratio}.

\noindent To evaluate magnetic properties of the thermal $e^{+}e^{-}$ pair plasma we take inspiration from Ch. 9 of Melrose's treatise on magnetized plasmas~\cite{melrose2008quantum}. We focus on the bulk properties of thermalized plasmas in (near) equilibrium.\index{plasma!magnetization}

We consider a homogeneous magnetic field\index{magnetic!field} domain defined along the $z$-axis as
\begin{gather}
 \label{homoB:1}
 \bb{B}=(0,\,0,\,B)\,,
\end{gather}
with magnetic field\index{magnetic!fields} magnitude $|\bb{B}|=B$. Following \cite{Steinmetz:2018ryf}, we reprint the microscopic energy of the charged relativistic fermion within a homogeneous magnetic field given by
\begin{align}
 \label{cosmokgp}
 E^{n}_{\sigma,s}(p_{z},{B})=\sqrt{m_{e}^{2}+p_{z}^{2}+e{B}\left(2n+1+\frac{g}{2}\sigma s\right)}\,,
\end{align}
where $n\in0,1,2,\ldots$ is the Landau orbital quantum number\index{Landau levels}, $p_{z}$ is the momentum parallel to the field axis and the electric charge is $e\equiv q_{e^{+}}=-q_{e^{-}}$. The index $\sigma$ in \req{cosmokgp} differentiates electron $(e^{-};\ \sigma=+1)$ and positron $(e^{+};\ \sigma=-1)$ states. The index $s$ refers to the spin along the field axis: parallel $(\uparrow;\ s=+1)$ or anti-parallel $(\downarrow;\ s=-1)$ for both particle and antiparticle species.

\begin{figure}
 \centering
 \includegraphics[width=0.8\linewidth]{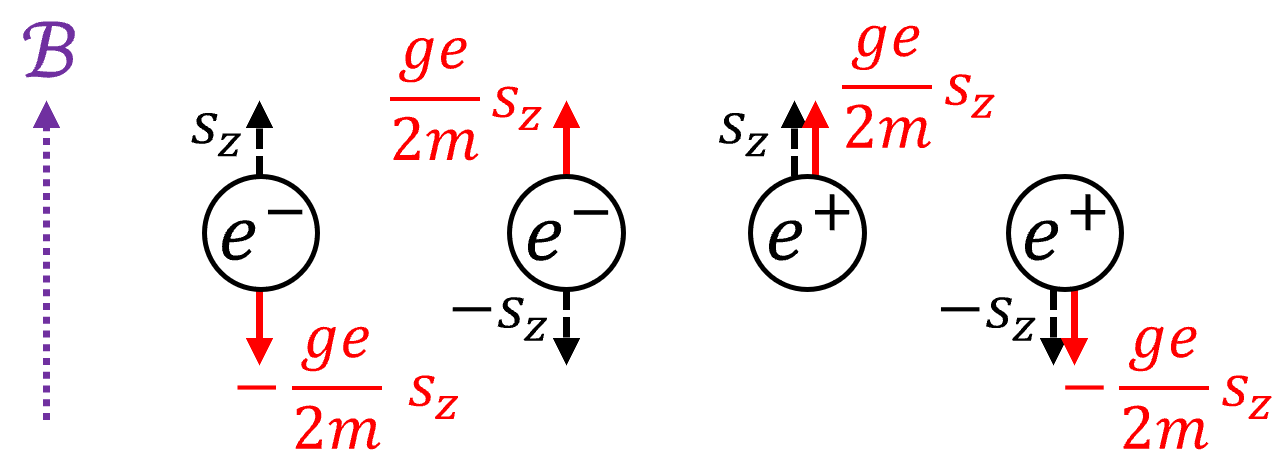}\Bstrut\\
 \begin{tabular}{ r|c|c| }
 \multicolumn{1}{r}{}
 & \multicolumn{1}{c}{aligned: $s=+1$}
 & \multicolumn{1}{c}{anti-aligned: $s=-1$} \\
 \cline{2-3}
 electron: $\sigma=+1$ & $U_{\rm Mag}>0$ & $U_{\rm Mag}<0$ \TBstrut\\
 \cline{2-3}
 positron: $\sigma=-1$ & $U_{\rm Mag}<0$ & $U_{\rm Mag}>0$ \TBstrut\\
 \cline{2-3}
 \end{tabular}\\
 \caption{Organizational schematic of matter-antimatter $(\sigma)$ and polarization $(s)$ states with respect to the sign of the nonrelativistic magnetic dipole energy $U_{\rm Mag}$ obtainable from~\req{cosmokgp}. \cccite{Steinmetz:2023nsc}}
 \label{fig:schematic}
\end{figure}

The reason \req{cosmokgp} distinguishes between electrons and positrons is to ensure the correct nonrelativistic limit for the magnetic dipole energy is reached. Following the conventions found in \cite{Tiesinga:2021myr}, we set the gyro-magnetic\index{g-factor} factor $g\equiv g_{e^{+}}=-g_{e^{-}}>0$ such that electrons and positrons have opposite $g$-factors and opposite magnetic moments relative to their spin; see \rf{fig:schematic}.

We recall the conventions established in \rsec{sec:flrw}. As the Universe undergoes the isotropic expansion, the temperature gradually decreases as $T\propto1/a(t)$, where $a(t)$ represents the scale factor\index{scale factor}. The assumption is made that the magnetic flux\index{magnetic!flux} is conserved over comoving surfaces, implying that the primordial relic field is expected to dilute as $B\propto1/a(t)^{2}$~\cite{Rafelski:2023emw}. Conservation of magnetic flux requires that the magnetic field\index{magnetic!fields} through a comoving surface $L_{0}^{2}$ remain unchanged. The magnetic field strength under expansion~\cite{Durrer:2013pga} starting at some initial time $t_{0}$ is then given by\index{magnetic!cosmic field scaling}
\begin{gather}
 \label{bscale}
 B(t)=B_{0}\frac{a^{2}_{0}}{a^{2}(t)}\rightarrow B(z)=B_{0}\left(1+z\right)^{2}\,,
\end{gather}
where $B_{0}$ is the comoving value obtained from the contemporary value of the magnetic field today. Magnetic fields in the cosmos generated through mechanisms such as dynamo or astrophysical sources do not follow this scaling~\cite{Pomakov:2022cem}. It is only in deep intergalactic space where matter density is low that magnetic fields are preserved (and thus uncontaminated) over cosmic time.

From \req{tscale} and \req{bscale} there emerges a natural ratio of interest which is conserved over cosmic expansion 
\begin{gather}
 \label{tbscale}
 b\equiv\frac{e{B}(t)}{T^{2}(t)}=\frac{e{B}_{0}}{T_{0}^{2}}\equiv b_0={\rm\ const.}\\
 10^{-3}>b_{0}>10^{-11}\,,
\end{gather}
given in natural units\index{natural units} ($c=\hbar=k_{B}=1$). We computed the bounds for this cosmic magnetic scale ratio by using the present day IGMF observations given by \req{igmf} and the present CMB\index{CMB} temperature $T_{0}=2.7{\rm\ K}\simeq2.3\times10^{-4}\eV$~\cite{Planck:2018vyg}.

\para{Eigenstates of magnetic moment in cosmology}
\label{sec:protection}
As statistical properties depend on the characteristic Boltzmann factor $E/T$, another interpretation of \req{tbscale} in the context of energy eigenvalues (such as those given in \req{cosmokgp}) is the preservation of magnetic moment energy relative to momentum under adiabatic cosmic expansion. The Boltzmann statistical factor is given by
\begin{alignat}{1}
 \label{Boltz} x\equiv\frac{E}{T}\,.
\end{alignat}
We can explore this relationship for the magnetized system explicitly by writing out \req{Boltz} using the KGP energy eigenvalues written in \req{cosmokgp} as
\begin{alignat}{1}
 \label{XExplicit} x_{\sigma,s}^{n} = \frac{E_{\sigma,s}^{n}}{T} = \sqrt{\frac{m_{e}^{2}}{T^{2}}+\frac{p_{z}^{2}}{T^{2}}+\frac{eB}{T^{2}}\left(2n+1+\frac{g}{2}\sigma s\right)}\,.
\end{alignat}

Introducing the expansion scale factor $a(t)$ via \req{tscale}, \req{bscale} and \req{tbscale}. The Boltzmann factor can then be written as
\begin{alignat}{1}
 \label{xscale:1} x_{\sigma,s}^{n}(a(t)) = \sqrt{\frac{m_{e}^{2}}{T^{2}(t_{0})}\frac{a(t)^{2}}{a_{0}^{2}}+\frac{p_{z,0}^{2}}{T_{0}^{2}}+\frac{eB_{0}}{T_{0}^{2}}\left(2n+1+\frac{g}{2}\sigma s\right)}\,.
\end{alignat}
This reveals that only the mass contribution is dynamic over cosmological time. The constant of motion $b_{0}$ defined in \req{tbscale} is seen as the coefficient to the Landau and spin portion of the energy. For any given eigenstate, the mass term drives the state into the nonrelativistic limit while the momenta and magnetic contributions are frozen by initial conditions. 

In comparison, the Boltzmann factor for the DP energy eigenvalues are given by
\begin{alignat}{1}
 \label{xscaledp:1} x_{\sigma,s}^{n}\vert_\mathrm{DP} = \sqrt{\left(\sqrt{\frac{m_{e}^{2}}{T^{2}}+\frac{eB}{T^{2}}\left(2n+1+\sigma s\right)}+\frac{eB}{2m_{e}T}\left(\frac{g}{2}-1\right)\sigma s\right)^{2}+\frac{p_{z}^{2}}{T^{2}}}\,,
\end{alignat}
which scales during FLRW\index{cosmology!FLRW} expansion as
\begin{equation}
 \label{xscaledp:2} x_{\sigma,s}^{n}(a(t))\vert_\mathrm{DP} =  \sqrt{\left(\sqrt{\frac{m_{e}^{2}}{T_{0}^{2}}\frac{a(t)^{2}}{a_{0}^{2}}+\frac{eB_{0}}{T_{0}^{2}}\left(2n+1+\sigma s\right)}+\frac{eB_{0}}{2m_{e}T_{0}}\frac{a_{0}}{a(t)}\left(\frac{g}{2}-1\right)\sigma s\right)^{2}+\frac{p_{z,0}^{2}}{T_{0}^{2}}}\,.
\end{equation}
While the above expression is rather complicated, we note that the KGP~\req{xscale:1} and DP~\req{xscaledp:1} Boltzmann factors both reduce to the Schroedinger-Pauli limit as $a(t)\rightarrow\infty$, thereby demonstrating that the total magnetic moment is protected under the adiabatic expansion of the primordial Universe.

Higher order non-minimal magnetic contributions can be introduced to the Boltzmann factor such as $\sim(e/m)^{2}B^{2}/T^{2}$. The reasoning above suggests that these terms are suppressed over cosmological time driving the system into minimal electromagnetic coupling with the exception of the anomalous magnetic moment. It is interesting to note that cosmological expansion then serves to `smooth out' the characteristics of more complex electrodynamics, erasing them from a statistical perspective in favor of minimal-like dynamics.

\para{Magnetized fermion partition function}
\label{sec:partition}
\noindent To obtain a quantitative description of the above evolution, we study the bulk properties of the relativistic charged/magnetic gases in a nearly homogeneous and isotropic primordial Universe via the thermal Fermi-Dirac or Bose distributions\index{Bose!distribution}\index{Fermi!distribution} .

The grand partition function\index{partition function} for the relativistic Fermi-Dirac distribution is given by the standard definition~\cite{Elze:1980er}
\begin{align}
 \label{part:1} \ln\mathcal{Z}_\mathrm{total} &= \sum_{\alpha}\ln\left(1+\Upsilon_{\alpha_{1}\ldots\alpha_{m}}\exp\left(-\frac{E_{\alpha}}{T}\right)\right)\,,\\
 \Upsilon_{\alpha_{1}\ldots\alpha_{m}} &= \lambda_{\alpha_{1}}\lambda_{\alpha_{2}}\ldots\lambda_{\alpha_{m}}\,,
\end{align}
where we are summing over the set all relevant quantum numbers $\alpha=(\alpha_{1},\alpha_{2},\ldots,\alpha_{m})$. We note here the generalized the fugacity\index{fugacity} $\Upsilon_{\alpha_{1}\ldots\alpha_{m}}$ allowing for any possible deformation caused by pressures effecting the distribution of any quantum numbers. In general, $\Upsilon=1$ represents the maximum entropy and corresponds to the normal Fermi distribution. The deviation of $\Upsilon\neq1$ represents the configurations of reduced entropy without pulling the system off a thermal temperature. Inhomogeneity can arise from the influence of other forces on the gas such as gravitational forces. This is precisely the kind of behavior that may arise in the $e^{\pm}$ epoch as the dominant photon thermal bath keeps the Fermi gas in thermal equilibrium while spatial nonequilibria could spontaneously develop.

In the case of the Landau\index{Landau levels} problem, there is an additional summation over $\widetilde{G}$, which represents the occupancy of Landau states~\cite{greiner2012thermodynamics}, which are matched to the available phase space within $\Delta p_{x}\Delta p_{y}$. If we consider the orbital Landau quantum number $n$ to represent the transverse momentum $p_{T}^{2}=p_{x}^{2}+p_{y}^{2}$ of the system, then the relationship that defines $\widetilde{G}$ is given by
\begin{alignat}{1}
 \label{phase:1} \frac{L^{2}}{(2\pi)^{2}}\Delta p_{x}\Delta p_{y}=\frac{eBL^{2}}{2\pi}\Delta n\,,\qquad\widetilde{G}=\frac{eBL^{2}}{2\pi}\,.
\end{alignat}
The summation over the continuous $p_{z}$ is replaced with an integration and the double summation over $p_{x}$ and $p_{y}$ is replaced by a single sum over Landau orbits
\begin{alignat}{1}
 \label{phase:2}
 \sum_{p_{z}}\rightarrow\frac{L}{2\pi}\int^{+\infty}_{-\infty}dp_{z}\,,\qquad\sum_{p_{x}}\sum_{p_{y}}\rightarrow\frac{eBL^{2}}{2\pi}\sum_{n}\,,
\end{alignat}
where $L$ defines the boundary length of our considered volume $V=L^{3}$.

The partition function of the $e^{+}e^{-}$ plasma can be understood as the sum of four gaseous species
\begin{align}
 \label{partition:0} 
 \ln\mathcal{Z}_{e^{+}e^{-}}=\ln\mathcal{Z}_{e^{+}}^{\uparrow}+\ln\mathcal{Z}_{e^{+}}^{\downarrow}+\ln\mathcal{Z}_{e^{-}}^{\uparrow}+\ln\mathcal{Z}_{e^{-}}^{\downarrow}\,,
\end{align}
of electrons and positrons of both polarizations $(\uparrow\downarrow)$. The change in phase space written in \req{phase:2} modifies the magnetized $e^{+}e^{-}$ plasma partition function from \req{part:1} into
\begin{gather}
 \label{partition:1}
 \ln\mathcal{Z}_{e^{+}e^{-}}=\frac{e{B}V}{(2\pi)^{2}}\sum_{\sigma}^{\pm1}\sum_{s}^{\pm1}\sum_{n=0}^{\infty}\int_{-\infty}^{\infty}\mathrm{d}p_{z}\left[\ln\left(1+\lambda_{\sigma}\xi_{\sigma,s}\exp\left(-\frac{E_{\sigma,s}^{n}}{T}\right)\right)\right]\,\\
 \label{partition:2} 
 \Upsilon_{\sigma,s} =\lambda_{\sigma}\xi_{\sigma,s} = \exp{\frac{\mu_{\sigma}+\zeta_{\sigma,s}}{T}}\,,
\end{gather}
where the energy eigenvalues $E_{\sigma,s}^{n}$ are given in \req{cosmokgp}. The index $\sigma$ in \req{partition:1} is a sum over electron and positron states while $s$ is a sum over polarizations. The index $s$ refers to the spin along the field axis: parallel $(\uparrow;\ s=+1)$ or anti-parallel $(\downarrow;\ s=-1)$ for both particle and antiparticle species.

We are explicitly interested in small asymmetries such as baryon excess over antibaryons, or one polarization over another. These are described by \req{partition:2} as the following two fugacities\index{fugacity!polarization}:
\begin{itemize}
 \item[(a)] Chemical fugacity $\lambda_{\sigma}$
 \item[(b)] Polarization fugacity $\xi_{\sigma,s}$
\end{itemize}
For matter $(e^{-};\ \sigma=+1)$ and antimatter $(e^{+};\ \sigma=-1)$ particles, a nonzero relativistic chemical potential\index{chemical potential} $\mu_{\sigma}=\sigma\mu$ is caused by an imbalance of matter and antimatter. While the primordial electron-positron plasma era was overall charge neutral\index{charge neutrality}, there was a small asymmetry in the charged leptons (namely electrons) from baryon asymmetry~\cite{Fromerth:2012fe,Canetti:2012zc}\index{baryon!asymmetry} in the primordial Universe. Reactions such as $e^{+}e^{-}\leftrightarrow\gamma\gamma$ constrains the chemical potential of electrons and positrons~\cite{Elze:1980er} as 
\begin{align}
 \label{cpotential}
 \mu\equiv\mu_{e^{-}}=-\mu_{e^{+}}\,,\qquad
 \lambda\equiv\lambda_{e^{-}}=\lambda_{e^{+}}^{-1}=\exp\frac{\mu}{T}\,,
\end{align}
where $\lambda$ is the chemical fugacity of the system.

We can then parameterize the chemical potential of the $e^{+}e^{-}$ plasma as a function of temperature $\mu\rightarrow\mu(T)$ via the charge neutrality of the primordial Universe which implies
\begin{align}
 \label{chargeneutrality}
 n_{p}=n_{e^{-}}-n_{e^{+}}=\frac{1}{V}\lambda\frac{\partial}{\partial\lambda}\ln\mathcal{Z}_{e^{+}e^{-}}\,.
\end{align}
In \req{chargeneutrality}, $n_{p}$ is the observed total number density of protons in all baryon species. The chemical potential defined in \req{cpotential} is obtained from the requirement that the positive charge of baryons (protons, $\alpha$ particles, light nuclei produced after BBN) is exactly and locally compensated by a tiny net excess of electrons over positrons.

We then introduce a novel polarization fugacity\index{fugacity!polarization} $\xi_{\sigma,s}$ and polarization potential $\zeta_{\sigma,s}=\sigma s\zeta$. We propose the polarization potential follows analogous expressions as seen in \req{cpotential} obeying
\begin{align}
 \label{spotential}
 \zeta\equiv\zeta_{+,+}=\zeta_{-,-}\,,\quad\zeta=-\zeta_{\pm,\mp}\,,\quad\xi_{\sigma,s}\equiv\exp{\frac{\zeta_{\sigma,s}}{T}}\,.
\end{align}
An imbalance in polarization within a region of volume $V$ results in a nonzero polarization potential $\zeta\neq0$: {\color{black}Please note that $\zeta$ in this Chapter should not be confused with $\zeta_\gamma$, the ratio of baryons to photons, and the interaction strength parameter introduced in \req{etaCTY}, both $\zeta$-s used elsewhere in this review but not in this Chapter}. Conveniently since antiparticles have opposite signs of charge and magnetic moment, the same magnetic moment is associated with opposite spin orientations. A completely particle-antiparticle symmetric magnetized plasma will have therefore zero total angular momentum.

\para{Euler-Maclaurin integration}
\label{sec:eulermac}
\noindent Before we proceed with the Boltzmann distribution\index{Boltzmann!distribution} approximation which makes up the bulk of our analysis, we will comment on the full Fermi-Dirac\index{Fermi!distribution} distribution analysis. The Euler-Maclaurin\index{Euler-Maclaurin integration} formula~\cite{abramowitz1988handbook} is used to convert the summation over Landau levels $n$ into an integration given by
\begin{equation}
 \label{eulermaclaurin}\sum^{b}_{n=a}f(n)-\int^{b}_{a}f(x)dx = \frac{1}{2}\left(f(b)+f(a)\right) 
 +\sum_{i=1}^{j}\frac{b_{2i}}{(2i)!}\left(f^{(2i-1)}(b)-f^{(2i-1)}(a)\right)+R(j)\,,
\end{equation}
where $b_{2i}$ are the Bernoulli numbers and $R(j)$ is the error remainder defined by integrals over Bernoulli polynomials. The integer $j$ is chosen for the level of approximation that is desired. Euler-Maclaurin integration is rarely convergent, and in this case serves only as an approximation within the domain where the error remainder is small and bounded; see~\cite{greiner2012thermodynamics} for the nonrelativistic case. In this analysis, we keep the zeroth and first order terms in the Euler-Maclaurin formula. We note that regularization of the excess terms in \req{eulermaclaurin} is done in the context of strong field QED~\cite{greiner2008quantum} though that is outside our scope.

Using \req{eulermaclaurin} allows us to convert the sum over $n$ quantum numbers in \req{partition:1} into an integral. Defining
\begin{alignat}{1}
 \label{Func} f_{\sigma,s}^{n}=\ln\left(1+\Upsilon_{\sigma,s}\exp\left(-\frac{E_{\sigma,s}^{n}}{T}\right)\right)\,,
\end{alignat}
\req{partition:1} for $j=1$ becomes
\begin{equation}
 \label{PartFuncTwo} \ln\mathcal{Z}_{e^{+}e^{-}} = \frac{e{B}V}{(2\pi)^{2}}\sum_{\sigma,s}^{\pm1}\int_{-\infty}^{+\infty}dp_{z}
 \left(\int_{0}^{+\infty}dn f_{\sigma,s}^{n} + \frac{1}{2}f_{\sigma,s}^{0} + \frac{1}{12}\frac{\partial f_{\sigma,s}^{n}}{\partial n}\bigg\rvert_{n=0} + R(1)\right)
\end{equation}
It will be useful to rearrange \req{cosmokgp} by pulling the spin dependency and the ground state Landau orbital into the mass writing
\begin{gather}
 \label{effmass:1}
 E^{n}_{\sigma,s}={\tilde m}_{\sigma,s}\sqrt{1+\frac{p_{z}^{2}}{{\tilde m}_{\sigma,s}^{2}}+\frac{2e{B}n}{{\tilde m}_{\sigma,s}^{2}}}\,,\\
 \label{effmass:2}
 \varepsilon_{\sigma,s}^{n}(p_{z},{B})=\frac{E_{\sigma,s}^{n}}{{\tilde m}_{\sigma,s}}\,,\qquad{\tilde m}_{\sigma,s}^{2}=m_{e}^{2}+e{B}\left(1+\frac{g}{2}\sigma s\right)\,,
\end{gather}
where we introduced the dimensionless energy $\varepsilon^{n}_{\sigma,s}$ and effective polarized mass ${\tilde m}_{\sigma,s}$ which is distinct for each spin alignment and is a function of magnetic field strength ${B}$. The effective polarized mass ${\tilde m}_{\sigma,s}$ allows us to describe the $e^{+}e^{-}$ plasma with the spin effects almost wholly separated from the Landau characteristics of the gas when considering the plasma's thermodynamic properties.

With the energies written in this fashion, we recognize the first term in \req{PartFuncTwo} as mathematically equivalent to the free particle fermion partition function with a re-scaled mass $m_{\sigma,s}$. The phase-space relationship between transverse momentum and Landau orbits in \req{phase:1} and \req{phase:2} can be succinctly described by
\begin{gather}
 p_{T}^{2} \sim 2eBn\,,\qquad2p_{T}dp_{T} \sim 2eBdn\,,\qquad d\bb{p}^{3}=2\pi p_{T}dp_{T}dp_{z}\,,\\
 \frac{eBV}{(2\pi)^{2}}\int_{-\infty}^{+\infty}dp_{z}\int_{0}^{+\infty}dn \rightarrow \frac{V}{(2\pi)^{3}}\int d\bb{p}^{3}\,,
\end{gather}
which recasts the first term in \req{PartFuncTwo} as
\begin{align}
 \label{FreePart}
 \ln\mathcal{Z}_{e^{+}e^{-}} = \frac{V}{(2\pi)^{3}}\sum_{\sigma,s}^{\pm1}\int d\bb{p}^{3}\ln\left(1+\Upsilon_{\sigma,s}\exp{\left(-\frac{m_{\sigma,s}\sqrt{1+p^{2}/m_{\sigma,s}^{2}}}{T}\right)}\right)+\ldots\,.
\end{align}
As we will see in the following section, this separation of the `free-like' partition function can be reproduced in the Boltzmann distribution limit as well. This marks the end of the analytic analysis without approximations.

\para{Boltzmann approach to electron-positron plasma}
\label{sec:boltzmann}
\noindent Since we address the temperature interval $200\keV>T>20\keV$ where the effects of quantum Fermi statistics on the $e^{+}e^{-}$ pair plasma\index{plasma!electron-positron} are relatively small, but the gas is still considered relativistic, we will employ the Boltzmann approximation\index{Boltzmann!approximation} to the partition function in \req{partition:1}. However, we extrapolate our results for presentation completeness up to $T\simeq 4m_{e}$.

\begin{table}[ht]
 \centering
 \begin{tabular}{ r|c|c| }
 \multicolumn{1}{r}{}
 & \multicolumn{1}{c}{aligned: $s=+1$}
 & \multicolumn{1}{c}{anti-aligned: $s=-1$} \\
 \cline{2-3}
 electron: $\sigma=+1$ & $+\mu+\zeta$ & $+\mu-\zeta$ \TBstrut\\
 \cline{2-3}
 positron: $\sigma=-1$ & $-\mu-\zeta$ & $-\mu+\zeta$ \TBstrut\\
 \cline{2-3}
 \end{tabular}\\\,\Bstrut\\
 \caption{Organizational schematic of matter-antimatter $(\sigma)$ and polarization $(s)$ states with respect to the chemical $\mu$ and polarization $\zeta$ potentials as seen in~\req{partitionpower:2}. Companion to \rt{fig:schematic}.}
 \label{fig:org}
\end{table}

To simplify the partition function, we consider the expansion of the logarithmic function
\begin{align}
\ln\left(1+x\right)=\sum^{\infty}_{k=1}\frac{(-1)^{k+1}}{k}x^k,\qquad\mathrm{for}\,|x|<1\,.
\end{align}
The partition function shown in equation \req{partition:1} can be rewritten removing the logarithm as
\begin{gather}
\label{partitionpower:1}
\ln{\mathcal{Z}_{e^{+}e^{-}}}=\frac{e{B}V}{(2\pi)^{2}}\sum_{\sigma,s}^{\pm1}\sum_{n=0}^{\infty}\sum_{k=1}^{\infty}\int_{-\infty}^{+\infty}\mathrm{d}p_{z}
\frac{(-1)^{k+1}}{k}\exp\left({k\frac{\sigma\mu+\sigma s\zeta-{\tilde m}_{\sigma,s}\varepsilon^{n}_{\sigma,s}}{T}}\right)\,,\\
\label{bapprox} 
\sigma\mu+\sigma s\zeta-{\tilde m}_{\sigma,s}\varepsilon_{\sigma,s}^{n}<0\,,
\end{gather}
which is well behaved as long as the factor in \req{bapprox} remains negative. We evaluate the sums over $\sigma$ and $s$ as
\begin{align}
 \label{partitionpower:2}
 \notag\ln{\mathcal{Z}_{e^{+}e^{-}}}&=\frac{e{B}V}{(2\pi)^{2}}\sum_{n=0}^{\infty}\sum_{k=1}^{\infty}\int_{-\infty}^{+\infty}\mathrm{d}p_{z}\frac{(-1)^{k+1}}{k}\times\\
 \notag&\,\left(\exp\left(k\frac{+\mu+\zeta}{T}\right)\exp\left(-k\frac{{\tilde m}_{+,+}\varepsilon_{+,+}^{n}}{T}\right)\right.
 +\exp\left(k\frac{+\mu-\zeta}{T}\right)\exp\left(-k\frac{{\tilde m}_{+,-}\varepsilon_{+,-}^{n}}{T}\right)\qquad\\
 &+\exp\left(k\frac{-\mu-\zeta}{T}\right)\exp\left(-k\frac{{\tilde m}_{-,+}\varepsilon_{-,+}^{n}}{T}\right)
 +\left.\exp\left(k\frac{-\mu+\zeta}{T}\right)\exp\left(-k\frac{{\tilde m}_{-,-}\varepsilon_{-,-}^{n}}{T}\right)\right)\,.
\end{align}
We note from \rf{fig:schematic} that the first and fourth terms and the second and third terms share the same energies via
\begin{align}
 \label{partitionpower:3}
 \varepsilon_{+,+}^{n}=\varepsilon_{-,-}^{n}\,,\qquad
 \varepsilon_{+,-}^{n}=\varepsilon_{-,+}^{n}\,.\qquad
 \varepsilon_{+,-}^{n}<\varepsilon_{+,+}^{n}\,.
\end{align}
\req{partitionpower:3} allows us to reorganize the partition function with a new magnetization quantum number $s'$, which characterizes paramagnetic flux increasing states $(s'=+1)$ and diamagnetic flux decreasing states $(s'=-1)$. This recasts \req{partitionpower:2} as
\begin{equation}
 \label{partitionpower:4}
 \ln{\mathcal{Z}_{e^{+}e^{-}}}=\frac{e{B}V}{(2\pi)^{2}}\sum_{s'}^{\pm1}\sum_{n=0}^{\infty}\sum_{k=1}^{\infty}\int_{-\infty}^{+\infty}\mathrm{d}p_{z}\frac{(-1)^{k+1}}{k} \left[2\xi_{s'}\cosh\frac{k\mu}{T}\right]\exp\left(-k\frac{{\tilde m}_{s'}\varepsilon_{s'}^{n}}{T}\right)\,,
\end{equation}
with dimensionless energy $\varepsilon_{s'}^{n}$, polarization mass $\tilde{m}_{s'}$, and polarization $\zeta_{s'}$ redefined in terms of the moment orientation quantum number $s'$
\begin{gather}
 {\tilde m}_{s'}^{2}=m_{e}^{2}+e{B}\left(1-\frac{g}{2}s'\right)\,,\\
 \zeta\equiv\zeta_{+}=-\zeta_{-}\qquad\xi\equiv\xi_{+}=\xi_{-}^{-1}\,,\qquad\xi_{s'}=\xi^{\pm1}=\exp\left(\pm\frac{\zeta}{T}\right)\,.
\end{gather}

We introduce the modified Bessel function\index{Bessel function} $K_{\nu}$ (see Ch. 10 of~\cite{Letessier:2002ony}) of the second kind
\begin{gather}
\label{besselk}
K_{\nu}\left(\frac{m}{T}\right)=\frac{\sqrt{\pi}}{\Gamma(\nu-1/2)}\frac{1}{m}\left(\frac{1}{2mT}\right)^{\nu-1}
\int_{0}^{\infty}\mathrm{d}p\,p^{2\nu-2}\exp\left({-\frac{m\varepsilon}{T}}\right)\,,\\
\nu>1/2\,,\qquad\varepsilon=\sqrt{1+p^{2}/m^{2}}\,,
\end{gather}
allowing us to rewrite the integral over momentum in \req{partitionpower:4} as
\begin{align}
 \label{besselkint}
 \frac{1}{T}\int_{0}^{\infty}\!\!\mathrm{d}p_{z}\exp\!\left(\!{-\frac{k{\tilde m}_{s'}\varepsilon_{s'}^{n}}{T}}\!\right)\!=\!W_{1}\!\!\left(\frac{k{\tilde m}_{s'}\varepsilon_{s'}^{n}(0,{B})}{T}\right)\,.
\end{align}
The function $W_{\nu}$ serves as an auxiliary function of the form $W_{\nu}(x)=xK_{\nu}(x)$. The notation $\varepsilon(0,{B})$ in \req{besselkint} refers to the definition of dimensionless energy found in \req{effmass:2} with $p_{z}=0$.

Summation over the auxillary function $W_{\nu}$ can be replaced via Euler-Maclaurin\index{Euler-Maclaurin integration} integration~\req{eulermaclaurin} as
\begin{align}
\sum^{\infty}_{n=0}W_1(n)=\int^\infty_0\!\!dn\,W_1(n)&+\frac{1}{2}\bigg[W_1(\infty)+W_1(0)\bigg]
+\frac{1}{12}\bigg[\left.\frac{\partial W_1}{\partial n}\right|_{\infty}-\left.\frac{\partial W_1}{\partial n}\right|_{0}\bigg]+R(2)\,.
\end{align}
Using the properties of Bessel function we have
\begin{align}
\frac{\partial W_1(s',n)}{\partial n}=-\frac{k^2eB}{T^2}K_0\left({\frac{k}{T}\sqrt{\tilde{m}^2_{s'}+2eBn}}\right),\qquad W_1(\infty)=0,\qquad
\int^\infty_a\!\!dx\,x^2K_1(x)=a^2K_2(a)\,.
\end{align}
This yields
\begin{align}
 \sum^{\infty}_{n=0}W_1(s',n)
 &=\left(\frac{T^2}{k^2eB}\right)\left[\left(\frac{k\tilde{m}_{s'}}{T}\right)^2K_2\left(\frac{k\tilde m_{s'}}{T}\right)\right]+\frac{1}{2}\left[\left(\frac{k\tilde{m}_{s'}}{T}\right)K_1\left(\frac{k\tilde m_{s'}}{T}\right)\right]
 +\frac{1}{12}\left[\left(\frac{k^2eB}{T^2}\right)K_0\left(\frac{k\tilde m_{s'}}{T}\right)\right].
\end{align}

The standard Boltzmann distribution\index{Boltzmann!distribution} is obtained by summing only $k=1$ and neglecting the higher order terms. Therefore we can integrate the partition function over the summed Landau levels. After truncation of the series and error remainder (up to the first derivative $j=2$), the partition function \req{partitionpower:1} can then be written in terms of modified Bessel $K_{\nu}$ functions of the second kind and cosmic magnetic scale $b_{0}$, yielding
\begin{gather}
 \label{boltzmann}
 \ln\mathcal{Z}_{e^{+}e^{-}}\simeq\frac{T^{3}V}{\pi^{2}}\sum_{s'}^{\pm1}\left[\xi_{s'}\cosh{\frac{\mu}{T}}\right]
 \left(x_{s'}^{2}K_{2}(x_{s'})+\frac{b_{0}}{2}x_{s'}K_{1}(x_{s'})+\frac{b_{0}^{2}}{12}K_{0}(x_{s'})\right)\,,\\
 \label{xfunc}
 x_{s'}=\frac{{\tilde m}_{s'}}{T}=\sqrt{\frac{m_{e}^{2}}{T^{2}}+b_{0}\left(1-\frac{g}{2}s'\right)}\,.
\end{gather}
The latter two terms in \req{boltzmann} proportional to $b_{0}K_{1}$ and $b_{0}^{2}K_{0}$ are the uniquely magnetic terms present in powers of magnetic scale \req{tbscale} containing both spin and Landau orbital influences in the partition function. These are magnetic effects to, the orders $\mathcal{O}(eB)$ and $\mathcal{O}(eB)^2$ respectively. The $K_{2}$ term is analogous to the free Fermi gas~\cite{greiner2012thermodynamics} being modified only by spin effects.

This `separation of concerns' can be rewritten as
\begin{gather}
 \label{spin}
 \ln\mathcal{Z}_\mathrm{S}=\frac{T^{3}V}{\pi^{2}}\sum_{s'}^{\pm1}\left[\xi_{s'}\cosh{\frac{\mu}{T}}\right]\left(x_{s'}^{2}K_{2}(x_{s'})\right)\,,\\
 \label{spinorbit}
 \ln\mathcal{Z}_\mathrm{SO}=\frac{T^{3}V}{\pi^{2}}\sum_{s'}^{\pm}\left[\xi_{s'}\cosh{\frac{\mu}{T}}\right]
 \left(\frac{b_{0}}{2}x_{s'}K_{1}(x_{s'})+\frac{b_{0}^{2}}{12}K_{0}(x_{s'})\right)\,, 
\end{gather}
where the spin (S) and spin-orbit (SO) partition functions can be considered independently. When the magnetic scale $b_{0}$ is small, the spin-orbit term \req{spinorbit} becomes negligible, leaving only paramagnetic effects in \req{spin} due to spin. In the nonrelativistic limit, \req{spin} reproduces a quantum gas whose Hamiltonian is defined as the free particle (FP) Hamiltonian plus the magnetic dipole (MD) Hamiltonian which span two independent Hilbert spaces $\mathcal{H}_\mathrm{FP}\otimes\mathcal{H}_\mathrm{MD}$. The nonrelativistic limit will be further discussed in the following.

Writing the partition function as \req{boltzmann} instead of \req{partitionpower:1} has the additional benefit that the partition function remains finite in the free gas $({B}\rightarrow0)$ limit. This is because the free Fermi gas and \req{spin} are mathematically analogous to one another. As the Bessel $K_{\nu}$ functions are evaluated as functions of $x_{\pm}$ in \req{xfunc}, the `free' part of the partition $K_{2}$ is still subject to spin magnetization\index{magnetization} effects. In the limit where ${B}\rightarrow0$, the free Fermi gas is recovered in both the Boltzmann approximation\index{Boltzmann!approximation} $k=1$ and the general case $\sum_{k=1}^{\infty}$.

\para{Nonrelativistic limit of the magnetized partition function}
While we label the first term in \req{FreePart} as the `free' partition function, this is not strictly true as the partition function is dependent on the magnetic-mass we defined in \req{effmass:2}. When determining the magnetization of the quantum Fermi gas, derivatives of the magnetic field $B$ will not fully vanish on this first term which will resulting in an intrinsic magnetization which is distinct from the Landau levels.

This represents magnetization that arises from the spin magnetic energy rather than orbital contributions. To demonstrate this, we will briefly consider the weak field limit for $g=2$. The effective polarized mass for electrons is then
\begin{align}
 \label{MagMassPlus}
 \tilde{m}_{+}^{2}&=m_{e}^{2}\,,\\
 \label{MagMassMinus}
 \tilde{m}_{-}^{2}&=m_{e}^{2}+2eB\,,
\end{align}
with energy eigenvalues
\begin{align}
 \label{EPlus}
 E_{n}^{+}&=\sqrt{p_{z}^{2}+m_{e}^{2}+2eBn}\,,\\
 \label{EMinus}
 E_{n}^{-}&=\sqrt{\left(E_{n}^{+}\right)^{2}+2eB}\,.
\end{align}
The spin anti-aligned states in the nonrelativistic (NR) limit reduce to
\begin{align}
 \label{EMinusNR} E_{n}^{-}\vert_\mathrm{NR}\approx E_{n}^{+}\vert_\mathrm{NR}+\frac{eB}{m_{e}}\,.
\end{align}
This shift in energies is otherwise not influenced by summation over Landau quantum number $n$, therefore we can interpret this energy shift as a shift in the polarization potential from \req{spotential}. The polarization potential is then
\begin{align}
 \label{SpinChem} \zeta_{e}^{\pm}=\zeta_{e}\pm\frac{eB}{2m_{e}}\,,
\end{align}
allowing us to rewrite the partition function in \req{partitionpower:1} as
\begin{gather}
 \label{PartTotalNR} \ln\mathcal{Z}_{e^{-}}\vert_{NR}=\frac{eBV}{(2\pi)^{2}}\sum_{s'}^{\pm}\sum_{n=0}^{\infty}\sum_{k=1}^{\infty}\int_{-\infty}^{+\infty}dp_{z}\frac{(-1)^{k+1}}{k}2\cosh(k\beta\zeta_{e}^{s'})\lambda^{k}\exp(-k\epsilon_{n}/T)\,,\\
 \label{NREnergy} \epsilon_{n}=m_{e}+\frac{p_{z}^{2}}{2m_{e}}+\frac{eB}{2m_{e}}\left(n+1\right)\,.
\end{gather}

\req{PartTotalNR} is then the traditional NR quantum harmonic oscillator partition function with a spin dependant potential shift differentiating the aligned and anti-aligned states. We note that in this formulation, the spin contribution is entirely excised from the orbital contribution. Under Euler-Maclaurin integration\index{Euler-Maclaurin integration}, the now spin-independent Boltzmann factor can be further separated into `free' and Landau quantum parts as was done in \req{FreePart} for the relativistic case. We note however that the inclusion of anomalous magnetic moment spoils this clean separation.

 \para{Electron-positron chemical potential}
\index{chemical potential!electron}
\noindent We refer to developments present \rsec{section:electron}: Considering the temperature after neutrino freeze-out\index{neutrino!freeze-out}, the charge neutrality\index{charge neutrality} condition can be written as
\begin{align}
 \label{density_proton}
 \left(n_{e^{-}}-n_{e^{+}}\right)=n_{p}=X_p\,\left(\frac{n_{B}}{\sigma_{\gamma,e}}\right)\,\sigma_{\gamma,e},\qquad X_p\equiv\frac{n_p}{n_B}\,,
\end{align}
where $n_{p}$ and $n_B$ is the number density of protons and baryons\index{baryon!entropy ratio} respectively. The parameter $s_{\gamma,e}$ is the entropy density\index{entropy!density} which is primarily dominated by photons and electron(positrons) in this era. Due to the adiabatic expansion of the primordial Universe, the comoving entropy density is conserved, making the ratio
\begin{align}
 \frac{n_{B}}{\sigma_{\gamma,e}} = \mathrm{const.}\,,
\end{align}
a constant which can be measured today from the entropy content of the CMB\index{CMB} today~\cite{Fromerth:2012fe}. The proton-to-baryon\index{baryon!per proton ratio} ratio is slightly offset by the presence of neutrons.

In presence of a magnetic field\index{magnetic!fields} in the Boltzmann approximation, the charge neutrality condition \req{chargeneutrality} and \req{density_proton} becomes
\begin{gather}
 \label{chem}
 \sinh\frac{\mu}{T}=n_{p}\frac{\pi^{2}}{T^{3}}
 \left[\sum_{s'}^{\pm1}\xi_{s'}\!\left(\!x_{s'}^{2}K_{2}(x_{s'})\!+\!\frac{b_{0}}{2}x_{s'}K_{1}(x_{s'})\!+\!\frac{b_{0}^{2}}{12}K_{0}(x_{s'}\!)\!\right)\!\right]^{-1}\!.
\end{gather}
\req{chem} is fully determined by the right-hand side expression if the polarization fugacity\index{fugacity!polarization} is set to unity $\zeta=0$ implying no external bias to the number of polarizations except as a consequence of the difference in energy eigenvalues. In practice, the latter two terms in \req{chem} are negligible to chemical potential in the bounds of the primordial $e^{+}e^{-}$ plasma considered and only becomes relevant for extreme (see \rf{fig:chemicalpotential}) magnetic field strengths well outside our scope.

\begin{figure}
 \centering
\includegraphics[width=0.8\linewidth]{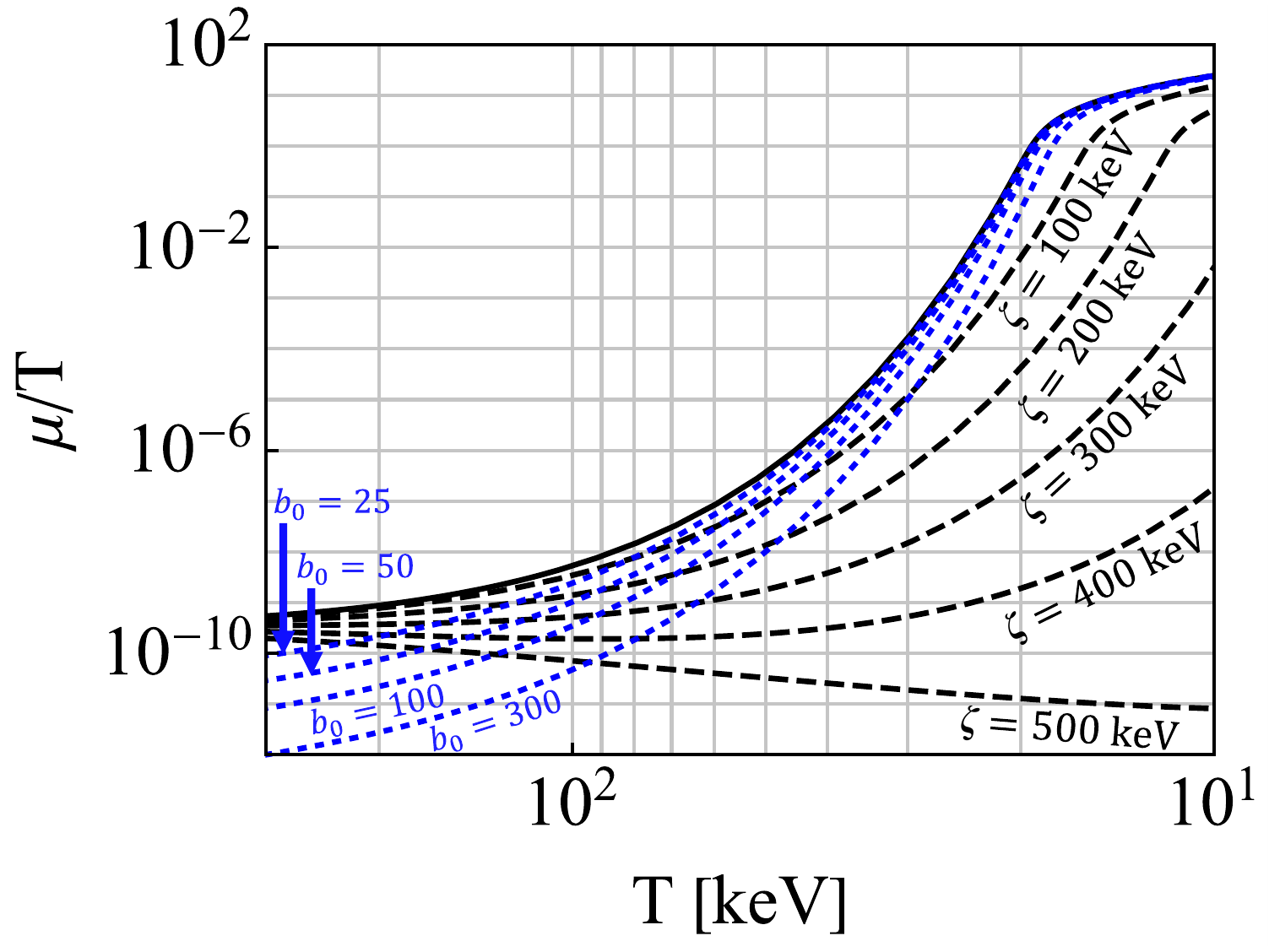}
 \caption{The chemical potential over temperature $\mu/T$ is plotted as a function of temperature with differing values of spin potential $\zeta$ and magnetic scale $b_{0}$. \cccite{Steinmetz:2023ucp}. \radapt{Steinmetz:2023nsc}}
 \label{fig:chemicalpotential}
\end{figure}

If there is no external magnetic field $b_{0}=0$ \req{chem} simplifies into
\begin{align}
 \label{simpchem:1}
 \sinh\frac{\mu}{T}=n_{p}\frac{\pi^{2}}{T^{3}}\left[2\cosh\frac{\zeta}{T}\left(\frac{m_{e}}{T}\right)^{2}K_{2}\left(\frac{m_{e}}{T}\right)\right]^{-1}\,.
\end{align}
In \rf{fig:chemicalpotential} we plot the chemical potential $\mu/T$ in \req{chem} and \req{simpchem:1} that characterizes the importance of the charged lepton asymmetry\index{lepton!asymmetry} as a function of temperature. Since the baryon (and thus charged lepton) asymmetry remains fixed, the suppression of $\mu/T$ at high temperatures indicates a large pair density which is seen explicitly in \rf{fig:densityratio}. The black line corresponds to the $b_{0}=0$ and $\zeta=0$ case. 

The para-diamagnetic contribution from \req{spinorbit} does not appreciably influence $\mu/T$ until the magnetic scales involved become incredibly large well outside the observational bounds defined in \req{igmf} and \req{tbscale} as seen by the dotted blue curves of various large values $b_{0}=\{25,\ 50,\ 100,\ 300\}$. The chemical potential is also insensitive to forcing by the spin potential until $\zeta$ reaches a significant fraction of the electron mass $m_{e}$ in size. The chemical potential for large values of spin potential $\zeta=\{100,\ 200,\ 300,\ 400,\ 500\}\,\keV$ are also plotted as dashed black lines with $b_{0}=0$.

It is interesting to note that there are crossing points where a given chemical potential can be described as either an imbalance in spin-polarization or presence of external magnetic field. While spin potential suppresses the chemical potential at low temperatures, external magnetic fields only suppress the chemical potential at high temperatures.

The profound insensitivity of the chemical potential to these parameters justifies the use of the free particle chemical potential (black line) in the ranges of magnetic field strength considered for cosmology. Mathematically this can be understood as $\xi$ and $b_{0}$ act as small corrections in the denominator of \req{chem} if expanded in powers of these two parameters.

\subsection{Relativistic paramagnetism of electron-positron gas}
\label{sec:magnetization}
\noindent The total magnetic flux within a region of space can be written as the sum of external fields and the magnetization of the medium via
\begin{align}
 \label{totalmag}
 {B}_\mathrm{total} = {B} + \mathcal{M}\,.
\end{align}
For the simplest mediums without ferromagnetic\index{ferromagnetism} or hysteresis considerations, the relationship can be parameterized by the susceptibility\index{magnetic!susceptibility} $\chi$ of the medium as
\begin{align}
 \label{susceptibility}
 {B}_\mathrm{total} = (1+\chi){B}\,,\qquad \mathcal{M} = \chi{B}\,,\qquad \chi\equiv\frac{\partial\mathcal{M}}{\partial{B}}\,,
\end{align}
with the possibility of both paramagnetic materials $(\chi>1)$ and diamagnetic materials $(\chi<1)$. The $e^{+}e^{-}$ plasma however does not so neatly fit in either category as given by \req{spin} and \req{spinorbit}. In general, the susceptibility of the gas will itself be a field dependant quantity.

In our analysis, the external magnetic field always appears within the context of the magnetic scale $b_{0}$, therefore we can introduce the change of variables
\begin{align}
 \frac{\partial b_{0}}{\partial{B}}=\frac{e}{T^{2}}\,.
\end{align}
The magnetization of the $e^{+}e^{-}$ plasma described by the partition function in \req{boltzmann} can then be written as
\begin{align}
 \label{defmagetization}
 \mathcal{M}\equiv\frac{T}{V}\frac{\partial}{\partial{B}}\ln{\mathcal{Z}_{e^{+}e^{-}}} = \frac{T}{V}\left(\frac{\partial b_{0}}{\partial{B}}\right)\frac{\partial}{\partial b_{0}}\ln{\mathcal{Z}_{e^{+}e^{-}}}\,.
\end{align}
Magnetization\index{magnetization} arising from other components in the cosmic gas (protons, neutrinos, etc.) could in principle also be included. Neutrinos for example may have a non-zero magnetic moment and participate in the magnetization of the cosmic plasma. Additionally, a more complete understanding of magnetic dynamics is required; for classical and semi-classical efforts in this regime see~\cite{Rafelski:2017hce,Formanek:2019cga,Formanek:2021mcp}. Localized inhomogeneities of matter evolution are often non-trivial and generally be solved numerically using magneto-hydrodynamics (MHD)~\cite{melrose2008quantum,Vazza:2017qge,Vachaspati:2020blt} or with a suitable Boltzmann-Vlasov transport equation. An extension of our work would be to embed magnetization into transport theory~\cite{Formanek:2021blc}. In the context of MHD, primordial magnetogenesis from fluid flows in the electron-positron epoch was considered in~\cite{Gopal:2004ut,Perrone:2021srr}.

We introduce dimensionless units for magnetization ${\mathfrak M}$ by defining the critical field strength
\begin{align}
 {B}_{C}\equiv\frac{m_{e}^{2}}{e}\,,\qquad{\mathfrak M}\equiv\frac{\mathcal{M}}{{B}_{C}}\,.
\end{align}
The scale ${B}_{C}$ is where electromagnetism is expected to become subject to non-linear effects, though luckily in our regime of interest, electrodynamics should be linear. We note however that the upper bounds of IGMFs in \req{igmf} (with $b_{0}=10^{-3}$; see \req{tbscale}) brings us to within $1\%$ of that limit for the external field strength in the temperature range considered.

The total magnetization ${\mathfrak M}$ can be broken into the sum of magnetic moment parallel ${\mathfrak M}_{+}$ and magnetic moment anti-parallel ${\mathfrak M}_{-}$ contributions
\begin{align}
\label{g2mag}
{\mathfrak M}&={\mathfrak M}_{+}+{\mathfrak M}_{-}\,.
\end{align}

We note that the expression for the magnetization simplifies significantly for $g\!=\!2$ which is the `natural' gyro-magnetic\index{g-factor} factor~\cite{Evans:2022fsu,Rafelski:2022bsv} for Dirac particles. For illustration, the $g\!=\!2$ magnetization from \req{defmagetization} is then
\begin{align}
 \label{g2magplus}
 {\mathfrak M}_{+}&=\frac{e^{2}}{\pi^{2}}\frac{T^{2}}{m_{e}^{2}}\xi\cosh{\frac{\mu}{T}}\left[\frac{1}{2}x_{+}K_{1}(x_{+})+\frac{b_{0}}{6}K_{0}(x_{+})\right]\,,\\
 \label{g2magminus}
 -{\mathfrak M}_{-}&=\frac{e^{2}}{\pi^{2}}\frac{T^{2}}{m_{e}^{2}}\xi^{-1}\cosh{\frac{\mu}{T}}
 \left[\left(\frac{1}{2}+\frac{b_{0}^{2}}{12x_{-}^{2}}\right)x_{-}K_{1}(x_{-})+\frac{b_{0}}{3}K_{0}(x_{-})\right]\,,\\
 x_{+}&=\frac{m_{e}}{T}\,,\qquad
 x_{-}=\sqrt{\frac{m_{e}^{2}}{T^{2}}+2b_{0}}\,.
\end{align}
As the $g$-factor of the electron is only slightly above two at $g\simeq2.00232$~\cite{Tiesinga:2021myr}, the above two expressions for ${\mathfrak M}_{+}$ and ${\mathfrak M}_{-}$ are only modified by a small amount because of anomalous magnetic moment (AMM) and would be otherwise invisible on our figures.\index{magnetization!g-factor dependence}

\para{Evolution of electron-positron magnetization}
\label{sec:paramagnetism}
\noindent In \rf{fig:magnet}, we plot the magnetization\index{magnetization} as given by \req{g2magplus} and \req{g2magminus} with the spin potential set to unity $\xi=1$. The lower (solid red) and upper (solid blue) bounds for cosmic magnetic scale $b_{0}$ are included. The external magnetic field strength ${B}/{B}_{C}$ is also plotted for lower (dotted red) and upper (dotted blue) bounds. Since the derivative of the partition function governing magnetization may manifest differences between Fermi-Dirac and the here used Boltzmann limit more acutely, out of abundance of caution, we indicate extrapolation outside the domain of validity of the Boltzmann limit with dashes.

\begin{figure}
 \centering
 \hspace*{0.4cm}\includegraphics[width=0.8\linewidth]{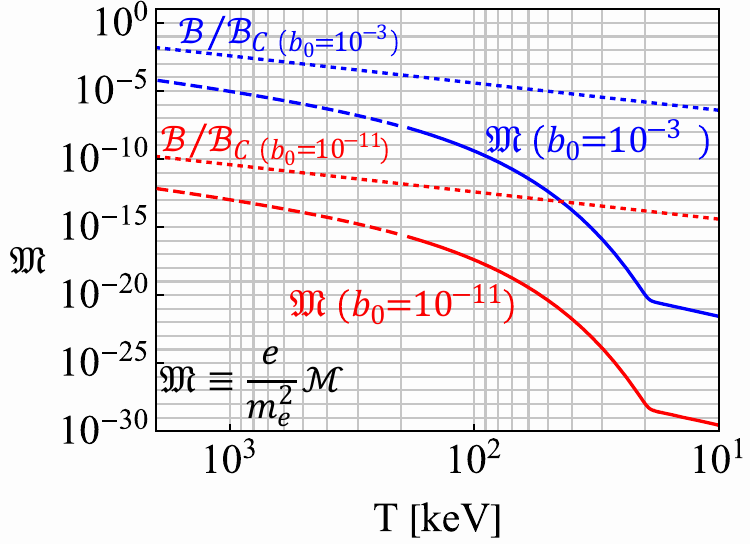}
 \caption{The magnetization ${\mathfrak M}$, with $g\!=\!2$, of the primordial $e^{+}e^{-}$ plasma is plotted as a function of temperature. \cccite{Steinmetz:2023ucp}. \radapt{Rafelski:2023emw,Steinmetz:2023nsc}}
 \label{fig:magnet} 
\end{figure}

We see in \rf{fig:magnet} that the $e^{+}e^{-}$ plasma is overall paramagnetic and yields a positive overall magnetization which is contrary to the traditional assumption that matter-antimatter plasma lack significant magnetic responses of their own in the bulk. With that said, the magnetization never exceeds the external field under the parameters considered which shows a lack of ferromagnetic behavior. 

The large abundance of pairs causes the smallness of the chemical potential\index{chemical potential!electron} seen in~\rf{fig:chemicalpotential} at high temperatures. As the primordial Universe expands and temperature decreases, there is a rapid decrease of the density $n_{e^{\pm}}$ of $e^{+}e^{-}$ pairs. This is the primary the cause of the rapid paramagnetic decrease seen in \rf{fig:magnet} above $T\!=\!21\keV$. At lower temperatures $T<21\keV$ there remains a small electron excess (see~\rf{fig:densityratio}) needed to neutralize proton charge. These excess electrons then govern the residual magnetization and dilutes with cosmic expansion.

An interesting feature of \rf{fig:magnet} is that the magnetization in the full temperature range increases as a function of temperature. This is contrary to Curie's law~\cite{greiner2012thermodynamics} which stipulates that paramagnetic susceptibility of a laboratory material is inversely proportional to temperature. However, Curie's law applies to systems with fixed number of particles which is not true in our situation.

A further consideration is possible hysteresis as the $e^{+}e^{-}$ density drops with temperature. It is not immediately obvious the gas's magnetization should simply `degauss' so rapidly without further consequence. If the very large paramagnetic susceptibility present for $T\simeq m_{e}$ is the origin of an overall magnetization of the plasma, the conservation of magnetic flux through the comoving surface ensures that the initial residual magnetization is preserved at a lower temperature by Faraday induced kinetic flow processes however our model presented here cannot account for such effects.

Primordial Universe conditions may also apply to some extreme stellar objects with rapid change in $n_{e^{\pm}}$ with temperatures above $T\!=\!21\keV$. Production and annihilation of $e^{+}e^{-}$ plasmas is also predicted around compact stellar objects~\cite{Ruffini:2009hg,Ruffini:2012it} potentially as a source of gamma-ray bursts.

\para{Dependency on g-factor}
\noindent As discussed at the end of \rsec{sec:magnetization}, the AMM of $e^{+}e^{-}$ is not relevant in the present model. However out of academic interest, it is valuable to consider how magnetization is affected by changing the $g$-factor\index{g-factor} significantly.

\begin{figure}
 \centering
 \hspace*{-0.8cm}\includegraphics[width=0.74\linewidth]{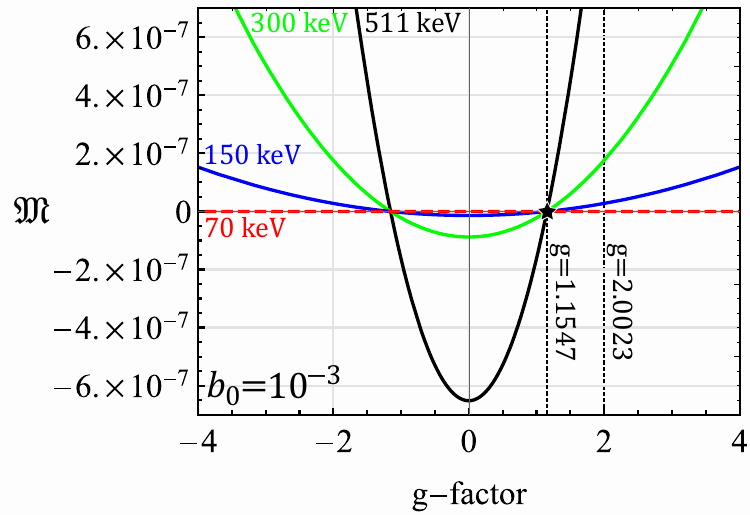}
 \caption{The magnetization $\mathfrak M$ as a function of $g$-factor plotted for several temperatures with magnetic scale $b_{0}=10^{-3}$ and polarization fugacity $\xi=1$. \cccite{Steinmetz:2023ucp}. \radapt{Steinmetz:2023nsc}}
 \label{fig:gfac} 
\end{figure}

The influence of AMM would be more relevant for the magnetization of baryon gases since the $g$-factor for protons $(g\approx5.6)$ and neutrons $(g\approx3.8)$ are substantially different from $g\!=\!2$. The influence of AMM on the magnetization of thermal systems with large baryon content (neutron stars, magnetars, hypothetical bose stars, etc.) is therefore also of interest~\cite{Ferrer:2019xlr,Ferrer:2023pgq}.

\req{g2magplus} and \req{g2magminus} with arbitrary $g$ reintroduced is given by
\begin{gather}
\label{arbg:1}
{\mathfrak M}=\frac{e^{2}}{\pi^{2}}\frac{T^{2}}{m_{e}^{2}}\sum_{s'}^{\pm1}\xi_{s'}\cosh{\frac{\mu}{T}}
\left[C^{1}_{s'}(x_{s'})K_{1}(x_{s'})+C^{0}_{s'}K_{0}(x_{s'})\right]\,,\\
\label{arbg:2}
C^{1}_{s'}(x_{\pm}) = \left[\frac{1}{2}-\left(\frac{1}{2}-\frac{g}{4}s'\right)\left(1+\frac{b^2_0}{12x^{2}_{s'}}\right)\right]x_{s'}\,,\qquad
C^{0}_{s'} = \left[\frac{1}{6}-\left(\frac{1}{4}-\frac{g}{8}s'\right)\right]b_0\,,
\end{gather}
where $x_{s'}$ was previously defined in \req{xfunc}. In \rf{fig:gfac}, we plot the magnetization as a function of $g$-factor between $4>g>-4$ for temperatures $T\!=\!\{511,\ 300,\ 150,\ 70\}\keV$. We find that the magnetization is sensitive to the value of AMM, revealing a transition point between paramagnetic $({\mathfrak M}>0)$ and diamagnetic gases $({\mathfrak M}<0)$. Curiously, the transition point was numerically determined to be around $g\simeq1.1547$ in the limit $b_{0}\rightarrow0$. The exact position of this transition point however was found to be both temperature and $b_{0}$ sensitive, though it moved little in the ranges considered.

It is not surprising for there to be a transition between diamagnetism and paramagnetism given that the partition function (see \req{spin} and \req{spinorbit}) contained elements of both. With that said, the transition point presented at $g\approx1.15$ should not be taken as exact because of the approximations used to obtain the above results. 

It is likely that the exact transition point has been altered by our taking of the Boltzmann approximation and Euler-Maclaurin\index{Euler-Maclaurin integration} integration steps. It is known that the Klein-Gordon-Pauli solutions to the Landau problem in \req{cosmokgp} have periodic behavior~\cite{Steinmetz:2018ryf,Evans:2022fsu,Rafelski:2022bsv} for $|g|=k/2$ (where $k\in1,2,3\ldots$).

These integer and half-integer points represent when the two Landau towers of orbital levels match up exactly. Therefore, we propose a more natural transition between the spinless diamagnetic gas of $g=0$ and a paramagnetic gas is $g=1$. A more careful analysis is required to confirm this, but that our numerical value is close to unity is suggestive.

\para{Magnetization per lepton}
Despite the relatively large magnetization\index{magnetization} seen in \rf{fig:magnet}, the average contribution per lepton\index{lepton}\index{magnetization!per lepton} is only a small fraction of its overall magnetic moment indicating the magnetization is only loosely organized. Specifically, the magnetization regime we are in is described by
\begin{align}
 \label{fractionalmagnetization}
 \mathcal{M}\ll\mu_{B}\frac{N_{e^{+}}+N_{e^{-}}}{V}\,,\qquad\mu_{B}\equiv\frac{e}{2m_{e}}\,,
\end{align}
where $\mu_{B}$ is the Bohr magneton and $N=nV$ is the total particle number in the proper volume V. To better demonstrate that the plasma is only weakly magnetized, we define the average magnetic moment per lepton given by along the field ($z$-direction) axis as
\begin{align}
 \label{momentperlepton}
 \vert\vec{m}\vert_{z}\equiv\frac{\mathcal{M}}{n_{e^{-}}+n_{e^{+}}}\,,\qquad\vert\vec{m}\vert_{x}=\vert\vec{m}\vert_{y}=0\,.
\end{align}
Statistically, we expect the transverse expectation values to be zero. We emphasize here that despite $|\vec{m}|_{z}$ being nonzero, this doesn't indicate a nonzero spin angular momentum as our plasma is nearly matter-antimatter symmetric. The quantity defined in \req{momentperlepton} gives us an insight into the microscopic response of the plasma.

\begin{figure}
 \centering
 \includegraphics[width=0.60\linewidth]{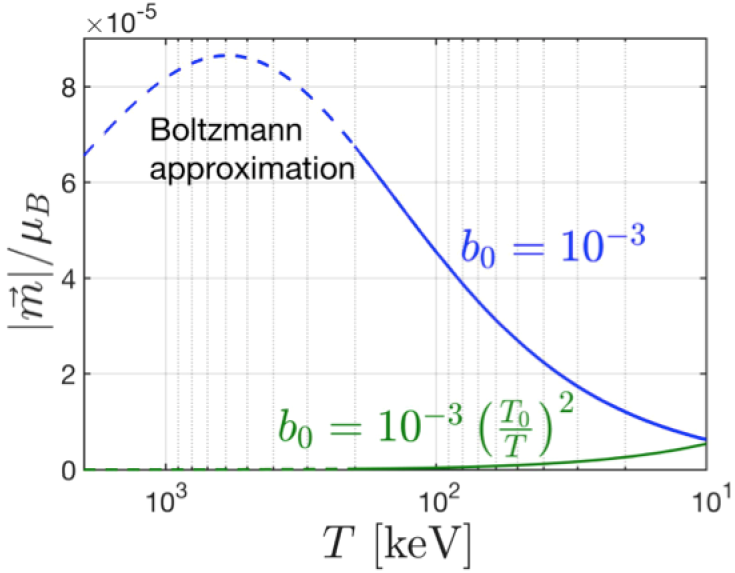}
 \caption{The magnetic moment per lepton $\vert\vec{m}\vert_{z}$ along the field axis as a function of temperature. \cccite{Steinmetz:2023nsc}}
 \label{fig:momentperlepton}
\end{figure}

The average magnetic moment $\vert\vec{m}\vert_{z}$ defined in \req{momentperlepton} is plotted in \rf{fig:momentperlepton} which displays how essential the external field is on the `per lepton' magnetization. The $b_{0}=10^{-3}$ case (blue curve) is plotted in the Boltzmann approximation. The dashed lines indicate where this approximation is only qualitatively correct. For illustration, a constant magnetic field\index{magnetic!fields} case (solid green line) with a comoving reference value chosen at temperature $T_{0}=10\keV$ is also plotted.

If the field strength is held constant, then the average magnetic moment per lepton is suppressed at higher temperatures as expected for magnetization satisfying Curie's law. The difference in \rf{fig:momentperlepton} between the non-constant (blue solid curve) case and the constant field (solid green curve) case demonstrates the importance of the conservation of primordial magnetic flux in the plasma, required by \req{bscale}. While not shown, if \rf{fig:momentperlepton} was extended to lower temperatures, the magnetization per lepton of the constant field case would be greater than the non-constant case which agrees with our intuition that magnetization is easier to achieve at lower temperatures. This feature again highlights the importance of flux conservation in the system and the uniqueness of the primordial cosmic environment.

\subsection{Polarization potential and ferromagnetism}
\label{sec:ferro}\index{ferromagnetism}
\noindent Up to this point, we have neglected the impact that a nonzero spin potential $\zeta\neq0$ (and thus $\xi\neq1$) would have on the primordial $e^{+}e^{-}$ plasma magnetization. In the limit that $(m_{e}/T)^2\gg b_0$ the magnetization given in \req{arbg:1} and \req{arbg:2} is entirely controlled by the polarization fugacity\index{fugacity!polarization} $\xi$ asymmetry generated by the spin potential $\zeta$ yielding up to first order $\mathcal{O}(b_{0})$ in magnetic scale
\begin{equation}
 \label{ferro}
 \lim_{m_{e}^{2}/T^{2}\gg b_0}{\mathfrak M}=\frac{g}{2}\frac{e^{2}}{\pi^{2}}\frac{T^{2}}{m_{e}^{2}}\sinh{\frac{\zeta}{T}}\cosh{\frac{\mu}{T}}\left[\frac{m_{e}}{T}K_{1}\left(\frac{m_{e}}{T}\right)\right] 
 +b_{0}\left(g^{2}-\frac{4}{3}\right)\frac{e^{2}}{8\pi^{2}}\frac{T^{2}}{m_{e}^{2}}\cosh{\frac{\zeta}{T}}\cosh{\frac{\mu}{T}}K_{0}\left(\frac{m_{e}}{T}\right)
 +\mathcal{O}\left(b_{0}^{2}\right)\,.
\end{equation}

Given \req{ferro}, we can understand the spin potential as a kind of `ferromagnetic'\index{ferromagnetism} influence on the primordial gas that allows for magnetization even in the absence of external magnetic fields. This interpretation is reinforced by the fact the leading coefficient is $g/2$. We suggest that a variety of physics could produce a small nonzero $\zeta$ within a domain of the gas. Such asymmetries could also originate statistically as while the expectation value of free gas polarization is zero, the variance is likely not.

As $\sinh{\zeta/T}$ is an odd function, the sign of $\zeta$ also controls the alignment of the magnetization. In the high temperature limit \req{ferro} with strictly $b_{0}=0$ assumes a form of to lowest order for brevity
\begin{align}
 \label{hiTferro}
 \lim_{m_{e}/T\rightarrow0}{\mathfrak M}\vert_{b_{0}=0}=\frac{g}{2}\frac{e^{2}}{\pi^{2}}\frac{T^{2}}{m_{e}^{2}}\frac{\zeta}{T}\,.
\end{align}

While the limit in \req{hiTferro} was calculated in only the Boltzmann limit, it is noteworthy that the high temperature (and $m\rightarrow0$) limit of Fermi-Dirac\index{Fermi!distribution} distributions only differs from the Boltzmann result by a proportionality factor. The natural scale of the $e^{+}e^{-}$ magnetization with only a small spin fugacity ($\zeta<1\eV$) fits easily within the bounds of the predicted magnetization during this era if the IGMF measured today was of primordial origin. The reason for this is that the magnetization\index{magnetization} seen in \req{g2magplus}, \req{g2magminus} and \req{ferro} are scaled by $\alpha{B}_{C}$ where $\alpha$ is the fine structure constant.

\index{ferromagnetism}
\para{Hypothesis of ferromagnetic self-magnetization}
\label{sec:self}
\noindent One exploratory model we propose is to fix the spin polarization asymmetry, described in \req{spotential}, to generate a homogeneous magnetic field which dissipates as the primordial Universe cools down. In this model, there is no external primordial magnetic field $({B}_\mathrm{PMF}=0)$ generated by some unrelated physics, but rather the $e^{+}e^{-}$ plasma itself is responsible for the field by virtue of spin polarization.

\begin{figure}
 \centering
 \includegraphics[width=0.8\linewidth]{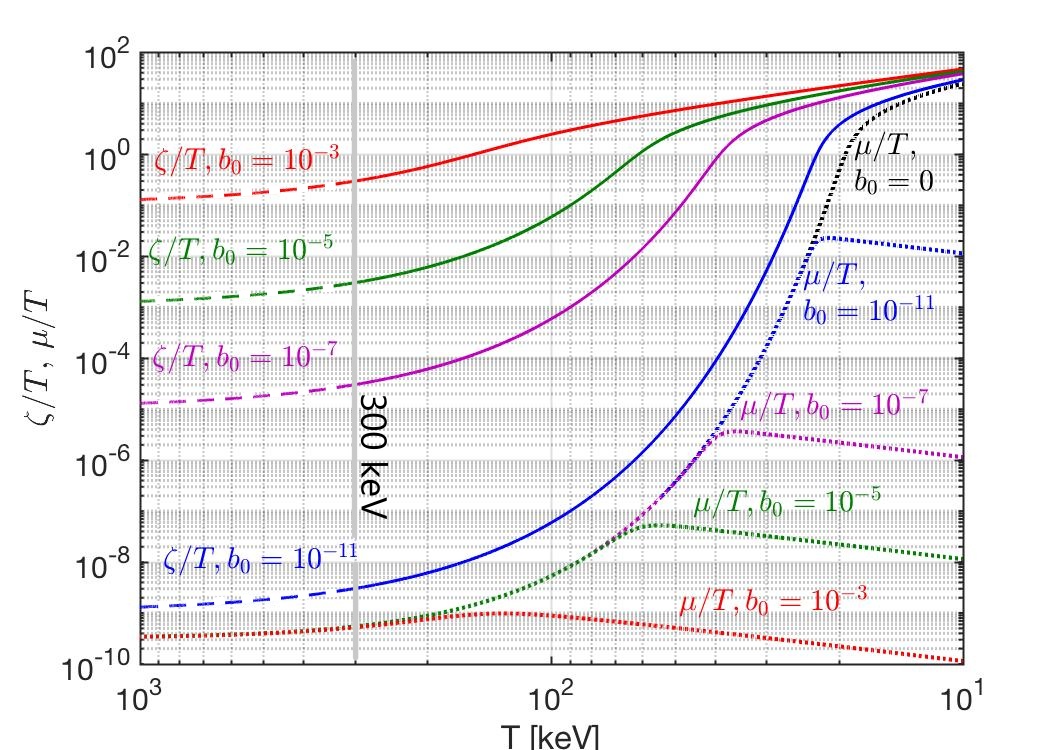}
 \caption{The spin potential $\zeta$ and chemical potential\index{chemical potential} $\mu$ are plotted under the assumption of self-magnetization through a nonzero spin polarization in bulk of the plasma. \radapt{Steinmetz:2023nsc}}
 \label{fig:self} 
\end{figure}

This would obey the following assumption of
\begin{align}
 \label{selfmag}
 {\mathfrak M}(b_{0})=\frac{\mathcal{M}(b_0)}{{B}_{C}}\longleftrightarrow\frac{B}{{B}_{C}}=b_{0}\frac{T^{2}}{m_{e}^{2}}\,,
\end{align}
which sets the total magnetization as a function of itself. The spin polarization described by $\zeta\rightarrow\zeta(b_{0},T)$ then becomes a fixed function of the temperature and magnetic scale. The underlying assumption would be the preservation of the homogeneous field would be maintained by scattering within the gas (as it is still in thermal equilibrium) modulating the polarization to conserve total magnetic flux.

The result of the self-magnetization assumption in \req{selfmag} for the potentials is plotted in \rf{fig:self}. The solid lines indicate the curves for $\zeta/T$ for differing values of $b_{0}=\{10^{-11},\ 10^{-7},\ 10^{-5},\ 10^{-3}\}$ which become dashed above $T\!=\!300\keV$ to indicate that the Boltzmann approximation is no longer appropriate though the general trend should remain unchanged.

\begin{figure}
 \centering
 \includegraphics[width=0.8\linewidth]{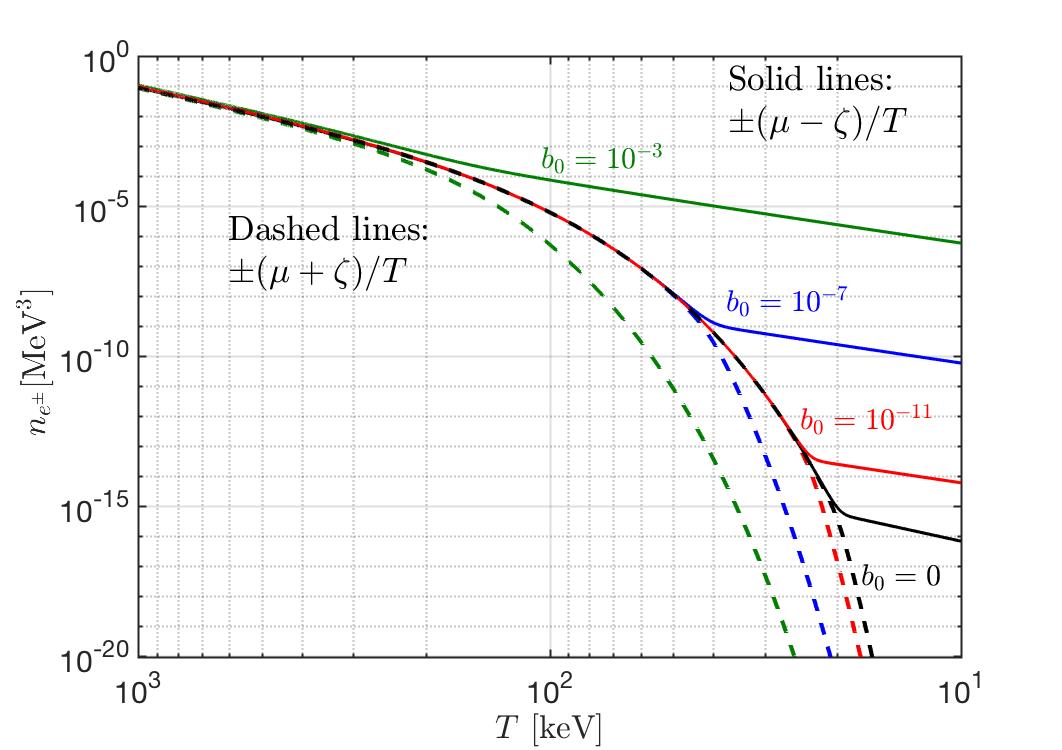}
 \caption{The number density $n_{e^{\pm}}$ of polarized electrons and positrons under the self-magnetization model for differing values of $b_{0}$. \cccite{Steinmetz:2023nsc}}
 \label{fig:polarswap} 
\end{figure}

The dotted lines are the curves for the chemical potential $\mu/T$. At high temperatures we see that a relatively small $\zeta/T$ is needed to produce magnetization owing to the large densities present. \rf{fig:self} also shows that the chemical potential does not deviate from the free particle case until the spin polarization becomes sufficiently high which indicates that this form of self-magnetization would require the annihilation of positrons to be incomplete even at lower temperatures.

This is seen explicitly in~\rf{fig:polarswap} where we plot the numerical density of particles as a function of temperature for spin aligned $(+\zeta)$ and spin anti-aligned $(-\zeta)$ species for both positrons $(-\mu)$ and electrons $(+\mu)$. Various self-magnetization strengths are also plotted to match those seen in~\rf{fig:self}. The nature of $T_{\rm split}$ changes under this model since antimatter and polarization states can be extinguished separately. Positrons persist where there is insufficient electron density to maintain the magnetic flux. Polarization asymmetry therefore appears physical only in the domain where there is a large number of matter-antimatter pairs.

\para{Matter inhomogeneities in the cosmic plasma}
\label{sec:inhomogeneous}
\noindent In general, an additional physical constraint is required to fully determine $\mu$ and $\zeta$ simultaneously as both potentials have mutual dependency (see \rsec{sec:ferro}). We note that spin polarizations are not required to be in balanced within a single species to preserve angular momentum.

The CMB\index{CMB}~\cite{Planck:2018vyg}\index{CMB} indicates that the primordial Universe was home to domains of slightly higher and lower baryon densities which resulted in the presence of galactic super-clusters, cosmic filaments, and great voids seen today. However, the CMB, as measured today, is blind to the localized inhomogeneities required for gravity to begin galaxy and supermassive black hole formation.

Such acute inhomogeneities distributed like a dust~\cite{Grayson:2023flr} in the plasma would make the proton density sharply and spatially dependant $n_{p}\rightarrow n_{p}(x)$ which would directly affect the potentials $\mu(x)$ and $\zeta(x)$ and thus the density of electrons and positrons locally. This suggests that $e^{+}e^{-}$ may play a role in the initial seeding of gravitational collapse. If the plasma were home to such localized magnetic domains, the nonzero local angular momentum within these domains would provide a natural mechanism for the formation of rotating galaxies today.

Recent measurements by the James Webb Space Telescope (JWST)~\cite{Yan:2022sxd,adams2023discovery,arrabal2023spectroscopic} indicate that galaxy formation began surprisingly early at large redshift values of $z\gtrsim10$ within the first 500 million years of the primordial Universe, requiring gravitational collapse to begin in a hotter environment than expected. The observation of supermassive black holes already present~\cite{CEERSTeam:2023qgy} in this same high redshift period (with millions of solar masses) indicates the need for local high density regions in the primordial Universe whose generation is not yet explained and likely need to exist long before the recombination epoch.

\section{Summary and Discussion}\label{ part6}
We have presented a compendium of theoretical models addressing the particle and plasma content of the primordial Universe. The Universe at a temperature above 10\keV\ is dominated by `visible' matter, and dependence on unknown dark matter and dark energy is minimal. However any underlying dark component will later surface, thus the understanding of this primordial epoch also as a source of darkness (including neutrinos background) in the present day Universe is among our objectives.

Select introductory material addressing kinetic theory, statistical physics, and general relativity has been presented. Kinetic and plasma theory are described in greater detail. Einstein's gravity theory, found in many other sources, is limited to the minimum required in the study of the primordial Universe within the confines of the FLRW cosmology model. 

This work connects several of our prior and ongoing studies of cosmic particle plasma in the primordial Universe. The three primary eras, radiation, matter, and dark energy dominance, can be recognized in terms of the acceleration parameter $q$. We introduce this tool in the cosmology primer \rsec{sec:flrw}, connecting these distinct epochs smoothly in \rsec{sec:dynamic}. Detailed results concerning time and temperature relation allowing for the reheating of the Universe were shown. Entropy transfer (reheating) inflates the Universe's expansion whenever the ambient temperature is too low to support the massive particle abundance.

In detailed studies, we explored particle abundances and plasma properties, which improved our comprehensive understanding of the Universe and its evolution. {\color{black} Many of the methods we presented and used in the study of the early Universe were developed for relativistic heavy-ion collision applications. However, the hot Universe environment differs from relativistic heavy-ion laboratory experiments, and this is best appreciated considering the evolution of particle inventory: In the evolving Universe, we allow for a full adjustment of particle yields to the ambient temperature of the dynamically expanding Universe, and we follow these yields as a function of temperature with chemical potential(s) constrained by the Universe baryon asymmetry. Each particle, irrespective of its interaction, has individual decoupling temperature, with unstable particles usually rapidly disappearing after decoupling, while stable decoupled particles free-streaming.} 

{\color{black}In contrast to this, in the laboratory heavy-ion experiments: a) Only strongly interacting degrees of freedom are typically observed; b) Particle yields follow from a single freeze-out near to QGP-hadron gas phase cross-over; c) Detailed particle yields allow us to measure the dynamically created chemical potentials; d) An analysis of laboratory experimental data creates a snap-shot image of the one high $T$ freeze-out instant. Clearly, a direct, detailed comparison between the early Universe particle inventory and laboratory experiments cannot be attempted. Similar remarks apply to all other observables: we learn in laboratory experiments the required methodology which we use in the study of the dynamic primordial Universe.}

{\color{black} This conceptual connection between heavy ion experiments and the early Universe has profound importance beyond the study of particle inventory. Known probes of the primordial Universe, such as the cosmic microwave background and primordial light nuclei abundances, are indirectly driven by macroscopic EM fields and transport properties in primordial plasma, including the QGP, which underpin response to primordial magnetic fields and eventual early structure formation. In~\rsec{chap:QCD}, we presented the electromagnetic properties of QGP in heavy ion collisions. This insight leads us to derive an analytic formula that predicts the freeze-out magnetic field that governs the micro-bang in the laboratory, potentially enabling experimental determination of the QGP electromagnetic conductivity, as was recently proposed~\cite{STAR:2023jdd}, which determines the  understanding of the primordial QGP.}

One important aspect of the laboratory study of the hot primordial Universe is the experimental access in ultra relativistic heavy-ion collision experiments to the process of melting of matter into constituent quarks at high enough temperature. The idea that one could recreate this Big-Bang condition in laboratory was indeed the beginning of the modern interest in better understanding the structure of the primordial Universe. 

We recalled the 50 years of effort that began with the recognition of novel structures in the primordial Universe beyond the Hagedorn temperature and the exploration of this high-temperature deconfined quark-gluon phase. Moreover, the study of the phase transformation between confined hadrons and deconfined quark-gluon plasma in the laboratory facilitates the understanding of the primordial Universe dating to the earliest instants after its birth, about 20-30\,$\mu$s after the Big-Bang. The question of how we can recognize the quark-gluon plasma observed in the laboratory to be different from the hadron Universe content was mentioned.

Many interesting phenomena in the primordial Universe depend on nonequilibrium conditions, and this topic is at the core of our theoretical interest. Nuance differences between kinetic and chemical equilibrium, dynamic but stationary detailed balance, and non-stationary phenomena recur as topics of interest in our discussion. For bottom quarks in \rsec{Bottom} we recognize in detail the deviations from thermal equilibrium, particle freeze-out, and detailed balance away from the thermal equilibrium condition and isolate the non-stationary components. These nonequilibrium concepts developed for more esoteric purposes are pivotal, in our opinion, in recognizing any remnant observable of the primordial Universe. 

The experimental study in the laboratory of the dynamic micro-bang stimulates the development of detailed models of the strongly interacting hadron era of the Universe. We use some of the tools created for laboratory experiment interpretation to study properties of hadronic matter in the Universe and strangeness flavor freeze-out in particular in \rsec{Strangeness}. 

These kinetic and dynamic insights drive our interest, leading beyond our interest in strangeness and bottom quarks to all heavy PP-SM particles. We question the potential that primordial QGP era harbors opportunity for baryogenesis, we look both for the bottom quarks and the Higgs particle induced reactions, \rsec{HiggsQGP}. This work will continue.

The different epochs in the Universe evolution are often considered as being distinctly separate. However, we have shown that this is not always the case. We note the `squeeze' of neutrino decoupling between: The electron-positron annihilation reheating of photons at the low temperature edge at about $T=1\MeV$; and heavy lepton (muon) disappearance on the high-$T$ edge at about $T=4.5\MeV$, \rsec{Electron}. 

This fine-tuning into a narrow available domain prompted our investigation of neutrino decoupling as a function of the magnitude of the governing natural constants, \rsec{ch:param:studies}. This characterization of neutrino freeze-out constrains the time variation of natural constants. We present in \rapp{ch:boltz:orthopoly} a novel computationally efficient moving-frame numerical method we developed to obtain the required results.

Our in-depth study of the neutrino background shows future potential to reconcile observational tensions that arise between the reported present day speed of Universe expansion $H_0$ (Hubble parameter in present epoch) and extrapolations from the recombination epoch. One can question how $H_0$ could depend on a better understanding of the dynamics of the free-streaming quantum neutrinos\index{neutrino!massive free-streaming quantum liquid} across mass thresholds. We recently laid a relevant theoretical foundation, allowing us to develop further this very intricate topic~\cite{Birrell:2024bdb}.

In~\rsec{sec:BoltzmannEinstein}, we provided background on the Boltzmann-Einstein equation, including proofs of the conservation laws and the Boltzmann's H-theorem for interactions between any number of particles; this is of interest as the evolution of the Universe often requires detailed balance involving more than two particle scattering. To our knowledge, proof for general numbers $m$, $n$ with $m\to n$-particle interactions is not available in other references on the subject.

Following on the neutrino decoupling we encounter in the temporal evolution of the Universe another example of two era overlap, this time potentially much more consequential: The era of electron-positron pair plasma annihilation begins immediate after neutrino decoupling and yet the primordial nucleosynthesis at a temperature that is 15 times lower proceeds amidst a dense $e^+e^-$-pair plasma background, which fades out well after BBN ends. 

This effect is clearly visible but maybe is not fully appreciated when inspecting in~\rf{fig:energy:frac}: We see that the line for the $e^+e^-$-component is a ``small'' $e^+e^-$-energy fraction during the marked BBN epoch. It seems that the $e^+e^-$-pair plasma is in process of disappearance and does not matter. This is, however, a wrong first impression: The $e^+e^-$-energy fraction is starting with a giant $10^9$ pair ratio over nucleon dust. Dropping by three orders of magnitude there remains a huge $e^+e^-$-pair abundance left with millions of pairs per each nucleon at the onset of the BBN era.

We studied the ratio of $e^+e^-$-pair abundance to baryon number in detail in \rf{fig:densityratio} (see also \rf{BBN:Electron} right ordinate): As a curious tidbit let us note that as long as there are more than a few thousand $e^+e^-$-pairs per nucleon the antimatter content in the primordial Universe is practically symmetric with the matter content in any applicable measure. The nuclear dust is not tilting the balance as matter are electrons and antimatter are positrons. Thus it is not entirely correct to consider the disappearance of of antibaryons, see \rf{Baryon:fig}, at $T\simeq 38.2\MeV$, as the end of antimatter epoch. It is instead correct to view the temperature\index{antimatter!disappearance} $T=30\keV$ as the onset of the antimatter disappearance which completes at $T=20.3\keV$, as is seen in~\rf{fig:densityratio}.

Investigation of the dense charged particle plasma background during BBN constitutes a major part of this work. In~\rsec{part4} we develop a covariant kinetic plasma theory to analyze the influence of $e^+e^-$-pair plasma polarization. We solve the dynamic phase space equations using linear response method considering both spatial and temporal dispersion. We are focusing our attention on the understanding how the covariant polarization tensor, which includes collisional damping, shapes the self-consistent electromagnetic fields within the medium. This approach allows us to elucidate the intricate dynamics introducing QED damping effects that characterize the behavior of the $e^+e^-$-pair plasma.

We explore the damped-dynamic screening effects between reacting nucleons and light elements in $e^+e^-$-pair plasma during the Big-Bang Nucleosynthesis (BBN). Our results indicate that the in plasma screening can modify inter nuclear potentials and thus also nuclear fusion reaction rates in an important manner. However, the effect during the accepted BBN temperature range is found to remain a minor correction to the usually used effective screening enhancement. Despite the significant perturbatively evaluated damping, and high temperatures characteristic of BBN, the enhancement in nuclear reaction rates remains relatively small, around $10^{-5}$, yet it provides a valuable refinement to our understanding of the primordial Universe's conditions. We also show a very significant impact of non-perturbative self-consistent evaluation of damping in \rsec{section:electron}. We have not yet had an opportunity to explore how the non-perturbative damping impacts BBN epoch fusion rates. 

The long lasting (in relative terms) antimatter $e^+e^-$-pair plasma offers an opportunity to consider a novel mechanism of magneto-genesis in primordial Universe: Extrapolating the intergalactic fields observed in the current era back in time to the $e^+e^-$-pair plasma era, magnetic field strengths are encountered which approach the strength of the surface magnetar fields~\rsec{sec:theory}.

This has prompted our interest to study the primordial $e^+e^-$-pair plasma as the source of Universe magnetization. We studied the temperature range of $2000\keV$ to $20\keV$ where all of space was filled with a hot dense electron-positron plasma (up to 450 million pairs per baryon) still present in primordial Universe within the first few minutes after the Big-Bang. We note that our chosen period also includes the BBN era.

We found that subject to a primordial magnetic field, the primordial Universe electron-positron plasma has a significant paramagnetic response, see~\rf{fig:magnet} due to magnetic moment polarization. We considered the interplay of charge chemical potential, baryon asymmetry, anomalous magnetic moment, and magnetic dipole polarization on the nearly homogeneous medium. We presented a simple model of self-magnetization of the primordial electron-positron plasma which indicates that only a small polarization asymmetry is required to generate significant magnetic flux when the primordial Universe was very hot and dense.

Our novel approach to high temperature magnetization, see Chapter~\ref{sec:mag:universe} shows that the $e^{+}e^{-}$-plasma paramagnetic response (see \req{g2magplus} and \req{g2magminus}) is dominated by the varying abundance of electron-positron pairs, decreasing with decreasing $T$ for $T\!<\!m_{e}c^2$. This is unlike conventional laboratory cases where the magnetic properties emerge with the number of magnetic particles being constant.  As the number of pairs depletes while the primordial Universe cools the electron-positron spin magnetization clearly cannot be maintained. However, once created magnetic fields want to persist. How the transit from Gilbertian to Amperian magnetism proceeds will be topic of future investigation: This presents an opportunity for understanding formation of space-time persistent  induced currents helping to facilitate magnetic and potentially matter inhomogeneity in the primordial Universe. 
 
Outside of the scope of our report we can also check for era overlaps at temperature below 10\keV: Inspecting \rf{fig:energy:frac} one can wonder about the coincidental multiple crossing of different visible energy components in the Universe seen near to $T=0.25\meV$. This means at condition of recombination there is an unexpected component coincidence. This special situation depends directly on the interpretation of our current era in terms of specific matter and darkness components. The analysis of cosmic background microwave (CBM) data which underpins this, is not retold here. However, the present day conditions propagate on to the primordial times in the particles and plasma Universe and provide for the era overlaps we reported in regard of earlier eras.

Skeptics could interpret the appearance of several such coincidences as indicative of a situation akin to pre-Copernican epicycles. Are we seeing odd `orbits' because we do not use the `solar' centered model? We note that current standard model of cosmology is being challenged by Fulvio Melia~\cite{Melia:2022itm} ``One cannot avoid the conclusion that the standard model needs a complete overhaul in order to survive.'' or by the same author~\cite{Melia:2024rzy} ``\ldots the timeline in $\Lambda$CDM is overly compressed at $z\ge 6$, while strongly supporting the expansion history in the early Universe predicted by\ldots" the Melia model of cosmology. 

This well could be the case. However, we believe that in order to argue for or against different models of primordial cosmology we need first to establish the Universe particles and plasma model properties very well as we presented in coherent fashion for the first time in the wide $130\GeV\le T\le 10\keV$ range. Without this any declarations about the cosmological context of particles and plasma Universe based on a few atomic, molecular, stellar phenomena observed at in comparison tiniest imaginable redshift $z=6\simeq7$ are not compelling. Similarly we view with some hesitance the many hypothesis about the properties of the Universe prior to the formation of the PP-SM particles with properties we have explored in laboratory.

The search to understand the grand properties of the primordial Universe without understanding is particle and plasma content has a much longer historical backdrop, which we noted and which had to evolve: Before about year 1971 there was no inkling about the particle physics standard model; we were attempting to understand the primordial Universe based on a thermal hadron model. Hagedorn's bootstrap approach~\cite{Rafelski:2016hnq} was particularly welcome as the exponential mass spectrum of hadronic resonances generated divergent energy density for point-sized hadrons. This well known result allowed the hypothesis that there is a maximum (Hagedorn) temperature in the Universe. 

This argument had excellent and convincing footing and yet it was not lasting: We needed to accommodate the energy content we observe in the infinite Universe. A divergence of energy at a singular starting point converts to a divergence, inflation in space size. However, as soon as experiments in laboratory clarified our understanding of fundamental particle physics, this narrative collapsed within weeks as one of us (JR) saw in late 70s at CERN, working with Hagedorn in his office long hours developing non-divergent models of hadrons. 

The outcome of more than 50 years of ensuing effort is seen in these pages, and yet with certainty this is just a tip of an iceberg. We presented here the primordial Universe within the realm of the known laws of physics. As the reader will note while  turning pages, there are many `loose' ends: we show and tell clearly about any and all we recognize. We cannot tell as yet what happened `before' our PP-SM begins at $T\simeq 130\GeV$. Many further key dynamic details characterizing evolution before recombination at $T=0.25\eV$ need to be resolved. The particles and plasma Universe based on PP-SM spans a 12 orders of magnitude temperature window $ 130\GeV > T > 0.25\eV $. And, there is the need to understand the ensuing atomic and molecular Universe which presents another challenge we did not mention. We believe that there is a lot more work to do, which will be much helped by gaining better insights into the riddles of the present day Universe dynamics.

\section*{Declarations}
\textbf{Author Contributions} All authors participated in every stage of the development of this work.\\
\textbf{Data Availability} No datasets were generated or analyzed during the current study.\\
\textbf{Competing Interests} The authors declare no competing interests.

\appendix
\section{Connecting Prior Works: Note to the Reader}\label{list:Works}
In this review we have expanded considerably both in scope and content our recent review:
\begin{enumerate}
\item ``A Short Survey of Matter-Antimatter evolution in the Primordial Universe'' by \allcite{Rafelski:2023emw}, which focused on the role of antimatter in the primordial Universe. 
\end{enumerate}
The scope is widened by including in an edited and re-sequenced manner select material from the contents of four Ph.D. theses completed at the Department of Physics, The University of Arizona. {\color{black} These individual projects were written by authors of this review and address distinctly different events in the evolution of the primordial Universe:}
\begin{enumerate}
\setcounter{enumi}{1}
\item ``Non-Equilibrium Aspects of Relic Neutrinos: From Freeze-out to the Present Day'' by \aucite{Birrell:2014ona} studies the evolution of the relic (or cosmic) neutrino distribution from neutrino freeze-out at $T=\mathcal{O}(1)\MeV$ through the free-streaming era up to today. {\color{black} This work impacts our understanding of the values of natural constants in this primordial epoch. It is of great relevance to the ongoing discussion regarding speed of Universe expansion in the current epoch.}
\item ``Dense Relativistic Matter-Antimatter Plasmas'' by \aucite{Grayson:2024okq} explores dense electron-positron and quark-gluon plasma (QGP)\index{QGP!quark-gluon plasma} with strong electromagnetic fields generated during heavy-ion collisions and prevalent in extreme astrophysical environments. {\color{black} These methods also allow us to understand the long range cosmological magnetic fields if originating in the QGP phase.}
\item ``Modern topics in relativistic spin dynamics and magnetism'' by \aucite{Steinmetz:2023ucp} explores spin and magnetic moments in \emph{relativistic} mechanics from both
quantum and classical perspectives. A model of primordial magnetization in the Universe is presented, originating during the hot dense electron-positron plasma epoch. {\color{black} The methods developed in this work are fully portable to the study of magnetization originating in the earlier QGP epoch.}
\item ``Elementary Particles and Plasma in the First Hour of the Early Universe'' by \aucite{Yang:2024ret} deepens the understanding of the primordial composition of the Universe in the temperature range $300 \MeV>T>0.02\MeV$, which transits from quark-gluon plasma to hadron matter. {\color{black} Equilibrium and non-equilibrium particle abundances in the primordial Universe are explored and different primordial Universe epochs connected.}
\end{enumerate}

Due to graduation time constraints some of this presented material is only found in follow-up publications and in reports yet to be readied for publication. Many of these results are also included in this work. We further report in a large part on our research papers and related reports. Thus beyond the four theses, the key input too this review includes:
\begin{enumerate}
\setcounter{enumi}{5}
%
\item ``Matter-antimatter origin of cosmic magnetism'' by \allcite{Steinmetz:2023nsc} proposes a model of spin para-magnetization driven by the large matter-antimatter (electron-positron) content of the primordial Universe. {\color{black} The theoretical framework presented is general and allows us to seek the origins of Universe magnetization in all epochs containing high antimatter components, including the other most interesting case, the QGP.}
\item ``Electron-positron plasma in BBN: Damped-dynamic screening'' by \allcite{Grayson:2023flr} employs the linear response theory to describe the inter-nuclear potential screened by in electron-positron pair plasma in the BBN epoch. This work includes also the computation of the chemical potential and plasma damping rate required in semi-analytical study of the relativistic Boltzmann equation in the context of the linear response theory. {\color{black} The methods of linear response theory presented in this work lay the foundation for many yet to come studies of interacting particles and plasma in the primordial Universe, example is above Ref.\,\cite{Grayson:2024uwg}.}
\item ``Dynamic magnetic response of the quark-gluon plasma to electromagnetic fields'' by \allcite{Grayson:2022asf} applies linear response method to characterize the quark-gluon plasma environment in the presence of strong magnetic fields. {\color{black} Future study of magnetization phenomena in the primordial QGP can rely on results obtained in this work.}
\item ``Cosmological Strangeness Abundance'' by \allcite{Yang:2021bko}[Rafelski] presents strange particle abundances in the expanding primordial Universe including determination of strange particle freeze-out temperatures. {\color{black} Considerable difference with experimental relativistic heavy ion environment arises due to importance of reactions absent at the time scale of laboratory experiments, \eg\ lepton fusion $l^+ +l^-\to\phi(1020)\to \mathrm{K}+\overline{\mathrm K}$ strangeness producing reactions.}
\item ``Current-conserving Relativistic Linear Response for Collisional Plasmas'' by \allcite{Formanek:2021blc} develops relativistic linear response plasma theory implementing conservation laws, obtaining general solutions and laying the foundation for applications to primordial Universe plasma conditions. {\color{black} Of great relevance to both laboratory and cosmological environments is the allowance for the damping term.}
\item ``The Muon Abundance in the Primordial Universe'' by \allcite{Rafelski:2021aey}[Yang] is a conference proceedings paper dedicated to exploration of muon abundance and its persistence temperature in the primordial Universe. {\color{black} One can wonder if it is a mere coincidence that muons disappear from the Universe particle inventory when their ratio to net baryon number is nearly unity.}
\item ``Reactions Governing Strangeness Abundance in Primordial Universe'' by \allcite{Rafelski:2020ajx}[Yang] is a conference proceeding paper that lays the groundwork for the study of strangeness reactions in the primordial Universe. {\color{black} This work is a prequel to the Ref.\,\cite{Yang:2021bko} described above. It is more accessible in some of technical details.}
\item ``Possibility of bottom-catalyzed matter genesis near to primordial QGP hadronization'' by \allcitep{Yang:2020nne}[Rafelski] was our fist study of the bottom flavor abundance in the primordial Universe. {\color{black} We show unexpected appearance of chemical nonequilibrium behavior near to QGP hadronization. Follow-up investigations of allowed us to recognize the presence of non-stationary heavy particle abundances in the primordial QGP required for baryogenesis, a topic which awaits our near term attention.}
\item ``Lepton Number and Expansion of the Universe'' by \allcitep{Yang:2018oqg} proposes a model of large lepton asymmetry and explores how this large cosmological lepton yield relates to the effective number of (Dirac) neutrinos. {\color{black} This work explores several cosmological Universe properties freed from the $B-L=0$ constraint: The conservation of baryon minus lepton number. This is highly relevant to the present day Hubble parameter $H_0$ value as additional neutrino pressure wanes with time due to emergence of neutrino mass.}
\item ``Temperature Dependence of the Neutron Lifespan'' by \allcitep{Yang:2018qrr} is a study of neutron lifespan in primordial cosmic plasma with Fermi-blocking of the decay electrons and neutrinos. {\color{black} The widely discussed dual value of neutron lifespan depending on measurement method stimulated our interest in understanding primordial neutron response to ambient conditions. A study of the influence of strong magnetic fields in both cosmological and laboratory environment remains on our to-do list.}
\item ``Strong fields and neutral particle magnetic moment dynamics'' by \allcite{Formanek:2017mbv} is an overview of our research group's early efforts to understand neutral particle dynamics in the presence of (strong) electromagnetic fields. {\color{black} We explore the dynamics as function of the $g$-factor which can assume arbitrary values. The cosmic neutrino response to strong magnetic fields remains on to-do list.}
\item ``The hot Hagedorn Universe'' by \allcite{Rafelski:2016cho}[Birrell] {\color{black} recounts and updates the impact of Hagedorn's work on phase transformation at Hagedorn temperature in the primordial Universe. Hagedorn had a pivotal impact on the cosmic paradigm as his statistical bootstrap model of point-sized hadrons allowed a singular energy density at a temperature near to pion mass equivalent. This paradigm was dramatically expanded allowing for hadrons of finite size.}
\item ``Relic Neutrino Freeze-out: Dependence on Natural Constants'' by \allcite{Birrell:2014uka} is a study of neutrino freeze-out temperature as a function of standard model parameters and its application on the effective number of neutrinos. {\color{black} The presence of the a background neutrino radiation is impacting the speed of Universe expansion today. The magnitude of unseen neutrino component depends on our belief in the constancy of fundamental constants near to onset of neutrino free-streaming. In this work we investigate, in a quantitative manner, this relation showing  that the Weinberg angle and a particular combination of several other natural constants determine the present day Hubble parameter $H_0$. This work provides all neutrino-matter weak interaction matrix elements required for the Boltzmann transport code developed and needed to obtain these result.}
\item ``Quark–gluon Plasma as the Possible Source of Cosmological Dark Radiation'' by \allcite{Birrell:2014cja}[Rafelski] {\color{black} explores the role of hypothetical dark radiation particles decoupling at the time of QGP hadronization. The presence of dark radiation is often invoked to fine-tune models of Universe dynamics. Our contribution argues that the freezing of color degrees of freedom could liberate dark particles. We evaluate the effect of a single `dark' particle and show how it is reduced in impact due to reheating of visible particles. This work was dedicated to the study of nearly massless particles; a similar origin and study could be developed to model unobserved massive dark matter component in the Universe. }
\item ``Boltzmann Equation Solver Adapted to Emergent Chemical Non-equilibrium'' by \allcite{Birrell:2014gea} addresses the transport theory tools we developed to characterize the slow in time freeze-out of neutrinos in the primordial Universe. {\color{black} There are several species of neutrinos that undergo in the temperature interval $1\MeV<T<4\MeV$ a gradual decoupling from the electron-positron background: This manuscript presents the blueprint of a novel dynamical solution of the decoupling (freeze-out) using moving basis method.}
\item ``Proposal for Resonant Detection of Relic Massive Neutrinos'' by \allcite{Birrell:2014qna}[Rafelski] characterizes the primordial neutrino flux spectrum today and explores experimental approaches for experimental observations. {\color{black} The ambient cosmic microwave background (CMB) temperature today is $kT_0=0.235\meV$, well below the mass of two of the three neutrinos which thus move relatively slowly compared to the cosmic Earth motion. This situation is offering unusual detection strategy opportunities. Future work is required to reach definitive conclusions about primordial neutrino background detection.}
\item ``Traveling Through the Universe: Back in Time to the Quark-Gluon Plasma Era'' by \allcite{Rafelski:2013yka}[Birrell] presents a first study of the connection between quark-gluon plasma and neutrino freeze-out epochs. {\color{black} This is a decade old prequel to this work, comparing to it we see enormous progress made since.}
\item ``Connecting QGP-Heavy Ion Physics to the Early Universe'' by \aucite{Rafelski:2013qeu} explores the properties of the primordial Universe at QGP hadronization and connects to the ongoing experimental heavy-ion effort to study the hadronization process. {\color{black} This detailed look at the primordial QGP hadronization helped define several projects which are described in this report.}
\item ``Fugacity and Reheating of Primordial Neutrinos'' by \allcite{Birrell:2013gpa} is a study of neutrino fugacity as a function of neutrino kinetic freeze-out temperature. This short work includes {\color{black} neutrino interaction matrix elements and paves the way for the evaluation of neutrino relaxation time. This is a conference report driven prequel to later detailed work on neutrino freeze-out.}
\item ``Relic Neutrinos: Physically Consistent Treatment of Effective Number of Neutrinos and Neutrino Mass'' by \allcite{Birrell:2012gg} is a model-independent study of the neutrino momentum distribution at freeze-out, treating the freeze-out temperature as a free parameter. {\color{black} The speed of Universe expansion depends on correct accounting of the free-streaming massive neutrinos which this work offers.}
\item ``From Quark-Gluon Universe to Neutrino Decoupling: $200 < T < 2\MeV$'' by \allcite{Fromerth:2012fe} {\color{black} connects the Quark-Hadron phase transformation with neutrino decoupling as a function of the current era cosmological properties. A conference report addressing ongoing effort to connect primordial QGP to neutrino decoupling eras.}
\item ``Unstable Hadrons in Hot Hadron Gas in Laboratory and in the Early Universe'' by \allcite{Kuznetsova:2010pi}[Rafelski] shows that some unstable hadrons may persist in the evolution of the primordial Universe as the detailed balance condition is never broken due to strong coupling to the photon background. {\color{black} This is preparatory work providing methods and initial results for the study of hadrons and heavy leptons in the primordial Universe.}
\item ``Hadronization of the Quark Universe'' by \allcite{Fromerth:2002wb}[Rafelski] is {\color{black} a path-breaking work which established the tools needed to describe the primordial quark-hadron phase of matter.} It includes a first detailed study of chemical potentials and conditions of hadronization of QGP in the primordial Universe. 
\item {\color{black}``Hadrons and Quark–Gluon Plasma''~\allcite{Letessier:2002ony}[Rafelski] is a textbook where the introductory chapters connect the properties of the primordial Universe with the relativistic heavy-ion collision experimental program. This was a first effort to explain the importance of the ongoing experimental effort in the context of creating understanding of the primordial Universe as described in this report.}
\end{enumerate}
Additionally, material adapted from Refs.~\cite{Rafelski:2019twp,Rafelski:2016hnq,Rafelski:2015cxa} has been included. This lets us address strong interactions and quark-gluon plasma (QGP)\index{QGP!quark-gluon plasma} hadronization\index{QGP!hadronization} in the Universe: (i) Deconfined states of hot quarks and gluons, the QG; and (ii) Hot hadronic phase of matter, also called hadronic gas\index{hadrons!gas phase}, as applicable to the context of the primordial Universe. 

\section{Geometry Background: Volume Forms on Submanifolds}\label{ch:vol:forms}
In this appendix we develop the geometric machinery that will be used to derive computationally efficient formulas for the scattering integrals. This facilitates the study of the neutrino freeze-out using the Boltzmann-Einstein equation in \rsec{ch:param:studies}. This appendix is much more mathematical than the main text and, when standard, we use geometrical language and notation here without further explanation; see, e.g., \cite{o1983semi,lee2003introduction,lee1997riemannian}. We found this formalism to be useful for our development of an improved method for computing scattering integrals, as presented in \rapp{ch:coll:simp}. However, if one is content with simply using the results then this appendix is non-essential. See also \cite{Birrell:2014uka}.

\subsection{Inducing volume forms on submanifolds}

Given a Riemannian manifold $(M,g)$ with volume form $dV_g$ and a hypersurface $S$, the standard Riemannian hypersurface area form, $dA_g$, is defined on $S$ as the volume form of the pullback metric tensor on $S$. Given vectors $v_1,...,v_k$ we define the interior product\index{interior product} (i.e. contraction) operator acting on a form $\omega$ of degree $n\geq k$ as the $n-k$ form 
\begin{equation}
i_{(v_1,...,v_k)}\omega=\omega(v_1,...,v_k,\cdot)\,.
\end{equation}
 With this notation, the hypersurface area form\index{hypersurface area form} can equivalently be computed as
\begin{equation}
dA_g=i_v dV_g\,,
\end{equation}
where $v$ is a unit normal vector to $S$. This method extends to submanifolds of codimension greater than one as well as to semi-Riemannian manifolds, as long as the metric restricted to the submanifold is non-degenerate. 

However, there are many situations where one would like to define a natural volume form on a submanifold that is induced by a volume form in the ambient space, but where the above method is inapplicable, such as defining a natural volume form on the light cone or other more complicated degenerate submanifolds in general relativity. In this section, we will describe a method for inducing volume forms on regular level sets of a function that is applicable in cases where there is no metric structure and show its relation to more widely used semi-Riemannian case. We prove analogues of the coarea formula\index{coarea formula} and Fubini's theorem\index{Dirac delta!Fubini's theorem} in this setting. 

Let $M$, $N$ be smooth manifolds, $c$ be a regular value of a smooth function $F:M\rightarrow N$, and $\Omega^M$ and $\Omega^N$ be volume forms on $M$ and $N$ respectively. Using this data, we will be able to induce a natural volume form on the level set $F^{-1}(c)$. The absence of a metric on $M$ is made up for by the additional information that the function $F$ and volume form $\Omega^N$ on $N$ provide. The following theorem makes our definition precises and proves the existence and uniqueness of the induced volume form\index{induced volume form}.

\begin{theorem}\label{inducedVolForm}
Let $M$, $N$ be $m$ (resp. $n$)-dimensional smooth manifolds with volume forms $\Omega^M$ (resp. $\Omega^N$). Let $F:M\rightarrow N$ be smooth and $c$ be a regular value. Then there is a unique volume form $\omega$ (also denoted $\omega^M$) on $F^{-1}(c)$ such that $\omega_x=i_{(v_1,...,v_n)}\Omega^M_x$ whenever $v_i\in T_xM$ are such that 
\begin{equation}\label{unitVolume}
\Omega^N(F_*v_1,...,F_* v_n)=1\,.
\end{equation}
We call $\omega$ the {\bf volume form induced by $F:(M,\Omega^M)\rightarrow (N,\Omega^N)$}\index{induced volume form}.
\end{theorem}
\begin{proof}
$F_*$ is onto $T_{F(x)}N$ for any $x\in F^{-1}(c)$. Hence there exists $\{v_i\}_1^n\subset T_xM$ such that $\Omega^N(F_*v_1,...,F_* v_n)=1$. In particular, $F_* v_i$ is a basis for $T_{F(x)} N$. Define $\omega_x=i_{(v_1,...,v_n)}\Omega_x$. This is obviously a nonzero $m-n$ form on $T_xF^{-1}(c)$ for each $x\in F^{-1}(c)$. We must show that this definition is independent of the choice of $v_i$ and the result is smooth.

 Suppose $F_*v_i$ and $F_*w_i$ both satisfy \req{unitVolume}. Then $F_*v_i=A_i^jF_*w_j$ for $A\in SL(n)$. Therefore $v_i-A_i^jw_j\in \ker F_{*x}$. This implies
\begin{equation}
i_{(v_1,...,v_n)}\Omega^M_x=\Omega^M_x(A_1^{j_1}w_{j_1},...,A_n^{j_n}w_{j_n},\cdot)\,,
\end{equation}
since the terms involving $\ker F_*$ will vanish on $T_x F^{-1}(c)=\ker F_{*x}$. Therefore
\begin{align}\label{indOfvProof}
i_{(v_1,...,v_n)}\Omega^M_x&=A_1^{j_1}...A_n^{j_n}\Omega^M_x(w_{j_1},...,w_{j_n},\cdot)
=\sum_{\sigma\in S_n} \pi(\sigma)A_1^{\sigma(1)}...A_n^{\sigma(n)}\Omega^M_x(w_1,...,w_n,\cdot)\notag\\
&=\det(A)i_{(w_1,...,w_n)}\Omega^M_x
=i_{(w_1,...,w_n)}\Omega^M_x\,.
\end{align}
This proves that $\omega$ is independent of the choice of $v_i$. If we can show $\omega$ is smooth, then we are done. We will do better than this by proving that for any $v_i\in T_xM$ the following holds
\begin{equation}
i_{(v_1,...,v_n)}\Omega^M_x=\Omega^N(F_*v_1,...,F_*v_n)\omega_x\,.
\end{equation}
To see this, take $w_i$ satisfying \req{unitVolume}. Then $F_*v_i=A_i^j F_*w_j$. This determinant can be computed from
\begin{align}
\Omega^N(F_*v_1,...,F_*v_n)=\det(A)\Omega^N(F_*w_1,...,F_*w_n)=\det(A)\,.
\end{align}
Therefore, the same computation as \req{indOfvProof} gives
\begin{align}
i_{(v_1,...,v_n)}\Omega^M_x=\det(A)\omega_x=\Omega^N(F_*v_1,...,F_*v_n)\omega_x\,,
\end{align}
as desired. To prove that $\omega$ is smooth, take a smooth basis of vector fields $\{V_i\}_1^m$ in a neighborhood of $x$. After relabeling, we can assume $\{F_*V_i\}_1^n$ are linearly independent at $F(x)$ and hence, by continuity, they are linearly independent at $F(y)$ for all $y$ in some neighborhood of $x$. In that neighborhood, $\Omega^N(F_*V_1,...,F_*V_n)$ is non-vanishing and therefore
\begin{equation}
\omega=(\Omega^N(F_*V_1,...,F_*V_n))^{-1}i_{(V_1,...,V_n)}\Omega\,,
\end{equation} 
which is smooth.
\end{proof}

\begin{corollary}\label{inducedVolEq}
For any $v_i\in T_xM$ the following holds
\begin{equation}\label{volFormula1}
i_{(v_1,...,v_n)}\Omega^M_x=\Omega^N(F_*v_1,...,F_*v_n)\omega_x\,.
\end{equation}
\end{corollary}

\begin{corollary}
If $\phi:M\rightarrow\mathbb{R}$ is smooth and $c$ is a regular value then by equipping $\mathbb{R}$ with its canonical volume form we have 
\begin{equation}
\omega_x=i_v\Omega^M_x\,,
\end{equation}
where $v\in T_xM$ is any vector satisfying $d\phi(v)=1$.
\end{corollary}

It is useful to translate \req{volFormula1} into a form that is more readily applicable to computations in coordinates. Choose arbitrary coordinates $y^i$ on $N$ and write $\Omega^N=h^N(y) dy^n$. Choose coordinates $x^i$ on $M$ such that $F^{-1}(c)$ is the coordinate slice
\begin{equation}
F^{-1}(c)=\{x:x^1=...=x^n=0\}
\end{equation}
and write $\Omega^M=h^M(x)dx^m$. The coordinate vector fields $\partial_{x^i}$ are transverse to $F^{-1}(c)$ and so
\begin{equation}
\Omega^N(F_*\partial_{x^1},...,F_*\partial_{x^n})=h^N(F(x))\det \left(\frac{\partial F^i}{\partial x^j}\right)_{i,j=1..n}
\end{equation}
and
\begin{equation}
i_{(\partial_{x^1},...,\partial_{x^n})}\Omega^M=h^M(x) dx^{n+1}...dx^m\,.
\end{equation}
Therefore we obtain
\begin{equation}\label{volFormCoords}
\omega_x=\frac{h^M(x)}{h^N(F(x))}\det \left(\frac{\partial F^i}{\partial x^j}\right)^{-1}_{i,j=1..n}dx^{n+1}...dx^m\,.
\end{equation}

Just like in the (semi)-Riemannian case, the induced measure allows us to prove a coarea formula where we break integrals over $M$ into slices. In this theorem and the remainder of the section, we consider integration with respect to the density defined by any given volume form, i.e., we ignore the question of defining consistent orientations.
\begin{theorem}[Coarea formula]\label{volFormCoarea}\index{coarea formula}
Let $M$ be a smooth manifold with volume form $\Omega^M$, $N$ a smooth manifold with volume form $\Omega^N$ and $F:M\rightarrow N$ be a smooth map. If $F_*$ is surjective at a.e. $x\in M$ then for $f\in L^1(\Omega^M)\bigcup L^+(M)$ we have
\begin{equation}\label{coareaFormula}
\int_Mf(x) \Omega^M(dx)=\int_{N}\int_{F^{-1}(z)} f(y)\omega^M_z(dy) \Omega^N(dz)\,,
\end{equation}
where $\omega^M_z$ is the volume form induced on $F^{-1}(z)$ as in Lemma \ref{inducedVolForm}.
\end{theorem}
\begin{proof}
First suppose $F$ is a submersion. By the rank theorem there exists a countable collection of charts $(U_i,\Phi_i)$ that cover $M$ and corresponding charts $(V_i,\Psi_i)$ on $N$ such that 
\begin{align}
\Psi_i\circ F\circ \Phi_i^{-1}(y^1,...,y^{m-n},z^1,...,z^n)=(z^1,...,z^n)\,.
\end{align}
Let $\sigma_i$ be a partition of unity subordinate to $U_i$. For each $i$ and $z$ we have $\Phi_i(U_i\cap F^{-1}(z))=\left(\mathbb{R}^{m-n}\times\{\Psi_i(z)\}\right)\cap \Phi_i(U_i)$. We can assume that the $\Phi_i(U_i)=U_i^1\times U_i^2\subset \mathbb{R}^{m-n}\times \mathbb{R}^n$ and therefore each $\Phi_i$ is a slice chart for $F^{-1}(z)$ for all $y$ such that $F^{-1}(z)\cap U_i\neq \emptyset$. In other words, $\Phi_i(U_i\cap F^{-1}(z))= U_i^1\times \{\Psi(z)\}$. This lets us compute the left and right-hand sides of \req{coareaFormula} for $f\in L^+(M)$:
\begin{align}
&\int_Mf(x) \Omega^M(dx)=\sum_i\int_{U_i}(\sigma_if)(x) \Omega^M(dx)
=\sum_i\int_{\Phi_i(U_i)}(\sigma_if)\circ \Phi^{-1}(y,z) \Phi^{-1*}\Omega^M(dy,dz)\notag\\
&=\sum_i\int_{\Phi_i(U_i)}(\sigma_if)\circ \Phi^{-1}(y,z)|g^M(y,z)| dy^{m-n}dz^n
=\sum_i\int_{U_i^2}\left[\int_{U_i^1}(\sigma_if)\circ \Phi^{-1}(y,z)|g^M(y,z)| dy^{m-n}\right]dz^n\\
&\text{where \hspace{2cm}}\Omega^M=g^M dy^1\wedge...\wedge dy^{m-n}\wedge dz^1\wedge...\wedge dz^n\,,\notag
\end{align}
and
\begin{align}
\int_{N}\int_{F^{-1}(z)} f(y)\omega^M_z(dy) \Omega^N(dz)=&\sum_i \int_{N}\left[\int_{\Phi_i(U_i\cap F^{-1}(z))} (\sigma_if)\circ\Phi_i^{-1}(y,\Psi(z))\Phi_i^{-1*}\omega^M_z(dy)\right] \Omega^N(dz)\notag\\
=&\sum_i \int_{V_i}\left[\int_{\Phi_i(U_i\cap F^{-1}(z))} (\sigma_if)\circ\Phi_i^{-1}(y,\Psi(z))\Phi_i^{-1*}\omega^M_z(dy)\right] \Omega^N(dz)\notag\\
=&\sum_i \int_{\Psi_i(V_i)}\left[\int_{\Phi_i(U_i\cap F^{-1}(\Psi^{-1}(z))} (\sigma_if)\circ\Phi_i^{-1}(y,z)\Phi_i^{-1*}\omega^M_z(dy)\right] \Psi^{-1*}\Omega^N(dz)\notag\\
=&\sum_i \int_{U_i^2}\left[\int_{U_i^1\times \{z\}} (\sigma_if)\circ\Phi_i^{-1}(y,z)|g^M_z(y)| dy^{m-n}\right] |g^N(z)| dz^n\,,\\
&\text{\hspace{-5cm}where \hspace{2cm}}\omega^M_z=g^M_z dy^1\wedge...\wedge dy^{m-n} \text{ and }\Omega^N=g^N dz^1\wedge...\wedge dz^n \text{ for } g_1^M,g_N>0\,.\notag
\end{align}
Therefore, if we can show $|g^M(y,z)|=|g_z^M(y)g^N(z)|$ on $U_i^1\times U_i^2$ we are done. From Corollary \ref{inducedVolEq} we have
\begin{align}
&(-1)^{n(m-n)} g^M(y,z)
=\Omega^M(\partial_{z^1},...,\partial_{z^n},\partial_{y^1},...,\partial_{y^{m-n}})=\Omega^N(F_*\partial_{z^n},...,F_*\partial_{z^n})g_z^M(y)\,.
\end{align}
Since $\Psi\circ F\circ\Phi^{-1}=\pi_2$ we have $F_*\partial_{z^j}=\partial_{z_j}$ and so $\Omega^N(F_*\partial_{z^n},...,F_*\partial_{z^n})=g^N$ which completes the proof in the case where $F$ is a submersion. The generalization to the case where $F_*$ is surjective a.e. follows from Sard's theorem and the fact that the set of $x\in M$ at which $F_*$ is surjective is open.
\end{proof}

\para{Comparison to Riemannian coarea formula}
We now recall the classical coarea formula for semi-Riemannian metrics, see, e.g., \cite{chavel1995riemannian}, and give its relation to Theorem \ref{volFormCoarea}.
\begin{definition}
Let $F:(M,g)\rightarrow (N,h)$ be a smooth map between semi-Riemannian manifolds. The {\bf normal Jacobian}\index{normal Jacobian} of $F$ is
\begin{equation}
NJF(x)=|\det(F_*|_x(F_*|_x)^T)|^{1/2}\,,
\end{equation}
where $(F_*|_x)^T$ denotes the adjoint map $T_xN\rightarrow T_xM$ obtained pointwise from the pullback $T^*N\rightarrow T^*M$ combined with the tangent-cotangent bundle isomorphisms defined by the metrics.
\end{definition}

\begin{lemma}
The normal Jacobian has the following properties.
\begin{itemize}
\item $(F_*|_x)^T:T_{F(x)}N\rightarrow (\ker F_*|_x)^\perp$.
\item If $F_*|_x$ is surjective then $(F_*|_x)^T$ is 1-1.
\item In coordinates
\begin{equation}
NJF(x)=\left|\det\left(h_{ik}(F(x))\frac{\partial F^k}{\partial x^l}(x)g^{lm}(x)\frac{\partial F^j}{\partial x^m}(x)\right)\right|^{1/2}\,.
\end{equation}
\item If $F_*|_x$ is surjective and $g$ is nondegenerate on $ker F_*|_x$ then $F_*|_x(F_*|_x)^T$ is invertible.
\item If $c\in N$ is a regular value of $F$ and $g$ is nondegenerate on $F^{-1}(c)$ then $NJF(x)$ is non-vanishing and smooth on $F^{-1}(c)$.
\end{itemize}
\end{lemma}

Combining these lemmas with the rank theorem, one can prove the standard semi-Riemannian coarea formula\index{coarea formula}
\begin{theorem}[Coarea formula]\index{coarea formula!semi-Riemannian}
Let $F:(M,g)\rightarrow (N,h)$ be a smooth map between semi-Riemannian manifolds such that $F_*$ is surjective at a.e. $x\in M$ and $g$ is nondegenerate on $F^{-1}(c)$ for a.e $c\in N$. Then for $\phi\in L^1(dV_g)$ we have
\begin{equation}
\int_M\phi(x)dV_g=\int_{y\in N}\int_{x\in F^{-1}(y)}\frac{\phi(x)}{NJF(x)}dA_g dV_h\,,
\end{equation}
where $dA_g$ is the volume measure induced on $F^{-1}(y)$ by pulling back the metric $g$. In particular, if $N=\mathbb{R}$ with its canonical metric then $NJF=|\nabla F|$ and 
\begin{equation}
\int_M \phi dV_g=\int_\mathbb{R}\int_{F^{-1}(r)}\frac{\phi(x)}{|\nabla F(x)|} dA_g dr\,.
\end{equation}
\end{theorem}

The relation between the Riemannian coarea formula and Theorem \ref{volFormCoarea} follows from the following theorem.
\begin{theorem}
Let $F:(M,g)\rightarrow (N,h)$ be a smooth map between semi-Riemannian manifolds and $c$ be a regular value. Suppose $g$ is nondegenerate on $F^{-1}(c)$. Let $\omega$ be the volume form on $F^{-1}(c)$ induced by $F:(M,dV_g)\rightarrow (N,dV_h)$. Then
\begin{equation}
\omega=NJF^{-1}dA_g
\end{equation}
as densities.
\end{theorem}
\begin{proof}
By Corollary \ref{inducedVolEq}, for any $v_i\in T_xM$ we have
\begin{equation}
i_{(v_1,...,v_n)}\Omega^M_x=dV_h(F_*v_1,...,F_*v_n)\omega_x\,.
\end{equation}
If we let $v_i$ be an orthonormal basis of vectors orthogonal to $F^{-1}(c)$ at $x$ then $F_*v_i$ are linearly independent and so
\begin{align}
\omega=&(dV_h(F_*v_1,...,F_*v_n))^{-1}i_{(v_1,...,v_n)}dV_g=(dV_h(F_*v_1,...,F_*v_n))^{-1}dA_g\,.
\end{align}
Choose coordinates about $x$ and $F(x)$ so that $\partial_{x^i}=v_i$ for $i=1...n$, $\{\partial_{x^i}\}_{n+1}^m$ span $\ker F_*$, and $\partial_{y_i}$ are orthonormal. Then 
\begin{align}
dV_h(F_*v_1,...,F_*v_n)&=\sqrt{|\det(h)|}\frac{\partial F^{j_1}}{\partial x^1}...\frac{\partial F^{j_n}}{\partial x^n}dy^1\wedge...\wedge dy^n(\partial_{y^{j_1}},...,\partial_{y^{j_n}})=\det\left(\frac{\partial F^{j}}{\partial x^i}\right)_{i,j=1}^n\,.
\end{align}
$F_*\partial_{x^i}=0$ for $i=n+1...m$ and so $\frac{\partial F^j}{\partial x_i}=0$ for $i=n+1...m$. Letting $\eta=\diag(\pm 1)$ be the signature of $g$, we find
\begin{align}
NJF(x)=&\left|\det\left(h_{ik}(F(x))\frac{\partial F^k}{\partial x^l}(x)g^{lm}(x)\frac{\partial F^j}{\partial x^m}(x)\right)\right|^{1/2}=\left|\det\left(\sum_{l,m=1}^n\frac{\partial F^k}{\partial x^l}(x)\eta^{lm}(x)\frac{\partial F^j}{\partial x^m}(x)\right)\right|^{1/2}\notag\\
=&\left|\det\left(\frac{\partial F^k}{\partial x^l}\right)_{k,l=1}^n\det(\eta^{lm})_{l,m=1}^n\det\left(\frac{\partial F^j}{\partial x^m}\right)_{j,m=1}^n\right|^{1/2}=\left|\det\left(\frac{\partial F^k}{\partial x^l}\right)_{k,l=1}^n\right|=|dV_h(F_*v_1,...,F_*v_n)|\,.
\end{align}
Therefore 
\begin{equation}
\omega=NJF^{-1}dA_g
\end{equation}
as densities.
\end{proof}
In particular, this shows that even though $NJF$ and $dA_g$ are undefined individually when $g$ is degenerate on $F^{-1}(c)$, one can make sense of their ratio in this situation as the induced volume form\index{induced volume form} $\omega$.

\para{Delta function supported on a level set}
 The induced measure defined above allows for a coordinate independent definition of a delta function supported on a regular level set. Such an object is of great use in performing calculations in relativistic phase space. We give the definition and prove several properties that justify several common formal manipulations that one would like to make with such an object.
\begin{definition}
Motivated by the coarea formula, we define the composition of the {\bf Dirac delta function}\index{Dirac delta} supported on $c\in N$ with a smooth map $F:M\rightarrow N$ such that $c$ is a regular value of $F$ by
\begin{equation}\label{deltaDef}
 \delta_c(F(x))\Omega^M \equiv \omega^M
\end{equation}
on $F^{-1}(c)$. This is just convenient shorthand, but it commonly used in the physics literature (typically without the justification presented above or in the following results). For $f\in L^1(\omega^M)$ we will write 
\begin{equation}
\int_M f(x)\delta_c(F(x))\Omega^M(dx)
\end{equation} 
in place of 
\begin{equation}
\int_{F^{-1}(c)} f(x) \omega^M(dx)\,.
\end{equation}

More generally, if the subset of $F^{-1}(c)$ consisting of critical points, a closed set whose complement we call $U$, has $\dim M-\dim N$ dimensional Hausdorff measure zero in $M$ then we define
\begin{equation}
\int_M f(x)\delta_c(F(x))\Omega^M(dx)=\int_{F|_U^{-1}(c)} f(x)\omega^M\,.
\end{equation}
This holds, for example, if $U^c$ is contained in a submanifold of dimension less than $\dim M-\dim N$. 

Equivalently, we can replace $U$ in this definition with any {\bf open} subset of $U$ whose complement still has $\dim M-\dim N$ dimensional Hausdorff measure zero. In this situation, we will say $c$ is a regular value except for a lower dimensional exceptional set. Note that while Hausdorff measure depends on a choice of Riemannian metric on $M$, the measure zero subsets are the same for each choice.
\end{definition}

Using \req{volFormCoords}, along with the coordinates described there, we can (at least locally) write the integral with respect to the delta function in the more readily usable form
\begin{equation}\label{deltaIntegralCoords}
\int_M f(x)\delta_c(F(x))\Omega^M=\int_{F^{-1}(c)} f(x)\frac{h^M(x)}{h^N(F(x))}\bigg|\det \left(\frac{\partial F^i}{\partial x^j}\right)^{-1}\bigg|dx^{n+1}...dx^m\,.
\end{equation}
The absolute value comes from the fact that we use $\delta_c(F(x))\Omega^M$ to define the orientation on $F^{-1}(c)$.

As expected, such an operation behaves well under diffeomorphisms.
\begin{lemma}\label{diffeoProperty}
Let $c$ be a regular value of $F:M\rightarrow N$ and $\Phi:M^{'}\rightarrow M$ be a diffeomorphism. Then the delta functions induced by $F:(M,\Omega^M)\rightarrow (N,\Omega^N)$ and $F\circ\Phi:(M^{'},\Phi^*\Omega^M)\rightarrow (N,\Omega^N)$ satisfy
\begin{equation}
\delta_c(F\circ\Phi)(\Phi^*\Omega^M)=\Phi^*(\delta_c(F)\Omega^M)\,.
\end{equation}
\end{lemma}

\begin{lemma}
Let $c$ be a regular value of $F:(M,\Omega^M)\rightarrow (N,\Omega^N)$ and $\Phi:N\rightarrow (N^{'},\Omega^{N^{'}})$ be a diffeomorphism where $\Phi^*\Omega^{N^{'}}=\Omega^N$. Then the delta functions induced by $F:(M,\Omega^M)\rightarrow (N,\Omega^N)$ and $\Phi\circ F:(M,\Omega^M)\rightarrow (N^{'},\Omega^{N^{'}})$ satisfy
\begin{equation}
\delta_c(F)\Omega^M=\delta_{\Phi(c)}(\Phi\circ F)\Omega^M\,.
\end{equation}
\end{lemma}

We also have a version of Fubini's theorem.
\begin{theorem}[Fubini's Theorem for Delta functions]\index{Dirac delta!Fubini's theorem}
Let $M_1,M_2,N$ be smooth manifolds with volume forms $\Omega_1,\Omega_2, \Omega^N$. Let $M\equiv M_1\times M_2$ and $\Omega\equiv \Omega_1\wedge\Omega_2$. Suppose that the set of $(x,y)\in F^{-1}(c)$ such that $F|_{M_1\times\{y\}}$ is not regular at $x$ has $\dim M_1+\dim M_2-\dim N$ dimensional Hausdorff measure zero in $M_1\times M_2$ (we denote the complement of this closed set by $U$). Then for $f\in L^1(\omega)\bigcup L^+(F^{-1}(c))$ we have
\begin{equation}\label{FubiniEq}
\int_Mf(x,y)\delta_c(F(x,y)) \Omega(dx,dy)=\int_{M_2}\left[\int_{U^y} f(x,y) \delta_c(F(x,y))\Omega_1(dx) \right]\Omega_2(dy)\,,
\end{equation}
where $U^y=\{x\in M_1:(x,y)\in U\}$.
\end{theorem}
\begin{proof}
Our assumption about $F|_{M_1\times\{y\}}$ implies that $c$ is a regular value of $F:M_1\times M_2\rightarrow N$ except for the lower dimensional exceptional set $U^c$ and for $y\in M_2$, $c$ is also a regular value of $F|_{U^y\times\{y\}}$, hence both sides of \req{FubiniEq} are well defined. Rewriting \req{FubiniEq} without the delta function, we then need to show that 
\begin{equation}
\int_{F|_U^{-1}(c)} f(x,y) d\omega=\int_{M_2}\left[\int_{F|_{U^y\times\{y\}}^{-1}(c)} f(x,y) \omega^1_{c,y}(dx)\right]\Omega_2(dy)\,,
\end{equation}
where $\omega^1_{c,y}$ is the induced volume form\index{induced volume form} on $F|_{U^y\times\{y\}}^{-1}(c)$. 

Consider the projection map restricted to the $c$-level set, $\pi_2:F|_U^{-1}(c)\rightarrow M_2$. By assumption, $F|_{M_1\times\{y\}}$ is regular at $x$ for all $(x,y)\in F|_U^{-1}(c)$. For such an $(x,y)$, take a basis $w_i\in T_yM_2$. Since $F|_{M_1\times\{y\}}$ has full rank at $x$, for each $i$ there exists $v_i\in T_xM_1$ such that $F(\cdot,y)_*v_i=F_*(0,w_i)$. Therefore $(-v_i,w_i)\in \ker F_*|_{(x,y)}=T_{(x,y)}F|_U^{-1}(c)$. Hence $w_i\in\pi_{2*} T_{(x,y)}F^{-1}(c)$ and so $\pi_2:F|_U^{-1}(c)\rightarrow M_2$ is regular at $(x,y)$. 

Since $\pi_2$ is regular for all $(x,y)\in F|_U^{-1}(c)$ the coarea formula\index{coarea formula} applies, giving
\begin{align}
\int_{F|_U^{-1}(c)}f d\omega=&\int_{M_2}\left[\int_{\pi_2^{-1}(y)}f\tilde{\omega}_{c,y}^1\right]\Omega_2(dy)
\end{align}
for all $f\in L^1(\omega)\bigcup L^+(F^{-1}(c))$, where $\tilde{\omega}_{c,y}^1$ is the volume form on $\pi_2^{-1}(y)$ induced by $\pi_2:(F|_U^{-1}(c),\omega)\rightarrow (M_2,\Omega_2)$.

As a point set, $\pi_2^{-1}(y)=F|_{ U^y\times\{y\}}^{-1}(c)$ and both are embedded submanifolds of $M_1\times M_2$ for a.e. $y\in M_2$, hence are equal as manifolds. So if we can show $\tilde{\omega}_{c,y}^1=\omega^1_{c,y}$ as densities whenever $F|_{M_1\times\{y\}}$ is regular at $x$ for some $(x,y)$ then we are done. 

Given any such $(x,y)$, take $v_i\in T_xM_1$ such that $\Omega^N(F(\cdot,y)_*v_i)=1$. By definition, $\omega_{c,y}^1=i_{(v_1,...,v_n)}\Omega_1$. We also have $(v_i,0)\in T_{(x,y)}M_1\times M_2$ and $\Omega^N(F_*(v_i,0))=1$. Hence 
\begin{align}
\omega=&i_{((v_1,0),...,(v_n,0))}(\Omega_1\wedge\Omega_2)
=(i_{((v_1,0),...,(v_n,0))}\Omega_1)\wedge\Omega_2\,.
\end{align}
Let $w_i\in T_y M_2$ such that $\Omega_2(w_1,...,w_{m_2})=1$. By the same argument as above, there exists $\tilde{v}_i\in T_xM_1$ such that $(\tilde{v}_i,w_i)\in \ker F_*=T_{(x,y)}F^{-1}(c)$. $\pi_{2*}(\tilde{v}_i,w_i)=w_i$ and $\Omega_2(w_1,...,w_{m_2})=1$ so by definition,
\begin{equation}
\tilde{\omega}_{c,y}^1=i_{((\tilde{v}_1,w_1),...,(\tilde{v}_{m_2},w_{m_2}))}\omega\,.
\end{equation}
Since any term containing $\Omega_2$ will vanishes on $T_F(\cdot,y)^{-1}(c)\subset T M_1$, we have 
\begin{align}
\tilde{\omega}_{c,y}^1=&(-1)^{m_1-n}i_{((v_1,0),...,(v_n,0))}\Omega_1=(-1)^{m_1-n}\omega_{c,y}^1\wedge\left(i_{((\tilde{v}_1,w_1),...,(\tilde{v}_{m_2},w_{m_2}))}\Omega_2\right)=(-1)^{m_1-n}\omega_{c,y}^1\,.
\end{align}
As we are integrating with respect to the densities defined by $\omega_{c,y}^1$ and $\tilde{\omega}_{c,y}^1$ we are done. 
\end{proof}

Before moving on, we give a few more useful identities.

\begin{theorem}\label{delta:associative}
Let $(c_1,c_2)$ be a regular value of $F\equiv F_1\times F_2:(M,\Omega^M)\rightarrow (N_1\times N_2,\Omega^{N_1}\wedge\Omega^{N_2})$. Then $c_2$ is a regular value of $F_2$, $c_1$ is a regular value of $F_1|_{F_2^{-1}(c_2)}$ and we have
\begin{equation}
\delta(F)\Omega^M=\delta(F_1)(\delta(F_2)\Omega^M)\,.
\end{equation}
\end{theorem}
\begin{proof}
$(c_1,c_2)$ is a regular value of $F$, hence there exists $v_i$, $w_i$ such that $F_* v_i=(\tilde{v}_i,0)$, $F_* w_i=(0,\tilde{w}_i)$ satisfy 
\begin{equation}
\Omega^{N_1}\wedge\Omega^{N_2}( (\tilde{v}_1,0),..., (0,\tilde{w}_1),...)=1\,.
\end{equation}
After rescaling, we can assume
\begin{equation}
\Omega^{N_1}( \tilde{v}_1,...,\tilde{v}_{n_1})=1,\hspace{2mm} \Omega^{N_2}(\tilde{w}_1,...,\tilde{w}_{n_2})=1\,.
\end{equation}
Therefore $c_2$ is a regular value of $F_2$ and 
\begin{equation}
\delta(F_2)\Omega^M=i_{w_1,...,w_n}\Omega^M\,.
\end{equation}
The tangent space to $F_2^{-1}(c_2)$ is $\ker (F_2)_*$ which contains $v_i$. Hence $c_1$ is a regular value of $F_1|_{F_2^{-1}(c_2)}$ and 
\begin{align}
\delta(F_1)(\delta(F_2)\Omega^M)=&i_{v_1,...,v_n}\delta(F_2)\Omega^M=\pm i_{v_1,...,v_n,w_1,...,w_n}\Omega^M\,,
\end{align}
therefore they agree as densities.
\end{proof}

\begin{theorem}
Let $c_i\in N_i$ be regular values of $F_i:M_i\rightarrow N_i$ and define $F=F_1\times F_2:M_1\times M_2\rightarrow N_1\times N_2$, $c=(c_1,c_2)$. If $\Omega^{M_i}$ and $\Omega^{N_i}$ are volume forms on $M_i$ and $N_i$ respectively then 
\begin{equation}
\delta_c( F) \left(\Omega^{M_1}\wedge\Omega^{M_2}\right)=\left(\delta_{c_1}( F_1)\Omega^{M_1}\right)\wedge\left(\delta_{c_2}( F_2)\Omega^{M_2}\right)
\end{equation}
as densities.
\end{theorem}
\begin{proof}
Our assumptions ensure that both sides are $m_1+m_2-n_1-n_2$-forms on $F_1^{-1}(c_1)\times F_2^{-1}(c_2)$. Choose $v_i^j\in TM_i$ that satisfy $\Omega^{N_i}(F_{i*}v^1_i,...,F_{i*}v^{n_i}_i)=1$ then
\begin{align}
&\Omega^{N_1}\wedge \Omega^{N_2}(F_*(v_1^1,0),...,F_*(v_1^{n_1},0),F_*(0,v_2^1),...,F_*(0,v_2^{n_2}))\\
=&\Omega^{N_1}\wedge \Omega^{N_2}(F_{1*}v_1^1,...,F_{2*}v_2^{n_2})\notag\\
=&\Omega^{N_1}(v_1^1,...,v_1^{n_1})\Omega^{N_2}(v_2^1,...,v_2^{n_2})=1\,.\notag
\end{align}
Therefore, by definition
\begin{align}
\delta_c\circ F \left(\Omega^{M_1}\wedge\Omega^{M_2}\right)=&i_{(v_1^1,0),...,(v_1^{n_1},0),(0,v_2^1),...,(0,v_2^{n_2})}\left(\Omega^{M_1}\wedge\Omega^{M_2}\right)\\
=&(-1)^{n_2}\left(i_{v_1^1,...,v_1^{n_1}}\Omega^{M_1}\right)\wedge\left(i_{v_2^1,...,v_2^{n_2}}\Omega^{M_2}\right)\notag\\
=&(-1)^{n_2}\left(\delta_{c_1}\circ F_1\right)\wedge\left(\delta_{c_2}\circ F_2\right)\,.\notag
\end{align}
Therefore they agree as densities.
\end{proof}

\begin{theorem}\label{deltaProduct}
Let $F_i:M_i\rightarrow N_i$ and $g:N_1\times N_2\rightarrow K$ be smooth. Let $\Omega^{M_i}$, $\Omega^{N_1}$, $\Omega^K$ be volume forms on $M_i$, $N_1$, $K$ respectively. Suppose $c$ is a regular value of $F_1$ and $d$ is a regular value of $g(c,F_2)$ and of $g\circ F_1\times F_2$. Then 
\begin{equation}
\delta_c(F_1)\left[\delta_d( g\circ F_1\times F_2)\left(\Omega^{M_1}\wedge\Omega^{M_2}\right)\right]=\left(\delta_c(F_1)\Omega^{M_1}\right)\wedge\left(\delta_d(g(c, F_2))\Omega^{M_2}\right)\,.
\end{equation}
\end{theorem}
\begin{proof}
 Let $(x,y)\in (f\circ F_1\times F_2)^{-1}(d)$ with $x\in F^{-1}(c)$. For any $w\in T_c N_1$ there exists $v\in T_x M_1$ such that $F_{1*}v=w$. $d$ is a regular value of $g(c,F_2)$ hence there exists $\tilde{v}$ such that $g(c,F_2)_*\tilde{v}=(g\circ F_1\times F_2)_*(v,0)$. Therefore $(g\circ F_1\times F_2)_*(v,-\tilde{v})=0$ and $F_1*(v,-\tilde{v})=w$. This proves $c$ is a regular value of $F_1$ on $(g\circ F_1\times F_2)^{-1}(d)$. This proves both sides are defined and are forms on $F^{-1}(c)\times g(c,F_2)^{-1}(d)$.

 Let $x\in F^{-1}(c)$ and $y\in g(c,F_2)^{-1}(d)$ and choose $v_i$, $w_j$ such that
\begin{equation}
\Omega^{N_1}(F_{1*}v_1,...,F_{1*}v_{n_1})=1\,, \hspace{2mm} \Omega^K(g(c,F_2)_*w_1,...,g(c,F_2)_*w_k)=1\,.
\end{equation}
Then 
\begin{equation}
\Omega^K((g\circ F_1\times F_2)_*(0,w_1),...,(g\circ F_1\times F_2)_*(0,w_k))=1
\end{equation}
and so 
\begin{align}
\delta_d( g\circ F_1\times F_2)\left(\Omega^{M_1}\wedge\Omega^{M_2}\right)&=i_{(0,w_1),...,(0,w_k)}\left(\Omega^{M_1}\wedge\Omega^{M_2}\right)\\
&=\Omega^{M_1}\wedge\left(i_{w_1,...,w_k}\Omega^{M_2}\right)\notag\\
&=\Omega^{M_1}\wedge\left(\delta_d(g(c,F_2))\Omega^{M_2}\right)\,.\notag
\end{align}
By the same argument as above, we get $\tilde{v}_i$ such that $(v_i,\tilde{v}_i)\in T_{(x,y)} (g\circ F_1\times F_2)^{-1}(d)$. Hence
\begin{equation}
\delta_c(F_1)\left[\delta_d( g\circ F_1\times F_2)\left(\Omega^{M_1}\wedge\Omega^{M_2}\right)\right]=i_{(v_1,\tilde{v}_1),...,(v_{n_1},\tilde{v}_{n_1})}\left[\Omega^{M_1}\wedge\left(i_{w_1,...,w_k}\Omega^{M_2}\right)\right]\,.
\end{equation}
The only non-vanishing term is
\begin{equation}
\left(i_{(v_1,\tilde{v}_1),...,(v_{n_1},\tilde{v}_{n_1})}\Omega^{M_1}\right)\wedge\left(i_{w_1,...,w_k}\Omega^{M_2}\right)=\left(i_{v_1,...,v_{n_1}}\Omega^{M_1}\right)\wedge\left(i_{w_1,...,w_k}\Omega^{M_2}\right)
\end{equation}
since the other terms all contain a $m_1-n_1+l$ form on the $m_1-n_1$-dimensional manifold $F^{-1}(c)$ for some $l>0$. This proves the result.
\end{proof}

Sometimes it is convenient to use the delta function to introduce ``dummy integration variables", by which we mean utilizing the following simple corollary of the coarea formula.
\begin{corollary}\label{dummyInt}
Let $\Omega^M$ be a volume form on $M$, $F:M\rightarrow (N,\Omega^N)$ be smooth, and $f:N\times M\rightarrow \mathbb{R}$ such that $f(F(\cdot),\cdot)\in L^1(\Omega^M)\bigcup L^+(M)$. If $F_*$ is surjective at a.e. $x\in M$ then
\begin{equation}
\int_M f(F(x),x)\Omega^M(dx)=\int_N\int_{F^{-1}(z)} f(z,x)\delta_z(F)\Omega^M(dx) \Omega^N(dz)\,.
\end{equation}
\end{corollary}

\subsection{Applications}
\para{Relativistic volume element}
We now discuss an application of the above results to the single particle phase space volume element. We first define it in the massive case, where the semi-Riemannian method of defining volume forms is applicable. The massless case is often handled via a limiting argument \cite{tsamparlis}. We will show that our method is able to handle both the massive and massless case in a unified manner.

 Given a time oriented $n+1$ dimensional semi-Riemannian manifold $(M,g)$, there is a natural induced metric $\tilde{g}$ on the tangent bundle, called the diagonal lift. At a given point $(x,p)\in TM$ its coordinate independent definition is
\begin{align}
\tilde{g}_{(x,p)}(v,w)=g_x(\pi_{*} v,\pi_{*} w)+g_x(D_t \gamma_v, D_t \gamma_w)\,,
\end{align}
where $\gamma_v$ is any curve in $TM$ with tangent $v$ at $x$, $\pi:TM\longrightarrow M$ is the projection, and $D_t\gamma_v$ is the covariant derivative of $\gamma_v$, treated as a vector field along the curve $\pi\circ\gamma_v$, and similarly for $\gamma_w$, see, e.g., \cite{pettini}. The result can be shown to be independent of the choice of curves. In a coordinate system on $M$ where the first coordinate is future timelike and the Christoffel symbols are $\Gamma^\beta_{\sigma\eta}$, consider the induced coordinates $(x^\alpha,p^\alpha), \hspace{2mm}\alpha=0,...,n$ on $TM$. In these coordinates we have 
\begin{equation}
\tilde{g}_{(x^\alpha,p^\alpha)}=g_{\beta,\delta}(x^\alpha)dx^\beta\otimes dx^\delta +g_{\beta,\delta}(x^{\alpha})\epsilon^\beta\otimes \epsilon^\delta, \hspace{2mm} \epsilon^\beta=dp^\beta+p^\sigma\Gamma^\beta_{\sigma\eta}(x^{\alpha})dx^\eta\,.
\end{equation}
The vertical and horizontal subspaces are spanned by
\begin{equation}\label{horizontalSubspace}
V_\alpha=\partial_{p^\alpha}, \hspace{2mm} H_\alpha=\partial_{x^\alpha}-p^\sigma\Gamma_{\sigma\alpha}^\beta\partial_{p^\beta}\,,
\end{equation}
respectively. The horizontal vector fields satisfy
\begin{equation}
\tilde{g}(H_\alpha,H_\beta)=g_{\alpha\beta}\,.
\end{equation}
For any manifold (oriented or not), the tangent bundle has a canonical orientation. With this orientation, the volume form on $TM$ induced by $\tilde{g}$ is
\begin{equation}
\widetilde{dV}_{(x^\alpha,p^{\alpha})}=|g(x^\alpha)|dx^0\wedge...\wedge dx^n\wedge dp^0\wedge...\wedge dp^n\,,
\end{equation}
where $|g(x^\alpha)|$ denotes the absolute value of the determinant of the component matrix of $g$ in these coordinates.

Of primary interest in kinetic theory for a particle of mass $m\geq 0$ is the mass shell bundle
\begin{equation}
P_m=\left\{p\in TM :g(p,p)=m^2,\hspace{1mm} p\text{ future directed}\right\}
\end{equation}
and it will be necessary to have a volume form on $P_m$. $P_m$ is a connected component of the zero set of the of the smooth map 
\begin{equation}\label{definingFunction}
h:TM\setminus \{0_x:x\in M\}\longrightarrow \mathbb{R},\hspace{2mm} h(x,p)= \frac{1}{2}(g_x(p,p)-m^2)\,. 
\end{equation} 
We remove the image of the zero section to avoid problems when $m=0$. Its differential is
\begin{equation}\label{dh}
dh=\frac{1}{2}\frac{\partial g_{\sigma\delta}}{\partial x^\alpha}p^\sigma p^\delta dx^\alpha+g_{\sigma\delta}p^\sigma dp^\delta=g_{\sigma\delta}p^\sigma\epsilon^\delta\,.
\end{equation}
$g$ is nondegenerate, so for $p=p^{\alpha}\partial_{x^\alpha}\in TM_x\setminus{\{0_x\}}$ there is some $v=v^\alpha\partial{x^\alpha}\in TM_x$ with $g(v,p)\neq 0$. Therefore
\begin{equation}
dh_{(x,p)}(v^\alpha\partial_{p^\alpha})=g(v,p)\neq 0\,.
\end{equation}
This proves $P_m$ is a regular level set of $h$, and hence is a closed embedded hypersurface of $TM\setminus \{0_x:x\in M\}$. For $m\neq 0$ it is also closed in $TM$, but for $m=0$ every zero vector is a limit point of $P_m$.\\

\noindent{\bf Massive Case:}\\
For $m\neq 0$, we will show that $P_m$ is a semi-Riemannian hypersurface in $TM$ and hence inherits a volume form from $TM$. This is the standard method of inducing a volume form, as presented in \cite{tsamparlis}. 

The normal to $P_m$ is 
\begin{equation}
\grad h=\tilde{g}^{-1}(dh)=p^\alpha\partial_{p^\alpha}
\end{equation}
which has norm squared 
\begin{equation}
\tilde{g}(\grad h,\grad h)=g(p,p)=m^2\,.
\end{equation}
Therefore, for $m\neq 0$, $P_m$ has a unit normal $N=\grad h/m$ and so it is a semi-Riemannian hypersurface with volume form
\begin{equation}
\widetilde{dV}_m=i_N \widetilde{dV}=\frac{|g|}{m}dx^0\wedge...\wedge dx^n\wedge\left(\sum_\alpha (-1)^\alpha p^\alpha dp^0\wedge...\wedge\widehat{dp^\alpha}\wedge...\wedge dp^n\right)\,,
\end{equation}
where $i_N$ denotes the interior product\index{interior product} (or contraction) and a hat denotes an omitted term. We are also interested in the volume form on $P_{m,x}$ the fiber of $P_m$ over a point $x\in M$. We obtain this by contracting $\widetilde{dV}$ with an orthonormal basis of vector fields normal to $P_{m,x}$. Such a basis is composed of $N$ together with an orthonormalization of the basis of horizontal fields, $W_\alpha=\Lambda^\beta_\alpha H_\beta$, where $H_\beta$ are defined in \req{horizontalSubspace}. Therefore we have
\begin{equation}
\widetilde{dV}_{m,x}=i_{W_0}...i_{W_n}\widetilde{dV}_m\,.
\end{equation}
 We can simplify these expressions by defining a coordinate system on the momentum bundle, writing $p^0$ as a function of the $p^i$. The details, which are standard, are carried out in 
 at the end of this Appendix. The results are
\begin{equation}
\widetilde{dV}_m=\frac{m|g|}{p_0}dx^0\wedge...\wedge dx^n\wedge dp^1\wedge...\wedge dp^n\,,
\end{equation}
\begin{equation}
\widetilde{dV}_{m,x}=\frac{m|g|^{1/2}}{p_0}dp^1\wedge...\wedge dp^n\,.
\end{equation}
We define $\pi$ and $\pi_x$ by
\begin{equation}
\pi=\frac{1}{m}\widetilde{dV}_m=\frac{|g|}{p_0}dx^0\wedge...\wedge dx^n\wedge dp^1\wedge...\wedge dp^n\,,
\end{equation}
\begin{equation}\label{pi:x}
\pi_x=\frac{1}{m}\widetilde{dV}_{m,x}=\frac{|g|^{1/2}}{p_0}dp^1\wedge...\wedge dp^n\,.
\end{equation}
We will typically omit the subscript $x$ and let the context distinguish whether we are integrating over the full momentum bundle (i.e. both over spacetime and momentum variables) or just momentum space at a single point in spacetime. \\

\noindent{\bf Massless Case:}\\
When $m=0$ the above construction fails. However, we can use Theorem \ref{inducedVolForm} to induce a volume form using the map \req{definingFunction} defined above. Here we carry out the construction for the induced volume\index{induced volume form} form on $P_{m,x}$ for any $m\geq 0$. The volume form on each tangent space $T_xM$ is
\begin{equation}
\tilde{dV}_x=|g(x)|^{1/2}dp^0\wedge...\wedge dp^n\,.
\end{equation}
We assume that the coordinates are chosen so that the vector field $\partial_{p^0}$ is timelike. By \req{dh} we find
\begin{equation}
dh(\partial_{p^0})=g_{\alpha 0}p^\alpha\neq 0
\end{equation}
on $P_{m,x}$. Therefore, by Corollary \ref{inducedVolEq} the induced volume form is
\begin{align}\label{massShellVol}
\omega=&\frac{1}{dh(\partial_{p^0})} i_{\partial_{p^0}} \tilde{dV}_x
=\frac{|g|^{1/2}}{p_0}dp^1\wedge...\wedge dp^n\,.
\end{align}
We can also pull this back under the coordinate chart on $P_{m,x}$ defined at the end of this Appendix
and obtain the same expression in coordinates. This result agrees with our prior definition of \req{pi:x} in the case where $m>0$ but is also able to handle the massless case in a uniform manner, without resorting to a limiting argument as $m\rightarrow 0$.

We also point out another convention in common use where $h$ is replaced by $2h$. This leads to an additional factor of $1/2$ in the volume element, distinguishing this definition from the one based on semi-Riemannian geometry. However, the convention
\begin{equation}
\omega=\frac{|g|^{1/2}}{2p_0}dp^1\wedge...\wedge dp^n
\end{equation}
 is in common use and will be employed in the scattering integral computations in \rapp{ch:coll:simp}.

\para{Relativistic phase space}
Here we justify several manipulations that are useful for working with relativistic phase space integrals.

\begin{lemma}\label{parallel:lemma}
Let $V$ be an $n$-dimensional vector space. The subset of $\prod_1^N V\setminus\{0\}$ consisting of $N$-tuples of parallel vectors is an $n+N-1$ dimensional closed submanifold of $\prod_1^N V\setminus\{0\}$.
\end{lemma}
\begin{proof}
The map $V\times \mathbb{R}^{N-1}\rightarrow \prod_1^N V\setminus\{0\}$ given by
\begin{equation}
F(p,a^2,...,a^N)=(p,a^2p,...,a^{N}p)
\end{equation}
is an injective immersion and maps onto the desired set.
\end{proof}
For reactions converting $k$ particles to $l$ particles, the relevant phase space is $3(k+l)-4$ dimensional and so for $k+l\geq 4$ (in particular for $2$-$2$ reactions), the set of parallel $4$-momenta is lower dimensional and can be ignored. This will be useful as we proceed.

\begin{lemma}
Let $N\geq 4$. Then
\begin{equation}
\prod_i \delta(p_i^2-m_i^2)d^4p_i=\left(\prod_i \delta(p_i^2-m_i^2)\right)\prod_i d^4p_i
\end{equation}
 and 
\begin{equation}
\delta(\Delta p)\left[\left(\prod_i \delta(p_i^2-m_i^2)\right)\prod_i d^4p_i\right]=\left(\delta(\Delta p)\prod_i \delta(p_i^2-m_i^2)\right)\prod_i d^4p_i\,,
\end{equation}
where each $d^4p_i$ is the standard volume form on future directed vectors, $\{p:p^2\geq 0, p^0>0\}$, we give $\mathbb{R}$ its standard volume form, and $\Delta p=a^ip_i$, $a^i=1$, $i=1,...,l$, $a^i=-1$, $i=l,...,N$. 
\end{lemma}
\begin{proof}
Let $F_1(p_i)=(p_1^2,...,p_N^2)$ and $F_2(p_i)=(\Delta p,F_1(p_i))$. We need to show that $(m_1^2,...,m_N^2)$ is a regular value of $F_1$ and $(0,m_1^2,...,m_k^2)$ is a regular value of $F_2$. The result then follows from Theorem \ref{delta:associative}.

It holds for $F_1$ since each $p_i\neq 0$. For $F_2$, the differential is
\begin{equation}
(F_2)*=\left( \begin{array}{cccc}
a^{1}I&a^{2}I&...& a^{N}I \\
2 \eta_{ij}p^j_1&0&...&0\\
\vdots&&&\vdots\\
0&...&0&2 \eta_{ij}p^j_N\\
\end{array} \right)\,,
\end{equation}
where $I$ is the $4$-by-$4$ identity. The fact that $(F_1)_*$ is onto means that we need only show $(F_2)_*$ maps onto $\mathbb{R}^4\times(0,...,0)$. 

By Lemma \ref{parallel:lemma} we assume there exists $i,j$ such that $p_i,p_j$ are not parallel. We are done if for each standard basis vector $e_k\in\mathbb{R}^4$ there exists $q\in\mathbb{R}^4$ such that
\begin{equation}
p_i\cdot q=\frac{1}{a^j}p_i\cdot e_k,\hspace{2mm} p_j\cdot q=0\,.
\end{equation}
If $p_j$ is null then there is a $c$ such that $q=c p_j$ satisfies these conditions. If $p_j$ is non-null then complete it to an orthonormal basis. $p_i$ must have a component along the orthogonal complement of $p_j$ and we can take $q$ to be proportional to that component.

\end{proof}

\para{Volume form in coordinates}
Here we derive a useful formula for the volume form on the momentum bundle in a simple coordinate system. We begin in a coordinate system $x^\alpha$ on $U\subset M$ and the induced coordinates $p^\alpha$ on $TM$ where our only assumption is that the $0$'th coordinate direction is future timelike, and so $g_{00}>0$. For any $v^i\in \mathbb{R}^n$, let $v^0=-g_{0i}v^i/g_{00}$. $v^\alpha$ is orthogonal to the $0$'th coordinate direction, and therefore spacelike. Hence 
\begin{equation}
0\geq g_{\alpha \beta}v^\alpha v^\beta=-(g_{0i}v^i)^2/g_{00}+g_{ij}v^iv^j\,.
\end{equation}
and is zero iff $v^\alpha=0$. Therefore, the following map is well defined
\begin{align}
(x^\alpha,p^j)&\longrightarrow (x^\alpha,p^0(x^\alpha,p^j),p^1,...,p^n), \hspace{2mm} \alpha=0...n, \hspace{1mm} j=1...n \,,\notag\\
 p^0&=-g_{0j}p^j/g_{00}+\left((g_{0j}p^j/g_{00})^2+(m^2-g_{ij}p^ip^j)/g_{00}\right)^{1/2}\,,
\end{align}
and is smooth on $\mathbb{R}^{n+1}\times\mathbb{R}^n$ if $m\neq 0$, and on $\mathbb{R}^{n+1}\times\left(\mathbb{R}^n\setminus{0}\right)$ if $m=0$. We also have $g_{00}p^0+g_{0j}p^j>0$ under either of these cases, and so the resulting element of $TM$ is future directed and has squared norm $m^2$, so it maps into $P_m$. It is a bijection and has full rank, hence it is a coordinate system on $P_m$. In these coordinates, the volume form is
\begin{align}
\widetilde{dV}_m=&\frac{|g|}{m}dx^0\wedge...\wedge dx^n\wedge\left(p^0dp^1\wedge...\wedge dp^n+\sum_j (-1)^j p^j dp^0\wedge...\wedge\widehat{dp^j}\wedge...\wedge dp^n\right)\notag\\
dp^0=&\partial_{x^\alpha} p^0dx^\alpha+\partial_{p^j}(p^0) dp^j\,.
\end{align}
The terms in $dp^0$ involving $dx^\alpha$ drop out once they are wedged with $dx^0\wedge...\wedge dx^n$, hence
\begin{align}
\widetilde{dV}_m
=&\frac{|g|}{m}dx^0\wedge...\wedge dx^n\wedge\left(p^0dp^1\wedge...\wedge dp^n+\sum_{i,j} (-1)^j p^j \partial_{p^i}p^0 dp^i\wedge...\wedge\widehat{dp^j}\wedge...\wedge dp^n\right)\notag\\
=&\frac{|g|}{m}\left(p^0-\sum_{j}p^j \partial_{p^j}(p^0) \right)dx^0\wedge...\wedge dx^n\wedge dp^1\wedge...\wedge dp^n\,,
\end{align}
where
\begin{align}
 \hspace*{1cm}\left(p^0-\sum_jp^j\partial_{p^j}(p^0)\right) =& p^0+g_{0j}p^j/g_{00}-\frac{(g_{0j}p^j/g_{00})^2-g_{ij}p^ip^j/g_{00}}{\left((g_{0j}p^j/g_{00})^2+(m^2-g_{ij}p^ip^j)/g_{00}\right)^{1/2}}\notag\\
=&\frac{1}{p_0}\left(\frac{1}{g_{00}}(g_{00}p^0+g_{0,j}p^j)^2-(g_{0j}p^j)^2/g_{00}+g_{ij}p^ip^j\right)=\frac{m^2}{p_0}\,. 
\end{align}
Therefore
\begin{equation}
\widetilde{dV}_m=\frac{m|g|}{p_0}dx^0\wedge...\wedge dx^n\wedge dp^1\wedge...\wedge dp^n\,.
\end{equation}

To compute the volume form on $P_{m,x}$, recall that 
\begin{equation}\label{contractHoriz}
\widetilde{dV}_{m,x}=i_{W_0}...i_{W_n}\widetilde{dV}_m\,,
\end{equation}
where $W_i$ is an orthonormalization of the basis of horizontal fields, $W_\alpha=\Lambda^\beta_\alpha H_\beta$, and where $H_\beta$ are defined in \req{horizontalSubspace}. All of the contractions in \req{contractHoriz} that involve the $dp^\alpha$'s will be zero when restricted to $P_{m,x}$ since the $dx^\alpha$ are zero there. Hence we obtain
\begin{align}\label{dVx}
\widetilde{dV}_{m,x}=&\frac{|g|}{m}\left(p^0-\sum_{j}p^j \partial_{p^j}(p^0) \right)dx^0\wedge...\wedge dx^n\left(W_0,...,W_n)\right) dp^1\wedge...\wedge dp^n\\
=&\frac{|g|\det(\Lambda)}{m}\left(p^0-\sum_{j}p^j \partial_{p^j}(p^0) \right)dx^0\wedge...\wedge dx^n\left(H_0,...,H_n)\right) dp^1\wedge...\wedge dp^n\notag\\
=&\frac{|g|^{1/2}}{m}\left(p^0-\sum_{j}p^j \partial_{p^j}(p^0) \right) dp^1\wedge...\wedge dp^n\,,\notag
\end{align}
where we used $\det(\Lambda^\sigma_\alpha g_{\sigma\delta}\Lambda^\delta_\beta)=1$.
 In the coordinate system on $P_{m,x}$,
\begin{align}
(p^j)&\longrightarrow (p^0(x^\alpha,p^j),p^1,...,p^n)\,,\\
 p^0&=-g_{0j}(x)p^j/g_{00}(x)+\left((g_{0j}(x)p^j/g_{00}(x))^2+(m^2-g_{ij}(x)p^ip^j)/g_{00}(x)\right)^{1/2}\,.\notag
\end{align}
The above calculation gives the formula
\begin{equation}
\widetilde{dV}_{m,x}=\frac{m|g|^{1/2}}{p_0}dp^1\wedge...\wedge dp^n\,.
\end{equation}

\section{Boltzmann-Einstein Equation Solver Adapted to Emergent Chemical Nonequilibrium}\label{ch:boltz:orthopoly}
Having completed the geometrical background in \rapp{ch:vol:forms}, we now proceed to develop a numerical method for the Boltzmann-Einstein equation in an FLRW\index{cosmology!FLRW} Universe.  This will allow us to efficiently study  nonequilibrium aspects of neutrino freeze-out\index{neutrino!freeze-out}. The analysis in \rsec{sec:model:ind} was based on exact chemical and kinetic equilibrium and sharp freeze-out transitions at $T_{ch}$ and $T_k$, but these are  only approximations.  The  Boltzmann-Einstein equation is a more precise model of the dynamics of the freeze-out process and furthermore, given the collision dynamics it is capable of capturing in a {\em quantitative manner} the non-thermal distortions from equilibrium, for example the emergence of actual distributions and the approximate values  of $T_{ch}$, $T_k$, and $\Upsilon$.  Indeed,  in  such a dynamical description no hypothesis about the presence of kinetic or chemical (non) equilibrium needs to be made, as the distribution close to \req{kinetic:equilibrium} with   $\Upsilon\ne  1$ emerges naturally as the outcome of collision processes, even when the particle system approaches the freeze-out temperature domain  in chemical equilibrium\index{chemical equilibrium}.

Considering the natural way in which chemical nonequilibrium emerges from chemical equilibrium during freeze-out, it is striking that the literature on Boltzmann solvers does not reflect on the accommodation of emergent chemical nonequilibrium into the method of solution. For an all-numerical solver this may not be a necessary step as long as there are no constraints that preclude development of a general nonequilibrium solution. However, when strong chemical nonequilibrium is present either in the intermediate time period or/and at the end of the evolution, a brute force approach can be very costly in computer time. Motivated by this circumstance and past work with physical environments in which chemical nonequilibrium arises,  we introduce here a  spectral method for solving the Boltzmann-Einstein equation that utilizes a dynamical basis of orthogonal polynomials, which is adapted to the case of emerging chemical nonequilibrium. We validate our method via a  model problem  that captures the essential physical characteristics of interest and use it to highlight the type of situation where this new method exhibits its advantages.

In the cosmological context, the Boltzmann-Einstein equation has been used to study neutrino freeze-out in the early Universe and has been successfully solved using both discretization in momentum space \cite{Hannestad:1995rs,Dolgov:1997mb,Dolgov:1998sf,Gnedin:1997vn,Mangano:2005cc} and a spectral method based on a fixed basis of orthogonal polynomials \cite{Esposito:2000hi,Mangano:2001iu}.    In Refs.\cite{Wilkening,Wilkening2} the nonrelativistic Boltzmann equation was solved via a spectral method similar in  one important mathematical idea to the approach we present here.  For near equilibrium solutions, the spectral methods have the advantage of requiring a relatively small number of modes to obtain an accurate solution, as opposed to momentum space discretization, which in general leads to a large highly coupled nonlinear system of odes irrespective of the near equilibrium nature of the system.  

The efficacy of the spectral method used in \cite{Esposito:2000hi,Mangano:2001iu} can largely be attributed to the fact that, under the conditions considered there, the true solution is very close to a chemical equilibrium distribution, \req{equilibrium}, where the temperature is controlled by the dilution of the system. However, as we have discussed, the Planck CMB results \cite{Planck:2013pxb} indicate the possibility that neutrinos participated in reheating to a greater degree than previously believed, leading to a more pronounced chemical nonequilibrium and reheating. Efficiently obtaining this emergent chemical nonequilibrium within the realm of kinetic theory motivates the development of a new numerical method that is adapted to this  circumstance.

In  \rsec{sec:orthopolyApp}  we give important general background on moving frames of orthogonal polynomials, deriving several formulas and properties that will be needed in our method for solving the Boltzmann-Einstein equation. In \rsec{sec:theMethod} we develop the details of our method. We start with a basic overview of the Boltzmann-Einstein equation in an FLRW\index{cosmology!FLRW} Universe, then we recall the orthogonal polynomial basis used in \cite{Esposito:2000hi,Mangano:2001iu} and compare this with our modified basis moving frame method.  We use the Boltzmann-Einstein equation to derive the dynamics of the mode coefficients and identify physically motivated evolution equations for the effective temperature and fugacity. In \rsec{validation} we validate the method using a model problem. This section is adapted from \cite{Birrell:2014ona,Birrell:2014gea,Birrell:2014uka}.

\subsection{Orthogonal polynomials}\label{sec:orthopolyApp}
In this section we give details regarding the construction of the moving frame of orthogonal polynomials\index{orthogonal polynomials} that will be required for our Boltzmann-Einstein equation solver.
\para{Generalities}
Let $w:(a,b)\rightarrow [0,\infty)$ be a weight function where $(a,b)$ is a (possibly unbounded) interval and consider the Hilbert space $L^2(w(x) dx)$.   We will consider weights such that $x^n\in L^2(w(x) dx)$ for all $n\in\mathbb{N}$. We denote the inner product by $\langle\cdot,\cdot\rangle$, the norm by $||\cdot||$, and for a vector $\psi\in L^2$ we let $\hat{\psi}\equiv \psi/||\psi||$.  The classical three term recurrence formula can be used to define a set of orthonormal polynomials $\hat{\psi}_i$ using this weight function, see, e.g.,  \cite{Olver},
\begin{align}\label{polyRecursion}
&\psi_0=1\,, \hspace{2mm} \psi_1=||\psi_0||(x-\langle x\hat\psi_0,\hat{\psi}_0\rangle)\hat{\psi}_0\,,\\
&\psi_{n+1}=||\psi_n||\left[\left(x-\langle x\hat\psi_n,\hat\psi_n\rangle\right)\hat\psi_n-\langle x\hat\psi_n,\hat\psi_{n-1}\rangle\hat\psi_{n-1}\right]\,.\notag
\end{align}
One can also derive recursion relations for the derivatives of $\psi_n$ with respect to $x$, denoted with a prime,
\begin{align}\label{derivRec}
&\psi_0^{'}=0, \hspace{2mm} \hat{\psi}_1^{'}=\frac{||\psi_0||}{||\psi_1||}\hat\psi_0\,,\\
&\hat{\psi}_{n+1}^{'}=\frac{||\psi_n||}{||\psi_{n+1}||}\left[\hat\psi_n+\left(x-\langle x\hat\psi_n,\hat\psi_n\rangle\right)\hat{\psi}_n^{'}-\langle x\hat\psi_n,\hat\psi_{n-1}\rangle\hat{\psi}_{n-1}^{'}\right]\,.\notag
\end{align}
Since $\hat{\psi}_n^{'}$ is a degree $n-1$ polynomial, we have the expansion 
\begin{equation}
\hat{\psi}_n^{'}=\sum_{k<n} a_n^k \hat{\psi}_k\,.
\end{equation}
Using \req{derivRec} we obtain a recursion relation for the $a_n^k$
\begin{align}
a_{n+1}^k=&\frac{||\psi_n||}{||\psi_{n+1}||}\left(\delta_{n,k}-\langle x\hat\psi_n,\hat\psi_{n}\rangle a_n^k-\langle x\hat\psi_n,\hat\psi_{n-1}\rangle a_{n-1}^k+\sum_{l<n}^la_n^l\langle x\hat\psi_l,\hat\psi_k\rangle\right)\,,\notag\\
a_1^0=&\frac{||\psi_0||}{||\psi_1||}\,.\notag
\end{align}

\para{Parametrized families of orthogonal polynomials}
Our method requires not just a single set of orthogonal polynomials, but rather a parametrized family of orthogonal polynomials that are generated by a weight function $w_t(x)$ that is a $C^1$ function of both $x\in(a,b)$ and the parameter $t$. The corresponding time-dependent basis of orthogonal polynomials, also called a moving frame, is used to define the spectral method for solving the  Boltzmann-Einstein equation as outlined in \rsec{dynamicsSec}.  To emphasize the time dependence, in this section we write $g_t(\cdot,\cdot)$ for the inner product $\langle\cdot,\cdot\rangle$ (not to be confused with the spacetime metric tensor).  We will assume that $\partial_t w$ is dominated by some $L^1(dx)$ function of $x$ only that decays exponentially as $x\rightarrow\pm\infty$ (if the interval is unbounded). In particular, this holds for the weight function \req{weight}.

Given the above assumption about the decay of $\partial_t w$, the dominated convergence theorem implies that $\langle p,q\rangle$ is a $C^1$ function of $t$ for all polynomials $p$ and $q$ and justifies  differentiation under the integral sign. By induction, it also implies implies that the $\hat\psi_i$ have coefficients that are $C^1$ functions of $t$. Therefore, for any polynomials $p$, $q$ whose coefficients are $C^1$ functions of $t$, we have
\begin{equation}
\frac{d}{dt}g_t( p,q)=\dot{g}_t(p,q)+g_t(\dot{p},q)+g_t( p,\dot{q})\,,
\end{equation}
where a dot denotes differentiation with respect to $t$ and we use $\dot{g}_t(\cdot,\cdot)$ to denote the inner product with respect to the weight $\dot{w}$.  

\req{bEq} for the mode coefficients requires us to compute $g(\dot{\hat\psi}_i,\hat\psi_j)$.  Differentiating the relation
\begin{equation}
\delta_{ij}=g_t(\hat\psi_i,\hat\psi_j)
\end{equation}
yields
\begin{equation}\label{orthoDerivEq}
0=\dot g_t(\hat\psi_i,\hat\psi_j)+g_t(\dot{\hat\psi}_i,\hat\psi_j)+g_t(\hat\psi_i,\dot{\hat\psi}_j)\,.
\end{equation}
For $i=j$ we obtain
\begin{equation}\label{normDerivEq}
g_t(\dot{\hat\psi}_i,\hat\psi_i)=-\frac{1}{2}\dot{g}_t(\hat\psi_i,\hat\psi_i)\,.
\end{equation}
For $i<j$, $\dot{\hat\psi}_i$ is a degree $i$ polynomial and so it is orthogonal to $\hat\psi_j$. Therefore \req{orthoDerivEq} simplifies to
\begin{equation}
g_t(\dot{\hat\psi}_i,\hat\psi_j)=-\dot{g}_t(\hat\psi_i,\hat\psi_j),\hspace{2mm} i\neq j\,.
\end{equation}

\para{Proof of lower triangularity}
Here we prove that the matrices that define the dynamics of the mode coefficients $b^k$ are lower triangular. This fact reduces the number of integrals that must be computed in practice.  Recall the definitions
\begin{align}\label{eq:lowerTri1}
A^k_i(\Upsilon)\equiv&\langle\frac{z}{f_\Upsilon }\hat\psi_i\partial_zf_\Upsilon ,\hat\psi_k\rangle+\langle z\partial_z \hat\psi_i,\hat\psi_k\rangle\,,\\
B^k_i(\Upsilon)\equiv &\Upsilon\left(\langle\frac{1}{f_\Upsilon }\frac{\partial f_\Upsilon }{\partial\Upsilon}\hat\psi_i,\hat\psi_k\rangle+\langle\frac{\partial\hat{\psi}_i}{\partial \Upsilon},\hat\psi_k\rangle\right)\,.\notag
\end{align}
Using integration by parts, we see that
\begin{equation}\label{eq:lowerTri2}
A^k_i=-3\langle\hat\psi_i,\hat\psi_k\rangle-\langle \hat \psi_i,z\partial_z\hat\psi_k\rangle\,.
\end{equation}
Since $\hat\psi_i$ is orthogonal to all polynomials of degree less than $i$ we have $A^k_i=0$ for  $k<i$.  

$B^k_i$ can be simplified as follows.  First differentiate 
\begin{equation}
\delta_{ik}=\langle \hat\psi_i,\hat\psi_j\rangle
\end{equation}
with respect to $\Upsilon$ to obtain
\begin{align}
0=&\int \hat\psi_i\hat\psi_k\partial_{\Upsilon}wdz+\langle \partial_{\Upsilon}\hat\psi_i,\hat\psi_k\rangle+\langle \hat\psi_i,\partial_{\Upsilon}\hat\psi_k\rangle\\
=&\langle\frac{\hat\psi_i}{f_\Upsilon}\partial_{\Upsilon}f_\Upsilon,\hat\psi_k \rangle+\langle\partial_{\Upsilon}\hat\psi_i,\hat\psi_k\rangle+\langle \hat\psi_i,\partial_{\Upsilon}\hat\psi_k\rangle\,.\notag
\end{align}
Therefore 
\begin{equation}\label{eq:lowerTri5}
B^k_i=-\Upsilon\langle\hat\psi_i,\partial_{\Upsilon}\hat\psi_k\rangle\,.
\end{equation}
$\partial_\Upsilon \hat\psi_k$ is a degree $k$ polynomial, hence $B_i^k=0$ for $k<i$ as desired.

\subsection{Spectral method for Boltzmann-Einstein equation  in an FLRW Universe}\label{sec:theMethod}
\para{Boltzmann-Einstein equation  in an FLRW Universe}
Recall the  Boltzmann-Einstein equation in a general spacetime\index{cosmology!FLRW}, as introduced in \rsec{sec:BoltzmannEinstein},
\begin{equation}
p^\alpha\partial_{x^\alpha}f-\Gamma^j_{\mu\nu}p^\mu p^\nu\partial_{p^j}f=C[f]\,.
\end{equation}
As discussed above, the left-hand side expresses the fact that particles undergo geodesic motion in between point collisions. The term $C[f]$ on the right-hand side of the Boltzmann-Einstein equation is called the collision operator and models the short range scattering processes that cause deviations from geodesic motion. For $2\leftrightarrow 2$ reactions between fermions, such as neutrinos and $e^\pm$, the collision operator takes the form
\begin{align}\label{coll}
C[f_1]=&\frac{1}{2}\int F(p_1,p_2,p_3,p_4) S |\mathcal{M}|^2(2\pi)^4\delta(\Delta p)\prod_{i=2}^4\delta_0(p_i^2-m_i^2)\frac{d^4p_i}{(2\pi)^3}\,,\\
F=&f_3(p_3)f_4(p_4)f^1(p_1)f^2(p_2)-f_1(p_1)f_2(p_2)f^3(p_3)f^4(p_4)\,,\notag\\
f^i=&1- f_i\,.\notag
\end{align}
Here $|\mathcal{M}|^2$ is the process amplitude or matrix element, $S$ is a numerical factor that incorporates symmetries and prevents over-counting, $f^i$ are the Fermi blocking factors, $\delta(\Delta p)$ enforces four-momentum conservation in the reactions, and the $\delta_0(p_i^2-m_i^2)$ restrict the four momenta to the future timelike mass shells.

The matrix element for a $2\leftrightarrow2$ reaction is some function of the Mandelstam variables $s, t, u$, of which only two are independent, defined by\index{Mandelstam variables}
\begin{align}\label{Mandelstam}
&s=(p_1+p_2)^2=(p_3+p_4)^2\,,\\
&t=(p_3-p_1)^2=(p_2-p_4)^2\,,\notag\\
&u=(p_3-p_2)^2=(p_1-p_4)^2\,,\notag\\
&s+t+u=\sum_i m_i^2\,.\notag
\end{align}
We will provide a detailed study of $2$-$2$ scattering kernels for neutrino processes in \rapp{ch:coll:simp}. In this section, when testing the numerical method presented below, we will use a simplified scattering model to avoid any application specific details.

We now restrict our attention to  systems of fermions under the assumption of homogeneity and isotropy. We assume that the particle are effectively massless,  i.e. the temperature is much greater than the mass scale.  Homogeneity and isotropy imply that the distribution function of each particle species under consideration has the form $f=f(t,p)$, where $p$ is the magnitude of the spacial component of the four momentum.  In a flat FLRW Universe the Boltzmann-Einstein equation reduces to
\begin{equation}\label{boltzmann:p}
\partial_t f-pH \partial_p f=\frac{1}{E}C[f]\,,\hspace{2mm} H=\frac{\dot{a}}{a}\,.
\end{equation}

The Boltzmann-Einstein equation, \req{boltzmann:p}, can be simplified by the method of characteristics. Writing $f(p, t)=g(a(t)p,t)$ and reverting back to call the new distribution $g\to f$, the 2nd term in \req{boltzmann:p} cancels out and the evolution in time can be studied directly.  Using the formulas for the moments of $f$ (equivalent to \req{energy_density}, \req{Pressure_density}, \req{number_density}, \req{entropy_density})
\begin{align}\label{moments}
\rho=&\frac{g_p}{(2\pi)^3}\int f(t,p)Ed^3p\,, \hspace{2mm} E=\sqrt{m^2+p^2}\,,\\
P=&\frac{g_p}{(2\pi)^3}\int f(t,p)\frac{p^2}{3E}d^3p\,,\\
n=&\frac{g_p}{(2\pi)^3}\int f(t,p) d^3p\,,\\
\sigma=&-\frac{g_p}{(2\pi)^3}\int (f\ln(f)\pm(1\mp f)\ln(1\mp f)) d^3p\,,
\end{align}
this transformation implies for the rate of change in the   number density and energy density
\begin{align}\label{n:div}
\frac{1}{a^3}\frac{d}{dt}(a^3n_1)=&\frac{g_p}{(2\pi)^3}\int C[f_1] \frac{d^3p}{E}\,,\\
\label{rho:div}
\frac{1}{a^4}\frac{d}{dt}(a^4\rho_1)=&\frac{g_p}{(2\pi)^3}\int C[f_1] d^3p \,.
\end{align} 
For free-streaming particles the vanishing of the collision operator implies conservation of comoving particle number of the particle species. From the associated powers of $a$ in \req{n:div} and \req{rho:div} we see that the energy per free streaming particle as measured by an observer scales as $1/a$, a manifestation or redshift.

\para{Orthogonal polynomials for systems close to kinetic and chemical equilibrium}
Here we outline the approach for solving \req{aVars} used in\index{chemical equilibrium}
 \cite{Esposito:2000hi,Mangano:2001iu} in order to contrast it with our approach as presented below.  As just discussed, the Boltzmann-Einstein equation equation  is a first order partial differential equation and can be reduced using a new variable $y=a(t)p$  via the method of characteristics and exactly solved in the collision free ($C[f]=0)$ limit.   This motivates a change of variables from $p$ to $y$, which eliminates the momentum derivative, leaving the simplified equation
\begin{equation}\label{aVars}
\partial_tf=\frac{1}{E} C[f]\,.
\end{equation}

We let $\hat\chi_i$ be the orthonormal polynomial basis on the interval $[0,\infty)$ with respect to the weight function
\begin{equation}\label{freeStreamWeight}
f_{ch}=\frac{1}{e^y+1}\,,
\end{equation}
constructed as in  \rsec{sec:orthopolyApp}. $f_{ch}$ is the Fermi-Dirac chemical equilibrium distribution for massless fermions and with temperature $T=1/a$.  Therefore this ansatz is well suited to distributions that are manifestly in chemical equilibrium ($\Upsilon=1$) or remain close and with $T\propto 1/a$, which we call dilution temperature scaling.  Assuming that $f$ is such a distribution, one is   motivated to decompose the distribution function as
\begin{equation}\label{freeStreamAnsatz}
f=f_{ch}\chi\,,\qquad \chi=\sum_i d^i\hat\chi_i
\end{equation}
and derive evolution equations for the coefficients, leading to a spectral method\index{spectral method} for the Boltzmann-Einstein equation in a FLRW Universe.

Using this ansatz \req{aVars} becomes
\begin{equation}\label{eq:ChemEquilibMethod}
\dot{d}^k=\int_0^\infty\frac{1}{E}\hat{\chi}_k C[f]dy\,.
\end{equation}
We call this the chemical equilibrium method.

One also have the following expressions for the particle number density and energy density
\begin{align}\label{freeStreamMoments}
n&=\frac{g_p}{2\pi^2 a^3}\sum_0^2 d^i\int_0^\infty f_{ch}\hat\chi_i y^2dy\,,\\
\rho&=\frac{g_p}{2\pi^2a^4}\sum_0^3 d^i\int_0^\infty f_{ch}\hat\chi_i y^3dy\,.\notag
\end{align}

Note that the sums truncate at $3$ and $4$ terms respectively, due to the fact that $\hat\chi_k$ is orthogonal to all polynomials of degree less than $k$. This implies that in general, at least four modes are required to capture both the particle number and energy flow. More modes are needed if the non-thermal distortions are large and the back reaction of higher modes on lower modes is significant.

\para{Polynomial basis for systems   far from chemical equilibrium}
Our primary interest is in solving \req{Tvars} for systems close to the kinetic equilibrium\index{kinetic equilibrium} distribution, \req{kinetic:equilibrium}, but not necessarily in chemical equilibrium, a task for which the method in the previous section is not well suited. For a general kinetic equilibrium distribution, the temperature does not necessarily scale as $T\propto 1/a$ i.e. the temperature is not controlled solely by dilution.  For this reason, we will find it more useful to make the change of variables $z=p/T(t)$ rather than the scaling used in \req{aVars}.  Here $T(t)$ is to be viewed as the time dependent effective temperature of the distribution $f$, a notion we will make precise later.  With this change of variables, the Boltzmann-Einstein equation becomes
\begin{equation}\label{TBoltzmann}
\partial_t f-z\left(H+\frac{\dot T}{T}\right)\partial_z f=\frac{1}{E}C[f]\,.
\end{equation}

To model a distribution close to kinetic equilibrium at temperature $T$ and fugacity\index{fugacity} $\Upsilon$, we assume
\begin{equation}\label{kineticApprox}
f(t,z)= f_\Upsilon (t,z)\psi(t,z)\,,\hspace{2mm} f_\Upsilon(z)=\frac{1}{\Upsilon^{-1}e^z+1}\,,
\end{equation}
where the kinetic equilibrium distribution $f_\Upsilon $ depends on $t$ because we are assuming $\Upsilon$ is time dependent (with dynamics to be specified later). 

We will solve \req{TBoltzmann} by expanding $\psi$ in the basis of orthogonal polynomials generated by the parameterized weight function
\begin{equation}\label{weight}
w(z)\equiv w_\Upsilon(z)\equiv z^2f_\Upsilon (z)=\frac{z^2}{\Upsilon^{-1} e^z+1}
\end{equation}
on the interval $[0,\infty)$. See  \rsec{sec:orthopolyApp} for details on the construction of these polynomials and their dependence on the parameter $\Upsilon$. This choice of weight is physically motivated by the fact that we are interested in solutions that describe massless particles not too far from kinetic equilibrium, but (potentially) far from chemical equilibrium. We refer to the resulting spectral method as the chemical nonequilibrium method.

We emphasize that we have made three important changes as compared to  the chemical equilibrium method:
\begin{enumerate}
\item  We allow a general time dependence of the effective temperature parameter $T$, i.e., we do not assume dilution temperature scaling $T=1/a$.
\item We have replaced the chemical  equilibrium distribution in the weight, \req{freeStreamWeight},  with a chemical nonequilibrium distribution  $f_\Upsilon $, i.e., we introduced $\Upsilon$.
\item We have introduced an additional factor of $z^2$ to the functional form of the weight as proposed in a different context in Refs.\cite{Wilkening,Wilkening2}. 
\end{enumerate} 
We note that the authors of \cite{Esposito:2000hi} did consider the case of fixed chemical potential imposed as an initial condition. This is not the same as an emergent chemical nonequilibrium, i.e. time dependent $\Upsilon$, that we study here, nor do they consider a $z^2$ factor in the weight. We borrowed the idea for the $z^2$ prefactor from   Ref.\cite{Wilkening2}, where it was found that including a $z^2$ factor along with the nonrelativistic chemical equilibrium distribution in the weight improved the accuracy of their method. Fortuitously,  this will also allow us to capture the particle number and energy flow with fewer terms than required by the chemical equilibrium method. 

\begin{figure}
\centerline{ 
\includegraphics[width=0.48\linewidth]{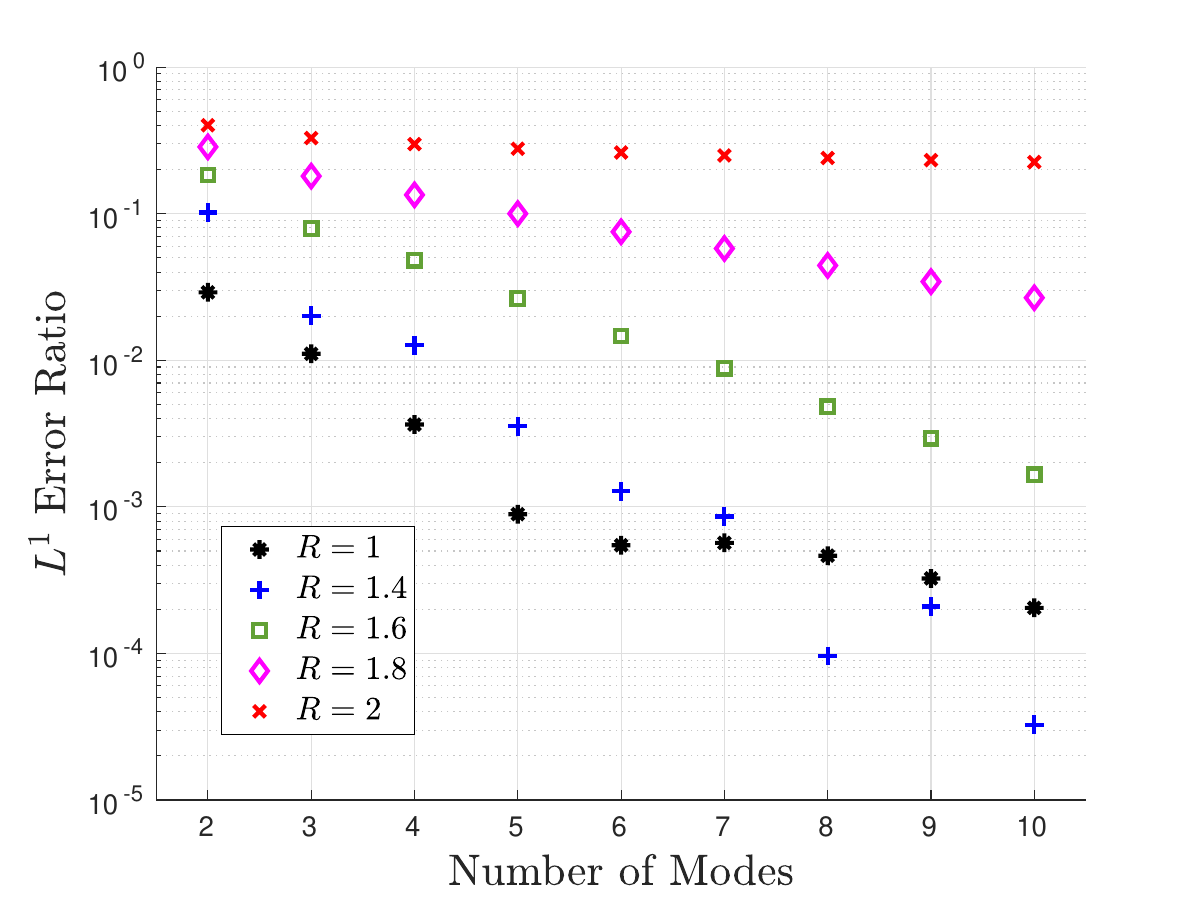} \includegraphics[width=0.48\linewidth]{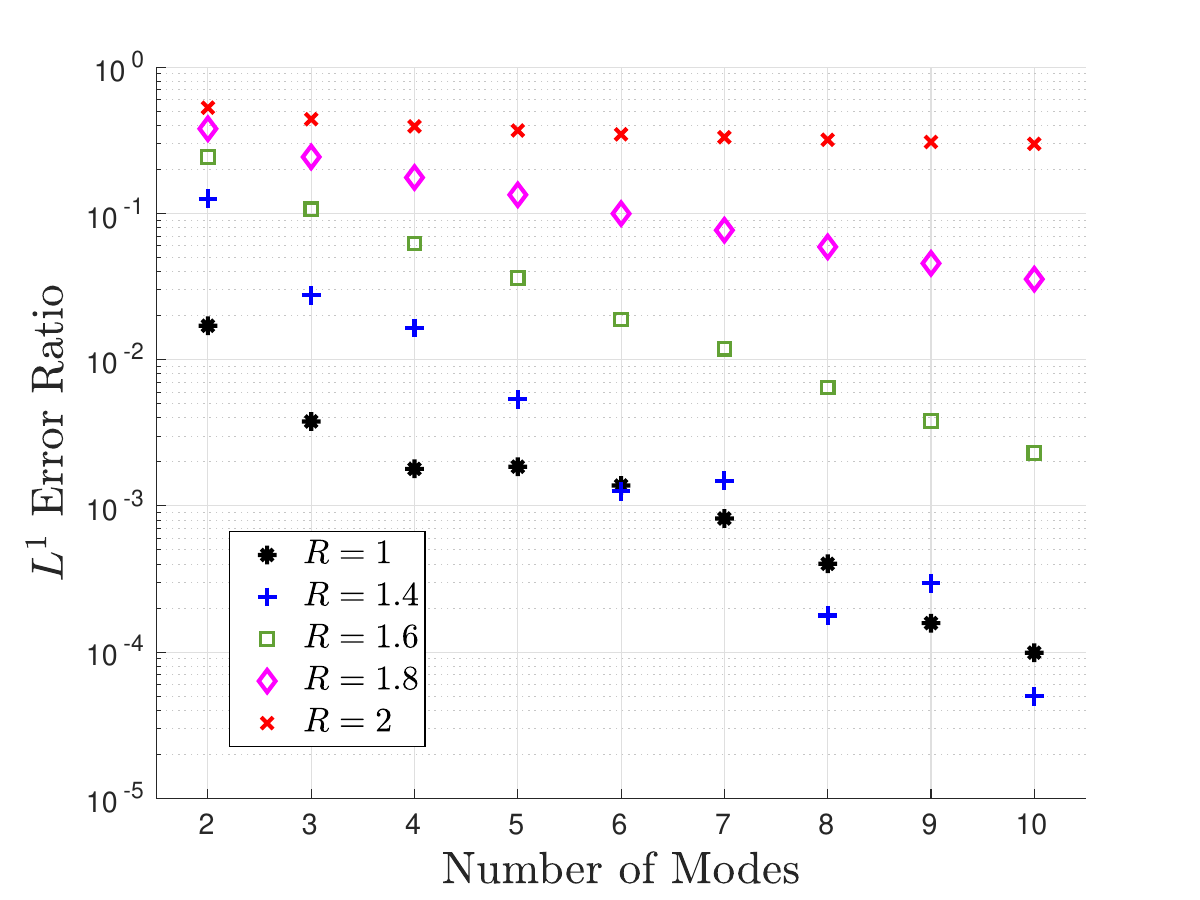}}
\caption{Errors in expansion of \req{zerothApprox} as a function of number of modes, left-hand side for $\Upsilon=0.5$ and right-hand side for $\Upsilon=1.5$. \radapt{Birrell:2014gea}.}\label{fig:freeStreamf0approxUps5}
\end{figure}

\para{Comparison of bases}
Before deriving the dynamical equations for the method outlined in previous part of this appendix,
we illustrate the error inherent in approximating the chemical nonequilibrium distribution, \req{kinetic:equilibrium},  with a  chemical equilibrium distribution, \req{equilibrium}, whose temperature is $T=1/a$.   Given a chemical nonequilibrium distribution 
\begin{equation}\label{zerothApprox}
f_\Upsilon (y)=\frac{1}{\Upsilon^{-1}e^{y/(aT)}+1}\,,
\end{equation}
 we can attempt to write it as a perturbation of the chemical equilibrium distribution,  
\begin{equation}\label{chiDef}
f_\Upsilon=f_{ch}\chi,
\end{equation} as we would need to when using the method \req{eq:ChemEquilibMethod}.  We expand $\chi=\sum_i d^i\hat\chi_i$ in the orthonormal basis generated by $f_{ch}$ and, using $N$ terms, form the $N$-mode approximation $f_\Upsilon^N$ to $f_\Upsilon$.  The $d^i$ are obtained by taking the $L^2(f_{ch}dy)$ inner product of $\chi$ with the basis function $\hat\chi_i$,
\begin{equation}
d^i=\int\hat\chi_i \chi f_{ch}dy=\int\hat\chi_i  f_\Upsilon dy\,.
\end{equation}
 Figure~\ref{fig:freeStreamf0approxUps5} 
 shows for two values of $\Upsilon=0.5,\ \Upsilon=1.5$ the normalized $L^1(dx)$ errors between $f_\Upsilon^N$ and $f_\Upsilon$, computed via
\begin{equation}
\text{error}_N=\frac{\int_0^\infty |f_\Upsilon -f_\Upsilon ^N|dy}{\int_0^\infty |f_\Upsilon |dy}\,.
\end{equation}

We note the appearance of the reheating ratio
\begin{equation}\label{reheat}
 R\equiv aT  
\end{equation}
in the denominator of \req{zerothApprox}, which comes from changing variables from $z=p/T$ in \req{weight} to $y=ap$ in order to compare with \req{freeStreamWeight}.  Physically, $R$ is the ratio of the physical temperature $T$ to the dilution controlled temperature scaling  of $1/a$.   In physical situations, including cosmology, $R$ can vary from unity when dimensioned energy scales influence dynamical equations for $a$. From the error plots we see that for $R$ sufficiently close to $1$, the approximation performs well with a small number of terms, even with $\Upsilon\neq 1$.  

In the case of large reheating, we find that when $R$ approaches and surpasses $2$, large spurious oscillations begin to appear in the expansion and they persist even when a large number of terms are used, as seen in   \rf{fig:freeStreamf0approxUps1Tr185}, 
where we compare $f_\Upsilon/f_{ch}^{1/2}$ with $f_{\Upsilon}^N/f_{ch}^{1/2}$ for $\Upsilon=1$ and $N=20$. See Ref.~\cite{Birrell:2014gea} for further discussion of the origin of these oscillations. This demonstrates that the chemical equilibrium method with dilution temperature scaling will  perform extremely poorly in situations that experience a large degree of reheating. For $R\approx 1$, the benefit of including fugacity is not as striking, as the chemical equilibrium basis is able to approximate \req{zerothApprox} reasonably well.  However, for more stringent error tolerances including $\Upsilon$ can reduce the number of required modes in cases, where the degree of chemical nonequilibrium is large.
 
\begin{figure}
\centerline{\includegraphics[width=0.5\linewidth]{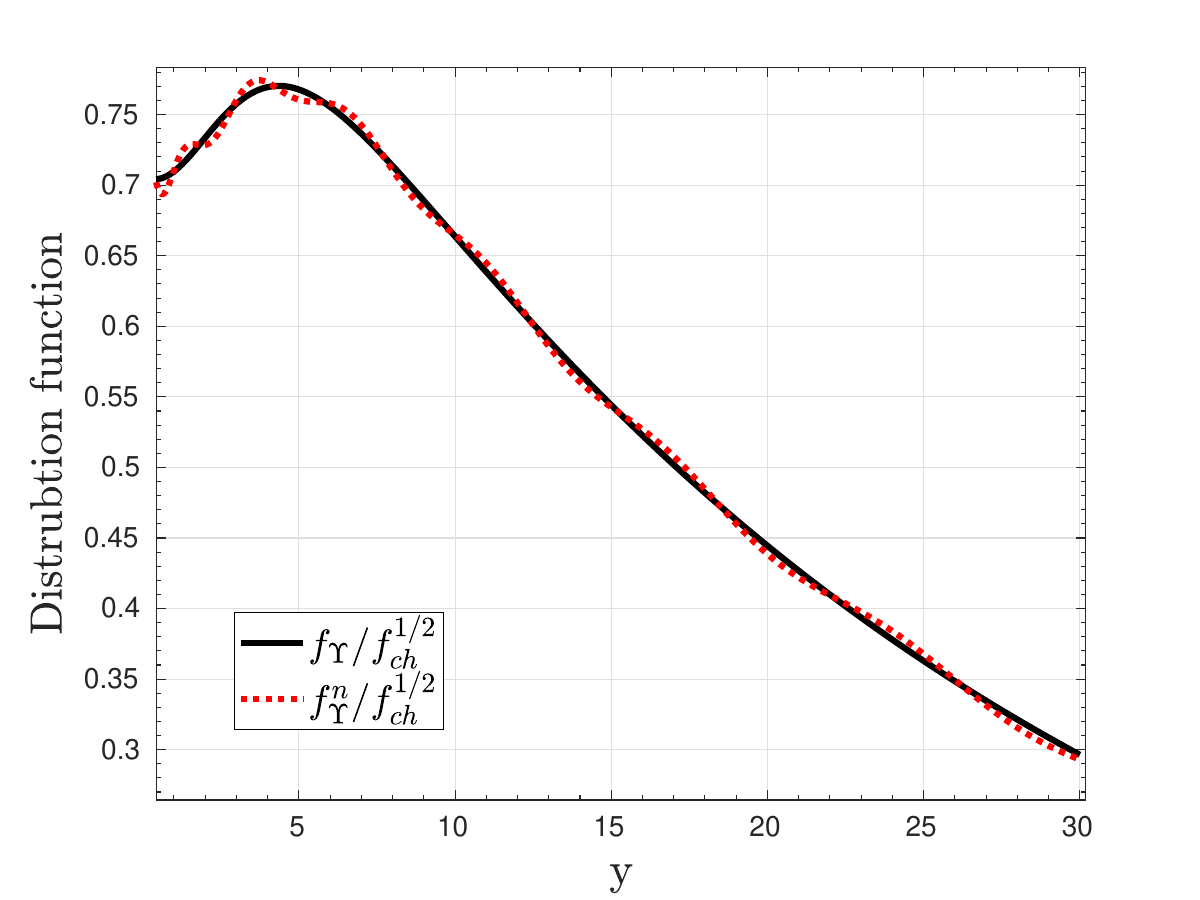}\hspace{-0.6cm}
\includegraphics[width=0.5\linewidth]{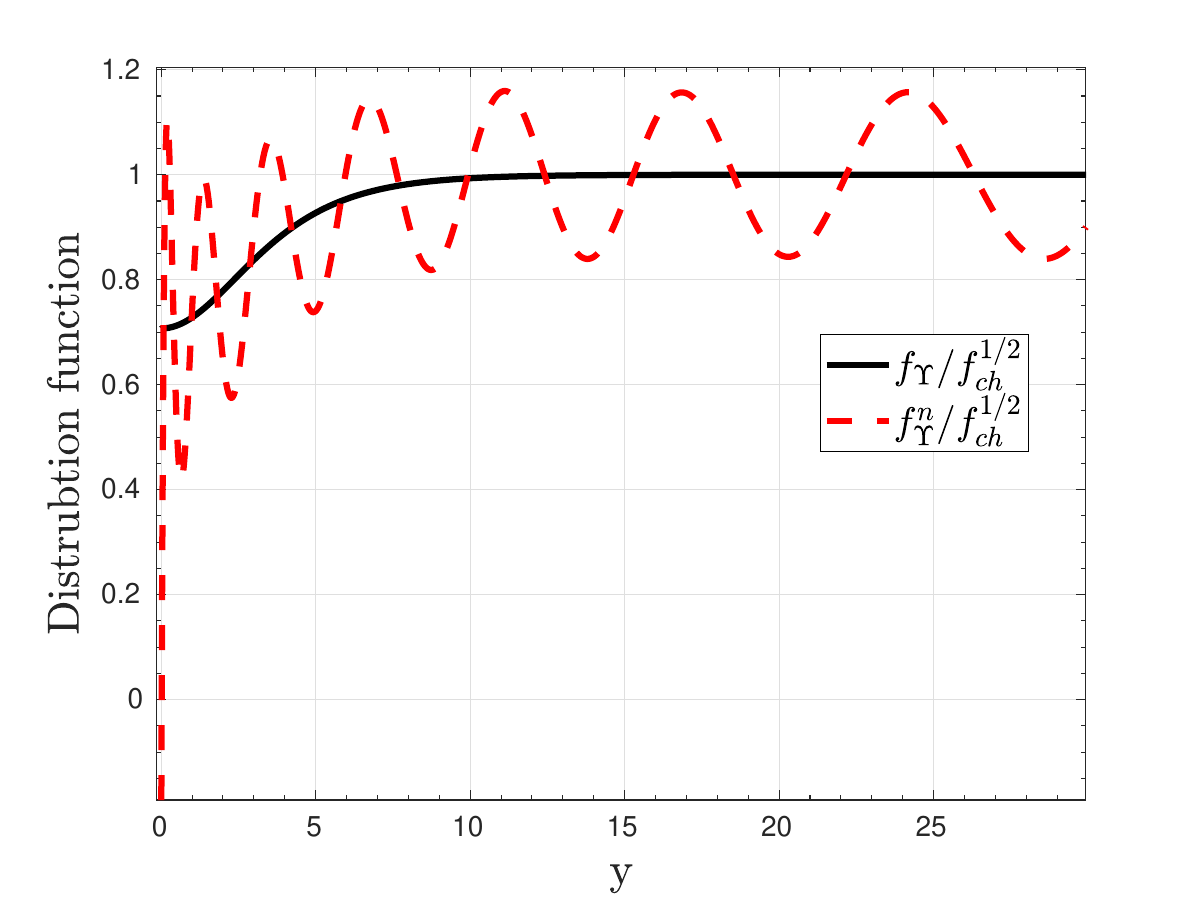}}
\caption{Approximation to \req{zerothApprox} for $\Upsilon=1$ with $R=1.85$ (on left-hand side) and $R=2$ (on right-hand side)  using the first $20$ basis elements generated by \req{freeStreamWeight}. \radapt{Birrell:2014gea}.}\label{fig:freeStreamf0approxUps1Tr185}
\end{figure}

\para{Nonequilibrium dynamics}\label{dynamicsSec}
In this section we derive the dynamical equations for the chemical nonequilibrium method. In particular, we identify physically motivated dynamics for the effective temperature and fugacity. Using \req{TBoltzmann} and the definition of $\psi$ from \req{kineticApprox} we have
\begin{align}\label{nearEquilibEq}
\partial_t \psi+\frac{1}{f_\Upsilon }\frac{\partial f_\Upsilon }{\partial\Upsilon}\dot\Upsilon\psi-\frac{z}{f_\Upsilon }\left(H+\frac{\dot{T}}{T}\right)\left(\psi\partial_zf_\Upsilon +f_\Upsilon \partial_z \psi\right)=\frac{1}{f_\Upsilon E}C[f_\Upsilon \psi]\,.
\end{align}
Denote the monic orthogonal polynomial basis generated by the weight, \req{weight}, by $\psi_n$, $n=0,1,...$, where $\psi_n$ is degree $n$ and call the normalized versions  $\hat{\psi}_n$. Recall that $\hat\psi_n$ depend on $t$ due to the $\Upsilon$ dependence of the weight function used in the construction; therefore the method developed here is a moving-frame spectral method. Consider the space of polynomial of degree less than or equal to $N$, spanned by $\hat\psi_n$, $n=0,...,N$.   For $\psi$ in this subspace, we expand $\psi=\sum_{j=0}^Nb^j\hat\psi_j$ and use \req{nearEquilibEq}  to obtain
\begin{align}\label{Tvars}
\sum_i \dot{b}^i\hat\psi_i=&\sum_ib^i\frac{z}{f_\Upsilon }\left(H+\frac{\dot{T}}{T}\right)\left(\partial_z(f_\Upsilon )\hat\psi_i+f_\Upsilon \partial_z\hat\psi_i\right)\\
&-\sum_ib^i\left(\dot{\hat{\psi}}_i+\frac{1}{f_\Upsilon }\frac{\partial f_\Upsilon }{\partial\Upsilon}\dot\Upsilon\hat\psi_i\right)+\frac{1}{f_\Upsilon E}C[f]\,.
\notag
\end{align}
From this we see  that  projecting the Boltzmann-Einstein equation onto the finite dimensional subspace gives
\begin{align}
\dot b^k=& \sum_i b^i\left(H+\frac{\dot{T}}{T}\right)\left(\langle\frac{z}{f_\Upsilon }\hat\psi_i\partial_zf_\Upsilon ,\hat\psi_k\rangle+\langle z\partial_z \hat\psi_i,\hat\psi_k\rangle\right) \\
&-\sum_i b^i\dot{\Upsilon}\left(\langle\frac{1}{f_\Upsilon }\frac{\partial f_\Upsilon }{\partial\Upsilon}\hat\psi_i,\hat\psi_k\rangle+\langle\frac{\partial\hat{\psi}_i}{\partial \Upsilon},\hat\psi_k\rangle\right)+\langle\frac{1}{f_\Upsilon E}C[f],\hat\psi_k\rangle\notag\,,
\end{align}
where $\langle\cdot,\cdot\rangle$ denotes the inner product defined by the weight function, \req{weight},
\begin{equation}
\langle h_1,h_2\rangle=\int_0^\infty h_1(z)h_2(z)w_\Upsilon(z)dz\,.
\end{equation}
The collision term contains polynomial nonlinearities when multiple coupled distribution are being modeled using a $2$-$2$ collision operator, \req{coll}, while the other terms are linear.  

To isolate the linear part, we define matrices
\begin{align}\label{ABmatrices}
A^k_i(\Upsilon)\equiv&\langle\frac{z}{f_\Upsilon }\hat\psi_i\partial_zf_\Upsilon ,\hat\psi_k\rangle+\langle z\partial_z \hat\psi_i,\hat\psi_k\rangle\,,\\
B^k_i(\Upsilon)\equiv& C_i^k(\Upsilon)+D_i^k(\Upsilon),\hspace{2mm} C_i^k\equiv\Upsilon\langle\frac{1}{f_\Upsilon }\frac{\partial f_\Upsilon }{\partial\Upsilon}\hat\psi_i,\hat\psi_k\rangle,\hspace{2mm} D_i^k\equiv\Upsilon\langle\frac{\partial\hat{\psi}_i}{\partial \Upsilon},\hat\psi_k\rangle\,. \notag
\end{align}
 With these definitions, the equations for the $b^k$ become
\begin{align}\label{bEq}
\dot b^k=& \left(H+\frac{\dot{T}}{T}\right)\sum_i A_i^k(\Upsilon)b^i-\frac{\dot{\Upsilon}}{\Upsilon}\sum_i B_i^k(\Upsilon)b^i+\langle\frac{1}{f_\Upsilon E}C[f],\hat\psi_k\rangle\,.
\end{align}
 See  \rsec{sec:orthopolyApp} for details on how to recursively construct the $\partial_z\hat\psi_i$. We also showed how to compute the inner products $\langle\hat\psi_k,\partial_{\Upsilon}\hat\psi_k\rangle$.
 In  \req{eq:lowerTri1}-\req{eq:lowerTri5} we proved that that both $A$ and $B$ are lower triangular and show that the only inner products involving the $\partial_\Upsilon\hat{\psi}_i$ that are required in order to compute $A$ and $B$ are those the above mentioned diagonal elements, $\langle\hat\psi_k,\partial_{\Upsilon}\hat\psi_k\rangle$.

We fix the dynamics of $T$ and $\Upsilon$ by imposing the conditions
\begin{equation}\label{bIcs}
b^0(t)\hat\psi_0(t)=1\,,\hspace{2mm}b^1(t)=0\,.
\end{equation}
In other words,
\begin{equation}
f(t,z)=f_\Upsilon (t,z)\left(1+\phi(t,z)\right),\hspace{2mm} \phi=\sum_{i=2}^N b^i\hat\psi_i\,.
\end{equation}
This reduces the number of degrees of freedom in \req{bEq} from $N+3$ to $N+1$.  In other words, after enforcing \req{bIcs}, \req{bEq} constitutes $N+1$ equations for the remaining $N+1$ unknowns, $b^2,...,b^N$, $\Upsilon$, and $T$.  We will call $T$ and $\Upsilon$ the first two ``modes", as their dynamics arise from imposing the conditions \req{bIcs} on the zeroth and first order coefficients in the expansion. We will solve for their dynamics explicitly below.

To see the physical motivation for the choices \req{bIcs}, consider the particle number density and energy density.  Using the orthonormality of the $\hat\psi_i$ and \req{bIcs} we have
\begin{align}
n=&\frac{g_pT^3}{2\pi^2}\sum_ib^i\int_0^\infty f_\Upsilon  \hat\psi_i z^2 dz=\frac{g_pT^3}{2\pi^2}\sum_ib^i\langle \hat\psi_i ,1\rangle\\
=&\frac{g_pT^3}{2\pi^2}b^0\langle \hat\psi_0 ,1\rangle=\frac{g_pT^3}{2\pi^2}\langle 1 ,1\rangle\,,\notag\\
\rho=&\frac{g_pT^4}{2\pi^2}\sum_ib^i\int_0^\infty f_\Upsilon  \hat\psi_i z^3 dz=\frac{g_pT^4}{2\pi^2}\sum_ib^i\langle\hat\psi_i, z\rangle\\
=&\frac{g_pT^4}{2\pi^2}\left(b^0\langle\hat\psi_0, z\rangle+b^1\langle\hat\psi_1, z\rangle\right)=
\frac{g_pT^4}{2\pi^2}\langle 1,z\rangle\,.\notag
\end{align}
These, together with the definition of the weight function, \req{weight}, imply
\begin{align}\label{thEqMoments}
n=&\frac{g_pT^3}{2\pi^2}\int_0^\infty f_\Upsilon  z^2dz\,,\\
\label{thEqMoments2}
\rho=&\frac{g_pT^4}{2\pi^2}\int_0^\infty f_\Upsilon  z^3dz\,.
\end{align}
Equations (\ref{thEqMoments}) and (\ref{thEqMoments2}) show  that the first two modes, $T$ and $\Upsilon$, with time evolution fixed by \req{bIcs} cause the chemical nonequilibrium distribution $f_\Upsilon $ to capture the number density and energy density of the system exactly.  This fact is very significant, as it implies that within the chemical nonequilibrium approach as long as the back-reaction from the non-thermal distortions is small (meaning that the evolution of $T(t)$ and $\Upsilon(t)$ is not changed significantly when more modes are included), {\em all the effects relevant to the computation of  particle and energy flow are modeled by the time evolution of $T$ and $\Upsilon$ alone} and no further modes are necessary.  This gives a clear separation between the averaged physical quantities, characterized by $f_\Upsilon $, and the momentum dependent non-thermal distortions as captured by 
\begin{equation}
\phi=\sum_{i=2}^N b^i\hat\psi_i\,.
\end{equation}

One should contrast this chemical nonequilibrium behavior  with the chemical equilibrium method, where a minimum of four modes is required to describe the number and energy densities, as shown in \req{freeStreamMoments}.   Moreover we will show that convergence to the desired precision is faster in the chemical nonequilibrium approach as compared to chemical equilibrium. Due to the high cost of numerically integrating realistic collision integrals of the form \req{coll}, this fact can be very significant in applications. We remark that the relations \req{thEqMoments} are the physical motivation for including the $z^2$ factor in the weight function. All three modifications we have made in constructing our new method, the introduction of an effective temperature, i.e., $R\ne 1$, the generalization to chemical nonequilibrium $f_\Upsilon $, and the introduction of $z^2$ to the weight, \req{reheat}, were needed to obtain the properties, \req{thEqMoments}, but it is the introduction of $z^2$ that reduces the number of required modes and hence reduces the computational cost. 

With $b^0$ and $b^1$ fixed as in \req{bIcs} we can solve the equations for $\dot b^0$ and $\dot b^1$ from \req{bEq} for $\dot\Upsilon$ and $\dot T$ to obtain

\begin{align}\label{UpsTEqs}
\dot{\Upsilon}/{\Upsilon}=&\frac{(Ab)^1\langle\frac{1}{f_\Upsilon E}C[f],\hat\psi_0\rangle-(Ab)^0\langle\frac{1}{f_\Upsilon E}C[f],\hat\psi_1\rangle }{[\Upsilon\partial_\Upsilon \langle1,1\rangle/(2||\psi_0||)+(Bb)^0](Ab)^1-(Ab)^0(Bb)^1}\,,\\[0.5cm]
\dot{T}/T
=&\frac{(Bb)^1\langle\frac{1}{f_\Upsilon E}C[f],\hat\psi_0\rangle-\langle\frac{1}{f_\Upsilon E}C[f],\hat\psi_1\rangle[\Upsilon\partial_\Upsilon \langle1,1\rangle/(2||\psi_0||)+(Bb)^0]}{[\Upsilon\partial_\Upsilon \langle1,1\rangle/(2||\psi_0||)+(Bb)^0](Ab)^1-(Ab)^0(Bb)^1}-H\notag\\[0.3cm]
=&\frac{1}{(Ab)^1}\left((Bb)^1\dot{\Upsilon}/\Upsilon-\langle\frac{1}{f_\Upsilon E}C[f],\hat\psi_1\rangle\right)-H\,.\label{Teq}
\end{align}

Here $(Ab)^n=\sum_{j=0}^NA^n_jb^j$ and similarly for $B$ and $||\cdot||$ is the norm induced by $\langle\cdot,\cdot\rangle$. In deriving this, we used
\begin{equation}
\dot{b}^0=\frac{1}{2||\psi_0||}\dot\Upsilon\partial_\Upsilon \langle1,1\rangle\,, \hspace{4mm} \partial_\Upsilon \langle1,1\rangle=\int_0^\infty \frac{z^2}{(e^{z/2}+ \Upsilon e^{-z/2})^2}dz\,,
\end{equation}
which comes from differentiating \req{bIcs}. 
 
It is easy to check that when the collision operator vanishes, then the above system is solved by 
\begin{equation}\label{freeStreamSol}
\Upsilon=\text{constant}\,,\hspace{4mm} \frac{\dot T}{T}=-H\,,\hspace{2mm}  b^n=\text{constant}\,,\hspace{1mm} n>2\,,
\end{equation}
i.e., the fugacity and non-thermal distortions are `frozen' into the distribution and the temperature satisfies dilution scaling $T\propto 1/a$.

When the collision term becomes small, \req{freeStreamSol} motivates another change of variables. Letting $T=(1+\epsilon)/a$  gives the equation
\begin{equation}
\dot\epsilon=\frac{1+\epsilon}{(Ab)^1}\left((Bb)^1\dot{\Upsilon}/\Upsilon-\langle\frac{1}{f_\Upsilon E}C[f],\hat\psi_1\rangle\right).
\end{equation}
Solving this in place of \req{Teq} when the collision terms are small avoids having to numerically track the free-streaming evolution. In particular this will ensure conservation of comoving particle number, which equals a function of $\Upsilon$ multiplied by $(aT)^3$, to much greater precision in this regime as well as resolve the freeze-out temperatures more accurately.\\

\noindent{\bf Projected Dynamics are Well-defined:}\\
The following calculation shows that, for a distribution initially in kinetic equilibrium, the determinant factor in the denominator of \req{UpsTEqs} is nonzero and hence the dynamics for $T$ and $\Upsilon$, as well as the remainder of the projected system, are well-defined, at least for sufficiently small times. 

Kinetic equilibrium\index{kinetic equilibrium} implies the initial conditions $b^0=||\psi_0||$, $b^i=0$, $i>0$.  Therefore we have
\begin{align}
K\equiv& (\Upsilon\partial_\Upsilon \langle 1,1\rangle/(2||\psi_0||)+(Bb)^0)(Ab)^1-(Ab)^0(Bb)^1\\[0.3cm]
=&(C^0_0A^1_0-A^0_0C^1_0)(b^0)^2+\left[(D^0_0A^1_0-A^0_0D^1_0)(b^0)^2+\Upsilon\partial_\Upsilon \langle 1,1\rangle/(2||\psi_0||)A^1_0b^0\right]\notag\\[0.3cm]
\equiv & K_1+K_2\,.\notag
\end{align}
\begin{align}
K_1=&\langle \frac{1}{1+\Upsilon e^{-z}},1\rangle\langle \frac{-z}{1+\Upsilon e^{-z}}\hat\psi_1,\hat\psi_0\rangle-\langle\frac{-z}{1+\Upsilon e^{-z}},\hat\psi_0\rangle\langle\frac{1}{1+ \Upsilon e^{-z}}\hat\psi_1,1\rangle\,.
\end{align}
Inserting the formula for $\hat\psi_1$ from \req{polyRecursion}, we find
\begin{align}
K_1=&-\frac{1}{||\psi_1||\,||\psi_0||}\left[\langle\frac{1}{1+ \Upsilon e^{-z}},\hat\psi_0\rangle\langle\frac{z^2}{1+\Upsilon e^{-z}},\hat\psi_0\rangle-\langle\frac{z}{1+\Upsilon e^{-z}},\hat\psi_0\rangle^2\right]\,.
\end{align}
The Cauchy-Schwarz inequality  applied to the inner product with weight function
\begin{equation}
\tilde{w}=\frac{w}{1+\Upsilon e^{-z}}\hat\psi_0
\end{equation}
together with linear independence of $1$ and $z$ implies that the term in brackets is positive and so $K_1<0$ at $t=0$.  For the second term, noting that $D^1_0=0$ by orthogonality and using \req{normDerivEq}, we have
\begin{align}
K_2=&[\langle\partial_\Upsilon\hat\psi_0,\hat\psi_0\rangle||\psi_0||+\partial_\Upsilon \langle 1,1\rangle/(2||\psi_0||)]\Upsilon A_0^1||\psi_0||=0\,.
\end{align}
This proves that $K$ is nonzero at $t=0$.\\

\subsection{Validation}\label{validation}
We will validate our numerical method on an exactly solvable model problem
\begin{equation}\label{toyEq}
\partial_t f-pH \partial_p f=M\left(\frac{1}{\Upsilon^{-1}e^{p/T_{eq}}+1}-f(p,t)\right)\,, \hspace{2mm} f(p,0)=\frac{1}{e^{p/T_{eq}(0)}+1}\,,
\end{equation}
where $M$ is a constant with units of energy and we choose units in which it is equal to $1$. This model describes a distribution that is attracted to a given equilibrium distribution at a prescribed time dependent temperature $T_{eq}(t)$ and fugacity\index{fugacity} $\Upsilon$. This type of an idealized scattering operator, without fugacity, was first introduced in \cite{Anderson:1974nyl}. By changing coordinates $y=a(t)p$ we find
\begin{equation}\label{freeStreamToy}
\partial_tf(y,t)=\frac{1}{\Upsilon^{-1}\exp[y/(a(t)T_{eq}(t))]+1}-f(y,t)\,,
\end{equation}
 which has as solution
\begin{equation}\label{exactSol}
f(y,t)=\int_0^t\frac{e^{s-t}}{\Upsilon^{-1}\exp[y/(a(s)T_{eq}(s))]+1}ds+\frac{e^{-t}}{\exp[y/(a(0)T_{eq}(0))]+1}\,.
\end{equation}
We now transform to $z=p/T(t)$, where the temperature $T$ of the distribution $f$ is defined as in \rsec{dynamicsSec}. Therefore, we have the exact solution to
\begin{equation}\label{kEqToy}
\partial_tf-z\left(H+\frac{\dot{T}}{T}\right)\partial_zf=\frac{1}{\Upsilon^{-1}e^{zT/T_{eq}}+1}-f(z,t)\,,
\end{equation}
given by
\begin{align}
f(z,t)=&\int_0^t\frac{e^{s-t}}{\Upsilon^{-1}\exp[a(t)T(t)z/(a(s)T_{eq}(s))]+1}ds\\
&+\frac{e^{-t}}{\exp[a(t)T(t)z/(a(0)T_{eq}(0))]+1}\,.\notag
\end{align}
We use this to test the chemical equilibrium and chemical nonequilibrium methods under two different conditions. 

\para{Reheating test}
First we compare the chemical equilibrium\index{chemical equilibrium} and nonequilibrium methods in a scenario that exhibits reheating.  Motivated by applications to cosmology, we choose a scale factor evolving as in the radiation dominated era, a fugacity $\Upsilon=1$, and choose an equilibrium temperature that exhibits reheating like behavior with $aT_{eq}$ increasing for a period of time,
\begin{align}\label{aTDef}
a(t)=\left(\frac{t+b}{b}\right)^{1/2}\,,\ \  \ \
T_{eq}(t)=\frac{1}{a(t)}\left(1+\frac{1-e^{-t}}{e^{-(t-b)}+1}(R-1)\right)\,,
\end{align}
where $R$ is the desired reheating ratio. Note that $(aT_{eq})(0)=1$ and $(aT_{eq})(t)\rightarrow R$ as $t\rightarrow\infty$. Qualitatively, this is reminiscent of the dynamics of neutrino freeze-out, but the range of reheating ratio for which we will test our method is larger than found there.

We solved \req{freeStreamToy} and \req{kEqToy} numerically using the chemical equilibrium and chemical nonequilibrium methods respectively for $t\in[0,10]$ and $b=5$ and the cases $R=1.1$, $R=1.4$, as well as the more extreme ratio of $R=2$.  The bases of orthogonal polynomials were generated numerically using the recursion relations from \rsec{sec:orthopolyApp}.  For the applications we are considering, where the solution is a small perturbation of equilibrium, only a small number of terms are required and so the numerical challenges associated with generating a large number of such orthogonal polynomials are not an issue.\\

\noindent{\bf Chemical equilibrium method}\\
We solved \req{freeStreamToy} using the chemical equilibrium method, with the orthonormal basis defined by the weight function \req{freeStreamWeight} for $N=2,...,10$ modes (mode numbers $n=0,...,N-1$) and prescribed  single step relative and absolute error tolerances of $10^{-13}$ for the numerical integration, and with asymptotic reheating ratios of $R=1.1$, $R=1.4$, and $R=2$ and will here compare various aspects of the numerical solution with the  exact solution, \req{exactSol}.

In  \rf{fig:freeStreamNumErr} 
we show the maximum relative error in the number densities and energy densities respectively over the time interval $[0,10]$ for various numbers of computed modes.  The particle number density and energy density are accurate, up to the integration tolerance level, for $3$ or more and $4$ or more modes respectively. This is consistent with \req{freeStreamMoments}, which shows the number of modes required to capture each of these quantities. However, fewer modes than these minimum values lead to a large error in the corresponding moment of the distribution function.

\begin{figure}
\centerline{\includegraphics[width=0.5\linewidth]{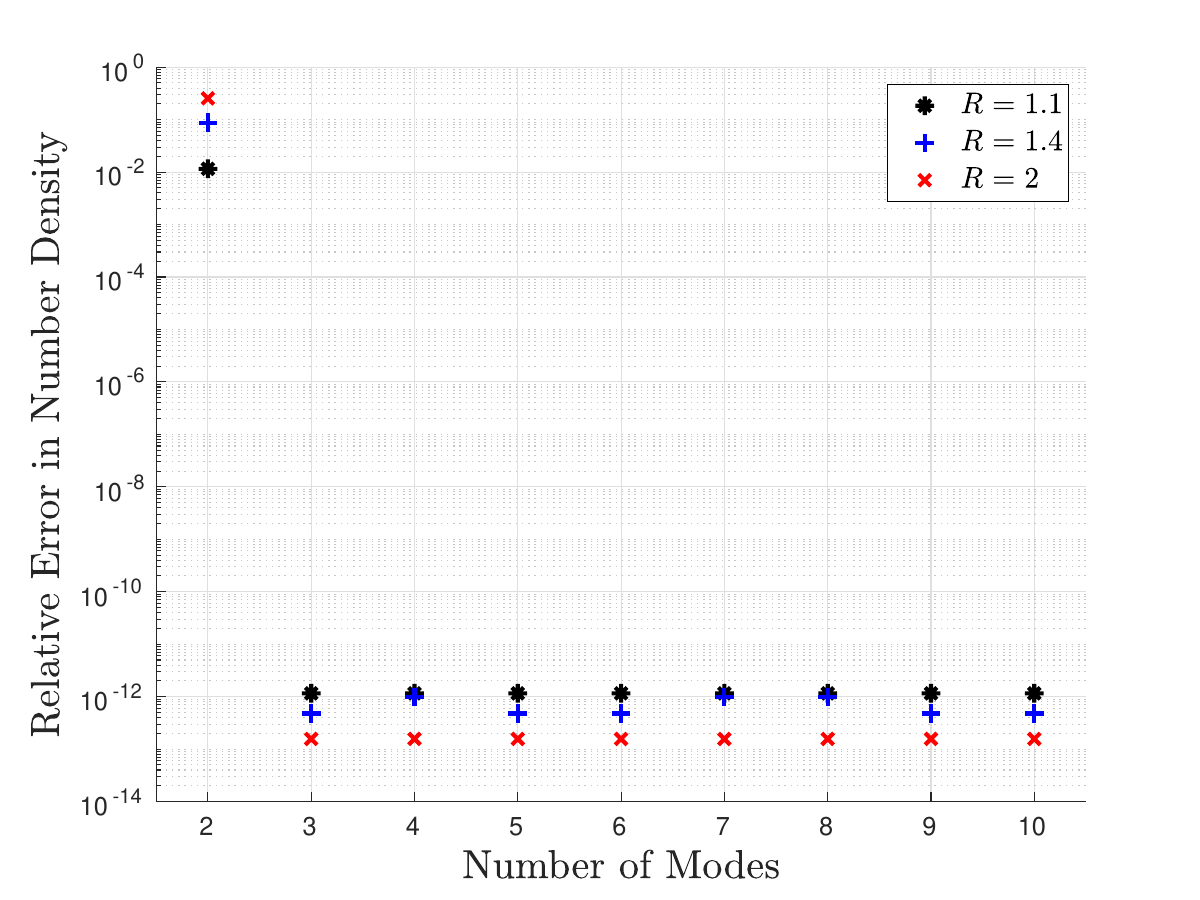}\hspace*{-0.5cm}
\includegraphics[width=0.5\linewidth]{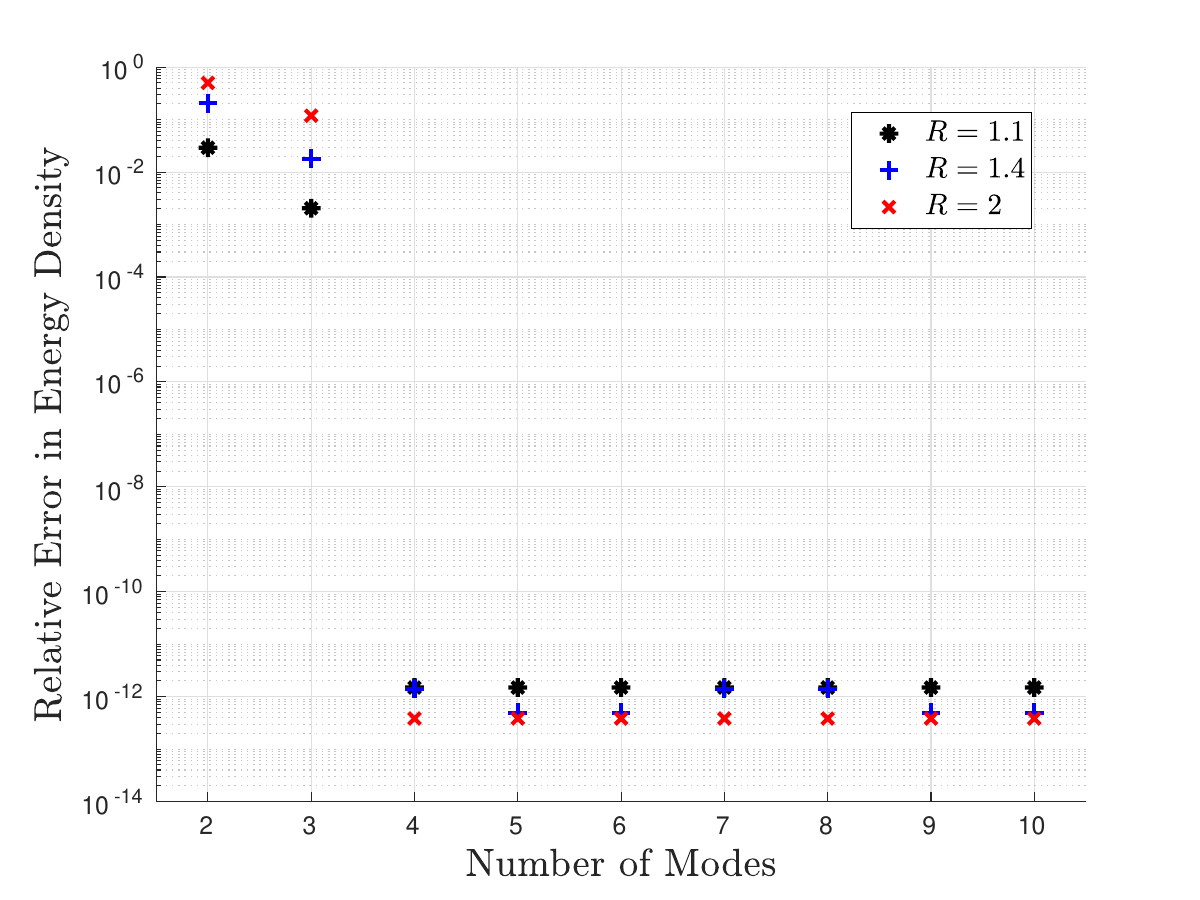}}
\caption{Maximum relative error, on left-hand side in particle number density and on right-hand side in energy density for various numbers of computed modes. \radapt{Birrell:2014gea}.}\label{fig:freeStreamNumErr}
\end{figure}

To show that the numerical integration accurately captures the mode coefficients, in \rf{fig:freeStreambErr} on the left we show the difference between the computed coefficients and actual coefficients, denoted by $\tilde b_n$ and $b_n$ respectively, as measured by the $L^\infty$-error 
\begin{equation}\label{modeErrDef}
\text{error}_{L^\infty}(n)=\max_{t} |\tilde{b}_n(t)-b_n(t)|\,.
\end{equation}
The error is shown as a function of mode number $n$; here the evolution of the system was computed using a total of $N=10$ modes.

\begin{figure}
\centerline{\includegraphics[width=0.5\linewidth]{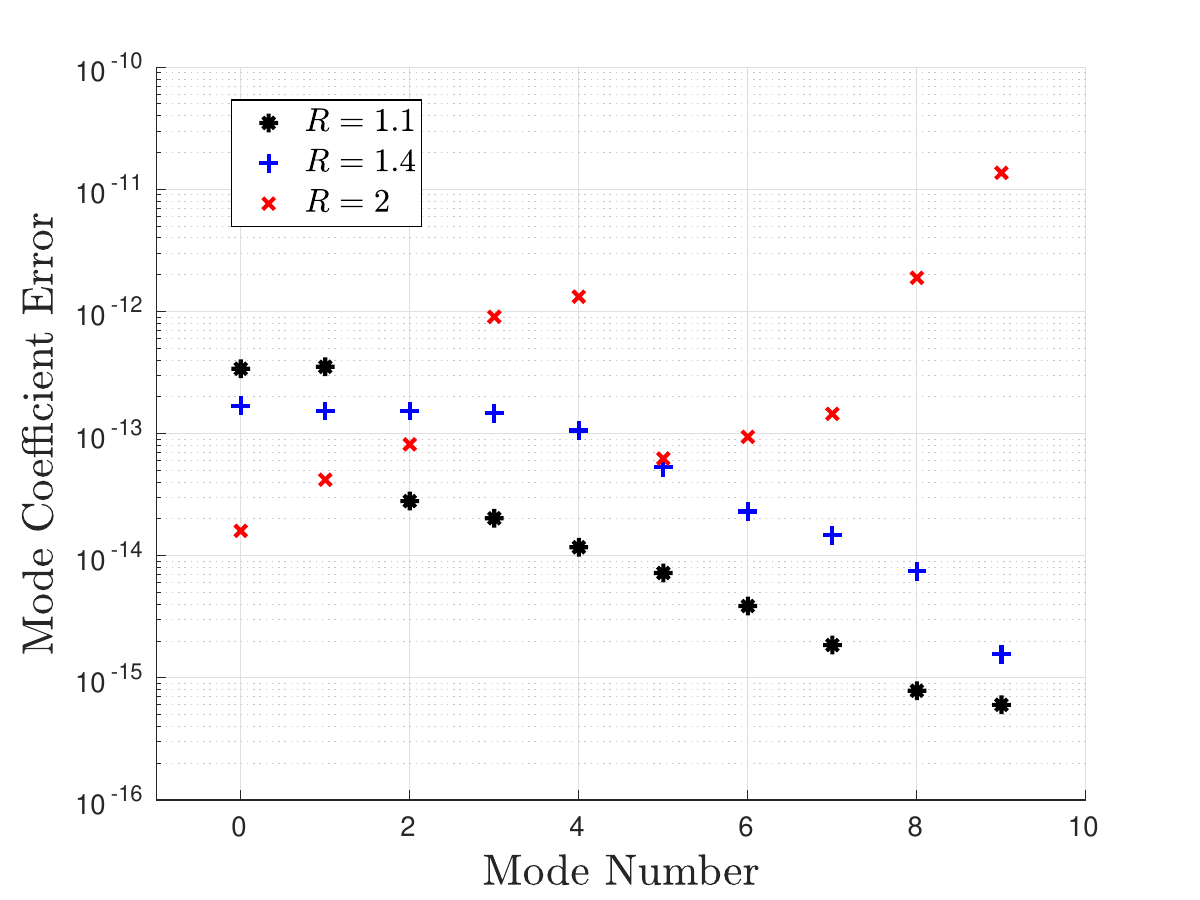}\hspace*{-0.5cm}
\includegraphics[width=0.5\linewidth]{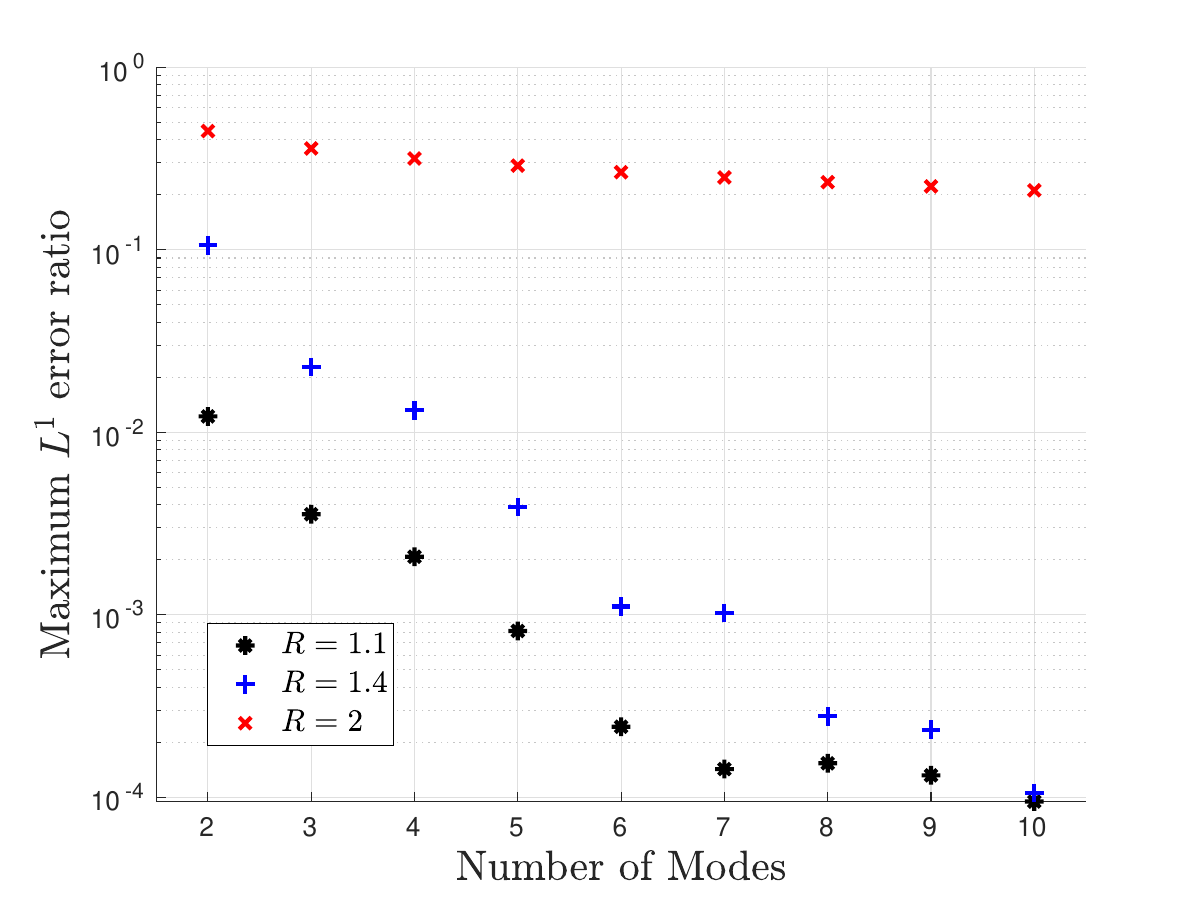}}
\caption{On left: Mode coefficient $L^\infty$-error. On right: Maximum ratio  of $L^1$ error between computed and exact solutions to $L^1$ norm of the exact solution. \radapt{Birrell:2014gea}.}\label{fig:freeStreambErr}
\end{figure}

In  \rf{fig:freeStreambErr} on  the right we show the difference between the exact solution $f$, and the numerical solution $f^N$ computed using a total number of $N=2,...,10$ modes, as measured by the maximum relative $L^1$ error  
\begin{equation}\label{fErr}
\text{error}_{L^1}(N)=\max_{t} \frac{\int |f-f^N|dy}{\int |f|dy}\,.
\end{equation}

For $R=1$ and $R=1.4$ we see that the chemical equilibrium method works reasonably well (as long as the number of modes is at least 4, so that the energy and number densities are properly captured) but for $R=2$ the  $L^1$ error  is large (see  \rf{fig:freeStreambErr}, right) and the approximate solution exhibits spurious oscillations as  seen in \rf{fig:freeStreamApproxTr2}  (left). This behavior is expected based on the results we did present earlier in this Appendix.
Further clarifying the behavior, in  \rf{fig:freeStreamApproxTr2} on the right-hand side
we show the $L^1$ error ratio as a function of time for $N=10$ modes. In the $R=2$ case we see that the error increases as the reheating ratio approaches its asymptotic value of $R=2$ as $t\rightarrow\infty$.  As we will see, our methods achieves a much higher accuracy for a small number of terms in the case of large reheating ratio due to the replacement of dilution temperature scaling with the dynamical effective temperature $T$.\\ 

\begin{figure}
\centerline{\includegraphics[width=0.5\linewidth]{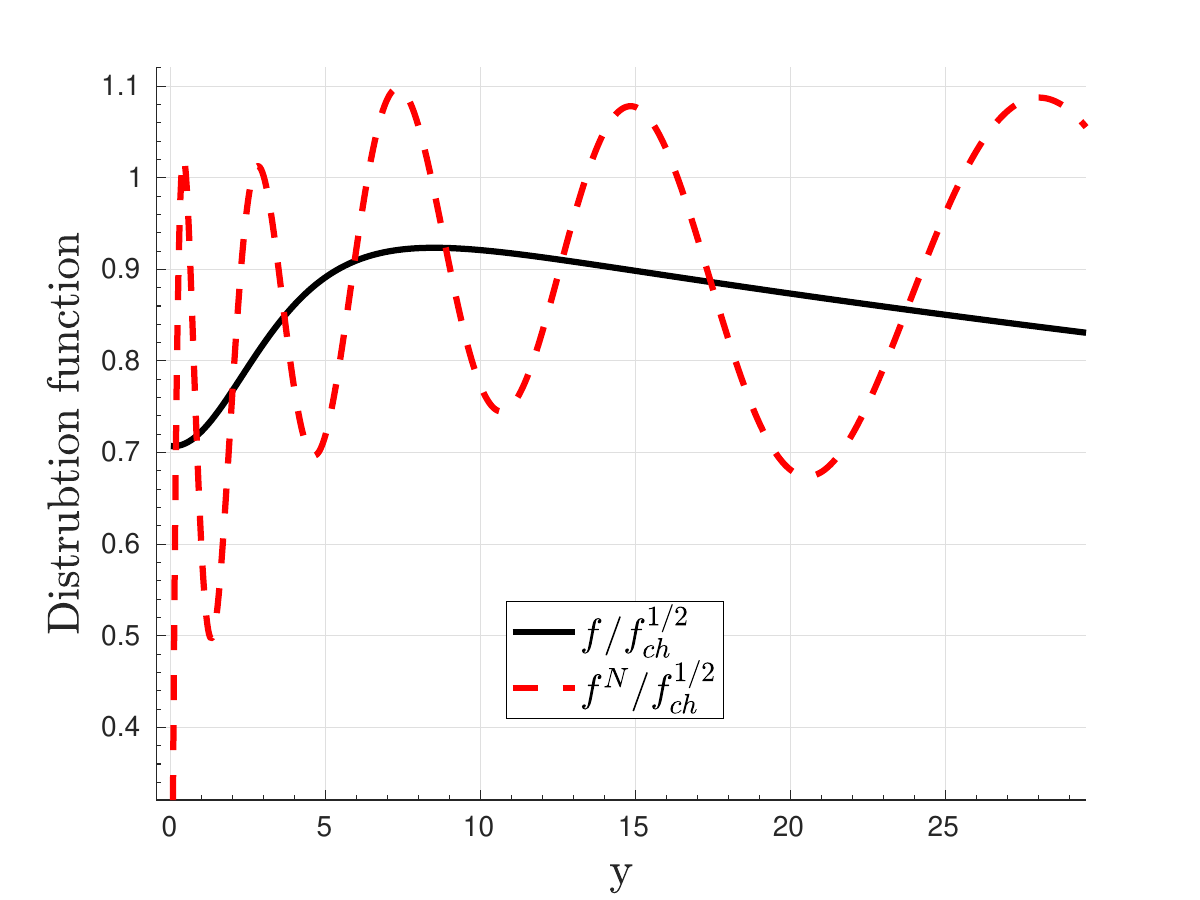}\hspace{-0.5cm}
\includegraphics[width=0.5\linewidth]{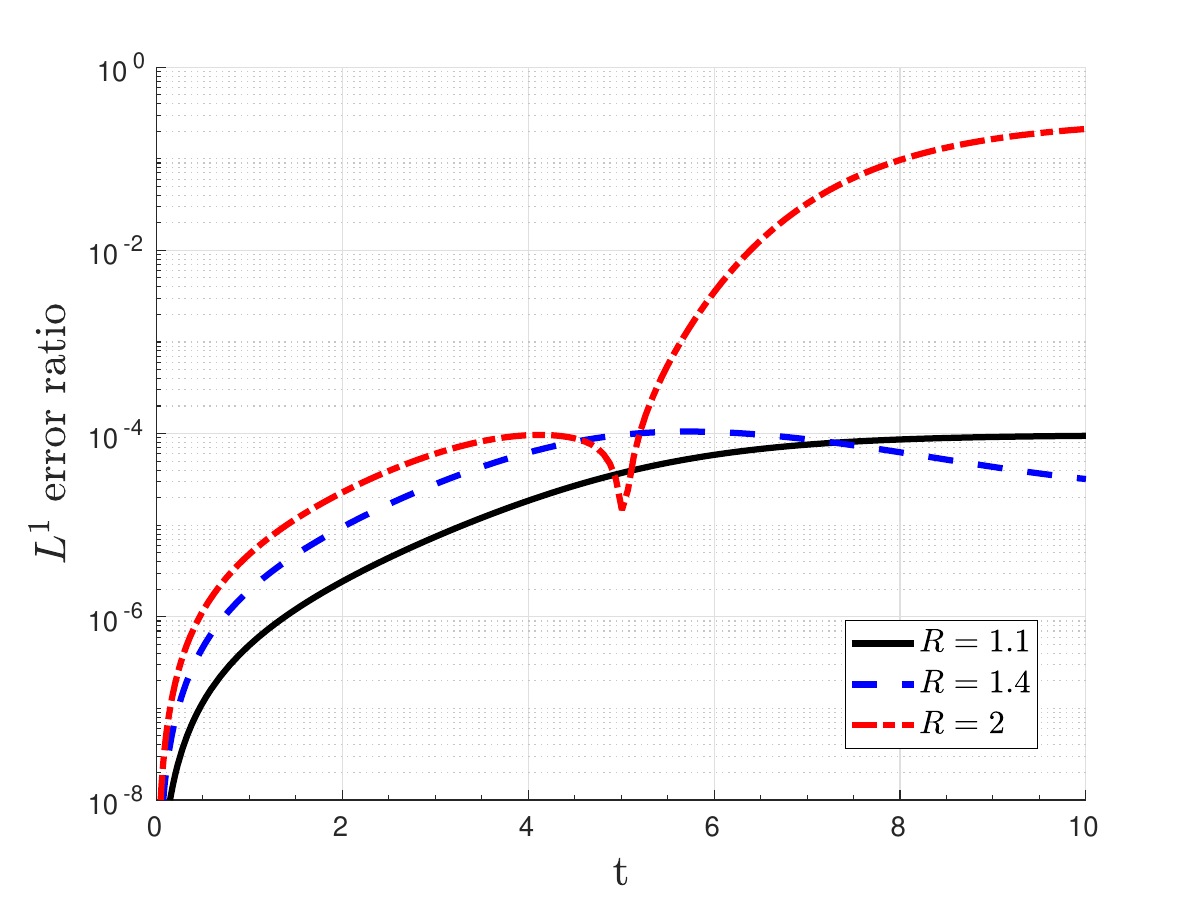}}
\caption{On left: Approximate and exact solution for a reheating ratio $R=2$ and $N=10$ modes. On right: $L^1$ error ratio as a function of time for $N=10$ modes. \radapt{Birrell:2014gea}.}\label{fig:freeStreamApproxTr2}
 \end{figure}

\noindent{\bf Chemical non-equilibrium method}\\
We now solve  \req{freeStreamToy} using the chemical nonequilibrium method, with the orthonormal basis defined by the weight function, \req{weight}, for $N=2,...,10$ modes, a prescribed numerical integration tolerance of $10^{-13}$, and asymptotic reheating ratios of $R=1.1$, $R=1.4$, and $R=2$ and will again compare with the exact solution, \req{exactSol}.  Recall that we are referring to $T$ and $\Upsilon$ as the first two modes ($n=0$ and $n=1$).

In  \rf{fig:keqNumErr} 
we show the maximum relative error over the time interval $[0,10]$ in the number densities and energy densities respectively for various numbers of computed modes. Even for only $2$ modes, the number and energy densities are accurate up to the integration tolerance level. This is in agreement with the analytical expressions in \req{thEqMoments}.

\begin{figure} 
\centerline{\includegraphics[width=0.5\linewidth]{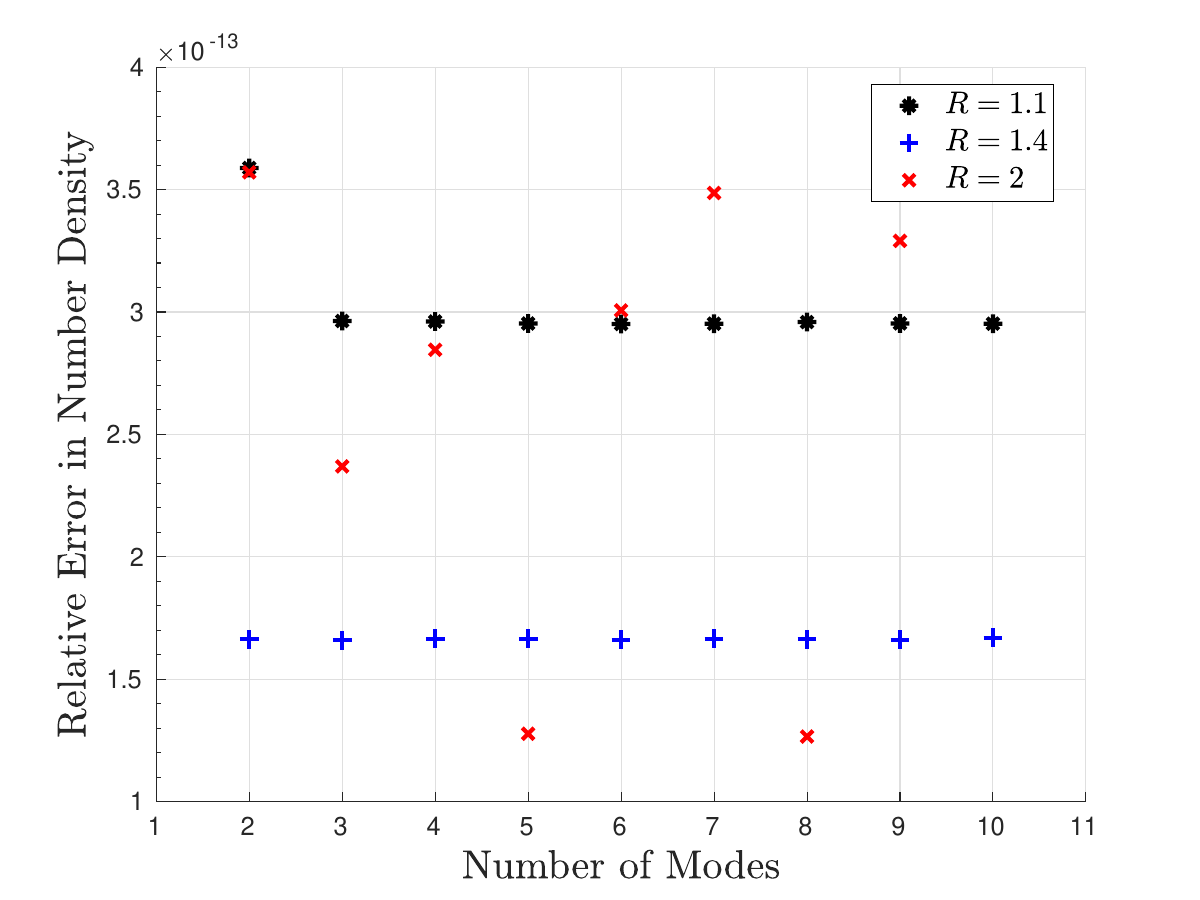}\hspace*{-0.5cm}
\includegraphics[width=0.5\linewidth]{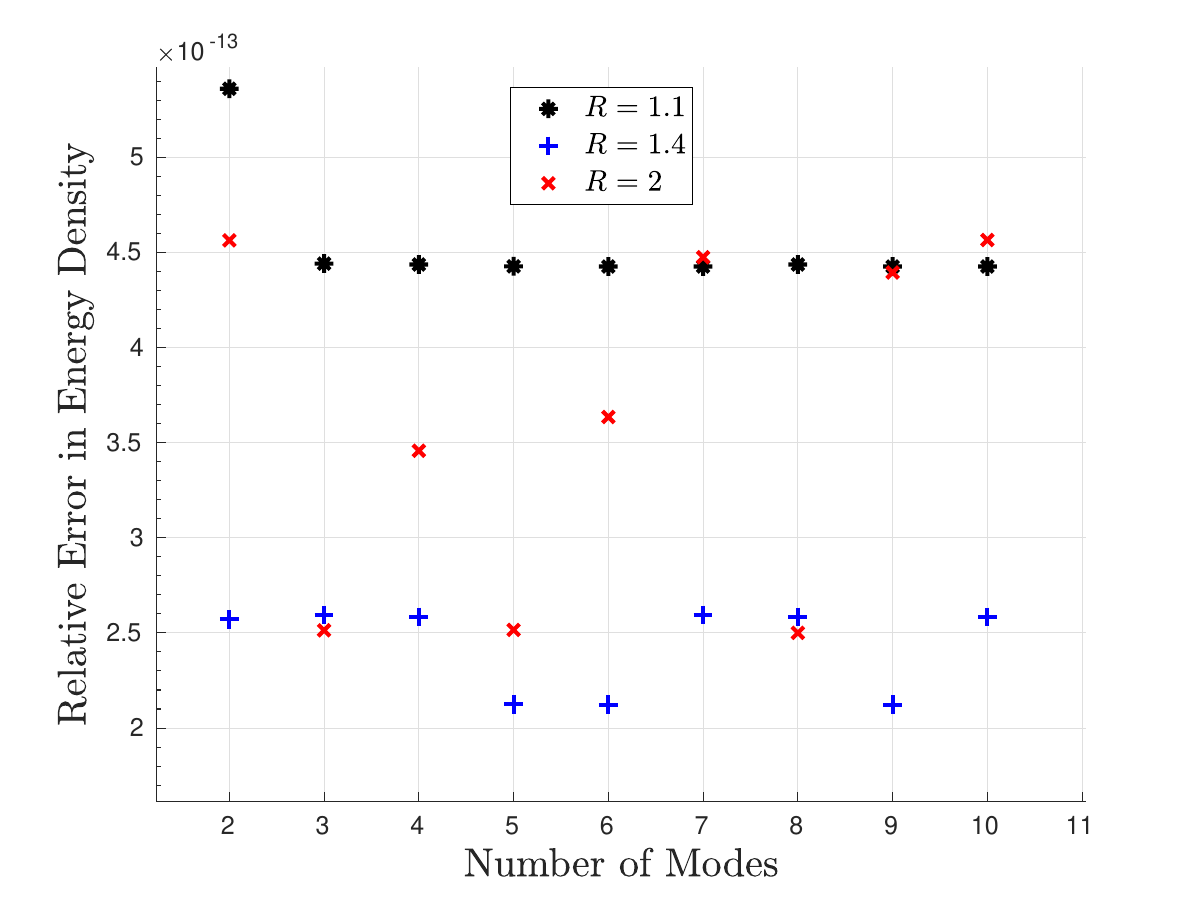}}
\caption{Maximum relative error, on left in particle number density and on right in energy density. \radapt{Birrell:2014gea}.}\label{fig:keqNumErr}
\end{figure}

To show that the numerical integration accurately captures the mode coefficients, in \rf{fig:keqL1Err} on left-hand side 
we show the error \req{modeErrDef}, where the evolution of the system was computed using a total of $N=10$ modes. In \rf{fig:keqL1Err} on right-hand side we show the error between the approximate and exact solutions, computed as in \req{fErr} for $N=2,...,10$ and $R=1.1$, $R=1.4$, and $R=2$ respectively.  For most mode numbers and $R$ values, the error using $2$ modes is substantially less than the error from the chemical equilibrium\index{chemical equilibrium} method using $4$ modes.  The result is most dramatic for the case of large reheating, $R=2$, where the spurious oscillations from the chemical equilibrium solution are absent in our method, as seen in  \rf{fig:keqApproxTr2} on the left-hand side, as compared to the chemical equilibrium method in \rf{fig:freeStreamApproxTr2} on left-hand side.  Note that due to the relation $z=y/R$ as discussed in before in this Appendix we plot from $z\in [0,15]$ in comparison to $y\in[0,30]$ in  \rf{fig:keqApproxTr2}.

\begin{figure}
\centerline{ \includegraphics[width=0.5\linewidth]{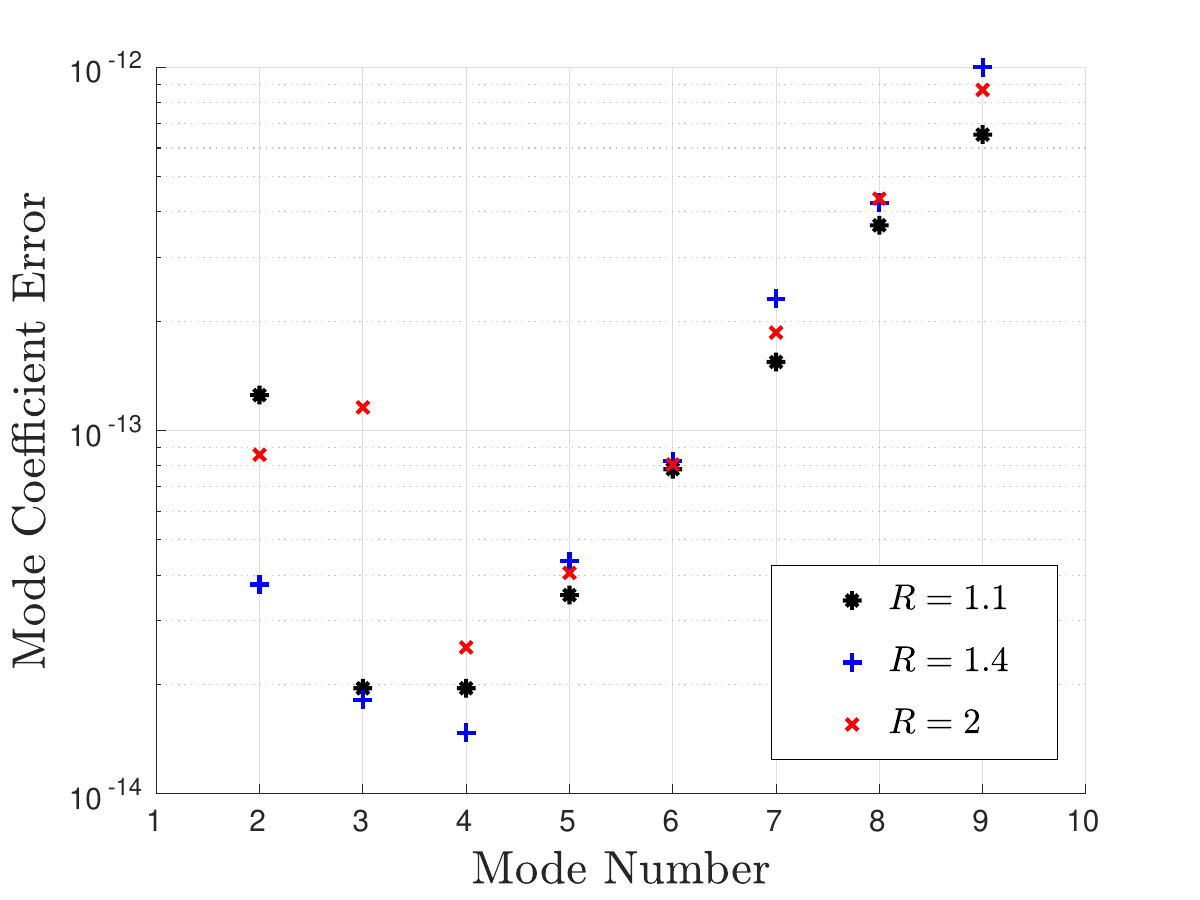}\hspace*{-0.5cm}
\includegraphics[width=0.5\linewidth]{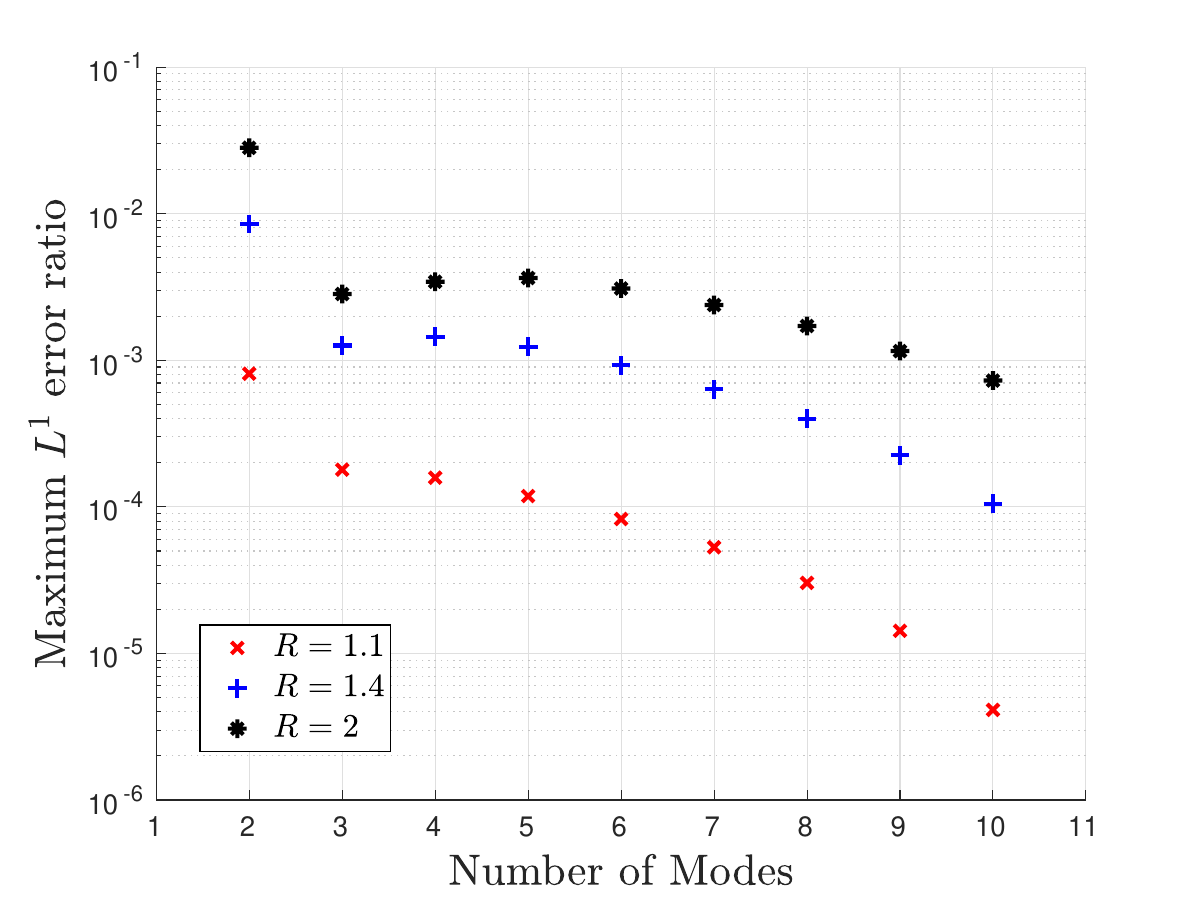} }
\caption{On left: Maximum error in mode coefficients. On right: Maximum ratio of $L^1$ error between computed and exact solutions to $L^1$ norm of the exact solution.  \radapt{Birrell:2014gea}.}\label{fig:keqL1Err}
\end{figure}

Additionally, the error no longer increases as $t\rightarrow\infty$, as it did for the chemical equilibrium method, see \rf{fig:keqApproxTr2} right-hand side.
In fact it decreases since the exact solution approaches chemical equilibrium at a reheated temperature and hence can be better approximated by $f_\Upsilon$.

\begin{figure}
\centerline{\includegraphics[width=0.5\linewidth]{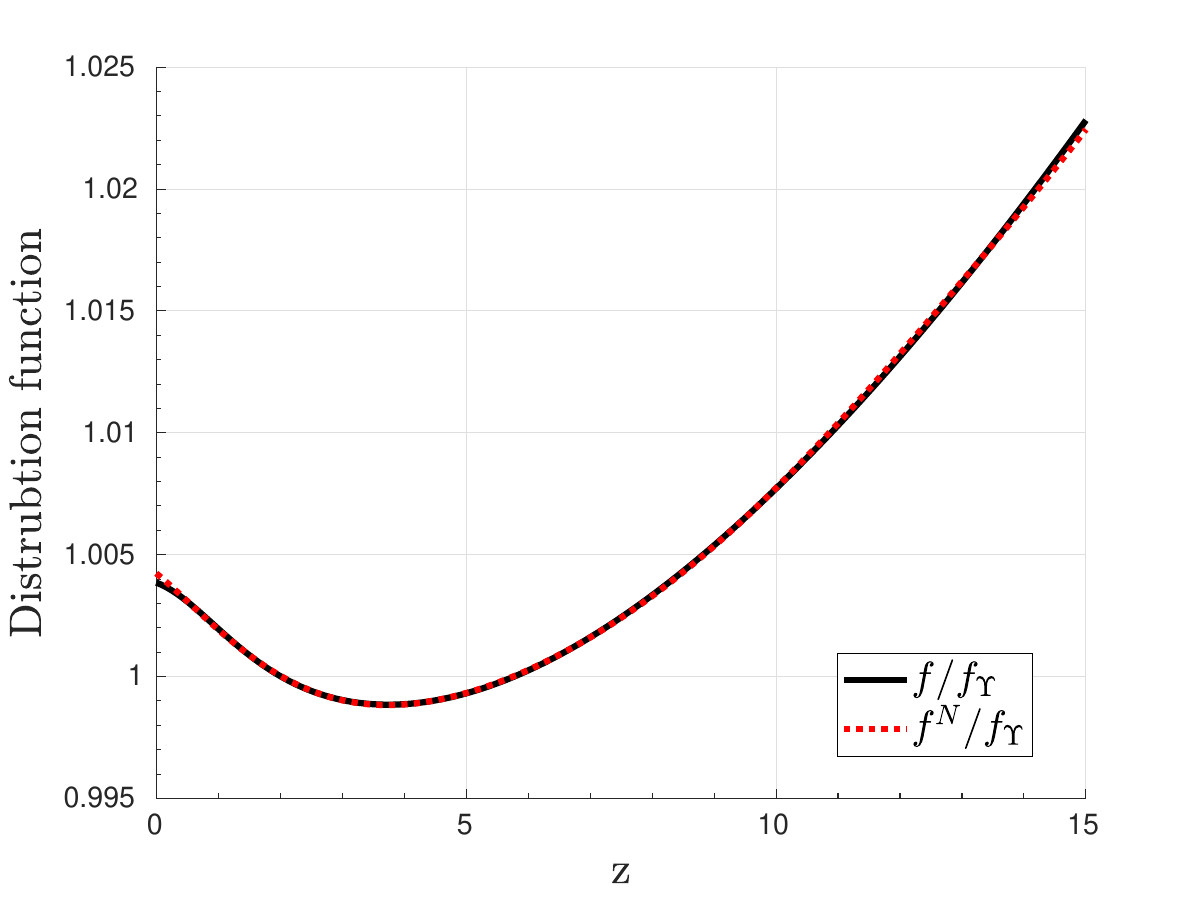}\hspace*{-0.5cm}
\includegraphics[width=0.5\linewidth]{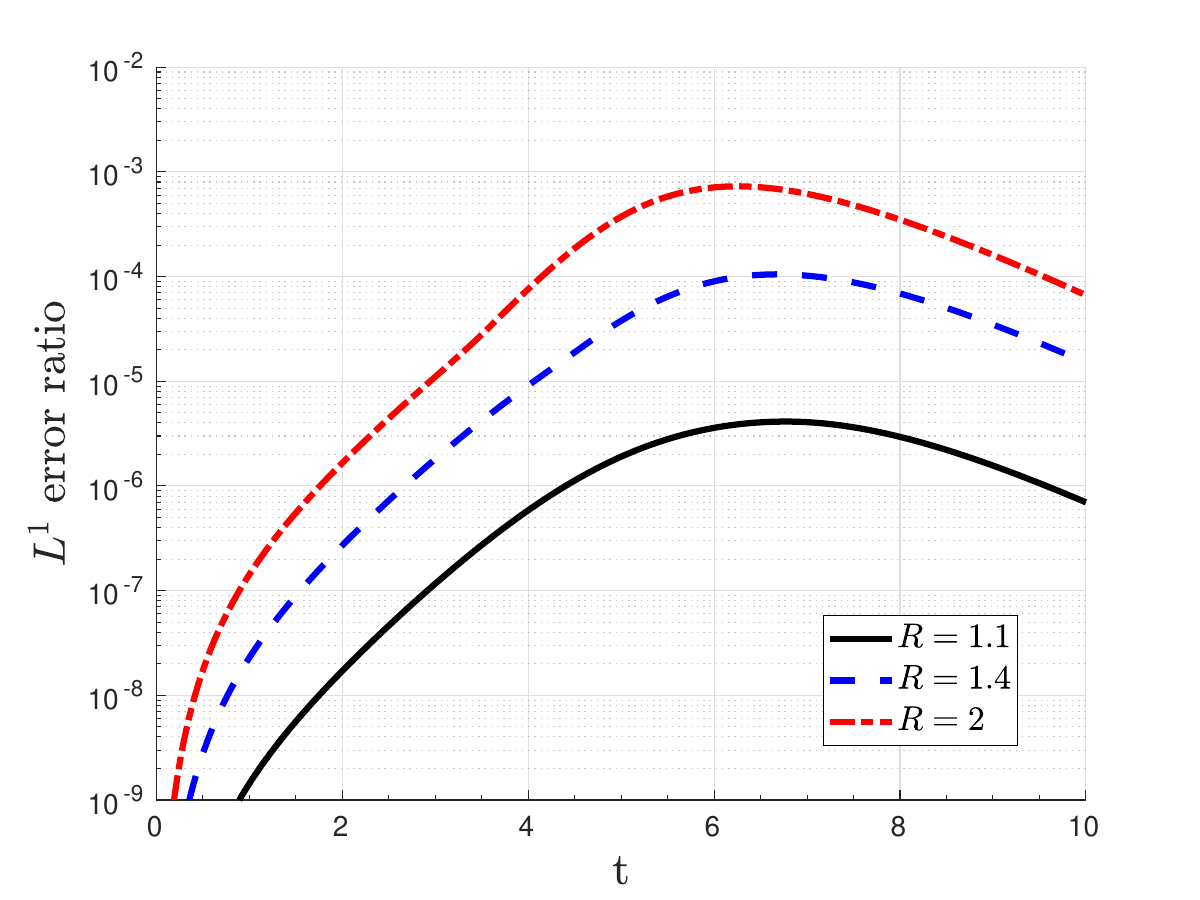}}
\caption{On left: Approximate and exact solution for $R=2$ obtained with two modes. On right: $L^1$ error ratio as a function of time for $N=10$ modes.  \radapt{Birrell:2014gea}.}\label{fig:keqApproxTr2}
\end{figure}

In summary, in addition to the reduction in the computational cost when going from $4$ to $2$ modes, our method also reduces the error as compared to the chemical equilibrium method, all while accurately capturing the number and energy densities. 

\section{Neutrino Collision Integrals}\label{ch:coll:simp}
\subsection{Collision integral inner products}
Having detailed our method for solving the Boltzmann-Einstein equation in \rsec{ch:boltz:orthopoly},  we address the computation of collision integrals\index{collision operator} for neutrino processes; see also \cite{Birrell:2014uka}. To solve for the mode coefficients using \req{bEq}, we must evaluate the collision operator inner products
\begin{align}\label{collision:integrals}
R_k\equiv&\langle\frac{1}{f_\Upsilon E_1}C[f_1],\hat\psi_k\rangle=\int_0^\infty \hat\psi_k(z_1)C[f_1](z_1) \frac{z_1^2}{E_1}dz_1\\
=&\frac{1}{2}\int \hat\psi_k(z_1)\int\left[f_3(p_3)f_4(p^4)f^1(p_1)f^2(p_2)-f_1(p_1)f_2(p_2)f^3(p_3)f^4(p^4)\right]\notag\\
&\hspace{30mm}\times S |\mathcal{M}|^2(s,t)(2\pi)^4\delta(\Delta p)\prod_{i=2}^4\frac{d^{3}p_i}{2(2\pi)^3E_i}\frac{z_1^2}{E_1}dz_1\,,\notag\\
=&\frac{2(2\pi)^3}{8\pi}T_1^{-3}\int G_k(p_1,p_2,p_3,p_4)S |\mathcal{M}|^2(s,t)(2\pi)^4\delta(\Delta p)\prod_{i=1}^4\frac{d^{3}p_i}{2(2\pi)^3E_i}\,,\notag\\
=&2\pi^2T_1^{-3}\int G_k(p_1,p_2,p_3,p_4)S |\mathcal{M}|^2(s,t)(2\pi)^4\delta(\Delta p)\prod_{i=1}^4 \delta_0(p_i^2-m_i^2)\frac{d^4p_i}{(2\pi)^3}\,,\notag\\
G_k=&\hat\psi_k(z_1)\left[f_3(p_3)f_4(p_4)f^1(p_1)f^2(p_2)-f_1(p_1)f_2(p_2)f^3(p_3)f^4(p_4)\right],\hspace{2mm} f^i=1- f_i\,.\notag
\end{align}
Note that $R_k$ only uses information about the distributions at a single spacetime point, and so we can work in a local orthonormal basis for the momentum.  Among other things, this implies that $p^2=p^\alpha p^\beta\eta_{\alpha\beta}$ where $\eta$ is the Minkowski metric
\begin{equation}
\eta_{\alpha\beta}=\diag(1,-1,-1,-1)\,.
\end{equation}

From \req{collision:integrals}, we see that a crucial input into the chemical nonequilibrium spectral method with $2 \leftrightarrow 2$ reactions is the ability to efficiently compute a numerical approximation to integrals of the form
\begin{align}\label{coll:Ip}
&M\equiv\int G(p_1,p_2,p_3,p_4) S |\mathcal{M}|^2(s,t) (2\pi)^4\delta(\Delta p)\prod_{i=1}^4 \delta_0(p_i^2-m_i^2)\frac{d^4p_i}{(2\pi)^3}\,,\\
&G(p_1,p_2,p_3,p_4)=g_1(p_1)g_2(p_2)g_3(p_3)g_4(p_4)\,,\notag
\end{align}
for some functions $g_i$. Even after eliminating the delta functions in \req{coll:Ip}, we are still left with an $8$ dimensional integral.  To facilitate numerical computation, we must analytically reduce this expression down to fewer dimensions.  Fortunately, the systems we are interested in have a large amount of symmetry that  can be utilized for this purpose.  

The distribution functions we are concerned with are isotropic in some frame defined  by a unit timelike vector $U$, i.e. they depend only on the four-momentum only through $p_i\cdot U$.  The same is true of the basis functions $\hat\psi_k$, and hence the  $g_i$ depend only on $p_i\cdot U$ as well.  In \cite{Hannestad:1995rs,Dolgov:1997mb,Dolgov:1998sf}, approaches are outlined that reduce integrals of this type down to $3$ dimensions.  We outline the method from \cite{Dolgov:1997mb,Dolgov:1998sf}, as applied to our spectral method solver, in \rapp{app:dogovMethod}. However, the integrand one obtains from these methods is only piecewise smooth or has an integration domain with a complicated geometry.  This presents difficulties for the integration routine we employ, which utilizes adaptive mesh refinement to ensure the desired error tolerance.  We take an alternative approach that, for the scattering kernels found in $e^\pm$, neutrino interactions, reduces the problem nested integrals  of depth three while also resulting in an integrand with better smoothness properties.  In our comparison with the method in \cite{Dolgov:1997mb,Dolgov:1998sf}, the resulting formula evaluates significantly faster under the numerical integration scheme we used.   The derivation presented here expands on what is found in \cite{Letessier:2002ony}.

\para{Simplifying the collision integral}
Our strategy for simplifying the collision integrals is as follows.  We first make a change of variables designed to put the 4-momentum conserving delta function in a particularly simple form, allowing for the integral to be reduced from $16$ to $12$ dimensions.  The remaining four delta functions, which impose the mass shell constraints, are then seen to reduce to integration over a product of spheres.  The simple form of the submanifold that these delta function restrict us to allows us to use the method in chapter \ref{ch:vol:forms} to analytically evaluate all four of the remaining delta functions simultaneously.  During this process, the isotropy of the system in the frame given by the 4-vector $U$ allows for further reduction of the dimensionality  by analytically evaluating several of the angular integrals. 

The change of variables that simplifies the 4-momentum conserving delta function is given by
\begin{equation}
p=p_1+p_2\,,\hspace{2mm} q=p_1-p_2\,, \hspace{2mm} p^{\prime}=p_3+p_4\,, \hspace{2mm} q^{\prime}=p_3-p_4\,.
\end{equation}
The Jacobian of this transformation is $1/2^{8}$.  Therefore, using Lemma \ref{diffeoProperty}, we find
\begin{align}\label{Meq1}
M=&\frac{1}{256(2\pi)^8 }\int   G((p+q)\cdot U/2,(p-q)\cdot  U/2,(p^{\prime}+q^{\prime})\cdot U/2, (p^{\prime}-q^{\prime})\cdot U/2)\notag\\ 
&\times S |\mathcal{M}|^2  \delta(p-p^{\prime})\delta((p+q)^2/4-m_1^2)\delta((p-q)^2/4-m_2^2)\delta((p^{\prime}+q^{\prime})^2/4-m_3^2)\notag\\
&\times\delta((p^{\prime}-q^{\prime})^2/4-m_4^2)1_{p^0>|q^0|} 1_{(p^{\prime})^0>|(q^{\prime})^0|}d^4pd^4qd^4p^{\prime}d^4q^{\prime}\,.
\end{align}
Next, we eliminate the integration over $p^{\prime}$ using $\delta(p-p^{\prime})$
 and then use Fubini's theorem\index{Dirac delta!Fubini's theorem} to write
\begin{align}\label{useFubini}
M=\frac{1}{256(2\pi)^8 }\int &\bigg[ \int G((p+q)\cdot U/2,(p-q)\cdot  U/2,(p^{\prime}+q^{\prime})\cdot U/2, (p^{\prime}-q^{\prime})\cdot U/2)\notag\\ 
&\times  1_{p^0>|q^0|} 1_{p^0>|(q^{\prime})^0|}S |\mathcal{M}|^2 \delta((p+q)^2/4-m_1^2)\delta((p-q)^2/4-m_2^2)\notag\\
&\times \delta((p+q^{\prime})^2/4-m_3^2)\delta((p-q^{\prime})^2/4-m_4^2)d^4qd^4q^{\prime}\bigg]d^4p\,.
\end{align}
Subsequent computations will justify this use of Fubini's theorem.

Since $p^0>0$ we have $dp\neq 0$ and so we can use Corollary \ref{dummyInt} of the coarea formula\index{coarea formula} to decompose this into an integral over the center of mass energy $s=p^2$,
\begin{align}\label{Meq2}
&M=\frac{1}{256(2\pi)^8 }\!\int_{s_0}^\infty\!\!\int \!\bigg[\int 1_{p^0>|q^0|} 1_{p^0>|(q^{\prime})^0|}  S |\mathcal{M}|^2  F(p,q,q^{\prime}) \delta((p+q)^2/4-m_1^2)\\
&\!\times\delta((p-q)^2/4-m_2^2)\delta((p+q^{\prime})^2/4-m_3^2)\delta((p-q^{\prime})^2/4-m_4^2)d^4qd^4q^{\prime}\bigg]\delta(p^2-s) d^4pds\,,\notag\\
&F(p,q,q^{\prime})=G((p+q)\cdot U/2,(p-q)\cdot  U/2,(p+q^{\prime})\cdot U/2, (p-q^{\prime})\cdot U/2)\,,\notag\\
&s_0=\max\{(m_1+m_2)^2,(m_3+m_4)^2\}\,.\notag
\end{align}
The lower bound on $s$ comes from the fact that both $p_1$ and $p_2$ are future timelike and hence 
\begin{equation}
p^2=m_1^2+m_2^2+2p_1\cdot p_2\geq m_1^2+m_2^2+2m_1m_2=(m_1+m_2)^2\,.
\end{equation}
The other inequality is obtained by using $p=p^{\prime}$. 

Note that the integral in brackets in \req{Meq2} is invariant under $SO(3)$ rotations of $p$ in the frame defined by $U$.  Therefore we obtain
\begin{align}\label{Kdef}
M=&\frac{1}{256(2\pi)^8 }\int_{s_0}^\infty\int_0^\infty K(s,p)\frac{4\pi |\vec p|^2}{2p^0}d|\vec p|ds\,,\hspace{2mm} p^0=p\cdot U=\sqrt{|\vec p|^2+s}\,,\\
K(s,p)=&\int\! 1_{p^0>|q^0|} 1_{p^0>|(q^{\prime})^0|}  S |\mathcal{M}|^2  F(p,q,q^{\prime}) \delta((p+q)^2/4-m_1^2)\delta((p-q)^2/4-m_2^2)\notag\\
&\times\delta((p+q^{\prime})^2/4-m_3^2)\delta((p-q^{\prime})^2/4-m_4^2)d^4qd^4q^{\prime}\,,\notag
\end{align}
where $|\vec p|$ denotes the norm of the spacial component of $p$ and in the formula for $K(s,p)$, $p$ is any four vector whose spacial component has norm $|\vec p|$ and timelike component $\sqrt{|\vec p|^2+s}$. Note that in integrating over $\delta(p^2-s)dp^0$, only the positive root was taken, due to the indicator functions in the $K(s,p)$.

We now simplify $K(s,p)$ for fixed but arbitrary $p$ and $s$ that satisfy $p^0=\sqrt{|\vec p|^2+s}$ and $s>s_0$.  These conditions imply $p$ is future timelike, hence we can we can change variables in  $q,q^{\prime}$ by an element of $Q\in SO(1,3)$ so that 
\begin{equation}
Qp=(\sqrt{s},0,0,0)\,, \hspace{2mm} QU=(\alpha,0,0,\delta)\,,
\end{equation}
where
\begin{equation}
\alpha=\frac{p\cdot U}{\sqrt{s}}\,, \hspace{2mm} \delta=\frac{1}{\sqrt{s}}\left((p\cdot U)^2-s \right)^{1/2}\,.
\end{equation}
Note that the delta functions in the integrand imply $p\pm q$ is  timelike (or null if the corresponding mass is zero).  Therefore $p^0>\pm q^0$ iff $p\mp q$ is future timelike (or null).  This condition is preserved by $SO(1,3)$ hence $p^0>|q^0|$ in one frame iff it holds in every frame.  Similar comments apply to $p^0>|(q^{\prime})^0|$ and so $K(s,p)$ has the same formula in the transformed frame as well.

We now evaluate the measure that is induced by the delta functions, using the method given in chapter \ref{ch:vol:forms}.  We have the constraint function
\begin{equation}
\Phi(q,q^{\prime})=((p+q)^2/4-m_1^2,(p-q)^2/4-m_2^2,(p+q^{\prime})^2/4-m_3^2,(p-q^{\prime})^2/4-m_4^2)
\end{equation}
and must compute the solution set $\Phi(q,q^{\prime})=0$. Adding and subtracting the first two components and the last two respectively, we have the equivalent conditions
\begin{align}
\frac{s+q^2}{2}=m_1^2+m_2^2\,,\hspace{2mm} p\cdot q=m_1^2-m_2^2\,, \hspace{2mm}\frac{s+(q^{\prime})^2}{2}=m_3^2+m_4^2\,,\hspace{2mm} p\cdot q^{\prime}=m_3^2-m_4^2\,.
\end{align}
If we let $(q^0,\vec{q})$, $((q^{\prime})^0,\vec{q}^{\prime})$ denote the spacial components in the frame defined by $p=(\sqrt{s},0,0,0)$ we have another set of equivalent conditions
\begin{align}\label{coordConditions}
&q^{0}=\frac{m_1^2-m_2^2}{\sqrt{s}}\,,\hspace{2mm} |\vec{q}|^2=\frac{(m_1^2-m_2^2)^2}{s}+s-2(m_1^2+m_2^2)\,,\\
&(q^{\prime})^{0}=\frac{m_3^2-m_4^2}{\sqrt{s}}\,, \hspace{2mm} |\vec{q}^{\prime}|^2=\frac{(m_3^2-m_4^2)^2}{s}+s-2(m_3^2+m_4^2)\,.\notag
\end{align}
Note that if these hold then using $s\geq s_0$, we obtain
\begin{equation}
\frac{|q^0|}{p^0}\leq \frac{|m_1^2-m_2^2|}{(m_1+m_2)^2}<1
\end{equation}
and similarly for $q^{\prime}$.  Hence the conditions in the indicator functions are satisfied and we can drop them from the formula for $K(s,p)$.

The conditions \req{coordConditions} imply that our solution set is a product of spheres in $\vec{q}$ and $\vec{q}^{\prime}$, as long as the conditions are consistent i.e. so long as $|\vec{q}|,|\vec{q}^{\prime}|>0$. To see that this holds for almost every $s$, first note
\begin{equation}
\frac{d}{ds}|\vec{q}|^2=1-\frac{(m_1^2-m_2^2)^2}{s^2}>0
\end{equation}
since $s\geq (m_1+m_2)^2$.  At $s=(m_1+m_2)^2$, $|\vec{q}|^2=0$.  Therefore, for $s>s_0$ we have $|\vec{q}|>0$ and similarly for $q^{\prime}$.  Hence we have the result
\begin{equation}
\Phi^{-1}(0)=\{q^{0}\}\times B_{|\vec{q}|}\times \{(q^{\prime})^{0}\}\times B_{|\vec{q}^{\prime}|}\,,
\end{equation}
where $B_r$ denotes the radius $r$ ball centered at $0$.  We will parametrize this by spherical angular coordinates in $q$ and $q^{\prime}$. 

 We now compute the induced volume form\index{induced volume form}. First consider the differential 
\begin{equation}
 D\Phi=\left( \begin{array}{c}
\frac{1}{2}(q+p)^\alpha\eta_{\alpha\beta}dq^\beta \\
\frac{1}{2}(q-p)^\alpha\eta_{\alpha\beta}dq^\beta\\
\frac{1}{2}(q^{\prime}+p)^\alpha\eta_{\alpha\beta}dq^{'^\beta}  \\
\frac{1}{2}(q^{\prime}-p)^\alpha\eta_{\alpha\beta}dq^{'^\beta}  \end{array} \right)\,.
\end{equation}
Evaluating this on the coordinate vector fields $\partial_{q^0}$, $\partial_r$, we obtain
\begin{equation}
 D\Phi(\partial_{q^0})=\left( \begin{array}{c}
\frac{1}{2}(q^0+\sqrt{s}) \\
\frac{1}{2}(q^0-\sqrt{s}) \\
0\\
0 \end{array} \right), \hspace{2mm}  D\Phi(\partial_{r})=\left( \begin{array}{c}
-\frac{1}{2}|\vec{q}| \\
-\frac{1}{2}|\vec{q}| \\
0\\
0 \end{array} \right)=\left( \begin{array}{c}
-\frac{1}{2}r \\
-\frac{1}{2}r \\
0\\
0 \end{array} \right)\,.
\end{equation}
Similar results hold for $q^{\prime}$.  Therefore we have the determinant
\begin{equation}
\det\left( \begin{array}{cccc}
D\Phi(\partial_{q^0}) & D\Phi(\partial_{r}) & D\Phi(\partial_{(q^{\prime})^0}) & D\Phi(\partial_{r^{\prime}}) \end{array} \right)=\frac{s}{4}rr^{\prime}.
\end{equation}
Note that this determinant being nonzero implies that our use of Fubini's theorem\index{Dirac delta!Fubini's theorem} in \req{useFubini} was justified.

By \req{volFormCoords} and \req{deltaDef}, the above computations imply that the induced volume measure is
\begin{align}
&\delta((p+q)^2/4-m_1^2)\delta((p-q)^2/4-m_2^2)\delta((p+q^{\prime})^2/4-m_3^2)\delta((p-q^{\prime})^2/4-m_4^2)d^4qd^4q^{\prime}\\
&=\frac{4}{srr^{\prime}}i_{(\partial_{q^0},\partial_{r},\partial _{(q^{\prime})^0},\partial_{r^{\prime}})}\left(r^2\sin(\phi)dq^0drd\theta d\phi\right)\wedge\left((r^{\prime})^2\sin(\phi^{\prime})d(q^{\prime})^0dr^{\prime}d\theta^{\prime}d\phi^{\prime}\right)\notag\\
&=\frac{4rr^{\prime}}{s}\sin(\phi)\sin(\phi^{\prime})d\theta d\phi d\theta^{\prime}d\phi^{\prime}\,,\notag
\end{align}
where
\begin{align}
r=&\frac{1}{\sqrt{s}}\sqrt{(s-(m_1+m_2)^2)(s-(m_1-m_2)^2)}\,,\\
r^{\prime}=&\frac{1}{\sqrt{s}}\sqrt{(s-(m_3+m_4)^2)(s-(m_3-m_4)^2)}\,.\notag
\end{align}

Consistent with our interest in the Boltzmann equation, we assume $F$ factors as
\begin{align}
 F(p,q,q^{\prime})=&F_{12}((p+q)\cdot U/2,(p-q)\cdot U/2)F_{34}((p+q^{\prime})\cdot U/2,(p-q^{\prime})\cdot U/2)\\
\equiv &G_{12}(p\cdot U,q\cdot U)G_{34}(p\cdot U,q^{\prime}\cdot U)\,.\notag
\end{align}
For now we suppress the dependence on $p$, as it is not of immediate concern. In our chosen coordinates where $U=(\alpha,0,0,\delta)$, we have
\begin{equation}
q\cdot U=q^0\alpha-r\delta\cos(\phi)
\end{equation}
and similarly for $q^{\prime}$.
To compute
\begin{align}\label{KAngular1}
K(s,p)=\frac{4rr^{\prime}}{s}\int \left[\int S |\mathcal{M}|^2 (s,t) G_{34}\sin(\phi^{\prime})d\theta^{\prime} d\phi^{\prime}\right] G_{12}\sin(\phi)d\theta d\phi\,,
\end{align}
first recall 
\begin{align}
t=&(p_1-p_3)^2=\frac{1}{4}(q- q^{\prime})^2=\frac{1}{4}(q^2+(q^{\prime})^2-2(q^0(q^{\prime})^0-\vec{q}\cdot \vec{q}^{\prime}))\,,\\
\vec{q}\cdot\vec{q}^{\prime}&=rr^{\prime}(\cos(\theta-\theta^{\prime})\sin(\phi)\sin(\phi^{\prime})+\cos(\phi)\cos(\phi^{\prime}))\,.\notag
\end{align}
Together, these imply that the integral in brackets in  \req{KAngular1} equals
\begin{align}
&\int_0^\pi\int_0^{2\pi} S |\mathcal{M}|^2 (s,t(\cos(\theta-\theta^{\prime})\sin(\phi)\sin(\phi^{\prime})+\cos(\phi)\cos(\phi^{\prime})))\\
&\hspace{15mm}\times G_{34}((q^{\prime})^0\alpha-r^{\prime}\delta\cos(\phi^{\prime}))\sin(\phi^{\prime})d\theta^{\prime} d\phi^{\prime}\notag\\
=&\int_{-1}^1\int_0^{2\pi} S |\mathcal{M}|^2 (s,t(\cos(\psi)\sin(\phi)\sqrt{1-y^2}+\cos(\phi)y)) G_{34}((q^{\prime})^0\alpha-r^{\prime}\delta y)d\psi dy\,.\notag
\end{align}

Therefore 
\begin{align}
K(s,p)=&\frac{8\pi rr^{\prime}}{s}\int_{-1}^1 \left[\int_{-1}^1\left(\int_0^{2\pi} S |\mathcal{M}|^2 (s,t(\cos(\psi)\sqrt{1-y^2}\sqrt{1-z^2}+yz))d\psi\right)\right.\\
&\hspace{26mm}\times G_{34}((q^{\prime})^0\alpha-r^{\prime}\delta y) dy\bigg] G_{12}(q^0\alpha-r\delta z)dz\,,\notag
\end{align}
where
\begin{align}
t(x)=&\frac{1}{4}((q^0)^2-r^2+((q^{\prime})^0)^2-(r^{\prime})^2-2q^0(q^{\prime})^0+2rr^{\prime}x)
= \frac{1}{4}((q^0-(q^{\prime})^0)^2-r^2-(r^{\prime})^2+2rr^{\prime}x)\,. 
\end{align}

\subsection{Electron and neutrino collision integrals}\label{nu:matrix:elements}
In this section, we further simplify the various integrals of the scattering matrix element that appear in the scattering kernels for processes involving $e^\pm$ and neutrinos.  For reference, we collect the important prior results 
on evaluation of the scattering kernel integrals \req{collision:integrals}, where we have changed notation from $|\vec p|$ to $p$.
\begin{align}\label{M:simp}
M=\frac{1}{256(2\pi)^7 }\int_{s_0}^\infty&\int_0^\infty K(s,p)\frac{ p^2}{p^0}dpds\,,
\end{align}
\begin{align}\label{matrixElemInt}
K(s,p)=&\frac{8\pi rr^{\prime}}{s}\int_{-1}^1 \left[\int_{-1}^1\left(\int_0^{2\pi} S |\mathcal{M}|^2 (s,t(\cos(\psi)\sqrt{1-y^2}\sqrt{1-z^2}+yz))d\psi\right)\right.\\
&\hspace{26mm}\times G_{34}((q^{\prime})^0\alpha-r^{\prime}\delta y) dy\bigg] G_{12}(q^0\alpha-r\delta z)dz\,,\notag
\end{align}
where
\begin{align}\label{tDef}
p^0=&\sqrt{p^2+s}\,,\hspace{2mm} \alpha=\frac{p^0}{\sqrt{s}}\,, \hspace{2mm} \delta=\frac{p}{\sqrt{s}}\,,\hspace{2mm}q^{0}=\frac{m_1^2-m_2^2}{\sqrt{s}}\,,\hspace{2mm}(q^{\prime})^{0}=\frac{m_3^2-m_4^2}{\sqrt{s}}\,,\\
r=&\frac{1}{\sqrt{s}}\sqrt{(s-(m_1+m_2)^2)(s-(m_1-m_2)^2)}\,,\notag\\
 r^{\prime}=&\frac{1}{\sqrt{s}}\sqrt{(s-(m_3+m_4)^2)(s-(m_3-m_4)^2)}\,,\notag\\
t(x)=&\frac{1}{4}((q^0-(q^{\prime})^0)^2-r^2-(r^{\prime})^2+2rr^{\prime}x)\,,\notag\\
s_0=&\max\{(m_1+m_2)^2,(m_3+m_4)^2\}\,,\notag
\end{align}
and
\begin{align}
 F(p,q,q^{\prime})=&F_{12}((p+q)\cdot U/2,(p-q)\cdot U/2)F_{34}((p+q^{\prime})\cdot U/2,(p-q^{\prime})\cdot U/2)\\
\equiv &G_{12}(p\cdot U,q\cdot U)G_{34}(p\cdot U,q^{\prime}\cdot U)\notag\,.
\end{align}

This is as far as we can simplify the collision integrals without more information about the form of the matrix elements.  The  matrix elements for weak force scattering processes involving neutrinos and $e^\pm$ in the limit $|p|\ll M_W,M_Z$, taken from \cite{Dolgov:1997mb,Dolgov:1998sf}, are as follows
\begin{table}[ht]
\centering 
\begin{tabular}{|c|c|}
\hline
Process &$S|\mathcal{M}|^2$  \\
\hline
$\nu_e+\bar\nu_e\rightarrow\nu_e+\bar\nu_e$ & $128G_F^2(p_1\cdot p_4)(p_2\cdot p_3)$\\
\hline
$\nu_e+\nu_e\rightarrow\nu_e+\nu_e$ & $64G_F^2(p_1\cdot p_2)(p_3\cdot p_4)$\\
\hline
$\nu_e+\bar\nu_e\rightarrow\nu_j+\bar\nu_j$&$32G_F^2(p_1\cdot p_4)(p_2\cdot p_3)$\\
\hline
$\nu_e+\bar\nu_j\rightarrow\nu_e+\bar\nu_j$ & $32G_F^2(p_1\cdot p_4)(p_2\cdot p_3)$\\
\hline
$\nu_e+\nu_j\rightarrow\nu_e+\nu_j$&$32G_F^2(p_1\cdot p_2)(p_3\cdot p_4)$\\
\hline
$\nu_e+\bar\nu_e\rightarrow e^++e^-$ & $128G_F^2[g_L^2(p_1\cdot p_4)(p_2\cdot p_3)+g_R^2(p_1\cdot p_3)(p_2\cdot p_4)+g_Lg_Rm_e^2(p_1\cdot p_2)]$\\
\hline
$\nu_e+e^-\rightarrow\nu_e+e^-$ & $128G_F^2[g_L^2(p_1\cdot p_2)(p_3\cdot p_4)+g_R^2(p_1\cdot p_4)(p_2\cdot p_3)-g_Lg_Rm_e^2(p_1\cdot p_3)]$\\
\hline
$\nu_e+e^+\rightarrow\nu_e+e^+$ & $128G_F^2[g_R^2(p_1\cdot p_2)(p_3\cdot p_4)+g_L^2(p_1\cdot p_4)(p_2\cdot p_3)-g_Lg_Rm_e^2(p_1\cdot p_3)]$\\
\hline
\end{tabular}
\caption{Matrix elements for electron neutrino processes where $j=\mu,\tau$,  $g_L=\frac{1}{2}+\sin^2\theta_W$, $g_R=\sin^2\theta_W$, $\sin^2(\theta_W)\approx 0.23$ is the Weinberg angle, and $G_F=1.16637\times 10^{-5}\text{GeV}^{-2}$ is Fermi's constant.}
\label{table:nu:e:reac}
\end{table}

\begin{table}[ht]
\centering 
\begin{tabular}{|c|c|}
\hline
Process &$S|\mathcal{M}|^2$  \\
\hline
$\nu_i+\bar\nu_i\rightarrow\nu_i+\bar\nu_i$ & $128G_F^2(p_1\cdot p_4)(p_2\cdot p_3)$\\
\hline
$\nu_i+\nu_i\rightarrow\nu_i+\nu_i$ & $64G_F^2(p_1\cdot p_2)(p_3\cdot p_4)$\\
\hline
$\nu_i+\bar\nu_i\rightarrow\nu_j+\bar\nu_j$&$32G_F^2(p_1\cdot p_4)(p_2\cdot p_3)$\\
\hline
$\nu_i+\bar\nu_j\rightarrow\nu_i+\bar\nu_j$ & $32G_F^2(p_1\cdot p_4)(p_2\cdot p_3)$\\
\hline
$\nu_i+\nu_j\rightarrow\nu_i+\nu_j$&$32G_F^2(p_1\cdot p_2)(p_3\cdot p_4)$\\
\hline
$\nu_i+\bar\nu_i\rightarrow e^++e^-$ & $128G_F^2[\tilde{g}_L^2(p_1\cdot p_4)(p_2\cdot p_3)+g_R^2(p_1\cdot p_3)(p_2\cdot p_4)+\tilde{g}_Lg_Rm_e^2(p_1\cdot p_2)]$\\
\hline
$\nu_i+e^-\rightarrow\nu_i+e^-$ & $128G_F^2[\tilde{g}_L^2(p_1\cdot p_2)(p_3\cdot p_4)+g_R^2(p_1\cdot p_4)(p_2\cdot p_3)-\tilde{g}_Lg_Rm_e^2(p_1\cdot p_3)]$\\
\hline
$\nu_i+e^+\rightarrow\nu_i+e^+$ & $128G_F^2[g_R^2(p_1\cdot p_2)(p_3\cdot p_4)+\tilde{g}_L^2(p_1\cdot p_4)(p_2\cdot p_3)-\tilde{g}_Lg_Rm_e^2(p_1\cdot p_3)]$\\
\hline
\end{tabular}
\caption{Matrix elements for $\mu$ and $\tau$ neutrino processes where $i=\mu,\tau$, $j=e,\mu,\tau$, $j\neq i$,  $\tilde{g}_L=g_L-1=-\frac{1}{2}+\sin^2\theta_W$, $g_R=\sin^2\theta_W$, $\sin^2(\theta_W)\approx 0.23$ is the Weinberg angle, and $G_F=1.16637\times 10^{-5}\text{GeV}^{-2}$ is Fermi's constant.}
\label{table:nu:mu:reac}
\end{table}
In the following subsections, we will analytically simplify \req{M:simp} for each of these processes.

\para{Neutrino-neutrino scattering} Using \req{Mandelstam}, the matrix elements for neutrino-neutrino scattering $\nu\nu\rightarrow\nu\nu$ can be simplified to
\begin{align}
\label{TA002}
S|\mathcal{M}|^2=C(p_1\cdot p_2)(p_3\cdot p_4)=C\frac{s^2}{4}\,,
\end{align}
 where the coefficient $C$ is given in table \ref{table:nu:nu:coeff}.

\begin{table}[ht]
\centering 
\begin{tabular}{|c|c|}
\hline
Process &$C$ \\
\hline
$\nu_i+\nu_i\rightarrow\nu_i+\nu_i,\hspace{2mm} i\in\{e,\mu,\tau\}$& $64 G_F^2$\\
\hline
$\nu_i+\nu_j\rightarrow\nu_i+\nu_j,\hspace{2mm} i\neq j, \hspace{1mm} i,j\in\{e,\mu,\tau\}$& $32 G_F^2$\\
\hline
\end{tabular}
\caption{Matrix element coefficients for neutrino neutrino scattering processes.}
\label{table:nu:nu:coeff}
\end{table}
From here we obtain
\begin{align}
K(s,p)=&\frac{8\pi rr^{\prime}}{s}\int_{-1}^1 \left[\int_{-1}^1\left(\int_0^{2\pi}S|\mathcal{M}|^2 (s,t(\cos(\psi)\sqrt{1-y^2}\sqrt{1-z^2}+yz))d\psi\right)\right.\\
&\hspace{26mm}\times G_{34}(p^0,(q^{\prime})^0\alpha-r^{\prime}\delta y) dy\bigg] G_{12}(p^0,q^0\alpha-r\delta z)dz\notag\\
=& 4\pi^2 Crr^{\prime}s \int_{-1}^1 G_{12}(p^0,q^0\alpha-r\delta z)dz \int_{-1}^1G_{34}(p^0,(q^{\prime})^0\alpha-r^{\prime}\delta y) dy\,.\notag
\end{align}
Therefore
\begin{align}
M_{\nu\nu\rightarrow\nu\nu}=&\frac{C}{256(2\pi)^5 } T^8\!\!\!\int_{0}^\infty\!\!\!\tilde{s}^2\!\!\int_0^\infty  \left[\int_{-1}^1 \tilde{G}_{12}(\tilde p^0,-\tilde{p} z)dz \int_{-1}^1\tilde{G}_{34}(\tilde p^0,-\tilde{p} y) dy\right]\frac{\tilde{p}^2}{\tilde{p}^0}d\tilde{p}d\tilde{s}\,,
\end{align}
where the tilde quantities are obtained by non-dimensionalizing via scaling by $T$ and we have re-introduced the dependence of $G_{i,j}$ on $p^0$. If we want to emphasize the role of $C$ then we write $M_{\nu\nu\rightarrow\nu\nu}(C)$.

\para{Neutrino-antineutrino scattering} Using \req{Mandelstam}, the matrix elements for neutrino antineutrino scattering $\nu\bar{\nu}\rightarrow\nu\bar{\nu}$ can be simplified to
\begin{align}
S|\mathcal{M}|^2=C\left(\frac{s+t}{2}\right)^2\,,
\end{align}
where the coefficient $C$ is given in table \ref{table:nu:nubar:coeff}.

\begin{table}[ht]
\centering 
\begin{tabular}{|c|c|}
\hline
Process &$C$  \\
\hline
$\nu_i+\bar\nu_i\rightarrow\nu_i+\bar\nu_i,\hspace{2mm} i\in\{e,\mu,\tau\}$& $128 G_F^2$\\
\hline
$\nu_i+\bar\nu_i\rightarrow\nu_j+\bar\nu_j,\hspace{2mm} i\neq j, \hspace{1mm} i,j\in\{e,\mu,\tau\}$& $32 G_F^2$\\
\hline
$\nu_i+\bar\nu_j\rightarrow\nu_i+\bar\nu_j,\hspace{2mm} i\neq j, \hspace{1mm} i,j\in\{e,\mu,\tau\}$& $32 G_F^2$\\
\hline
\end{tabular}
\caption{Matrix element coefficients for neutrino neutrino scattering processes.}
\label{table:nu:nubar:coeff}
\end{table}
Using this we find
 \begin{align}
\int_0^{2\pi} S |\mathcal{M}|^2 (s,t(\cos(\psi)\sqrt{1-y^2}\sqrt{1-z^2}+yz))d\psi 
=& \frac{\pi C}{16} s^2(3+4 yz-y^2-z^2+3y^2z^2)
\equiv\frac{\pi C}{16} s^2q(y,z)\,, \\
K(s,p)=&\frac{\pi^2C}{2}s^2\int_{-1}^1 \left[\int_{-1}^1q(y,z)G_{34}(p^0,-p y) dy\right] G_{12}(p^0,-p z)dz\,.\notag
\end{align}
Therefore
\begin{align}\label{eq:M:nu:nubar}
M_{\nu\bar\nu\rightarrow\nu\bar\nu}=\frac{C}{2048(2\pi)^5 }T^8\! \int_0^\infty\!\!\int_0^\infty\!\!\! \tilde{s}^2\bigg[\int_{-1}^1\!\int_{-1}^1q(y,z)\tilde{G}_{34}(\tilde p^0,-\tilde{p} y) 
\tilde{G}_{12}(\tilde p^0,-\tilde{p} z)dydz\bigg]\frac{\tilde{p}^2}{\tilde{p}^0}d\tilde{p}d\tilde{s}\,.
\end{align}
 If we want to emphasize the role of $C$ then we write $M_{\nu\bar\nu\rightarrow\nu\bar\nu}(C)$. Note that due to the polynomial form of the matrix element integral, the double integral in brackets breaks into a linear combination of products of one dimensional integrals, meaning that the nesting of integrals is again only three deep in  practice.

\para{Neutrino-antineutrino annihilation to electron-positrons}
Using \req{Mandelstam}, the matrix elements for leptonic neutrino antineutrino annihilation $\nu\bar{\nu}\rightarrow e^+e^-$ can be simplified to
\begin{align}
S|\mathcal{M}|^2=A\left(\frac{s+t-m_e^2}{2}\right)^2+B\left(\frac{m_e^2-t}{2}\right)^2+Cm_e^2\frac{s}{2}\,,
\end{align}
where the coefficients $A,B,C$ are given in table \ref{table:nu:nubar:ee:coeff}.

\begin{table}[ht]
\centering 
\begin{tabular}{|c|c|c|c|}
\hline
Process &$A$&$B$&$C$  \\
\hline
$\nu_e+\bar\nu_e\rightarrow e^++e^-$&$128G_F^2g_L^2$&$128G_F^2g_R^2$&$128G_F^2g_Lg_R$\\
\hline
$\nu_i+\bar\nu_i\rightarrow e^++e^-,\hspace{2mm} i\in\{\mu,\tau\}$&$128G_F^2\tilde g_L^2$&$128G_F^2g_R^2$&$128G_F^2\tilde g_Lg_R$\\
\hline
\end{tabular}
\caption{Matrix element coefficients for neutrino neutrino annihilation into $e^\pm$.}
\label{table:nu:nubar:ee:coeff}
\end{table}

The integral of each of these terms is
 \begin{align}
\int_0^{2\pi}\frac{(s+t(\psi)-m_e^2)^2}{4}d\psi=&\frac{\pi}{16}s(3s-4m_e^2)+\frac{\pi}{4}s^{3/2}\sqrt{s-4m_e^2}yz
-\frac{\pi}{16}s(s-4m_e^2)(y^2+z^2)+\frac{3\pi}{16}s(s-4m_e^2)y^2z^2\,,\\
\int_0^{2\pi} \frac{(m_e^2-t(\psi))^2}{4}d\psi=&\frac{\pi}{16}s(3s-4m_e^2)-\frac{\pi}{4}s^{3/2}\sqrt{s-4m_e^2}yz
-\frac{\pi}{16}s(s-4m_e^2)(y^2+z^2)+\frac{3\pi}{16}s(s-4m_e^2)y^2z^2\,,\notag\\
\int_0^{2\pi} m_e^2\frac{s}{2} d\psi=&\pi m_e^2s\,.\notag
\end{align}
Therefore 
\begin{align}
\int_0^{2\pi} S |\mathcal{M}|^2 (s,t(\psi))d\psi
=&\frac{\pi}{16}s[3s(A+B)+4m_e^2(4C-A-B)]+\frac{\pi}{4}s^{3/2}\sqrt{s-4m_e^2}(A-B)yz\\
&-\frac{\pi}{16}s(s-4m_e^2)(A+B)(y^2+z^2)+\frac{3\pi}{16}s(s-4m_e^2)(A+B)y^2z^2
\equiv \pi q(m_e,s,y,z)\,,\notag
\end{align}
and hence
\begin{align}
&M_{\nu\bar\nu\rightarrow e^+e^-}\\
=&\frac{1}{128(2\pi)^5 }\int_{4m_e^2}^\infty\int_0^\infty\!\!\!\sqrt{1-4m_e^2/s}\left[\int_{-1}^1\int_{-1}^1q(s,y,z,m_e)G_{34}(p^0,-(\sqrt{1-4m_e^2/s})p y)\right.\notag\\
&\hspace{68mm}\times G_{12}(p^0,-p z)dydz\bigg]\frac{ p^2}{p^0}dpds,\notag\\
=&\frac{T^8}{128(2\pi)^5 }\int_{4\tilde m_e^2}^\infty\int_0^\infty\!\!\!\sqrt{1-4\tilde m_e^2/\tilde s}\left[\int_{-1}^1 \int_{-1}^1q(\tilde s,y,z,\tilde m_e)\tilde G_{34}(\tilde p^0,-(\sqrt{1-4\tilde{m}_e^2/\tilde s})\tilde p y)\right.\notag\\
&\hspace{68mm}\times\tilde G_{12}(\tilde p^0,-\tilde p z)dydz\bigg] \frac{ \tilde p^2}{\tilde p^0}d\tilde pd\tilde s\,,\notag
\end{align}
where $\tilde{m_e}=m_e/T$.  If we want to emphasize the role of $A,B,C$ then we write $M_{\nu\bar\nu\rightarrow e^+e^-}(A,B,C)$.  Note that this expression is linear in $(A,B,C)\in\mathbb{R}^3$. Also note that, under our assumptions that the distributions of $e^+$ and $e^-$ are the same,  the $G_{ij}$ terms that contain the product of $e^\pm$ distributions are even functions. Hence the term involving the integral of $yz$ vanishes by antisymmetry.

\para{Neutrino-electron(positron) scattering}
Using \req{Mandelstam}, the matrix elements for neutrino $e^\pm$ scattering $\nu e^\pm\rightarrow \nu e^\pm$ can be simplified to
\begin{align}
S|\mathcal{M}|^2=A\left(\frac{s-m_e^2}{2}\right)^2+B\left(\frac{s+t-m_e^2}{2}\right)^2+Cm_e^2\frac{t}{2}\,,
\end{align}
 where the coefficients $A,B,C$ are given in table \ref{table:nu:e:coeff}.

\begin{table}[ht]
\centering 
\begin{tabular}{|c|c|c|c|}
\hline
Process &$A$&$B$&$C$  \\
\hline
$\nu_e+e^-\rightarrow \nu_e+e^-$&$128G_F^2g_L^2$&$128G_F^2g_R^2$&$128G_F^2g_Lg_R$\\
\hline
$\nu_i+e^-\rightarrow \nu_i+e^-,\hspace{2mm} i\in\{\mu,\tau\}$&$128G_F^2\tilde g_L^2$&$128G_F^2g_R^2$&$128G_F^2\tilde g_Lg_R$\\
\hline
$\nu_e+e^+\rightarrow \nu_e+e^+$&$128G_F^2g_R^2$&$128G_F^2g_L^2$&$128G_F^2g_Lg_R$\\
\hline
$\nu_i+e^+\rightarrow \nu_i+e^+,\hspace{2mm} i\in\{\mu,\tau\}$&$128G_F^2 g_R^2$&$128G_F^2\tilde g_L^2$&$128G_F^2\tilde g_Lg_R$\\
\hline
\end{tabular}
\caption{Matrix element coefficients for neutrino $e^\pm$ scattering.}
\label{table:nu:e:coeff}
\end{table}

  The integral of each of these terms is
 \begin{align}
&\int_0^{2\pi}\frac{(s-m_e^2)^2}{4} d\psi=\pi\frac{(s-m_e^2)^2}{2},\\
&\int_0^{2\pi}\frac{(s+t(\psi)-m_e^2)^2}{4}d\psi=\frac{\pi}{16s^2}(s-m_e^2)^2(3m_e^4+2m_e^2s+3s^2)\notag\\
&\qquad+\frac{\pi}{4s^2}(s-m_e^2)^3(s+m_e^2)yz
-\frac{\pi}{16s^2}(s-m_e^2)^4(y^2+z^2)+\frac{3\pi}{16s^2}(s-m_e^2)^4y^2z^2\,,\notag\\
&\int_0^{2\pi} m_e^2\frac{t(\psi)}{2}d\psi=-\frac{\pi}{2s}m_e^2(s-m_e^2)^2(1-yz)\,.\notag
\end{align}
Therefore we have
\begin{align}
\int_0^{2\pi} S |\mathcal{M}|^2 (s,t(\psi))d\psi=&\pi\left[\frac{A}{2}+\frac{B}{16s^2}(3m_e^4+2m_e^2s+3s^2)-\frac{C}{2s}m_e^2\right](s-m_e^2)^2\notag\\
&+\pi\left[\frac{B}{4s^2}(s-m_e^2)(s+m_e^2)+\frac{C}{2s}m_e^2\right](s-m_e^2)^2yz\notag\\
&-B\frac{\pi}{16s^2}(s-m_e^2)^4(y^2+z^2)+B\frac{3\pi}{16s^2}(s-m_e^2)^4y^2z^2\notag\\
\equiv& \pi q(m_e,s,y,z)
\end{align}
and
\begin{align}
K(s,p)=&\frac{8\pi^2 rr^{\prime}}{s}\!\!\int_{-1}^1 \!\left[\int_{-1}^1 \! q(m_e,s,y,z) G_{34}(p^0,(q^{\prime})^0\alpha-r^{\prime}\delta y) dy\right] \!G_{12}(p^0,q^0\alpha-r\delta z)dz\,,\\
r=r^{\prime}=&\frac{s-m_e^2}{\sqrt{s}}\,,\hspace{2mm} q^0=(q^{\prime})^0=-\frac{m_e^2}{\sqrt{s}}\,,\hspace{2mm} \delta=\frac{p}{\sqrt{s}}\,,\hspace{2mm}  \alpha=\frac{p^0}{\sqrt{s}}\,.\notag
\end{align}
This implies
\begin{align}
M_{\nu e\rightarrow\nu e}=&\frac{1}{128(2\pi)^5 }\int_{m_e^2}^\infty\!\int_0^\infty (1-m_e^2/s)^2\left(\int_{-1}^1 \int_{-1}^1 q(m_e,s,y,z) G_{34}(p^0,(q^{\prime})^0\alpha-r^{\prime}\delta y)\right.\\
&\hspace{50mm}\times  G_{12}(p^0,q^0\alpha-r\delta z)dydz\bigg)\frac{ p^2}{p^0}dpds\,.\notag
\end{align}
As above, after scaling all masses by $T$, we obtain a prefactor of $T^8$. If we want to emphasize the role of $A,B,C$ then we write $M_{\nu e\rightarrow\nu e}(A,B,C)$.  Note that this expression is also linear in $(A,B,C)\in\mathbb{R}^3$.

\para{Total collision integral}
We now give the total collision integrals for neutrinos.    In the following, we indicate which distributions are used in each of the four types of scattering integrals discussed above by using the appropriate subscripts. For example, to compute $M_{\nu_e\bar\nu_\mu\rightarrow\nu_e\bar\nu_\mu}$  we set $G_{1,2}=\hat\psi_jf^1f^2$, $G_{3,4}=f_3f_4$, $f_1= f_{\nu_e}$, $f_3=f_{\nu_e}$, and $f_2=f_4=f_{\bar\nu_\mu}$ in the expression \req{eq:M:nu:nubar} for $M_{\nu\bar\nu\rightarrow\nu\bar\nu}$  and then, to include the reverse direction of the process, we must {\emph subtract}  the analogous expression whose only difference is $G_{1,2}=\hat\psi_jf_1f_2$, $G_{3,4}=f^3f^4$.
With this notation the collision integral for $\nu_e$ is
\begin{align}\label{Mtot}
M_{\nu_e}=&[M_{\nu_e\nu_e\rightarrow\nu_e\nu_e}+M_{\nu_e\nu_\mu\rightarrow\nu_e\nu_\mu}+M_{\nu_e\nu_\tau\rightarrow\nu_e\nu_\tau}]\\
&+[M_{\nu_e\bar\nu_e\rightarrow\nu_e\bar\nu_e}+M_{\nu_e\bar\nu_e\rightarrow\nu_\mu\bar\nu_\mu}+M_{\nu_e\bar\nu_e\rightarrow\nu_\tau\bar\nu_\tau}+M_{\nu_e\bar\nu_\mu\rightarrow\nu_e\bar\nu_\mu}+M_{\nu_e\bar\nu_\tau\rightarrow\nu_e\bar\nu_\tau}]\notag\\
&+M_{\nu_e\bar\nu_e\rightarrow e^+e^-}+[M_{\nu_e e^-\rightarrow\nu_e e^-}+M_{\nu_e e^+\rightarrow\nu_e e^+}]\,.\notag
\end{align}

Symmetry among the interactions implies that the distributions of $\nu_\mu$ and $\nu_\tau$ are equal.  We also neglect the small matter anti-matter asymmetry and so we take the distribution of each particle to be equal to that of the corresponding antiparticle.  Therefore there are only three independent distributions, $f_{\nu_e}$, $f_{\nu_\mu}$, and $f_e$. This allows us to combine several of the terms in \req{Mtot} to obtain
\begin{align}
M_{\nu_e}=&M_{\nu_e\nu_e\rightarrow\nu_e\nu_e}(64G_F^2)+M_{\nu_e\nu_\mu\rightarrow\nu_e\nu_\mu}(2\times 32 G_F^2)+M_{\nu_e\bar\nu_e\rightarrow\nu_e\bar\nu_e}(128G_F^2)  \\
&+M_{\nu_e\bar\nu_e\rightarrow\nu_\mu\bar\nu_\mu}(2\times 32G_F^2)+M_{\nu_e\bar\nu_\mu\rightarrow\nu_e\bar\nu_\mu}(2\times 32 G_F^2)\notag\\
&+M_{\nu_e\bar\nu_e\rightarrow e^+e^-}(128G_F^2g_L^2,128G_F^2g_R^2,128G_F^2g_Lg_R)\notag\\
&+M_{\nu_e e\rightarrow\nu_e e}(128 G_F^2( g_L^2+g_R^2),128 G_F^2 (g_L^2+ g_R^2),256G_F^2g_Lg_R)\notag\,.
\end{align}
Introducing one more piece of notation, we use a subscript $k$ to denote the orthogonal polynomial basis element that multiplies $f_1$ or $f^1$ in the inner product.  The inner product of the $k$th basis element with the total scattering operator for electron neutrinos is therefore 
\begin{align}
R_k=&2\pi^2T^{-3} M_{k,\nu_e}\,.
\end{align}
Under these same assumptions and conventions, the total collision integral for the combined $\nu_\mu$, $\nu_\tau$ distribution (which we label $\nu_\mu$) is
\begin{align}
M_{\nu_\mu}=&M_{\nu_\mu\nu_\mu\rightarrow\nu_\mu\nu_\mu}(64G_F^2+32G_F^2)+M_{\nu_\mu\nu_e\rightarrow\nu_\mu\nu_e}(32 G_F^2)\\&+M_{\nu_\mu\bar\nu_\mu\rightarrow\nu_\mu\bar\nu_\mu}(128G_F^2+32G_F^2+32G_F^2) \notag \\
&+M_{\nu_\mu\bar\nu_\mu\rightarrow\nu_e\bar\nu_e}(32G_F^2)+M_{\nu_\mu\bar\nu_e\rightarrow\nu_\mu\bar\nu_e}( 32 G_F^2)\notag\\
&+M_{\nu_\mu\bar\nu_\mu\rightarrow e^+e^-}(128G_F^2\tilde g_L^2,128G_F^2g_R^2,128G_F^2\tilde g_Lg_R)\notag\\
&+M_{\nu_\mu e\rightarrow\nu_\mu e}(128 G_F^2(\tilde  g_L^2+g_R^2),128 G_F^2 (\tilde g_L^2+ g_R^2),256G_F^2\tilde g_Lg_R)\,,\notag\\
R_k=&2\pi^2T^{-3} M_{k,\nu_\mu}\,.
\end{align}

\para{Neutrino freeze-out test}
Now that we have the above expressions for the neutrino scattering integrals, we can compare the chemical equilibrium\index{chemical equilibrium} and nonequilibrium methods on the problem of neutrino freeze-out\index{neutrino!freeze-out} using the full $2$-$2$ scattering kernels for neutrino processes.  We solve the Boltzmann-Einstein equation, \req{boltzmann}, for both the electron neutrino distribution and the combined $\mu$, $\tau$ neutrino distribution, including all of the  processes outlined above in the scattering operator, together with the Hubble\index{Hubble!equation} equation for $a(t)$, \req{Hubble:eq}.  The total energy density  appearing in the Hubble equation consists of the contributions from both independent neutrino distributions as well as chemical equilibrium $e^\pm$ and photon distributions at some common temperature $T_\gamma$, all computed using \req{moments}.  The dynamics of $T_\gamma$ are fixed by the divergence freedom condition of the total stress energy tensor\index{stress-energy tensor} implied by Einstein's equations.  In addition, we include the QED corrections to the $e^\pm$ and photon equations of state from  \rsec{ch:param:studies}\index{QED!Corrections EOS}.

To compare our results with Ref.~\cite{Mangano:2005cc}, where neutrino freeze-out was simulated using $\sin^2(\theta_W)=0.23$ and $\eta=\eta_0$, in table \ref{table:method:comp} we present $N_\nu$ together with the following quantities
\begin{align}
 z_{fin}=T_\gamma a,\hspace{2mm}  \rho_{\nu 0}=\frac{7}{120}\pi^2a^{-4}, \hspace{2mm}  \delta\bar\rho_{\nu}= \frac{\rho_\nu}{\rho_{\nu 0}}-1.
\end{align}
This quantities were introduced in Ref.~\cite{Mangano:2005cc}, but some additional discussion of their significance is in order.  The normalization of the scale factor $a$ is chosen so that at the start of the computation $T_\gamma=1/a$.  This means that $1/a$ is the temperature of a (hypothetical) particle species that is completely decoupled throughout the computation. Here we will call it the free-streaming\index{free-streaming} temperature. $z_{fin}$ is the ratio of photon temperature to the free-streaming temperature.  It is a measure of the amount of reheating that photons underwent due to the annihilation of $e^\pm$.  For completely decoupled neutrinos, whose temperature is the free-streaming temperature, the well known value can be computed from conservation of entropy
\begin{equation}
z_{fin}=(11/4)^{1/3}\approx 1.401.
\end{equation}
For coupled neutrinos, one expects this value to be slightly reduced, due to the  transfer of some entropy from annihilating $e^\pm$ into neutrinos. This is reflected in Table \ref{table:method:comp}.

$\rho_{\nu0}$ is the energy density of a massless fermion with two degrees of freedom and temperature equal to the free-streaming temperature.  In other words, it is the energy density of a single neutrino species, assuming it decoupled before reheating. Consequently, $\delta\bar\rho_\nu$ is the fractional increase in the energy density of a coupled neutrino species, due to its participation in reheating.

We compute the above using both the chemical equilibrium and nonequilibrium methods. For the following results, we used $\sin^2(\theta_W)=0.23$ and $\eta=\eta_0$. 
\begin{table}[ht]\label{table:method:comp}
\centering 
\begin{tabular}{|c|c|c|c|c|c|}
\hline
Method &Modes&$z_{fin}$ & $\delta\bar\rho_{\nu_e}$&   $\delta\bar\rho_{\nu_{\mu,\tau}}$ & $N_{\nu}$  \\
\hline
Chemical Eq& 4 &1.39785 &0.009230 &0.003792 &3.044269\\
\hline
Chemical Non-Eq& 2&1.39784 &0.009269 & 0.003799&3.044383 \\
\hline
Chemical Non-Eq& 3&1.39785&0.009230 & 0.003791&3.044264 \\
\hline
\end{tabular}
\caption{Equilibrium and nonequilibrium values for relevant parameters in different modes.}
\end{table}
We see that $\Delta N_\nu\equiv N_\nu-3$ agrees to $2$ digits and $4$ digits when using $2$ and $3$ modes respectively for the chemical nonequilibrium method, and similar behavior holds for the other quantities. Due to the reduction in the required number of modes, the chemical nonequilibrium method with the minimum number of required modes ($2$ modes) is more than $20\times$ faster than the chemical equilibrium method with its minimum number of required modes ($4$ modes), a very significant speed-up when the minimum number of modes meets the required precision.  The value of $N_\nu$ we obtain agrees with that found by \cite{Mangano:2005cc}, up to their cited error tolerance of $\pm 0.002$.

\para{Conservation laws and scattering integrals}
For some processes, various of the $R_k$'s vanish exactly, as we now show. First consider processes in which $f_1=f_3$ and $f_2=f_4$, such as in kinetic scattering processes. Since $m_1=m_3$ and $m_2=m_4$ we have $r=r^{\prime}$, $q^0=(q^{\prime})^0$.  The scattering terms are all two dimensional integrals of some function of $s$ and $p$ multiplied by the quantity
\begin{align}
&I_k
\equiv\int_{-1}^1\! \left[\int_{-1}^1\int_0^{2\pi}\!\!S|\mathcal{M}|^2 (s,t(\cos(\psi)\sqrt{1-y^2}\sqrt{1-z^2}+yz))d\psi f_1(h_1(y))f_2(h_2(y)) dy\right]\\
&\hspace{26mm}\times f_k^1(h_1(z))f^2(h_2(z))dz\notag\\
&-\int_{-1}^1 \!\left[\int_{-1}^1\int_0^{2\pi}\!\!S|\mathcal{M}|^2 (s,t(\cos(\psi)\sqrt{1-y^2}\sqrt{1-z^2}+yz))d\psi f^1(h_1(y))f^2(h_2(y)) dy\right] \notag\\
&\hspace{26mm}\times f_{1,k}(h_1(z))f_2(h_2(z))dz\,,\notag\\
&h_1(y)=(p^0+(q^{\prime})^0\alpha-r^{\prime}\delta y)/2\,,\hspace{2mm} h_2(y)=(p^0-q^0\alpha+r\delta y)/2\,,\notag\\
&f_{1,k}=\hat\psi_k f_1\,, f^1_k=\hat\psi_k f^1\,.\notag
\end{align}
Note that for $k=0$, $\hat\psi_0$ is constant.  After factoring it out of $I_k$, the result is clearly zero and so $R_0=0$.  

We further specialize to a distribution scattering from itself i.e. $f_1=f_2=f_3=f_4$.  Since $m_1=m_2$ and $m_3=m_4$ we have $q^0=(q^{\prime})^0=0$ and
\begin{equation}
h_1(y)=(p^0-r^{\prime}\delta y)/2,\hspace{2mm} h_2(y)=(p^0+r\delta y)/2\,.
\end{equation}
 By the above, we know that $R_0=0$.  $\hat\psi_1$ appears in $I_1$ in the form $\hat\psi_1(h_1(z))$, a degree one polynomial in $z$.  Therefore $R_1$ is a sum of two terms, one which comes from the degree zero part and one from the degree one part.  The former is zero, again by the above reasoning.  Therefore, to show that $R_1=0$ we need only show $I_1=0$, except with $\hat\psi_1(h_1(z))$ replaced by  $z$.  Since $h_1(-y)=h_2(y)$, changing variables  $y\rightarrow -y$ and $z\rightarrow -z$ in the following shows that this term is equal to its own negative, and hence is zero
\begin{align}
&\int_{-1}^1 \left[\int_{-1}^1\int_0^{2\pi}S|\mathcal{M}|^2 (s,t(\cos(\psi)\sqrt{1-y^2}\sqrt{1-z^2}+yz))d\psi f_1(h_1(y))f_1(h_2(y)) dy\right] \notag\\
&\hspace{26mm}\times  zf^1(h_1(z))f^1(h_2(z))dz\\
&-\int_{-1}^1 \left[\int_{-1}^1\int_0^{2\pi}S|\mathcal{M}|^2 (s,t(\cos(\psi)\sqrt{1-y^2}\sqrt{1-z^2}+yz))d\psi f^1(h_1(y))f^1(h_2(y)) dy\right] \notag\\
&\hspace{26mm}\times zf_{1}(h_1(z))f_1(h_2(z))dz\,.\notag
\end{align}
We note that the corresponding scattering integrals do not vanish for the chemical equilibrium spectral method.  This is another advantage of the method developed in \rapp{ch:boltz:orthopoly} and leads to a further reduction in cost of the method, beyond just the reduction in minimum number of modes.

Finally, we point out how the vanishing of these inner products is a reflection of certain conservation laws. From \req{n:div}, \req{collision:integrals}, and the fact that $\hat\psi_0,\hat\psi_1$ span the space of polynomials of degree $\leq 1$, we have the following expressions for the change in number density and energy density of a massless particle
\begin{align}
\frac{1}{a^3} \frac{d}{dt}(a^3n)= \frac{g_p}{2\pi^2}\int \frac{1}{E}C[f]p^2dp=c_0 R_0\,,\qquad
\frac{1}{a^4}\frac{d}{dt}(a^4\rho)= \frac{g_p}{2\pi^2}\int C[f] p^2dp=d_0R_0+d_1R_1\,, 
\end{align}
for some $c_0,d_0,d_1$. Therefore, the vanishing of $R_0$ is equivalent to conservation of comoving particle number.  The vanishing of $R_0$ and $R_1$ implies $\rho\propto 1/a^4$ i.e. that the reduction in energy density is due entirely to redshift; energy is not lost from the distribution due to scattering.  These findings match the situations above where we found one or both of $R_0=0$, $R_1=0$.  $R_0$ vanished for all kinetic scattering processes and we know that all such processes conserve comoving particle number.  Both $R_0$ and $R_1$ vanished for a distribution scattering from itself and in such a process  there is no energy loss energy  from the distribution by scattering; energy is only redistributed among the particles corresponding to that distribution.

\subsection{Comparison with an alternative method for computing scattering integrals}\label{app:dogovMethod}
As a comparison and consistency check for our method of computing the scattering integrals, in this appendix we analytically reduce the collision integral down to $3$ dimensions by a method adapted from \cite{Dolgov:1997mb,Dolgov:1998sf}.  The only difference between our treatment in this section and theirs being that they solved the Boltzmann equation numerically on a grid in momentum space and not via a spectral method.  Therefore we must take an inner product of the collision operator with a basis function and hence we are integrating over all particle momenta, whereas they integrate over all momenta except that of particle one.  For completeness we give a detailed discussion of their method.

 Writing the conservation of four-momentum enforcing delta function
\begin{equation}
\delta(\Delta p)=\frac{1}{(2\pi)^3}\delta(\Delta E)e^{i\vec z\cdot \Delta \vec p}d^3z\,,
\end{equation}
where the arrow denoted the spatial component, we can simplify the collision integral as follows
\begin{align}
R\equiv&\int G(E_1,E_2,E_3,E_4) S|\mathcal{M}|^2(s,t)(2\pi)^4\delta(\Delta p)\prod_{i=1}^4\frac{d^3p_i}{2(2\pi)^3 E_i}\\
=&\frac{1}{16(2\pi)^{11}}\int G(E_i) S|\mathcal{M}|^2(s,t)\delta(\Delta E)e^{i\vec z\cdot\Delta p}\prod_{i=1}^4\frac{d^3p_i}{ E_i}d^3z\notag\\
=&\frac{2}{(2\pi)^{6}}\int G(E_i)K(E_i) \delta(\Delta E)\prod_{i=1}^4\frac{p_i}{E_i}dp_i z^2dz\,,\notag\\
K\equiv&\frac{p_1p_2p_3p_4}{(4\pi)^5}\int S|\mathcal{M}|^2(s,t)e^{i\vec z\cdot\Delta\vec p}\prod_{i=1}^4d\Omega_id\Omega_z\,.
\end{align}
We can change variables from $p_i$ to $E_i$ in the outer integrals and use the delta function to eliminate the integration over $E_4$ to obtain
\begin{align}
R=&\frac{2}{(2\pi)^{6}}\int1_{E_1+E_2-E_3>m_4}G(E_i)\left[\int_0^\infty K(z,E_i)z^2dz\right]dE_1dE_2dE_3\,,\\
p_i=&\sqrt{E_i^2-m_i^2},\hspace{2mm} E_4=E_1+E_2-E_3\,.\notag
\end{align}
From Tables \ref{table:nu:e:reac} and \ref{table:nu:mu:reac} we see that the matrix elements for weak scattering involving neutrinos are linear combinations of the terms
\begin{equation}
p_1\cdot p_2,\hspace{2mm} p_1\cdot p_3,\hspace{2mm}(p_1\cdot p_4)(p_2\cdot p_3), \hspace{2mm} (p_1\cdot p_2)(p_3\cdot p_4),\hspace{2mm} (p_1\cdot p_3)(p_2\cdot p_4).
\end{equation}
Therefore we must compute the angular integral term $K$ with $S|\mathcal{M}|^2$ replaced by elements from the following list
\begin{align}\label{matrixElementPieces}
&1,\hspace{2mm}\vec p_1 \cdot\vec p_2,\hspace{2mm}\vec p_1 \cdot\vec p_3,\hspace{2mm}\vec p_1 \cdot\vec p_4 \,,\hspace{2mm}\vec p_2\cdot\vec p_3,\hspace{2mm}\vec p_2\cdot\vec p_4 \,,\hspace{2mm}\vec p_3\cdot\vec p_4 \,,\\
& (\vec p_1 \cdot\vec p_2)(\vec p_3\cdot\vec p_4 ),\hspace{2mm}(\vec p_1 \cdot\vec p_4 )(\vec p_2\cdot\vec p_3),\hspace{2mm} (\vec p_1 \cdot\vec p_3)(\vec p_2\cdot\vec p_4 )\,,\notag
\end{align}
producing $K_0$, $K_{12}$, $K_{13}$,...,$K_{1324}$.  All of these are rotationally invariant, and so we can always rotate coordinates so that $\vec z=z\hat z$.  This allows us to evaluate the $z$ angular integral
\begin{equation}
K=\frac{p_1p_2p_3p_4}{(4\pi)^4}\int S|\mathcal{M}|^2(s,t)e^{iz \hat z\cdot\Delta\vec p}\prod_{i=1}^4d\Omega_i\,.
\end{equation}

The remaining angular integrals are straightforward to evaluate analytically for each expression in \req{matrixElementPieces}
\begin{align}
&K_0=\prod_{i=1}^4\frac{\sin(p_iz)}{z}\,,\\
&K_{12}=-\frac{(\sin(p_1z)-p_1z\cos(p_1z))(\sin(p_2z)-p_2z\cos(p_2z))\sin(p_3z)\sin(p_4z)}{z^6}\,,\notag\\
&K_{13}=\frac{(\sin(p_1z)-p_1z\cos(p_1z))\sin(p_2z)(\sin(p_3z)-p_3z\cos(p_3z))\sin(p_4z)}{z^6}\,,\notag\\
&K_{14}=\frac{(\sin(p_1z)-p_1z\cos(p_1z))\sin(p_2z)\sin(p_3z)(\sin(p_4z)-p_4z\cos(p_4z))}{z^6}\,,\notag\\
&K_{23}=\frac{\sin(p_1z)(\sin(p_2z)-p_2z\cos(p_2z))(\sin(p_3z)-p_3z\cos(p_3z))\sin(p_4z)}{z^6}\,,\notag\\
&K_{24}=\frac{\sin(p_1z)(\sin(p_2z)-p_2z\cos(p_2z))\sin(p_3z)(\sin(p_4z)-p_4z\cos(p_4z))}{z^6}\,,\notag\\
&K_{34}=-\frac{\sin(p_1z)\sin(p_2z)(\sin(p_3z)-p_3z\cos(p_3z))(\sin(p_4z)-p_4z\cos(p_4z))}{z^6}\,,\notag\\
&K_{1234}=K_{1423}=K_{1324}=\prod_{i=1}^4\frac{(\sin(p_iz)-p_iz\cos(p_iz))}{z^2}\,.\notag
\end{align}

To compute $\int_0^\infty K(z) z^2 dz$ we need to evaluate the following three integrals
\begin{align}
D_1=&\int_0^\infty \frac{\sin(p_1z)\sin(p_2z)\sin(p_3z)\sin(p_4z)}{z^2}dz\,,\\
D_2=&\int_0^\infty\frac{\sin(p_1z)\sin(p_2z)(\sin(p_3z)-p_3z\cos(p_3z))(\sin(p_4z)-p_4z\cos(p_4z))}{z^4}dz\,,\notag\\
D_3=&\int_0^\infty\frac{\prod_{i=1}^4(\sin(p_iz)-p_iz\cos(p_iz))}{z^6}dz\,.\notag
\end{align}
These expressions are symmetric under $1\leftrightarrow 2$ and $3\leftrightarrow 4$ and so without loss of generality we can assume $p_1\geq p_2$, $p_3\geq p_4$. We require $p_1\leq p_2+p_3+p_4$ (and cyclic permutations) by conservation of energy.  In the case where the above conditions all hold, we separate the computation into four additional cases in which the integrals can be evaluated analytically, as  in \cite{Dolgov:1997mb,Dolgov:1998sf}:\\
${\bf p_1+p_2>p_3+p_4\text{, \hspace{1mm} }p_1+p_4>p_2+p_3}${\bf :}
\begin{align}
D_1=&\frac{\pi}{8}(p_2+p_3+p_4-p_1)\,,\qquad
D_2=\frac{\pi}{48}((p_1-p_2)^3+2(p_3^3+p_4^3)-3(p_1-p_2)(p_3^2+p_4^2)\,, \\
D_3=&\frac{\pi}{240}(p_1^5-p_2^5+5p_2^3(p_3^2+p_4^2)-5p_1^3(p_2^2+p_3^2+p_4^2)-(p_3+p_4)^3(p_3^2-3p_3p_4+p_4^2)\notag\\
&\hspace{7mm}+5p_2^2(p_3^3+p_4^3)+5p_1^2(p_2^3+p_3^3+p_4^3))\,.\notag
\end{align}
${\bf p_1+p_2<p_3+p_4\text{, \hspace{1mm} }p_1+p_4>p_2+p_3}${\bf :}
\begin{align}
D_1= \frac{\pi }{4}p_2\,,\qquad
D_2= \frac{\pi }{24}p_2(3(p_3^2+p_4^2-p_1^2)-p_2^2)\,,\qquad 
D_3= \frac{\pi}{120}p_2^3(5(p_1^2+p_3^2+p_4^2)-p_2^2)\,. 
\end{align}
${\bf p_1+p_2>p_3+p_4\text{, \hspace{1mm} }p_1+p_4<p_2+p_3}${\bf :}
\begin{align}
D_1= \frac{\pi }{4}p_4\,,\qquad 
D_2= \frac{\pi}{12} p_4^3\,,\qquad 
D_3= \frac{\pi }{120}p_4^3(5(p_1^2+p_2^2+p_3^2)-p_4^2)\,. 
\end{align}
${\bf p_1+p_2<p_3+p_4\text{, \hspace{1mm} }p_1+p_4<p_2+p_3}${\bf :}
\begin{align}
D_1=&\frac{\pi}{8}(p_1+p_2+p_4-p_3)\,,\qquad
D_2= \frac{\pi}{48}(-(p_1+p_2)^3-2p_3^3+2p_4^3+3(p_1+p_2)(p_3^2+p_4^2))\,, \\
D_3=&\frac{\pi}{240}(p_3^5-p_4^5-(p_1+p_2)^3(p_1^2-3p_1p_2+p_2^2)+5(p_1^3+p_2^3)p_3^2-5(p_1^2+p_2^2)p_3^3\notag\\
&\hspace{7mm}+5(p_1^3+p_2^3-p_3^3)p_4^2+5(p_1^2+p_2^2+p_3^2)p_4^3)\,.\notag
\end{align}
We computed the remaining integrals numerically in several test cases for each of the reaction types in \rsec{nu:matrix:elements} and obtained agreement between this method and ours, up to the integration tolerance used.  However, the method we have developed in this Appendix has the distinct advantage of resulting in  smooth integrand.  The expressions obtained here are only piecewise smooth and therefore much costlier to integrate numerically.  Since the cost of numerically solving the Boltzmann equation is dominated by the cost of computing the collision integrals, we find that our approach constitutes a  significant optimization in practice.


\addcontentsline{toc}{section}{References}
\bibliographystyle{sn-aps}
\bibliography{00CosmoThesis.bib}
\addcontentsline{toc}{section}{Index}
\printindex
\end{document}